\documentclass[12pt, double, phd]{osudiss-2}
\usepackage{tocloft}

\usepackage{amsmath, amssymb, amsthm,mathtools}
\usepackage{blkarray}
\usepackage{mathdots}
\usepackage{graphicx} 
\usepackage{mwe}
\usepackage{float}
\usepackage{rotating} 
\usepackage{multirow} 
\usepackage{caption}
\usepackage{subcaption}
\usepackage{bm,bbm} 
\usepackage{booktabs} 
\usepackage{dsfont}
\usepackage{bbm}
\usepackage{bookmark} 
\hypersetup{colorlinks=true,linkcolor=blue, linktoc=page} 

\usepackage{cite} 

\usepackage{amsmath,amsfonts,enumerate,array}
\usepackage{graphicx}
\usepackage[usenames]{color}
\usepackage[english]{babel}
\usepackage{fancybox}
\usepackage{chngpage} 

\usepackage{blindtext}
\usepackage[utf8]{inputenc}

\newcommand{\captionfonts}{\small}

\makeatletter  
\long\def\@makecaption#1#2{%
  \vskip\abovecaptionskip
  \sbox\@tempboxa{{\captionfonts #1: #2}}%
 \ifdim \wd\@tempboxa >\hsize
    {\captionfonts #1: #2\par}
  \else
    \hbox to\hsize{\hfil\box\@tempboxa\hfil}%
  \fi
  \vskip\belowcaptionskip}
\makeatother   

\usepackage{mciteplus}

\usepackage[all]{hypcap}     

\numberwithin{equation}{section}

\mathchardef\mhyphen="2D

\makeatletter
\newcommand{\contraction}[5][1ex]{%
  \mathchoice
    {\contraction@\displaystyle{#2}{#3}{#4}{#5}{#1}}%
    {\contraction@\textstyle{#2}{#3}{#4}{#5}{#1}}%
    {\contraction@\scriptstyle{#2}{#3}{#4}{#5}{#1}}%
    {\contraction@\scriptscriptstyle{#2}{#3}{#4}{#5}{#1}}}%
\newcommand{\contraction@}[6]{%
  \setbox0=\hbox{$#1#2$}%
  \setbox2=\hbox{$#1#3$}%
  \setbox4=\hbox{$#1#4$}%
  \setbox6=\hbox{$#1#5$}%
  \dimen0=\wd2%
  \advance\dimen0 by \wd6%
  \divide\dimen0 by 2%
  \advance\dimen0 by \wd4%
  \vbox{%
    \hbox to 0pt{%
      \kern \wd0%
      \kern 0.5\wd2%
      \contraction@@{\dimen0}{#6}%
      \hss}%
    \vskip 0.2ex%
    \vskip\ht2}}

\newcommand{\contraction@@}[3][0.06em]{%
  \hbox{%
    \vrule width #1 height 0pt depth #3%
    \vrule width #2 height 0pt depth #1%
    \vrule width #1 height 0pt depth #3%
    \relax}}
\makeatother

\newcommand{\be}{\begin{equation}} 
\newcommand{\ee}{\end{equation}} 
\newcommand{\bea}{\begin{eqnarray}\displaystyle}
\newcommand{\eea}{\end{eqnarray}}
\newcommand{\bt}{\begin{tabular}}
\newcommand{\et}{\end{tabular}}
\newcommand{\bs}{\begin{split}}
\newcommand{\es}{\end{split}}
\newcommand{\pa}{\partial}

\newcommand{\nsnsket}{|0_{NS}\rangle^{(1)} |0_{NS}\rangle^{(2)}}					
\newcommand{\nsnsbra}{{}^{(1)}\langle 0_{NS}| {}^{(2)}\langle 0_{NS}|}		
\newcommand{\nstket}{|0_{NS}\rangle_t}																										
\newcommand{\nstbra}{{}_t\langle 0_{NS}|}																								
\newcommand{\rmmket}{|0_R^-\rangle^{(1)}|0_R^-\rangle^{(2)}}							
\newcommand{\rmket}{|0_R^-\rangle}																									
\newcommand{\rmbra}{\langle 0_{R,-}|}																								
\newcommand{\rpket}{|0_R^+\rangle}																									
\newcommand{\rpbra}{\langle 0_{R,+}|}																								
\newcommand{\rpmket}{| 0_R^+\rangle^{(1)} \newotimes | 0_R^-\rangle^{(2)}}					
\newcommand{\rpmbra}{{}^{(1)}\langle 0_{R,+}| \newotimes {}^{(2)}\langle 0_{R,-}|}	

\newcommand{\sh}{\,\hat\sigma\,} 

\newcommand{\nstclose}{0_{NS}\rangle}

\newcommand{\rmutvket}{|0_R^-\rangle^{(1)}\otimes |0_R^-\rangle^{(2)}}
 
\newcommand{\rmtvket}{|0_R^-\rangle}

\newcommand{\rptvket}{|0_R^+\rangle}

\renewcommand{\a}{\alpha}	
\renewcommand{\b}{\beta}
\newcommand{\g}{\gamma}		
\newcommand{\G}{\Gamma}
\renewcommand{\d}{\delta}
\newcommand{\D}{\Delta}
\renewcommand{\c}{\chi}			

\newcommand{\p}{\psi}			

\newcommand{\s}{\sigma}		

\renewcommand{\t}{\tau}		
\newcommand{\e}{\epsilon}
\newcommand{\n}{\nu}
\newcommand{\m}{\mu}
\renewcommand{\r}{\rightarrow}

\newcommand{\nn}{\nonumber\\} 		
\newcommand{\newotimes}{}  				
\newcommand{\diff}{\,\text{d}}		
\newcommand{\h}{{1\over2}}				

\def\h{\mathbb{h}}

\def\D{\Delta}

\def\h{\frac{1}{2}}

\def\one{{\hbox{\kern+.5mm 1\kern-.8mm l}}}
\def\zero{{\hbox{0\kern-1.5mm 0}}}
\def\eq#1{(\ref{#1})}

\def\sqi{{1\over \sqrt{2}}}

\newcommand{\ac}[2]{\ensuremath{\{ #1, #2 \}}}
\renewcommand{\ell}{l}

\newcommand{\bra}[1]{{\langle {#1} |\,}}
\newcommand{\ket}[1]{{\,| {#1} \rangle}}
\newcommand{\braket}[2]{\ensuremath{\langle #1 | #2 \rangle}}

\def\b{\bigskip}

\author{Shaun Hampton}
\title{Understanding Black Hole Formation in String Theory}
\authordegrees{B.S., M.Sc.}  
\unit{The Physics Department\vspace{-1em}}
\advisorname{Samir. D. Mathur}
\member{Yuri Kovchegov}
\member{Junko Shigemitsu}
\member{Brian Winer}

\begin{document}
\frontmatter

\renewcommand{\contentsname}{\centerline {Table of Contents}}   
\renewcommand{\cftaftertoctitle}{\hfill}

\begin{abstract}
The strongly coupled dynamics of black hole formation in bulk AdS is conjectured to be dual to the thermalization of a weakly interacting CFT on the boundary for low $N$ which, for $N\to\infty$, becomes strongly coupled. We search for this thermalization effect by utilizing the D1D5 CFT to compute effective string interactions for $N=2$. This is done by turning on a marginal deformation of the theory which twists together or untwists effective strings. For a system to thermalize, the initial state, which is far from thermal, must redistribute it's energy via interactions until a thermal state is achieved. In our case, we consider excited states of the effective strings. We compute splitting amplitudes for 1) one excitation going to three excitations and 2) two excitations going to four excitations using two insertions of the deformation. Scenario 1) corresponds to a single particle moving in AdS. Scenario 2) corresponds to two particles moving and colliding in AdS. We find that the `1 to 3' amplitude has terms which oscillate with time, $t$, where $t$ is the duration of the two deformations. We find that the `2 to 4' amplitude has similar oscillatory terms as well as secular terms which grow like $t^2$. For this case the growth implies that for large $t$ the excitations in the initial state, which carry a given energy, prefer to redistribute themselves amongst lower energy modes in the final state. This is a key feature of thermalization. Albeit in a simplified setting, we therefore argue that we have identified the thermalization vertex in the D1D5 CFT, which after repeated applications, should lead to thermalization. This ultimately maps to two particles colliding and forming a black hole in AdS, which in our case, is a fuzzball.
\end{abstract}

%
%
%


%
%

\dedication{To Robert Hampton, my precious grandfather who has been added to the great cloud of witnesses in heaven, to Rebecca T. Hampton my precious grandmother, and to Beverly Winston, my precious Godmother who has been added to the great cloud of witnesses in heaven.}
%
%

\begin{acknowledgements}
First and foremost, and I can't stress that enough, I want to thank God my Father, my Lord and Savior Jesus Christ and the precious Holy Spirit, my comforter, master teacher, and best friend,
for helping me get to this point. I know without a doubt, that without you, I would not be getting my Ph.D in physics. You have sustained me, carried me, lifted me, held me, protected me, strengthened me, and most importantly, loved me through everything. Unfortunately, words can not express my truest gratitude towards you.


Secondly, I want to thank my advisor, Professor Samir D. Mathur. I am truly grateful to God for placing me with you as your student. You've taught me how to become a better theoretical physicist. You've demonstrated, time and time again, how to think deeply and physically about problems. You've shown me how to interpret the mathematics, extracting the important details regarding the physics. You've taught me so much during my time as a graduate student. I want to thank Professors Yuri Kovchegov, Junko Shigemitsu and Brian Winer for serving on my candidacy and dissertation committee. Thank you for your patience and suggestions during my oral exams. I also want to thank Professors Samir Mathur, Stuart Raby, Linda Carpenter, Junko Shigemitsu, and Ulrich Heinz for teaching me many things about numerous topics in theoretical physics in their classes.  I want to thank my former advisor at Clemson University, Professor Dieter Hartmann. Thank you for training and developing me in the area of physics. Your support and insight were critical in my early years of graduate school in physics. I also want to thank Professor Catalina Marinescu for her her advice and encouragement also during my early years in physics. I want to thank Zaq Carson, David Turton, Bin Guo, and Ida Zadeh whom I've collaborated over the years. Thank you for all of your hard work in our collaborations and for providing stimulating discussions. I want to thank the physics administrative staff. In particular, I want to thank Kris Dunlap and Professor John Pelz for tirelessly helping me in many ways. 
I also want to thank other colleagues such as Bowen Shi, Zijie Poh, Abhishek Mohapatra, Joe McEwan, Hudson Smith, Chuck Bryant, Russell Colburn, Dennis Bazow, Alex Dyhdalo, and Hong Zhang for helpful discussions and for providing a stimulating environment. 

Next, I want to express my sincerest gratitude to my grandparents, Robert and Rebecca Hampton, my first spiritual parents. You are a blessing from God. I can't express in words what you truly mean to me. 
I know that if it had not been for your love, support, impartation, correction, example, character and stability throughout the years, I would not be where I am today. 
You raised me in the admonition of the Lord. For this I am truly and eternally grateful. 
I love you dearly. 
I would also like to thank my mom and dad, Kathy Cline and David Hampton. I am truly grateful to have you as my parents and for the support that you have shown over the years. You both are blessing from the Lord. Thank you for your love and sacrifice.
I want to thank my Godmother Beverly Winston who poured into me in an incredible way despite our spatial separation. You and my grandfather have both been added to the great cloud of witnesses in heaven. I want to thank my Aunt Brigitte for all of the support that she has shown over the years. Though I am your nephew, I am like your own child. You have provided for me in many ways and for that I am truly grateful. I want to thank my step mom Sonya Hampton. You have also been like a mother to me. I am truly grateful for your support and care over the years. I want to thank my Uncle Robbie, my Aunt Maribel and my Uncle John for their support and love. I am grateful to have you as my aunt and uncles. I want to thank my siblings, Nekia Cline, Natalia Cline, Isaac Hampton, and Triniti Hampton for their support and for motivating me to be an example to them. I want to thank all of my great grandparents, great aunts and uncles, grandperents, aunts, uncles, cousins, nieces and nephews for their love, support and encouragement. I want to thank J.R. Girardi for his lasting friendship and brotherhood throughout everything. I love you all dearly.

Next, I want to thank my church community, my family in the Body of Christ and in the Kingdom of our Heavenly Father of which my natural family is also a part. I first want to thank my spiritual parents for raising me in the gospel of Jesus Christ and imparting the Word of the Lord to me. I am so blessed to have been given several spiritual parents by the Lord over the years. 
I want to thank Pastors Floyd and Sharon Dodson, Apostles Dale and Vicki Sides, and Apostles Eric and Carolyn Warren. Each of you 
have contributed to my life in significant ways and have catapulted me to reach my destiny in God. Thank you all for the love, impartation, prayers, correction, encouragement, and support that you have provided along this journey. I am truly grateful to God for providing me with each of you. I want to thank Pastor and Prophetess Best, and Pastor David Judah for their continued prayers and encouragement. I love you all.
I want to thank my church family, The Church, The Body of Christ in Pilot Mountain, NC, Lovell's Chapel Church International in Pilot Mountain, NC, Kings Park International Church in Durham, NC, New Life Church in Central SC, Liberating Ministries for Christ International in Bedford, Virginia, Equippers City Church in Columbus, OH, and Vintage Grounds Church in Columbus, OH. I want to thank my spiritual family near and far for all of the help, support, meals, fellowship, hospitality, encouragement, and prayers throughout this process. 
I want to thank Felix Meyer for mentoring me over years. Thank you for being a big brother to me and for giving me advice and wisdom on many things. I want to thank Pat Williams for also being a big brother to me and supporting me over the years. I want to thank Nancy Blatnik for her encouragement and prayer throughout this process. You have been a blessing. I want to thank Pastors Michael and Joy Bivens for the incredible hospitality that they have shown to me during my time in Columbus. Thank you for opening up your home to me and treating me like part of the family. I am truly grateful. I want to thank Pastor Jason Wilson for supporting and believing in me. Thank you for all of the fruitful discussions and insight regarding numerous topics which came to mind. I want to thank Abby Boateng for being a sister and a friend to me and encouraging me in many ways. To my Equippers family, thank you. Thank you for helping me to make it through the process. You've embraced me as part of the family. For that I am grateful. I love you all dearly.

To my Columbus family, thank you for being here to encourage me and help lift me up during the tough times. To the members of BGSPC, SACNAS, and NSBP, thank you for your friendship and support. Thank you for persevering and pushing right along with me. 

To my Clemson family, thank you for the lasting friendships and support over the years. Thank you for being there at such a critical time in my life. I appreciate you.

To my Chapel Hill family, my first college family, thank you for beginning this journey with me over a decade ago and for maintaining the relationships that we all have. 

To all of my friends and family from Pilot Mountain, NC, thank you for the continued friendship and support over many years. I am grateful for you. 

To everyone that has contributed in some way to my journey up to this point whom I did not specifically name, I truly thank God for you and for what you have given to me. Know that you are appreciated and I am grateful for you.

This work was supported by the Presidential Fellowship awarded by the graduate school at The Ohio State University.


\end{acknowledgements}

\begin{vita}

\dateitem{October 26, 1986}{Born - Asheville North Carolina, USA}

\dateitem{2010}{B.S. Chemistry}

\dateitem{2013}{M.S. Physics}

\dateitem{2013-2015}{Fowler Fellow,\\
			 The Ohio State University.}
			 
\dateitem{2015-2016}{Graduate Teaching Assistant,\\
			 The Ohio State University.}
			 
\dateitem{2017}{Graduate Research Assistant,\\
			 The Ohio State University.}
			 
\dateitem{2017}{Graduate Teaching Assistant,\\
			 The Ohio State University.}	
			 
\dateitem{2018}{Presidential Fellow,\\
			 The Ohio State University.}

\begin{publist}


\pubitem{S.~Hampton, S., D., Mathur, I., G.,  Zadeh
\newblock``Lifting of D1-D5-P states".\\
\newblock {\em arXiv:}1804.10097, (2018)}

\pubitem{B.~Guo, S. Hampton, S., D., Mathur
\newblock``Can we observe fuzzballs or firewalls".\\
\newblock {\em arXiv:}1711.01617, (2017)}

\pubitem{Z.~Carson, S. Hampton, S., D., Mathur
\newblock``Full action of two deformation operators in the D1D5 CFT''.
\newblock {\em Journal of High Energy Physics} 11 096 (2017)}

\pubitem{Z.~Carson, S. Hampton, S., D., Mathur
\newblock``One loop transition amplitudes in the D1D5 CFT''.
\newblock {\em Journal of High Energy Physics} 01 006 (2017)}

\pubitem{Z.~Carson, S. Hampton, S., D., Mathur
\newblock``Second order effect of twist deformations in the D1D5 CFT''.
\newblock {\em Journal of High Energy Physics} 04 115 (2016)}

\pubitem{Z.~Carson, S. Hampton, S., D., Mathur, D. Turton
\newblock``Effect of the deformation operator in the D1D5 CFT''.
\newblock {\em Journal of High Energy Physics} 1501 071 (2014)}

\pubitem{Z.~Carson, S. Hampton, S., D., Mathur, D. Turton
\newblock``Effect of the twist operator in the D1D5 CFT''.
\newblock {\em Journal of High Energy Physics} 08 064 (2014)}

%

\end{publist}

\begin{fieldsstudy}


 \majorfield{Physics}



\end{fieldsstudy}

\end{vita}

\tableofcontents
\newpage
\listoffigures
\newpage
\listoftables

\mainmatter
\doublespacing
\chapter{Introduction}
Black holes are some of the most fascinating objects in the universe. They are predicted by Einstein's theory of general relativity \cite{einstein} and classically, they are regions in spacetime from which light can't even escape. The simplest of these solutions, called the Schwarzschild solution, was found by Karl Schwarzschild \cite{schwarzschild} in 1916 and independently by Johannes Droste around the same time. It is an uncharged, nonrotaing, spherically symmetric vacuum solution to Einstein's equations. In 1974, the traditional view that nothing could escape a black hole was modified by Stephen Hawking where he showed, using quantum mechanics, that black holes actually have a temperature and can radiate particles \cite{hawking}. While this was a remarkable breakthrough, it created a problem. Black holes will slowly evaporate away, but in a way that is forbidden by quantum mechanics. As a result, the information stored within the black hole can  not be recovered by the radiation that evaporated away. This problem is famously called 'the information paradox' and was put forth by Hawking after his discovery. Fortunately, string theory provides a framework to address this paradox and other fundamental questions about black holes. 


String theory says that all matter and energy is described by tiny vibrating strings. For a brief list of textbooks and some early works on the subject, see \cite{gsw1,gsw2,polchinski1,polchinski2,zwiebach,bbs,kiritsis}. As a result of having a theory of strings instead of particles, spacetime is constrained to be 10 dimensional. This constraint specifically comes from including both bosons and fermions into the theory yielding a supersymmetric theory. This describes superstring theory. Furthuremore, string theory is  theory of quantum gravity where the graviton, the fundamental particle that transmits the gravitational interaction, is described by a closed string. 


Since string theory is formulated in ten dimensions and our world is four dimensional (three spatial dimensions and time), the extra dimensions must be extremely small or `compact' so that they are hidden from our world. The strings can then be `wound` around these compact dimensions. Even though the string is now hidden from the perspective of the large dimensions, it's presence is felt through its gravitational field. This produces an object with a large mass in a small region which are exactly the right conditions needed to make a black hole. Furthermore, a string can carry transverse vibrations, allowing it to fill a finite region within the large dimensions. This finite region resembles the region enclosed by a traditional black hole. One key difference between these black holes and traditional black holes is the absence of an event horizon. This new perspective of black holes was pioneered by Samir Mathur and Oleg Lunin \cite{fuzzball}. These black holes are referred to as `fuzzballs.' Since the construction of the first solution, much work has been done in this area and many solutions have been constructed. A list of works and review articles pertaining to the subject, is given here \cite{fuzzballs_i, fuzzballs_ii, fuzzballs_iii, fuzzballs_iv, fuzzballs_v, Lunin:2001dt, Bena:2005ay, Jejjala:2005yu,Taylor:2005db, Mathur:2005ai, Giusto:2005ag, Emparan:2001ux, Kruczenski:2002mn, Lunin:2002fw, Lunin:2002iz,Mathur:2008kg}. This list is certainly not exhaustive. However, one remaining question is ``How do they form?". This is the question that I seek to address. 


In general, black hole formation is a very difficult problem to solve. When matter is extremely close together, a necessary condition to form a black hole, each particle interacts strongly with the other particles through gravity. Therefore, the number and strength of the interactions are large. In order to understand the full formation process, one would need to compute each interaction. There is, however, a simplification one can make. According to a conjecture made by Juan Maldecena \cite{adscft}, a theory of gravity containing strings that live in a certain number of dimensions, called the bulk, can be mapped to a quantum theory of `effective strings' without gravity living in one less dimension, called the boundary. This conjecture is termed the AdS/CFT correspondence.  The bulk gravity theory is an AdS (Anti De Sitter) background which is a negatively curved solution to Einstein's equations. CFT stands for Conformal Field Theory which is characterized by conformal symmetry and it lives on the boundary of AdS space. This conformal field theory is a quantum theory of effective strings. Computations difficult to perform in the bulk are much more tractable on the boundary. We can therefore map our problem of black hole formation in the AdS bulk to a quantum theory of `effective' strings on the boundary.

The boundary is divided into a compact dimension and several large dimensions. We wind the strings around the compact dimension and allow them to interact by joining and splitting. When two strings interact on the boundary, we directly know what is happening in the bulk. Because black holes have a temperature, as found by Hawking, we expect that their formation will manifest on the boundary as a process that evolves into a final state characterized by a temperature \cite{witten one}. This evolution is called thermalization and it is this thermalization process we seek. Other works on thermalization and black hole formation in the standard case (standard meaning geometries which contain horizons) are given in \cite{Max1,Max2,formation,Balasubramanian:2014cja,Arefeva:2013uta,Aoki:2015uha,Hubeny:2013dea,Arefeva:2012jp,Das:2011nk,Arefeva:2012bly,Garfinkle:2011tc,Garfinkle:2011hm,Erdmenger:2011jb,Ebrahim:2010ra,Das:2010yw,Bhattacharyya:2009uu,kaplan}. We are particularly interested in processes which describe tunneling into fuzzball states. It is in this context by which we refer to black formation in this work.


In order to identify thermalization we must know what to look for. To gain some intuition about the thermalization process on a microscopic level let's consider a simple scenario. Imagine that you have a box of particles. Consider that one of the particles in the box has a very high energy while all the other particles have very low energies. Now if the high energy particle never interacts with the low energy particles then a thermal state can never by achieved. This is because the energy of the high energy particle is never transfered to the low energy particles and therefore thermalization can't happen. The interaction between particles is therefore crucial to achieving a thermal state. Now let us apply these ideas to the problem of thermalization in a theory of `effective strings' which is described by a CFT. In the CFT, imagine that we have strings wrapped around a compact dimension. The only interactions between these strings are that they can join together into a single multiwound string and then split back into two strings. This describes the string interactions we will consider. Our goal is to compute the fundamental interaction vertex in the CFT. Computing this vertex is the first step in proving thermalization. We will consider two scenarios: 1) two strings of arbitrary winding joined into one multiwound string with winding equal to the addition of the two initial windings, and 2) two singly wound strings joined into one doubly wound string and split back into two singly strings. The scenario of primary importance is the second scenario but we compute both scenarios. Also, since the second scenario is technically more difficult than the first scenario, we restrict our computation to singly wound strings in the initial state. The simplicity of the first scenario enabled us to use multiwound initial states. Continuing with the second scenario, imagine that we have two singly wound strings with an initial excitation along one of the strings.  We want to see what happens to this initial excitation if we join and split the two strings. Does the initial excitation split into lower energy excitations. The initial excitation corresponds to a scalar particle moving in the AdS dual. Next we want to know what happens when we apply the interaction to either two initial excitations along one of the strings or an initial excitation on each string. We want to address whether or not there is a fundamental difference in the splitting amplitude if we start with two excitations as opposed to one. In chapter 9 we show that there is a difference in the amplitude for splitting between the case with one excitation and two excitations. The amplitude with one excitation in the initial state has a term which oscillates with time, $t$, where $t$ measures the duration of the interaction. The amplitude with two excitations in the initial state has components which oscillate with $t$ and another component which grows like $t^2$. 
The amplitudes are related to the probabilities that those processes will happen. After a long time, $t$, the probability that two excitations split into four excitations is significantly larger than the probability that one excitation splits into three excitations. While we found that the two processes differ, more work is needed to prove thermalization. A thermal state has a large entropy compared to the initial state from which it evolved. We have only computed one splitting event and one splitting combination for various initial and final states. We must first compute multiple splitting combinations for a single splitting event. We must then argue that multiple splitting events will move us toward a thermal state in the CFT, which corresponds to black hole formation in the bulk. In our case, this formation process is to be thought of as an initial state tunneling into a fuzzball state. This is left for future work.



\chapter{The D1D5 System and the twist interaction}\label{D1D5 system}
In this chapter we describe the D1D5 system. We first describe the nature of D-branes. We then describe the D1D5 supergravity solution. Lastly and most importantly, we describe the dual D1D5 CFT and the twist interaction. This will help us to understand string interactions. 
\section{D branes}
D branes are essential objects in string theory. They were discovered by Dai, Leigh and Polchinksi, and independently by Horava in 1989 \cite{Dbranes1}. They are hypersurfaces on which the end points of open strings could end. One of the great advantages of this feature is that the D-branes are useful in defining gauge theories. Since the open string has two endpoints the field can be written as $\lambda_{ij} X$, where for now $\lambda_{ij}$ is an arbitrary matrix, the labels $i,j$, indicating on which D-brane the open string endpoints lie. These indices are called Chan-Paton indices \cite{chan paton}. If there is just one D-brane present then the open string endpoints must both lie on the same D-brane. If there are two D-branes present then the open string can have both endpoints lie along one of the branes, both endpoints lie along the other brane, or one endpoint on one brane and one endpoint on the other brane. One can continue adding more D-branes giving more combinations on which open string points can end. It is useful to group the endpoint locations into matrices. Now the open string can be represented as a matrix. 
On the worldsheet of the string $X$, defined by coordinates $\s$ and $\t$, invariance under transformations of this matrix is a global symmetry. However, in spacetime, this becomes a gauge symmetry because one can make different matrix transformations at separate points. This symmetry then defines a gauge theory of open strings. D-branes can extend along an arbitrary number of spatial dimensions in which case we label them as D$p$-branes where $p$ represents the number of spatial dimensions of the brane. It was discovered by Polchinski \cite{Dbranes} that Dp-branes carry what's called $RR$ (Ramond Ramond) charges which couple to $p+1$ dimensional gauge fields the same way that the electric charge, $e$, couples to the $U(1)$ gauge field, $A_{\mu}$ in Quantum Electrodynamics. These gauge fields appear as states in the spectrum of Type II superstring theories. For more references on D branes see \cite{D brane notes}.
\section{Gravity Dual}
To describe the D1D5 system we first consider the gravitational dual. We start with Type IIB string theory in 10 spacetime dimensions compactified in the following way
\bea
M_{9,1}\rightarrow M_{4,1}\times S^1\times T^4
\eea 
where $M_{4,1}$ is a large 5 dimensional manifold and $S^1\times T^4$ our compactified manifolds. We then wrap $N_1$ D1 branes on $S^1$, $N_5$ D5 branes on $S^1\times
T^4$, and allow the D1D5 system to rotate with parameter of $a$.
This corresponds to the 10D supergravity solution described by the metric \cite{fuzzball two}
\bea
ds^2 \!\!&=\!\!&-{1\over h}(dt^2-dy^2)+hf\bigg(d\theta^2 + {dr^2\over r^2 +a^2}\bigg)\cr
&&-{2 a \sqrt{Q_1Q_5}\over hf}\big(\cos^2\theta dy d\psi + \sin^2\theta dtd\phi\big)\cr
&&+h\bigg[\bigg(r^2+{a^2Q_1Q_5\cos^2\theta\over h^2f^2}\bigg)\cos^2\theta d\psi^2 + \bigg(r^2 + a^2 - {a^2 Q_1Q_5\sin^2\theta\over h^2f^2}\bigg)\sin^2\theta d\phi^2\bigg]\cr
&&+\sqrt{Q_1 +f\over Q_5 +f}\sum_{a=1}^4dz_adz_a
\label{10D metric}
\eea
where
\bea
f=r^2 + a^2\cos^2\theta,\qquad h = \bigg[\bigg(1+{Q_1\over f}\bigg)\bigg(1+{Q_5\over f}\bigg)\bigg]^{1\over2}\cr
\eea
and the charges $Q_1$ and $Q_5$ are proportional to their integer winding numbers $n_1$ and $n_5$ in a nontrivial way \cite{fuzzball two}.
Here $x_1,x_2,x_3,x_4$ parameterize the torus with volume $(2\pi)^4V$, and $y$ is compactified on a circle with circumference $2\pi R$. 
 At spatial infinity the geometry is flat. We note there is an $S^3$ in the geometry spanned by the coordinates $\theta, \phi,\psi$. The size of $S^3$ settles down to a constant radius for $ r<<(Q_1Q_5)^{1\over4}$. We can see this explicitly as follows. In this limit $f$ and $h$ become
 \bea
 f &=& r^2 + a^2\cos^2\theta\cr
 h &\approx& {\sqrt{Q_1Q_5}\over r^2+a^2\cos^2\theta}
 \eea
 Inserting this into the metric (\ref{10D metric}) we get
 \bea
ds^2 \!\!&=\!\!&-{(r^2+a^2\cos^2\theta)\over \sqrt{Q_1Q_5}}(dt^2-dy^2)+\sqrt{Q_1Q_5} {dr^2\over r^2 +a^2}\cr
&&-2 a\sqrt{Q_1Q_5}\big(\cos^2\theta dy d\psi + \sin^2\theta dtd\phi\big)\cr
&&+ \sqrt{Q_1Q_5}\bigg[d\theta^2+\cos^2\theta d\psi^2 +\sin^2\theta d\phi^2\bigg] \cr
&&+\sqrt{Q_1\over Q_5 }\sum_{a=1}^4dx_adx_a
\eea
where
\bea
d\theta^2+\cos^2\theta d\psi^2 +\sin^2\theta d\phi^2 = d\Omega^2_{S^3}
\eea
We call the the region $r< (Q_1Q_5)^{1\over4}$ the throat where the start of the throat is at $r\sim(Q_1Q_5)^{1\over4}$. Now as $r$ approaches $a$ the space ends in a cap. This is because the $y$ circle shrinks to zero at this distance down the throat. Let's rewrite the above metric in the following way (we write $r'=r/a$)
\bea
ds^2 \!\!&=\!\!&-(r'^2+1){a^2\over \sqrt{Q_1Q_5}}dt^2+r'^2{a^2\over \sqrt{Q_1Q_5}}dy^2+\sqrt{Q_1Q_5} {dr'^2\over r'^2 +1}\cr
&&+\cos^2\theta{a^2\over \sqrt{Q_1Q_5}}dy^2+\sin^2\theta{a^2\over \sqrt{Q_1Q_5}}dt^2\cr
&&-2 a\sqrt{Q_1Q_5}\big(\cos^2\theta dy d\psi + \sin^2\theta dtd\phi\big)\cr
&&+ \sqrt{Q_1Q_5}\bigg[d\theta^2+\cos^2\theta d\psi^2 +\sin^2\theta d\phi^2\bigg] \cr
&&+\sqrt{Q_1\over Q_5 }\sum_{a=1}^4dx_adx_a
\cr
\cr
ds^2\!\!&=\!\!&-(r'^2+1){a^2\over \sqrt{Q_1Q_5}}dt^2+r'^2{a^2\over \sqrt{Q_1Q_5}}dy^2+\sqrt{Q_1Q_5} {dr'^2\over r'^2 +1}\cr
&&+ \sqrt{Q_1Q_5}\bigg[d\theta^2+\cos^2\theta \bigg(d\psi - {ady\over\sqrt{Q_1Q_5}}\bigg)^2 +\sin^2\theta \bigg(d\phi - {adt\over\sqrt{Q_1Q_5}}\bigg)^2\bigg] \cr
&&+\sqrt{Q_1\over Q_5 }\sum_{a=1}^4dx_adx_a
\eea
Now transforming our angular coordinates by
\bea
\psi' &=& \psi - {a\over Q_1Q_5}y\cr
\phi'&=& \phi - {a\over Q_1Q_5}t
\eea 
Our metric becomes
\bea
ds^2\!\!&=\!\!&\underbrace{\sqrt{Q_1Q_5}\bigg[-(r'^2+1){dt^2\over R^2}+r'^2{dy^2\over R^2}+ {dr'^2\over r'^2 +1}\bigg]}_{AdS_3}\cr
&&+ \underbrace{\sqrt{Q_1Q_5}\bigg[d\theta^2+\cos^2\theta d\psi'^2 +\sin^2\theta d\phi'^2 \bigg]}_{S^3} \cr
&&+\underbrace{\sqrt{Q_1\over Q_5 }\sum_{a=1}^4dx_adx_a}_{T^4}
\eea
where
\bea
R = {\sqrt{Q_1Q_5}\over a}
\eea
is the radius of $y$ circle. We see that in the limit of $r<<(Q_1Q_5)^{1\over4}$ the space becomes $AdS_3\times S^3\times T^4$. The CFT will live at the boundary of $AdS_3$. We note that $S^1\subset AdS_3$.
\section{D1D5 CFT}

The dual description of the D1-D5 supergravity solution is described by a field theory with $n_1$ D1 branes wrapping $S^1$ and $n_5$ D5 branes wrapping $S^1\times T^4$ where the $S^1$ is labeled by the coordinate $y$. This gives a theory of open strings whose endpoints can begin and end on the $D1$ and the $D5$ branes. We think of the $S^1$ as being large compared to the $T^4$ ($T^4 \sim l_s$), so that at low energies we only look for open string excitations along the direction of $S^1$.  This low energy limit gives a Conformal Field Theory (CFT) on the circle $S^1$. CFT's are special classes of quantum field theories which are   invariant under conformal transformations. They provide a framework for which to study the behavior of strings. String theory contains a variety of moduli or parameters which characterize the theory such as the string coupling $g$, the shape and size of the torus, the values of flat connections for gauge fields etc. These parameters can be continuously varied. These variations move us to different points in the moduli space or parameter space of the CFT. It has been conjectured that we can move to a point called the `orbifold point' where the CFT is particularly simple \cite{orbifold}. At this orbifold point the CFT is a 1+1 dimensional sigma model. We will work in the Euclidean theory, where the base space is a cylinder spanned by the coordinates 
\be
\tau, \sigma: ~~~0\le \sigma<2\pi, ~~~-\infty<\tau<\infty
\ee
The target space of the sigma model is the symmetrized product of
$N_1N_5$ copies of $T^4$,
\be
(T_4)^{N_1N_5}/S_{N_1N_5},
\ee
with each copy of $T^4$ giving 4 bosonic excitations which we call $X^1, X^2, X^3, X^4$ for left movers and $\bar{X}^1, \bar{X}^2, \bar{X}^3, \bar{X}^4$. It also gives 4 fermionic excitations, which we call $\psi^1, \psi^2, \psi^3, \psi^4$ for the left movers, and $\bar\psi^1, \bar\psi^2,\bar\psi^3,\bar\psi^4$ for the right movers. These are the four directions in which the string can oscillate away from $S^1$.  The central charge of the theory with fields $X^i, \psi^i, ~i=1\dots 4$ is $c=6$. The total central charge of the entire system is thus $6 N_1N_5$. In the CFT there are left moving excitations (holomorphic sector) and right moving excitations (antiholomorphic sector). In many computations, we only consider the left movers explicitly since the right movers function almost identically.

Since the different copies are symmetrized, the operator content includes twist operators. A twist operator $\sigma_n$, which we call the bare twist, takes $n$ different copies of the CFT and links them into a single copy living on a circle that is $n$ times longer. We call any such linked set of copies a `component string'. Let's illustrate this process below. Imagine that we have n singly wound copies 
\bea
X^1, X^2, X^3,\ldots,  X^n
\eea

Now by applying $\sigma_n$, these $n$ copies are permuted into each other as follows 
\bea
X_1 \to X_2 \to X_3 \to \ldots X_{n-1}\to X_n \to X_1
\eea
The action of $\sigma_n$ changes the boundary conditions on each of the singly wound copies so that after going around the $\s$ direction by $2\pi$ the endpoint of the first string is identified with the second string and then going around $2\pi$ again identifies the endpoint of the second string with the third string and so on:
\bea
X^1(\s + 2\pi) &=& X^2(\s),\quad X^2(\s + 2\pi)= X^3(\s),\ldots, X^n(\s +2\pi)=X_1(\s)\nn
\eea
We have now created a single `multiwound' copy of length $n$. The fermions are twisted together in a similar manner but with an additional feature. The twist must now carry a charge. Along with the bare twist, $\s_n$, we must include something called a spin field $S^{\pm}$. The full twist operator is therefore defined as \cite{acm1}
\bea
\s^{\pm\pm}_n=S^{\pm\pm}\s_n
\eea 
Fermionic strings can exist within two sectors: the Ramond sector, which corresponds to periodic boundary    conditions as one goes around the string, and the $NS$ sector which correspond to antiperiodic boundary conditions. The spin field, $S^\pm$, changes the boundary condition of the string from the $NS$ sector to the the positive or negative Ramond sector. Lets discuss the vacuum states in more detail.

\section{NS and R vacua}
Consider a single copy of the $c=6$ CFT. The lowest energy state of the left-moving sector is the NS vacuum, 
\be
|0_{NS}\rangle \,, \qquad\quad h = 0, \quad m = 0 \,
\ee
where $h$ denotes the $L_0$ eigenvalue.
However, we study the CFT in the R sector. In particular we are interested in the R vacua denoted by\footnote{There are two other left-moving R ground states, which may be obtained by acting with fermion zero modes. Including right-movers then gives a total of 16 R-R ground states; see e.g.~\cite{Avery:2010qw}.}
\be
|0_{R}^{\pm}\rangle \,, 
\qquad\quad h=\frac14, \quad m=\pm\frac12 \,.
\ee
We can relate the NS and R sectors using spectral flow.  In particular, spectral flow by $\a = \pm 1$ in the left-moving sector produces the transformations
\bea
\alpha= 1 : && \ket{0_R^-} \rightarrow \ket{0_{NS}} \,, 
\quad \ket{0_{NS}}  \rightarrow \ket{0_R^+}  \cr
\alpha= -1 : && \ket{0_R^+} \rightarrow \ket{0_{NS}} \,, 
\quad \ket{0_{NS}}  \rightarrow \ket{0_R^-}  \label{eq:sfstates}
\eea

Similar relations hold for the right-moving sector, which we denote with bars.  In many places we will write expressions only for the left-moving sector, as the two sectors are on an equal footing in this chapter.  However, in certain places we will write expressions involving the full vacuum of both sectors.  We therefore introduce the notation
\bea
|0_R^{--}\rangle &\equiv& |0_R^-\rangle |\bar{0}_R^-\rangle \,. \label{eq:0r--}
\eea
Following the notation of \cite{acm2}, we denote the hermitian conjugate of $\ket{0_R^{--}}$ as 
\bea
\bra{0_{R,--}} \,.
\eea

Now that we've discussed the vacuum structure of the theory, lets understand more about the twist interaction. The free CFT is characterized by multiwound copies of various winding including singly wound copies.
The `free' theory at the orbifold point has been surprisingly successful in reproducing many aspects of black hole physics. The free CFT yields exact agreement with the properties of   near-extremal black holes; for example the entropy and greybody factors are reproduced exactly \cite{sv,dmcompare}. 
However, the full dynamics of black holes is not given by the CFT at the orbifold point; we have to deform away from this `free' theory by an operator which corresponds to turning on the coupling constant of the orbifold theory \cite{acm2}. In particular the process of black hole formation is described in the CFT by the thermalization of an initially non-thermal  state; such a process requires nontrivial interactions in the CFT.
The deformation operator describing these interactions has the form of a twist operator $\sigma_2$, dressed with a supercharge: $\hat O\sim G_{-{1\over2}} \sigma_2$. As just discussed, since we are dealing with fermions as well as bosons the twist must carry charge and since the deformation should be charge neutral, the $G$ must come with the opposite charge. Schematically deformation should look something like $\hat O\sim G^-_{-{1\over2}} \sigma^+_2$. The effect of the twist is depicted in Fig.\;\ref{fone}. Since the $G$ is a contour around the twist operator we can deform that contour into two parts. An upper component where the supercharge is acting after the twist and a lower component where the supercharge is acting before the twist.

In \cite{acm2} it was shown that we can write the deformation operator as
\be
\hat O_{\dot A\dot B}(w_0)=\Big [{1\over 2\pi i} \int _{w_0} dw G^-_{\dot A} (w)\Big ]\Big [{1\over 2\pi i} \int _{\bar w_0} d\bar w \bar G^-_{\dot B} (\bar w)\Big ]\sigma_2^{++}(w_0)
\label{pert}
\ee

Only considering the left moving sector, we stretch the $G$ contours to locations above and below the twist. This is illustrated in figure. We therefore write the two components of the contour. The component which exist above the twist insertion is:

\begin{eqnarray}
\frac{1}{2\pi i}\int\limits_{w=\tau_0+\e}^{\tau_0+\e+2\pi i(M+N)}G_{\dot{A}}^{-}\left(w\right)dw &=& G_{\dot{A},0}^{-}
\label{G0_above}
\end{eqnarray}
and the lower 
 part  of the contour gives
\begin{eqnarray}
- \frac{1}{2\pi i} \left[\int\limits_{w=\tau_0-\e}^{\tau_0-\e+2\pi iM}G_{\dot{A}}^{(1)-}
\left(w\right)dw +\int\limits_{w=\tau_0-\e}^{\tau_0-\e+2\pi iN}G_{\dot{A}}^{(2)-}
\left(w\right)dw \right]
&=& - \left( G_{\dot{A},0}^{(1)-} +G_{\dot{A},0}^{(2)-} \right) . \qquad\quad
\label{G0_below-1}
\end{eqnarray}
Thus we obtain
\be
\hat{O}_{\dot A} = G^{-}_{\dot A,0}\s_2^+ - \s_2^+ \left ( G^{(1)-}_{\dot A,0} + G^{(2)-}_{\dot A,0}\right )
\ee
and similarly for the right sector.  When acting upon a state, we can use the fact that the supercharge annihilates the vacuum to simplify our expressions.  For the vacuum, we have
\bea
\hat{O}_{\dot A}|0_R^-\rangle^{(1)} |0_R^-\rangle^{(2)} & = & G^{-}_{\dot A,0}\s_2^+|0_R^-\rangle^{(1)} |0_R^-\rangle^{(2)}
\eea
while for some operator $\hat{Q}$, we have
\bea
\hat{O}_{\dot A}\hat{Q}|0_R^-\rangle^{(1)}  |0_R^-\rangle^{(2)} & = & G^{-}_{\dot A,0}\s_2^+\hat{Q}|0_R^-\rangle^{(1)}  |0_R^-\rangle^{(2)} - \s_2^+ \left [ \left ( G^{(1)-}_{\dot A,0} + G^{(2)-}_{\dot A,0}\right ),\hat{Q} \right ]|0_R^-\rangle^{(1)}  |0_R^-\rangle^{(2)} \nn
&=& G^{-}_{\dot A,0}\s_2^+\hat{Q}|0_R^-\rangle^{(1)}  |0_R^-\rangle^{(2)} - \s_2^+ \left \{ \left ( G^{(1)-}_{\dot A,0} + G^{(2)-}_{\dot A,0}\right ),\hat{Q} \right \}|0_R^-\rangle^{(1)}  |0_R^-\rangle^{(2)} \nn
\eea
The commutation/anticommutation relations of the supercharge with basic operators such as the creation and annihilation operators for the bosonic and fermionic fields are known.  As such, the bulk of our computations focus on the effects of the twist.

Note that the interaction term only contains $\s^+_2$ and not higher orders of $\s^+_n$. $\s^+_2$ is a chiral primary and therefore the only twist that one could combine with a supercharge $G^-$ along with it's antiholomorphic counterpart to produce a dimension two operator in the Lagrangian.

Let's discuss the boson and fermion field content of the CFT.

We have 4 real left moving fermions $\psi_1, \psi_2, \psi_3, \psi_4$ which are grouped into doublets $\psi^{\alpha A}$ as follows:
\be
\begin{pmatrix}
\psi^{++} \cr \psi^{-+}
\end{pmatrix}
={1\over\sqrt{2}}
\begin{pmatrix}
\psi_1+i\psi_2 \cr \psi_3+i\psi_4
\end{pmatrix}
\ee
\be
\begin{pmatrix}
\psi^{+-} \cr \psi^{--}
\end{pmatrix}
={1\over\sqrt{2}}
\begin{pmatrix}
\psi_3-i\psi_4 \cr -(\psi_1-i\psi_2)
\end{pmatrix}.
\ee
The index $\alpha=(+,-)$ corresponds to the subgroup $SU(2)_L$ of rotations on $S^3$ and the index $A=(+,-)$ corresponds to the subgroup $SU(2)_1$ from rotations in $T^4$. The 2-point functions read
\be
\langle\psi^{\alpha A}(z)\psi^{\beta B}(w)\rangle=-\epsilon^{\alpha\beta}\epsilon^{AB}{1\over z-w}
\ee
where we have 
\be
\epsilon_{12}=1, ~~~\epsilon^{12}=-1
\ee

We also have four real bosons $X^1,X^2,X^3,X^4$ but we find it convenient to group them into four complex bosons 
\be
X_{A\dot A}= \sqi X_i \sigma_i=\sqi\begin{pmatrix}  X_3+iX_4& X_1-iX_2\\ X_1+iX_2&-X_3+iX_4 \end{pmatrix}
\ee
where $\sigma_i=\sigma_1, \sigma_2, \sigma_3, iI$. The 2-point functions are
\be
<\partial X_{A\dot A}(z) \partial X_{B\dot B}(w)>~=~{1\over (z-w)^2}\epsilon_{AB}\epsilon_{\dot A\dot B}\,.
\label{ope}
\ee


\section{Symmetries of the CFT}
The D1D5 CFT has a (small) $\mathcal{N}=4$ superconformal symmetry in both the left and right-moving sectors. The generators and their OPEs are given in Appendix~\ref{app_cft}. 

Each superconformal algebra contains an R symmetry $SU(2)$.  Thus we have the global symmetry $SU(2)_L\times SU(2)_R$, with quantum numbers:
\be\label{ExternalQuantumNumbers}
SU(2)_L: ~(j, m);~~~~~~~SU(2)_R: ~ (\bar j, \bar m).
\ee
In addition there is an $SO(4) \simeq SU(2)\times SU(2)$ symmetry, coming from rotations in the four directions of the $T^4$, which is broken by the fact that we have compactified these directions into a torus.  However, the symmetry still provides a useful organizing tool for states. We label this symmetry by
\be
SU(2)_1\times SU(2)_2.
\ee

We use indices $\alpha, \dot\alpha$ for $SU(2)_L$ and $SU(2)_R$ respectively, and indices $A, \dot A$ for $SU(2)_1$ and $SU(2)_2$ respectively.  The 4 real fermion fields of the left sector are grouped into complex fermions $\psi^{\alpha A}$. The right fermions are grouped into fermions $\bar{\psi}^{\dot\alpha A}$. The boson fields $X^i$ are a vector in $T^4$ and have no charge under $SU(2)_L$ or $SU(2)_R$, so are grouped as $X_{A \dot A}$. Different copies of the $c=6$ CFT are denoted with a copy label in brackets, e.g.~
\bea
X^{(1)} \,, ~ X^{(2)} \,,~ \cdots \,, ~X^{(N_1N_5)} \,.
\eea
Further details about the CFT are given in Appendix~\ref{app_cft}.

\section{Nature of the twist interaction}

\begin{figure}[t]
\begin{center}
\includegraphics[scale=.55]{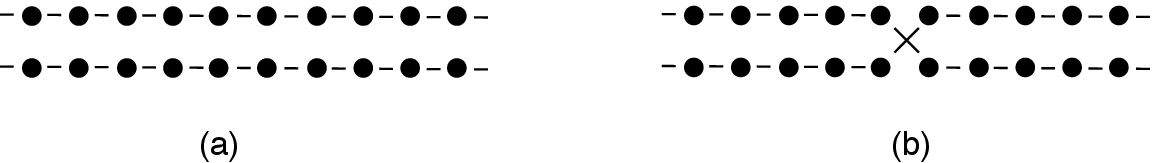}
\caption{(a) The scalar field on the component strings modeled by point masses joined by springs. (b) The twist operator $\sigma_2$ changes the springs so that the masses are linked in a different way.}
\label{ftwo}
\end{center}
\end{figure}

To understand the effect of such a twist $\sigma_2$ in a more physical way, consider a discretization of a 1+1 dimensional bosonic free field $X$.  We can model this field by a collection of point masses joined by springs. This gives a set of coupled harmonic oscillators, and the     oscillation amplitude of the masses then gives the field $X(\tau,  \sigma)$. Consider such a collection of point masses on two different circles, and let the state in each case be the ground state of the coupled oscillators (Fig.\;\ref{ftwo}(a)). At time $\tau_0$ and position $\sigma_0$, we insert a twist $\sigma_2$. The effect of this twist is to connect the masses with a different set of springs, so that the masses make a single chain of longer  length (Fig.\;\ref{ftwo}(b)).

To see what we might expect of such an interaction, consider a single harmonic oscillator,  which starts in its ground state. The Hamiltonian can be expressed in terms of annihilation and creation operators $\hat a,  {\hat a}^\dagger$, and the ground state  $|0\rangle_a$ is given by $\hat a|0\rangle_a=0$.  At time $\tau=\tau_0$, imagine changing the spring constant to a different value. The wavefunction does not change at this instant, and the Lagrangian remains quadratic. But the ground state with this new spring constant is a different state $|0\rangle_b$, and the operator $\hat {a}$ can be expressed as a linear combination of the new annihilation and creation operators $\hat b, {\hat b}^\dagger$:
\be
\hat a = \alpha \hat b +\beta {\hat b}^\dagger
\ee
The wavefunction after this change of coupling can be reexpressed as
\be
|0\rangle_a=Ce^{\h \gamma {\hat b}^\dagger {\hat b}^\dagger}|0\rangle_b
\label{exp}
\ee
where $C$ is a constant.
If we had a single initial excitation before the twist, it will give a single excitation after the twist, but with a nontrivial coefficient\footnote{We have $f=\alpha^*+\beta^*\gamma$.} $f$
\be
\hat a ^\dagger |0\rangle_a=f \, {\hat b}^\dagger \, Ce^{\h \gamma {\hat b}^\dagger {\hat b}^\dagger}|0\rangle_b \,.
\label{ff}
\ee

Now let us return to the CFT. Regarding the scalar field on the component strings as a set of coupled harmonic oscillators, we note that the twist interaction changes the coupling matrix between the oscillators but not the wavefunction itself. Thus the effect of the twist is captured by reexpressing the state before the twist in terms of the natural oscillators after the twist. In analogy to (\ref{exp})  the state
 after the twist will then have the form
\bea
|\chi\rangle&\equiv& \sigma_2(w_0)|0^{(1)}\rangle\otimes |0^{(2)}\rangle\nn
&=&C(w_0)e^{\sum_{ k\ge 1, l\ge 1}\gamma^B_{kl}(-\alpha_{++, -k}\alpha_{--, -l}+\alpha_{-+, -k}\alpha_{+-, -l})}
|0\rangle \,.
\label{pfive}
\eea
The index structure on the $\alpha$ oscillators is arranged to obtain a singlet under the group of rotations in the $T^4$. This is explained in detail in  \cite{acm2}, where we also fix 
the normalization $C(w_0)$  to unity after we include the fermions \cite{acm2}. 


An initial excitation on one of the component strings will transform to a linear combination of excitations on the final component string above the state $|\chi\rangle$. Analogous to (\ref{ff}), we will have
\bea
&&\sigma_2(w_0)\alpha^{(i)}_{A\dot A, -m}|0\rangle^{(1)}\otimes |0\rangle^{(2)}=\nn
&&\qquad\qquad\sum_{k\ge 1}~f^{B(i)}_{mk}~\alpha_{A\dot A, -k} ~
 e^{\sum_{ k'\ge 1, l'\ge 1}\gamma^B_{k'l'}(-\alpha_{++, -k'}\alpha_{--, -l'}+\alpha_{-+, -k'}\alpha_{+-, -l'})}|0\rangle \,.\qquad\quad
 \label{wfourqq}
\eea
where $i=1,2$ for the initial component strings with windings $M,N$ respectively. We will find the $f^{B(i)}_{mk}$. In the next chapter we show how one computes such states. In subsequent chapter, we show how one computes these quantities for two twists.

\chapter{Effect of the twist operator on a general vacuum state of winding M and winding N}\label{sec:gamma}
In this chapter we discuss what a twist $\s_2$ does to the vacuum state of two initial strings, one of winding $M$ and the other of winding $N$. We compute the bogoliubov coefficients which characterize the twisted state on an $M+N$ wound copy of the string.
\section{Introduction}
\begin{figure}[t]
\begin{center}
\includegraphics[scale=.50]{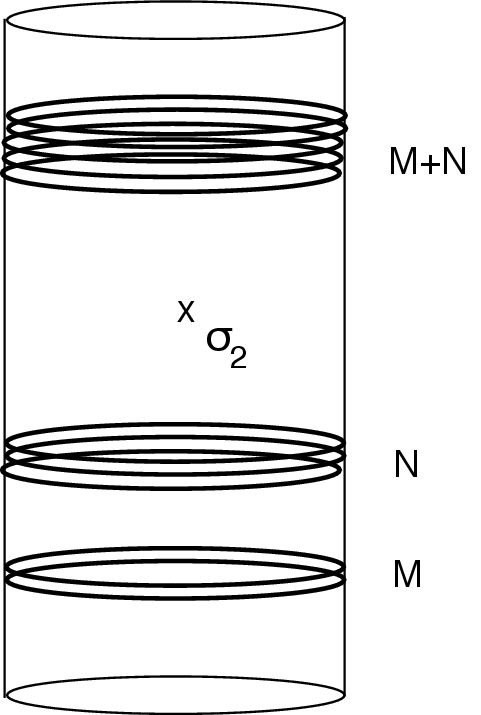}
\caption{The cylinder with coordinate $w$. The state before the twist has component strings with windings $M,N$. The twist operator $\sigma_2$  links these into a single component string of winding $M+N$.}
\label{fone}
\end{center}
\end{figure}

Consider the process depicted in Fig.\;\ref{fone}. We insert the twist operator $\sigma_2$ at a location 
\be
w_0=\tau_0+i\sigma_0
\ee
The twist operator changes the Lagrangian of the theory from one free Lagrangian (that describes  free CFTs on circles of length $2\pi M$ and $2\pi N$) to another free Lagrangian (that for a free CFT on a circle of length $2\pi(M+N)$). As explained in \cite{acm2}, in this situation the vacuum state of the initial theory does not go over to the vacuum of the new theory. But the excitations must take the special form of a Gaussian (\ref{pfive}), so our goal is to find the coefficients $\gamma^B_{kl}$. 

The steps we follow are analogous to those in \cite{acm2}. We first map the cylinder $w$ to the complex plane through $z=e^w$. The CFT field $X$ will be multivalued in the $z$ plane, due to the presence of twist operators.  The initial component strings with  windings $M,N$ are created by twist operators $\sigma_M, \sigma_N$. The interaction is another twist operator $\sigma_2$. The point at infinity has a twist of order $M+N$ corresponding to the component string in the final state.

To handle the twist operators, we go to a covering space $t$ where $X$ is single valued. The twist operators become simple punctures in the $t$ plane, with no insertions at these punctures. We can therefore trivially close these punctures. The nontrivial physics is now encoded in the definition of oscillator modes on the $t$ plane -- the creation operators on the cylinder are linear combinations of creation and annihilation operators on the $t$ plane.  Performing appropriate Wick contractions, we obtain the $\gamma^B_{kl}$. Finally, we change notation to a form that will be more useful in the situation where $k\gg 1, l\gg 1$.

\section{The exponential ansatz}

We begin our computations by working firstly with the twist operator $\s^{++}_2$.  After determining the effects of the twist operator, we use the known commutation relations of the supercharge to determine the effects of the full deformation operator.  We consider separately the effects of this operator on the vacuum, and on a state with a single bosonic or fermionic excitation.

Let us now record the exponential ansatz with all notation explicit:
\begin{eqnarray}
\ket{\boldsymbol{\chi}} &\equiv& \sigma_{2}^{++}(w_{0})
|0_R^{--}\rangle^{(1)}|0_R^{--}\rangle^{(2)} 
\cr
&=&
C_{MN} \left[
\exp\left(\sum_{m\geq 1,n\geq 1}
\gamma_{mn}^{B}\left[-\alpha_{++,-m}\alpha_{--,-n}
+\alpha_{+-,-m}\alpha_{-+,-n}\right]\right)
\right. \cr
&& \qquad \qquad  \left.
\times\exp\left(\sum_{m\geq 0,n\geq 1}
\gamma_{mn}^{F}
\left[d_{-m}^{++}d_{-n}^{--}-d_{-m}^{+-}d_{-n}^{-+}
\right]\right)
\times [\mathrm{antihol.}] \right]
\ket{0_{R}^{--}} \,. \qquad\nn
\label{eq:expans}
\end{eqnarray}
This ansatz is a generalization of that made in \cite{acm2} in the case of $M=N=1$, where the exponential ansatz was verified up to fourth order in the bosonic oscillators.

The main novel feature of the above ansatz relative to the $M=N=1$ case is the non-trivial overall factor $C_{MN}$. A priori, this could be a function of $w_0$, but will turn out to be a pure c-number independent of $w_0$. We observe that
\bea
C_{MN}&=&\bra{0_{R,--}}
\sigma_{2}^{++}\left(w_{0}\right)
\ket{0_{R}^{--}}^{\left(1\right)}\ket{0_{R}^{--}}^{\left(2\right)} \,.
\label{eq:cmn}
\eea
In the case of $M=N=1$, the normalization of the twist $\sigma_2^{++}$ implies that $C_{11}=1$ \cite{acm2}. 
We compute the general factor $C_{MN}$ in Section \ref{sec:CMN}.

\subsection{Bosonic modes on the cylinder $w$}

Let us begin by defining operator modes on the cylinder.
Below the twist insertion ($\tau<\tau_0$)
we  have modes $\alpha^{(1)}_{A\dot A, m}$ on the component string of winding $M$ and modes $\alpha^{(2)}_{A\dot A, n}$ on the component string of winding $N$:
\be
\alpha^{(1)}_{A\dot A, m}= {1\over 2\pi} \int\limits_{\sigma=0}^{2\pi M} \partial_w X^{(1)}_{A\dot A}(w) e^{{m\over M}w} dw
\ee
\be
\alpha^{(2)}_{A\dot A, n}= {1\over 2\pi} \int\limits_{\sigma=0}^{2\pi N} \partial_w X^{(2)}_{A\dot A}(w) e^{{n\over N}w} dw
\ee
From (\ref{ope}), we find that the commutation relations are
\be
[\alpha^{(i)}_{A\dot A, m}, \alpha^{(j)}_{B\dot B, n}]=-\epsilon_{AB}\epsilon_{\dot A\dot B}\delta^{ij} m \delta_{m+n,0}
\ee

Above this twist insertion ($\tau>\tau_0$) we have a single component string of winding $M+N$. The modes are
\be
\alpha_{A\dot A, k}= {1\over 2\pi} \int\limits_{\sigma=0}^{2\pi (M+N)} \partial_w X_{A\dot A}(w) e^{{k\over M+N}w} dw
\label{qaone}
\ee
The commutation relations are
\be\label{bcommtwist}
[\alpha_{A\dot A, k}, \alpha_{B\dot B, l}]=-\epsilon_{AB}\epsilon_{\dot A\dot B}\,  k\,  \delta_{k+l,0}\;.
\ee

\subsection{Bosonic modes on the $z$ plane}

We wish to go to a covering space where the field $X_{A\dot A}$ will be single valued.  As a preparatory step, it is convenient to map the cylinder with coordinate $w$ to the plane with coordinate $z$,
\be
z=e^w
\ee
Under this map the operator modes change as follows. Before the insertion of the twist ($|z|<e^{\tau_0}$) we have,  using a contour circling ${z=0}$
\be
\alpha^{(1)}_{A\dot A, m}\rightarrow {1\over 2\pi} \int\limits_{{\rm arg}(z)=0}^{2\pi M} \partial_z X^{(1)}_{A\dot A}(z) z^{m\over M} dz
\ee
\be
\alpha^{(2)}_{A\dot A, n}\rightarrow {1\over 2\pi} \int\limits_{{\rm arg}(z)=0}^{2\pi N} \partial_z X^{(2)}_{A\dot A}(z) z^{n\over N} dz
\ee

After the twist ($|z|>e^{\tau_0}$) we have, using a contour circling ${z=\infty}$
\be
\alpha_{A\dot A, k}\rightarrow {1\over 2\pi} \int\limits_{{\rm arg}(z)=0}^{2\pi (M+N)} \partial_z X_{A\dot A}(z) z^{{k\over M+N}} dz \,.
\ee

\subsection{Bosonic modes on the covering space $t$}

We now proceed to the covering space $t$ where $X_{A\dot A}$ will be single-valued. Consider the map
\be
z=t^M(t-a)^N
\label{coverone}
\ee
The various operator insertions map as follows:

(i) The initial component strings were at $w\r-\infty$ on the cylinder, which corresponds to $z=0$ on the $z$ plane. In the $t$ plane, the component string of winding $M$ maps to $t=0$, while the component string with winding $N$ maps to $t=a$.

(ii) The final component string state is at $w\r\infty$ on the cylinder, which corresponds to $z=\infty$. This maps to $t=\infty$.

(iii) The twist operator $\sigma_2$ is at $w_0$ on the cylinder, which corresponds to $e^{w_0}$ on the $z$ plane. To find its location on the covering space $t$, we note that ${dz\over dt}$ should vanish at the location of every twist, since these are ramification points of map (\ref{coverone}) to the covering space. We find that apart from the ramification points at $t=0,a, \infty$, the function  ${dz\over dt}$ also vanishes at
\be
t={a M\over M+N}
\ee
which corresponds to the following value of $z$:
\be
z_0={a^{M+N}} {M^MN^N\over (M+N)^{M+N}}(-1)^N \,.
\ee
To solve for $a$, we must specify how to deal with the fractional exponent. We do this by choosing
\be
z_0={a^{M+N}} {M^MN^N\over (M+N)^{M+N}}e^{i\pi N} \,
\ee
which determines the quantity $a$ in terms of $z_0=e^{w_0}$, as
\be
a=e^{-i\pi \frac{N}{M+N}}\left( {z_0\over M^MN^N}\right)^{1\over M+N}(M+N)\,.
\label{az}
\ee

Now let us consider the modes in the $t$ plane. Before the twist we have
\be
\alpha^{(1)}_{A\dot A, m}\to {1\over 2\pi} \oint\limits_{t=0} dt \, \partial_t X_{A\dot A}(t)  \left ( t^M(t-a)^N \right ) ^{m\over M}
\ee
\be
\alpha^{(2)}_{A\dot A, n}\to {1\over 2\pi} \oint\limits_{t=a} dt \, \partial_t X_{A\dot A}(t)  \left ( t^M(t-a)^N \right ) ^{n\over N}
\ee
After the twist we have
\be
\alpha_{A\dot A, k}\to {1\over 2\pi}\oint\limits_{t=\infty} dt \, \partial_t X_{A\dot A}(t)  \left ( t^M(t-a)^N \right ) ^{k\over M+N}
\ee
We also define mode operators that are natural to the $t$ plane, as follows:
\be
\tilde\alpha_{A\dot A, m}\equiv  {1\over 2\pi} \oint\limits_{t=0} dt \, \partial_t X_{A\dot A}(t) t^m 
\label{qathree}
\ee
The commutation relations are
\be
[\tilde{\alpha}_{A \dot A, k}, \tilde{\alpha}_{B \dot B, l}]=-\epsilon_{A B} \epsilon_{\dot A \dot B} \, k\, \delta_{k+l, 0}
\label{pttwo}
\ee

\subsection{Method for finding  the  $\gamma^B_{kl}$}

Let us consider the amplitude
\bea
{\cal A}_1&=&\langle 0|\sigma_2(w_0)|0\rangle^{(1)}\otimes |0\rangle^{(2)}\nn
&=&C(w_0)\langle 0|e^{\sum_{ k\ge 1, l\ge 1}\gamma^B_{kl}[-\alpha_{++, -k}\alpha_{--, -l}+\alpha_{-+, -k}\alpha_{+-, -l}]}
|0\rangle\nn
&=& C(w_0)
\eea
where we assume that the vacuum is normalized as $\langle 0 | 0 \rangle=1$.

We compare this to the amplitude
\bea
{\cal A}_2&=&\langle 0|\Big (\alpha_{++,_l}\alpha_{--,_k}\Big )\sigma_2(w_0)|0\rangle^{(1)}\otimes |0\rangle^{(2)}\nn
&=&C(w_0)\langle 0|\Big (\alpha_{++,_l}\alpha_{--,_k}\Big )e^{\sum_{ k'\ge 1, l'\ge 1}\gamma^B_{k'l'}[-\alpha_{++, -k'}\alpha_{--, -l'}+\alpha_{-+, -k'}\alpha_{+-, -l'}]}
|0\rangle\nn
&=& -C(w_0)~kl\gamma^B_{kl}~\langle 0|0\rangle=-C(w_0)~kl\gamma^B_{kl}
\eea
Thus we see that
\be
\gamma^B_{kl}=-{1\over kl} {{\cal A}_2\over {\cal A}_1} \,.
\ee

To compute ${\cal A}_1$ we map the cylinder $w$ to the plane $z$ and then to the cover $t$. In this cover the locations of the twist operators are just punctures with no insertions. Thus these punctures can be closed, making the $t$ space just a sphere. Closing the punctures involves normalization factors, so we write
\be
{\cal A}_1=\langle 0|\sigma_2(w_0)|0\rangle^{(1)}\otimes |0\rangle^{(2)}=D(z_0)~{}_t\langle 0 |0\rangle_t
\ee
Factors like $D(z_0)$ were computed in \cite{lm1}, but here we do not need to compute $D(z_0)$ since it will cancel in the ratio ${\cal A}_2/{\cal A}_1$. We have
\be
{\cal A}_2=\langle 0|\Big (\alpha_{++,_l}\alpha_{--,_k}\Big )\sigma_2(w_0)|0\rangle^{(1)}\otimes |0\rangle^{(2)}=D(z_0)~{}_t\langle 0| \Big (\alpha'_{++,_l}\alpha'_{--,_k}\Big )|0\rangle_t
\ee
where the primes on the operators on the RHS signify the fact that these operators arise from the unprimed operators by the various maps leading to the $t$ plane description. Thus we have
\be
\gamma^B_{kl}=-{1\over kl} {{\cal A}_2\over {\cal A}_1}=-{1\over kl} ~{{}_t\langle 0| \Big (\alpha'_{++,_l}\alpha'_{--,_k}\Big )|0\rangle_t\over {}_t\langle 0 |0\rangle_t}\,.
\ee

\section{Computing the $\gamma^B_{kl}$}

The operators $\alpha'$ are given by contour integrals at large $t$:
\bea
&&\hskip-.25in{}_t\langle 0| \Big (\alpha'_{++,_l}\alpha'_{--,_k}\Big )|0\rangle_t=\nn
&&\hskip-.25in{}_t\langle 0 |\Big ({1\over 2\pi }\int_{} dt_1 \partial_t X_{++}(t_1) \left ( t_1^M(t_1-a)^N \right ) ^{l\over M+N}\Big )\Big(
{1\over 2\pi }\int_{} dt_2 \partial_t X_{--}(t_2) \left ( t_2^M(t_2-a)^N \right ) ^{k\over M+N}\Big )
|0\rangle_t\nn
\label{ptfive}
\eea
with $|t_1|>|t_2|$. We have\footnote{The symbol ${}^nC_m$ is the binomial coefficient, also written $\begin{pmatrix} n\\ m\\ \end{pmatrix}$.}
\be
\left ( t_1^M(t_1-a)^N \right ) ^{l\over M+N}=t_1^l(1-a t_1^{-1})^{lN\over M+N}=t_1^l\sum_{p\ge 0} {}^{lN\over M+N}C_p (-a)^p t_1^{-p}=\sum_{p\ge 0} {}^{lN\over M+N}C_p (-a)^p t_1^{l-p}
\ee
\be
\left ( t_2^M(t_2-a)^N \right ) ^{k\over M+N}=t_2^k(1-a t_2^{-1})^{kN\over M+N}=t_2^k\sum_{q\ge 0} {}^{kN\over M+N}C_q (-a)^q t_2^{-q}=\sum_{q\ge 0} {}^{kN\over M+N}C_q (-a)^q t_2^{k-q}
\ee
Thus
\bea
{1\over 2\pi }\int_{} dt_1 \pa_t X_{++}(t_1) \left ( t_1^M(t_1-a)^N \right ) ^{l\over M+N}&=&\sum_{p\ge 0} {}^{lN\over M+N}C_p (-a)^p \t\alpha_{++, {l-p}}\nn
{1\over 2\pi }\int_{} dt_2 \pa_t X_{--}(t_2) \left ( t_2^M(t_2-a)^N \right ) ^{k\over M+N}&=&\sum_{q\ge 0} {}^{kN\over M+N}C_q (-a)^q \t\alpha_{--, {k-q}}
\eea
We then find
\be
\gamma^B_{kl}=-{1\over kl}\sum_{p\ge 0} \sum_{q\ge 0}~{}^{lN\over M+N}C_p ~{}^{kN\over M+N}C_q (-a)^{p+q}~ {{}_t\langle 0|\t \alpha_{++,l-p}~ \t \alpha_{--,k-q}|0\rangle_t\over {}_t\langle 0|0\rangle_t}
\ee
Using the commutation relations (\ref{pttwo}) we get
\be
k-q=-(l-p) ~\Rightarrow ~q=(k+l)-p
\ee
This gives
\be
\gamma^B_{kl}={(-a)^{k+l}\over kl}\sum_{p\ge 0} ~{}^{lN\over M+N}C_p ~{}^{kN\over M+N}C_{(k+l)-p} ~(l-p)
\ee
Note that in order to give a non-zero contribution, $\t\alpha_{++,l-p}$ needs to be an annihilation operator, so we require $p\le l$. Thus we have
\be
\gamma^B_{kl}={(-a)^{k+l}\over kl}\sum_{p= 0}^l ~{}^{lN\over M+N}C_p ~{}^{kN\over M+N}C_{(k+l)-p} ~(l-p)
\ee
Evaluating this sum gives
\be
\gamma^B_{kl}=-{(-a)^{k+l}\sin[{\pi  Mk\over M+N}]\sin[{\pi Ml\over M+N}]\over \pi^2}{MN\over (M+N)^2}{1\over (k+l)}{\Gamma[{Mk\over M+N}]\Gamma[{Nk\over M+N}]\over \Gamma[k]}{\Gamma[{Ml\over M+N}]\Gamma[{Nl\over M+N}]\over \Gamma[l]}\,.
\label{gammapre}
\ee

\subsection{Expressing $\gamma^B_{kl}$ in final form}

 It will be convenient to write the expression for $\gamma^B_{kl}$ in a slightly different notation. We make the following changes:

 (i) We can replace the parameter $a$ by the variable $z_0=e^{ w_0}$ through (\ref{az}).

(ii) In addressing the continuum limit it is useful to use fractional mode numbers  defined as
\be
s={k\over M+N}, ~~~s'={l\over M+N}
\ee
The parameters $s,s'$ directly give the physical wavenumbers of the modes on the cylinder with coordinate $w$. 
When using $s,s'$ in place of $k,l$, we will write
\be
\gamma^B_{kl}\r \tilde\gamma^B_{ss'}\,.
\ee

(iii) We define the useful shorthand notation
\be
1-e^{2\pi i Ms}=\mu_s \,.
\ee
Note that
\be
(-1)^k\sin[{\pi  Mk\over M+N}]= (-1)^{(M+N)s}\sin(\pi Ms)={i\over 2}(-1)^{Ns}(1-e^{2\pi i Ms})=  {i\over 2}(-1)^{Ns}\mu_s \,.
\ee

With these changes of notation we find
\bea
\tilde\gamma^B_{ss'}&=&{1\over 4\pi^2}\,z_0^{s+s'}\,
{\mu_s\mu_{s'}\over  s+s'}\,{MN\over (M+N)^3} 
\left ({(M+N)^{M+N}\over M^MN^N}\right )^{s+s'} 
{\Gamma[{Ms}]\Gamma[{Ns}]\over \Gamma[(M+N)s]}~{\Gamma[{Ms'}]\Gamma[{Ns'}]\over \Gamma[(M+N)s']}\,.\nn
\label{gammafinal}
\eea

\subsection{The case $M=N=1$}

In \cite{acm2} the $\gamma^B_{ss'}$ were computed for the case $M=N=1$. Let us check that our general result reduces to the result in \cite{acm2} for these parameters.

Our general expression for $\gamma^B_{ss'}$ is given in (\ref{gammafinal}). 
For $M=N=1$ we have
\be
\Gamma[(M+N) s]=\Gamma[2s]={2^{2s-{1\over2}}\over (2\pi)^\h} \Gamma[s]\Gamma[s+{1\over2}]
\label{gammahalf}
\ee
Substituting in (\ref{gammafinal}) we get
\bea
\tilde\gamma^B_{ss'}={z_0^{s+s'}\over 2\pi (s+s')}\,
{\Gamma[{s}]\over \Gamma[s+{1\over2}]}\,
{\Gamma[{s'}]\over \Gamma[s'+{1\over2}]}
\eea
which agrees with the result in \cite{acm2}. 

Now we look at computing $\g^F$

\section{Computing the Bogoliubov coefficients $\gamma^{F}$}
In this section we compute $\gamma_{kl}^{F}$. First we compute the expressions for the fermionic modes.

\subsection{Mode Expansions on the cylinder} \label{sec:expansions}

The mode expansion for the fermionic strings are given below. 

Before the twist, the mode expansions are
\begin{eqnarray}
d_m^{(1)\alpha A} &=& \frac{1}{2\pi i\sqrt M} \int\limits_{\sigma=0}^{2\pi M} \psi^{(1)\alpha A} (w) e^{\frac{m}{M}w} dw
\label{fermM} \\
d_m^{(2)\alpha A} &=&  \frac{1}{2\pi i\sqrt N} \int\limits_{\sigma=0}^{2\pi N} \psi^{(2)\alpha A} (w) e^{\frac{m}{N}w} dw \,.
\label{fermN}
\eea
From the two-point functions, the commutation/anticommutation relations are
\bea
\left\{d_m^{(i)\alpha A}, d_n^{(j)\beta B}\right\} &= &-\varepsilon ^{\alpha\beta} \varepsilon^{AB}\delta^{ij}\delta_{m+n,0} \,.
\label{commrelation}
\eea
After the twist, there is a single component string of length $ M + N $. We thus have modes
\bea
d_k^{\alpha A} &=& \frac{1}{2\pi i\sqrt{M+N}} \!\! \int\limits_{\sigma=0}^{2\pi (M+N)} \psi^{\alpha A} (w) e^{\frac{k}{M+N}w} dw
\label{fermMN}
\eea
with commutation relations
\bea
\left\{d_k^{\alpha A}, d_l^{\beta B}\right\} &=& -\varepsilon^{\alpha\beta} \varepsilon^{AB} \delta_{k+l,0}  \,.
\eea

Now we must map these modes to the $t$ plane. Since the fermions carry charge they will be affected by the presence of spin fields which will appear in the $t$ plane. Next we show show how handle these spin fields.

\subsection{Spectral flowing to an empty covering plane} \label{sec:sf}

In the $t$ plane, there are no twist insertions.  However, we have spin field insertions at each of the bifurcation points:
\begin{eqnarray}
S^- (t=0) \,,\quad S^+ (t = \tfrac{M}{M+N}a)\,,\quad S^- (t=a) \,, \quad S^+ (t=\infty)\,
\end{eqnarray}
along with appropriate normalization factors. 
To compute $\gamma^F$, 
we shall take a ratio of amplitudes just as we did for the bosons, and so it it will not be necessary to keep account of the normalization factors of the spin fields at this point; such factors will cancel out. 
However, these normalization factors will later be important when we come to calculate  the overall coefficient $C_{MN}$.

Our method of computing $\gamma^F$ is to perform a sequence of spectral flows and coordinate transformations to
remove the spin fields, and thus to map the problem to an empty $ t $ plane.

The sequence is as follows (we write only the left-moving part, the right-moving part is identical):
\begin{enumerate}
\item[i)] Spectral flow by $\alpha = 1$ in the $t$ plane.

\item[ii)] Change coordinate to $t' = t - \tfrac{M}{M+N} a$.

\item[iii)] Spectral flow by $\alpha = -1$ in the $t'$ plane.

\item[iv)]  Change coordinate to $\hat t = t' - \frac{N}{M+N}a = t-a$.

\item[v)]  Spectral flow by $\alpha = 1$ in the $\hat t$ plane.

\item[vi)] Change coordinate back to $t = \hat t + a$ (when necessary).
\end{enumerate}

From the action of spectral flow on the R ground states in \eq{eq:sfstates} one sees that this sequence indeed maps the problem to one with no spin field insertions. 

The action of spectral flow on operators is straightforward for those operators where the fermion content may be expressed as a simple exponential in the language in which the fermions are bosonized. For such operators with charge $j$, spectral flow with parameter $\alpha$ gives rise to the transformation
$$\hat O_j (t)\to t^{-\alpha j}\hat O_j (t).$$
The fermion fields and spin fields are of this form.

Since we take a ratio of amplitudes to compute $\gamma^F$, we will not need to take account of the transformation of the spin fields (again, these will however be important for the computation of $C_{MN}$).

By contrast, an essential part of the computation of $\gamma^F$ is to follow the transformation of the modes of the fermion fields through the above sequence of spectral flows and coordinate transformations. 
Here we present only the final expressions for these modes; details of the derivation are given in Appendix \ref{sec:mapping}.

In terms of the $\hat t$-plane, the cylinder modes \eq{fermM}, \eq{fermN} and \eq{fermMN} map to the following expressions.  For the modes before the twist, we have
\begin{eqnarray}
    d_{m}^{\left(1\right)+A}&\to& \hat{d}_{m}'^{\left(1\right)+A}
		~=~\frac{\sqrt{M+N}}{2\pi i\sqrt{M}}
		\oint\limits_{\hat{t}=-a}d\hat{t}\,\psi^{\left(1\right)+A}\left(\hat{t}\;\!\right)\left[\left(\hat{t}+a\right)^{m-1}\left(\hat{t}+\tfrac{Na}{M+N}\right)\hat{t}^{\frac{Nm}{M}-1}\right]\cr
     d_{m}^{\left(2\right)+A}&\to& \hat{d}_{m}'^{\left(2\right)+A}
		~=~\frac{\sqrt{M+N}}{2\pi i\sqrt{N}}
		\oint\limits_{\hat{t}=0}d\hat{t}\,\psi^{\left(2\right)+A}\left(\hat{t}\;\!\right)\left[\left(\hat{t}+a\right)^{\frac{Mm}{N}-1}\left(\hat{t}+\tfrac{Na}{M+N}\right)\hat{t}^{m-1}\right]\cr
     d_{m}^{\left(1\right)-A}&\to& \hat{d}_{m}'^{\left(1\right)-A}
		~=~\frac{\sqrt{M+N}}{2\pi i\sqrt{M}}
		\oint\limits_{\hat{t}=-a}d\hat{t}\,\psi^{\left(1\right)-A}\left(\hat{t}\;\!\right)\left[\left(\hat{t}+a\right)^{m}\hat{t}^{\frac{Nm}{M}}\right]\cr
     d_{m}^{\left(2\right)-A}&\to& \hat{d}_{m}'^{\left(2\right)-A}
		~=~\frac{\sqrt{M+N}}{2\pi i\sqrt{N}}
		\oint\limits_{\hat{t}=0}d\hat{t}\,\psi^{\left(2\right)-A}\left(\hat{t}\;\!\right)\left[\left(\hat{t}+a\right)^{\frac{Mm}{N}}\hat{t}^{m}\right]
\end{eqnarray}
After the twist, we obtain
\begin{eqnarray}
   d_{k}^{+A}&\to&\hat{d}_{k}'^{+A}~=~\frac{1}{2\pi i}\oint\limits_{\hat{t}=\infty}d\hat{t}\,\psi^{+A}\left(\hat{t}\;\!\right)\left[\left(\hat{t}+a\right)^{\tfrac{Mk}{M+N}-1}\left(\hat{t}+\tfrac{N}{M+N}a\right)\hat{t}^{\frac{Nk}{M+N}-1}\right]\cr
   d_{k}^{-A}&\to&\hat{d}_{k}'^{-A}~=~\frac{1}{2\pi i}\oint\limits_{\hat{t}=\infty}d\hat{t}\,\psi^{-A}\left(\hat{t}\;\!\right)\left[\left(\hat{t}+a\right)^{\tfrac{Mk}{M+N}}\hat{t}^{\frac{Nk}{M+N}}\right]
\label{dhat}
\end{eqnarray}

For later use, let us also rewrite these modes in the $t$ plane.  The modes before the twist then become
\begin{eqnarray}
    d_{m}^{\left(1\right)+A}&\to& d_{m}'^{\left(1\right)+A}~=~\frac{\sqrt{M+N}}{2\pi i\sqrt{M}}\oint\limits_{t=0}dt\,\psi^{\left(1\right)+A}\left(t\right)\left[t^{m-1}\left(t-\tfrac{M}{M+N}a\right)\left(t-a\right)^{\frac{Nm}{M}-1}\right]\cr
     d_{m}^{\left(2\right)+A}&\to& d_{m}'^{\left(2\right)+A}~=~\frac{\sqrt{M+N}}{2\pi i\sqrt{N}}\oint\limits_{t=a}dt\,\psi^{\left(2\right)+A}\left(t\right)\left[t^{\frac{Mm}{N}-1}\left(t-\tfrac{M}{M+N}a\right)\left(t-a\right)^{m-1}\right]\cr
     d_{m}^{\left(1\right)-A}&\to& d_{m}'^{\left(1\right)-A}~=~\frac{\sqrt{M+N}}{2\pi i\sqrt{M}}\oint\limits_{t=0}dt\,\psi^{\left(1\right)-A}(t)\left[t^{m}\left(t-a\right)^{\frac{Nm}{M}}\right]\cr
      d_{m}^{\left(2\right)-A}&\to& d_{m}'^{\left(2\right)-A}~=~\frac{\sqrt{M+N}}{2\pi i\sqrt{N}}\oint\limits_{t=a}dt\,\psi^{\left(2\right)-A}(t)\left[t^{\frac{Mm}{N}}\left(t-a\right)^{m}\right]
\label{d_t(i)_modes}
\end{eqnarray}
and after the twist we obtain
\begin{eqnarray}
   d_{k}^{+A}&\to&d_{k}'^{+A}~=~\frac{1}{2\pi i}\oint\limits_{t=\infty}dt\,\psi^{+A}\left(t\right)\left[t^{\tfrac{Mk}{M+N}-1}\left(t-\tfrac{M}{M+N}a\right)\left(t-a\right)^{\frac{Nk}{M+N}-1}\right]\cr
   d_{k}^{-A}&\to&d_{k}'^{-A}~=~\frac{1}{2\pi i}\oint\limits_{t=\infty}dt\,\psi^{-A}\left(t\right)\left[t^{\tfrac{Mk}{M+N}}\left(t-a\right)^{\frac{Nk}{M+N}}\right] .
 \label{d_t_modes}
\end{eqnarray}
For later use, we also define modes natural to the $t$ and $\hat{t}$ planes,
\begin{eqnarray}
\tilde{d}_{r}^{\alpha A}&=&\frac{1}{2\pi i}\oint\limits_{t=0}\psi^{\alpha A}(t)t^{r-\frac{1}{2}} \label{nat_t_modes} \\
\hat{d}_{r}^{\alpha A}
&=&\frac{1}{2\pi i}\oint\limits_{\hat t=0}
\psi^{\alpha A}\left(\hat{t}\;\!\right){\hat t}^{r-\frac{1}{2}} \,.
\label{nat_that_modes}
\end{eqnarray}
The anticommutation relations for these modes are
\begin{eqnarray}
\lbrace \tilde{d}^{\alpha A}_{r},\tilde{d}^{\beta B}_{s}\rbrace~=~\lbrace \hat{d}^{\alpha A}_{r},\hat{d}^{\beta B}_{s}\rbrace&=&-\epsilon^{\alpha\beta}\epsilon^{AB}\delta_{r+s,0} \,.
\label{anti_comm}
\end{eqnarray}

\subsection{Computing expressions for $\gamma_{kl}^{F}$}
We now compute $\gamma_{kl}^{F}$. Let us consider the ratio of amplitudes
\begin{eqnarray}
\frac{\mathcal{A}_{2}}{\mathcal{A}_{1}}&=&\frac{\bra{0_{R,--}}\left(d^{++}_{l}d^{--}_{k}\right)\sigma_{2}^{+}\left(w_{0}\right)
\ket{0_{R}^{--}}^{\left(1\right)}
\ket{0_{R}^{--}}^{\left(2\right)}}
{\bra{0_{R,--}}\sigma_{2}^{+}\left(w_{0}\right)\ket{0_{R}^{--}}^{\left(1\right)}\ket{0_{R}^{--}}^{\left(2\right)}} \,.
\label{amp}
\end{eqnarray}
From the exponential ansatz \eq{eq:expans} we observe that
\begin{eqnarray}
\mathcal{A}_{2}&=&\bra{0_{R,--}}d^{++}_{l}d^{--}_{k}\ket{\boldsymbol{\chi}} \cr
&=& C_{MN} \bra{0_{R,--}}d^{++}_{l}d^{--}_{k} 
\exp\left(\sum_{m\geq 0,n\geq 1}
\gamma_{mn}^{F}
\left[d_{-m}^{++}d_{-n}^{--}-d_{-m}^{+-}d_{-n}^{-+}
\right]\right)
\ket{0_{R}^{--}} \cr
&=& C_{MN} \, \gamma_{kl}^{F}
\end{eqnarray}
and so using \eq{eq:cmn} we see that
\begin{eqnarray}
\gamma_{kl}^{F}&=&\frac{\mathcal{A}_{2}}{\mathcal{A}_{1}} \,.
\label{amp_ratio}
\end{eqnarray}
We now map this ratio of amplitudes to the empty $\hat t$ plane using the sequence of spectral flows and coordinate transformations discussed in Section \ref{sec:sf}, obtaining
\begin{eqnarray}
\gamma_{kl}^{F} ~=~ \frac{\bra{0_{R,--}}\left(d^{++}_{l}d^{--}_{k}\right)\sigma_{2}^{+}\left(w_{0}\right)\ket{0_{R}^{--}}^{\left(1\right)}\ket{0_{R}^{--}}^{\left(2\right)}}
{\bra{0_{R,--}}\sigma_{2}^{+}\left(w_{0}\right)\ket{0_{R}^{--}}^{\left(1\right)}\ket{0_{R}^{--}}^{\left(2\right)}}
&=&\frac{{}_{\hat{t}}\bra{0_{NS}}\hat{d}'^{++}_{l}\hat{d}'^{--}_{k}\ket{0_{NS}}_{\hat{t}}}
{{}_{\hat{t}}\langle {} 0_{NS}|0_{NS}\rangle_{\hat{t}}}\nn
\label{amp-2}
\end{eqnarray}
where $\hat{d}'^{++}_l$ and $\hat{d}'^{--}_k$ are the modes after all spectral flows and coordinate transformations given in \eq{dhat}. 

Let us now evaluate the $\hat{t}$ plane amplitude in the numerator of (\ref{amp-2}). 
To do so, we expand the transformed modes $\hat{d}'$ in terms of the natural modes $\hat{d}$ on the $\hat{t}$ plane \eq{nat_that_modes}, whose commutation relations are known. From \eq{dhat} we have 
\begin{eqnarray}
\hat{d}_{l}'^{++}~=~\frac{1}{2\pi i}\oint\limits_{\hat{t}=\infty}d\hat{t}\,\psi^{++}\left(\hat{t}\;\!\right)\left[\left(\hat{t}+a\right)^{\tfrac{Ml}{M+N}-1}\left(\hat{t}+\tfrac{N}{M+N}a\right)\hat{t}^{\frac{Nl}{M+N}-1}\right]
\end{eqnarray}
Since we are at large $\hat{t}$ we expand
\begin{eqnarray}
\left(\hat{t}+a\right)^{\frac{Ml}{M+N}-1}\hat{t}^{\frac{Nl}{M+N}-1}
&=&
\hat{t}^{l-2}\left(1+a\hat{t}^{-1}\right)^{\frac{Ml}{M+N}-1}
~=~
\sum_{q\geq 0}{}^{\frac{Ml}{M+N}-1}C_{q} \, a^{q} \, \hat{t}^{l-q-2}\nn
\label{binom_exp}
\end{eqnarray}
where ${}^x C_y$ is the binomial coefficient.  This gives
\begin{eqnarray}
\hat{d}_{l}'^{++}~=~
\sum_{q\geq 0}{}^{\frac{Ml}{M+N}-1}C_{q} 
\left( a^{q} \, \hat{d}_{l-q-\frac{1}{2}}^{++} 
+ \tfrac{N}{M+N} \, a^{q+1} \, \hat{d}_{l-q-\frac{3}{2}}^{++} \right) \,.
\end{eqnarray}
Similarly, for $\hat{d}'^{--}_k$ we obtain the relation
\begin{eqnarray}
\hat{d}_{k}'^{--}~=~
\sum_{p\geq 0}{}^{\frac{Mk}{M+N}}C_{p} \,
 a^{p} \, \hat{d}_{k-p+\frac{1}{2}}^{--} \,.
\end{eqnarray}
We then obtain 
\begin{eqnarray}
&&{}_{\hat{t}}\bra{0_{NS}}d'^{++}_{l}d'^{--}_{k}\ket{0_{NS}}_{\hat{t}}\cr
\cr
&&{}\qquad=~\sum_{p\geq 0}\sum_{q\geq 0}a^{p+q}\,{}^{\frac{Mk}{M+N}}C_{p}{}^{\frac{Ml}{M+N}-1}C_{q}\,\,{}_{\hat{t}}\bra{0_{NS}}\hat{d}^{++}_{l-q-\frac{1}{2}}\hat{d}^{--}_{k-p+\frac{1}{2}}\ket{0_{NS}}_{\hat{t}}\cr
&&{}\qquad\qquad+\frac{N}{M+N}\sum_{p\geq 0}\sum_{q\geq 0}a^{p+q+1}\,{}^{\frac{Mk}{M+N}}C_{p}{}^{\frac{Ml}{M+N}-1}C_{q}\,\,{}_{\hat{t}}\bra{0_{NS}}\hat{d}^{++}_{l-q-\frac{3}{2}}\hat{d}^{--}_{k-p+\frac{1}{2}}\ket{0_{NS}}_{\hat{t}} \qquad\quad\nn
\label{t_hat_amp_binom}
\end{eqnarray}
We have two terms. For the first term, utilizing commutation relations \eq{anti_comm} we obtain
\begin{eqnarray}
l-q-\frac12=-\left(k-p+\frac12\right)&\Rightarrow&p=k+l-q \,.
\end{eqnarray}
Similarly, for the second term we obtain
\begin{eqnarray}
p=k+l-q-1 \,.
\end{eqnarray}
Note that in the first term, $\hat{d}^{++}_{l-q-\frac{1}{2}}$ must be an annihilation operator, so we have nonzero contributions only from modes with
\begin{eqnarray}
q \leq l-1
\end{eqnarray}
and similarly for the second term we have nonzero contributions only from
\begin{eqnarray}
q \leq l-2 \,.
\end{eqnarray}
Then (\ref{t_hat_amp_binom}) becomes
\begin{eqnarray}
&&{\phantom{\rangle}}_{\hat{t}}\bra{0_{NS}}d'^{++}_{l}d'^{--}_{k}\ket{0_{NS}}_{\hat{t}}
~=~ \cr
&& \quad
-a^{k+l} \left[
\sum_{q=0}^{l-1}{}^{\frac{Ml}{M+N}-1}C_{q}{}^{\frac{Mk}{M+N}}C_{l+k-q} 
+\frac{N}{M+N}\sum_{q=0}^{l-2}{}^{\frac{Mk}{M+N}-1}C_{q}{}^{\frac{Ml}{M+N}}C_{l+k-q-1}
\right]\cr
&&\quad{}_{\hat{t}}\langle {} 0_{NS}|0_{NS}\rangle_{\hat{t}} \qquad\qquad\qquad 
\end{eqnarray}
which gives
\begin{eqnarray}
\gamma_{kl}^{F}&=&
-a^{k+l} \left[
\sum_{q=0}^{l-1}{}^{\frac{Ml}{M+N}-1}C_{q}{}^{\frac{Mk}{M+N}}C_{l+k-q} 
+\frac{N}{M+N}\sum_{q=0}^{l-2}{}^{\frac{Mk}{M+N}-1}C_{q}{}^{\frac{Ml}{M+N}}C_{l+k-q-1}
\right]. \qquad \nn
\end{eqnarray}
Evaluating the sums in $Mathematica$ we find 
\begin{eqnarray}
\gamma_{kl}^{F}&=&\frac{a^{k+l}
}
{\pi^{2}}\sin\left[\tfrac{N\pi k}{M+N}\right]
\sin\left[\tfrac{N\pi l}{M+N}\right]
\frac{MN}{(M+N)^{2}}\frac{k}{k+l}\frac{\Gamma\left[\frac{Mk}{M+N}\right]\Gamma\left[\frac{Nk}{M+N}\right]}{\Gamma\left[k\right]}\frac{\Gamma\left[\frac{Ml}{M+N}\right]\Gamma\left[\frac{Nl}{M+N}\right]}{\Gamma\left[l\right]} \,. \cr &&
\end{eqnarray}

\subsubsection{Writing $\gamma^{F}$ in final form}

We next write $\gamma^{F}$ in final form by making the following changes. 
\begin{enumerate}
\item[(i)] We replace the parameter $a$ by $z_{0}=e^{w_0}$, so that our result is expressed in terms of the insertion point of the twist, using \eq{az}.
\item[(ii)] We define fractional modes:
\begin{eqnarray}
s=\frac{k}{M+N},&&s'=\frac{l}{M+N} \,.
\end{eqnarray}
The parameters $s$ and $s'$ then give directly the physical wavenumbers of the modes on the cylinder with coordinate $w$.
\item[(iii)]We define the shorthand notation
\begin{eqnarray}
\mu_{s} &\equiv& 1-e^{2\pi i Ms} \,.  
\end{eqnarray}
 Note that
 \begin{eqnarray}
 \sin\left[\frac{\pi N k}{M+N}\right]
=\sin\left(\pi Ns\right)
=-\frac{i}{2}e^{i\pi Ns}\left(1-e^{-2\pi i Ns}\right)
=-\frac{i}{2}e^{i\pi Ns}\mu_{s} \,.
 \end{eqnarray}
 \end{enumerate}
In this notation, $\gamma^F$ becomes 
\begin{eqnarray}
\tilde{\gamma}_{ss'}^{F}&=&-\frac{1}{4\pi^{2}}z_{0}^{s+s'}\mu_{s}\mu_{s'}\frac{s}{s+s'}\frac{MN}{(M+N)^{2}}\left(\frac{(M+N)^{M+N}}{M^{M}N^{N}}\right)^{s+s'}\cr
&&\quad\times\frac{\Gamma\left[Ms\right]\Gamma\left[Ns\right]}{\Gamma\left[(M+N)s\right]}\frac{\Gamma\left[Ms'\right]\Gamma\left[Ns'\right]}{\Gamma\left[(M+N)s'\right]}. 
\label{gen_gamma}
\end{eqnarray}
In order to verify that in the case of $M=N=1$ this expression agrees with that computed in \cite{acm2}, note that due to different choices in normalization of the modes of the fermions, the expression $\frac{1}{2}\gamma^{F}$ in this chapter should agree with the $\gamma^{F}$ in \cite{acm2}. One can check that this is indeed the case.

\section{Computing the coefficient $C_{MN}$}

\subsection{Computing the overall prefactor $C_{MN}$} \label{sec:CMN}

We next compute the overall prefactor $C_{MN}$. Recall that from the exponential ansatz \eq{eq:expans} we have 
\bea
C_{MN}&=&\bra{0_{R,--}}
\sigma_{2}^{++}\left(w_{0}\right)
\ket{0_{R}^{--}}^{\left(1\right)}\ket{0_{R}^{--}}^{\left(2\right)} \,.
\label{eq:cmn-2}
\eea
We use the methods developed in \cite{lm1,lm2} to compute this correlator.
The full calculation is somewhat lengthy and is presented in Appendix \ref{app:cmn}; here we summarize the main steps.  
\begin{enumerate}
\item[(i)] We first lift to the $z$ plane. Since $\sigma_{2}^{++}$ has weight (1/2,1/2), we obtain the Jacobian factor contribution
\begin{eqnarray}
\left|\frac{dz}{dw}\right|_{z=z_{0}}&=&|a|^{M+N}\frac{M^{M}N^{N}}{(M+N)^{M+N}}
\end{eqnarray}
\item[(ii)] We compute the above correlator of spin-twist fields following the method of \cite{lm1,lm2} as follows. We work in a path integral formalism, and we define regularized spin-twist operators by cutting circular holes in the $ z $ plane.  
We lift to the covering space $ t $ where the fields become single-valued. 
In the $ t $ plane there is a non-trivial metric;  we take account of this by defining
a fiducial metric and computing the Liouville action. The Liouville action terms give
\begin{eqnarray}
2^{-\frac{5}{4}}
|a|^{-\frac{3}{4}\left(M+N\right)+\frac{1}{2}\left(\frac{M}{N}+\frac{N}{M}+1\right)}M^{-\frac{3}{4}M-\frac{1}{4}}N^{-\frac{3}{4}N-\frac{1}{4}}(M+N)^{\frac{3}{4}(M+N)-\frac{1}{4}} \,.
\end{eqnarray}
\item[(iii)] In the covering space, we must insert spin fields, each within an appropriate normalization to take account of the local form of the map near each insertion, $\left(z-z_{*}\right)\approx b_*\left(t-{t_{*}}\right)^{n}$ \cite{lm2}. These normalization factors give the contribution
\begin{eqnarray}
&&|b_{\infty}|^{-\frac{1}{2(M+N)}}|b_{t_{0}}|^{-\frac{1}{4}}|b_{a}|^{-\frac{1}{2N}}|b_{0}|^{-\frac{1}{2M}}=\cr
&&\qquad\qquad
2^{\frac{1}{4}}|a|^{-\frac{1}{2}\left(\frac{M}{N}+\frac{N}{M}+\frac{M}{2}+\frac{N}{2}-1\right)}M^{-\frac{1}{4}(M-1)}N^{-\frac{1}{4}(N-1)}(M+N)^{\frac{1}{4}(M+N-3)} \,. \qquad\qquad 
\end{eqnarray}
\item[(iv)] Finally, the correlator of the spin fields in the $t$ plane gives
\bea
\frac{\langle S^{+}(\infty)S^{+}(t_{0})S^{-}(a)S^{-}(0)\rangle}{\langle S^{+}(\infty)S^{-}(0)\rangle}&=&\frac{(M+N)^2}{MN} \frac{1}{|a|}\,.
\label{eq:spinfieldans1}
\end{eqnarray}
\end{enumerate}
These four results combine to give the final result
\begin{eqnarray}
C_{MN}=\frac{M+N}{2MN} \,.
\end{eqnarray}

In order to write the full expression for the effect of the twist operator 
$\sigma_{2}^{++}$ on the state $\ket{0_{R}^{--}}^{\left(1\right)} \ket{0_{R}^{--}}^{\left(2\right)}$, let us recall the expression for $\gamma^{B}$ computed in (\ref{gammapre}), 
\bea
\gamma^B_{kl}=-
{(-a)^{k+l}\over \pi^2}
\sin\left[
\tfrac{\pi  Mk}{M+N} \right]
\sin\left[
\tfrac{\pi Ml}{M+N} \right]
{MN\over (M+N)^2}
{1\over k+l}
{\Gamma[{Mk\over M+N}]
\Gamma[{Nk\over M+N}]\over \Gamma[k]}
{\Gamma[{Ml\over M+N}]
\Gamma[{Nl\over M+N}]\over \Gamma[l]} \cr &&
\eea
which since $k$ and $l$ are integers can be rewritten using 
\begin{eqnarray}
(-1)^{k}\sin\left[\frac{\pi M k}{M+N}\right]&=&-\sin\left[\frac{\pi N k}{M+N}\right],
\label{sin}
\end{eqnarray}
which gives
\bea
\gamma^B_{kl}=-
{a^{k+l}\over \pi^2}
\sin\left[
\tfrac{\pi  Nk}{M+N} \right]
\sin\left[
\tfrac{\pi Nl}{M+N} \right]
{MN\over (M+N)^2}
 \, {1\over k+l} \,
{\Gamma[{Mk\over M+N}]
\Gamma[{Nk\over M+N}]\over \Gamma[k]}
{\Gamma[{Ml\over M+N}]
\Gamma[{Nl\over M+N}]\over \Gamma[l]} . \quad
\label{eq:gammaB}
\eea
Therefore, the full effect of the twist operator 
$\sigma_{2}^{++}(w_{0},\bar{w}_{0})$ 
on the state $\ket{0_{R}^{--}}^{\left(1\right)} \ket{0_{R}^{--}}^{\left(2\right)}$ is
\begin{eqnarray}
\ket{\boldsymbol{\chi}} &\equiv& \sigma_{2}^{++}(w_{0})
|0_R^{--}\rangle^{(1)}|0_R^{--}\rangle^{(2)} 
\cr
&=&
C_{MN} \left[
\exp\left(\sum_{m\geq 1,n\geq 1}
\gamma_{mn}^{B}\left[-\alpha_{++,-m}\alpha_{--,-n}
+\alpha_{+-,-m}\alpha_{-+,-n}\right]\right)
\right. \cr
&& \qquad \qquad  \left.
\times\exp\left(\sum_{m\geq 0,n\geq 1}
\gamma_{mn}^{F}
\left[d_{-m}^{++}d_{-n}^{--}-d_{-m}^{+-}d_{-n}^{-+}
\right]\right)
\times [\mathrm{antihol.}] \right]
\ket{0_{R}^{--}} \qquad\nn
\label{eq:expans-2}
\end{eqnarray}
where
\bea
C_{MN}&=&\frac{M+N}{2MN} \cr
\gamma^B_{kl}&=&-
{a^{k+l}\over \pi^2}
\sin\left[
\tfrac{\pi  Nk}{M+N} \right]
\sin\left[
\tfrac{\pi Nl}{M+N} \right]
{MN\over (M+N)^2}
 \, {1\over k+l} \,
{\Gamma[{Mk\over M+N}]
\Gamma[{Nk\over M+N}]\over \Gamma[k]} \,
{\Gamma[{Ml\over M+N}]
\Gamma[{Nl\over M+N}]\over \Gamma[l]} \cr
\gamma_{kl}^{F}&=& \phantom{-}
\frac{a^{k+l}}{\pi^{2}}
\sin\left[
\tfrac{\pi  Nk}{M+N} \right]
\sin\left[
\tfrac{\pi Nl}{M+N} \right]
\frac{MN}{(M+N)^{2}}
 \, \frac{k}{k+l} \,
{\Gamma[{Mk\over M+N}]
\Gamma[{Nk\over M+N}]\over \Gamma[k]} \,
{\Gamma[{Ml\over M+N}]
\Gamma[{Nl\over M+N}]\over \Gamma[l]}  \,. \cr &&
\label{gammab gammaf}
\eea

\section{The continuum limit}\label{sec:cont}

While we have obtained the exact expressions for $\gamma^B_{kl}$, $\gamma^F_{kl}$, the nature of the physics implied by these expressions may not be immediately clear because the expressions look somewhat involved. We will comment on the structure of these expressions in the discussion section below. But first we note that these expressions simplify considerably in the limit where the arguments $s, s'$ are much larger than unity. We call the resulting    approximation the `continuum limit', since large mode numbers correspond to short wavelengths, and at short wavelength the physics is not sensitive to the finite length of the component string. Thus the expressions in this continuum limit describe the results obtained in the limit of  {\it infinite} component strings. Such expressions are useful for the following reason. When the D1D5 system is used to describe a black hole, then the total winding $N_1N_5$ of the component strings is very large, and so one expects the individual component strings to have large winding as well. Having long component strings ($M, N$ much larger than unity) is approximately  equivalent to holding $M,N$ fixed and taking $s,s'$ large.

\subsection{Continuum limit for the $\gamma^B_{kl}$}\label{sec:gammacont}

Let us start by looking at $\gamma^B_{kl}$. We have the exact expression
\bea
\tilde\gamma^B_{ss'}&=&{1\over 4\pi^2}\,z_0^{s+s'}\,
{\mu_s\mu_{s'}\over  s+s'}\,{MN\over (M+N)^3} 
\left ({(M+N)^{M+N}\over M^MN^N}\right )^{s+s'} \cr
&&\times{\Gamma[{Ms}]\Gamma[{Ns}]\over \Gamma[(M+N)s]}~{\Gamma[{Ms'}]\Gamma[{Ns'}]\over \Gamma[(M+N)s']}\,.\nn
\label{wnine}
\eea
We wish to find an approximation for this expression when
\be
s\gg 1, ~~~s'\gg 1
\ee
We have the basic identity, for positive integer $K$:
\be
\Gamma[x]\Gamma[x+{1\over K}]\Gamma[x+{2\over K}]\dots \Gamma[x+{K-1\over K}]=(2\pi)^{K-1\over 2}K^{\h-Kx}\Gamma[Kx]
\ee
Using this identity, we get
\be
\Gamma[{Ms}]={1\over (2\pi)^{M-1\over 2}M^{\h-M s}}
\Gamma[s]\Gamma[s+{1\over M}]\dots \Gamma[s+{M-1\over M}]
\ee
\be
\Gamma[{Ns}]={1\over (2\pi)^{N-1\over 2}N^{\h-N s}}
\Gamma[s]\Gamma[s+{1\over N}]\dots \Gamma[s+{N-1\over N}]
\ee
\be
\Gamma[(M+N)s]={1\over (2\pi)^{M+N-1\over 2}(M+N)^{\h-(M+N) s}}
\Gamma[s]\Gamma[s+{1\over M+N}]\dots \Gamma[s+{M+N-1\over M+N}]
\ee
We have, for $s\gg 1$, $x\ll s$
\be
{\Gamma[s+x]\over \Gamma[s]}\approx s^x
\ee
which gives
\be
{\Gamma[s+x]}\approx {\Gamma[s]} s^x
\ee
We use the above approximations in the expression
\be
{\Gamma[{Ms}]\Gamma[{Ns}]\over \Gamma[(M+N)s]}
\ee
There are an equal number of factors ${\Gamma[s]}$ at the numerator and denominator, so they cancel out. We can now collect the powers of $s$. In the numerator we have the power
\be
\left ( {1\over M}+{2\over M}+\dots {M-1\over M} \right ) + \left ({1\over N}+{2\over N}+\dots {N-1\over N}\right ) ={M+N-2\over 2}
\ee
In the denominator we have the power
\be
{1\over N+M}+{2\over N+M}+\dots {N+M-1\over N+M}={(M+N-1)\over 2}
\ee
Thus overall we get
\be
{\Gamma[{Ms}]\Gamma[{Ns}]\over \Gamma[(M+N)s]}\approx {(2\pi)^\h \left ({M^MN^N\over (M+N)^{M+N}}\right )^s \sqrt{M+N\over MN}}\, s^{-\h}
\label{wthree}
\ee
Similarly, we get
\be
{\Gamma[{Ms'}]\Gamma[{Ns'}]\over \Gamma[(M+N)s']}\approx {(2\pi)^\h \left ({M^MN^N\over (M+N)^{M+N}}\right )^{s'} \sqrt{M+N\over MN}}\, {s'}^{-\h}
\ee
Using these approximations in (\ref{wnine}), we find
\be
\tilde\gamma^B_{ss'}\approx{1\over 2\pi}{1\over(M+N)^2}\,z_0^{s+s'}{\mu_s\mu_{s'}\over \sqrt{ss'}}\,{1\over s+s'}\,.
\ee

\subsection{Continuum Limit for $\gamma_{kl}^{F}$}

We start by taking the continuum limit of $\gamma^{F}$. From \eq{gen_gamma} we have 
\begin{eqnarray}
\tilde{\gamma}_{ss'}^{F}=-\frac{1}{4\pi^{2}}z_{0}^{s+s'}\mu_{s}\mu_{s'}\frac{s}{s+s'}\frac{MN}{(M+N)^{2}}\left(\frac{(M+N)^{M+N}}{M^{M}N^{N}}\right)^{s+s'}\frac{\Gamma\left[Ms\right]\Gamma\left[Ns\right]}{\Gamma\left[(M+N)s\right]}\frac{\Gamma\left[Ms'\right]\Gamma\left[Ns'\right]}{\Gamma\left[(M+N)s'\right]}\cr
\label{final_gamma}
\end{eqnarray}
We wish to find the approximation to this expression when 
\begin{eqnarray}
s >> \frac{1}{M+N},\quad s>>\frac{1}{M},\quad s>>\frac{1}{N}
\end{eqnarray}
and likewise for $s'$.
We use Stirling's formula, 
\begin{eqnarray}
\Gamma[x]\sim\sqrt{\frac{2\pi}{x}}\left(\frac{x}{e}\right)^{x}
\end{eqnarray}
for $x>>1$.
Using this, we get
\begin{eqnarray}
\Gamma[Ms]&\sim&\sqrt{\frac{2\pi}{Ms}}\left(\frac{Ms}{e}\right)^{Ms} \,. 
\end{eqnarray}
For the Gamma function terms of (\ref{final_gamma}) with variable $s$, we then have
\begin{eqnarray}
\frac{\Gamma\left[Ms\right]\Gamma\left[Ns\right]}
{\Gamma\left[(M+N)s\right]}
\approx
\sqrt{\frac{M+N}{MN}}\sqrt{\frac{2\pi}{s}}\left(\frac{M^{M}N^{N}}{(M+N)^{M+N}}\right)^{s}
\label{cont_limit}
\end{eqnarray}
and likewise for the terms with variable $s'$.

Inserting these approximations in (\ref{final_gamma}), we find
\begin{eqnarray}
\tilde{\gamma}_{ss'}^{F}&\approx&-\frac{1}{2\pi}z_{0}^{s+s'}\frac{\mu_{s}\mu_{s'}}{(M+N)}\frac{1}{s+s'}\sqrt{\frac{s}{s'}}\,.
\end{eqnarray}

\section{Discussion}\label{sec:disc}

We have considered the effect of the twist operator $\sigma_2$ when it links together component strings of windings $M,N$ into a single component string of length $M+N$. In the bosonic sector we have found the final state in the case where the initial state on the component strings was the vacuum $|0\rangle^{(1)}\otimes |0\rangle^{(2)}$. In the fermion sector we have found the final state in the case where the initial state on the component strings was the vacuum $|0^-_R\rangle^{(1)}\otimes |0^-_R\rangle^{(2)}$.

While we have discussed this problem in the context of the D1D5 CFT, we note that this is a problem that could arise in other areas of physics. Each component string describes a free field theory on a circle, and the twist interaction joins these circles. In this process the vacuum state of the initial field theory goes over to a `squeezed state' of  the final theory; the coefficients $\gamma^B_{ss'},\gamma^F_{ss'}$ describe this squeezed state.

It is interesting to analyze the structure of the results that we have found. Consider the expression for $\gamma^B_{kl}$ and $\gamma^F_{kl}$  given in (\ref{gammab gammaf}). This contains a factor
\be
{1\over \Gamma[k]}{1\over \Gamma[l]}
\ee
which vanishes when either of the integers $k,l$ is zero or negative.  From (\ref{eq:expans}) we see that this implies that the $\gamma^B_{kl}$ and $\gamma^F_{kl}$ will multiply only creation operators.

A second observation is that the expression for $\gamma^B_{ss'}$ and $\gamma^F_{ss'}$  almost separates into a product of terms corresponding to $s$ and $s'$. Only the term ${1\over s+s'}$ fails to separate in this manner. The part corresponding to $s$ contains the beta function ${\Gamma[{Ms}]\Gamma[{Ns}]\over \Gamma[(M+N)s]}$
and the part for $s'$ contains the beta function ${\Gamma[{Ms'}]\Gamma[{Ns'}]\over \Gamma[(M+N)s']}$.

We have considered the vacuum    state for bosons and fermions and found the corresponding bogoluibov coefficients characterizing the final state. We should next consider bosonic and fermionic excitations in the initial state. Then we should apply the supercharge $G^-$ in (\ref{pert}) to the overall state for bosons and fermions. These steps were carried out for the case $M=N=1$ in \cite{acm2,acm3}; we compute these quantities for the case of general $M,N$ in the next chapter.

The expressions for $\gamma^B_{ss'}, \gamma^F_{ss'}$ simplify considerably in the continuum limit. This  limit  may be more useful for obtaining the qualitative dynamics of thermalization, which is expected to be dual to the process of black hole formation. It would therefore be helpful to have a way of obtaining the continuum limit expressions directly, without having to obtain the exact expressions first. We hope to return to this issue elsewhere.

In general, it is hoped that by putting together knowledge of the fuzzball construction 
(which gives the gravity description  of individual black hole microstates) and  dynamical processes in the interacting CFT (which include black hole formation), we will arrive at a deeper understanding of black hole dynamics.

\chapter{Effect of the twist operator on an initial excitation}\label{f function}

In this chapter we compute what happens when we apply the full deformation operator to the state where one of the initial component strings has an oscillator excitation. We do this in two steps. We must first understand what happens when we just apply the twist, $\s^+$ without the supercharge. We compute the state where a single twist is applied to the state where one of the initial component strings has an oscillator excitation. We then compute the action of the supercharge on this state. 
\section{Introduction}
In the case of an initial bosonic or fermionic excitation, exponential ansatz in (\ref{eq:expans}) implies \cite{acm1,cmt} that the twist operator converts an initial excitation into a linear combination of excitations above the state $\ket{\boldsymbol{\chi}}$, which we write as 
\bea
\sigma_{2}^{++}(w_{0})\alpha^{(i)}_{A\dot A,-m} |0^{--}_R\rangle^{(1)} |0^{--}_R\rangle^{(2)} ~&=&~\sum\limits_k f^{B(i)}_{mk}\,  \alpha_{A\dot A,-k}|\boldsymbol{\chi}\rangle \,, \qquad i=1,2
\label{eq:fbdef} \\
\sigma_{2}^{++}(w_{0})d^{(i)\pm A}_{-m} |0^{--}_R\rangle^{(1)} |0^{--}_R\rangle^{(2)} ~&=&~\sum\limits_k f^{F(i)\pm}_{mk}\,  d^{(i)\pm A}_{-k}|\boldsymbol{\chi}\rangle \,, \qquad i=1,2 \,.
\label{eq:ffdef}
\eea

We therefore want to compute the functions $f^{B(i)}_{mk},f^{F(i)\pm}_{mk}$. We note that in our subsequent computations we only compute the holomorphic part. The antiholomorphic computation is identical. Let us begin with $f^{B(i)}_{mk}$.

\section{Computing $f^{B(i)}_{mk}$}
In this section we compute the expressions for $f^{B(i)}_{mk}$. We denote this excitation by $\alpha^{(1)}_{A\dot A, -m}$ for the component string with winding $M$ and by $\alpha^{(2)}_{A\dot A, -m}$ for the component string with winding $N$. 
We write
\bea
&&\sigma_2(w_0)\alpha^{(1)}_{A\dot A, -m}|0\rangle^{(1)}\otimes |0\rangle^{(2)}=\nn
&&\qquad\qquad\sum_{k\ge 1}~f^{B(1)}_{mk}~\alpha_{A\dot A, -k} ~
 e^{\sum_{ k'\ge 1, l'\ge 1}\gamma^B_{k'l'}(-\alpha_{++, -k'}\alpha_{--, -l'}+\alpha_{-+, -k'}\alpha_{+-, -l'})}|0\rangle \qquad
 \label{wfour}
\eea
\bea
&&\sigma_2(w_0)\alpha^{(2)}_{A\dot A, -m}|0\rangle^{(1)}\otimes |0\rangle^{(2)}=\nn
&&\qquad\qquad\sum_{k\ge 1}~f^{B(2)}_{mk}~\alpha_{A\dot A, -k} ~
 e^{\sum_{ k'\ge 1, l'\ge 1}\gamma^B_{k'l'}(-\alpha_{++, -k'}\alpha_{--, -l'}+\alpha_{-+, -k'}\alpha_{+-, -l'})}|0\rangle \qquad
 \label{wfourq}
\eea
In this section we find the functions $f^{B(i)}_{mk}$.

\section{Method for finding the $f^{B(i)}_{mk}$}

Analogously to the computation of $\gamma^B_{kl}$, let us consider the amplitude
\bea
{\cal A}_3&=&\langle 0|\alpha_{--,_k}\, \sigma_2(w_0)\, \alpha^{(1)}_{++, -m}|0\rangle^{(1)}\otimes |0\rangle^{(2)}\nn
&=&C(w_0)\sum_{l\ge 1}~f^{B(1)}_{ml}~\langle 0|\alpha_{--,_k}\, ~\alpha_{++, -l} ~
 e^{\sum_{ k'\ge 1, l'\ge 1}\gamma^B_{k'l'}(-\alpha_{++, -k'}\alpha_{--, -l'}+\alpha_{-+, -k'}\alpha_{+-, -l'})}|0\rangle\nn
  &=& C(w_0)\sum_{l\ge 1}~f^{B(1)}_{ml} \, (-k) \delta_{k l} \nn
&=& -C(w_0)\, k\, f^{B(1)}_{mk}\,.
\eea
In the second step above, we note that there is also a contribution when $\alpha_{--,_k}$ contracts with the terms in the exponential, but this contribution  consists of $\alpha_{++, -k'}$ with $k'>0$. Such oscillators annihilate the vacuum $\langle 0|$, and so this contribution in fact vanishes. 

Thus we see that
\be
f^{B(1)}_{mk}=-{1\over k} {{\cal A}_3\over {\cal A}_1}\,.
\ee
We have
\be
{\cal A}_3=\langle 0|\alpha_{--,_k}\sigma_2(w_0)\alpha^{(1)}_{++, -m}|0\rangle^{(1)}\otimes |0\rangle^{(2)}=D(z_0)~{}_t\langle 0| \alpha'_{--,_k}\alpha'^{(1)}_{++, -m}|0\rangle_t
\ee
Thus we obtain
\be
f^{B(1)}_{mk}=-{1\over k} {{\cal A}_3\over {\cal A}_1}=-{1\over k}~{{}_t\langle 0| \alpha'_{--,_k}\alpha'^{(1)}_{++, -m}|0\rangle_t\over {}_t\langle 0 |0\rangle_t}\,.
\label{wsix}
\ee

\subsection{Computing  $f^{B(1)}_{mk}$}

Let us now carry out the details of the  computation we outlined above. 

The operator $\alpha'_{--,k}$ is applied at $w=\infty$, and is thus given in the $t$ plane by a contour at large $t$. The operator $\alpha'^{(1)}_{++, -m}$ on the other hand is applied at $w=-\infty$ to the component string with winding $M$, and is thus given in the $t$ plane by a contour around $t=0$. Thus we get
\bea
&&{}_t\langle 0|\alpha'_{--,_k}\alpha'^{(1)}_{++, -m}|0\rangle_t=
{}_t\langle 0 |\Big ({1\over 2\pi }\int\limits_{t_1=\infty} dt_1 \p_t X_{--}(t_1) \left ( t_1^M(t_1-a)^N \right ) ^{k\over M+N}\Big )
\nn
&&\qquad\qquad \qquad\qquad\qquad\quad\times ~~
\Big(
{1\over 2\pi }\int\limits_{t_2=0} dt_2 \p_t X_{++}(t_2) \left ( t_2^M(t_2-a)^N \right ) ^{-{m\over M}}\Big )
|0\rangle_t  \;. \qquad\quad
\label{ptfiveq}
\eea
Since $t_1$ is large, we expand as
\be
\left ( t_1^M(t_1-a)^N \right ) ^{k\over M+N}=\sum_{p'\ge 0} {}^{Nk\over M+N} C_{p'} (-a)^{p'} t_1^{k-p'}
\ee
Thus we find
\be
\alpha'_{A\dot A, k}=\sum_{p'\ge 0} {}^{Nk\over M+N} C_{p'} (-a)^{p'}\t\alpha_{A\dot A, k-p'}\;.
\ee
Next, since $t_2\approx 0$  we  expand as
\bea
\left ( t_2^M(t_2-a)^N \right ) ^{-{m\over M}}&=&t_2^{-m}(t_2-a)^{-{mN\over M}}=t_2^{-m}(-a)^{-{mN\over M}}(1-{t_2\over a})^{-{mN\over M}}\nn
&=&t_2^{-m}(-a)^{-{mN\over M}}\sum_{p\ge 0} {}^{-{mN\over M}}C_p\, (-a)^{-p}t_2^{p}=\sum_{p\ge 0} {}^{-{mN\over M}}C_p\, (-a)^{-{mN\over M}-p}t_2^{p-m}\nn
\eea
Thus we find
\be
\alpha'^{(1)}_{A\dot A, -m}= \sum_{p\ge 0} {}^{-{mN\over M}}C_p\, (-a)^{-{mN\over M}-p}~ \t\alpha_{A\dot A, p-m}
\ee
Since $\tilde\alpha_{A\dot A, p-m}|0\rangle_t=0$ for $p-m\geq 0$, we get a contribution only from  $p<m$ in the above sum. Thus we have
\be
\alpha'^{(1)}_{A\dot A, -m}= \sum_{p= 0}^{m-1} {}^{-{mN\over M}}C_p\, (-a)^{-{mN\over M}-p}~ \tilde\alpha_{A\dot A, p-m}
\label{wone}
\ee
Using these expansions, we obtain
\bea
&&{}_t\langle 0 |\alpha'_{--,_k}\alpha'^{(1)}_{++, -m}|0\rangle_t\nn
&&=\sum_{p= 0}^{m-1} \sum_{p'\ge 0}~{}^{-{mN\over M}}C_p\,{}^{Nk\over M+N} C_{p'}~ (-a)^{-{mN\over M}-p+p'}~{}_t\langle 0 |\, \t\alpha_{--, k-p'}\, \t\alpha_{++, p-m}\,  |0\rangle_t\nn
&&=\sum_{p= 0}^{m-1} \sum_{p'\ge 0}~{}^{-{mN\over M}}C_p\,{}^{Nk\over M+N} C_{p'}~ (-a)^{-{mN\over M}-p+p'}~(p'-k)\delta_{k-p'+p-m,0}~{}_t\langle 0|0\rangle_t\nn
&&=(-a)^{k-{(M+N)m\over M}}\sum_{p= {\max(m-k, 0)}}^{m-1}~{}^{-{mN\over M}}C_p\,{}^{Nk\over M+N} C_{k+p-m} (p-m)~{}_t\langle 0|0\rangle_t\nn
&&=-{(-1)^m k\sin(\pi {Mk\over M+N})\over \pi (M+N)}\;
{(-a)^{k-{(M+N)m\over M}}\over {k\over M+N}-{m\over M}}\;
{\Gamma[{(M+N) m\over M}]\over \Gamma[m]\Gamma[{Nm\over M}]}\,
{\Gamma[{Nk\over M+N}]\Gamma[{Mk\over M+N}]\over \Gamma[k]}
~{}_t\langle 0|0\rangle_t\nn
\label{wseven}
\eea
and using (\ref{wsix}), we obtain
\be
f^{B(1)}_{mk}~=~{(-1)^m \sin(\pi {Mk\over M+N})\over \pi (M+N)}\;
{(-a)^{k-{(M+N)m\over M}}\over {k\over M+N}-{m\over M}}\;
{\Gamma[{(M+N) m\over M}]\over \Gamma[m]\Gamma[{Nm\over M}]}\,
{\Gamma[{Nk\over M+N}]\Gamma[{Mk\over M+N}]\over \Gamma[k]}
\;.
\label{fpre}
\ee
Note that in the above expression, when we have 
\be
{k\over M+N}={m\over M} \,,
\ee
both numerator and denominator vanish, since
\be
\sin(\pi {Mk\over M+N}) ~=~ \sin(\pi {m}) ~=~ 0 \,.
\ee
Since the above expression for $f^{B(1)}_{mk}$ is indeterminate in this situation, we return to the sum in the third  line of (\ref{wseven}), and take parameter values 
 \be
 m=M c, ~~~k=(M+N) c \,.
 \ee
 Here $c={j\over Y}$, where $j$ is a positive integer and $Y=\gcd(M,N)$. Then we have
 \bea
 {}_t\langle 0 |\alpha'_{--,_k}\alpha'^{(1)}_{++, -m}|0\rangle_t&=&(-a)^{-{(M+N)m\over M}+k}\sum_{p= {\max(m-k, 0)}}^{m-1}~{}^{-{mN\over M}}C_p\,{}^{Nk\over M+N} C_{k+p-m} (p-m)\nn
 &=&\sum_{p=0}^{m-1}~{}^{-{Nc}}C_p\,{}^{Nc} C_{Nc+p} (p-Mc)\,.
 \eea
 We note that since $Nc$ is a positive integer,
 \be
 {}^{Nc} C_{Nc+p}=0, ~~p>0
 \ee
 Thus only the $p=0$ term survives in the above sum, and we get
 \be
 {}_t\langle 0 |\alpha'_{--,_k}\alpha'^{(1)}_{++, -m}|0\rangle_t=-Mc
 \ee
 Thus
 \be
 f^{B(1)}_{mk}={Mc\over k}={Mc\over (M+N)c}={M\over M+N} \,.
 \ee
So for this special case, we write
\be \label{fpre2}
f^{B(1)}_{mk}|_{{k\over M+N}={m\over M}}~=~{M\over M+N} \,.
\ee

Using the gamma function identity (\ref{gammahalf}), one can check that the results (\ref{fpre}), (\ref{fpre2}) agree with the corresponding expression obtained in \cite{acm3} for the case $M=N=1$. 
 
\subsection{Expressing $f^{B(1)}_{mk}$ in final form}

We make the same changes of notation that we did for the $\gamma^B_{lk}$\;:

(i) We replace $a$ by $z_0$.

(ii) We use fractional modes
\be
q={m\over M}, ~~~s={k\over M+N} \,.
\ee
 When expressing our result using $q$ and $s$ in place of $m$ and $k$, we will write
 \be
 f^{B(1)}_{mk}\rightarrow \tilde f^{B(1)}_{qs}
 \ee

(iii) We again use the shorthand $ \mu_s=(1-e^{2\pi i Ms})$. In doing this we note that
 \be
(-1)^m\sin(\pi  {Mk\over M+N})= (-1)^{Mq}\sin(\pi M s)={i\over 2}e^{-i\pi M(s-q)}(1-e^{2\pi i Ms}) \,.
 \ee
With these changes of notation we get, for $s \ne q$,
 \bea
\tilde f^{B(1)}_{qs}&=&
{i\over 2\pi}
z_0^{s-q}
{\mu_s\over s-q}\,
{1\over (M+N)}\left( {(M+N)^{M+N}\over M^MN^N}\right)^{(s-q)}
{\Gamma[(M+N)q]\over \Gamma[Mq]\Gamma[Nq]}\,
 {\Gamma[Ms]\Gamma[Ns]\over \Gamma[(M+N)s]} \nn
 \label{exactft}
 \eea
 and for $s =q$ we have
 \be
 \tilde f^{B(1)}_{qs}|_{q=s}={M\over M+N}\,.
 \ee

\subsection{Computing  $f^{B(2)}_{mk}$}

The expression for $f^{B(2)}_{mk}$ can be obtained by a similar computation. There are only a few changes, which we mention here.
In place of (\ref{ptfiveq}) we have
\bea
&&{}_t\langle 0|\alpha'_{--,_k}\alpha'^{(2)}_{++, -m}|0\rangle_t=
{}_t\langle 0 |\Big ({1\over 2\pi }\int\limits_{t_1=\infty} dt_1 \p_t X_{--}(t_1) \left ( t_1^M(t_1-a)^N \right ) ^{k\over M+N}\Big )
\nn
&&\qquad\qquad \qquad\qquad\qquad\quad\times ~~
\Big(
{1\over 2\pi }\int\limits_{t_2=a} dt_2 \p_t X_{++}(t_2) \left ( t_2^M(t_2-a)^N \right ) ^{-{m\over N}}\Big )
|0\rangle_t\qquad\quad
\label{ptfiveqq}
\eea
The power ${m\over M}$ has been replaced by ${m\over N}$, and the $t_2$ contour is now around $t_2=a$ instead of around $t_2=0$. We define a shifted coordinate in the $t$ plane
\be
t'=t-a
\ee
which gives
\bea
&&{}_t\langle 0|\alpha'_{--,_k}\alpha'^{(2)}_{++, -m}|0\rangle_t=
{}_t\langle 0 |\Big ({1\over 2\pi }\int\limits_{t'_1=\infty} dt'_1 \p_t X_{--}(t'_1) \left ( {t'_1}^N(t'_1+a)^M \right ) ^{k\over M+N}\Big )
\nn
&&\qquad\qquad \qquad\qquad\qquad\quad\times ~~
\Big(
{1\over 2\pi }\int\limits_{t'_2=0} dt'_2 \p_t X_{++}(t'_2) \left ( {t'_2}^N(t'_2+a)^M \right ) ^{-{m\over N}}\Big )
|0\rangle_t \;.\qquad\qquad
\label{ptfiveqq2}
\eea
This expression is the same as (\ref{ptfiveq}) with the replacements
\be
M\to N ~~~N\to M~~~ a\to -a \,.
\ee
Thus $f^{B(2)}_{mk}$ can be obtained from (\ref{exactft}) with the above replacements.  For ${k \over M+N} \neq {m\over N}$ this gives:
\be
f^{B(2)}_{mk}~=~
{(-1)^m \sin(\pi {Nk\over M+N})\over \pi (M+N)}\;
{a^{k-{(M+N)m\over N}}\over {k\over M+N}-{m\over N}}\;
{\Gamma[{(M+N) m\over N}]\over \Gamma[m]\Gamma[{Mm\over N}]}\,
{\Gamma[{Nk\over M+N}]\Gamma[{Mk\over M+N}]\over \Gamma[k]}
\label{fpreq}
\ee
and we have the special case
\be
f^{B(2)}_{mk}|_{{k\over M+N}={m\over N}}~=~{N\over M+N} \,.
\ee
We can again use fractional modes
\be
r={m\over N}, ~~~s={k\over M+N}
\ee
 When expressing our result using $r$ and $s$ in place of $m$ and $k$, we will write
 \be
 f^{B(2)}_{mk}\rightarrow \tilde f^{B(2)}_{rs}\,.
 \ee
We then have for $r \neq s$:
 \bea
\tilde f^{B(2)}_{rs}&=&
-{i\over 2\pi}
z_0^{s-r}
{\mu_s\over s-r}\,
{1\over  (M+N)}\left( {(M+N)^{M+N}\over M^MN^N}\right)^{(s-r)}
{\Gamma[(M+N)r]\over \Gamma[Mr]\Gamma[Nr]}\,
 {\Gamma[Ms]\Gamma[Ns]\over \Gamma[(M+N)s]}\nn
 \label{exactf2}
\eea
while for $r = s$:
\be
\tilde f^{B(2)}_{rs}|_{r=s} = {N \over M+N}\,.
\ee

\subsection{ Summarizing the result for $\tilde f^{B(1)}_{qs}$, $\tilde f^{B(2)}_{rs}$}
 
For convenient reference, we record the result for $\tilde f^{B(1)}_{qs}$, $\tilde f^{B(2)}_{rs}$ for general $M,N$:
\bea
\tilde f^{B(1)}_{qs} & = & \begin{cases}
{M \over M+N} & q = s \\
{i\over 2\pi}z_0^{s-q}
{\mu_s \over s-q}\,
{1\over  (M+N)}\left( {(M+N)^{M+N}\over M^MN^N}\right)^{(s-q)}
{\Gamma[(M+N)q]\over \Gamma[Mq]\Gamma[Nq]}\,
{\Gamma[Ms]\Gamma[Ns]\over \Gamma[(M+N)s]}
\qquad\quad & q \neq s 
\end{cases}\qquad\quad
\eea

\bea
\tilde f^{B(2)}_{rs} & = & \begin{cases}
{N \over M+N} & r = s \\
-{i\over 2\pi}z_0^{s-r}
{\mu_s\over s-r}\,
{1\over  (M+N)}\left( {(M+N)^{M+N}\over M^MN^N}\right)^{(s-r)}
{\Gamma[(M+N)r]\over \Gamma[Mr]\Gamma[Nr]}\,
 {\Gamma[Ms]\Gamma[Ns]\over \Gamma[(M+N)s]} \qquad\quad & r \neq s
\end{cases}\qquad\quad
\eea

\subsection{Continuum limit for the $f^{B(i)}_{qs}$}\label{sec:fcont}

Let us also obtain the continuum limit for the $f^{B(i)}_{qs}$. 
We have the exact expression
 \bea
\tilde f^{B(1)}_{qs}&=&
{i\over 2\pi}
z_0^{s-q}
{\mu_s\over s-q}\,
{1\over (M+N)}\left( {(M+N)^{M+N}\over M^MN^N}\right)^{(s-q)}
{\Gamma[(M+N)q]\over \Gamma[Mq]\Gamma[Nq]}\,
 {\Gamma[Ms]\Gamma[Ns]\over \Gamma[(M+N)s]} \,. \nn
 \label{exactftq}
 \eea
 We have, from (\ref{wthree})
 \be
{\Gamma[{Ms}]\Gamma[{Ns}]\over \Gamma[(M+N)s]}\approx {(2\pi)^\h \left ({M^MN^N\over (M+N)^{M+N}}\right )^s \sqrt{M+N\over MN}}\, s^{-\h}
\ee
Similarly,
\be
{\Gamma[(M+N)q]\over \Gamma[{Mq}] \Gamma[{Nq}]}\approx {(2\pi)^{-\h} \left ({M^MN^N\over (M+N)^{M+N}}\right )^{-q} \sqrt{MN\over M+N}}\, q^{\h}
\ee
Thus we obtain
\be
\tilde f^{B(1)}_{qs}\approx {i\over 2\pi}{1\over (M+N)}z_0^{s-q}
\mu_s\sqrt{q\over s}\,{1\over s-q}\,.
\ee
Similarly, for $\tilde f^{B(2)}_{rs}$ we obtain
\be
\tilde f^{B(2)}_{rs}\approx -{i\over 2\pi}{1\over (M+N)}z_0^{s-r}
\mu_s\sqrt{r\over s}\,{1\over s-r}\,.
\ee
Next we compute the fermion, $f^{F}$'s,

\section{Fermion {$f^F$}'s} \label{sec:f}
We now consider the case where one of the initial component strings has an initial fermionic oscillator excitation. We thus compute the $f$ functions for a fermion in the initial state, $f^{F,\pm}$. From \eq{eq:ffdef} we recall the definition
\bea
\sigma_{2}^{++}(w_{0})d^{(i)\pm A}_{-m} |0^{--}_R\rangle^{(1)} |0^{--}_R\rangle^{(2)} ~&=&~\sum\limits_k f^{F(i)\pm}_{mk}\,  d^{(i)\pm A}_{-k}|\boldsymbol{\chi}\rangle \,, \qquad i=1,2 \,.
\label{eq:ffdef-2}
\eea
In this section we compute $f_{mk}^{F(i)\pm}$.

\subsection{Computing $f_{mk}^{F(1)+}$}

Let us start with $f_{mk}^{F(1)+}$. Consider the ratio of amplitudes
\begin{eqnarray}
\frac{\mathcal{A}_{3}}{\mathcal{A}_{1}}&=&\frac{
\bra{0_{R,--}}d_{k}^{--}\sigma_{2}^{++}\left(w_{0}\right)d_{-m}^{(1)++}\ket{0_{R}^{--}}^{(1)} \ket{0_{R}^{--}}^{(2)}
}
{
\bra{0_{R,--}}\sigma_{2}^{++}\left(w_{0}\right)\ket{0_{R}^{--}}^{\left(1\right)}\ket{0_{R}^{--}}^{\left(2\right)}} \,.
\label{amp-3}
\end{eqnarray}
From \eq{eq:ffdef-2} we observe that
\begin{eqnarray}
\mathcal{A}_{3}&=&\bra{0_{R,--}}d_{k}^{--}\sigma_{2}^{++}\left(w_{0}\right)d_{-m}^{(1)++}\ket{0_{R}^{--}}^{(1)} \ket{0_{R}^{--}}^{(2)}\cr
\cr
&=&\sum_{l\geq 1}f_{ml}^{F(1)+}\bra{0_{R,--}}d_{k}^{--}d_{-l}^{++}\ket{\boldsymbol{\chi}}\cr
&=&-C_{MN}f_{mk}^{F(1)+}
\end{eqnarray}
So we have
\begin{eqnarray}
f_{mk}^{F(1)+}=-\frac{\mathcal{A}_{3}}{\mathcal{A}_{1}}=-\frac{{}_{t}\bra{0_{NS}}d'^{--}_{k}d'^{(1)++}_{-m}\ket{0_{NS}}_{t}}{{}_{t}\langle 0_{NS}|0_{NS}\rangle_{t}}
\label{f_amp}
\end{eqnarray}
where on the RHS, the t-plane amplitude is after all spectral flows and coordinate shifts have been carried out for the copy 1 quantities $f^{F(1)\pm}$. For this calculation, we use the $t$ coordinate in which copy 1 is at the origin. When we calculate the copy 2 quantities $f^{F(2)\pm}$ we will use the $\hat{t}$ coordinate. 

We now compute the empty $t$ plane amplitude
\begin{eqnarray}
-\mathcal{A}_{3}&=&-{}_{t}\bra{0_{NS}}d'^{--}_{k}d'^{(1)++}_{-m}\ket{0_{NS}}_{t}
\end{eqnarray}
using the mode expansions in (\ref{d_t(i)_modes}) and (\ref{d_t_modes}). Let us first expand $d'^{--}_{k}$ around $t=\infty$. From (\ref{d_t_modes}) we have
\begin{eqnarray}
d'^{--}_{k}&=&\frac{1}{2\pi i}\oint\limits_{t=\infty}dt \,\psi^{--}(t)\left[t^{\frac{Mk}{M+N}}\left(t-a\right)^{\frac{Nk}{M+N}}\right].
\label{d_prime_f1}
\end{eqnarray}
We therefore expand
\begin{eqnarray}
t^{\frac{Mk}{M+N}}\left(t-a\right)^{\frac{Nk}{M+N}}&=&t^{k}\left(1-at^{-1}\right)^{\frac{Nk}{M+N}}=\sum_{p\geq 0}{}^{\frac{Nk}{M+N}}C_{p}(-a)^{p}t^{k-p} \,.
\end{eqnarray}
For our purposes, the sum is truncated by the requirement that $\tilde{d}_{k-p+\frac{1}{2}}^{--}$ be an annihilation operator.
So in terms of modes natural to the $t$-plane we find
\begin{eqnarray}
d'^{--}_{k}&\to&\sum_{p=0}^{k}{}^{\frac{Nk}{M+N}}C_{p}(-a)^{p}\tilde{d}^{--}_{k-p+\frac{1}{2}} \,.
\end{eqnarray}
Next, using (\ref{d_t(i)_modes}), we expand
\begin{eqnarray}
d'^{(1)++}_{-m}=\frac{\sqrt{M+N}}{2\pi i\sqrt{M}}\oint\limits_{t=0}dt\,\psi^{(1)++}(t)\left[t^{-m-1}\left(t-\tfrac{Ma}{M+N}\right)\left(t-a\right)^{-\frac{Nm}{M}-1}\right]
\label{d1_prime_f1}
\end{eqnarray}
We find
\begin{eqnarray}
&&t^{-m-1}\left(t-\tfrac{M}{M+N}a\right)\left(t-a\right)^{-\frac{Nm}{M}-1}\cr
&&\,\,\,\,\,\,\,\,\,\,\,\,\,\,\,=(-a)^{-\frac{Nm}{M}-1}\left(t^{-m}-\tfrac{M}{M+N}at^{-m-1}\right)\sum_{p'\geq 0}{}^{-\frac{Nm}{M}-1}C_{p'}(-a)^{-p'}t^{p'}\cr
&&\,\,\,\,\,\,\,\,\,\,\,\,\,\,\,=(-a)^{-\frac{Nm}{M}-1}\left[\sum_{p'\geq 0}{}^{-\frac{Nm}{M}-1}C_{p'}(-a)^{-p'}t^{p'-m}+\sum_{p'\geq 0}{}^{-\frac{Nm}{M}-1}C_{p'}\tfrac{M}{M+N}(-a)^{1-p'}t^{p'-m-1}\right]\nn
\end{eqnarray}
This gives
\begin{eqnarray}
d'^{(1)++}_{-m}&\to&\frac{\sqrt{M+N}}{\sqrt{M}}\sum_{p'=0}^{m-1}{}^{-\frac{Nm}{M}-1}C_{p'}(-a)^{-\frac{Nm}{M}-p'-1}\tilde{d}^{++}_{p'-m+\frac{1}{2}}\cr
&& {} + \frac{\sqrt{M}}{\sqrt{M+N}}\sum_{p'=0}^{m}{}^{-\frac{Nm}{M}-1}C_{p'}(-a)^{-\frac{Nm}{M}-p'}\tilde{d}^{++}_{p'-m-\frac{1}{2}}
\label{exp}
\end{eqnarray}
where again the upper limits on the sums are determined by the requirement that the operators on the RHS be creation operators. We now use the mode expansions to compute 
\begin{eqnarray}
-{}_{t}\bra{0_{NS}}d'^{++}_{k}d'^{(1)--}_{-m}\ket{0_{NS}}_{t}
\end{eqnarray} 
Since the expansion of $d'^{(1)++}_{-m}$ involves two separate terms, each involving a sum, we will separately calculate the contributions from these terms and add the resulting expressions to find the above amplitude. The contribution to the amplitude from the first term in the expansion of $d'^{(1)++}_{m}$ in (\ref{exp}) is
\begin{eqnarray}
-\sum_{p=0}^{k}\sum_{p'=0}^{m-1}\frac{\sqrt{M+N}}{\sqrt{M}}{}^{\frac{Nk}{M+N}}C_{p}{}^{-\frac{Nm}{M}-1}C_{p'}(-a)^{-\frac{Nm}{M}+(p-p')-1}{}_{t}\bra{0_{NS}}\tilde{d}^{--}_{k-p+\frac{1}{2}}\tilde{d}^{++}_{p'-m+\frac{1}{2}}\ket{0_{NS}}_{t}
\end{eqnarray}
where we have used the modes natural to the $t$ plane given in (\ref{nat_t_modes}). Using the anticommutation relations in (\ref{anti_comm}) we have the following constraints on the mode numbers
\begin{eqnarray}
k-p+\frac{1}{2}&=&-\left(p'-m+\frac{1}{2} \right)
\end{eqnarray}
This gives
\begin{eqnarray}
&&\frac{\sqrt{M+N}}{\sqrt{M}}(-a)^{k-\frac{m\left(M+N\right)}{M}}\sum_{p'=\max(m-k-1,0)}^{m-1}{}^{\frac{Nk}{M+N}}C_{k-m+p'+1}{}^{{}^{-\frac{Nm}{M}-1}}C_{p'}\,{}_{t}\langle 0_{NS}|0_{NS}\rangle_{t}\quad\nn
\label{1st_term}
\end{eqnarray}
For the second term we find 
\begin{eqnarray}
-\frac{\sqrt{M}}{\sqrt{M+N}}\sum_{p=0}^{k}\sum_{p'=0}^{m}{}^{\frac{Nk}{M+N}}C_{p}{}^{-\frac{Nm}{M}-1}C_{p'}(-a)^{-\frac{Nm}{M}+(p-p')}{}_{t}\bra{0_{NS}}\tilde{d}^{--}_{k-p+\frac{1}{2}}\tilde{d}^{++}_{p'-m-\frac{1}{2}}\ket{0_{NS}}_{t} \,\nn
\end{eqnarray}
which upon using the commutation relations becomes
\begin{eqnarray}
\frac{\sqrt{M}}{\sqrt{M+N}}(-a)^{k-\frac{m\left(M+N\right)}{M}}\sum_{p'=\max(m-k,0)}^{m}{}^{\frac{Nk}{M+N}}C_{k-m+p'}{}^{-\frac{Nm}{M}-1}C_{p'}\;{}_{t}\langle 0_{NS}|0_{NS}\rangle_{t} \,.
\label{2nd_term}
\end{eqnarray}
Adding together (\ref{1st_term}) and (\ref{2nd_term}),  and using (\ref{f_amp}), we find
\begin{eqnarray}
f_{mk}^{F(1)+}
&=&\frac{\sqrt{M+N}}{\sqrt{M}}(-a)^{k-\frac{m(M+N)}{M}}\sum_{p'=\max(m-k-1,0)}^{m-1}{}^{\frac{Nk}{M+N}}C_{k-m+p'+ 1}{}^{{}^{-\frac{Nm}{M}-1}}C_{p'}\cr
&&{}+\frac{\sqrt{M}}{\sqrt{M+N}}(-a)^{k-\frac{m\left(M+N\right)}{M}}\sum_{p'=\text{max}(m-k,0)}^{m}{}^{\frac{Nk}{M+N}}C_{k-m+p'}{}^{-\frac{Nm}{M}-1}C_{p'}\nn
\label{f_sum}
\end{eqnarray}
Evaluating the sums, we obtain
\begin{eqnarray}
&&\!\!\!\!f_{mk}^{F(1)+}\cr
&&=~\frac{(-1)^{m}\sqrt{M}\sin\left(\pi\frac{Mk}{M+N}\right)}{\pi\left(M+N\right)^{\frac{3}{2}}}\frac{(-a)^{k-\frac{m(M+N)}{M}}}{\frac{k}{M+N}-\frac{m}{M}}\frac{k}{m}\frac{\Gamma\left[\frac{Mk}{M+N}\right]\Gamma\left[\frac{Nk}{M+N}\right]}{\Gamma\left[k\right]}\frac{\Gamma\left[\frac{(M+N)m}{M}\right]}{\Gamma\left[m\right]\Gamma\left[\frac{Nm}{M}\right]}\nn
\,. \quad
\label{f_F1+}
\end{eqnarray}
Note that in the above expression, when we have
\begin{eqnarray}
\frac{k}{M+N}=\frac{m}{M}
\end{eqnarray}
both numerator and denominator vanish, since
\begin{eqnarray}
\sin\left(\pi\tfrac{Mk}{M+N}\right)=\sin\left(\pi m\right)=0 \,.
\end{eqnarray}
Since the above expression for $f_{mk}^{F(1)+}$ is indeterminate in this situation, we return to the sum in (\ref{f_sum}), and take parameter values
\begin{eqnarray}
m~=~Mc \,,\qquad k&=&(M+N)c \,, \qquad c~=~\frac{j}{Y}
\end{eqnarray} 
where $j$ is a positive integer and $Y=\text{gcd}(M,N)$. Since $m<k$, we have 
\begin{eqnarray}
f_{mk}^{F(1)+}&=&\frac{\sqrt{M+N}}{\sqrt{M}}(-a)^{k-\frac{m(M+N)}{M}}\sum_{p'= 0}^{m-1}{}^{\frac{Nk}{M+N}}C_{k-m+p'+1}{}^{{}^{-\frac{Nm}{M}-1}}C_{p'}\cr
&& {}+\frac{\sqrt{M}}{\sqrt{M+N}}(-a)^{k-\frac{m\left(M+N\right)}{M}}\sum_{p'=0}^{m}{}^{\frac{Nk}{M+N}}C_{k-m+p'}{}^{-\frac{Nm}{M}-1}C_{p'}\cr
&=&\frac{\sqrt{M+N}}{\sqrt{M}}\sum_{p'= 0}^{Mc-1}{}^{Nc}C_{Nc+p'+1}{}^{-Nc-1}C_{p'}
+\frac{\sqrt{M}}{\sqrt{M+N}}\sum_{p'=0}^{Mc}{}^{Nc}C_{Nc+p'}{}^{-Nc-1}C_{p'}\cr
&=&\frac{\sqrt{M}}{\sqrt{M+N}} \,.
\label{s=q}
\end{eqnarray}

\subsection{Computing $f_{mk}^{F(1)-}$}

Next we compute $f^{F(1)-}$. The method is entirely analogous; this time, we take the amplitude in the numerator to be 
\begin{eqnarray}
\mathcal{A}_4&=&\bra{0_{R,--}}d_{k}^{++}\sigma_{2}^{+}\left(w_{0}\right)d'^{(1)--}_{-m}\ket{0_{R}^{--}}^{(1)}\ket{0_{R}^{--}}^{(2)}\cr
&=&-C_{MN}f_{mk}^{F(1)-} \,,
\end{eqnarray}
so we have 
\begin{eqnarray}
f_{mk}^{F(1)-}=-\frac{\mathcal{A}_{4}}{\mathcal{A}_{1}}=-\frac{{}_{t}\bra{0_{NS}}d'^{++}_{k}d'^{(1)--}_{-m}\ket{0_{NS}}_t}{{}_t\langle 0_{NS}|0_{NS}\rangle_{t}} \,
\label{f-amp}
\end{eqnarray}
where again on the RHS the amplitudes are in the empty $t$ plane after all spectral flows. 

Expanding the modes as before, we obtain
\begin{eqnarray}\label{f-sum}
f_{mk}^{F(1)-}&=&\frac{1}{\sqrt{M(M+N)}}(-a)^{k-\frac{(M+N)m}{M}}\\
&&\times\left[(M+N)\sum_{p'=\text{max}(m-k,0)}^{m-1}{}^{\frac{Nk}{M+N}-1}C_{k-m+p'}{}^{-\frac{Nm}{M}}C_{p'} \right. \cr
&&{}\left.
+M\sum_{p'=\text{max}(m-k+1,0)}^{m-1}{}^{\frac{Nk}{M+N}-1}C_{k-m+p'-1}{}^{-\frac{Nm}{M}}C_{p'}\right] . \nonumber 
\end{eqnarray}
Evaluating the sums, we find
\begin{eqnarray}
f_{mk}^{F(1)-}=\frac{(-1)^{m}\sin\left(\pi\frac{Mk}{M+N}\right)}{\pi\sqrt{M\left(M+N\right)}}\frac{(-a)^{k-\frac{m(M+N)}{M}}}{\frac{k}{M+N}-\frac{m}{M}}\frac{\Gamma\left[\frac{Mk}{M+N}\right]\Gamma\left[\frac{Nk}{M+N}\right]}{\Gamma\left[k\right]}\frac{\Gamma\left[\frac{(M+N)m}{M}\right]}{\Gamma\left[m\right]\Gamma\left[\frac{Nm}{M}\right]} .
\label{final_f1_-}
\end{eqnarray}
As before, for the special case of
\begin{eqnarray}
\frac{k}{M+N}=\frac{m}{M}
\end{eqnarray} we have an indeterminate expression for $f_{mk}^{F(1)-}$ and using (\ref{f-sum}) we find 
\begin{eqnarray}
f_{mk}^{F(1)-}&=&\frac{1}{\sqrt{M(M+N)}}\cr
&&\times\left[(M+N)\sum_{p'=0}^{Mc-1}{}^{Nc-1}C_{Nc+p'}{}^{-Nc}C_{p'}+M\sum_{p'=0}^{Mc-1}{}^{Nc-1}C_{Nc+p'-1}{}^{-Nc}C_{p'}\right]\cr
&=&\frac{\sqrt{M}}{\sqrt{M+N}} \,.
\label{s=q_2-}
\end{eqnarray}

\subsection{Computing $f_{mk}^{F(2)+}$ and $f_{mk}^{F(2)-}$}
For an initial excitation on copy 2, we calculate $f_{mk}^{F(2)+}$. Similar to the calculation of $f_{mk}^{F(1)+}$, we consider the amplitude
\begin{eqnarray}
\mathcal{A}_{5}&=&\bra{0_{R,--}}d_{k}^{--}\sigma_{2}^{++}\left(w_{0}\right)d_{-m}^{(2)++}\ket{0_{R}^{--}}^{(1)} \ket{0_{R}^{--}}^{(2)}\cr
\cr
&=&\sum_{l\geq 1}f_{ml}^{F(2)+}\bra{0_{R,--}}d_{k}^{--}d_{-l}^{++}\ket{\boldsymbol{\chi}}\cr
&=&-C_{MN}f_{mk}^{F(2)+}
\end{eqnarray}
So we have
\begin{eqnarray}
f_{mk}^{F(2)+}=-\frac{\mathcal{A}_{4}}{\mathcal{A}_{1}}=-\frac{{}_{t}\bra{0_{NS}}d'^{--}_{k}d'^{(2)++}_{-m}\ket{0_{NS}}_{t}}{{}_{t}\langle 0_{NS}|0_{NS}\rangle_{t}} \,.
\label{f_amp_2}
\end{eqnarray}
Let us now compute the empty $\hat{t}$ plane amplitude
\begin{eqnarray}
&&{}_{\hat{t}}\bra{0_{NS}}d'^{--}_{k}d'^{(2)++}_{-m}\ket{0_{NS}}_{\hat{t}}\cr
&&\quad={}_{\hat{t}}\bra{0_{NS}}\frac{1}{2\pi i}\oint\limits_{\hat{t}=\infty}d\hat{t}\psi^{--}(\hat{t})\left[(\hat{t}+a)^{\frac{Mk}{M+N}}\hat{t}^{\frac{Nk}{M+N}}\right]\cr
&&\quad\quad\times\frac{\sqrt{M+N}}{\sqrt{N}}\frac{1}{2\pi i}\oint\limits_{\hat{t}=0}d\hat{t}\psi^{++}(\hat{t})\left[(\hat{t}+a)^{-\frac{Mm}{N}-1}\left(\hat{t}+\frac{Na}{M+N}\right)\hat{t}^{-m-1}\right]\ket{0_{NS}}_{\hat{t}}\qquad\nn
\label{f_2_+}
\end{eqnarray}
Looking at (\ref{f_2_+}), (\ref{d_prime_f1}) and (\ref{d1_prime_f1}), we notice that if we make following interchanges
\begin{eqnarray}
M\leftrightarrow N,\quad a\leftrightarrow -a
\label{interchanges}
\end{eqnarray}
then
\begin{eqnarray} 
f^{F(2)+}_{mk}\leftrightarrow f^{F(1)+}_{mk}
\end{eqnarray}
Therefore we can use the results obtained for the expression $f^{F(1)+}_{mk}$ and make the interchanges given in (\ref{interchanges}) to obtain the expression for $f^{F(2)+}_{mk}$. The same applies for 
\begin{eqnarray} 
f^{F(2)-}_{mk}\leftrightarrow f^{F(1)-}_{mk} \,.
\end{eqnarray}
Then from (\ref{f_F1+}), (\ref{interchanges}) we find that for the case 
$
\frac{k}{M+N}\neq \frac{m}{N}
$
we have
\begin{eqnarray}
&&\!\!\!\!f_{mk}^{F(2)+}\cr
&&= \frac{(-1)^{m}\sqrt{N}\sin\left(\pi\frac{Nk}{M+N}\right)}{\pi\left(M+N\right)^{\frac{3}{2}}} \, \frac{a^{k-\frac{m(M+N)}{N}}}{\frac{k}{M+N}-\frac{m}{N}} 
\,\frac{k}{m} \,
\frac{\Gamma\left[\frac{Mk}{M+N}\right]\Gamma\left[\frac{Nk}{M+N}\right]}{\Gamma\left[k\right]}
\,
\frac{\Gamma\left[\frac{(M+N)m}{N}\right]}{\Gamma\left[m\right]\Gamma\left[\frac{Mm}{N}\right]}
\label{f_F2+} \qquad \\
&&\!\!\!\!f_{mk}^{F(2)-}\cr
&&=
\frac{(-1)^{m}\sin\left(\pi\frac{Nk}{M+N}\right)}{\pi\sqrt{N\left(M+N\right)}}
 \, \frac{a^{k-\frac{m(M+N)}{N}}}{\frac{k}{M+N}-\frac{m}{N}} \, 
\frac{\Gamma\left[\frac{Mk}{M+N}\right]\Gamma\left[\frac{Nk}{M+N}\right]}{\Gamma\left[k\right]}
\, \frac{\Gamma\left[\frac{(M+N)m}{N}\right]}{\Gamma\left[m\right]\Gamma\left[\frac{Mm}{N}\right]}
\end{eqnarray}
and for the case of
$
\frac{k}{M+N}=\frac{m}{N}
$
we have
\begin{eqnarray}
f_{mk}^{F(2)+} ~=~ f_{mk}^{F(2)-} &=& \frac{\sqrt{N}}{\sqrt{M+N}}\,.
\eea

\subsection{Expressing $f_{mk}^{F(i)\pm}$ in final form}

To express $f_{mk}^{F(i)\pm}$ in final form, we now make analogous changes of notation as done for $\gamma_{kl}^{F}$.
\begin{enumerate}
\item[(i)] Using \eq{az}, we replace the parameter $a$ by $z_{0}=e^{w_0}$.
\item[(ii)] For component string (1) we use fractional modes
\begin{eqnarray}
q=\frac{m}{M},&&s=\frac{k}{M+N}
\end{eqnarray}
and for component string  (2) we use fractional modes
\begin{eqnarray}
r=\frac{m}{N},&&s=\frac{k}{M+N} \,.
\end{eqnarray}
When expressing our result for component string (1) using indices $q$ and $s$, we write
\begin{eqnarray}
f_{mk}^{F(1)\pm}\to\tilde{f}_{qs}^{F(1)\pm}
\end{eqnarray} 
and for component string (2) using indices $r$ and $s$, we write
\begin{eqnarray}
f_{mk}^{F(2)\pm}\to\tilde{f}_{rs}^{F(2)\pm} \,.
\end{eqnarray}
\item[(iii)] We use the shorthand $\mu_{s}=(1-e^{2\pi i Ms})$. In doing this we note that
\begin{eqnarray}
\sin\left(\pi Ms\right)
=\frac{i}{2}e^{-i\pi Ms}\mu_s \,.
\end{eqnarray}
\end{enumerate}
With these changes in notation, for component string (1) with $s\neq q$ we have
\begin{eqnarray}
&&\!\!\!\!\!\!\tilde{f}_{qs}^{F(1)+}\cr
&&=\frac{i}{2\pi}z_{0}^{s-q}\frac{\mu_{s}}{\sqrt{M(M+N)}}\frac{1}{s-q}
 \, \frac{s}{q}\left(\frac{(M+N)^{M+N}}{M^{M}N^{N}}\right)^{s-q}
\frac{\Gamma\left[(M+N)q\right]}{\Gamma\left[Mq\right]\Gamma\left[Nq\right]}
\frac{\Gamma\left[Ms\right]\Gamma\left[Ns\right]}{\Gamma\left[(M+N)s\right]}
\cr
&&\!\!\!\!\!\!\tilde{f}_{qs}^{F(1)-}\cr
&&=\frac{i}{2\pi}z_{0}^{s-q}\frac{\mu_{s}}{\sqrt{M(M+N)}}\frac{1}{s-q}
\left(\frac{(M+N)^{M+N}}{M^{M}N^{N}}\right)^{s-q}
\frac{\Gamma\left[(M+N)q\right]}{\Gamma\left[Mq\right]\Gamma\left[Nq\right]}
\frac{\Gamma\left[Ms\right]\Gamma\left[Ns\right]}{\Gamma\left[(M+N)s\right]}
\nn
\label{f+}
\end{eqnarray}
For component string (2) with $s\neq r$ we obtain
\begin{eqnarray}
&&\!\!\!\!\!\!\tilde{f}_{rs}^{F(2)+}\cr
&&=-\frac{i}{2\pi}z_{0}^{s-r}\frac{\mu_{s}}{\sqrt{N(M+N)}}\frac{1}{s-r}
 \, \frac{s}{r}\left(\frac{(M+N)^{M+N}}{M^{M}N^{N}}\right)^{s-r}
\frac{\Gamma\left[(M+N)r\right]}{\Gamma\left[Mr\right]\Gamma\left[Nr\right]}
\frac{\Gamma\left[Ms\right]\Gamma\left[Ns\right]}{\Gamma\left[(M+N)s\right]}
\cr
&&\!\!\!\!\!\!\tilde{f}_{rs}^{F(2)-}\cr
&&=-\frac{i}{2\pi}z_{0}^{s-r}\frac{\mu_{s}}{\sqrt{N(M+N)}}\frac{1}{s-r}
\left(\frac{(M+N)^{M+N}}{M^{M}N^{N}}\right)^{s-r}
\frac{\Gamma\left[(M+N)r\right]}{\Gamma\left[Mr\right]\Gamma\left[Nr\right]}
\frac{\Gamma\left[Ms\right]\Gamma\left[Ns\right]}{\Gamma\left[(M+N)s\right]}
\nn
\label{f2+}
\end{eqnarray}
For $M=N=1$, given the normalizations chosen, the quantities $\frac{1}{\sqrt{2}}f^{F(i)+}$ should agree with the  $f^{F(i)+}$ computed in \cite{acm3}. One can check that this is indeed the case.

Let us also obtain the continuum limit for $f^{F(i)\pm}$. We need only consider the cases where $s\neq q$ for $f_{qs}^{F(1)\pm}$ and $s\neq r$ for $f_{qs}^{F(2)\pm}$ because for the cases $s=q$ and $s=r$ the continuum limit expression is identical to the exact expression.

For $f^{F(1)+}$, from \eq{f+} we have the exact expression
\begin{eqnarray}
\tilde{f}_{qs}^{F(1)+}&=&\frac{i}{2\pi}z_{0}^{s-q}
\frac{\mu_{s}}{\sqrt{M(M+N)}}\frac{1}{s-q}
 \, \frac{s}{q}\left(\frac{(M+N)^{M+N}}{M^{M}N^{N}}\right)^{s-q}\frac{\Gamma\left[Ms\right]\Gamma\left[Ns\right]}{\Gamma\left[(M+N)s\right]}\frac{\Gamma\left[(M+N)q\right]}{\Gamma\left[Mq\right]\Gamma\left[Nq\right]}\cr &&
\end{eqnarray}
thus we obtain
\begin{eqnarray}
\tilde{f}_{qs}^{F(1)+}\approx\frac{i}{2\pi}z_{0}^{s-q}
\frac{\mu_{s}}{\sqrt{M(M+N)}}\frac{1}{s-q}
\sqrt{\frac{s}{q}} \,. \qquad
\end{eqnarray}
Similarly, we obtain
\begin{eqnarray}
\tilde{f}_{qs}^{F(1)-}\approx\frac{i}{2\pi}z_{0}^{s-q}
\frac{\mu_{s}}{\sqrt{M(M+N)}}\frac{1}{s-q}
\sqrt{\frac{q}{s}}
\end{eqnarray}
and the component string  (2) quantities can be found in the same way.

\section{Applying the Supercharge}
Thus far we've only computed the action of the bare twist. We need to compute the action of the supercharge operator given in \cite{acm2}.
We now apply the supercharge to obtain the full effect of the deformation operator on the state $|0_R^{--}\rangle^{(1)}|0_R^{--}\rangle^{(2)} $.
For ease of notation, we introduce notation for the holomorphic and antiholomorphic parts of the state $\ket{\boldsymbol{\chi}}$,
\bea
\ket{\boldsymbol{\chi}} &=& \ket{\chi}\ket{\bar\chi},
\eea
where we divide the prefactor $C_{MN}$ equally between the holomorphic and antiholomorphic parts, i.e.
\bea
\ket{\chi} &=& 
\sqrt{C_{MN}} \left[
\exp\left(\sum_{m\geq 1,n\geq 1}
\gamma_{mn}^{B}\left[-\alpha_{++,-m}\alpha_{--,-n}
+\alpha_{+-,-m}\alpha_{-+,-n}\right]\right)
\right. \cr
&& \qquad \qquad  \left.
\times\exp\left(\sum_{m\geq 0,n\geq 1}
\gamma_{mn}^{F}
\left[d_{-m}^{++}d_{-n}^{--}-d_{-m}^{+-}d_{-n}^{-+}
\right]\right)
\right]
\ket{0_{R}^{-}}. \qquad
\label{eq:expans-3}
\end{eqnarray}
Similarly, we write the final state obtained by acting with holomorphic and antiholomorphic supercharges as 
\bea
\ket{\Psi} &=& \ket{\psi}\ket{\bar\psi}
\eea
where
\begin{eqnarray} \label{psistart}
\ket{\psi}=G_{\dot{A},0}^{-}\ket{\chi} \,
\end{eqnarray}
and similarly for the antiholomorphic part.
We now compute the state $ \ket{\psi}$.

From \eq{G0_above} we have
\begin{eqnarray}
G_{\dot{A},0}^{-} &=& \frac{1}{2\pi i}\int\limits_{w=\tau_0+\e}^{w=\tau_0+\e+2\pi i(M+N)}G_{\dot{A}}^{-}\left(w\right)dw ~=~ \frac{i}{\sqrt{M+N}}\sum_{n=-\infty}^{\infty}d_{n}^{-A}\alpha_{A\dot{A},-n} \,.
\label{G0_above-1}
\end{eqnarray}
We wish to write (\ref{psistart}) with only negative index modes acting on $\ket{0_{R}^{-}}$. We thus write
\begin{eqnarray}
G_{\dot{A},0}^{-}&=&\frac{i}{\sqrt{M+N}}\left (\sum_{l>0}^{\infty}d^{-A}_{-l}\alpha_{A\dot{A},l}+\sum_{l>0}^{\infty}d_{l}^{-A}\alpha_{A\dot{A},-l} + d^{-A}_0 \a_{A \dot A, 0} \right )
\end{eqnarray}
We note e.g.~from \eq{eq:gammaB} that $\gamma^{B}_{kl}$ is symmetric, so we have 
\begin{eqnarray}
\frac{i}{\sqrt{M+N}}\sum_{l\geq 1}^{\infty}d^{-A}_{-l}\alpha_{A\dot{A},l}\ket{\chi}&=&\frac{i}{\sqrt{M+N}}\sum_{l\geq 1}\sum_{l\geq 1}l\gamma^{B}_{kl}d^{-A}_{-l}\alpha_{A\dot{A},-k}\ket{\chi}\cr
\frac{i}{\sqrt{M+N}}\sum_{k\geq 1}^{\infty}d_{k}^{-A}\alpha_{A\dot{A},-k}\ket{\chi}&=&-\frac{i}{\sqrt{M+N}}\sum_{k\geq 1}\sum_{l\geq 1}\gamma_{kl}^{F}d_{-l}^{-A}\alpha_{A\dot{A},-k}\ket{\chi} \cr
\frac{i}{\sqrt{M+N}}d^{-A}_{0}\alpha_{A\dot{A}0}\ket{\chi}&=& 0
\end{eqnarray}
where we have used the commutation relations in (\ref{commrelation}). Thus
\begin{eqnarray}
G_{\dot{A},0}^{-}\ket{\chi}&=&\frac{i}{\sqrt{M+N}}\sum_{k\geq 1,l\geq 1}\left(l\gamma^{B}_{kl}-\gamma_{kl}^{F}\right)d_{-l}^{-A}\alpha_{A\dot{A},-k}\ket{\chi} \,.
\end{eqnarray}
We observe that the $l$ and $k$ sums factorize and we obtain
\begin{eqnarray}
\ket{\psi} ~=~G_{\dot{A},0}^{-}\ket{\chi}&=&-\frac{i}{\pi^{2}}\frac{MN}{(M+N)^{\frac{5}{2}}}\left(\sum_{l\geq 1}a^{l}\sin\left[\tfrac{N\pi l}{M+N}\right]\frac{\Gamma\left[\frac{Ml}{M+N}\right]\Gamma\left[\frac{Nl}{M+N}\right]}{\Gamma\left[l\right]}d_{-l}^{-A}\right)\cr
&& \qquad\times\left(\sum_{k\geq 1}a^{k}\sin\left[\tfrac{N\pi k}{M+N}\right]\frac{\Gamma\left[\frac{Mk}{M+N}\right]\Gamma\left[\frac{Nk}{M+N}\right]}{\Gamma\left[k\right]}\alpha_{A\dot{A},-k}\right)\ket{\chi} \,. \qquad\nn
\label{eq:psians}
\end{eqnarray}
One can check\footnote{In order to verify this, note that due to different choices in normalization of the modes of the fermions, one should replace $d^{-A}\to \frac{1}{\sqrt{2}}d^{-A}$; in addition, there is an overall minus sign due to the different directionality of the contour for $G^{-}_{\dot{A},0}$.} that in the case of $M=N=1$ this expression agrees with that computed in \cite{acm2}.

Analogous expressions hold for 
\begin{eqnarray}
\ket{\bar{\psi}}&=&\bar{G}_{\dot{B},0}^{-}\ket{\bar{\chi}} \,,
\end{eqnarray}
and the complete final state is given by
\begin{eqnarray}
\ket{\Psi}=\ket{\psi} \ket{\bar{\psi}} \,.
\end{eqnarray}

\subsection{Applying the supercharge for an initial bosonic excitation}

We now compute the state produced when the deformation operator acts on an initial bosonic excitation. 
In the same way done for an initial fermionic excitation, we introduce the notation
\begin{eqnarray}
\ket{\Psi^{B(i)}} &=& 
\hat{O}_{\dot{A}}(w_0)\, \a^{(i)}_{B\dot B,-m}\ket{0_{R}^{--}}^{(1)} \ket{0_{R}^{--}}^{(2)} ~=~ 
\ket{\psi^{B(i)}}\ket{\bar\psi^{B(i)}} \,.
\eea

\subsubsection{$\ket{\psi^{B(1)}}$}

We first compute the state $\ket{\psi^{B(1)}}$. We have
\begin{eqnarray}
\ket{\psi^{B(1)}} 
&=&
\left(G_{\dot{A},0}^{-}\sigma_{2}^{+}(w_{0})\alpha^{(1)}_{B\dot B,-m}-\sigma_{2}^{+}G_{\dot{A},0}^{(1)-}\alpha^{(1)}_{B\dot B,-m}\right)\ket{0_{R}^{-}}^{(1)} \ket{0_{R}^{-}}^{(2)}
\end{eqnarray}
The first term becomes
\begin{eqnarray}
G_{\dot{A},0}^{-}\sum_{k\geq 1}f_{mk}^{B(1)}\alpha_{B\dot{B},-k}\ket{\chi}&=&i\epsilon_{AB}\epsilon_{\dot{A}\dot{B}}\sum_{k\geq 1}\left(\frac{k \, f_{mk}^{B(1)}}{\sqrt{M+N}}\right)d_{-k}^{-A}\ket{\chi}+\sum_{k\geq 1}f_{mk}^{B(1)}\alpha_{B\dot{B},-k}\ket{\psi}\nonumber
\end{eqnarray}
For the second term, we find
\begin{eqnarray}
\sigma_{2}^{+}G_{\dot{A},0}^{(1)-}\alpha^{(1)}_{B\dot{B},-m}\ket{0_{R}^{-}}^{(1)} \ket{0_{R}^{-}}^{(2)}&=&i\epsilon_{AB}\epsilon_{\dot{A}\dot{B}}\frac{m}{\sqrt{M}}\sigma_{2}^{+}d_{-m}^{(1)-A}\ket{0_{R}^{-}}^{(1)} \ket{0_{R}^{-}}^{(2)}\cr
&=&i\epsilon_{AB}\epsilon_{\dot{A}\dot{B}}\sum_{k\geq 1}
\left(
\frac{m \, f_{mk}^{F(1)-}}{\sqrt{M}} \right)
d_{-k}^{-A}\ket{\chi}
\end{eqnarray}
Combining both terms, we obtain
\begin{eqnarray}
\ket{\psi^{B(1)}}&=&i\epsilon_{AB}\epsilon_{\dot{A}\dot{B}}
\sum_{k\geq 1}
\left(\frac{k \, f_{mk}^{B(1)}}{\sqrt{M+N}}-\frac{m \, f_{mk}^{F(1)-}}{\sqrt{M}}\right)
d_{-k}^{-A}\ket{\chi}+\sum_{k\geq 1}f_{mk}^{B(1)}\alpha_{B\dot{B},-k}\ket{\psi} \quad\nn
\label{final_f(1)}
\eea
where we note that\footnote{To compare to \cite{acm3} in the limit of $ M = N = 1 $, one should take into account the different conventions on fermion modes. Note also that there is a typo in the last term of equation (7.5) of \cite{acm3}; this term should resemble the last term in \ref{final_f(1)} above.}
\bea
&&\!\!\!\!\!\!\!\!\!\!\!\!\frac{k \, f_{mk}^{B(1)}}{\sqrt{M+N}}-\frac{m \, f_{mk}^{F(1)-}}{\sqrt{M}}\cr
&&=\frac{(-1)^{m}\sin\left(\pi\frac{Mk}{M+N}\right)}{\pi\sqrt{M+N}}
(-a)^{k-\frac{m(M+N)}{M}}
\frac{\Gamma\left[\frac{Mk}{M+N}\right]\Gamma\left[\frac{Nk}{M+N}\right]}{\Gamma\left[k\right]}
\times
\frac{\Gamma\left[\frac{(M+N)m}{M}\right]}{\Gamma\left[m\right]\Gamma\left[\frac{Nm}{M}\right]}.
\cr &&
\label{final_f(1)-2}
\end{eqnarray}

\subsubsection{$\ket{\psi^{B(1)}}$}

To find $\ket{\psi^{B(2)}}$ one simply modifies (\ref{final_f(1)}) by $M\leftrightarrow N$, $a\rightarrow -a$, and $(1) \rightarrow (2)$. 

\subsection{Summary of Results}
For convenient reference, here we record the results for $C_{MN}$ and $f^{B(i)},f^{F(i)\pm}$. For completeness we include also the bogoliubov coefficents $\g^B,\g^F$ which computed in the previous chapter.
\begin{eqnarray}
\tilde\gamma^B_{ss'}&=&{z_0^{s+s'}\over 4\pi^2}\,\mu_s\mu_{s'}\,
{1 \over  s+s'}\,{MN\over (M+N)^3} 
\left ({(M+N)^{M+N}\over M^MN^N}\right )^{s+s'} \cr
&&\qquad\qquad\qquad\qquad\qquad\qquad\times{\Gamma[{Ms}]\Gamma[{Ns}]\over \Gamma[(M+N)s]}~{\Gamma[{Ms'}]\Gamma[{Ns'}]\over \Gamma[(M+N)s']}\cr
\tilde\gamma_{ss'}^{F}&=&-\frac{z_{0}^{s+s'}}{4\pi^{2}}\mu_{s}\mu_{s'}\frac{s}{s+s'}\frac{MN}{(M+N)^{2}}\left(\frac{(M+N)^{M+N}}{M^{M}N^{N}}\right)^{s+s'}\cr
&&\qquad\qquad\qquad\qquad\qquad\qquad\times\frac{\Gamma\left[Ms\right]\Gamma\left[Ns\right]}{\Gamma\left[(M+N)s\right]}\frac{\Gamma\left[Ms'\right]\Gamma\left[Ns'\right]}{\Gamma\left[(M+N)s'\right]}\cr
C_{MN}&=&\frac{M+N}{2MN}
\end{eqnarray}
\bea
\tilde f^{B(1)}_{qs} & = & \begin{cases}
{M \over M+N} & q = s \\
{i\over 2\pi}z_0^{s-q}
{\mu_s \over s-q}\,
{1\over  (M+N)}\left( {(M+N)^{M+N}\over M^MN^N}\right)^{(s-q)}
{\Gamma[(M+N)q]\over \Gamma[Mq]\Gamma[Nq]}\,
{\Gamma[Ms]\Gamma[Ns]\over \Gamma[(M+N)s]}
\qquad\quad & q \neq s 
\end{cases}\qquad\quad \cr
\tilde f^{B(2)}_{rs} & = & \begin{cases}
{N \over M+N} & r = s \\
-{i\over 2\pi}z_0^{s-r}
{\mu_s\over s-r}\,
{1\over  (M+N)}\left( {(M+N)^{M+N}\over M^MN^N}\right)^{(s-r)}
{\Gamma[(M+N)r]\over \Gamma[Mr]\Gamma[Nr]}\,
 {\Gamma[Ms]\Gamma[Ns]\over \Gamma[(M+N)s]} \qquad\quad & r \neq s
\end{cases}\qquad\quad
\eea

\bea
\tilde{f}_{qs}^{F(1)+}&=&\begin{cases}
\frac{\sqrt{M}}{\sqrt{M+N}}& q=s\\
&\\
\frac{i}{2\pi}z_{0}^{s-q}\frac{\mu_{s}}{\sqrt{M(M+N)}}\frac{1}{s-q}
 \, \frac{s}{q}\left(\frac{(M+N)^{M+N}}{M^{M}N^{N}}\right)^{s-q}
\frac{\Gamma\left[(M+N)q\right]}{\Gamma\left[Mq\right]\Gamma\left[Nq\right]}
\frac{\Gamma\left[Ms\right]\Gamma\left[Ns\right]}{\Gamma\left[(M+N)s\right]}
 & q\neq s
\end{cases} \cr
\tilde{f}_{qs}^{F(1)-}&=&
\begin{cases}\frac{\sqrt{M}}{\sqrt{M+N}}& q=s\\
&\\
\frac{i}{2\pi}z_{0}^{s-q}\frac{\mu_{s}}{\sqrt{M(M+N)}}\frac{1}{s-q}
\left(\frac{(M+N)^{M+N}}{M^{M}N^{N}}\right)^{s-q}
\frac{\Gamma\left[(M+N)q\right]}{\Gamma\left[Mq\right]\Gamma\left[Nq\right]}
\frac{\Gamma\left[Ms\right]\Gamma\left[Ns\right]}{\Gamma\left[(M+N)s\right]}
&q\neq s
\end{cases}
\end{eqnarray}

\bea
\tilde{f}_{rs}^{F(2)+}&=&\begin{cases}
\frac{\sqrt{N}}{\sqrt{M+N}}& r=s\\
&\\
-\frac{i}{2\pi}z_{0}^{s-r}\frac{\mu_{s}}{\sqrt{N(M+N)}}\frac{1}{s-r}
 \, \frac{s}{r}\left(\frac{(M+N)^{M+N}}{M^{M}N^{N}}\right)^{s-r}
\frac{\Gamma\left[(M+N)r\right]}{\Gamma\left[Mr\right]\Gamma\left[Nr\right]}
\frac{\Gamma\left[Ms\right]\Gamma\left[Ns\right]}{\Gamma\left[(M+N)s\right]}
 & r\neq s
\end{cases} \cr
\tilde{f}_{rs}^{F(2)-}&=&
\begin{cases}\frac{\sqrt{N}}{\sqrt{M+N}}& r=s\\
&\\
-\frac{i}{2\pi}z_{0}^{s-r}\frac{\mu_{s}}{\sqrt{N(M+N)}}\frac{1}{s-r}
\left(\frac{(M+N)^{M+N}}{M^{M}N^{N}}\right)^{s-r}
\frac{\Gamma\left[(M+N)r\right]}{\Gamma\left[Mr\right]\Gamma\left[Nr\right]}
\frac{\Gamma\left[Ms\right]\Gamma\left[Ns\right]}{\Gamma\left[(M+N)s\right]}
&r\neq s
\end{cases}
\end{eqnarray}

\section{Discussion} \label{sec:disc}

In this chapter we have studied the effect of the deformation operator when it joins together two component strings of length $ M, N $ with an initial bosonic and fermionic excitation. We obtain obtain a single component string of winding $ M+N$ with a linear combination of excitations built on the single twist $|\chi\rangle$. 

Our results extend those from chapter 3, where we only computed expressions obtained from twisting the vacuum. In addition to these computations, we have also computed the overall prefactor on the final state. The calculation of this prefactor is quite nontrivial, but the answer    is compact: $C_{MN} = (M+N)/(2MN) $. If one considers the special case $ M=N $, this becomes $C_{MM} = 1/M$, which agrees with the fact that the twist operator $\s_2^{++}$ responsible for this coefficient has weight $(1/2,1/2)$.


We note an interesting structure for both $f^{B(i)}$ and $f^{F(i)\pm}$. We see that they both vanish unless there is a creation operator in the initial state, and a creation operator in the final state. 


In addition, both $f^{B(i)}$ and $f^{F(i)}$ share the property of `almost factorization' observed for the quantities $\gamma^B$, $\gamma^F$ in chapter 2. For $f^{B(1)}_{qs}$ and $f^{F(1)}_{qs}$, the only part which does not factorize in this way is the factor $\frac{1}{s-q}$.

For applications to black hole physics, one is interested in the limit of large $N_{1}N_{5}$. In this limit, component strings typically have parametrically large winding;
this is the main physical reason for studying the general $ M$, $ N $ problem.  Thus, as well as obtaining our exact results, we have extracted the behaviour of our results in the continuum limit just as we did for the $\g$'s in chapter 3. Again, the mode numbers are large compared to the spacing of modes on the component string. We found significant simplification to the various quantities, similar to that observed for $\g$'s. 

The results in this chapter further our understanding of the effect of the deformation operator in the D1D5 CFT. This work is based on the papers \cite{chmt1,chmt2}.  In the next chapter we analyze the two twist case.



\chapter{Effect of two twist operators on the vacuum state}\label{two twist gamma chapter}
Everything that we have done thus far has been for a single deformation operator. In the case of the single twist operator we are going from two singly wound strings to a doubly wound string. The size and number of component strings change. This makes any thermalization effects hard to identify because the box size is changing. It's hard to distinguish a thermalization effect from casimir energy effects. This causes us to consider two deformation operators. We separate this computation into three chapters. In this chapter we'll compute the Bogoliubov coefficients, $\g^F,\g^B$ which characterize the twisted vacuum. In the next chapter we'll compute the action of the twist on a initial excitation, giving us the transition ampitudes $f^{B,F}$. In the following chapter we'll compute the action of the supercharges. Due to the increased technicality and length of the computations at second order we devote an entire chapter to each. Let us begin 

\section{Outline of the calculations}\label{Outline}

As we've stated before, the deformation operator is composed of a supercharge contour acting on a twist operator. The action of the supercharge contour can be split off from the main computation, which involves the effects of the twist operators. In this chapter we will focus of the effect of the twist operators alone. 

Since we are looking at deformations to second order, we will have two twist-2 operators acting on our initial state. If $n$ copies of the CFT are twisted together, we say that we have an $n$-times wound `component string'. We will start with the simplest case of two singly wound component strings. The twist from the first deformation operator will change this to a single component string with winding 2, while the second twist will take this new component string back to two component strings of winding 1 each. Thus we are looking at a `1-loop' process in the interacting CFT. We believe that such second order effects will lead to the thermalization that we seek. 

Each of the initial component strings are taken to be in the negative Ramond vacuum state $\rmket$.  The final state after the application of two twists will be denoted by $|\c(w_1,w_2)\rangle$. Thus:
\bea
|\c(w_1,w_2)\rangle & \equiv & \s_2^+(w_2)\s_2^+(w_1) \rmmket.
\label{stateq}
\eea
Here $w$ is a coordinate on the cylinder, with
\be
w=\tau+i\sigma,
\ee
where $\sigma$ is an angular coordinate for the compact spacial dimension and $\tau$ is a Euclideanized time coordinate and $w_1,w_2$ correspond to the location of the first and second twist on the cylinder  In (\ref{stateq}), we assume that $\tau_2 > \tau_1$.

In the region $\t_1 < \t < \t_2$,  the two component strings are joined together to form a doubly wound componet string. The bosonic fields are now periodic only after an interval $\Delta \sigma = 4\pi$. The fermionic fields, a priori, can be periodic or antiperiodic after $\Delta \sigma = 4\pi$.  Outside the above interval of $\tau$ we have two singly wound component strings.  We will now map this configuration to a double cover of the cylinder.  To do this, we first map the cylinder into the complex plain with coordinate $z$ and then map the complex plane to a double cover of itself.

\begin{figure}[tbh]
\begin{center}
\includegraphics[width=0.5\columnwidth]{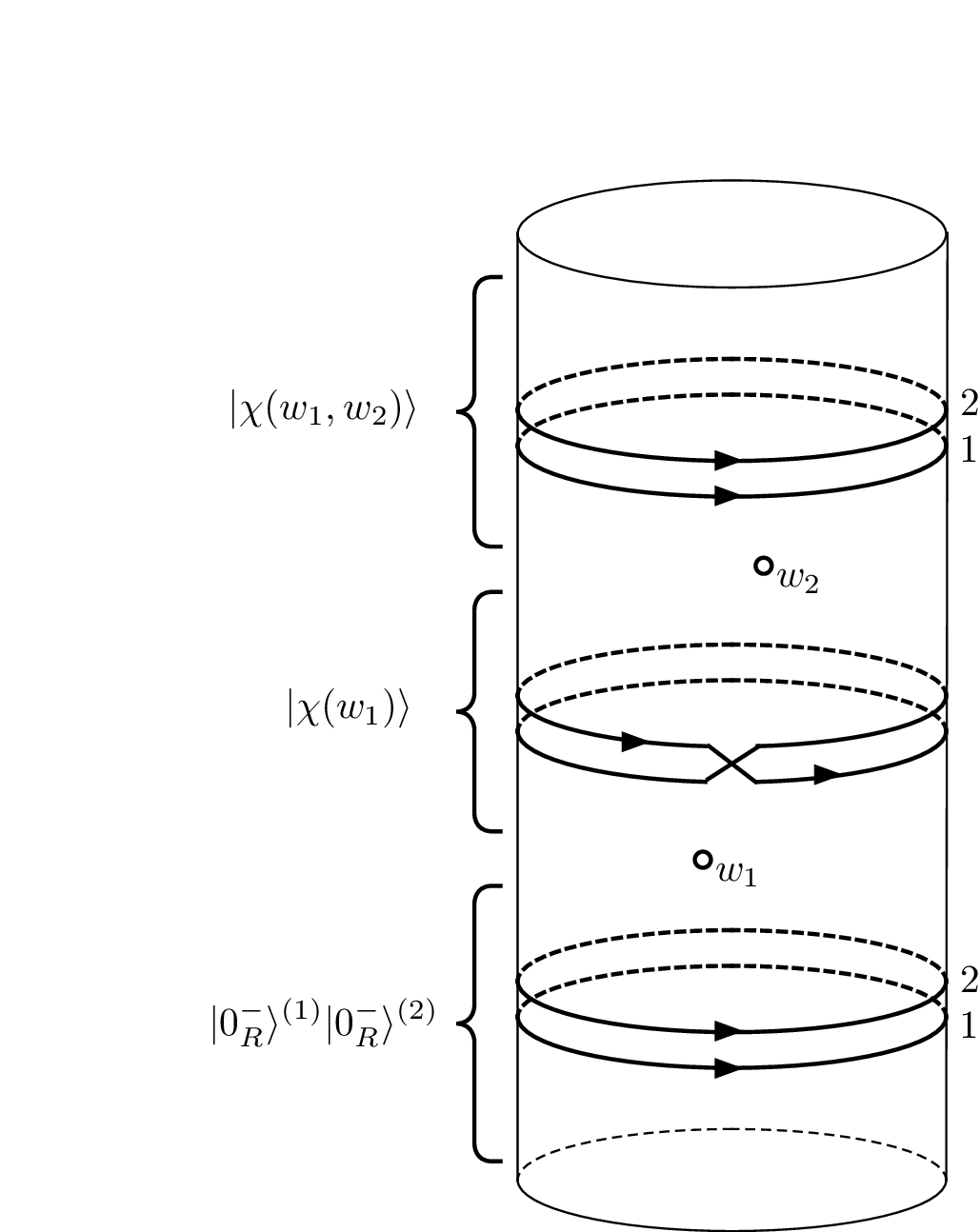}
\end{center}
\caption{The cylinder with twist insertions at $w_1$ and $w_2$.  Below the first twist we have the negative Ramond vacuum on each component string.  Above both twists we have the state $|\chi(w_1,w_2)\rangle$, which we will compute.  In the intermediate regions we have a single doubly-wound component string in the state $|\chi(w_1,w_2)\rangle$.  This state was computed in \cite{acm1} and is not used here.}
\label{figone}
\end{figure}

Around any point $z$ on the plane, a fermionic field can have either periodic or antiperiodic boundary conditions. If we insert only local operators at $z$, then the fermion will be periodic (i.e. it will have NS boundary conditions). If we wish to have instead the anti-periodic boundary condition (i.e., the R boundary condition) then we must insert a spin field $S^\pm$ at the point $z$. The spin field carries  a charge $j=\pm \h$ and has dimension $h={1\over 4}$. 

  The initial state (the `in' state) now has two singly wound component strings, each in the R sector. This brings in two spin fields, one from each component string.  We write
\bea
\rmmket = S^{(1)-}(\tau = -\infty)S^{(2)-}(\tau = -\infty)\nsnsket.
\eea
In addition, the twist operator $\s_2^+$ contains a spin field:
\bea
\s_2^+(w) = S^+(w) \s_2(w).
\eea
We will compute an amplitude by taking the inner product of the state we generate after the twists with some state of our choosing. This state with which we `cap' the cylinder may then bring additional spin fields of its own. 

Each of the above mentioned spin fields will have some position in our double cover.  We can remove a spin field by performing a spectral flow around that point, since a spectral flow can map an R sector state to an NS sector state. We will thus perform a series of spectral flows at various points     in the double cover in order to remove these spin field insertions.  Once all the spin fields have been removed, the locations where they were inserted will have just the local NS vacuum inserted there, and we can close the corresponding puncture at that location with no insertions. Any contour can then be smoothly deformed through such a location.  It is through these  smooth deformations that we will be able to map in states to out states and determine the nature of $|\c(w_1,w_2)\rangle$

We now divide the remainder of this section into four parts.  In the first part, we outline the coordinate changes used to map the cylinder into a double cover of the complex plane and identify the images of all critical points.  In the second part, we introduce the mode operators   on the cylinder.  In the third part, we present the general form of the state $|\c(w_1,w_2)\rangle$ in terms of these mode operators.  This general form motivates capping with certain types of states, which allows us to derive expressions for the parameters of our ansatz.  Finally, we determine the spectral flows needed for each type of capping state based on the images of the spin fields of those states.

\subsection{Coordinate maps}\label{CoordinateMaps}
\begin{figure}[tbh]
\begin{center}
\includegraphics[width=0.4\columnwidth]{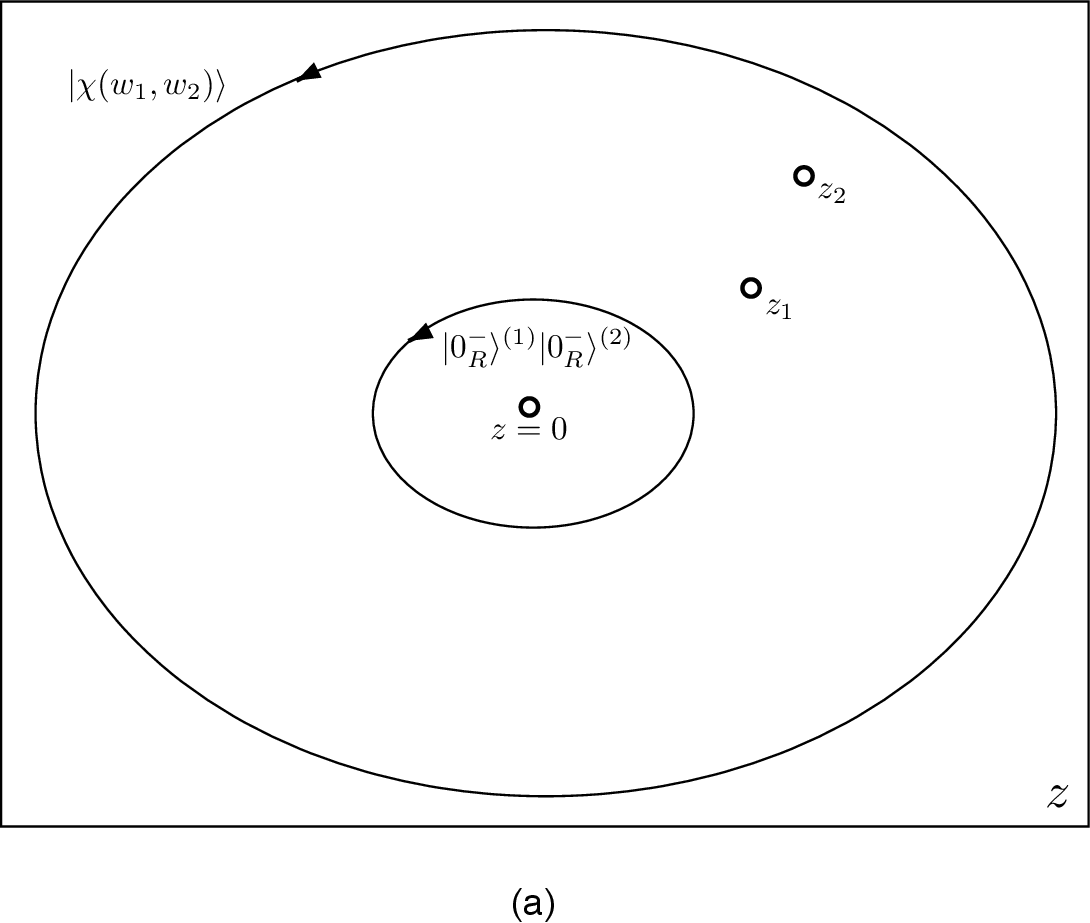} $\qquad\qquad$ \includegraphics[width=0.4\columnwidth]{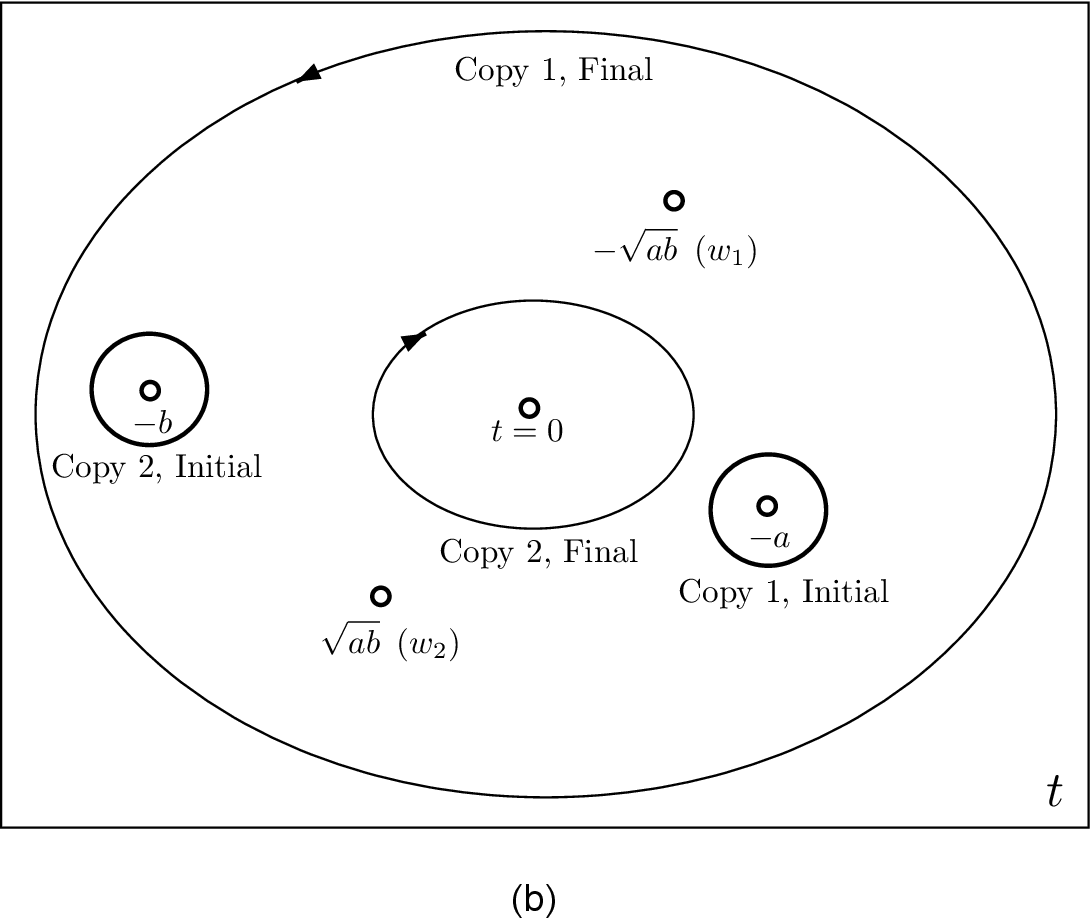}
\end{center}
\caption{The $z$ plane (a) and $t$ plane (b) with all the relevant image points labeled.  The intermediate state $|\chi(w_1)\rangle$ is not depicted.  In the $z$ plane the $\tau$ coordinate maps to the radial coordinate, while the $\s$ coordinate maps to the phase.  In the $t$ plane there are no simple directions corresponding to $\t$ and $\s$.}
\label{CoordinateMapFigure}
\end{figure}

The effects of our coordinate maps are illustrated in Figure \ref{CoordinateMapFigure}.  First we map the cylinder to the complex plane through the map:
\bea
z & = & e^w ~\equiv~ e^{\t + i\s}.
\eea
Here the in states at $\tau = -\infty$ map to the origin of the complex plane, while the out states at $\tau = +\infty$ map to $z = \infty$.  Our fields are still double-valued near these points, as there are two component strings in both the in and out states.  The timelike direction $\tau$ becomes the radial direction, while the spacelike $\s$ becomes the phase of the $z$ plane.  Because $\tau_2 > \tau_1$, we have:
\bea \label{zmap}
|z_2| &=& e^{\t_2} ~>~ |z_1| ~=~ e^{\t_1}.
\eea

Now we map the $z$ plane into a double cover of itself.  We need to ensure that this map will separate out the two distinct copies of our fields for the in and out states.  Thus we require that $z=0$ and $z=\infty$ both have two distinct image points.  It is also useful to ensure that near these points we have leading order relations of the form $z \sim t^{\pm 1}$.  Such a map can in general be written as:
\bea \label{tmap}
z &=& {(t+a)(t+b)\over t}.
\eea
Here, the point $z = \infty$ maps to both $t = \infty$ and $t = 0$.  Near $t=\infty$ we have the leading order behavior $z \sim t$, while near $t=0$ we have the behavior $z \sim t^{-1}$.  The point $z = 0$ also has two images, at $t= -a$ and $t = -b$.  This allows us to split up the images of the two copies for our in and out states, thus leaving us with only single-valued fields in the $t$ plane.

At this point, we have some choice of which component strings map to which regions in the $t$ plane.  The in state has images at $t = -a$ and $t = -b$, but the physics is independent of which component string we call copy 1 and which we call copy 2.  Indeed, the map (\ref{tmap}) is symmetric under the interchange $a \leftrightarrow b$.  Similarly, we may choose which component string of the out state maps to $t = \infty$ and which component string maps to $t = 0$.  Here we present the conventions used in this chapter:
\bea
\text{Copy 1 In State} & \to & t = -a \nn
\text{Copy 2 In State} & \to & t = -b \nn
\text{Copy 1 Out State} & \to & t = \infty \nn
\text{Copy 2 Out State} & \to & t = 0.
\eea

We also need to know the images of the twist insertions at $w_1$ and $w_2$.  These insertions create bifurcation points in the double cover, so we find the images by solving for these bifurcation points.
\bea
{\diff z \over \diff t} & = & 1 - {ab \over t^2} ~=~ 0.
\eea
This relation has two solutions, corresponding to our two twist insertions.
\bea \label{twistimages}
t_1 & = & -\sqrt{ab} \nn
t_2 & = & \sqrt{ab}.
\eea
This in turn gives us a relationship between the double cover map parameters $a,b$ and our twist insertion points $w_1,w_2$:
\bea
z_1 & = & a+b - 2\sqrt{ab} ~=~ e^{w_1} \nn
z_2 & = & a+b + 2\sqrt{ab} ~=~ e^{w_2},
\eea
where we choose the branch of $\sqrt{ab}$ so as to maintain $|z_2| > |z_1|$.  As expected, this relationship is unaffected by the interchange of $a$ and $b$.

\subsection{Modes on the cylinder}
In order to determine the final state $|\chi(w_1,w_2)\rangle$ we first introduce the bosonic and fermionic modes that live on the cylinder in the region $\t > \t_2$.  For the bosons, we have:
\bea\label{BosonCylinderMode}
\alpha_{A\dot A, n}^{(i)f} &=& {1\over 2\pi} \int\limits_{\sigma=0}^{2\pi} X_{A\dot A}^{(i)}(w) e^{nw} \diff w,
\eea
where $f$ stands for final modes, indicating that we are working in the region above both twists. The commutation relations are
\bea\label{BosonCommutator}
[\alpha_{A\dot A, n}^{(i)f}, \alpha_{B\dot B, m}^{(j)f}] &=& -n\epsilon_{AB}\epsilon_{\dot A\dot B}\, \delta^{ij} \delta_{n+m,0}\;.
\eea

For fermions, we can have two different types of modes depending on which sector we are in, NS or R.  In the NS sector, the modes are indexed by half-integers $r$:
\bea\label{FermionNSCylinderMode}
d^{(i)f,\a A}_{r} &=& {1\over 2\pi i} \int\limits_{\s=0}^{2\pi} \psi^{(i)\alpha A}(w) e^{rw}\diff w,
 \eea
with anticommutation relations
\bea
\left \{ d^{(i)f,\a A}_{r}, d^{(j)f,\b B}_{s} \right \} &=& -\e^{\a\b}\e^{AB}\d^{(i)(j)}\d_{r+s,0}\;.
\eea
Similarly, the modes in the Ramond sector are:
\bea\label{FermionRCylinderMode}
d^{(i)f,\a A}_{n} &=& {1\over 2\pi i} \int\limits_{\s=0}^{2\pi} \psi^{(i)\alpha A}(w) e^{nw}\diff w,
\eea
with integer $n$.  These modes have anticommutation relations
\bea
\left \{ d^{(i)f,\a A}_{n}, d^{(j)f,\b B}_{m} \right \} &=& -\e^{\a\b}\e^{AB}\d^{(i)(j)}\d_{n+m,0}\;.
\eea

In the Ramond sector there are fermion zero modes.  While bosonic zero modes annihilate all vacua, the fermionic zero modes do not always annihilate the vaccum.  Instead, one finds:
\be
d^{(i)f,+ A}_0 \rpket^{(i)} ~=~ d^{(i)f,- A}_0 \rmket^{(i)} ~=~ 0
\ee
\be
d^{(i)f,+ A}_0 \rmket^{(i)}~\neq~ 0, ~~~\, d^{(i)f,- A}_0 \rpket^{(i)} ~\neq~ 0,
\ee
where the copy index $(i)$ is not summed over.  For more details about the behavior of these zero modes, see Appendix \ref{RVN}.

One can of course construct modes which live before the two twist insertions, as well as modes which live between the two twists.  Since we begin with the vacuum state as our in state before the twists and we do not need the state between the two twists in an explicit way, we will not write the modes in these regions.   We do however require modes natural to the NS vacuum in the $t$ plane.  These are
\bea
\tilde{\a}_{A\dot A,n} &=& {1\over 2\pi} \oint\limits_{t=0} \partial_t X_{A\dot A}(t) t^n \diff t \label{BosontMode}\\
\tilde{d}^{\a A}_r &=& {1\over 2\pi i} \oint\limits_{t=0} \psi^{\a A}(t) t^{r-\h} \diff t, \label{FermiontMode}
\eea
with commutation relations
\begin{eqnarray}
\left[\tilde{\a}_{A\dot{A},m},\tilde{\a}_{A\dot{A},n}\right]=-\e_{AB}\e_{\dot{A}\dot{B}}m\delta_{n+m,0} \label{bosoncommutation}\\
\left\{\tilde{d}^{\a A}_m,\tilde{d}^{\a A}_n\right\}=-\e^{\a\b}\e^{AB}\delta_{n+m,0}\;. \label{fermioncommutation}
\end{eqnarray}

As a final remark, there is a subtle concern that needs to be addressed when we wish to map the cylinder modes into the $t$ plane.  Since we have chosen the final copy 1 to map to large $t$, copy 1 modes must always come to the left of copy 2 modes in the $t$ plane.  As such, when we apply annihilation modes to probe the deformed state we will always choose to place all copy 1 modes to the left of all copy 2 modes.  This has no effect on the bosonic calculation because bosonic annihilators commute, but it is important to account for this convention when working with fermions.

\subsection{The general form}  
Looking at the SU(2) charges of the twist operators and the initial vacuum state, it is clear that $|\chi(w_1,w_2)\rangle$ is overall neutral.   One may also expect the final state to contain only pairs of excitations in the form of an exponential similar to the one-twist result of \cite{acm1}.  However, this naive analysis seems to indicate that part of $|\chi(w_1,w_2)\rangle$ could live in the $NS$ sector, as the vacuum $\nsnsket$ is neutral.  We will perform the computations for both R and NS sectors in the final state, but argue at the end that only the R sector amplitudes are relevant for the physical system under study.

  In \cite{acm1} it was noted that the state produced after the twist had an exponential form, and a general argument was given for why this must be the case for the bosons of the theory.  This proof extends trivially to the use of arbitrarily many twists.  In Appendix \ref{GeneralFormAppendix} we present a similar proof for the fermionic contribution, which again extends to arbitrarily many twists and also applies to both the NS and Ramond sectors.  Keeping both sectors in the final state for the moment, we  note that $|\chi(w_1,w_2)\rangle$ can be written as:
\bea \label{generalform}
|\chi(w_1,w_2)\rangle & \!= \!& C_{NS}(w_1,w_2)\text{exp}\left [ \sum_{(i),(j)}\sum_{k,l > 0}\g^{B(i)(j)}_{kl} \left ( -\a^{(i)f}_{++,-k}\a^{(j)f}_{--,-l} + \a^{(i)f}_{+-,-k}\a^{(j)f}_{-+,-l} \right ) \right ] \nn
&&\!\!\!\!\!\!\!\!{}\times \text{exp}\left [ \sum_{(i),(j)}\sum_{r,s > 0}\g^{F(i)(j)}_{NS,rs} \left ( d^{(i)f,++}_{-r}d^{(j)f,--}_{-s} - d^{(i)f,+-}_{-r}d^{(j)f,-+}_{-s} \right ) \right ] \nsnsket \nn
&&\!\!\!\!\!\!\!\!{}+ C_{R+-}(w_1,w_2)\text{exp}\left [ \sum_{(i),(j)}\sum_{k,l > 0}\g^{B(i)(j)}_{kl} \left ( -\a^{(i)f}_{++,-k}\a^{(j)f}_{--,-l} + \a^{(i)f}_{+-,-k}\a^{(j)f}_{-+,-l} \right ) \right ] \nn
&&\!\!\!\!\!\!\!\!{}\times \text{exp}\left [ \sum_{(i),(j)}\sum_{k,l \geq 0}\g^{F(i)(j)}_{R+-,kl} \left ( d^{(i)f,++}_{-k}d^{(j)f,--}_{-l} - d^{(i)f,+-}_{-k}d^{(j)f,-+}_{-l} \right ) \right ] \rpmket, \nn
\label{chi state}
\eea
where the sums over the mode indices for the fermions in the Ramond sector include zero modes only when those modes do not annihilate the vacuum $\rpmket$.  Note that the various $\g$ coefficients also depend on the twist insertion points $w_1$ and $w_2$.

From here it is fairly straightforward to apply particular capping states with only a single pair of modes so as to pick out each specific $\g$ coefficient from $|\chi(w_1,w_2)\rangle$.  Taking a ratio of amplitudes, we can isolate these coefficients in a manner that is independent of other factors, such as the overall coefficients $C$.  We thus find:
\bea
\g^{B(i)(1)}_{mn} & = & -{1\over mn} {\nsnsbra \a^{(1)f}_{++,n}\a^{(i)f}_{--,m}|\chi(w_1,w_2)\rangle \over \nsnsbra \chi(w_1,w_2)\rangle} \label{BosonCylinderRatio1} \\
\g^{B(i)(2)}_{mn} & = & -{1\over mn} {\nsnsbra \a^{(i)f}_{--,m}\a^{(2)f}_{++,n}|\chi(w_1,w_2)\rangle \over \nsnsbra \chi(w_1,w_2)\rangle} \label{BosonCylinderRatio2} \\
\g^{F(i)(1)}_{NS,rs} & = & {\nsnsbra d^{(1)f,++}_{r}d^{(i)f,--}_{s}|\chi(w_1,w_2)\rangle \over \nsnsbra \chi(w_1,w_2)\rangle} \label{FermionNSCylinderRatio1} \\
\g^{F(i)(2)}_{NS,rs} & = & -{\nsnsbra d^{(i)f,--}_{s}d^{(2)f,++}_{r}|\chi(w_1,w_2)\rangle \over \nsnsbra \chi(w_1,w_2)\rangle} \label{FermionNSCylinderRatio2} \\
\g^{F(i)(1)}_{R+-,mn} & = & {\rpmbra d^{(1)f,++}_{n}d^{(i)f,--}_{m}|\chi(w_1,w_2)\rangle \over \rpmbra \chi(w_1,w_2)\rangle}\label{FermionRCylinderRatio1} \\
\g^{F(i)(2)}_{R+-,mn} & = & -{\rpmbra d^{(i)f,--}_{m}d^{(2)f,++}_{n}|\chi(w_1,w_2)\rangle \over \rpmbra \chi(w_1,w_2)\rangle},\label{FermionRCylinderRatio2}
\eea
where the minus signs in the fourth and sixth lines come from the fact that the fermion annihilators anticommute.  Note that the bosonic coefficients do not depend on the vacuum the state is built upon.  This is because the bosonic modes are unaffected by spectral flow, so they behave identically in both sectors.  Indeed, one can calculate these same bosonic coefficients by capping with a Ramond state instead.

From these expressions, we must perform the series of coordinate maps outlined in Section \ref{CoordinateMaps}.  This procedure removes all of the multivalued regions, though the Ramond vacua and twist insertions still introduce spin fields.  We deal with these spin fields in the next subsection.


\subsection{Spectral flows}
We have two sectors we wish to cap with, the NS sector and the R sector.  The behavior of the bosonic modes is independent of the sector they act on, and as such we do not have to worry about which spectral flows we perform when it comes to the bosonic coefficients.  Thus we only need to apply the coordinate maps to the right side of (\ref{BosonCylinderRatio1}) and (\ref{BosonCylinderRatio2}).  Doing so yields:
\bea
\g^{B(i)(1)}_{mn} & = & -{1\over mn} {\nstbra \a'^{(1)f}_{++,n}\a'^{(i)f}_{--,m}\nstket \over \nstbra \nstclose} \label{BosonCoverRatio1}\\
\g^{B(i)(2)}_{mn} & = & -{1\over mn} {\nstbra \a'^{(i)f}_{--,m}\a'^{(2)f}_{++,n}\nstket \over \nstbra \nstclose}, \label{BosonCoverRatio2}
\eea
where the primes denote that the modes have been altered by the coordinate shifts.

\begin{figure}[tbh]
\begin{center}
\includegraphics[width=0.4\columnwidth]{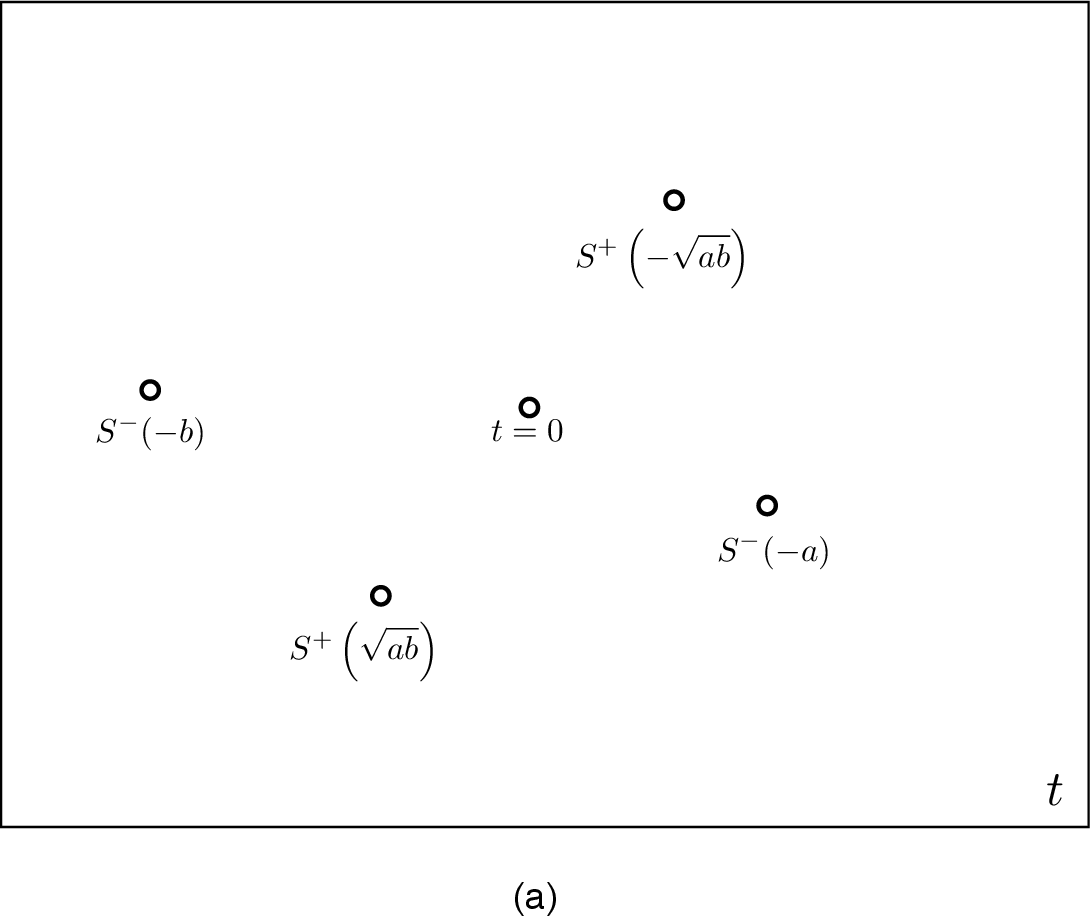} $\qquad\qquad$ \includegraphics[width=0.4\columnwidth]{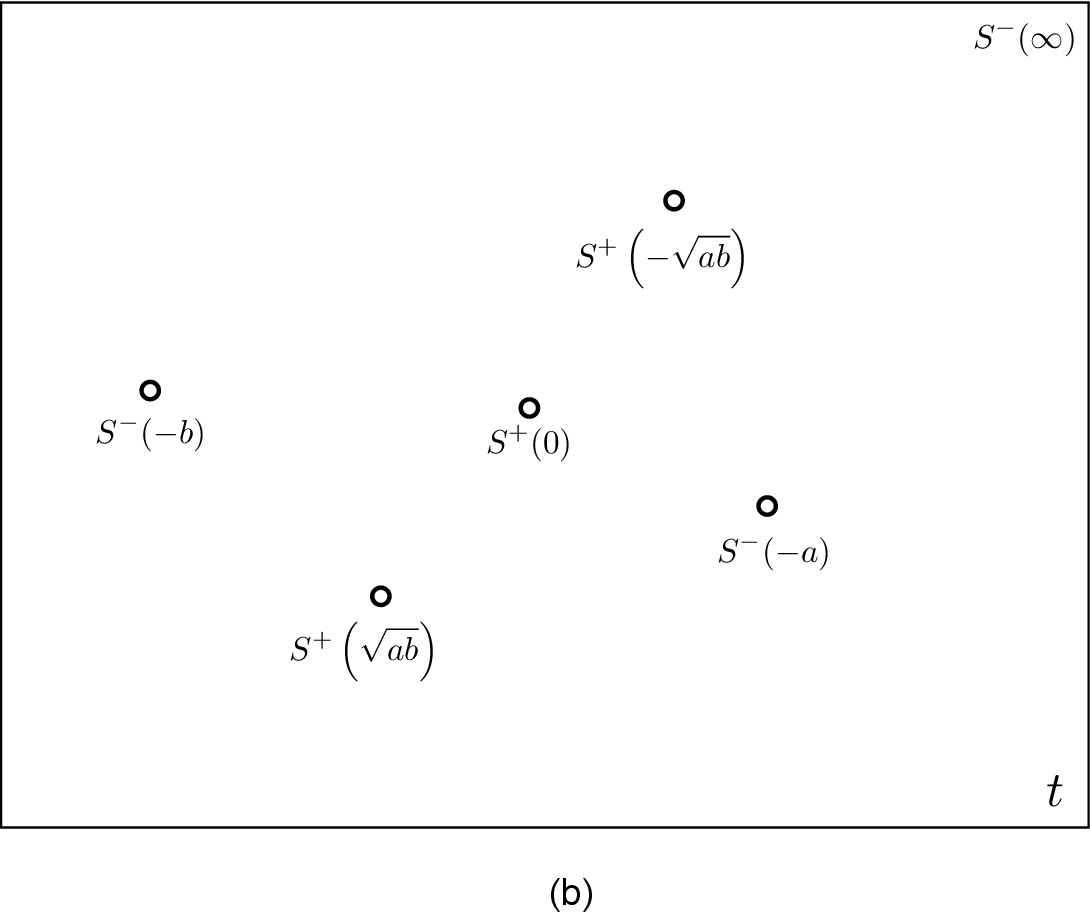}
\end{center}
\caption{Here we illustrate the insertions of all the spin fields in the $t$ plane when capping with the NS (a) and Ramond (b) sectors.}
\label{SpinInsertionFigure}
\end{figure}

For the fermionic coefficients, it is important to specify which sector we are building upon because the Ramond sector brings in additional spin fields.  Both cases are illustrated in Figure \ref{SpinInsertionFigure}.

When building upon the NS sector, the $t$-plane contains the following spin fields:
\bea\label{NSSpinLocations}
S^-(-a) &\text{from}& \rmket^{(1)} \nn
S^-(-b) &\text{from}& \rmket^{(2)} \nn
S^+(-\sqrt{ab}) &\text{from}& \s_2^+(w_1) \nn
S^+(\sqrt{ab}) &\text{from}& \s_2^+(w_2).
\label{spin field locations}
\eea
To remove these spin fields we must perform spectral flows at each of these points.  We spectral flow by $\a=+1$ at the locations of the $S^-$ spin fields, and by $\a=-1$ at the locations of the $S^+$ spin fields.  Since this combination contains two spectral flows in each direction all at finite points, there is no net effect at infinity.  This is important as we do not have any insertions at infinity.  After these spectral flows we can close all punctures in the $t$ plane, allowing us to smoothly deform the contours of the fermion modes.  Recalling (\ref{FermionNSCylinderRatio1}) and (\ref{FermionNSCylinderRatio2}), we find:
\bea
\g^{F(i)(1)}_{NS,rs} & = & {\nstbra d'^{(1)f,++}_{s}d'^{(i)f,--}_{r}\nstket \over \nstbra \nstclose} \label{FermionNSCoverRatio1}\\
\g^{F(i)(2)}_{NS,rs} & = & -{\nstbra d'^{(i)f,--}_{r}d'^{(2)f,++}_{s}\nstket \over \nstbra \nstclose}, \label{FermionNSCoverRatio2}
\eea
where here the primes denote that the modes have been altered by both the coordinate shifts and the four spectral flows given in (\ref{NSSpinLocations}).

When building upon the Ramond sector, we also have spin fields from the capping state $\rpmbra$.  Recall that for out states, copy 1 maps to $t=\infty$ while copy $2$ maps to the origin.  Also note that the lower-index charge indicated in the bra denotes that it is the conjugate of the ket of that charge, and thus has the opposite charge.  We now see that in addition to the spin fields encountered when building upon the NS vacuum, we also have:
\bea\label{RSpinLocations}
S^-(\infty) &\text{from}& {}^{(1)}\rpbra \nn
S^+(0) &\text{from}& {}^{(2)}\rmbra.
\eea
Both of these spin fields can be removed with a single spectral flow by $\a=-1$ units at the origin of the $t$ plane.  Recalling (\ref{FermionRCylinderRatio1}) and (\ref{FermionRCylinderRatio2}), we find:
\bea
\g^{F(i)(1)}_{R+-,mn} & = & {\nstbra \hat{d}^{(1)f,++}_{n}\hat{d}^{(i)f,--}_{m}\nstket \over \nstbra\nstclose}\label{FermionRCoverRatio1}\\
\g^{F(i)(2)}_{R+-,mn} & = & -{\nstbra \hat{d}^{(i)f,--}_{m}\hat{d}^{(2)f,++}_{n}\nstket \over \nstbra\nstclose},\label{FermionRCoverRatio2}
\eea
where the hats indicate that we have performed all of the coordinate maps and spectral flows required by (\ref{NSSpinLocations}) along with an additional spectral flow by $\a=-1$ units at the origin of the $t$ plane.

We now have useful expressions for the various $\g$ coefficients.  By applying the required coordinate shifts and spectral flows to the bosonic and fermionic modes and expanding the results in terms of modes natural to the $t$ plane, we can use the known commutation relations to evaluate equations (\ref{BosonCoverRatio1}) through (\ref{FermionRCoverRatio2}).  We perform these calculations in the following sections.

\section{Computation of $\g^{B(i)(j)}_{mn}$}\label{Boson}

Here we compute the bosonic $\g$ coefficients using the relations (\ref{BosonCoverRatio1}) and (\ref{BosonCoverRatio2}).  To use these expressions, we first need to compute the $\a'$ modes in the $t$ plane.  Since the bosonic modes are unaffected by spectral flow, we need    only apply the coordinate maps to the original modes on the cylinder.  We thus organize this section into three parts.  In the first part, we will apply the coordinate maps outlined in Section \ref{CoordinateMaps} to determine the expressions for the $\a'$ modes.  In the second part, we will expand the $\a'$ modes in terms of the modes natural to the $t$ plane.  In the final part, we will use these expressions in (\ref{BosonCoverRatio1}) and (\ref{BosonCoverRatio2}) to compute the bosonic $\g$ coefficients.

\subsection{Mapping the boson modes to the $t$ plane}
We wish to apply the various coordinate maps presented in Section \ref{CoordinateMaps} to the post-twist modes on the cylinder given in (\ref{BosonCylinderMode}).  Under a general coordinate transformation $w \to w'$, the bosonic field $X$ tranforms as:
\bea
X^{(i)}_{A\dot A}(w) &\to& {\diff w' \over \diff w} X^{(i)}_{A\dot A}\left ( w' \right ).
\eea
However, the jacobian brings a factor:
\bea\label{Jacobian}
\diff w &\to& \left ({\diff w' \over \diff w} \right )^{-1} \diff w'.
\eea
Thus the combination of the two behaves as:
\bea
X^{(i)}_{A\dot A}(w)\diff w &\to& X^{(i)}_{A\dot A}(w') \diff w'.
\eea

We first map the cylinder with coordinate $w$ to the plane with coordinate $z$ via
\be
z=e^w.
\ee
Since we do not make use of the initial or intermediate modes in this chapter, we just write the modes after the second twist insertion ($|z|>e^{\tau_2}$).  Here we have a contour circling ${z=\infty}$:
\be
\alpha_{A\dot A, n}^{(1)f}\to{1\over 2\pi} \oint_{z=\infty} X_{A\dot A}^{(1)f}(z) z^{n} \diff z
\ee
\be
\alpha_{A\dot A, n}^{(2)f}\to{1\over 2\pi} \oint_{z=\infty} X_{A\dot A}^{(2)f}(z) z^{n} \diff z . 
\ee

We now proceed to the covering space $t$, where $X_{A\dot A}$ will be single-valued. Using the map defined earlier,
\be
z=\frac{(t+a)(t+b)}{t}, \label{covertwo}
\ee
and again only considering the modes after the second twist insertion, we have
\be
\alpha^{(1)f}_{A\dot A, n}\to \a'^{(1)f}_{A\dot A ,n}~=~{1\over 2\pi} \oint_{t=\infty}X_{A\dot A}(t) \left(\frac{(t+a)(t+b)}{t}\right)^{n} \diff t
\label{a1f}
\ee
\be
\alpha^{(2)f}_{A\dot A, n}\to \a'^{(2)f}_{A\dot A ,n}~=~-{1\over 2\pi} \oint_{t=0}X_{A\dot A}(t) \left(\frac{(t+a)(t+b)}{t}\right)^{n} \diff t.
\label{a2f}
\ee
Note the minus sign in (\ref{a2f}).  As mentioned earlier, the copy 2 values at $z=\infty$ map to $t=0$.  This map has the leading-order behavior $z \sim \tfrac{ab}{t}$, which means $arg(z) \sim -arg(t)$.  Thus the contour changes direction when we map to the origin of the $t$ plane.  This is the source of the minus sign.

\subsection{Expanding the Bosonic Modes}
In order to expand our $\a'$ modes in terms of $\tilde{\a}$, it suffices to expand the integrand of the former in powers of $t$.  This expansion must be performed in the region where the mode lives, $t\sim\infty$ for the copy 1 modes and $t\sim0$ for the copy 2 modes.  We perform the relevant expansions for each copy below.
\subsubsection*{Copy 1, \textbf{$t\sim\infty$}}
\bea
\left(\frac{(t+a)(t+b)}{t}\right)^{n}&=&t^{n}\left(1+at^{-1}\right)^{n}\left(1+bt^{-1}\right)^{n}\nn
&=&t^{n}\sum_{j,j'\geq 0}{}^{n}C_{j}{}^{n}C_{j'}a^{j}b^{j'}t^{-j-j'}\nn
&=&\sum_{j,j'\geq 0}{}^{n}C_{j}{}^{n}C_{j'}a^{j}b^{j'}t^{n-j-j'}.
\eea
\subsubsection*{Copy 2, \textbf{$t\sim0$}}
\bea
\left(\frac{(t+a)(t+b)}{t}\right)^{n}&=&t^{-n}a^{n}b^{n}\left(1+ta^{-1}\right)^{n}\left(1+tb^{-1}\right)^{n}\nn
&=&t^{-n}a^{n}b^{n}\sum_{j,j'\geq 0}{}^{n}C_{j}{}^{n}C_{j'}a^{-j}b^{-j'}t^{j+j'}\nn
&=&\sum_{j,j'\geq 0}{}^{n}C_{j}{}^{n}C_{j'}a^{n-j}b^{n-j'}t^{j+j'-n}.
\eea
Thus (\ref{a1f}) and (\ref{a2f}) become 
\bea
\a'^{(1),f}_{A\dot{A},n}&=&\sum_{j,j'\geq 0} {}^{n}C_{j}{}^{n}C_{j'}a^{j}b^{j'} {1\over 2\pi}\oint_{t=\infty}X_{A\dot A}(t)t^{n-j-j'}\diff t\nn
&=&\sum_{j,j'\geq 0} {}^{n}C_{j}{}^{n}C_{j'}a^{j}b^{j'} \tilde{\a}_{A\dot{A},n-j-j'} \label{AlphaOneExpansion}\\
\a'^{(2),f}_{A\dot{A},n}&=&-\sum_{j,j'\geq 0} {}^{n}C_{j}{}^{n}C_{j'}a^{n-j}b^{n-j'}{1\over 2\pi}\oint_{t=\infty}X_{A\dot A}(t)t^{j+j'-n}\diff t \nn
&=&-\sum_{j,j'\geq 0} {}^{n}C_{j}{}^{n}C_{j'}a^{n-j}b^{n-j'}\tilde{\a}_{A\dot{A},j+j'-n}. \label{AlphaTwoExpansion}
\eea

\subsection{Computing $\g^{B(i)(j)}_{mn}$}
In this section we compute the $\g^{B(i)(j)}_{mn}$ coefficients. From (\ref{BosonCylinderRatio1}) and (\ref{AlphaOneExpansion}) we find: 
\bea
\g^{B(1)(1)}_{mn}=-\frac{1}{mn}\sum_{j,j',k,k'\geq 0}{}^{n}C_{j}{}^{n}C_{j'}{}^{m}C_{k}{}^{m}C_{k'}a^{j+k}b^{j'+k'}\frac{\nstbra\tilde{\a}_{++,n-j-j'}\tilde{\a}_{--,m-k-k'}\nstket}{\nstbra\nstclose}.\nn\label{Gamma11TripleSum}
\eea
It is clear that the summand vanishes except when
\bea
m+n-j-j'-k-k'=0~&\implies& ~k'=m+n-j-j'-k \nn
n-j-j'> 0~&\implies&~n-j > j'. \label{constraints1}
\eea
Furthermore,
\bea
j'\geq 0~&\implies& n > j\nn
k'\geq 0 ~&\implies& ~ m+n-j-j'\geq k. \label{constraints2}
\eea
We can thus rewrite (\ref{Gamma11TripleSum}) as
\bea
\g^{B(1)(1)}_{mn}=\frac{1}{mn}\sum_{j=0}^{n-1}\sum_{j'=0}^{n-j-1}\sum_{k=0}^{m+n-j-j'}(n-j-j'){}^{n}C_{j}{}^{n}C_{j'}{}^{m}C_{k}{}^{m}C_{m+n-j-j'-k}a^{j+k}b^{m+n-j-k}.\nn
\eea
Following these same steps for the other cases, we find:
\bea
\g^{B(1)(2)}_{mn}=-\frac{1}{mn}\sum_{j=0}^{n-1}\sum_{j'=0}^{n-j-1}\sum_{k=0}^{m-n+j+j'}(n-j-j'){}^{n}C_{j}{}^{n}C_{j'}{}^{m}C_{k}{}^{m}C_{m-n+j+j'-k}a^{n-j+k}b^{m+j-k}\nn
\label{gamma12boson}
\eea
\bea
\g^{B(2)(1)}_{mn}=-\frac{1}{mn}\sum_{j=0}^{n-1}\sum_{j'=0}^{n-j-1}\sum_{k=0}^{m-n+j+j'}(n-j-j'){}^{n}C_{j}{}^{n}C_{j'}{}^{m}C_{k}{}^{m}C_{m-n+j+j'-k}a^{m+j-k}b^{n-j+k}\nn
\label{gamma21boson}
\eea
\bea
\g^{B(2)(2)}_{mn}=\frac{1}{mn}\sum_{j=0}^{n-1}\sum_{j'=0}^{n-j-1}\sum_{k=0}^{m+n-j-j'}(n-j-j'){}^{n}C_{k}{}^{n}C_{m+n-j-j'-k}{}^{m}C_{j}{}^{m}C_{j'}a^{m+n-j-k}b^{j+k}.\nn
\label{gamma22boson}
\eea
The physics here is symmetric under the interchange $a\leftrightarrow b$. One can explicitly check that this symmetry holds in the above expressions, though this fact is not immediately manifest. Using this symmetry, we find
\bea
\g^{B(2)(2)}_{mn}=\frac{1}{mn}\sum_{j=0}^{n-1}\sum_{j'=0}^{n-j-1}\sum_{k=0}^{m+n-j-j'}(n-j-j'){}^{n}C_{j}{}^{n}C_{j'}{}^{m}C_{k}{}^{m}C_{m+n-j-j'-k}a^{j+k}b^{m+n-j-k},\quad\nn
\eea
and therefore
\bea
\g^{B(2)(2)}_{mn}=\g^{B(1)(1)}_{mn}.\label{BosonRelation1}
\eea
Similarly,
\bea
\g^{B(2)(1)}_{mn}=\g^{B(1)(2)}_{mn}.\label{BosonRelation2}
\eea

We will now bring the $\g^{B(i)(j)}_{mn}$ coefficients into a form that involves only one summation over a pair of hypergeometric functions, binomial coefficients, and $a$ and $b$ terms raised to the appropriate powers. We achieve this through a re-definition of the summation indices.  We present the computation for $\g^{B(1)(1)}_{mn}$ and simply skip to the results for the remaining cases.

For $\g^{B(1)(1)}_{mn}$, we begin with:
\bea
\g^{B(1)(1)}_{mn}=\frac{1}{mn}\sum_{j=0}^{n-1}\sum_{j'=0}^{n-j-1}\sum_{k=0}^{m+n-j-j'}(n-j-j'){}^{n}C_{j}{}^{n}C_{j'}{}^{m}C_{k}{}^{m}C_{m+n-j-j'-k}a^{j+k}b^{m+n-j-k}.\nn
\label{gamma11}
\eea
We now define a new summation index:
\bea
l=n-j-j'~&\implies&~j'=n-j-l,
\label{indexredefition}
\eea
where (\ref{constraints1}) and (\ref{constraints2}) now give
\bea
n-j-j' > 0 ~&\implies&~ l > 0 \nn
j'\geq 0 ~&\implies&~ n-l \geq j \nn
j,j'\geq 0 ~&\implies&~ n \geq l.
\eea
Applying these constraints, we find:
\bea
\g^{B(1)(1)}_{mn}=\frac{1}{mn}\sum_{l=1}^{n}lb^{m+n}\left(\sum_{j=0}^{n-l}{}^{n}C_{j}{}^{n}C_{n-j-l}a^{j}b^{-j}\right)\left(\sum_{k=0}^{m+l}{}^{m}C_{k}{}^{m}C_{m+l-k}a^{k}b^{-k}\right).
\label{Gamma11_2}
\eea
The $k$ and $j$ sums can be evaluated in \textit{Mathematica}\footnote{Mathematica sometimes has trouble dealing with the fact that some of the binomial coefficients vanish when the second argument is negative.  It is occasionally necessary to alter the range of the $j$ sum to explicitly remove these vanishing terms in order to obtain the full result.}, resulting in:\footnote{While it is not immediately apparent, these expressions are symmetric under $a \leftrightarrow b$, as expected.}
\bea
\g^{B(1)(1)}_{mn}&=&\frac{1}{mn}\sum_{l=1}^{n}l{}^{n}C_{l}{}^{m}C_{l}a^{l}b^{m+n-l}{}_2F_1\left(-n,l-n;l+1;\frac{a}{b}\right){}_2F_1\left(-m,l-m;l+1;\frac{a}{b}\right).\nn
\eea

The other $\g^{B(i)(j)}$ coefficients are calculated in the same manner, using the same index re-definition.  We present the results here.
\bea
\g^{B(1)(2)}_{mn}&=&-\frac{1}{mn}\sum_{l=1}^{n}l{}^{m}C_{l}{}^{n}C_{l}a^{n}b^{m}{}_2F_1\left(-m,l-m;l+1;\frac{a}{b}\right) {}_2F_1\left(-n,l-n;l+1;\frac{b}{a}\right)\nn\nn
\g^{B(2)(1)}_{mn}&=&-\frac{1}{mn}\sum_{l=1}^{n}l{}^{m}C_{l}{}^{n}C_{l}a^{m}b^{n}{}_2F_1\left(-m,l-m;l+1;\frac{b}{a}\right){}_2F_1\left(-n,l-n;l+1;\frac{a}{b}\right)\nn\nn
\g^{B(2)(2)}_{mn}&=&\frac{1}{mn}\sum_{l=1}^{n}l{}^{n}C_{l}{}^{m}C_{l} a^{m+n-l}b^{l}{}_2F_1\left(-m,l-m;l+1;\frac{b}{a}\right){}_2F_1\left(-n,l-n;l+1;\frac{b}{a}\right). \nn
\eea

In addition to (\ref{BosonRelation1}) and (\ref{BosonRelation2}), the above forms allow us to show the additional relation:
\bea
\g^{B(1)(1)}_{mn} &=& - \g^{B(1)(2)}_{mn}.
\eea

\section{Computation of $\g^{F(i)(j)}_{NS,mn}$}\label{sec:f}
Here we perform the computation of the non-physical $\g^{F(i)(j)}_{NS,mn}$ coefficients, where the $NS$ indicates that we are capping the squeezed state with the non-physical $NS$ vacuum for both copy $1$ and copy $2$. We proceed in a manner identical to the boson case, except that the fermion modes are affected by the spectral flows which were needed to remove the spin fields in (\ref{NSSpinLocations}).

\subsection{Mapping the fermion modes to the $t$ plane}
The NS sector fermion modes on the cylinder were given in (\ref{FermionNSCylinderMode}).  Here we map these modes to the $t$ plane by way of the $z$ plane.

Under a general coordinate transformation $w\to w'$, the fermion field $\psi$ transforms as:
\bea
\psi^{(i),\a A}(w) \to \left ( {\diff w' \over \diff w}\right )^{\h} \psi^{(i),\a A}(w'),
\eea
while the Jacobian again brings the factor described in (\ref{Jacobian}).  Thus the two together transform as:
\bea
\psi^{(i),\a A}(w)\diff w \to \left ( {\diff w' \over \diff w}\right )^{-\h} \psi^{(i),\a A}(w')\diff w'.
\eea
Thus when mapping to the $z$ plane via $z = e^w$, we have
\bea
\psi^{\alpha A}(w)\diff w \to z^{-1/2}\psi^{\alpha A}(z) \diff z,
\eea
and the fermionic modes become:
\bea
d_{r}^{(1)f,\a A}&\to&\frac{1}{2\pi i}\int_{\sigma=0}^{2\pi} 
\psi^{(1)f,\a A}(z)z^{r-1/2}\diff z\nn
d_{r}^{(2)f,\a A}&\to&\frac{1}{2\pi i}\int_{\sigma=0}^{2\pi}
\psi^{(2)f,\a A}(z)z^{r-1/2}\diff z.\nn
\eea

We now move to the covering $t$ plane in the same manner, using the map (\ref{tmap}).  We then find:  
\bea
d_{r}^{(1)f,\a A}&\to&\frac{1}{2\pi i}\int_{t=\infty}
\psi^{(1)f,\a A}(t)t^{-r-1/2}(t+\sqrt{ab})^{1/2}(t-\sqrt{ab})^{1/2}(t+a)^{r-1/2}(t+b)^{r-1/2}\diff t\nn
d_{r}^{(2)f,\a A}&\to&\!-\frac{1}{2\pi i}\int_{t=0}
\psi^{(2)f,\a A}(t)t^{-r-1/2}(t+\sqrt{ab})^{1/2}(t-\sqrt{ab})^{1/2}(t+a)^{r-1/2}(t+b)^{r-1/2}\diff t,\nn
\eea
where again the copy 2 mode gains a minus sign from the reversal of the direction of the contour as $z \sim t^{-1}$.

\subsection{Applying the spectral flows}
We now perform the spectral flows required to eliminate the spin fields in the $t$ plane (\ref{NSSpinLocations}).  Under a spectral flow by $\a$ units at a point $t_0$, the fermion field $\psi$ behaves as:
\bea
\psi^{\pm A}(t)\to(t-t_0)^{\mp \h \a}\psi^{\pm A}(t).
\eea
We perform each of the four spectral flows in turn.

\subsubsection*{\underline{$\a = 1$ at $t=-b$}}
\bea
d_{r}^{(1)f,+A}&\to&\frac{1}{2\pi i}\int_{t=\infty}\psi^{(1)f,+A}(t)t^{-r-1/2}(t+\sqrt{ab})^{1/2}(t-\sqrt{ab})^{1/2}(t+a)^{r-1/2}(t+b)^{r-1}\diff t\nn
d_{r}^{(1)f,-A}&\to&\frac{1}{2\pi i}\int_{t=\infty}\psi^{(1)f,-A}(t)t^{-r-1/2}(t+\sqrt{ab})^{1/2}(t-\sqrt{ab})^{1/2}(t+a)^{r-1/2}(t+b)^{r}\diff t\nn
d_{r}^{(2)f,+A}&\to&-\frac{1}{2\pi i}\int_{t=0}\psi^{(2)f,+A}(t)t^{-r-1/2}(t+\sqrt{ab})^{1/2}(t-\sqrt{ab})^{1/2}(t+a)^{r-1/2}(t+b)^{r-1}\diff t\nn
d_{r}^{(2)f,-A}&\to&-\frac{1}{2\pi i}\int_{t=0}\psi^{(2)f,-A}(t)t^{-r-1/2}(t+\sqrt{ab})^{1/2}(t-\sqrt{ab})^{1/2}(t+a)^{r-1/2}(t+b)^{r}\diff t.\nn
\eea

\subsubsection*{\underline{$\a = 1$ at $t=-a$}}
\bea
d_{r}^{(1)f,+A}&\to&\frac{1}{2\pi i}\int_{t=\infty}\psi^{(1)f,+A}(t)t^{-r-1/2}(t+\sqrt{ab})^{1/2}(t-\sqrt{ab})^{1/2}(t+a)^{r-1}(t+b)^{r-1}\diff t\nn
d_{r}^{(1)f,-A}&\to&\frac{1}{2\pi i}\int_{t=\infty}\psi^{(1)f,-A}(t)t^{-r-1/2}(t+\sqrt{ab})^{1/2}(t-\sqrt{ab})^{1/2}(t+a)^{r}(t+b)^{r}\diff t\nn
d_{r}^{(2)f,+A}&\to&-\frac{1}{2\pi i}\int_{t=0}\psi^{(2)f,+A}(t)t^{-r-1/2}(t+\sqrt{ab})^{1/2}(t-\sqrt{ab})^{1/2}(t+a)^{r-1}(t+b)^{r-1}\diff t\nn
d_{r}^{(2)f,-A}&\to&-\frac{1}{2\pi i}\int_{t=0}\psi^{(2)f,-A}(t)t^{-r-1/2}(t+\sqrt{ab})^{1/2}(t-\sqrt{ab})^{1/2}(t+a)^{r}(t+b)^{r}\diff t.\nn
\eea

\subsubsection*{\underline{$\a = -1$ at $t=-\sqrt{ab}$}}
\bea
d_{r}^{(1)f,+A}&\to&\frac{1}{2\pi i}\int_{t=\infty}\psi^{(1)f,+A}(t)t^{-r-1/2}(t+\sqrt{ab})(t-\sqrt{ab})^{1/2}(t+a)^{r-1}(t+b)^{r-1}\diff t\nn
d_{r}^{(1)f,-A}&\to&\frac{1}{2\pi i}\int_{t=\infty}\psi^{(1)f,-A}(t)t^{-r-1/2}(t-\sqrt{ab})^{1/2}(t+a)^{r}(t+b)^{r}\diff t\nn
d_{r}^{(2)f,+A}&\to&-\frac{1}{2\pi i}\int_{t=0}\psi^{(2)f,+A}(t)t^{-r-1/2}(t+\sqrt{ab})(t-\sqrt{ab})^{1/2}(t+a)^{r-1}(t+b)^{r-1}\diff t\nn
d_{r}^{(2)f,-A}&\to&-\frac{1}{2\pi i}\int_{t=0}\psi^{(2)f,-A}(t)t^{-r-1/2}(t-\sqrt{ab})^{1/2}(t+a)^{r}(t+b)^{r}\diff t.
\eea

\subsubsection*{\underline{$\a = -1$ at $t=\sqrt{ab}$}}
\bea
d_{r}^{(1)f,+A}&\to&d'^{(1)f,+A}_{r}~=~\frac{1}{2\pi i}\int_{t=\infty}\psi^{(1)f,+A}(t)t^{-r-1/2}(t^{2}-ab)(t+a)^{r-1}(t+b)^{r-1}\diff t\nn
d_{r}^{(1)f,-A}&\to&d'^{(1)f,-A}_{r}~=~\frac{1}{2\pi i}\int_{t=\infty}\psi^{(1)f,-A}(t)t^{-r-1/2}(t+a)^{r}(t+b)^{r}\diff t\nn
d_{r}^{(2)f,+A}&\to&d'^{(2)f,+A}_{r}~=~-\frac{1}{2\pi i}\int_{t=0}\psi^{(2)f,+A}(t)t^{-r-1/2}(t^{2}-ab)(t+a)^{r-1}(t+b)^{r-1}\diff t\nn
d_{r}^{(2)f,-A}&\to&d'^{(2)f,-A}_{r}~=~-\frac{1}{2\pi i}\int_{t=0}\psi^{(2)f,-A}(t)t^{-r-1/2}(t+a)^{r}(t+b)^{r}\diff t.\label{fermionmodes}
\eea
We have now removed all spin field insertions in the $t$-plane and can close all punctures with the state $\nstket$.  This allows us to smoothly deform our contours.

\subsection{Expanding the fermion modes}
The operators $d'^{(1)f,\pm A}_r$ are given by contour integrals at large $t$ while the operators $d'^{(2)f,\pm A}_r$ are given by contour integrals around $t=0$. We need to express these operators in terms of the fermion modes natural to the $t$ plane, which are contour integrals of powers of $t$.  We thus expand   the integrand of each mode in powers of $t$ around its appropriate region in the $t$-plane.

\subsubsection*{$\underline{d'^{(1)f,+A}_{r}}$}
\bea
t^{-r-1/2}(t^{2}-ab)(t+a)^{r-1}(t+b)^{r-1} &=& t^{r-5/2}(t^{2}-ab)\sum_{q,q'\geq 0}^{\infty} {}^{r-1}C_q{}^{r-1}C_{q'} a^q b^{q'} t^{-q-q'} \nn
&=&\sum_{q,q'\geq 0} {}^{r-1}C_q{}^{r-1}C_{q'} a^q b^{q'} t^{r-q-q'-1/2}\nn
&&-\sum_{q,q'\geq 0} {}^{r-1}C_q{}^{r-1}C_{q'} a^{q+1} b^{q'+1} t^{r-q-q'-5/2}. \nn
\eea

\subsubsection*{$\underline{d'^{(1)f,-A}_{r}}$}
\bea
t^{-r-1/2}(t+a)^{r}(t+b)^{r} &=& t^{r-1/2}\sum_{q,q'\geq 0} {}^{r}C_q{}^{r}C_{q'} a^{q} b^{q'} t^{-q-q'} \nn
&=&\sum_{q,q\geq 0} {}^{r}C_q{}^{r}C_{q'} a^{q} b^{q'} t^{r-q-q'-1/2}.
\eea
\subsubsection*{$\underline{d'^{(2)f,+A}_{r}}$}
\bea
&&\!\!\!\!\!\!t^{-r-1/2}(t^{2}-ab)(t+a)^{r-1}(t+b)^{r-1}\cr
&&=t^{-r-1/2}(t^{2}-ab)a^{r-1}b^{r-1}\sum_{q,q'\geq 0} {}^{r-1}C_q{}^{r-1}C_{q'} a^{-q} b^{-q'} t^{q+q'} \nn
&&=\sum_{q,q'\geq 0}{}^{r-1}C_q{}^{r-1}C_{q'} a^{r-q-1} b^{r-q'-1} t^{q+q'-r+3/2}\nn
&&-\sum_{q,q'\geq 0} {}^{r-1}C_q{}^{r-1}C_{q'} a^{r-q} b^{r-q'} t^{q+q'-r-1/2}. \nn
\eea
\subsubsection*{$\underline{d'^{(2)f,-A}_{r}}$}
\bea
t^{-r-1/2}(t+a)^{r}(t+b)^{r} &=& t^{-r-1/2}a^r b^r\sum_{q,q'\geq 0} {}^{r}C_q{}^{r}C_{q'} a^{-q} b^{-q'} t^{q+q'} \nn
&=&\sum_{q,q'\geq 0} {}^{r}C_q{}^{r}C_{q'} a^{r-q} b^{r-q'} t^{q+q'-r-1/2}.
\eea

With these expansions, (\ref{fermionmodes}) becomes:
\bea
d'^{(1)f,+A}_{r}&=&\sum_{q,q'\geq 0} {}^{r-1}C_q{}^{r-1}C_{q'} a^q b^{q'} \frac{1}{2\pi i}\int_{t=\infty}\psi^{(1)f,+A}(t)t^{r-q-q'-1/2}\diff t\nn 
&&-\sum_{q,q\geq 0} {}^{r-1}C_q{}^{r-1}C_{q'} a^{q+1} b^{q'+1}\frac{1}{2\pi i}\int_{t=\infty}\psi^{(1)f,+A}(t)t^{r-q-q'-5/2}\diff t\nn
d'^{(1)f,-A}_{r}&\to&\sum_{q,q'\geq 0} {}^{r}C_q{}^{r}C_{q'} a^{q} b^{q'}\frac{1}{2\pi i}\int_{t=\infty}\psi^{(1)f,-A}(t)t^{r-q-q'-1/2}\diff t\nn
d'^{(2)f,+A}_{r}&\to&-\sum_{q,q'\geq 0}{}^{r-1}C_q{}^{r-1}C_{q'} a^{r-q-1} b^{r-q'-1} \frac{1}{2\pi i}\int_{t=0}\psi^{(2)f,+A}(t)t^{q+q'-r+3/2}\diff t\nn
&&+\sum_{q,q'\geq 0} {}^{r-1}C_q{}^{r-1}C_{q'} a^{r-q} b^{r-q'}\frac{1}{2\pi i}\int_{t=0}\psi^{(2)f,+A}(t)t^{q+q'-r-1/2}\diff t\nn
d'^{(2)f,-A}_{r}&\to&-\sum_{q,q'\geq 0} {}^{r}C_q{}^{r}C_{q'} a^{r-q} b^{r-q'}\frac{1}{2\pi i}\int_{t=0}\psi^{(2)f,-A}(t)t^{q+q'-r-1/2}\diff t.
\eea
In terms of modes natural to the $t$-plane, we have:
\bea
d'^{(1)f,+A}_{r}&=&\sum_{q,q'\geq 0} {}^{r-1}C_q{}^{r-1}C_{q'} a^q b^{q'} \tilde{d}^{+A}_{r-q-q'}-\sum_{q,q'\geq 0} {}^{r-1}C_q{}^{r-1}C_{q'} a^{q+1} b^{q'+1}\tilde{d}_{r-q-q'-2}^{+A}\nn
d'^{(1)f,-A}_{r}&=&\sum_{q,q'\geq 0} {}^{r}C_q{}^{r}C_{q'} a^{q} b^{q'}\tilde{d}^{-A}_{r-q-q'}\nn
d'^{(2)f,+A}_{r}&=&-\sum_{q,q'\geq 0}{}^{r-1}C_q{}^{r-1}C_{q'} a^{r-q-1} b^{r-q'-1} \tilde{d}^{+A}_{q+q'-r+2}+\sum_{q,q'\geq 0} {}^{r-1}C_q{}^{r-1}C_{q'} a^{r-q} b^{r-q'}\tilde{d}^{+A}_{q+q'-r}\nn
d'^{(2)f,-A}_{r}&=&-\sum_{q,q'\geq 0} {}^{r}C_q{}^{r}C_{q'} a^{r-q} b^{r-q'}\tilde{d}^{-A}_{q+q'-r}.\label{tplanemodes}
\eea

\subsection{Computing $\g^{F(i)(j)}_{NS,rs}$}
Here we shall compute the $\g^{F(i)(j)}_{NS}$ coefficients. We present the computation for $\g^{F(1)(1)}_{NS}$ explicitly. The other computations are performed in the same way, and we give only the results. Using (\ref{FermionNSCoverRatio1}) and (\ref{tplanemodes}), we find:
\bea
\g^{F(1)(1)}_{NS,rs}&=&\sum_{p,p',q,q'\geq 0}{}^{s-1}C_q{}^{s-1}C_{q'}{}^{r}C_p{}^{r}C_{p'}a^{q+p}b^{q'+p'}~\frac{\nstbra\tilde{d}^{++}_{s-q-q'}\tilde{d}^{--}_{r-p-p'}\nstket}{\nstbra\nstclose_{t}}\nn
&&{}-\sum_{p,p',q,q'\geq 0}{}^{s-1}C_q{}^{s-1}C_{q'}{}^{r}C_p{}^{r}C_{p'}a^{q+p+1}b^{q'+p'+1}~\frac{\nstbra\tilde{d}_{s-q-q'-2}^{++}\tilde{d}^{--}_{r-p-p'}\nstket}{\nstbra\nstclose_{t}}.\nn
\label{gamma11NS}
\eea
For the first term, it is clear that the summand vanishes except when:
\bea
r+s-p-p'-q-q'=0~&\implies& ~p'=r+s-q-q'-p\nn
s-q-q' > 0 ~&\implies& ~ s-q > q'.
\eea
Furthermore,
\bea
q'\geq 0 ~&\implies&~ s > q \nn
p' \geq 0 ~&\implies&~ r+s-q-q' \geq p.
\eea
We can thus rewrite (\ref{gamma11NS}) as:
\bea
\g^{F(1)(1)}_{NS,rs}&=&-\sum_{q=0}^{\lfloor s\rfloor}\sum_{q'=0}^{\lfloor s\rfloor-q}\,\sum_{p=0}^{r+s-q-p}{}^{s-1}C_q{}^{s-1}C_{q'}{}^{r}C_p{}^{r}C_{r+s-q-q'-p}a^{q+p}b^{r+s-q-q'} \nn
&&{}-\sum_{p,p',q,q'\geq 0}{}^{s-1}C_q{}^{s-1}C_{q'}{}^{r}C_p{}^{r}C_{p'}a^{q+p+1}b^{q'+p'+1}~\frac{\langle 0_{NS}|\tilde{d}_{s-q-q'-2}^{++}\tilde{d}^{--}_{r-p-p'}|0_{NS}\rangle_{t}}{\nstbra\nstclose}.\nn
\label{gamma11NS_2}
\eea
Applying the same line of reasoning for the second term, we find:
\bea
r+s-p-p'-q-q'-2=0~&\implies& ~p'=r+s-q-q'-p-2\nn
s-q-q'-2 >0~&\implies& s-q-2> q'\nn
p'\geq 0 ~&\implies& ~ r+s-q-q'-2\geq p\nn
q'\geq 0~&\implies& s-2> q.
\eea
So (\ref{gamma11NS_2}) becomes
\bea
\g^{F(1)(1)}_{NS,rs}&=&-\sum_{q=0}^{\lfloor s\rfloor}\sum_{q'=0}^{\lfloor s\rfloor-q}\,\sum_{p=0}^{r+s-q-p}{}^{s-1}C_q{}^{s-1}C_{q'}{}^{r}C_p{}^{r}C_{r+s-q-q'-p}a^{q+p}b^{r+s-q-q'}\nn
&&+\sum_{q=0}^{\lfloor s\rfloor-2}\sum_{q'=0}^{\lfloor s\rfloor-q-2}\,\sum_{p=0}^{r+s-q-q'-2}{}^{s-1}C_q{}^{s-1}C_{q'}{}^{r}C_p{}^{r}C_{r+s-q-q'-p-2}a^{q+p+1}b^{r+s-p-q-1},\nn
\eea
where the second term is instead zero when the sums' ranges are invalid ($s \leq \tfrac{3}{2}$).  We shall adopt this convention generally.  Henceforth, all sums are to be treated as zero when the given range is invalid. 

The other cases proceed along the same lines.
\bea
\g^{F(1)(2)}_{NS,rs}&=&-\sum_{q=0}^{\lfloor s\rfloor-2}\sum_{q'=0}^{\lfloor s\rfloor-q-2}\,\sum_{p=0}^{r-s+q+q'+2}{}^{s-1}C_{q}{}^{s-1}C_{q'}{}^{r}C_{p}{}^{r}C_{r-s+q+q'-p+2}a^{s-q+p-1}b^{r+q-p+1}\nn
&&+\sum_{q=0}^{\lfloor s\rfloor}\sum_{q'=0}^{\lfloor s\rfloor-q}\,\sum_{p=0}^{r-s+q+q'}{}^{s-1}C_{q}{}^{s-1}C_{q'}{}^{r}C_{p}{}^{r}C_{r-s+q+q'-p}a^{s-q+p}b^{r+q-p}
\label{gamma12NS}
\eea
\bea
\g^{F(2)(1)}_{NS,rs}&=&-\sum_{q=0}^{\lfloor s\rfloor-2}\sum_{q'=0}^{\lfloor s\rfloor-q-2}\,\sum_{p=0}^{r-s+q+q'+2}{}^{s-1}C_{q}{}^{s-1}C_{q'}{}^{r}C_{p}{}^{r}C_{r-s+q+q'-p+2}a^{r+q-p+1}b^{s-q+p-1}\nn
&&+\sum_{q=0}^{\lfloor s\rfloor}\sum_{q'=0}^{\lfloor s\rfloor-q}\,\sum_{p=0}^{r-s+q+q'}{}^{s-1}C_{q}{}^{s-1}C_{q'}{}^{r}C_{p}{}^{r}C_{r-s+q+q'-p}a^{r+q-p}b^{s-q+p}
\label{gamma21NS}
\eea
\bea
\g^{F(2)(2)}_{NS,rs}&=&-\sum_{q=0}^{\lfloor s\rfloor}\sum_{q'=0}^{\lfloor s\rfloor-q}\,\sum_{p=0}^{r+s-q-q'}{}^{s-1}C_q{}^{s-1}C_{q'}{}^{r}C_p{}^{r}C_{r+s-q-q'-p}a^{r+s-q-p}b^{q+p}\nn
&&\,\,+\sum_{q=0}^{\lfloor s\rfloor-2}\sum_{q'=0}^{\lfloor s\rfloor-q-2}\,\sum_{p=0}^{r+s-q-q'-2}{}^{s-1}C_q{}^{s-1}C_{q'}{}^{r}C_p{}^{r}C_{r+s-q-q'-p-2}a^{r+s-p-q-1}b^{q+p+1}.\nn
\label{gamma22NS}
\eea
Using the interchange symmetry $a \leftrightarrow b$, we find:
\bea
\g^{F(2)(2)}_{NS,rs}&=&\g^{F(1)(1)}_{NS,rs} \\
\g^{F(1)(2)}_{NS,rs}&=&\g^{F(2)(1)}_{NS,rs}.
\eea

Just as in the bosonic case, we shall now bring the $\g^{F(i)(j)}_{NS}$ coefficients into a form that involves only one summation over a pair of hypergeometric functions, binomial coefficients, and $a$ and $b$ terms raised to the appropriate powers. We present the computation for $i=j=1$ and simply give the results for the remaining cases. We start with:
\bea
\g^{F(1)(1)}_{NS,rs}&=&-\sum_{q=0}^{\lfloor s\rfloor}\sum_{q'=0}^{\lfloor s\rfloor-q}\,\sum_{p=0}^{r+s-q-q'}{}^{s-1}C_q{}^{s-1}C_{q'}{}^{r}C_p{}^{r}C_{r+s-q-q'-p}a^{q+p}b^{r+s-q-q'}\nn
&&+\sum_{q=0}^{\lfloor s\rfloor-2}\sum_{q'=0}^{\lfloor s\rfloor-q-2}\,\sum_{p=0}^{r+s-q-q'-2}{}^{s-1}C_q{}^{s-1}C_{q'}{}^{r}C_p{}^{r}C_{r+s-q-q'-p-2}a^{q+p+1}b^{r+s-p-q-1}.\nn
\eea
We now redefine our summation indices:
\bea
&&\text{Term 1: }\, j\equiv s-q-q'\implies q'=s-q-j\cr
&&\text{Term 2: }\, j\equiv s-q-q'-2\implies q'=s-q-j-2,
\label{redefintionsNS}
\eea
where in both cases $j$ is a positive half-integer. Since $q$ and $q'$ are both non-negative, we find:
\bea
&&\text{Term 1: }\,q'\geq 0~\implies~ q\leq s-j, \qquad\quad~~~\! q\geq 0~\implies~ j\leq s\nn
&&\text{Term 2: }\,q'\geq 0~\implies~ q\leq s-j-2, \qquad q\geq 0~\implies~ j\leq s-2.
\eea
This gives:
\bea
\g^{F(1)(1)}_{NS,rs}&=&-\sum_{j=1/2}^{s}b^{r+s}\left(\sum_{q=0}^{s-j}{}^{s-1}C_q{}^{s-1}C_{s-q-j}a^{q}b^{-q}\right)\left(\sum_{p=0}^{r+j}{}^{r}C_p{}^{r}C_{r+j-p}a^{p}b^{-p}\right) \cr
&&+\sum_{j=1/2}^{s-2}ab^{r+s-1}\left(\sum_{q=0}^{s-j-2}{}^{s-1}C_q{}^{s-1}C_{s-q-j-2}a^{q}b^{-q}\right)\left(\sum_{p=0}^{r+j}{}^{r}C_p{}^{r}C_{r+j-p}a^{p}b^{-p}\right).\nn
\eea
Evaluating the $q$ and $p$ sums in \textit{Mathematica}, we find:
\bea
\g^{F(1)(1)}_{NS,rs}&=&-\sum_{j=\h}^{s}{}^{s-1}C_{s-j}{}^{r}C_{r+j}b^{r+s} \nn
&&\quad{}\times{}_2F_1\left(1-s,j-s;j;\frac{a}{b}\right) {}_2F_1\left(-r,-j-r;1-j;\frac{a}{b}\right)\nn
&&{}+\sum_{j=\h}^{s-2}{}^{s-1/2}C_{s-j-2} {}^{r}C_{r+j}a b^{r+s-1}\nn
&&\quad{}\times {}_2F_1\left(1-s,j-s+2;j+2;\frac{a}{b}\right){}_2F_1\left(-r,-j-r;1-j;\frac{a}{b}\right).\nn
\eea

The other $\g^{F(i)(j)}_{NS,rs}$ coefficients are handled in the same manner. We present the final result along with the appropriate index redefinition that was made in each case.  For $\g^{F(1)(2)}_{NS,rs}$, we find:
\bea
\g^{F(1)(2)}_{NS,rs}&=&-\sum_{j=\h}^{s-2}{}^{s-1}C_{s-j-2}{}^{r}C_{r-j}a^{s-1}b^{r+1}\nn
&&{}\times_2F_1\left(-r,j-r;j;\frac{a}{b}\right)\,_2F_1\left(1-s,j-s+2;j+2;\frac{b}{a}\right)\nn
&&{}+\sum_{j=\h}^{s}{}^{s-1}C_{s-j}{}^{r}C_{r-j}a^{s}b^{r}\nn
&&{}\times_2F_1\left(-r,j-r;j+1;\frac{a}{b}\right)\,_2F_1\left(1-s,j-s;j;\frac{b}{a}\right),\nn
\eea
where in (\ref{gamma12NS}), we substituted:
\bea
&&\text{Term 1: }\,q'=s+j-q-2\nn
&&\text{Term 2: }\,q'=s+j-q.
\eea
For $\g^{F(2)(1)}_{NS,rs}$, we find:
\bea
\g^{F(2)(1)}_{NS,rs}&=&\sum_{j=\h}^{s}{}^{s-1}C_{s-j}{}^{r}C_{r-j}a^{r}b^{s}\nn
&&{}\times{}_2F_1\left(-r,j-r;j+1;\frac{b}{a}\right){}_2F_1\left(1-s,j-s;j;\frac{a}{b}\right)\nn
&&{}-\sum_{j=\h}^{s-2}{}^{s-1}C_{s-j-2}{}^{r}C_{r-j}a^{r+1}b^{s-1}\nn
&&{}\times {}_2F_1\left(-r-,j-r;j+1;\frac{b}{a}\right){}_2F_1\left(1-s,j-s+2;j+2;\frac{a}{b}\right) ,
{}   \nn\nn
\eea
where in (\ref{gamma21NS}), we substituted:
\bea
&&\text{Term 1: }\,q'=s-q-j\nn
&&\text{Term 2: }\,q'=s-q-j-2.
\eea
For $\g^{F(2)(2)}_{NS,rs}$, we find:
\bea
\g^{F(2)(2)}_{NS,rs}&=&\sum_{j=\h}^{s-2}{}^{s-1}C_{s-j-2}{}^{r}C_{r+j}a^{r+s-1}b\nn
&&{}\times _2F_1\left(-r,-j-r;1-j;\frac{b}{a}\right)\,_2F_1\left(1-s,j-s+2;j+2;\frac{b}{a}\right)\nn
&&{}-\sum_{j=\h}^{s}{}^{s-1}C_{s-j}{}^{r}C_{r+j}a^{r+s}\nn
&&{}\times  _2F_1\left(-r,-j-r;1-j;\frac{b}{a}\right)\,_2F_1\left(1-s,j-s;j;\frac{b}{a}\right),\nn
\eea
where in (\ref{gamma22NS}), we substituted:
\bea
&&\text{Term 1: }\,q'=s+j-q-2\cr
&&\text{Term 2: }\,q'=s+j-q.
\eea

\section{Computation of $\g^{F(i)(j)}_{R+-}$}\label{Ramond}
Here we perform the computation of the $\g^{F(i)(j)}_{R+-,mn}$ coefficients, where the $R+-$ indicates that we are capping the squeezed state with the positive Ramond vacuum on copy $1$ and the negative Ramond vacuum on copy $2$. In order to obtain an empty t-plane from this capping state we must apply one additional spectral flow beyond those performed in the $NS$ case. We then expand the resulting modes in terms of modes natural to the $t$ plane in order to compute $\g^{F(i)(j)}_{R+-,mn}$.

\subsection{Modes in t-plane from additional spectral flow by $\a=-1$}
Here we obtain the $t$-plane modes after one additional spectral flow of $\a=-1$ around the point $t=0$. We remind the reader that for the $NS$ sector computation, we performed four spectral flows:
\bea
&&\a=1\text{ around }t=-b\nn
&&\a=1\text{ around }t=-a\nn
&&\a=-1\text{ around }t=-\sqrt{ab}\nn
&&\a=-1\text{ around }t=\sqrt{ab}.
\eea
The resulting modes were given by
\bea
d'^{(1)f,+A}_{r}&=&\frac{1}{2\pi i}\int_{t=\infty}\psi^{(1)f,+A}(t)t^{-r-1/2}(t^{2}-ab)(t+a)^{r-1}(t+b)^{r-1}\diff t\nn
d'^{(1)f,-A}_{r}&=&\frac{1}{2\pi i}\int_{t=\infty}\psi^{(1)f,-A}(t)t^{-r-1/2}(t+a)^{r}(t+b)^{r}\diff t\nn
d'^{(2)f,+A}_{r}&=&-\frac{1}{2\pi i}\int_{t=0}\psi^{(2)f,+A}(t)t^{-r-1/2}(t^{2}-ab)(t+a)^{r-1}(t+b)^{r-1}\diff t\nn
d'^{(2)f,-A}_{r}&=&-\frac{1}{2\pi i}\int_{t=0}\psi^{(2)f,-A}(t)t^{-r-1/2}(t+a)^{r}(t+b)^{r}\diff t.
\eea
We now apply one additional spectral flow of $\a=-1$ around the point, $t=0$. We also note that the Ramond sector has integer-indexed fermions, so we will switch from the half-integer mode index $r$ to the integer index $n$.

Under this last spectral flow, the fermion field changes as follows:
\bea
\psi^{(i)f,\pm A}(t)\to t^{\pm 1/2}\psi^{(i)f,\pm}(t).
\eea
The modes then become
\bea
d'^{(1)f,+A}_{n}&\to&\hat{d}^{(1)f,+A}_{n}~=~\frac{1}{2\pi i}\int_{t=\infty}\psi^{(1)f,+A}(t)t^{-n}(t^{2}-ab)(t+a)^{n-1}(t+b)^{n-1}\diff t\nn
d'^{(1)f,-A}_{n}&\to&\hat{d}^{(1)f,-A}_{n}~=~\frac{1}{2\pi i}\int_{t=\infty}\psi^{(1)f,-A}(t)t^{-n-1}(t+a)^{n}(t+b)^{n}\diff t\nn
d'^{(2)f,+A}_{n}&\to&\hat{d}^{(2)f,+A}_{n}~=~-\frac{1}{2\pi i}\int_{t=0}\psi^{(2)f,+A}(t)t^{-n}(t^{2}-ab)(t+a)^{n-1}(t+b)^{n-1}\diff t\nn
d'^{(2)f,-A}_{n}&\to&\hat{d}^{(2)f,-A}_{n}~=~-\frac{1}{2\pi i}\int_{t=0}\psi^{(2)f,-A}(t)t^{-n-1}(t+a)^{n}(t+b)^{n}\diff t.\nn
\label{Rsector_tmodes}
\eea 

We now expand the modes (\ref{Rsector_tmodes}) in terms of modes natural to the $t$ plane.  Since the additional spectral flow produced a factor of $t^{\pm\h}$, the binomials we wish to expand are identical to the ones found in the NS case.  We thus present only the results: 
\bea
\hat{d}_{n}^{(1)f,+A}&=&\sum_{j,j'\geq 0} {}^{n-1}C_j{}^{n-1}C_{j'} a^j b^{j'} \tilde{d}^{+A}_{n-j-j'+1/2}\nn
&&{}-\sum_{j,j'\geq 0} {}^{n-1}C_j{}^{n-1}C_{j'} a^{j+1} b^{j'+1}\tilde{d}_{n-j-j'-3/2}^{+A}\nn
\hat{d}_{n}^{(1)f,-A}&=&\sum_{j,j'\geq 0} {}^{n}C_j{}^{n}C_{j'} a^{j} b^{j'}\tilde{d}^{-A}_{n-j-j'-1/2}\nn
\hat{d}_{n}^{(2)f,+A}&=&-\sum_{j,j'\geq 0}{}^{n-1}C_j{}^{n-1}C_{j'} a^{n-j-1} b^{n-j'-1} \tilde{d}^{+A}_{j+j'-n+5/2}\nn
&&{}+\sum_{j,j'\geq 0} {}^{n-1}C_j{}^{n-1}C_{j'} a^{n-j} b^{n-j'}\tilde{d}^{+A}_{j+j'-n+1/2}\nn
\hat{d}_{n}^{(2)f,-A}&=&-\sum_{j,j'\geq 0} {}^{n}C_j{}^{n}C_{j'} a^{n-j} b^{n-j'}\tilde{d}^{-A}_{j+j'-n-1/2}.\nn
\label{tplanemodesR}
\eea

\subsection{Computing $\g^{F(i)(j)}_{R+-,mn}$}
Here we show the computation of the $\g^{F(i)(j)}_{R+-}$ coefficients.
We shall explicitly show the computation for the case $i=j=1$. The other cases are computed in the same way and we shall simply state their results. Using (\ref{FermionRCoverRatio1}) and (\ref{tplanemodesR}), we find:
\bea
\g^{F(1)(1)}_{R+-,mn}&=&\sum_{j,j',k,k'\geq 0}{}^{n-1}C_k{}^{n-1}C_{k'}{}^{m}C_j{}^{m}C_{j'}a^{k+j} b^{k'+j'}~\frac{\nstbra\tilde{d}^{++}_{n-k-k'+1/2}\tilde{d}^{--}_{m-j-j'-1/2}\nstket}{\nstbra\nstclose_{t}}\nn
&&{}-\sum_{j,j',k,k'\geq 0}{}^{n-1}C_k{}^{n-1}C_{k'}  {}^{m}C_j{}^{m}C_{j'}a^{k+j+1} b^{k'+j'+1}\nn
&&\quad{}\times\frac{{}_{t}\langle 0_{NS}|\tilde{d}_{n-k-k'-3/2}^{++}\tilde{d}^{--}_{m-j-j'-1/2}|0_{NS}\rangle_{t}}{\nstbra\nstclose_{t}}.\nn
\label{gamma11R+-}
\eea
Looking at the first term, it is clear that the summand vanishes except when:
\bea
m+n-j-j'-k-k'=0~&\implies&~ j'=m+n-k-k'-j\nn
n-k-k'+\h > 0 ~&\implies&~ n-k \geq k'.
\eea
Furthermore,
\bea
k' \geq 0 ~&\implies&~ n \geq k\nn
j' \geq 0 ~&\implies&~ m+n-k-k' \geq j.
\eea
We can thus re-write (\ref{gamma11R+-}) as:
\bea
\g^{F(1)(1)}_{R+-,mn}&=&\sum_{k=0}^{n}\sum_{k'=0}^{n-k}\,\sum_{j=0}^{m+n-k-k'}{}^{n-1}C_k{}^{n-1}C_{k'}{}^{m}C_j{}^{m}C_{m+n-k-k'-j}a^{k+j}b^{m+n-j-k}\nn
&&{}-\sum_{j,j',k,k'\geq 0}{}^{n-1}C_k{}^{n-1}C_{k'}  {}^{m}C_j{}^{m}C_{j'}a^{k+j+1}b^{k'+j'+1}\nn
&&\quad{}\times\frac{{}_{t}\langle 0_{NS}|\tilde{d}_{n-k-k'-3/2}^{++}\tilde{d}^{--}_{m-j-j'-1/2}|0_{NS}\rangle_{t}}{\nstbra\nstclose}.\nn
\label{gamma11R+-_2}
\eea
Applying the same line of reasoning for the second term, we see:
\bea
m+n-j-j'-k-k'-2=0 ~&\implies&~ j'=m+n-k-k'-j-2 \nn
n-k-k'-{3\over 2} > 0 ~&\implies&~ n +2-k \geq k' \nn
j'\geq 0 ~&\implies& ~ m+n-k-k'-2\geq j.\nn
\eea
So (\ref{gamma11R+-_2}) becomes
\bea
\g^{F(1)(1)}_{R+-,mn} &=&-\sum_{k=0}^{n}\sum_{k'=0}^{n-k}\,\sum_{j=0}^{m+n-k-k'}{}^{n-1}C_k{}^{n-1}C_{k'} {}^{m}C_j{}^{m}C_{m+n-k-k'-j}a^{j+k} b^{m+n-j-k}\nn
&&{}+\sum_{k=0}^{n-2}\sum_{k'=0}^{n-k-2}\,\sum_{j=0}^{m+n-k-k'-2}{}^{n-1}C_k{}^{n-1}C_{k'}{}^{m}C_j{}^{m}C_{m+n-k-k'-j-2}a^{j+k+1}b^{m+n-j-k-1}.\nn\label{gamma11R+-_f}
\eea

Following these same steps for the other cases, we obtain:
\bea
\g^{F(1)(2)}_{R+-,mn}&=&-\sum_{k=0}^{n-3}\sum_{k'= 0}^{n-k-3}\,\sum_{j= 0}^{m-n+k+k'+2}{}^{n-1}C_{k}{}^{n-1}C_{k'}{}^{m}C_{j}{}^{m}C_{m-n+k+k'-j+2}a^{n+j-k-1}b^{m-j+k+1}\nn
&&{}+\sum_{k= 0}^{n-1}\sum_{k'= 0}^{n-k-1}\,\sum_{j=0}^{m-n+k+k'}{}^{n-1}C_{k}{}{}^{n-1}C_{k'}{}^{m}C_{j}{}^{m}C_{m-n+k+k'-j}a^{n+j-k}b^{m-j+k}\nn\label{gamma12R+-}
\eea
\bea
\gamma_{R+-,mn}^{F(2)(1)}&=&\sum_{k=0}^{n}\sum_{k'=0}^{n-k}\sum_{j=0}^{m-n+k+k'}{}^{n-1}C_{k}{}^{n-1}C_{k'}{}^{m}C_{j}{}^{m}C_{m-n+k+k'-j}a^{m-j+k}b^{n+j-k}\nn
&&{}-\sum_{k=0}^{n-2}\sum_{k'=0}^{n-k-2}\,\sum_{j=0}^{m-n+k+k'+2}{}^{n-1}C_{k}{}^{n-1}C_{k'}{}^{m}C_{j}{}^{m}C_{m-n+k+k'-j+2}a^{m-j+k+1}b^{n+j-k-1}\nn\label{gamma21R+-}
\eea
\bea
\g^{F(2)(2)}_{R+-,mn}&=&\sum_{k=0}^{n-3}\sum_{k'=0}^{n-k-3}\,\sum_{j=0}^{m+n-k-k'-2}{}^{n-1}C_{k}{}^{n-1}C_{k'}{}^{m}C_{j}{}^{m}C_{m+n-k-k'-j-2}a^{m+n-j-k-1}b^{j+k+1}\nn
&&{}-\sum_{k=0}^{n-1}\sum_{k'=0}^{n-k-1}\,\sum_{j=0}^{m+n-k-k'}{}^{n-1}C_{k}{}^{n-1}C_{k'}{}^{m}C_{j}{}^{m}C_{m+n-k-k'-j}a^{m+n-j-k}b^{j+k}.\nn\label{gamma22R+-}
\eea
The physics is symmetric under Copy 1 $\leftrightarrow$ Copy 2.  This symmetry manifests in the following relations:
\bea
\g^{F(1)(1)}_{R+-;m,0}&=&-\g^{(2)(1)}_{R+-;m,0},\quad m>1\nn
\g^{F(1)(1)}_{R+-,mn}&=&\g^{F(2)(2)}_{R+-,mn},\quad m,n > 1\nn
\g^{F(1)(2)}_{R+-,mn}&=&\g^{F(2)(1)}_{R+-,mn},\quad m,n > 1 \nn
\g^{(2)(1)}_{R+-;0,n} &=& \d_{n,0} \nn
\g^{(2)(2)}_{R+-;0,n} &=& 0. \label{symmetryrequirements}
\eea
The manner in which these relations enforce the copy symmetry is examined in detail in Appendix \ref{GammaRelations}. The fact that these relations are satisfied will become evident when we re-express the coefficients in terms of hypergeometric functions. As usual we present the computation for $i=j=1$ and simply give the results for the remaining cases.

The symmetry requirements of (\ref{symmetryrequirements}) suggests that $\g^{F(1)(1)}_{R+-,mn}$ may behave differently for $n=0$. Indeed, it is convenient to treat the $n=0$ case separately.  Setting $n=0$ in (\ref{gamma11R+-_f}), we find:
\bea
\g^{F(1)(1)}_{R+-;m,0} &=& -\sum_{j=0}^m a^j b^{m-j} \,{}^m C_j {}^m C_{m-j} ~=~ -b^m {}_2 F_1 \left ( -m,-m,1,{a\over b}\right).
\eea
When $n>0$, we begin:
\bea
&&\left[\g^{F(1)(1)}_{R+-,mn}\right]_{n > 0}\nn
&&\quad = -\sum_{k=0}^{n}\sum_{k'=0}^{n-k}\,\sum_{j=0}^{m+n-k-k'}{}^{n-1}C_k{}^{n-1}C_{k'} {}^{m}C_j{}^{m}C_{m+n-k-k'-j}a^{j+k}b^{m+n-j-k}\nn
&&\quad \quad +\sum_{k=0}^{n-2}\sum_{k'=0}^{n-k-2}\,\sum_{j=0}^{m+n-k-k'-2}{}^{n-1}C_k{}^{n-1}C_{k'}{}^{m}C_j{}^{m}C_{m+n-k-k'-j-2}a^{j+k+1} b^{m+n-j-k-1}.\nn
\eea
We now make the following index redefinitions:
\bea
&&\text{Term 1: }\,p\equiv n-k-k'+1/2\implies k'=n-k-p+1/2\nn
&&\text{Term 2: }\,p\equiv n-k-k'-3/2\implies k'=n-k-p-3/2,
\label{redefintionsR+-}
\eea
where in both cases $p$ is a positive half-integer.  We then find:
\bea
&&\text{Term 1}:\quad k'\geq 0\implies k\leq n-p+1/2~~~~\text{and}~~~~k,k'\geq 0\implies p\leq n+1/2\nn
&&\text{Term 2}:\quad k'\geq 0\implies k\leq n-p-3/2~~~~\text{and}~~~~k,k'\geq 0\implies p\leq n-3/2.\nn
\eea
This gives:
\bea
\left[\g^{F(1)(1)}_{R+-,mn}\right]_{n > 0} &=&-\sum_{p=1/2}^{n+1/2}b^{m+n}\left(\sum_{k=0}^{n-p+1/2}{}^{n-1}C_k{}^{n-1}C_{n-k-p+1/2}a^{k}b^{-k}\right)\nn
&&\qquad {}\times \left(\sum_{j=0}^{m+p-1/2}{}^{m}C_j{}^{m}C_{m+p-j-1/2}a^{j}b^{-j}\right)\nn
&&{}+\sum_{p=1/2}^{n-3/2}ab^{m+n-1}\left(\sum_{k=0}^{n-p-3/2}{}^{n-1}C_k{}^{n-1}C_{n-k-p-3/2}a^{k}b^{-k}\right)\nn
&&\qquad {}\times \left(\sum_{j=0}^{m+p-1/2}{}^{m}C_j{}^{m}C_{m+p-j-1/2}a^{j} b^{-j}\right).\nn\label{hypergeometricintermediate}
\eea
Careful examination of (\ref{hypergeometricintermediate}) reveals that the $p=\h$ terms either independantly vanish or precisely cancel for all values of $n$.  We thus shift the lower bound of the sums to $p = \tfrac{3}{2}$.  Performing the $k$ and $j$ sums in $\textit{Mathematica}$, we are left with:
\bea
&&\!\!\!\!\!\!\left[\g^{F(1)(1)}_{R+-,mn}\right]_{n>0}\cr
&&= -\sum_{p = {3\over 2}}^{n+\h}{}^{n-1}C_{n-p+\h}{}^m C_{p-\h}a^{p-\h}b^{m+n-p+\h}\nn
&&\quad{}\times {}_2 F_1 \left ( 1-n, p-n-\h; p-\h;{a\over b}\right ) {}_2 F_1 \left ( p-m-\h,-m;p+\h;{a\over b}\right ) \nn
&&{}+\sum_{p = {3\over 2}}^{n-{3\over2}}{}^{n-1}C_{n-p-{3\over2}}{}^m C_{p-\h}a^{p+\h}b^{m+n-p-\h}\nn
&&\quad{}\times {}_2 F_1 \left( 1-n, p-n+{3\over2}; p+{3\over2};{a\over b}\right) {}_2 F_1 \left( p-m-\h,-m;p+\h;{a\over b}\right).\nn
\eea

The other cases are performed similarly, and we present only the results along with the index redefinitions used.
\bea
\g^{F(1)(2)}_{R+-,mn}&=&-\sum_{p=\h}^{\min\left(m-\h,n-{5\over2}\right)}{}^{n-1}C_{n-p-{5\over2}}{}^{m}C_{m-p-\h}a^{n-1} b^{m+1}\nn
&&{}_2F_1\left(p-m+\h,-m;p+{3\over2};\frac{a}{b}\right) {}_2F_1\left(1-n,p-n+{5\over2};p+{5\over2};\frac{b}{a}\right)\nn
&&{}+\sum_{p=\h}^{\min\left(m-\h,n-\h\right)}{}^{n-1}C_{n-p-\h}{}^{m}C_{m-p-\h}a^{n}b^{m}\nn
&&{}\times {}_2F_1\left(p-m+\h,-m;p+{3\over2};\frac{a}{b}\right) \,\, \,\,_2F_1\left(1-n,p-n+\h;p+\h;\frac{b}{a}\right),\nn
\label{gammaF12}
\eea
where in (\ref{gamma12R+-}), we substituted:
\bea
&&\text{Term 1: }\,k'\equiv n+p-k-{5\over2} \nn
&&\text{Term 2: }\,k'\equiv n+p-k-\h.
\eea
\bea
\left[\g^{F(2)(1)}_{R+-,mn}\right]_{n=0}&=& \sum_{j=0}^m{}^m C_j{}^m C_{m-j}a^{m-j}b^j~=~a^m{}_2F_1\left(-m,-m,1,{b\over a}\right)\nn
\left[\g^{F(2)(1)}_{R+-,mn}\right]_{n>0}&=& \sum_{p={1\over 2}}^{\min\left(m-\h,n-\h\right)}{}^{n-1}C_{n-p-\h}{}^m C_{m-p-\h}a^m b^n\nn
&&{}\times {}_2 F_1 \left(1-n,p-n+\h;p+\h;{a\over b}\right){}_2 F_1 \left(p-m+\h,-m;p+\frac{3}{2};{b\over a}\right) \nn
&&{}-\sum_{p={1\over2}}^{\min\left(m-\h,n-{5\over2}\right)}{}^{n-1}C_{n-p-{5\over2}}{}^m C_{m-p-\h}a^{m+1}b^{n-1}\nn
&&{}\times {}_2 F_1 \left(1-n,p-n+{5\over2};p+{5\over2};{a\over b}\right){}_2 F_1 \left(p-m+\h,-m;p+{3\over2};{b\over a}\right),\nn
\label{gammaF21}
\eea
where in (\ref{gamma21R+-}), we substituted:
\bea
&&\text{Term 1: }\,k'\equiv n-p-k-\h\nn
&&\text{Term 2: }\,k'\equiv n-p-k-{5\over 2}.
\eea
\bea
\g^{F(2)(2)}_{R+-,mn} &=& \sum_{p={3\over2}}^{n+\h}{}^{n-1}C_{p-{3\over 2}}{}^mC_{m-p+\h}a^{m+n-p+\h}b^{p-\h}\nn
&&{}\times{}_2F_1\left(1-n,p-n-\h;p-\h;{b\over a}\right)\,{}_2F_1\left(p-m-\h,-m;p+\h;{b\over a}\right)\nn
&&{}- \sum_{p={3\over 2}}^{n-\frac{3}{2}}{}^{n-1}C_{p+\h}{}^mC_{m-p+\h}a^{m+n-p-\h}b^{p+\h}\nn
&&{}\times{}_2F_1\left(1-n,p-n+{3\over2};p+{3\over2};{b\over a}\right) {}_2F_1\left(p-m-\h,-m;p+\h;{b\over a}\right),\nn
\eea
where in (\ref{gamma22R+-}), we substituted:
\bea
&&\text{Term 1: }\,k'\equiv n+p-k-5/2\nn
&&\text{Term 2: }\,k'\equiv n+p-k-1/2.
\eea
Using the $a\leftrightarrow b$ symmetry, it is clear that the relations (\ref{symmetryrequirements}) are satisfied.  We also show in Appendix \ref{GammaRelations} the additional relationship:
\bea
\left[\g^{F(1)(1)}_{R+-,mn}\right]_{n>0} &=& -\left[\g^{F(2)(1)}_{R+-,mn}\right]_{m,n>0}.
\eea
We thus have only a single linearly independent fermion coefficient for the non-zero modes (along with coefficients for zero modes).

\section{NS vs R sectors in the final state}\label{NS-R Sectors}

We have started with two singly wound component strings, both in the R sector. This sector was chosen because the physical D1D5 black hole has R periodicity for its fermions. We applied two twists to this initial state, and then tried to determine the final state by taking an inner product with all possible states.

Here we found the following. We can get a nonzero inner product by taking a final state   which has two singly wound strings, each in the R sector. But we  also find a nonzero inner product if we take a final state which has two singly wound strings, each in the NS sector. Does this mean that the initial R sector state can evolve to a state in the NS sector?

Such an evolution would be surprising, since it is known that all states of the D1D5 system can be accounted for by the different choices of R sector states. So why are we finding a nonzero inner product with NS sector states? To understand this, we consider an analogy: the example of the Ising model.

 \begin{figure}[htbp]
\begin{center}
\includegraphics[scale=.52]{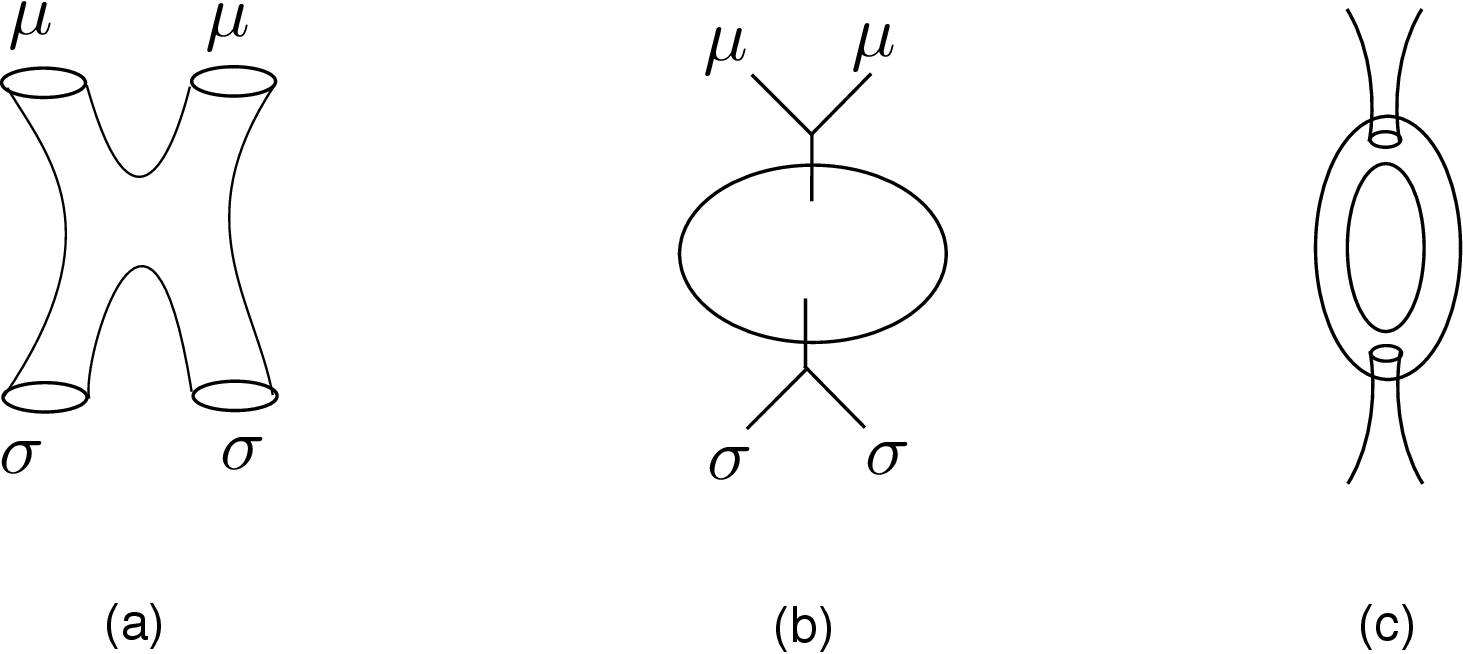}
\caption{{(a) An amplitude in the Ising model where we have put $\sigma$ states on two legs and $\mu$ states on the other two. (b) The same amplitude drawn differently, using conformal invariance. (c) A 1-loop amplitude; only a single set of complete fields should run in the loop.}}
\label{amplitude}
\end{center}
\end{figure}

The Ising model has 3 primary operators: $I, \sigma, \epsilon$. However it has a dual representation as well, where the operator order parameter $\sigma$ is replaced by the disorder parameter $\mu$. 

Let us use $\sigma$ as our basic variable; not the dual variable $\mu$. Now consider the CFT  amplitude shown in fig.\ref{amplitude}(a). The initial state has two copies of the CFT on a circle, each created by the application of a $\sigma$ operator to the identity. The final state again    has two copies of the CFT on a circle, but each state is generated by the application of $\mu$ to the identity state. Is this transition amplitude nonzero?

We can redraw this amplitude in the form shown in fig.\ref{amplitude}(b). Now it looks like the amplitude should be nonzero" the two $\sigma$ operators fuse to the identity, and the two $\mu$ operators do as well. The insertion of two identity operators into the sphere gives a nonzero amplitude. 

So should we conclude that an initial state made of two $\sigma$ operators can evolve to a state containing two $\mu$ operators? If this were true, then we would have to allow $\mu$ operators to run in loops like the one shown in fig.\ref{amplitude}(c), along with the $\sigma$ operator that can run in the loop. 

But at this stage we note that this looks like double counting: a complete set of operators is given by $I, \sigma, \epsilon$, without the inclusion of $\mu$. The operator $\mu$ does not need to be included in loops. This shows up in the fact that $\mu$ is not local with respect to $\sigma$. 

In a similar way we note that the NS and R sectors are different. The R sector state is created by applying a spin field to the NS sector state, and this spin field is not local with respect to the fermions in the theory. We can start with the NS sector, and stay in this sector all along, or we can spectral flow to the R sector, and stay in the R sector. But we should not include both sectors. The formalism we have used started with the states in the R sector, thus we should keep only the inner products we computed for transitions to the R sector, and not the ones for the NS sector.

\section{Analysis}\label{Analysis}

In the single twist case, we were able to write down explicit closed form expressions for $\gamma^B_{mn}, \gamma^F_{mn}$. In the 2-twist case, 
we do not have such closed form expressions.  In the single twist case the dependence on the position of the twist was trivial; it was just an overall phase. In the 2-twist case we expect a nontrivial dependence on the      separation between the two twists $\Delta w$. We would also like to see how the $\gamma^B_{mn}, \gamma^F_{mn}$ fall off for large values of $m,n$. While it is not possible to see these features from the form of these functions as finite sums, numerical analysis turns out to reveal the information we seek. In this section we will explore the functions $\gamma^B_{mn}, \gamma^F_{mn}$ by computing them using Mathematica.

\subsection{Wick Rotation and Origin Choice}

Note that the physical problem of the black hole needs us to do all computations for Minkowski signature on the cylinder. Thus we first rotate Euclidean time to Minkowski time. 
\bea
\t \to it ~\implies~ w = it + i\s.
\eea
Here $w$ is purely imaginary.

Now that we have returned to Minkowski time, we make the convenient choice of setting our origin midway between $w_1$ and $w_2$.  We thus have:
\bea
t_2 = -t_1 = {\D t \over 2}, \quad \s_2 = -\s_1 = {\D\s\over 2}.
\eea
With the choice of coordinates, we make several  observations that will simplify our algebra:
\bea
z_2 &=& e^{w_1} ~=~ e^{i{\D t + \D s \over 2}} ~=~ a+b + 2\sqrt{ab}\nn
z_1 &=& e^{w_1} ~=~ e^{-i{\D t + \D s \over 2}} ~=~ a+b - 2\sqrt{ab},
\eea
from which we find:
\bea
a+b &=& {z_2 + z_1 \over 2}\nn
\sqrt{ab} &=& {z_2 - z_1 \over 4},
\eea
and thus\footnote{We choose to have $b\to0$ as $\D w\to0$ so that the initial and final Copy 2 strands have the same image in this limit.}:
\bea
\sqrt{a} &=& \h\left(\sqrt{z_2} + \sqrt{z_1}\right) ~=~ \cos\left( {\D t + \D \s \over 4}\right)\nn
\sqrt{b} &=& \h\left(\sqrt{z_2} - \sqrt{z_1}\right) ~=~ i\sin\left( {\D t + \D \s \over 4}\right).\nn
\eea
This gives
\bea
a &=& \cos^2\left ({\D w \over 4i}\right)\nn
b &=& -\sin^2\left ({\D w \over 4i}\right) ~=~ 1-a,
\eea
eliminating the parameter $b$.  This also makes it clear that our $\g$ parameters will be periodic in $\D w$, with a period of $4\pi i$.

\subsection{Numerical Analysis}
Now that we have reduced our $\g$ coefficients to a single spacetime parameter, we can numerically calculate their values for various choices of $m$, $n$, and $\D w$.  We present here several figures plotting the behavior of the bosonic and fermionic $\g$ coefficients as we vary these parameters.

\begin{figure}
\includegraphics[width=0.5\columnwidth]{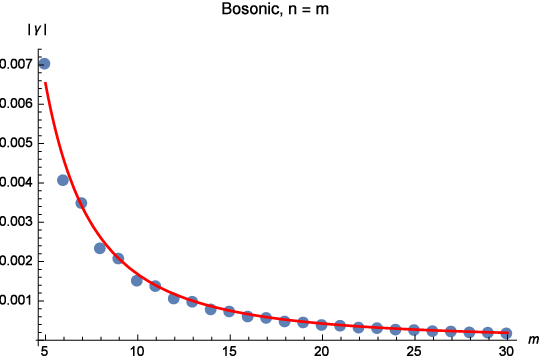}
\includegraphics[width=0.5\columnwidth]{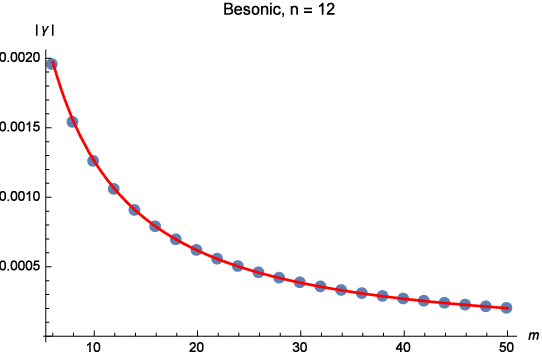}
\caption{Numerical values of $\left | \g^{B,(1)(1)}_{mn}\right | $ for $\D w = i\pi$.  For $n=m$ (left), a fit of the form $\G^2(m+\tfrac{1}{2})/[m\G^2(m+1)]$ is shown.  For $n=12$ (right), a fit of the form $\G(m+\tfrac{1}{2})/[\G(m+1)(m+12)]$ is shown.  The coefficient clearly behaves as $\G(m+\tfrac{1}{2})\G(n+\tfrac{1}{2})/[\G(m+1)\G(n+1)(m+n)]$.  Both plots show only $m>4$, as these simple fits do not work well for very low values of $m$ and $n$.  The plot on the right shows only even $m$, as the coefficient vanishes at $\D w = i\pi$ when $m+n$ is odd.}
\label{Bosonic_Fixed_w}
\end{figure}

\begin{figure}
\includegraphics[width=0.5\columnwidth]{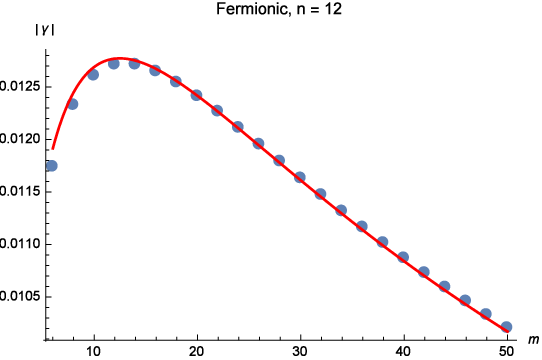}
\includegraphics[width=0.5\columnwidth]{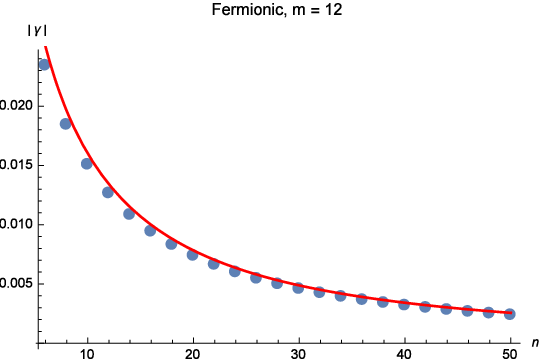}
\caption{Numerical values of $\left | \g^{F,(1)(1)}_{mn}\right | $ for $\D w = i\pi$.  For $n=12$ (left), a fit of the form $\G(m+\tfrac{1}{2})/[\G(m)(m+12)$ is shown.  For $m=12$ (right), a fit of the form $\G(n+\tfrac{1}{2})/[\G(n+1)(n+12)]$ is shown.  The coefficient clearly behaves as $\G(m+\tfrac{1}{2})\G(n+\tfrac{1}{2})/[\G(m)\G(n+1)(m+n)]$.  Both plots show only even $m,n>4$, as these simple fits do not work well for very low values of $m$ and $n$ and this coefficient also vanishes at $\D w = i\pi$ when $m+n$ is odd.}
\label{Fermionic_Fixed_w}
\end{figure}

Figures \ref{Bosonic_Fixed_w} and \ref{Fermionic_Fixed_w} show the behavior of the bosonic and fermionic $\g^{(1)(1)}_{mn}$ as $m$ or $n$ are varied.  The value of $\Delta w$ is held fixed. By considering a set of such graphs, ranging over different values of $\Delta w$, we can conjecture a simple approximate dependence of the $\gamma$ on $m,n$:
\bea
\g^{B,(1)(1)}_{mn} &\sim& {1\over (m+n)}{\G(m+\h)\G(n+\h)\over \G(m+1)\G(n+1)}\nn
\g^{F,(1)(1)}_{mn} &\sim& {1\over (m+n)}{\G(m+\h)\G(n+\h)\over \G(m)\G(n+1)}.
\eea
This leads us to guess a simple relationship for the magnitudes of the bosonic and fermionic $\g$ coefficients.  Fixing the sign, we have:
\bea
\g^{F,(1)(1)}_{mn} &=& -m\g^{B,(1)(1)}_{mn}.\label{GBFRelation}
\eea
We prove this relationship analytically in Appendix \ref{boson fermion relation}.  The relationship also exists for the first-order $\g$ coefficients, though this was not noted in \cite{acm1}.

For large $m$ and $n$, (\ref{GBFRelation}) simplifies to:
\bea
\g^{B,(1)(1)}_{mn} &\sim& {1\over \sqrt{mn}(m+n)}\nn
\g^{F,(1)(1)}_{mn} &\sim& \sqrt{m\over n}{1\over (m+n)},
\eea
which is exactly the same behavior as the single-twist $\g$.  Thus for any specific separation, each two-twist $\g$ looks just like a re-scaled single-twist $\g$.

\begin{figure}[h]
\includegraphics[width=0.5\columnwidth]{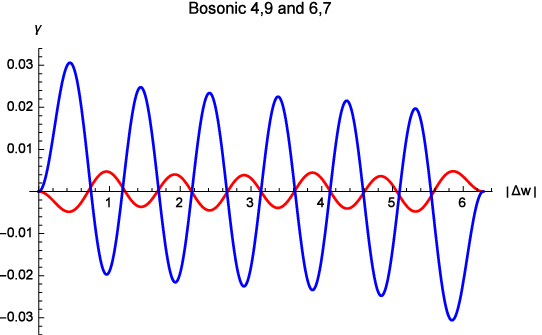}
\includegraphics[width=0.5\columnwidth]{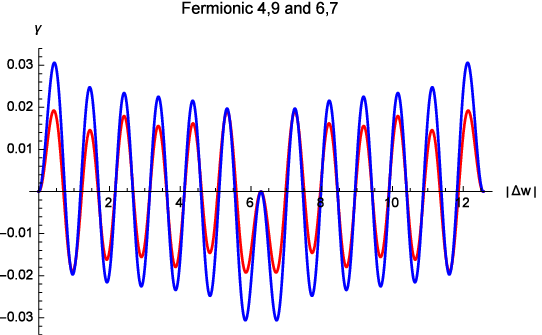}
\caption{Numerical values of $\left | \g^{F,(1)(1)}_{mn}\right | $ for $m=4,n=9$ (red) and $m=6,n=7$ (blue).  The $\D w$ dependence is clearly an oscillatory function.  The number of peaks is determined by $m+n$.}
\label{wPlot}
\end{figure}

Next we fix values for $m$ and $n$ while varying $\D w$.  As seen in Figure \ref{wPlot}, this analysis reveals that the $\D w$ dependance is an oscillation with a  frequency set by $m+n$. The amplitude of the oscillations, however does not remain constant over the full range of $\Delta w$. 

Combined with the previous result we reach a remarkable conclusion: The two-twist $\g$ coefficients have the exact same large $m,n$ behavior as their one-twist counterparts, except for the addition of an oscillation in $\D w$.

\section{Discussion}

In the preceding chapters, we had worked out the effect of one insertion of the deformation operator. Such an insertion takes two singly wound component strings, for example, to one doubly wound component string. Though this is a nontrivial interaction, creating excitations on the doubly wound string, it is hard to see evidence of thermalization at this level. One reason is that the vacuum state of the doubly wound string is different from the vacuum state of the two initial singly wound strings, so it is hard to separate the    effect of this vacuum change from any changes due to interactions among the excitations on the string. In this chapter we looked at a 1-loop effect where we have two insertions of $O$. The initial state has two singly wound strings. The first twist converts this to a doubly wound string, and the second twist returns us to two singly wound strings. Thus the vacuum structure of the initial and final states is the same, and any changes to the state can be seen as actual excitations that have been generated on the strings. 

In the single twist case, which we presented in chapters \ref{sec:gamma} and  \ref{f function}, we learned that the effect of a twist operator  generated a squeezed state, described by Bogoliubov coefficients $\gamma^B, \gamma^F$. We have a similar structure for the case of two twists. We have in this paper focused on the effect of the twists $\sigma_2^+$, and left out the action of the supercharge contours that surround these twist insertions. Though these supercharge insertions are essential to the form of the final state after the deformation, they can be split off from the twist insertions and written as supercharge actions on the initial and final states. This computation is straightforward but messy, so we hope to return to it elsewhere and in this paper have  focused only on the more essential physics: the computation of the Bogoliubov coefficients arising from the twists. 

In the single twist case one could find closed form expressions for the Bogoliubov coefficients. In the two twist case the best we could do was write the Bogoliubov coefficients as a finite sum of hypergeometric functions. This form does not make the properties of the coefficients obvious. In particular, we expect the coefficients $\gamma^B_{mn}, \gamma^F_{mn}$. to fall at large values of $m,n$. This is not obvious from any term in the sums. But numerical work showed that the coefficients do fall off as desired. Further, there was a very simple relation between the coefficients with one twist and the coefficients with two twists that can be obtained as a numerical approximation in the limit of large $m,n$. The two twist case also brings in a new feature that was not present in the 1-twist case: the coefficients depend on the separation $\Delta w$ between the twists. We found that there was a simple oscillatory behavior of the coefficients with $\Delta w$.

To obtain a complete analysis of the deformed theory at second order, we must use the Bogoluibov coefficients obtained here, but we must also allow excitations above the vacuum in the initial state, and  apply the  supercharge contours arising from the deformation operators. This work is based on the paper \cite{chm1}.

\chapter{Effect of two twist operators on an initial excitation}\label{two twist f}
In this chapter we compute the transition amplitudes, $f^B$, $f^{\pm F}$ for the two twist case which carries through in a similar manner as in the one twist case. However, the computations are more complicated. Lets discuss the notation we will use.

\section{Copy Notation}
Here we introduce some notation for handling the copy-identification indices in our computations.

For this chapter, we will label an arbitrary CFT copy with the index $(k)$.  If the CFT is specifically before the twist operators of interest, we'll use the index $(i)$ instead.  The index $(j)$ will be used if the CFT is specifically after the twisting. In short:
\bea
(i) \implies \text{initial copy}, \quad (j) \implies \text{final copy}, \quad (k) \implies \text{any copy}.
\eea
If we need to specify the copy number explicitly, we will use a prime for copies located after the twists.  So $(1)$ means Copy 1 before the twists, while $(1')$ means Copy 1 after the twists.  We will never need to talk about copies in-between twist operators of interest.

Each copy $(k)$ can have its own winding number, denoted as $N_{(k)}$.  By the nature of the twist interactions, total winding number is conserved.  We express this as:
\bea
\sum_{(j)}N_{(j)} &=& \sum_{(i)}N_{(i)}.
\eea

We will henceforth use the symbol $\hat\sigma$ to denote whatever arbitrary combinations of pairwise twists we wish to apply.  The initial vacuum will be a tensor product of Ramond vacuua across all initial copies.  In general, we will denote such a vacuum (before and after the twist) by:
\bea
|\emptyset\rangle &\equiv& \prod_{(i)}|0^*_R\rangle^{(i)}\nn
|\emptyset'\rangle &\equiv& \prod_{(j)}|0^*_R\rangle^{(j)},
\eea
where the notation $|0^*_R\rangle$ is used to indicate an unspecified type of R vacuum.  The four possibilities were are recorded below
\bea
|0_R^{\pm}\rangle, &&\qquad h={1\over4}, \quad m = \pm \h\nn
|0_R\rangle,|\tilde 0_R\rangle, &&\qquad h={1\over4}, \quad m=0.\label{RamondVacuua}
\eea
 In general, the type of R vacuum need not be the same between the different copies.  The twisted vacuum is then:
\bea
\hat\s\,|\emptyset\rangle &\equiv& |\chi\rangle.
\eea
where
\bea
\hat\s = \s^+_{2}(w_2)\s^+_2(w_1)
\eea
Our bosonic and fermionic fields are defined in a manner consistent with \cite{acm1, acm2}. Let us record the cylinder modes both before and after twist. Before the twist we have
\subsection*{$|\t|<|\tau_1|$}

\bea\label{BosonCylinderMode before}
\alpha_{A\dot A, n}^{(i)} &=& {1\over 2\pi} \int\limits_{\sigma=0}^{2\pi} \partial X_{A\dot A}(w) e^{nw} \diff w,
\eea

For fermions, we can have two different types of modes depending on which sector we are in, NS or R.  In the NS sector, the modes are indexed by half-integers $r$:
\bea\label{FermionNSCylinderMode before}
d^{(i)f,\a A}_{r} &=& {1\over 2\pi i} \int\limits_{\s=0}^{2\pi} \psi^{(i)\alpha A}(w) e^{rw}\diff w,
 \eea
Similarly, the modes in the Ramond sector are:
\bea\label{FermionRCylinderMode before}
d^{(i)f,\a A}_{n} &=& {1\over 2\pi i} \int\limits_{\s=0}^{2\pi} \psi^{(i)\alpha A}(w) e^{nw}\diff w,
\eea
with integer $n$.

\subsection*{$|\t| > |\t_2|$}
For modes after both twists we have the following
\bea\label{BosonCylinderMode after}
\alpha_{A\dot A, n}^{(j)} &=& {1\over 2\pi} \int\limits_{\sigma=0}^{2\pi} \partial X^{(j)}_{A\dot A}(w) e^{nw} \diff w,
\eea

For fermions, we can have two different types of modes depending on which sector we are in, NS or R.  In the NS sector, the modes are indexed by half-integers $r$:
\bea\label{FermionNSCylinderMode after}
d^{(j)f,\a A}_{r} &=& {1\over 2\pi i} \int\limits_{\s=0}^{2\pi} \psi^{(j)\alpha A}(w) e^{rw}\diff w,
 \eea

with $r$ half integer. Similarly, the modes in the Ramond sector are:
\bea\label{FermionRCylinderMode after}
d^{(j)f,\a A}_{n} &=& {1\over 2\pi i} \int\limits_{\s=0}^{2\pi} \psi^{(j)\alpha A}(w) e^{nw}\diff w,
\eea
with $n$ integer. This gives the following (anti)commutation relations for before and after the two twists:
\bea
\left[\a_{A\dot A,m}^{(k)},\a_{B\dot B,n}^{(k')}\right] &=& -m\e_{AB}\e_{\dot A\dot B}\d^{(k)(k')}\d_{m+n,0}\nn
\text{NS sector}:~~~ \left\{d_{m}^{(k),\a A},d_{n}^{(k'),\b B}\right\} &=& -N_{(k)}\e^{\a\b}\e^{AB}\d^{(k)(k')}\d_{m+n,0}\cr
\text{R sector}: ~~~\left\{d_{r}^{(k),\a A},d_{s}^{(k'),\b B}\right\} &=& -N_{(k)}\e^{\a\b}\e^{AB}\d^{(k)(k')}\d_{r+s,0} \label{CommutationRelations}
\eea

We also have a general supercurrent that can be used to form supercharge modes.  For each copy (k), the modes are defined in terms of a supercurrent contour.  However, we can also write the supercharge modes in terms of bosonic and fermionic modes:
\bea
G^{(k),\a}_{\dot A,n} &=& -{i\over N_{(k)}}\sum_m d^{(k),\a A}_m\a_{A\dot A,n-m}^{(k)}.
\eea
We will also make use of the `full $G$' operator, which is defined such that:
\bea
G^{\a}_{\dot A,n} &=& \begin{cases}
\sum\limits_{(i)}G^{(i),\a}_{\dot A,n} & \text{before twists}\\
\sum\limits_{(j)}G^{(j),\a}_{\dot A,n} & \text{after twists}.
\end{cases}
\eea
While this operator has been stripped away from the twist operators that we consider in this chapter, the supercurrent is still useful for proving certain transition amplitude relationships in full generality.

Lastly, we define the transition amplitudes $f$ as follows:
\bea
\hat \s \, \a^{(i)}_{A\dot A,-n}|\emptyset\rangle &=& \sum_{(j)}\sum_{p} f^{B,(i)(j)}_{np}\a^{(j)}_{A\dot A,-p}|\chi\rangle\nn
\hat \s \, d^{(i),\a A}_{-n}|\emptyset\rangle &=& \sum_{(j)}\sum_{p} f^{F\a,(i)(j)}_{np}d^{(j),\a A}_{-p}|\chi\rangle.
\label{eqone}
\eea
The mode number $p$ also runs over integers.
Here the $(j)$ sum spans all final copies while the $p$ sum spans all values for which the corresponding operator does not annihilate the particular R vacuum upon which $|\chi\rangle$ is built.  We anticipate two distinct cases for the fermions since their SU(2) R charge $\a$ is the same type of charge carried by the twist operators.  All other group indices should be symmetric under the twists.

\section{Relations between transition amplitudes}\label{Relations}

Our interest is in computing the following kind of amplitude. We start with either the vacuum state or a state containing some oscillator excitations. We then apply a certain number of twist operators $\hat\sigma$ at definite locations. We then ask for the final state generated by this procedure. The actual    deformation operator taking the CFT away from the orbifold point also contains contours of the supercharge $G$, but as mentioned before, these contours can be pulled away to act on the initial and final states, so a principal nontrivial part of the computation involves the effect of the twists $\hat \sigma$.

We have already mentioned some general properties of the states generated by the twists. Acting on the vacuum $|0\rangle$, the action of any number of twists is given by the form $\exp[\gamma^B\alpha\alpha+\gamma^F dd] |0\rangle$, where the $\alpha$ and $d$ are bosonic and fermionic oscillatirs respectively, and $\gamma^B, \gamma^F$ are Bogoliubov coefficients which we have computed in the previous chapter. If we had a single oscilaltor in the intial state, then we get the exponential as before, but the oscillator is changed to a linear combination of operators (\ref{eqone}). If there is more than one oscillator in the initial state, then one gets the same behavior for individual oscillators, but in addition one finds Wick contractions between pairs of operators present in the initial state; examples of this were computed in \cite{acm2}.

In this section, we will derive some general relationships governing the action of twist operators. These relations reduce the effort involved in computing the relevant amplitudes, as they relate some amplitudes to others.  

\subsection{Transpose Relation for Bosons}\label{TransposeRelations}
Consider a situation in which the winding configuration of the final state are identical to that of the initial state.  That is, we have the relation:
\bea
(i) = (j) &\implies& N_{(i)} = N_{(j)},
\eea
for all $(i),(j)$.  Now consider the amplitude:
\bea
\mathcal{A} &\equiv& \langle\emptyset'| \a^{(j)}_{++,n}\sh\a^{(i)}_{--,-m}|\emptyset\rangle.
\eea
Passing the initial boson through the twists, we find:
\bea
\mathcal{A} &=& \sum_{(j'),p}f^{B,(i)(j')}_{mp}\langle\emptyset'| \a^{(j)}_{++,n}\a^{(j')}_{--,-p}\sh|\emptyset\rangle.
\eea
We now apply the commutation relations (\ref{CommutationRelations}) to obtain:
\bea
\mathcal{A} &=& -n f^{B,(i)(j)}_{mn}\langle\emptyset'|\sh|\emptyset\rangle.\label{Step1}
\eea

Now the bosonic operators have no SU(2) R charge.  This means their behavior is in general independent of both the choice of particular R vacuua as well a the charge of the twists. The first independence allows us to    choose any combination of R vacuua for both $|\emptyset\rangle$ and $\langle\emptyset'|$ without altering the amplitude $\mathcal{A}$.  Since both cover the same winding configurations, one possible choice is to swap the two vacuum choices.  We thus find:
\bea
\mathcal{A} &=& \sum_{(j'),p}f^{B,(i)(j')}_{mp}\langle\emptyset| \a^{(i)}_{++,n}\sh\a^{(j')}_{--,-p}|\emptyset'\rangle.\label{ReversedAmp}
\eea

We now make use of the independence of the twist operator's SU(2) R charge.  This means that for bosons $\hat\s^{\dagger}=\hat\s^{-1}$, the reverse twisting process.\footnote{In general one would need also apply a global interchange of SU(2) R charges for this relation.}  We thus take the conjugate of both sides in Equation (\ref{ReversedAmp}) to obtain:
\bea
\mathcal{A}^* &=& \langle\emptyset'| \a^{(i)}_{++,m}\,\hat{\s}^{-1}\,\a^{(j)}_{++,-n}|\emptyset\rangle\nn
&=& -m\left( \tilde{f}^{B,(j)(i)}_{nm}\right)^*\langle\emptyset'|\sh|\emptyset'\rangle,\label{GeneralTransposeRelation}
\eea
where here the tilde denotes the transition amplitude for the reversed process.  Comparing to Equation (\ref{Step1}), we find:
\bea
n f^{B,(i)(j)}_{mn}&=& m \left(\tilde f^{B,(j)(i)}_{nm}\right)^*
\eea

For the particular one-loop case that we examine later, the twisting process is its own reversal (for bosons) and the transition amplitudes are real-valued (for convenient Minkowski coordinates). Furthermore, swapping the copy indices on $f$ amounts to either a global copy redefinition or nothing, neither of which has any physical effect.  Thus (\ref{GeneralTransposeRelation}) becomes:
\bea
nf^{B,(i)(j)}_{mn} &=& mf^{B,(i)(j)}_{nm}.\label{TwoTwistTransposeRelation}
\eea

\subsection{Supersymmetry Relations}
Here we find relations between the bosonic and fermionic transition amplitudes for arbitrary twisting.  To show these relations, we will make use of the fact that a $G^+$ current wrapped around any $\s_2^+$ operator with no extra weight in the integrand vanishes:
\bea
\oint_{w_0} G^{+}_{\dot A}(w)\,\s_2^+(w_0)\diff w &=& 0.
\eea
This can be seen by mapping to an appropriate covering plane, as shown in appendix \ref{GPlusProof}.  At the same time, one can deform the contour to obtain:
\bea
\oint_{w_0} G^{+}_{\dot A}(w)\,\s_2^+(w_0)\diff w &=& G^+_{\dot A,0}\,\s_2^+(w_0) -\s_2^+(w_0)G^+_{\dot A,0}\nn
&=& \left[G^+_{\dot A,0},\s_2^+(w_0)\right]~=~0.
\eea
Thus the full $G^+_{\dot A,0}$ mode commutes with the basic twist operator $\s_2^+$.  Since any twist operation of interest can be constructed from a combination of $\s_2^+$ operators, we find:
\bea
G^+_{\dot A,0}\,\hat\s &=& \hat\s\, G^+_{\dot A,0}.
\eea

Now for each copy, that copy's $G^+$ zero mode annihilates  the Ramond vacuua; thus in particular it annihilates the negative Ramond vacuum.  We thus have:
\bea
G^+_{\dot A,0}|\chi\rangle &=& G^+_{\dot A,0}\sh|\emptyset\rangle\nn
&=& \hat\s\,G^+_{\dot A,0}|\emptyset\rangle\nn
&=& 0.
\eea

We will now apply these relations to the two cases of possible SU(2) R charges for fermion modes.

\subsubsection{Fermions with negative SU(2) R charge}
Let us here consider the state with an initial negative-charge fermion mode and act on it with a $G^+$ zero mode.  Since this zero mode commutes with all twists, we have for instance:
\bea
G^+_{+,0}\sh d^{(i),--}_{-n}|\emptyset\rangle &=& \hat\s\,G^+_{+,0}d^{(i),--}_{-n}|\emptyset\rangle.\label{MinusRelationStart}
\eea

For $n=0$ the relationship is trivial: both sides vanish.  Setting aside such zero modes, we proceed with (\ref{MinusRelationStart}) in two ways. Starting with the left-hand side, we find:
\bea
G^+_{+,0}\sh d^{(i),--}_{-n}|\emptyset\rangle &=&\sum_{p,(k)} G^+_{+,0}f^{F-,(i)(j)}_{np}d^{(j),--}_{-p}|\chi\rangle\nn
&=& \sum_{p,(j)}f^{F-,(i)(j)}_{np}\left\{G^+_{+,0},d^{(j),--}_{-p}\right\}|\chi\rangle\nn
&=& \sum_{p',p,(j),(j'),A}\!\!\!\!\!\!\!f^{F-,(i)(j)}_{np}\left({-i\over N_{(j')}}\right)\left\{d^{(j'),+A}_{p'},d^{(j),--}_p\right\}\a_{A+,-p'}^{(j')}|\chi\rangle\nn
&=& -i\sum_{p,(j)}{1\over N_{(j)}}f^{F-,(i)(j)}_{np}\left(-N_{(j)}\right)\e^{+-}\e^{+-}\a^{(j)}_{++,-p}|\chi\rangle\nn
&=& i\sum_{p,(j)}f^{F-,(i)(j)}_{np}\a^{(j)}_{++,-p}|\chi\rangle.\label{NegativeLeftPart}
\eea

Now we turn to the right-hand side of (\ref{MinusRelationStart}) and perform a similar manipulation:
\bea
\s^+G^+_{+,0}d^{(i),--}_{-n}|\emptyset\rangle &=& \hat\s\,\left\{G^+_{+,0},d^{(i),--}_{-n}\right\}|\emptyset\rangle\nn
&=& \hat\s\,\sum_{(i'),n'}\left({-i\over N_{(i')}}\right)\left\{d^{(i'),+A}_{n'},d^{(i),--}_{-n}\right\}\a_{A+,-n'}^{(i')}|\emptyset\rangle\nn
&=& -{i\over N_{(i)}}\s^+\left(-N_{(i)}\right)\e^{+-}\e^{+-}\a_{++,-n}^{(i)}|\emptyset\rangle\nn
&=& i\sh\a_{++,-n}|\emptyset\rangle\nn
&=& i\sum_{p,(j)}f^{B,(i)(j)}_{np}\a_{++,-p}^{(j)}|\chi\rangle.\label{NegativeRightPart}
\eea
Combining this with (\ref{NegativeLeftPart}) we find that for any arbitrary twisting:
\bea
f^{B,(i)(j)}_{np} &=& f^{F-,(i)(j)}_{np},\qquad n,p>0.\label{NegativeRelation}
\eea

\subsubsection{Fermion with positive SU(2) R charge}
Let us now consider the state with an initial boson mode and act on it with a $G^+_0$ mode.  Since this zero mode commutes with all twists, we have for instance:
\bea
G^+_{+,0}\sh\a_{--,-n}^{(i)}|\emptyset\rangle &=& \hat\s\,G^+_{+,0}\a_{--,-n}^{(i)}|\emptyset\rangle.\label{PositiveRelationStart}
\eea

Again we set aside the $n=0$ case since it vanishes trivially.  We proceed with (\ref{PositiveRelationStart}) in the same two ways. Starting with the left-hand side, we find:
\bea
G^+_{+,0}\sh\a_{--,-n}^{(i)}|\emptyset\rangle &=& \sum_{p,(j)} G^+_{+,0}f^{B,(i)(j)}_{np}\a_{--,-p}^{(j)}|\chi\rangle\nn
&=& \sum_{p,(k)}f^{B,(i)(j)}_{np}\left[G^+_{+,0},\a_{--,-p}^{(j)}\right]|\chi\rangle\nn
&=& \sum_{p',p,(i),(j'),A}f^{B,(i)(j)}_{np}\left({-i\over N_{(j')}}\right)\left[\a_{A+,p'}^{(j')},\a_{--,-p}^{(j)}\right]d^{(j'),+A}_{-p'}|\chi\rangle\nn
&=& i\sum_{p,(j)}\left({-i\over N_{(j)}}\right)f^{B,(i)(j)}_{np}\left(-p\right)\e_{+-}\e_{+-}d^{(j),++}_{-p}|\chi\rangle\nn
&=& i\sum_{p,(j)}{p\over N_{(j)}}f^{B,(i)(j)}_{np}d^{(j),++}_{-p}|\chi\rangle.\label{PositiveLeftPart} 
\eea

Now we turn to the right-hand side of (\ref{PositiveRelationStart}) and perform a similar manipulation:
\bea
\hat\s\,G^+_{+,0}\a_{--,-n}^{(i)}|\emptyset\rangle &=& \hat{\s}\left[G^+_{+,0},\a_{--,-n}^{(i)}\right]|\emptyset\rangle\nn
&=& \hat\s\,\sum_{(i'),n'}\left({-i\over N_{(i')}}\right)\left[\a_{A+,n'}^{(i')},\a_{--,-n}^{(i)}\right]d^{(i'),+A}_{-n'}|\emptyset\rangle\nn
&=& -{i\over N_{(i)}}\sh\left(-n\right)\e_{+-}\e_{+-}d^{(i),++}_{-n}|\emptyset\rangle\nn
&=& i{n\over N_{(i)}}\sh d^{(i),++}_{-n}|\emptyset\rangle\nn
&=& i{n\over N_{(i)}}\sum_{p,(j)}f^{F+,(i)(j)}_{np}d^{(j),++}_{-n}|\chi\rangle.\label{PositiveRightPart}
\eea
Combining this with (\ref{PositiveLeftPart}) we find that for any arbitrary twisting:
\bea
{p\over N_{(j)}}f^{B,(i)(j)}_{np} &=& {n\over N_{(i)}}f^{F+,(i)(j)}_{np}, \qquad n,p>0.\label{PositiveRelation}
\eea

\subsection{The Capping States}
Let us first re-present Equation (\ref{eqone}):
\bea
\hat \s \, \a^{(i)}_{A\dot A,-n}|\emptyset\rangle &=& \sum_{(j)}\sum_{p} f^{B,(i)(j)}_{np}\a^{(j)}_{A\dot A,-p}|\chi\rangle\nn
\hat \s \, d^{(i),\a A}_{-n}|\emptyset\rangle &=& \sum_{(j)}\sum_{p} f^{F\a,(i)(j)}_{np}d^{(j),\a A}_{-p}|\chi\rangle.\label{GeneralForm2}
\eea
In  \cite{chm1} it was found that the physical part of $|\chi\rangle$ can be written as:
\bea
|\chi\rangle &=& C(w_1,w_2)\,e^{\hat{Q}(w_1,w_2)}\,\rpmket\nn
&=& C(w_1,w_2)\left(1 + \hat{Q}(w_1,w_2) + \left[\hat{Q}(w_1,w_2)\right]^2 + \ldots\right)\rpmket\label{ChiExpansion}
\eea
where the operator $\hat Q$ contains pairs of excitations.  Thus each term in in Equation (\ref{ChiExpansion}) contains an even number of excitations on the relevant R vacuum.  Inserting this expansion into Equation (\ref{GeneralForm2}), we find:
\bea
\hat \s \, \a^{(i)}_{A\dot A,-m}|\emptyset\rangle &=& \sum_{(j)}\sum_{n} f^{B,(i)(j)}_{mn}\a^{(j)}_{A\dot A,-n}C(w_1,w_2)\left(1 + \hat{Q}(w_1,w_2) + \ldots\right)\rpmket\nn
\hat \s \, d^{(i),\a A}_{-m}|\emptyset\rangle &=& \sum_{(j)}\sum_{n} f^{F\a,(i)(j)}_{mn}d^{(j),\a A}_{-n}C(w_1,w_2)\left(1 + \hat{Q}(w_1,w_2) + \ldots\right)\rpmket,\nn\label{ExpandedGeneralForm}
\eea
where each term now contains an odd number of excitations.

In order to avoid the messy details of the $\hat{Q}$ operator, we should choose a capping state that has a nonzero overlap with only the $0^{\text{th}}$  order term in Equation (\ref{ExpandedGeneralForm}).  We first note that this state should be built upon the same vacuum as $|\chi\rangle$, namely:
\bea
\langle\emptyset'| &\equiv& \rpmbra.
\eea
We will also need a single mode with appropriate charges to contract with the existing excitation from the $0^{\text{th}}$ order term in Equation (\ref{ExpandedGeneralForm}).  This yields, for example:
\bea
\langle\emptyset'|\a^{(j)}_{++,n}\sh\a^{(i)}_{--,-m}|\emptyset\rangle &=& -pf^{B,(i)(j)}_{mn}C(w_1,w_2)\langle\emptyset'|\emptyset'\rangle ~=~ -pf^{B,(i)(j)}_{mn}\langle\emptyset'|\sh|\emptyset\rangle\nn
\langle\emptyset'|d^{(j),++}_n\sh d^{(i),--}_{-m}|\emptyset\rangle &=& -f^{F-,(i)(j)}_{mn}C(w_1,w_2)\langle\emptyset'|\emptyset'\rangle ~=~ -pf^{B,(i)(j)}_{mn}\langle\emptyset'|\sh|\emptyset\rangle\nn
\langle\emptyset'|d^{(j),--}_n\sh d^{(i),++}_{-m}|\emptyset\rangle &=& -f^{F+,(i)(j)}_{mn}C(w_1,w_2)\langle\emptyset'|\emptyset'\rangle ~=~ -pf^{B,(i)(j)}_{mn}\langle\emptyset'|\sh|\emptyset\rangle.\nn\quad\label{CylinderRelations}
\eea
Equivalent relations can be obtained for other choices of the SU(2) indices.  Since we do not here determine the coefficient $C(w_1,w_2)$, we will not make use of the middle expressions.

\subsection{Spectral Flows}
Now that we know the capping states, we can determine the type and location of the spin fields they bring when mapped to the $t$ plane.  Recalling the relations in (\ref{spin field locations}), we have:
\bea
|0_{R,-}\rangle^{(1}) &\to& S^-(t=-a)\nn
|0_{R,-}\rangle^{(2}) &\to& S^-(t=-b)\nn
{}^{(1')}\langle0_{R,+}| &\to& S^-(t=\infty)\nn
{}^{(2')}\langle0_{R,-}| &\to& S^+(t=0).
\eea
We also list the spin fields accompanying the twists.  We remind the reader that the locations of these spin fields are the points that satisfy
\bea
{dz\over dt}={t^2 -ab\over t^2}=0
\eea
They occur at
\bea
\s_2^+(w_1) &\to& S^+\left(t=-\sqrt{ab}\right)\nn
\s_2^+(w_2) &\to& S^+\left(t=\sqrt{ab}\right).
\eea

We must now remove these six spin fields via local spectral flows just as we did when computing $\g^F_{mn}$'s.  A spectral flow by $\a=+1$ units removes an $S^-$ spin field while also applying an effect at infinity: Either removing an $S^+$ at infinity or applying an $S^-$ at infinity if there is no $S^+$ to be removed.  Table \ref{spectralflowtplane} tracks the behavior of the $t$ plane both locally and at infinity through five spectral flows.  The net effect is the elimination of all six spin fields in the $t$ plane.  We can now close all punctures with the local NS vacuum.  This vacuum is all that remains at the twist insertions, while the in and out states may carry additional contours from initial and capping excitations.  Thus the coordinate maps in conjunction with our spectral flows produces the following transformations:
\bea
|\emptyset\rangle &\to& \nstket \nn
\langle\emptyset'| &\to& \nstbra \nn
\sh &\to& \one.
\eea
There are also transformations on the bosonic and fermionic modes.  These transformations are the primary focus of the next two sections.  For now, we simply denote the transformed modes with primes:
\bea
\a^{(k)}_{A\dot A,n} &\to& \a'^{(k)}_{A\dot A,n}\nn
d^{(k),\a A}_{n} &\to& d'^{(k),\a A}_{n}.
\eea

\begin{table}[hbt]
\begin{center}
\begin{tabular}{|>{$}l<{$~}l>{~$}l<{$}>{$}c<{$}>{$}l<{$}c>{$}l<{$}|}\hline
\a = -1 & at & t=0 & \to & S^+(t=0) \text{ removed} & \& & S^-(t=\infty) \text{ removed}\\
&&&&&&\\
\a = -1 & at & t=\sqrt{ab} & \to & S^+\left(t=\sqrt{ab}\right) \text{ removed} & \& & S^+(t=\infty) \text{ added}\\
&&&&&&\\
\a = +1 & at & t=-a & \to & S^-(t=-a) \text{ removed} & \& & S^+(t=\infty) \text{ removed}\\
&&&&&&\\
\a = +1 & at & t=-b & \to & S^-(t=-b) \text{ removed} & \& & S^-(t=\infty) \text{ inserted}\\
&&&&&&\\
\a = -1 & at & t=-\sqrt{ab} & \to & S^+\left(t=-\sqrt{ab}\right) \text{ removed} & \& & S^-(t=\infty) \text{ removed}\\
\hline
\end{tabular}
\end{center}
\caption{One possible ordering of spectral flows.  While all orders have the same overall effect, we have presented here an order that never brings any complicated operators to the point at infinity.  The net result is the removal of all six spin insertions in the $t$ plane.}
\label{spectralflowtplane}
\end{table}

We now apply these transformations to the relations found in Equation (\ref{CylinderRelations}).  Naively, we obtain:
\bea
f^{B,(i)(j)}_{mn} &=& -{1\over n}{\nstbra\a'^{(j)}_{++,n}\a^{(i)}_{--,-m}\nstket\over\nstbra 0_{NS}\rangle_t}\nn
f^{F-,(i)(j)}_{mn} &=& -{\nstbra d'^{(j),++}_nd'^{(i),--}_{-m}\nstket\over\nstbra 0_{NS}\rangle_t} \nn
f^{F+,(i)(j)}_{mn}&=&-{\nstbra d'^{(j),--}_nd'^{(i),++}_{-m}\nstket\over\nstbra 0_{NS}\rangle_t\rangle}.
\eea
At this point there is one more subtlety to consider.  The out state modes on copy 2 are mapped to the origin of the $t$ plane before our in state modes are deformed to this region, so the in modes will have contours wrapped \emph{outside} of the out modes.  This means that we should write all copy 2 out modes to the left of other modes.  For fermions, this order-swapping will also bring a sign change.  Taking all of this into account, we find:
\bea
f^{B,(i)(1')}_{mn} &=& -{1\over n}{\nstbra\a'^{(1')}_{++,n}\a^{(i)}_{--,-m}\nstket\over\nstbra 0_{NS}\rangle_t}\nn
f^{B,(i)(2')}_{mn} &=& -{1\over n}{\nstbra\a^{(i)}_{--,-m}\a'^{(2')}_{++,n}\nstket\over\nstbra 0_{NS}\rangle_t}\nn
f^{F-,(i)(1')}_{mn} &=& -{\nstbra d'^{(1'),++}_nd'^{(i),--}_{-m}\nstket\over\nstbra 0_{NS}\rangle_t} \nn
f^{F-,(i)(2')}_{mn} &=& {\nstbra d'^{(i),--}_{-m}d'^{(2'),++}_n\nstket\over\nstbra 0_{NS}\rangle_t} \nn
f^{F+,(i)(1')}_{mn} &=& -{\nstbra d'^{(1'),--}_nd'^{(i),++}_{-m}\nstket\over\nstbra 0_{NS}\rangle_t} \nn
f^{F+,(i)(2')}_{mn}&=&{\nstbra d'^{(i),++}_{-m}d'^{(2'),--}_n\nstket\over\nstbra 0_{NS}\rangle_t\rangle}.\label{tPlaneRelations}
\eea

In the following two sections we will determine the integral expressions of the transformed modes.  We will then expand these modes in terms of modes natural to the $t$ plane in the appropriate regions.  This allows us to use the commutation relations found in Equations (\ref{bosoncommutation}) and (\ref{fermioncommutation}) to obtain an analytic expression for the transition amplitudes from the above relations.

\section{Calculating $f^B$}\label{Bosons}
Let us recall the expression for the bosonic transition amplitude from Equation (\ref{tPlaneRelations}).
\bea
f^{B,(i)(1')}_{mn} &=& -{1\over n}{\nstbra\a'^{(1')}_{++,n}\a^{(i)}_{--,-m}\nstket\over\nstbra 0_{NS}\rangle_t}\nn
f^{B,(i)(2')}_{mn} &=& -{1\over n}{\nstbra\a^{(i)}_{--,-m}\a'^{(2')}_{++,n}\nstket\over\nstbra 0_{NS}\rangle_t}.\label{BosonfRelation}
\eea

The rest of this session will consist primarily of writing these transformed modes in full detail.  Since bosonic modes are unaffected by spectral flows, we deal only with the coordinate maps.  We first expand the cylinder modes in terms of modes natural to the $t$ plane in the region of their image points.  After this we are free to drop any modes that annihilate the local NS vacuum (non-negative modes in this case).  We then deform the contours from our initial-state modes into the neighborhood of the capping mode.  From here we re-expand the initial-state mode in terms of $t$ plane modes natural to the new neighborhood.  We can then apply the commutation relations from Equation (\ref{bosoncommutation}) to obtain an expression for the transition amplitude. 

\subsection{The Boson Mode Expansions}
In \cite{chm1}, the transformed final modes were presented in full detail.  Here we merely give those results:
\bea
\a'^{(1')}_{A\dot A,n} &=& \sum_{j,j' = 0}^{\infty} {}^n C_j {}^n C_{j'} a^j b^{j'} \tilde{\a}_{A\dot A,n-j-j'}^{t\to\infty} \\
\a'^{(2')}_{A\dot A,n} &=& -\sum_{j,j' = 0}^{\infty} {}^n C_j {}^n C_{j'} a^{n-j} b^{n-j'} \tilde{\a}_{A\dot A,j+j'-n}^{t\to0}.
\eea

In keeping with the notation of \cite{chm1} we chose to have the initial Copy 1 map to the neighborhood around $t=a$ while the initial Copy 2 maps to the neighborhood around $t=b$.  Since the combination $X(w)\diff w$ has weight zero under coordinate transformations, we need only rewrite the $e^{nw}$ factor in terms of $t$.  Following the maps
\bea
e^{w}=z={(t+a)(t+b)\over t},
\eea
the factor is:
\bea
e^{nw} &=& z^n ~=~ \left( {(t+a)(t+b)\over t}\right)^n.
\eea
The initial boson modes are then:
\bea
\a'^{(1)}_{A\dot A,-m} &=& {1\over 2\pi} \oint_{t=-a} \partial X_{A\dot A}(t) \left ( {(t+a)(t+b) \over t} \right )^{-m} \diff t\nn
&=& {1\over 2\pi} \oint_{t'=0} \partial X_{A\dot A}(t) \left ( {t'(t'-a+b) \over t'-a} \right )^m \diff t' \label{1}\\
\a'^{(2)}_{A\dot A,-m} &=& {1\over 2\pi} \oint_{t=-b} \partial X_{A\dot A}(t') \left ( {(t+a)(t+b) \over t} \right )^{-m} \diff t\nn
&=&{1\over 2\pi} \oint_{t''=0} \partial X_{A\dot A}(t'') \left ( {(t''+a-b)t'' \over t''-b} \right )^{-m} \diff t'', \label{2}
\eea
where we have introduced shifted coordinates $t' = t-a$ and $ t'' = t-b$.  We shall deal with each copy in turn.

\subsubsection*{Copy 1 Initial Modes}
We now expand the integrand of (\ref{1}) in powers of $t'$.
\bea
\left ( {t'(t'-a+b) \over t'-a} \right )^{-m} &=& t'^{-m} \left ( t' + (b-a)\right )^{-m} (t'-a)^m\nn
&=& t'^{-m} \sum_{k=0}^{\infty} {}^{-m} C_k t'^k (b-a)^{-m-k} \sum_{k'=0}^{\infty} {}^{m} C_{k'} t'^{k'} (-a)^{m-k'}\nn
&=& \sum_{k,k' = 0}^{\infty} {}^{-m} C_k {}^{m}C_{k'} (b-a)^{-m-k} (-a)^{m-k'}t'^{k+k'-m} \nn
&\to&\sum_{k,k' = 0}^{k+k'<m} {}^{-m} C_k {}^{m}C_{k'} (b-a)^{-m-k} (-a)^{m-k'}t'^{k+k'-m}.  \label{1middle}
\eea
Here we have placed a limit on the sum to ensure that we work only with creation operators.  This means we have dropped those modes which annihilate the local NS vacuum.

We must now bring the contour to one of two different regions, depending on which copy of final modes we wish to interact with.  When interacting with Copy 1 final modes we must bring the contour out to infinity, thus expanding (\ref{1middle}) around $t=\infty$.  This will leave our contour \emph{inside} the final mode contour, so we expand in terms of creation operators.  When interacting with Copy 2 final modes we must instead bring the   contour to the origin of the $t$ plane, thus expanding (\ref{1middle}) around $t=0$.  This time our contour is \emph{outside} the final mode contour (indeed smaller $|z|$ map to larger $|t|$ in this region).  We must thus expand this region in terms of annihilation operators.  We also pick up an extra minus sign in this region to account from the contour reversing direction when it wraps around a finite point (the $t$ plane origin).

When expanding the Copy 1 initial modes for interaction with Copy 1 final modes, we find:
\bea
\left[ t'^{k+k'-m}\right]_{t \to \infty} &=& (t+a)^{k+k'-m}\nn
&=& t^{k+k'-m}\left(1+{a\over t}\right)^{k+k'-m}\nn
&=& t^{k+k'-m}\sum_{k''=0}^{\infty} {}^{k+k'-m} C_{k''}a^{k''}t^{-k''}\nn
&=& \sum_{k''=0}^{\infty} {}^{k+k'-m} C_{k''}a^{k''}t^{k+k'-m-k''},
\eea
where ${}^n C_m$ is the binomial coefficient of $m$ and $n$.
Plugging this into Equation (\ref{1middle}) gives:
\bea
\left ( {t'(t'-a+b) \over t'-a} \right )^{-m}_{t\to \infty}&\!\!\!=&\!\!\!\!\sum_{k,k',k'' = 0}^{\infty}\!\!\! {}^{-m}C_k {}^{m}C_{k'} {}^{k+k'-m} C_{k''} t^{k+k'-m-k''} \nn
&&\qquad {}\times(-1)^{m-k'}(b-a)^{-m-k} (a)^{m+k''-k'},
\eea
and thus:
\bea
\a'^{(1)}_{A\dot A,-m} \!\!&=& \!\!\!\!\sum_{k,k',k'' = 0}^{\infty} \!\!{}^{-m}C_k {}^{m}C_{k'} {}^{k+k'-m} C_{k''}(-1)^{m-k'}(b-a)^{-m-k} a^{m+k''-k'} \tilde{\a}_{A\dot A,k+k'-m-k''}^{t\to\infty},\nn \label{B1InitialLarge}
\eea
where we still have the constraint $k+k'<m$.

For interaction with the Copy 2 final modes, we instead have:
\bea
\left[t'^{k+k'-m}\right]_{t \to 0} &=& (t+a)^{k+k'-m}\nn
&=& a^{k+k'-m}\left(1+{t\over a}\right)^{k+k'-m}\nn
&=& a^{k+k'-m}\sum_{k''=0}^{\infty} {}^{k+k'-m} C_{k''}a^{-k''}t^{k''}\nn
&=& \sum_{k''=0}^{\infty} {}^{k+k'-m} C_{k''}a^{k+k'-m-k''}t^{k''}.
\eea
This gives:
\bea
\left ( {t'(t'-a+b) \over t'-a} \right )^{-m}_{t\to \infty}&\!\!\!\!=&\!\!\!\!\sum_{k,k',k'' = 0}^{\infty}\!\! {}^{-m}C_k {}^{m}C_{k'} {}^{k+k'-m} C_{k''}(-1)^{m-k'}(b-a)^{-m-k} (a)^{k-k''}t^{k''},\nn
\eea
and thus:
\bea
\a'^{(1)}_{A\dot A,-m} \!\!&=& \!\!\!\!-\sum_{k,k',k'' = 0}^{\infty}\!\!{}^{-m}C_k {}^{m}C_{k'}{}^{k+k'-m} C_{k''}(-1)^{m-k'}(b-a)^{-m-k} a^{k-k''}\tilde{\a}_{A\dot A,k''}^{t\to0},\nn \label{B1InitialSmall}
\eea
where again $k+k'<m$ and we have an extra minus sign from the change in direction of the contour.

\subsubsection*{Copy 2 Initial Modes}
In principle, we now expand the integrand of (\ref{2}) in powers of $t''$.  In practice we can shortcut this by noticing that this integrand can be obtained from the integrand of (\ref{1}) via the interchange $a \leftrightarrow b$, under which the reparameterization variables are also interchanged, $t' \leftrightarrow t''$.  In other words, $a \leftrightarrow b$ amounts to a redefinition of our initial copies $(1)\leftrightarrow(2)$.  We can thus apply $a \leftrightarrow b$ to (\ref{B1InitialLarge}) and (\ref{B1InitialSmall}), yielding:
\bea
\a'^{(2)}_{A\dot A,-m} \!\!&=&\!\!\!\! \sum_{k,k',k''=0}^{\infty}\!\! {}^{-m}C_k {}^m C_{k'} {}^{k+k'-m}C_{k''}(-1)^{m-k}(a-b)^{-m-k}b^{m+k''-k'}\tilde{\a}_{A\dot A,k+k'-m-k''}^{t\to\infty}\label{B2InitialLarge}\nn
\a'^{(2)}_{A\dot A,-m} \!\!&=&\!\!\!\! -\sum_{k,k',k''=0}^{\infty}\!\!{}^{-m}C_k {}^m C_{k'}{}^{k+k'-m}C_{k''}(-1)^{m-k}(a-b)^{-m-k}b^{k-k''}\tilde{\a}_{A\dot A,k''}^{t\to0}\label{B2InitialSmall},
\eea
with $k+k'<m$.

\subsubsection*{Summary}
For ease of reference, we group all of the expansions together.
\bea
\a'^{(1)}_{A\dot A,-m} \!\!&=& \!\!\!\!\sum_{k,k',k'' = 0}^{\infty} \!\!{}^{-m}C_k {}^{m}C_{k'} {}^{k+k'-m} C_{k''}(-1)^{m-k'}(b-a)^{-m-k} a^{m+k''-k'} \tilde{\a}_{A\dot A,k+k'-m-k''}^{t\to\infty}\nn
\!\!&=& \!\!\!\!-\sum_{k,k',k'' = 0}^{\infty} \!\!{}^{-m}C_k {}^{m}C_{k'} {}^{k+k'-m} C_{k''}(-1)^{m-k'}(b-a)^{-m-k} a^{k-k''} \tilde{\a}_{A\dot A,k''}^{t\to0}\label{1i}
\eea
\bea
\a'^{(2)}_{A\dot A,-m} \!\!&=&\!\!\!\! \sum_{k,k',k''=0}^{\infty}\!\! {}^{-m}C_k {}^m C_{k'} {}^{k+k'-m}C_{k''}(-1)^{m-k}(a-b)^{-m-k}b^{m+k''-k'}\tilde{\a}_{A\dot A,k+k'-m-k''}^{t\to\infty}\nn
\!\!&=&\!\!\!\!-\sum_{k,k',k''=0}^{\infty}\!\!{}^{-m}C_k {}^m C_{k'} {}^{k+k'-m}C_{k''}(-1)^{m-k}(a-b)^{-m-k}b^{k-k''}\tilde{\a}_{A\dot A,k''}^{t\to0}\label{2i}
\eea
\bea
\a'^{(1')}_{A\dot A,n} &=& \sum_{j,j' = 0}^{\infty} {}^n C_j {}^n C_{j'} a^j b^{j'} \tilde{\a}_{A\dot A,n-j-j'}^{t\to\infty} \label{1f}
\eea
\bea
\a'^{(2')}_{A\dot A,n} &=& -\sum_{j,j' = 0}^{\infty} {}^n C_j {}^n C_{j'} a^{n-j} b^{n-j'} \tilde{\a}_{A\dot A,j+j'-n}^{t\to0}\label{2f}.
\eea
For all initial-mode cases we have the additional constraint:
\bea
k+k' < m.
\eea

\subsection{Computing the $f^B$ Coefficients}
We can now compute the $f^B$ coefficients by plugging (\ref{1i} $-$ \ref{2f}) into (\ref{BosonfRelation}).  We'll split the cases into subsections.

\subsubsection{$(i)=(j)=1$}
Using the first line of (\ref{1i}) along with (\ref{1f}) and (\ref{BosonfRelation}), we find:
\bea
f^{B,(1)(1')}_{m,n} &=& \sum_{j,j',k,k',k''=0}^{\infty}{}^n C_j {}^n C_{j'}{}^{-m}C_k{}^m C_{k'}{}^{k+k'-m}C_{k''}(-1)^{m-k'}a^{m+k''-k'+j}b^{j'}(b-a)^{-m-k} \nn
&&{}\times{}\left (- {1\over n} {{}_t \langle 0|\tilde{\a}_{++,n-j-j'}^{t\to\infty}\tilde{\a}_{--,k+k'-m-k''}^{t\to\infty}|0\rangle_t\over {}_t \langle 0|0\rangle_t} \right ),
\eea
with the constraint:
\bea
k+k'<m.
\eea
We now apply the appropriate commutation relation from Equation (\ref{bosoncommutation}).  The amplitude ratio gives zero unless the following constraints are satisfied:
\bea
n-j-j' > 0 &\implies& j+j' < n \label{c1}\\
k+k'-m-k'' < 0 &\implies& k+k' < m+k'' \geq m \label{c2}\\
n-j-j' = -(k+k'-m-k'') &\implies& k'' = n+k+k'-m-j-j'.\geq 0\nn\label{c3}
\eea
When nonzero the commutator is simply $-(n-j-j')$.

Let us now look at our constraints in more detail.  Equation (\ref{c2}) is redundant with the earlier constraint $k+k' < m$.  Meanwhile, the right side of Equation (\ref{c3}) gives:
\bea
k'' = n+k+k'-m-j-j'\geq 0 &\implies& j+j' \leq n - m + k + k',
\eea
which is stricter than (\ref{c1}) since $m>0$.  From here, the fact that $j+j'\geq 0$ gives:
\bea
0\leq j+j' \leq n - m + k + k') &\implies& m-n \leq k+k',
\eea
which places another \emph{lower} bound on $k$ and $k'$.  We can now eliminate the $k''$ sum and use the constraints to set the limits.
\bea
\max(m-n-k,0) &\leq ~k'~ \leq& m-k-1 \nn
0 &\leq ~k~ \leq& m-1 \nn
0 &\leq ~j'~ \leq& n-(m-k-k')-j \nn
0 &\leq ~j~ \leq& n-(m-k-k'). \label{j=1limits}
\eea
This gives:
\bea
f^{B,(1)(1')}_{m,n} &\!\!=&\!\! \sum_{k=0}^{m-1}\,\sum_{k'=\max(m-n-k,0)}^{m-k-1}\,\sum_{j=0}^{n-m+k+k'}\,\sum_{j'=0}^{n-m+k+k'-j}{}^n C_j {}^n C_{j'} {}^{-m}C_k {}^m C_{k'}\nn
&&\quad{}\times {}^{k+k'-m}C_{n+k+k'-m-j-j'}{n-j-j' \over n}(-1)^{m-k'}a^{n+k-j'}b^{j'}(b-a)^{-m-k}.\nn\qquad\label{f11}
\eea

\subsubsection{$(i)=1,(j)=2$}
Here we use the second line of (\ref{1i}) along with (\ref{1f}) and (\ref{BosonfRelation}).  Noting that the initial mode maps to a contour \emph{outside} of the final mode due to $z \sim t^{-1}$, we find:
\bea
f^{B,(1)(2')}_{m,n} &=& \sum_{j,j',k,k',k''} {}^n C_j {}^n C_{j'}{}^{-m}C_k {}^m C_{k'} {}^{k+k'-m}C_{k''} (-1)^{m-k'} a^{k-k''+n-j} b^{n-j'}(b-a)^{-m-k} \nn
&&\quad{}\times{}\left (-{1\over n}{{}_t\langle 0|\tilde{a}_{--,k''}^{t\to0}\tilde{a}_{++, j+j'-n}^{t\to0}|0\rangle_t \over {}_t \langle 0| 0 \rangle_t} \right ),
\eea
with $k+k' < m$.  Here again we have constraints for the amplitude ratio to remain nonzero:
\bea
j+j'-n < 0 &\implies& j+j' < n \\
k'' = n-j-j' &\implies& \text{No extra limits}
\eea
When these conditions are met, the amplitude ratio gives a factor:
\bea
-k'' = -(n-j-j')
\eea
We can now eliminate the $k''$ sum and use the constraints to set the limits:
\bea
&0 \leq k' \leq m-k-1 \nn
&0 \leq k \leq m-1 \nn
&0 \leq j' \leq n-j-1 \nn
&0\leq j \leq n-1 \label{j=2limits}
\eea
\bea
f^{B,(1)(2')}_{m,n} &\!\!=& \!\!\sum_{k=0}^{m-1}\,\sum_{k'=0}^{m-k-1}\,\sum_{j=0}^{n-1}\,\sum_{j'=0}^{n-j-1}{}^n C_j {}^n C_{j'}{}^{-m}C_k {}^m C_{k'} {}^{k+k'-m}C_{n-j-j'}\nn
&&\qquad\qquad\qquad\qquad{}\times{}{n-j-j' \over n}(-1)^{m-k'}a^{k+j'} b^{n-j'}(b-a)^{-m-k}\nn\label{f12}
\eea

\subsubsection{The Other Cases by Symmetry}
The other cases can be obtained easily from the first two by applying the interchange $a \leftrightarrow b$, which swaps the initial copies while leaving the final copies unchanged.  We thus find:
\bea
f^{B,(2)(1')}_{m,n} &=& \left [ f^{B,(1)(1')}_{m,n}\right ]_{a\leftrightarrow b} \nn
&\!\!=&\!\! \sum_{k=\max(m-n,0)}^{m-1}\,\sum_{k'=\max(m-n-k,0)}^{m-k-1}\,\sum_{j=0}^{n-m+k+k'}\,\sum_{j'=0}^{n-m+k+k'-j}{}^n C_j {}^n C_{j'} {}^{-m}C_k {}^m C_{k'}\nn
&&\quad{}\times {}^{k+k'-m}C_{n+k+k'-m-j-j'}{n-j-j' \over n}(-1)^{m-k'}b^{n+k-j'}a^{j'}(a-b)^{-m-k}\nn\qquad\label{f21}\\\nn
f^{B,(2)(2')}_{m,n} &=& \left [ f^{B,(1)(2')}_{m,n} \right ]_{a\leftrightarrow b} \nn
&=& \!\!\sum_{k=0}^{m-1}\,\sum_{k'=0}^{m-k-1}\,\sum_{j=0}^{n-1}\,\sum_{j'=0}^{n-j-1}{}^n C_j {}^n C_{j'}{}^{-m}C_k {}^m C_{k'} {}^{k+k'-m}C_{n-j-j'}\nn
&&\qquad\qquad\qquad\qquad{}\times{}{n-j-j' \over n}(-1)^{m-k'}a^{n-j'}b^{k+j'}(a-b)^{-m-k}.\label{f22}
\eea

\subsection{Results}
Here we gather the results together for convenient reference.
\bea
f^{B,(1)(1')}_{m,n} &\!\!=&\!\! \sum_{k=\max(m-n,0)}^{m-1}\,\sum_{k'=\max(m-n-k,0)}^{m-k-1}\,\sum_{j=0}^{n-m+k+k'}\,\sum_{j'=0}^{n-m+k+k'-j}{}^n C_j {}^n C_{j'} {}^{-m}C_k {}^m C_{k'}\nn
&&\quad{}\times {}^{k+k'-m}C_{n+k+k'-m-j-j'}{n-j-j' \over n}(-1)^{m-k'}a^{n+k-j'}b^{j'}(b-a)^{-m-k}\qquad\nn\\
f^{B,(1)(2')}_{m,n} &\!\!=& \!\!\sum_{k=0}^{m-1}\,\sum_{k'=0}^{m-k-1}\,\sum_{j=0}^{n-1}\,\sum_{j'=0}^{n-j-1} (-1)^{m-k'}{n-j-j' \over n} a^{k+j'} b^{n-j'}(b-a)^{-m-k}\nn
&&\qquad\qquad\qquad\qquad{}\times{}^n C_j {}^n C_{j'}{}^{-m}C_k {}^m C_{k'} {}^{k+k'-m}C_{n-j-j'}\\\cr
f^{B,(2)(1')}_{m,n} &\!\!=&\!\! \sum_{k=\max(m-n,0)}^{m-1}\,\sum_{k'=\max(m-n-k,0)}^{m-k-1}\,\sum_{j=0}^{n-m+k+k'}\,\sum_{j'=0}^{n-m+k+k'-j}{}^n C_j {}^n C_{j'} {}^{-m}C_k {}^m C_{k'}\nn
&&\quad{}\times {}^{k+k'-m}C_{n+k+k'-m-j-j'}{n-j-j' \over n}(-1)^{m-k'}b^{n+k-j'}a^{j'}(a-b)^{-m-k}\qquad\nn\\
f^{B,(2)(2')}_{m,n} &\!\!=&\!\!\sum_{k=0}^{m-1}\,\sum_{k'=0}^{m-k-1}\,\sum_{j=0}^{n-1}\,\sum_{j'=0}^{n-j-1} (-1)^{m-k'}{n-j-j' \over n} a^{n-j'}b^{k+j'}(a-b)^{-m-k}\nn
&&\qquad\qquad\qquad\qquad{}\times{}^n C_j {}^n C_{j'}{}^{-m}C_k {}^m C_{k'} {}^{k+k'-m}C_{n-j-j'}.
\eea
Using the relationships from Equations (\ref{NegativeRelation}) and (\ref{PositiveRelation}), this also gives the fermion transition amplitude for all cases that do not involve a zero mode.

Because of global copy exchange symmetry we have the relations
\bea
f^{B,(1)(1')}_{mn}&=&f^{B,(2)(2')}_{mn}\cr
f^{B,(2)(1')}_{mn}&=&f^{B,(1)(2')}_{mn}
\label{copy symmetry relation}
\eea

\subsection{Numerical Analysis}
Let us now find a good approximation for our transition amplitudes when the mode numbers become large.  We proceed by plotting the exact values over a range of $\D w$ coordinates and then attempting to fit the resulting points.  We find a good, simple approximation for the bosonic transition amplitudes, which are themselves related to the fermionic transition amplitudes for nonzero modes through supersymmetry relations.  The fermion zero mode functions prove less tractable for this analysis.  This might be expected, as for such amplitudes at least one mode is necessarily well outside of the continuum limit.

\begin{figure}[bht]
\includegraphics[width=0.5\columnwidth]{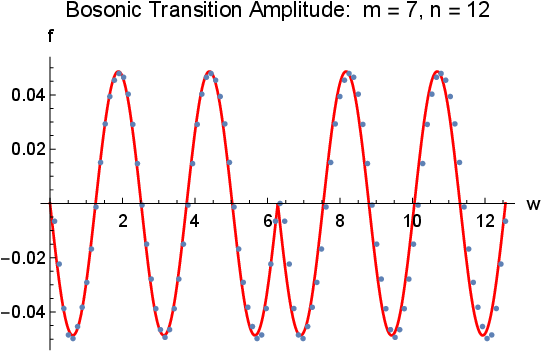} \includegraphics[width=0.5\columnwidth]{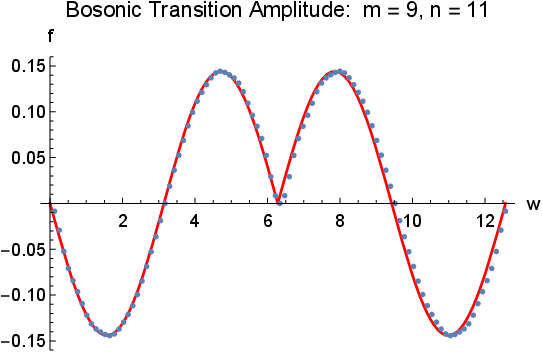}\\\\\\
\includegraphics[width=0.5\columnwidth]{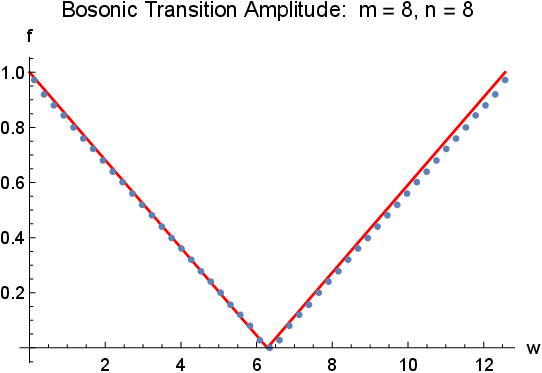} \includegraphics[width=0.5\columnwidth]{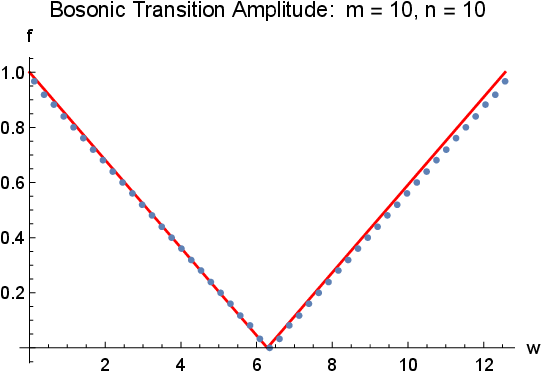}
\caption{The bosonic transition amplitudes $f^{B,(1)(1')}_{mn}$ for: $m=7$, $n=12$ (upper left); $m=9$, $n=11$ (upper right); $m=n=8$ (lower left); $m=n=10$ (lower right).  The exact values are plotted as blue dots, while the simple continuum limit approximation from Equation (\ref{CLApprox}) is plotted as a red line.  The axis labeled w is the real value $w/i$.}
\label{BosonGraphs}
\end{figure}

For the bosonic transition amplitude, large indices result in the approximate form (i.e., in the limit $m, n \gg 1$):
\bea
f^{B,(1)(1')}_{mn} &\approx& \begin{cases}
{1\over (m-n)\pi} \sqrt{m\over n}\sin\left((m-n){\D w\over 2i}\right)\text{sgn}\left({\D w \over 2\pi i}-1\right) & m\neq n\\
\left|{\D w \over 2\pi i}-1\right| & m=n.
\end{cases}\label{CLApprox}
\eea
This produces a mirroring about the point $\D w = 2\pi i$:
\bea
\left[f^{B,(1)(1')}_{mn}\right]_{\D w = 2\pi i + x} &=& \left[f^{B,(1)(1')}_{mn}\right]_{\D w = 2\pi i - x}.
\eea

Looking numerically at modes on the initial copy 2, we find a simple relationship:
\bea
f^{B,(1)(1')}_{mn} + f^{B,(2)(1')}_{mn} &=& \d_{mn}.\label{DeltaRelation}
\eea
While we have not proven this relationship in full generality, it has held for all cases we have checked including small $m$ and $n$.  There is also an analogous exact relationship for the single twist case:
\bea
f^{B,(1)}_{mn} + f^{B,(2)}_{mn} &=& \d_{mn}.
\eea

If we combine (\ref{DeltaRelation}) with (\ref{copy symmetry relation}), we find that a copy symmetric state is left unaffected by spectral flow. Schematically, 
\bea
\hat\sigma \big(\a^{(1)} + \a^{(2)}\big)|\emptyset\rangle = \big(\a^{(1)} + \a^{(2)}\big) \hat\sigma|\emptyset\rangle
\eea 
A similar relation holds for fermionic transition amplitudes with nonzero mode numbers.

\section{Fermion Zero Modes}\label{Fermions}

We have seen that the $f$ functions for fermions can be related to the $f$ functions for bosons. These relation    however do not hold when one or more of the modes involved in the relation is a fermion zero mode. In this case we need to be more careful, and computre the $f$ functions for fermions explicitly. In this section we will perform the relevant computations for the $f^{F(i)(j),\pm}$ when one or more of the indices represent a zero mode. .

We follow the same method used for the bosons, except that we must now account for the effects of our five spectral flows. These spectral flows were required to map the originital problem to a plane with no punctures. The bosons were not affected by the spectral flow, but the fermions will be. The full calculation of the spectral flow effects was performed in chapter \ref{two twist gamma} for capping modes, so we present only the results found there:
\bea
d'^{(1'),+A}_{n}&=&\sum_{j,j'\geq 0} {}^{n-1}C_j{}^{n-1}C_{j'} a^j b^{j'} \tilde{d}^{+A,t\to\infty}_{n-j-j'+1/2}\nn
&&{}-\sum_{j,j'\geq 0} {}^{n-1}C_j{}^{n-1}C_{j'} a^{j+1} b^{j'+1}\tilde{d}_{n-j-j'-3/2}^{+A,t\to\infty}\nn
d'^{(1'),-A}_{n}&=&\sum_{j,j'\geq 0} {}^{n}C_j{}^{n}C_{j'} a^{j} b^{j'}\tilde{d}^{-A,t\to\infty}_{n-j-j'-1/2}\nn
d'^{(2'),+A}_{n}&=&-\sum_{j,j'\geq 0}{}^{n-1}C_j{}^{n-1}C_{j'} a^{n-j-1} b^{n-j'-1} \tilde{d}^{+A,t\to0}_{j+j'-n+5/2}\nn
&&{}+\sum_{j,j'\geq 0} {}^{n-1}C_j{}^{n-1}C_{j'} a^{n-j} b^{n-j'}\tilde{d}^{+A,t\to0}_{j+j'-n+1/2}\nn
d'^{(2'),-A}_{n}&=&-\sum_{j,j'\geq 0} {}^{n}C_j{}^{n}C_{j'} a^{n-j} b^{n-j'}\tilde{d}^{-A,t\to0}_{j+j'-n-1/2}
\label{tplanemodesNS}
\eea
Since we are interested in the cases containing at least one zero mode, we evaluate (\ref{tplanemodesNS}) for such cases.
\bea
d'^{(1'),+A}_{0}&=&\sum_{j,j'\geq 0}(-1)^{j+j'} a^j b^{j'} \tilde{d}^{+A,t\to\infty}_{-j-j'+1/2}-\sum_{j,j'\geq 0} (-1)^{j+j'} a^{j+1} b^{j'+1}\tilde{d}_{-j-j'-3/2}^{+A,t\to\infty}\nn
d'^{(1'),-A}_{0}&=&\tilde{d}^{-A,t\to\infty}_{-1/2}\nn
d'^{(2'),+A}_{0}&=&-\sum_{j,j'\geq 0}(-1)^{j+j'} a^{-j-1} b^{-j'-1} \tilde{d}^{+A,t\to0}_{j+j'+5/2}+\sum_{j,j'\geq 0}(-1)^{j+j'} a^{-j} b^{-j'}\tilde{d}^{+A,t\to0}_{j+j'+1/2}\nn
d'^{(2'),-A}_{0}&=&-\tilde{d}^{-A,t\to0}_{-1/2}
\label{tplanezeromodesNS}
\eea

We now turn to modes of the in state.  We will work in three subsections.  First, we will follow the in states through the coordinate shifts and spectral flows.  Next we will express the transformed modes in terms of modes natural to the $t$ plane in the appropriate neighborhood, stretch the contours to the neighborhood of the out states, and then re-express the result in modes natural to the new neighborhoods.  Lastly we will plug the results into Equation (\ref{tPlaneRelations}) to obtain expressions for the fermion transition amplitudes.

\subsection{Coordinate maps and spectral flows}
We begin with the fermion modes on the cylinder before the twist insertions:
\bea
d^{(i),\a A}_n &=& {1\over 2\pi i}\int_{\t<\t_1,\s=0}^{\s=2\pi} \psi^{(i),\a A}(w)e^{nw}\diff w\nn
d^{(j),\a A}_n &=& {1\over 2\pi i}\int_{\t>\t_2,\s=0}^{\s=2\pi} \psi^{(j),\a A}(w)e^{nw}\diff w.\label{InModesCylinder}
\eea
The fermion field $\p$ has conformal weight $1/2$.  Combined with the jacobian, this gives:
\bea
\p^{(i),\a A}(x)\diff x &\to& \left({\diff x' \over \diff x}\right)^{-\h}\p^{(i),\a A}(x')\diff x'.
\eea

The derivatives of our two maps are as follows:
\bea
{\diff z \over \diff w} &=& e^w ~=~ z\\\nn
{\diff t \over \diff z} &=& \left({\diff z \over \diff t}\right)^{-1}\nn
&=& \left(1-{ab\over t^2}\right)^{-1}\nn
&=& t^2\left(t^2-ab\right)^{-1}.
\eea
Taking the two coordinate shifts together, we find:
\bea
\psi^{(i),\a A}(w)e^{nw}\diff w &\to& z^{n-\h}t^{-1}\left(t^2-ab\right)^{\h}\psi^{(i),\a A}(t)\diff t\label{FermionCoordinateResult}\\
&=& (t+a)^{n-\h}(t+b)^{n-\h}t^{-n-\h}\left(t^2-ab\right)^{\h}\psi^{(i),\a A}(t)\diff t.\nonumber
\eea

We must now apply our five spectral flows.  For each spectral flow by a unit $\a$ around a point $t_0$, a field with an SU(2) R charge q transforms as:
\bea
\psi^q(t) &\to& (t-t_0)^{-\a q}\psi^q(t).
\eea
We can determine the factors that come from the spectral flows outlined in Table \ref{spectralflowtplane}.  These factors are presented in Table \ref{SFResultsTable}.  Applying this to Equation (\ref{FermionCoordinateResult}) gives:
\bea
\psi^{(i),+A}(w)e^{nw}\diff w &\to& (t+a)^{n}(t+b)^{n-1}t^{-n-1}\left(t^2 - ab\right)\psi^{(i),+A}(t)\diff t\nn
\psi^{(i),-A}(w)e^{nw}\diff w &\to& t^{-n-1}(t+a)^{n}(t+b)^{n}\psi^{(i),+A}(t)\diff t.\label{FermionTransformations}
\eea

\begin{table}[hbt]
\begin{center}
\begin{tabular}{|>{$}c<{$~}|>{$}c<{$~}|>{~$}l<{$\quad}|>{$}l<{$}|}\hline
\a & t_0 & \psi^{+A}\text{ factor} & \psi^{-A}\text{ factor}\\
\hline
-1 & 0 & t^{\h} & t^{-\h}\\
-1 & \sqrt{ab} & \left(t-\sqrt{ab}\right)^{\h} & \left(t-\sqrt{ab}\right)^{-\h}\\
+1 & -a & (t+a)^{-\h} & (t+a)^{\h}\\
+1 & -b & (t+b)^{-\h} & (t+b)^{\h}\\
-1 & -\sqrt{ab} & \left(t+\sqrt{ab}\right)^{\h} & \left(t+\sqrt{ab}\right)^{-\h}\\
\hline
\text{All} && t^{\h}(t+a)^{-\h}(t+b)^{-\h}\left(t^2-ab\right)^{\h} & t^{-\h}(t+a)^{\h}(t+b)^{\h}\left(t^2-ab\right)^{-\h}\\
\hline
\end{tabular}
\end{center}
\caption{The factors obtained by the five spectral flows for the two possible fermion field charges. The last row groups all the factors together in a simplified form.}
\label{SFResultsTable}
\end{table}

Plugging Equation (\ref{FermionTransformations}) into (\ref{InModesCylinder}), we find:
\begin{eqnarray}
d^{(1),+A}_{n}&\to&\frac{1}{2\pi i}\oint_{t'=0}   
\psi^{+A}(t')((t' - a)^{2}-ab)(t'-a)^{-n}t'^{n-1}(t'+b-a)^{n-1}\diff t'\cr
d^{(1),-A}_{n}&\to&\frac{1}{2\pi i}\oint_{t'=0}  
\psi^{-A}(t')(t'-a)^{-n-1}t'^n(t'+b-a)^{n}\diff t'\cr
d^{(2),+A}_{n}&\to&\frac{1}{2\pi i}\oint_{t''=0}  
\psi^{+A}(t'')((t'' - b)^{2}-ab)(t''-b)^{-n}t''^{n-1}(t''+a-b)^{n-1}\diff t''\cr
d^{(2),-A}_{n}&\to&\frac{1}{2\pi i}\oint_{t''=0}  
\psi^{-A}(t'')(t''-b)^{-n-1}t''^n(t''+a-b)^{n}\diff t''.\label{tr_sf_modes}
\end{eqnarray}
Here we use the same notation as for the bosons, $t' = t-a$ and $t'' = t-b$.

\subsection{Natural modes on the $t$ plane}
We now expand the in state modes in terms of modes natural to the $t$-plane.  For both $+$ and $-$ cases,  we perform expansions first around $t'=0$ or $t''=0$ and then around $t=\infty$ or $t=0$.  In each case, we denote the integrand from Equation (\ref{tr_sf_modes}) as $I$ with appropriate indices.  We will also make use of the $a \leftrightarrow b$ interchange to swap the initial copies as we did for the bosons.

\subsubsection*{\underline{$d_{-m}^{(i),+A}$}}
\bea
I^{(1),+}&=&((t' - a)^{2}-ab)(t'-a)^{-n}t'^{n-1}(t'+b-a)^{n-1}\cr\cr
&=&(t'-a)^{m+2}t'^{-m-1}(t'+(b-a))^{-m-1}-ab(t'-a)^{m}t'^{-m-1}(t'+(b-a))^{-m-1}\cr\cr
&=&(-a)^{m+2}(b-a)^{-m-1}\left(1+\frac{t'}{-a}\right)^{m+2}t'^{-m-1}\left(1+\frac{t'}{b-a}\right)^{-m-1}\nn\nn
&&\qquad-ab(-a)^{m}(b-a)^{-m-1}\left(1+\frac{t'}{-a}\right)^{m}t'^{-m-1}\left(1+\frac{t'}{b-a}\right)^{-m-1}\nn\nn
&=&\sum_{k,k'\geq 0}{}^{m+2}C_{k}{}^{-m-1}C_{k'}(-a)^{m-k+2}(b-a)^{-m-k'-1}t'^{-m+k+k'-1}\nn
&&\qquad- \sum_{k,k'\geq 0}{}^{m}C_{k}{}^{-m-1}C_{k'}(-1)^{m-k}b\,a^{m-k+1}(b-a)^{-m-k'-1}t'^{-m+k+k'-1}\nn
&=& \sum_{k,k'\geq0}{}^{-m-1}C_{k'}\left({}^{m+2}C_{k}a-{}^{m}C_{k}b\right)(-1)^{m-k}a^{m-k+1}(b-a)^{-m-k'-1}t'^{-m+k+k'-1}\nn\nn
I^{(2),+}&=&I^{(1),+}\left(a\leftrightarrow b, t'\leftrightarrow t''\right)
\label{Integrand+}
\eea

\subsubsection*{\underline{$d_{-m}^{(i),-A}$}}
\bea
I^{(1),-}&=&(t'-a)^{m-1}t'^{-m}(t'+(b-a))^{-m}\nn
&=&(-a)^{m-1}(b-a)^{-m}\left(1+\frac{t'}{-a}\right)^{m-1}t'^{-m}\left(1+\frac{t'}{b-a}\right)^{-m}\nn
&=&\sum_{k,k'\geq 0}{}^{m-1}C_{k}{}^{-m}C_{k'}(-a)^{m-k-1}(b-a)^{-m-k'}t'^{-m+k+k'}\nn\nn
I^{(2),-}&=& I^{(1),-}\left(a\leftrightarrow b,t'\leftrightarrow t''\right)
\label{Integrand-}
\eea

We now insert these expansions back into Equation (\ref{tr_sf_modes}) and drop any powers of $t$ that give local annihilators.  This leaves us with:
\bea
d^{(1),+A}_{-m}&\!\!\to\!\!&\sum_{k=0}^{m}\sum_{k'=0}^{m-k}{}^{-m-1}C_{k'}\left({}^{m+2}C_{k}a-{}^{m}C_{k}b\right)(-1)^{m-k}a^{m-k+1}(b-a)^{-m-k'-1}\tilde{d}^{+A,t\to a}_{-m+k+k'-1/2}\nn
d^{(2),+A}_{-m}&\!\!\to\!\!&\sum_{k=0}^{m}\sum_{k'=0}^{m-k}{}^{-m-1}C_{k'}\left({}^{m+2}C_{k}b-{}^{m}C_{k}a\right)(-1)^{m-k}b^{m-k+1}(a-b)^{-m-k'-1}\tilde{d}^{+A,t\to b}_{-m+k+k'-1/2}\nn
d^{(1),-A}_{-m}&\!\!\to\!\!&\sum_{k=0}^{m-1}\sum_{k'=0}^{m-k-1}{}^{m-1}C_{k}{}^{-m}C_{k'}(-1)^{m-k-1}a^{m-k-1}(b-a)^{-m-k'}\tilde{d}^{-A,t\to a}_{-m+k+k'+1/2}\nn
d^{(2)i,-A}_{-m}&\!\!\to\!\!&\sum_{k=0}^{m-1}\sum_{k'=0}^{m-k-1}{}^{m-1}C_{k}{}^{-m}C_{k'}(-1)^{m-k-1}b^{m-k-1}(a-b)^{-m-k'}\tilde{d}^{-A,t\to b}_{-m+k+k'+1/2}
\label{initialmodestplane}
\eea

We must now expand these modes around $t=\infty$ and $t=0$.  
We do this by simply expanding the integrands hidden in the modes of (\ref{initialmodestplane}) in these regions. 
\subsubsection*{\underline{$t=\infty$}}
\bea
\tilde{d}^{(1)+A,t\to a}_{-m+k+k'-{1\over2}}&:&t'^{-m+k+k'-1}=(t+a)^{-m+k+k'-1}=\sum_{k''\geq 0}{}^{-m+k+k'-1}C_{k''}a^{k''}t^{-m+k+k'-k''-1}\nn
\tilde{d}^{(2)+A,t\to b}_{-m+k+k'-{1\over2}}&:&t''^{-m+k+k'-1}=(t+b)^{-m+k+k'-1}=\sum_{k''\geq 0}{}^{-m+k+k'-1}C_{k''}b^{k''}t^{-m+k+k'-k''-1}\nn
\tilde{d}^{(1)-A,t\to a}_{-m+k+k'+{1\over2}}&:&t'^{-m+k+k'}=(t+a)^{-m+k+k'}=\sum_{k''\geq 0}{}^{-m+k+k'}C_{k''}a^{k''}t^{-m+k+k'-k''}\nn
\tilde{d}^{(2)-A,t\to b}_{-m+k+k'+{1\over2}}&:&t''^{-m+k+k'}=(t+b)^{-m+k+k'}=\sum_{k''\geq 0}{}^{-m+k+k'}C_{k''}b^{k''}t^{-m+k+k'-k''}.
\eea
\subsubsection*{\underline{$t=0$}}
\bea
\tilde{d}^{(1),+A,t\to a}_{-m+k+k'-{1\over2}}&:&t'^{-m+k+k'-1}=(t+a)^{-m+k+k'-1}=\sum_{k''\geq 0}{}^{-m+k+k'-1}C_{k''}a^{-m+k+k'-k''-1}t^{k''}\nn
\tilde{d}^{(2),+A,t\to b}_{-m+k+k'-{1\over2}}&:&t''^{-m+k+k'-1}=(t+b)^{-m+k+k'-1}=\sum_{k''\geq 0}{}^{-m+k+k'-1}C_{k''}b^{-m+k+k'-k''-1}t^{k''}\nn
\tilde{d}^{(1),-A,t\to a}_{-m+k+k'+{1\over2}}&:&t'^{-m+k+k'}=(t+a)^{-m+k+k'}=\sum_{k''\geq 0}{}^{-m+k+k'}C_{k''}a^{-m+k+k'-k''}t^{k''}\nn
\tilde{d}^{(2),-A,t\to b}_{-m+k+k'+{1\over2}}&:&t''^{-m+k+k'}=(t+b)^{-m+k+k'}=\sum_{k''\geq 0}{}^{-m+k+k'}C_{k''}b^{-m+k+k'-k''}t^{k''}.
\eea
Equation (\ref{initialmodestplane}) now gives:
\subsubsection*{\underline{$t=\infty$}}
\bea
d'^{(1),+A}_{-m}&=&\sum_{k=0}^{m}\sum_{k'=0}^{m-k}\sum_{k''\geq0}{}^{-m-1}C_{k'}{}^{-m+k+k'-1}C_{k''}\left({}^{m+2}C_{k}a-{}^{m}C_{k}b\right)\nn
&&\quad{}\times(-1)^{m-k}a^{m-k+k''+1}(b-a)^{-m-k'-1}\tilde{d}^{+A,t\to\infty}_{-m+k+k'-k''-1/2}\nn
d'^{(1),-A}_{-m}&=&\sum_{k=0}^{m-1}\sum_{k'=0}^{m-k-1}\sum_{k''\geq0}{}^{m-1}C_{k}{}^{-m}C_{k'}{}^{-m+k+k'}C_{k''}\nn
&&\quad{}\times(-1)^{m-k-1}a^{m+k''-k-1}(b-a)^{-m-k'}\tilde{d}^{-A,t\to\infty}_{-m+k+k'-k''+1/2}\cr
d'^{(2),+A}_{-m}&=&\sum_{k=0}^{m}\sum_{k'=0}^{m-k}\sum_{k''\geq0}\left({}^{m+2}C_{k}b-{}^{m}C_{k}a\right){}^{-m-1}C_{k'}{}^{-m+k+k'-1}C_{k''}\nn
&&\quad{}\times(-1)^{m-k}b^{m-k+k''+1}(a-b)^{-m-k'-1}\tilde{d}^{+A,t\to\infty}_{-m+k+k'-k''-1/2}\nn
d'^{(2),-A}_{-m}&=&\sum_{k=0}^{m-1}\sum_{k'=0}^{m-k-1}\sum_{k''\geq0}{}^{m-1}C_{k}{}^{-m}C_{k'}{}^{-m+k+k'}C_{k''}\nn
&&\quad{}\times(-1)^{m-k-1}b^{m+k''-k-1}(a-b)^{-m-k'}\tilde{d}^{-A,t\to\infty}_{-m+k+k'-k''+1/2}.
\label{InitialModes t=infinity}
\eea

\subsubsection*{\underline{$t=0$}}
\bea
d'^{(1),+A}_{-m}&=&-\sum_{k=0}^{m}\sum_{k'=0}^{m-k}\sum_{k''\geq0}{}^{-m-1}C_{k'}{}^{-m+k+k'-1}C_{k''}\left({}^{m+2}C_{k}a-{}^{m}C_{k}b\right)\nn
&&\quad{}\times(-1)^{m-k}a^{k'-k''}(b-a)^{-m-k'-1}\tilde{d}^{+A,t\to0}_{k''+1/2}\nn\nn
d'^{(1),-A}_{-m}&=&-\sum_{k=0}^{m-1}\sum_{k'=0}^{m-k-1}\sum_{k''\geq0}{}^{m-1}C_{k}{}^{-m}C_{k'}{}^{-m+k+k'}C_{k''}\nn
&&\quad{}\times(-1)^{m-k-1}a^{k'-k''-1}(b-a)^{-m-k'}\tilde{d}^{-A,t\to0}_{k''+1/2}\nn
d'^{(2),+A}_{-m}&=&-\sum_{k=0}^{m}\sum_{k'=0}^{m-k}\sum_{k''\geq0}{}^{-m-1}C_{k'}{}^{-m+k+k'-1}C_{k''}\left({}^{m+2}C_{k}b-{}^{m}C_{k}a\right)\nn
&&\quad{}\times(-1)^{m-k}b^{k'-k''}(a-b)^{-m-k'-1}\tilde{d}^{+A,t\to0}_{k''+1/2}\nn\nn
d'^{(2),-A}_{-m}&=&-\sum_{k=0}^{m-1}\sum_{k'=0}^{m-k-1}\sum_{k''\geq0}{}^{m-1}C_{k}{}^{-m}C_{k'}{}^{-m+k+k'}C_{k''}\nn
&&\quad{}\times(-1)^{m-k-1}b^{k'-k''-1}(a-b)^{-m-k'}\tilde{d}^{+A}_{k''+1/2}.
\label{InitialModes t=0}
\eea

We also present the zero modes of (\ref{InitialModes t=infinity}) and (\ref{InitialModes t=0}) explicitly.

\subsubsection*{Zero Modes at $t=\infty$}
\bea
d'^{(1),+A}_{0}&=& -\sum_{k''\geq0}(-1)^{k''}a^{k''}\tilde{d}^{+A,t\to\infty}_{-k''-1/2}\nn
d'^{(1),-A}_{0}&=&0\nn
d'^{(2),+A}_{0}&=&-\sum_{k''\geq0}(-1)^{k''}b^{k''+1}\tilde{d}^{+A,t\to\infty}_{-k''-1/2}\nn
d'^{(2),-A}_{0}&=&0.
\label{InitialZeroModes t=infinity}
\eea
\subsubsection*{Zero Modes at $t=0$}
\bea
d'^{(1),+A}_{-m}&=&\sum_{k''\geq0}(-1)^{k''}a^{-k''}\tilde{d}^{+A,t\to0}_{k''+1/2}\nn
d'^{(1),-A}_{-m}&=&0\nn
d'^{(2),+A}_{-m}&=&\sum_{k''\geq0}(-1)^{k''} b^{-k''}\tilde{d}^{+A,t\to0}_{k''+1/2}\nn
d'^{(2),-A}_{-m}&=&0.
\label{InitialZeroModes t=0}
\eea
As expected, any expression that tries to start with a negative zero mode vanishes.

\subsection{Computing the transition amplitudes}
Here we compute all of the various $f^{F\pm,(i)(j)+}_{mn}$ factors that include at least one zero mode.  We will do this by applying the various expressions for the transformed modes to Equation (\ref{tPlaneRelations}).  We repeat the relavent parts of this equation here for convenience.
\bea
f^{F-,(i)(1')}_{mn} &=& -{\nstbra d'^{(1'),++}_nd'^{(i),--}_{-m}\nstket\over\nstbra 0_{NS}\rangle_t}\nn
f^{F-,(i)(2')}_{mn} &=& -{\nstbra d'^{(i),--}_{-m}d'^{(2'),++}_n\nstket\over\nstbra 0_{NS}\rangle_t} \nn
f^{F+,(i)(1')}_{mn}&=&-{\nstbra d'^{(1'),--}_nd'^{(i),++}_{-m}\nstket\over\nstbra 0_{NS}\rangle_t\rangle}\nn
f^{F+,(i)(2')}_{mn}&=&-{\nstbra d'^{(i),++}_{-m}d'^{(2'),--}_n\nstket\over\nstbra 0_{NS}\rangle_t\rangle}.\label{FermionRelation}
\eea

We will now work through the different copy and charge combinations in turn.

\subsubsection*{Positive SU(2) R charge, final copy 1}
We begin with $f^{F+,(1)(1')}_{mn}$.  Since our transition amplitude is defined specifically for modes that do not annihilate the vacuum on which $|\chi\rangle$ is built, there is no case $n=0$ for this term.  Indeed, if one tries to evaluate this transition amplitude for $n=0$ it is found to vanish.

For the case $m=0$, we insert Equations (\ref{tplanemodesNS}) and  (\ref{InitialZeroModes t=infinity}) into Equation (\ref{FermionRelation}).  This gives:
\bea
f^{F+,(1)(1')}_{0,n}&=&-\frac{{}_{t}\langle 0_{NS}|d'^{(1'),--}_{n}d'^{(1),++}_{0}|0_{NS}\rangle_{t}}{{}_{t}\langle 0_{NS}|0_{NS}\rangle_{t}}\nn
&=&\sum_{j,j',k''\geq 0} {}^{n}C_j{}^{n}C_{j'}(-1)^{k''}a^{j+k''+1} b^{j'}\frac{{}_{t}\langle 0_{NS}|\tilde{d}^{--}_{n-j-j'-1/2}\tilde{d}^{++}_{-k''-1/2}|0_{NS}\rangle_{t}}{{}_{t}\langle 0_{NS}|0_{NS}\rangle_{t}}\nn\quad \label{f11+ first}
\eea
Using the anticommutation relations from Equation (\ref{fermioncommutation}) along with creation/annihilation constraints, we find the following summation limits on the $j,j'$ sums:
\bea
n-j-j'-k''-1=0&\to&k'' = n-j-j'-1\nn
n-j-j'-1/2>0&\to&j'<n-j-1/2\nn
j'>0&\to&j<n-1/2.
\eea 
We then find:
\bea
\nn
f^{F+,(1)(1')}_{0,n}&=&-\sum_{j = 0}^{n-1} \sum_{j' = 0}^{n-j-1} {}^{n}C_j{}^{n}C_{j'}(-1)^{n-j-j'-1} a^{n-j'} b^{j'}.
\eea

We can now determine $f^{F+,(2)(1')}_{0,n}$ by appling the interchange $a\leftrightarrow b$.  This gives:
\bea
f^{F+,(2)(1')}_{0,n} &=& -\sum_{j = 0}^{n-1} \sum_{j' = 0}^{n-j-1} {}^{n}C_j{}^{n}C_{j'}(-1)^{n-j-j'-1} b^{n-j'} a^{j'}.
\eea

\subsubsection*{Positive SU(2) R charge, final copy 2}
We now turn to $f^{F+,(2)(2')}_{mn}$. Here both $m$ and $n$ are allowed to be zero.  We'll start wtih $m=0$.  Using (\ref{tplanezeromodesNS}),  (\ref{InitialZeroModes t=0}) and (\ref{FermionRelation}), we have:
\bea
f^{F+,(2)(2')}_{0,n} &=& \frac{{}_{t}\langle 0_{NS}|d'^{(2),++}_{0}d'^{(2'),--}_{n}|0_{NS}\rangle_{t}}{{}_{t}\langle 0_{NS}|0_{NS}\rangle_{t}}\label{F(f)22+_1}\\
&=& -\sum_{j,j',k''\geq 0}{}^{n}C_j{}^{n}C_{j'}(-1)^{k''}a^{n-j} b^{n-j'}b^{-k''}\frac{{}_{t}\langle 0_{NS}|\tilde{d}^{++,t\to0}_{k''+1/2}\tilde{d}^{--,t\to0}_{j+j'-n-1/2}|0_{NS}\rangle_{t}}{{}_{t}\langle 0_{NS}|0_{NS}\rangle_{t}}.\nonumber
\eea
We now use the anti commutation relations (\ref{fermioncommutation}) along with creation/annihilation constraints to limit the $j,j'$ sums:
\bea
k''+j+j'-n=0&\to& k''=n-j-j'\nn
j+j'-n-1/2< 0&\to& j'\leq n-j\nn
j'\geq 0&\to& j\leq n.
\eea
Using these constraints, (\ref{F(f)22+_1}) becomes:
\bea
f^{F+,(2)(2')}_{0,n}&=&-\sum_{j = 0}^{n}\sum_{j' = 0}^{n-j}{}^{n}C_j{}^{n}C_{j'}(-1)^{n-j-j'}a^{n-j} b^{j}.\label{fp220n}
\eea

Let us next examine the case $n=0$.  From (\ref{FermionRelation}) along with (\ref{tplanezeromodesNS}) and (\ref{InitialModes t=0}), we have:
\bea
f^{F+,(2)(2')}_{m,0} &=& \frac{{}_{t}\langle 0_{NS}|d'^{(2),++}_{m}d'^{(2'),--}_{0}|0_{NS}\rangle_{t}}{{}_{t}\langle 0_{NS}|0_{NS}\rangle_{t}}\nn
&=& \sum_{k=0}^{m}\sum_{k'=0}^{m-k}\sum_{k''\geq0}{}^{-m-1}C_{k'}{}^{-m+k+k'-1}C_{k''}\left({}^{m+2}C_{k}b-{}^{m}C_{k}a\right)\nn
&&\times(-1)^{m-k}b^{k'-k''}(a-b)^{-m-k'-1} { {}_{t}\langle 0_{NS}| \tilde{d}^{++,t\to0}_{k''+1/2}\tilde{d}^{--,t\to0}_{-1/2}|0_{NS}\rangle_{t} \over {}_{t}\langle 0_{NS}|0_{NS}\rangle_{t}} .\nn
\eea
The anti commutation relations (\ref{fermioncommutation}) now gives only $k''=0$.  Thus:
\bea
f^{F+,(2)(2')}_{m,0} &=&  -\sum_{k=0}^{m}\sum_{k'=0}^{m-k}{}^{-m-1}C_{k'}\left({}^{m+2}C_{k}b-{}^{m}C_{k}a\right)(-1)^{m-k}b^{k'}(a-b)^{-m-k'-1}.\nn
\eea
Naturally this expression should agree with (\ref{fp220n}) for the case $m=n=0$.  Indeed, both expressions yield:
\bea
f^{F+,(2)(2')}_{0,0} &=& 1.
\eea

We now assess the case of the initial mode living on copy 1 by applying the $a\leftrightarrow b$ interchange.  This gives:
\bea
f^{F+,(1)(2')}_{0,n} &=& -\sum_{j = 0}^{n}\sum_{j' = 0}^{n-j}{}^{n}C_j{}^{n}C_{j'}(-1)^{n-j-j'}b^{n-j} a^{j}\\
f^{F+,(1)(2')}_{m,0} &=&-\sum_{k=0}^{m}\sum_{k'=0}^{m-k}{}^{-m-1}C_{k'}\left({}^{m+2}C_{k}a-{}^{m}C_{k}b\right)(-1)^{m-k}a^{k'}(b-a)^{-m-k'-1}.\nonumber
\eea

\subsubsection*{Negative SU(2) R charge}
Here we analyze to the amplitudes $f^{F-,(i)(j)}_{mn}$.  These amplitudes vanish for $m=0$, since an initial negative zero mode on either copy annihilates the R vacuum upon which our in state is built.  Similarly, a negative zero mode on copy 2 above the twists annihilates the vacuum upon which the out state is built.  We are thus left with only two amplitudes to consider, $f^{F-,(1)(1')}_{m,0}$ and $f^{F-,(2)(1')}_{m,0}$.  Since these are related by the $a \leftrightarrow b$ interchange, we will explicitly calculate only the first amplitude.
\bea
f^{F-,(1)(1')}_{m,0} &=& -\frac{{}_{t}\langle 0_{NS}|d'^{(1'),++}_{0}d'^{(1),--}_{-m}|0_{NS}\rangle_{t}}{ {}_{t}\langle 0_{NS}|0_{NS}\rangle_{t}}.\label{fminusratio}
\eea
Let us now examine (\ref{tplanezeromodesNS}).  The relevant transformed mode contains two parts:
\bea
d'^{(1'),+A}_{0}&=&\sum_{j,j'\geq 0}(-1)^{j+j'} a^j b^{j'} \tilde{d}^{+A,t\to\infty}_{-j-j'+1/2}-\sum_{j,j'\geq 0} (-1)^{j+j'} a^{j+1} b^{j'+1}\tilde{d}_{-j-j'-3/2}^{+A,t\to\infty}.\nn
\eea
This is the leftmost mode, so we can drop any terms that annihilate $\nstbra$ on the left.  This includes the entirety of the second sum, as well as all terms from the first sum except $j=j'=0$.  Combining (\ref{InitialModes t=infinity}), Equation (\ref{fminusratio}) becomes:
\bea
f^{F-,(1)(1')}_{m,0} &=&-\sum_{k''\geq 0}\sum_{k=0}^{m-1}\sum_{k'=0}^{m-k-1}{}^{m-1}C_{k}{}^{-m}C_{k'}{}^{-m+k+k'}C_{k''}(-1)^{m-k-1}\\
&&\quad{}\times\left(a^{m+k''-k-1}(b-a)^{-m-k'}\frac{{}_{t}\langle 0_{NS}|\tilde{d}^{++,t\to\infty}_{1/2}\tilde{d}^{--,t\to\infty}_{-m+k+k'-k''+1/2}|0_{NS}\rangle_{t}}{ {}_{t}\langle 0_{NS}|0_{NS}\rangle_{t}}\right).\nonumber
\eea

We once again use the anticommutation relations (\ref{fermioncommutation}) to find the following constraints:
\bea
-m+k+k'-k''+1=0&\to&k''=-m+k+k'+1 \cr
 k''\geq 0&\to&k'\geq m-k-1
\eea
The $k'$ minimum here is already the maximum value it can take, so we can eliminate that sum in addition to eliminating the $k''$ sum.  We then find:
\bea
f^{F-,(1)(1')}_{m,0}&=&\sum_{k=0}^{m-1}{}^{m-1}C_{k}{}^{-m}C_{m-k-1}(-1)^{m-k-1}a^{m-k-1}(b-a)^{-2m+k+1}.
\eea
We now apply $a\leftrightarrow b$ to find:
\bea
f^{F-,(2)(1')}_{m,0}&=&\sum_{k=0}^{m-1}{}^{m-1}C_{k}{}^{-m}C_{m-k-1}(-1)^{m-k-1}b^{m-k-1}(a-b)^{-2m+k+1}.
\eea

\subsection{Summary of Results}
Here we gather all of the results for the fermionic transition amplitudes that involve at least one zero mode.
\bea
f^{F+,(1)(1')}_{0,n} &=&-\sum_{j = 0}^{n-1} \sum_{j' = 0}^{n-j-1} {}^{n}C_j{}^{n}C_{j'}(-1)^{n-j-j'-1} a^{n-j'} b^{j'}\nn
f^{F-,(1)(1')}_{m,0} &=&\sum_{k=0}^{m-1}{}^{m-1}C_{k}{}^{-m}C_{m-k-1}(-1)^{m-k-1}a^{m-k-1}(b-a)^{-2m+k+1}\nn
f^{F+,(2)(1')}_{0,n} &=& -\sum_{j = 0}^{n-1} \sum_{j' = 0}^{n-j-1} {}^{n}C_j{}^{n}C_{j'}(-1)^{n-j-j'-1} b^{n-j'} a^{j'}\nn
f^{F-,(2)(1')}_{m,0} &=&\sum_{k=0}^{m-1}{}^{m-1}C_{k}{}^{-m}C_{m-k-1}(-1)^{m-k-1}b^{m-k-1}(a-b)^{-2m+k+1}\nn
f^{F+,(2)(2')}_{m,0}&=& -\sum_{k=0}^{m}\sum_{k'=0}^{m-k}{}^{-m-1}C_{k'}\left({}^{m+2}C_{k}b-{}^{m}C_{k}a\right)(-1)^{m-k}b^{k'}(a-b)^{-m-k'-1}\cr
f^{F+,(2)(2')}_{0,n}&=&-\sum_{j = 0}^{n}\sum_{j' = 0}^{n-j}{}^{n}C_j{}^{n}C_{j'}(-1)^{n-j-j'}a^{n-j} b^{j}\nn
f^{F+,(1)(2')}_{m,0}&=&-\sum_{k=0}^{m}\sum_{k'=0}^{m-k}{}^{-m-1}C_{k'}\left({}^{m+2}C_{k}a-{}^{m}C_{k}b\right)(-1)^{m-k}a^{k'}(b-a)^{-m-k'-1}\nn
f^{F+,(1)(2')}_{0,n} &=& -\sum_{j = 0}^{n}\sum_{j' = 0}^{n-j}{}^{n}C_j{}^{n}C_{j'}(-1)^{n-j-j'}b^{n-j} a^{j}.
\eea

\section{Discussion}
In this chapter we computed the bosonic transition amplitudes, $f^B$, for the two twist case. Because of supersymmetry these expressions have a simple relation with the fermionic transition functions, $f^{F\pm}$, for nonzero mode numbers therefore saving us from having to compute the fermionic transition amplitudes directly. Additionally, however, we computed the relevant fermionic transition functions containing at least \textit{one} zero mode. This provided an additional complication as we had to use the $t$-plane spectral flowed modes to compute them. This is because the bosonic quantities contain no zero mode contributions and therefore no supersymmetry relations with the fermionic quantities. Also, because there is now a twist separation, $\Delta w$, we find interesting behavior for $f^B$ in the two twist case. An initial mode which starts on Copy 1 will produce a distribution of modes in the final state which peak around the initial energy for small twist separation. However, as the separation increases, the initial energy on Copy 1 one  transitions to Copy 2 and back to Copy 1 after a period of $4\pi$. Under the interchange of Copy 1 and Copy 2, this behavior is symmetric which is expected. Also for large mode numbers we find that the behavior of the two twist transition functions are very similar to the one twist case differing by an oscillating factor. 

Now that we have computed the action of the twist operators on an initial state we need to determine the action of the two supercharge operators. Since there are two twist's this time, we can't just stretch the $G$ contours to initial and final states on the cylinder. This is because the contours which end up below the second twist but above the first twist can not be removed from the cylinder. Therefore we will always obtain a term where one twist acts, then the combination of G contours act and then the second twist would act. This is problematic because this gives an infinite number of terms. The first twist produces a squeezed state which is an exponential of bilinear excitations, which contains an infinite number of terms when expanded out. Therefore the $G$ contours above the first twist would have to act on the infinite number of terms. Then the second twist would have to act on the result of that, which is again an infinite numbers of terms. We must therefore map our deformation operators to the $t$-plane, remove all spin fields in the $t$-plane through spectral flow, expand all remaining contours to initial and final states, and then map the final state back to the cylinder. We show how to do this next chapter. We note that this work is based on paper \cite{chm2}.

\chapter{Supercharge contours for the two twist case}\label{G contours}
So far, for the two twist case, we have computed the Bogoliubov coefficients, $\g^{B,F}$, and the transition amplitudes, $f^{B,F}$, both whose computations were analogous to the one twist case. However, in the one twist case, we were able to freely stretch the supercharge contours to initial and final states on the cylinder without obstruction. In the two twist case, this is not so simple. While we can stretch the contours on the cylinder to initial and final states, these are not the only contributions. We also obtain contours which lie between the two twists. This term is nearly impossible to compute directly. Therefore, instead of stretching the supercharge contours on the cylinder directly, we first map them to the covering space, remove all spin fields, stretch them to $t$ plane images corresponding to initial and final states on the cylinder, and then map the result back to the cylinder. We perform these computations in detail in this chapter.

\section{The Cylinder}

Now let us start with defining our cylinder coordinate
\bea
w=\t + i\s
\eea
where $\t$ is euclidean time and $\s$ is the spatial coordinate of our cylinder. The cylinder radius R has been included both coordinates so that $\s$ just becomes the angle. In our initial state at $\t=-\infty$ we have two singly wound copies of the CFT both in the negative Ramond vacuum which can be defined via spin fields applied to $NS$ vacuua for each copy:
\bea
\rmmket = S^{(1)-}(\t = -\infty)S^{(2)-}(\t = -\infty)\nsnsket\label{initial}
\eea
Each twist operator also contains a spin field:
\bea
\s^+_2(w_1)&=&S^+(w_1)\s_2(w_1)\cr
\s^+_2(w_2)&=&S^+(w_2)\s_2(w_2)
\eea
where we've taken $|w_2|>|w_1|$. At $\t \to \infty$ we have two singly wound copies of the CFT described by the state $|\chi(w_1,w_2)\rangle$ which was computed in (\ref{chi state}).
The locations of the initial and final copies of the CFT as well as the twist insertions are tabulated below:
\bea
w&=&-\infty + i\s,~~~~ \text{Copy 1 and 2 initial}\cr
w&=&\infty + i\s,~~~~ \text{Copy 1 and 2 final}\cr
w_1&=&\t_{1} + i\s_1,~~~~ \text{First twist insertion}\cr
w_2&=&\t_{2} + i\s_2,~~~~ \text{Second twist insertion}, \quad \t_2>\t_1
\eea
Next we describe the supercharge action on the cylinder.

\subsection{Supercharge Contours on the Cylinder}
Now we describe the supercharge action on the cylinder. We note that the two copies of the deformation operator will be taken to have $SU(2)_2$ indices $\dot A, \dot B$. We will write the supercharge action at each twist as a contour surrounding that twist. Our goal is to   remove the supercharge contours from these twists, and to move them to the initial and final states. This will relate the full amplitude for two deformation operators to an amplitude with just two bare twists.

The two deformation operators on the cylinder give the operator
\bea
\hat{O}_{\dot{B}}\hat{O}_{\dot{A}}&=&{1\over 2 \pi i}\oint_{w_2} dw G^-_{\dot{B}}\s_2^+(w_2){1\over 2 \pi i}\oint_{w_1} dw'G^-_{\dot{A}}(w')\s_2^+(w_1) 
\eea
where we have only written the holomorphic part. The antiholomorphic part is identical.

Let us first stretch the contour for the operator $G^-_{\dot B}$ away from the twist on which it is applied. We will get three different contributions corresponding to three different locations:
\begin{itemize}
\item $\dot{B}$ positive direction contour for both copies at $\t > \t_2$\\
\item $\dot{B}$ negative direction contour for both copies at $\t < \t_1$\\
\item $\dot{B}$ negative direction contour outside of $\dot{A}$ contour at $\t_1$
\end{itemize}

Stretching these contours therefore gives on the cylinder:
\bea
\hat{O}_{\dot{B}}\hat{O}_{\dot{A}} &=& {1\over 2 \pi i}\int_{\s =0,\t>\t_2}^{\s=2\pi}\diff w G^-_{\dot{B}}(w)\s_2^+(w_2){1\over 2 \pi i}\oint_{w_1} \diff w'G^-_{\dot{A}}(w')\s_2^+(w_1)\cr
&& -\s_2^+(w_2){1\over 2 \pi i}{1\over 2 \pi i} \oint_{w_1}\oint_{w_1,|w-w_1|>|w'-w_1|}  G^-_{\dot{B}}(w) G^-_{\dot{A}}(w')\s_2^+(w_1)\diff w \diff w'\cr\cr
&& + \s_2^+(w_2){1\over 2 \pi i}\oint_{w_1} \diff w'G^-_{\dot{A}}(w')\s_2^+(w_1) {1\over 2 \pi i}\int_{\s =0,\t<\t_1}^{\s=2\pi}\diff w G^-_{\dot{B}}(w)
\eea
Here the extra sign change in the second term comes from swapping the order of $G^-_{\dot{A}}$ and $G^-_{\dot{B}}$. The contours  above $\s_2^+(w_2)$ and below $\s_{2}^+(w_1)$ wrap around the cylinder and have no power of $e^w$ multiplying $G^-_{\dot B}(w)$. Thus we can rewrite our expression as 
\bea
\hat{O}_{\dot{B}}\hat{O}_{\dot{A}} &=&\big( G^{(1),-}_{\dot{B},0} + G^{(2),-}_{\dot{B},0}\big)\s_2^+(w_2){1\over 2 \pi i}\oint_{w_1} \diff w'G^-_{\dot{A}}(w')\s_2^+(w_1)\cr
&& -\s_2^+(w_2){1\over 2 \pi i}{1\over 2 \pi i} \oint_{w_1}\oint_{w_1,|w-w_1|>|w'-w_1|}  G^-_{\dot{B}}(w) G^-_{\dot{A}}(w')\s_2^+(w_1)\diff w \diff w'\cr\cr
&& + \s_2^+(w_2){1\over 2 \pi i}\oint_{w_1} \diff w'G^-_{\dot{A}}(w')\s_2^+(w_1)\big( G^{(1),-}_{\dot{B},0} + G^{(2),-}_{\dot{B},0}\big)
\cr
\cr
&\equiv&\big( G^{(1),-}_{\dot{B},0} + G^{(2),-}_{\dot{B},0}\big)\mathcal{I}_1 - \mathcal{I}_2 + \mathcal{I}_{1}\big( G^{(1),-}_{\dot{B},0} + G^{(2),-}_{\dot{B},0}\big)
\label{full deformation one}
\eea
where in the last line we have made the following definitions
\bea
\mathcal{I}_1&\equiv& \s_2^+(w_2){1\over 2 \pi i}\oint_{w_1} \diff w G^-_{\dot{A}}(w)\s_2^+(w_1)\cr
\mathcal{I}_2&\equiv&\s_2^+(w_2) {1\over 2 \pi i}{1\over 2 \pi i} \oint_{w_1}\oint_{w_1,|w-w_1|>|w'-w_1|}  G^-_{\dot{B}}(w) G^-_{\dot{A}}(w')\s_2^+(w_1)\diff w \diff w'
\eea

We now have to evaluate the terms $\mathcal{I}_1$ and $\mathcal{I}_2$. This will be one of the main steps of our computation. We will evaluate each of these terms by mapping from the cylinder $w$ to the a covering space described by a coordinate  $t$. We now turn to this map.

\subsection{Mapping from the Cylinder to the $t$ Plane}

As mentioned several times now, the map from the cylinder to the $z$ plane is given by:
\bea
z=e^{w}
\eea
The $z$ plane locations corresponding to initial and final copies of the CFT and the twist insertions are given by:
\bea
z&=&e^{-\infty+i\s}\to|z| = 0,~~~~\text{Copy 1 and 2 Final}\cr
z&=&e^{\infty+i\s}\to|z| = \infty,~~~~\text{Copy 1 and 2 Final}\cr
z_1&=&e^{w_1},~~~~ \text{First twist insertion}\cr
z_2&=&e^{w_2},~~~~ \text{Second twist insertion}
\label{w coor}
\eea
Mapping the measure combined with the supercharge as well as the twist operator to the $z$ plane gives
\bea
&&\diff w G^{-}_{\dot{A}}(w)\xrightarrow[]{w \to z} \diff z \bigg({\diff z\over \diff w} \bigg)^{1/2} G^{-}_{\dot{A}}(z) =\diff z z^{1/2} G^{-}_{\dot{A}}(z)\cr
&&\s^+_2(w_i)\xrightarrow[]{w \to z}\bigg({\diff z\over \diff w} \bigg)^{1/2}_{w=w_i}\s_2^+(z_i) =  z_i^{1/2}\s_2^{+}(z_i),\qquad i=1,2
\label{w to z}
\eea
Now let us map to the covering $t$ plane with the map:
\bea
z = {(t+a)(t+b)\over t}
\label{map2}
\eea
Since our twist operators carry spin fields we will have to compute the images of the spin fields in the $t$ plane. These are bifurcation points. To do this we take the derivative of our map in (\ref{map2}) and set it equal to zero:
\bea
{\diff z\over \diff t}={t^2-ab\over t^2}={(t-\sqrt{ab})(t+\sqrt{ab})\over t^2}=0
\eea
The solution to the above give the images of our spin fields to be:
\bea
t_1=-\sqrt{ab},\quad t_2 = \sqrt{ab} 
\label{twist image points}
\eea
Here we write the $t$ plane images corresponding to initial and final states on the cylinder where we use (\ref{w coor}):
\bea
z=0 \to  t&=&-a,~~~~ \text{Copy 1 Initial}\cr
z=0 \to t&=&-b ~~~~ \text{Copy 2 Initial}\cr
 z\to \infty \to t&=&\infty ~~~~ \text{Copy 1 Final}\cr
 z\to \infty\to t&=& 0  ~~~~ \text{Copy 2 Final}
\eea
where we have split the locations of the initial and final copies.
Now inserting the image points of our twist (\ref{twist image points}) into our map (\ref{map2}), we define the twist insertion points in the $z$ plane in terms of the images of our initial copies $a$ and $b$:
\bea
z_1&\equiv&{(t_1+a)(t_1+b)\over t_1}=(\sqrt{a}-\sqrt{b})^2\cr
z_2&\equiv&{(t_2+a)(t_2+b)\over t_2} = (\sqrt{a} +\sqrt{b})^2
\label{coordinates}
\eea
We note that our the measure and supercharge transform as:
\bea
\diff z ~z^{1/2} G^{-}_{\dot{A}}(z)&\xrightarrow[]{z \to t}& \diff t (t+a)^{1/2}(t+b)^{1/2}t^{-1/2} \bigg({\diff z \over \diff t}\bigg)^{-1/2} G^{-}_{\dot{A}}(t)\cr
&=&  \diff t (t+a)^{1/2}(t+b)^{1/2}t^{-1/2} (t-t_1)^{-1/2}(t-t_2)^{-1/2}t G^{-}_{\dot{A}}(t)\cr
&=&  \diff t (t+a)^{1/2}(t+b)^{1/2} (t-t_1)^{-1/2}(t-t_2)^{-1/2}t^{1/2} G^{-}_{\dot{A}}(t)
\label{G coor map}
\eea
Let us define the separation of the twists as $\Delta w \equiv w_2 - w_1$. We can therefore define $w_1 = - {\Delta w \over 2}$ and $w_2 = {\Delta w \over 2}$. Using (\ref{w coor}), (\ref{coordinates}) and our definitions of $w_1$ and $w_2$, we can rewrite our $a$ and $b$ coordinates in terms of the twist separation $\Delta w$:
\bea
\sqrt{a} &=& {1\over 2}(\sqrt{z_2}+\sqrt{z_1})={1\over 2}\big(e^{{\Delta w \over 4
}} + e^{-{\Delta w \over 4}}\big)~~\to~~ a = \cosh^2\bigg({\Delta w \over 4 }\bigg)\cr
\sqrt{b} &=& {1\over 2}(\sqrt{z_2}-\sqrt{z_1})={1\over 2}\big(e^{{\Delta w \over 4}} - e^{-{\Delta w \over 4}}\big)~~\to~~ b = \sinh^2\bigg({\Delta w \over 4 }\bigg)
\eea 
We see that $a,b$ are strictly positive so we are confident to take the positive branch to be $\sqrt{ab}$ and the negative branch to be $-\sqrt{ab}$.
\subsection{Spectral Flows}
Since we started on the cylinder in the Ramond sector, the $t$ plane contains spin fields coming from the initial state (\ref{initial}), the final state given by $|\chi(w_1,w_2)\rangle$, and two twist insertions. In the final state $|\chi(w_1,w_2)\rangle$, given in (\ref{chi state}), we have an exponential of bilinear boson and fermion operators built on the vacuum $\rpmket$. These bilinear operators are accompanied by bosonic and fermionic Bogoliubov coefficients which we computed in Chapter \ref{two twist gamma}. To obtain a nonzero amplitude it is necessary to cap any given final state with the vacuum $\rpmbra$ which is given by:
\bea 
\rpmbra = \nsnsbra S^{(1)-}(\t=\infty)S^{(2)+}(\t = \infty)
\eea
We therefore see that this capping state also brings in spin fields.
We then remove all of the above mentioned spin fields by spectral flowing them away. Below, we record the spin fields sitting at finite points in the $t$ plane:
\bea
|0_R^-\rangle^{(1)} &\to& S^{-}(t=-a)\cr
|0_R^-\rangle^{(2)} &\to& S^{-}(t=-b)\cr
z_1^{1/2}\s_2^+(z_1)&\to&z_1^{1/2} S^{+}(t=t_1)\cr
z_2^{1/2}\s_2^+(z_2)&\to& z_2^{1/2} S^{+}(t=t_2)\cr
{}^{(2)}\langle 0_{R,-}| &\to& S^{+}(t=0)
\eea
We also have a spin field at $t=\infty$ coming from ${}^{(1)}\rpbra$. As we spectral flow away the spin fields at finite points the spin field at infinity will also be removed.
We will spectral flow in a way that does not introduce any new operators such as $J$'s. Under spectral flow, our $G^-$ changes as follows
\bea
G^{-}_{\dot{A}}(t) \to (t-t_i)^{ {\a \over 2}} G^{-}_{\dot{A}}(t)
\eea
where $t_i$ is the location of the spectral flow. Performing the following spectral flows
\bea
\a=+1 ~~\text{around}~~ t_i = -a,-b\cr
\a=-1 ~~\text{around}~~ t_i = t_1,t_2,0\cr
\eea
transforms our supercharge as follows:
\bea
G^-_{\dot{A}}(t)\to (t-a)^{1/2}(t-b)^{1/2}(t-t_1)^{-1/2}(t-t_2)^{-1/2}t^{-1/2}G^-_{\dot{A}}(t)
\label{sf}
\eea
Combining our spectral flow in (\ref{sf}) with the coordinate map from the $z$ to $t$ plane given in (\ref{G coor map}), our modification of the integrand and measure is given by:
\bea
 &&\diff t (t+a)^{1/2}(t+b)^{1/2} (t-t_1)^{-1/2}(t-t_2)^{-1/2}t^{1/2} G^{-}_{\dot{A}}(t)\cr
 &&\qquad\qquad \xrightarrow[]{t \to sf ~ t}  \diff t (t+a)(t+b) (t-t_1)^{-1}(t-t_2)^{-1}G^{-}_{\dot{A}}(t)\nn
\label{z to t sf transformation}
\eea
So the full transformation from $w$ to spectral flowed $t$ plane.
\bea
\diff w G^{-}_{\dot{A}}(w)&\xrightarrow[]{w \to z}& \diff z ~z^{1/2} G^{-}_{\dot{A}}(z)\cr
 &\xrightarrow[]{z \to t}&\diff t (t+a)^{1/2}(t+b)^{1/2} (t-t_1)^{-1/2}(t-t_2)^{-1/2}t^{1/2} G^{-}_{\dot{A}}(t)\cr
 &\xrightarrow[]{t \to sf ~ t}&   \diff t (t+a)(t+b) (t-t_1)^{-1}(t-t_2)^{-1}G^{-}_{\dot{A}}(t)\nn
\label{full G transformation}
\eea
We will also obtain an over factor of $C$ when performing our spectral flows coming from the presence of other spin fields. This constant depends upon the order in which the spectral flows were performed. We won't worry about the value of $C$ because we will invert the spectral flows in exactly the opposite order which will remove it.

Our next goal is to compute the amplitudes $\mathcal{I}_1$ and $\mathcal{I}_2$.

\section{Computing $\mathcal{I}_1$}\label{computing I1}

Now let us compute integral expression $\mathcal{I}_1$.  We start with the expression $\mathcal{I}_1$.  We apply the transformation (\ref{full G transformation}) and note  the Jacobian factor $(z_1z_2)^{1/2}$ coming from (\ref{w to z}).   We obtain
\bea
\mathcal{I}_1&\to& C(z_1z_2)^{1/2}{1\over 2 \pi i}\oint_{t_1} \diff t (t+a)(t+b) (t-t_1)^{-1}(t-t_2)^{-1}G^{-}_{\dot{A}}(t)
\label{I two}
\eea
We have removed all spin fields but we still have a singularity at $t=t_2$. To remove the singular behavior from $t_2$ we expand $(t-t_2)^{-1}$ around $t_1$:
\bea
(t-t_2)^{-1}=(t-t_1 + t_1 - t_2)^{-1}&=&\sum_{k=0}^\infty{}^{-1}C_{k}(t_1-t_2)^{-k-1}(t-t_1)^{k}\cr
&=&(t_1 - t_2)^{-1} + \sum_{k=1}^\infty{}^{-1}C_{k}(t_1-t_2)^{-k-1}(t-t_1)^{k}\nn
\label{expansion}
\eea
Inserting this into (\ref{I two}) we see that the only nonvanishing term be will the $k=0$ term because all others will annihilate the vacuum at $t_1$. Let us show this in more detail. We first define supercharge modes natural to the $t$ plane:
\bea
\tilde{G}^{\a,t\to t_1}_{\dot{A} , r}={1\over 2 \pi i}\oint \diff t (t-t_1)^{r+1/2} G^{\a}_{\dot{A},r} 
\label{t plane G mode}
\eea
Then we have
\bea
\tilde{G}^{-,t\to t_1}_{\dot{A} , r}|0_{NS}\rangle_{t_1} = 0,\quad r>-3/2
\eea
Therefore as we stated previously, the only $\tilde{G}^-_{\dot{A}}$ mode that survives when acting locally at $t_1$ is the one corresponding to $k=0$ in the expansion given in (\ref{expansion}) which is $\tilde{G}^{-,t\to t_1}_{\dot{A},-{3 \over 2}}$.
Therefore (\ref{I two}) becomes
\bea
\mathcal{I}_1\to C(z_1z_2)^{1/2}(t_1-t_2)^{-1}{1\over 2 \pi i}\oint_{t_2} \diff t (t+a)(t+b) (t-t_1)^{-1}G^{-}_{\dot{A}}(t)
\eea
We now only have contours around $t=t_1$.
We can now expand $\mathcal{I}_1$ to points corresponding to initial and final states in the $t$ plane.
\bea
\mathcal{I}_1 &\to& - C(z_1z_2)^{1/2}{1\over (t_2-t_1)}\cr
&&~\times~{1\over 2 \pi i} \bigg(\oint_{t=\infty} -\oint_{t=0} - \oint_{t=-a} -\oint_{=-b}  \bigg)\diff t (t+a)(t+b) (t-t_1)^{-1} G^{-}_{\dot{A}}(t)
\cr
\cr
\cr
&&=   C(z_1z_2)^{1/2}{1\over (t_2-t_1)}
\cr
&&\quad\times~{1\over 2 \pi i} \bigg(-\oint_{t=\infty} +  \oint_{t=0} + \oint_{t=-a} +  \oint_{=-b}  \bigg)\diff t (t+a)(t+b) (t-t_1)^{-1} G^{-}_{\dot{A}}(t)\nn
\label{I three}
\eea
where the minus signs come from the fact that contours at finite points reverse their order. Let us now undo the spectral flows. Performing spectral flows in the reverse direction, the integrand and measure change as follows 
\bea
  G^-_{\dot{A}}(t)\diff t &\xrightarrow[]{ t ~ \to ~ rsf ~t}& (t-a)^{-1/2}(t-b)^{-1/2}(t-t_1)^{1/2}(t-t_2)^{1/2}t^{1/2}G^-_{\dot{A}}(t)\diff t
  \cr
  \cr
 \one (t_2)\one(t_1)&\to& S^{+}(t_2)S^{+}(t_1)
 \label{t to rsf t}
\eea
where we have restored all spin fields but have only explicitly written those at the twist insertion points. 
Now let us map back to the $z$ plane. The combination of the measure with $G^-_{\dot{A}}(t)$ along with the spin fields transform as:
\bea
 G^-_{\dot{A}}(t)\diff t&\xrightarrow[]{rsf ~ t\to z}&\bigg({\diff z \over \diff t}\bigg)^{1/2}G^-_{\dot{A}}(z)\diff z =(t-t_1)^{1/2}(t-t_2)^{1/2}t^{-1}G^-_{\dot{A}}(z)\diff z\cr
\cr
 S^{+}(t_2)S^{+}(t_1) &\xrightarrow[]{rsf ~ t\to z}& \s_{2}^+(z_2)\s_{2}^+(z_1)
\label{rsf t to z}
\eea
Combining (\ref{t to rsf t}) and (\ref{rsf t to z}) with (\ref{I three}) gives $\mathcal{I}_1$ to be:
\bea
\mathcal{I}_1 \!\!\!\!&\to&\!\!\!\!   -(z_1z_2)^{1/2}{1\over (t_2-t_1)} \bigg[\bigg({1\over 2 \pi i}\oint_{z=\infty}\diff z ~z^{1/2}(t-t_2) G^{(1)-}_{\dot{A}}(t)\cr
&& + {1\over 2 \pi i}\oint_{z=\infty}\diff z~z^{1/2} (t-t_2) G^{(2)-}_{\dot{A}}(z)\bigg)\s_2^+(z_2)\s_2^+(z_1)\cr
&&\qquad\qquad -\s_2^+(z_2)\s_2^+(z_1)\bigg( {1\over 2 \pi i}\oint_{z=0}\diff z  ~ z^{1/2}(t-t_2) G^{(1)-}_{\dot{A}}(z)\cr
&& +  {1\over 2 \pi i}\oint_{z=0}\diff z ~ z^{1/2} (t-t_2) G^{(2)-}_{\dot{A}}(z)\bigg) \bigg]\nn
\eea
where we have used our map $z=(t+a)(t+b)t^{-1}$. We note the sign reversal for the integral $\oint_{t=0}$ because for $t=0$ the map goes like $z\sim {1\over t}$. This reverses the contour direction in the $z$ plane.
Now mapping this result back to the cylinder with the transformation
\bea
\diff z G^-_{\dot{A}}(z)&\xrightarrow[]{z \to w} & \diff w z^{-1/2} G^-_{\dot{A}}(w)
\cr
\cr
\s_{2}^{+}(z_2)\s_{2}^{+}(z_1)&\xrightarrow[]{z \to w} &(z_1z_2)^{-1/2}\s_{2}^{+}(w_2)\s_{2}^{+}(w_1)
\eea
gives $\mathcal{I}_1$ to be:
\bea
\mathcal{I}_1 \!\!\!\!&\to&\!\!\!\!   -{1\over (t_2-t_1)} \bigg[{1\over 2 \pi i}\int_{\s = 0,\t>\t_2}^{\s = 2\pi}\diff w (t-t_2)\bigg( G^{(1)-}_{\dot{A}}(w) + G^{(2)-}_{\dot{A}}(w) \bigg) \s_2^+(w_2)\s_2^+(w_1)\cr
&&\qquad\qquad -\s_2^+(w_2)\s_2^+(w_1){1\over 2 \pi i}\int_{\s = 0,\t<\t_1}^{\s = 2\pi}\diff w (t-t_2)\bigg( G^{(1)-}_{\dot{A}}(w) + G^{(2)-}_{\dot{A}}(w) \bigg) \bigg]\nn
\label{I four prime}
\eea
Let us take $t_1=-t_2$. Inserting this into (\ref{I four prime}) gives $\mathcal{I}_1$ to be:
\bea
\mathcal{I}_1 \!\!\!\!&\to&\!\!\!\!   -{1\over 2t_2} \bigg[{1\over 2 \pi i}\int_{\s = 0,\t>\t_2}^{\s = 2\pi}\diff w (t-t_2)\bigg( G^{(1)-}_{\dot{A}}(w) + G^{(2)-}_{\dot{A}}(w) \bigg) \s_2^+(w_2)\s_2^+(w_1)\cr
&&\qquad\qquad -\s_2^+(w_2)\s_2^+(w_1){1\over 2 \pi i}\int_{\s = 0,\t<\t_1}^{\s = 2\pi}\diff w (t-t_2)\bigg( G^{(1)-}_{\dot{A}}(w) + G^{(2)-}_{\dot{A}}(w) \bigg) \bigg]\nn
\label{I four}
\eea
Looking at (\ref{I four}), we see that we have written $\mathcal{I}_1$ in terms of $G^-_{\dot{A}}$ contours at initial and final states on the cylinder. Later we will write our contours in terms of cylinder modes where we will have to expand the $t$ coordinate in terms $z=e^w$. 

As a side note, let us summarize the total transformation going from the $t$ plane all the way back to the cylinder. We note that it is exactly opposite from the transformation (\ref{full G transformation}):
\bea
\diff t G^{-}_{\dot{A}}(t) &\xrightarrow[]{t\, \to rsf ~ t}&    (t+a)^{-1/2}(t+b)^{-1/2} (t-t_1)^{1/2}(t-t_2)^{1/2}t^{1/2} \diff t G^{-}_{\dot{A}}(t)\cr
 &\xrightarrow[]{rsf\, t\to z}&\diff z z^{-1/2} t^{-1}(t-t_1)(t-t_2) G^{-}_{\dot{A}}(z)\cr
&\xrightarrow[]{w \to z}& \diff w ~ (t+a)^{-1}(t+b)^{-1}(t-t_1)(t - t_2)G^{-}_{\dot{A}}(w)
\label{t to w}
\eea
and for the spin fields coming from twist insertions
\bea
\one (t_2)\one (t_1)&\xrightarrow[]{ t\to rsf~t}& S^{+}(t_2)S^{+}(t_1)\cr
S^{+}(t_2)S^{+}(t_1) &\xrightarrow[]{rsf ~ t\to z}& \s_{2}^+(z_2)\s_{2}^+(z_1)
\cr
\s_{2}^{+}(z_2)\s_{2}^{+}(z_1)&\xrightarrow[]{z \to w} &(z_1z_2)^{-1/2}\s_{2}^{+}(w_2)\s_{2}^{+}(w_1)
\label{spin field t to w}
\eea
In the next section we compute the amplitude $\mathcal{I}_2$ following the same procedure as we did here for $\mathcal{I}_1$.

\section{Computing $\mathcal{I}_2$}\label{computing I2}

Now let us compute the integral expression $\mathcal{I}_2$. We start with the cylinder expression:
\bea
\mathcal{I}_2&\equiv&\s_2^+(w_2) {1\over 2 \pi i}{1\over 2 \pi i} \oint_{w_1}\oint_{w_1,|w-w_1|>|w'-w_1|}  G^-_{\dot{B}}(w) G^-_{\dot{A}}(w')\s_2^+(w_1)\diff w \diff w'
\eea
Let us map from the cylinder to the $t$ plane and perform the necessary spectral flows. In doing this we obtain a modification in our integral expression that is similar to $\mathcal{I}_{1}$ except now we have two integrals instead of one. Therefore inserting the result in (\ref{full G transformation}) for both the $t$ integral and the $t'$ integral gives
\bea
\mathcal{I}_{2} &\to& CC'(z_1z_2)^{1/2}{1\over 2 \pi i}{1\over 2 \pi i} \oint_{t_1}\oint_{t_1,|t-t_1|>|t'-t_1|}(t+a)(t+b) (t-t_1)^{-1}(t-t_2)^{-1} \cr
&&(t'+a)(t'+b) (t'-t_1)^{-1}(t'-t_2)^{-1} G^-_{\dot{B}}(t) G^-_{\dot{A}}(t')\diff t \diff t'
\label{GG one prime}
\eea
where the constants $C$ and $C'$ come from spectral flows in both the $t$ and $t'$ coordinates respectively. Both constants will again be removed later when spectral flowing in exactly the reverse order.
We again remove singular terms at $t_2$ by expanding them around $t_1$. Inserting the expansion given in (\ref{expansion}) into (\ref{GG one prime}) for both $(t-t_2)^{-1}$ and $(t'-t_2)^{-1}$ gives $\mathcal{I}_2$ to be:
\bea
\mathcal{I}_{2} &\equiv& CC'(z_1z_2)^{1/2}\sum_{k=0}^{\infty}(-1)^k(t_1-t_2)^{-k-2} {1\over 2 \pi i}{1\over 2 \pi i} \oint_{t_1}\oint_{t_1,|t-t_1|>|t'-t_1|}(t+a)(t+b)\cr
&& (t-t_1)^{k-1} (t'+a)(t'+b) (t'-t_1)^{-1} G^-_{\dot{B}}(t) G^-_{\dot{A}}(t')\diff t \diff t'
\label{GG one}
\eea
Here we make several comments about the nontrivial terms from our $k$ expansions.
Since the $\dot{A}$ contour is on the inside the only term which survives the $(t'-t_2)^{-1}$ expansion is the $k=0$ term which, as shown when computing $\mathcal{I}_1$, is $\tilde{G}^{-,t\to t_1}_{\dot{A},-{3\over 2}}$. For the $\dot{B}$ contour however we will have the $k=0$ term plus    additional terms corresponding to $k=1,2$ in the expansion of $(t-t_2)^{-1}$. These additional terms anticommute nontrivially with the $\dot{A}$ contour to produce $\tilde{J}^-$ terms. This can be seen when looking at the anticommutation relation:
\bea
\nn
\ac{\tilde{G}^{\alpha,t\to t_1}_{\dot{A},m}}{\tilde{G}^{\beta,t\to t_1}_{\dot{B},n}} &=& \hspace*{-4pt}\epsilon_{\dot{A}\dot{B}}\bigg[
   (m^2 - \frac{1}{4})\epsilon^{\alpha\beta}\delta_{m+n,0}
  + (m-n){(\sigma^{aT})^\alpha}_\gamma\epsilon^{\gamma\beta}\tilde{J}^{a,t\to t_1}_{m+n}
  + \epsilon^{\alpha\beta} \tilde{L}^{t\to t_1}_{m+n}\bigg]\quad\nn
  \label{anticommutator}
\eea 
where $\tilde{G}^{\a,t\to t_1}_{\dot{A},m}$ is defined in (\ref{t plane G mode}).
The $t$ plane $J^{-}$ terms are defined locally at $t_1$ as:
\bea
\tilde{J}^{-,t\to t_1}_{n} = {1\over 2 \pi i}\oint_{t_1}\diff t (t-t_1)^n J^-(t)
\label{t plane J}
\eea
However, instead of actually anticommuting the $\dot{B}$ mode for $k=1,2$ to produce $\tilde{J}^-$ terms, we leave our expression in terms of $\tilde{G}^-$'s. The reason is as follows: when we did anticommute the $\dot{B}$ modes through the $\dot{A}$ mode we obtained terms that were proportional to $\tilde{J}^{-,t\to t_1}_{-1}$ and $\tilde{J}^{-,t\to t_1}_{-2}$ with $J^{-,t\to t_1}_n$ defined in (\ref{t plane J}). We found that for the $\tilde{J}^{-,t\to t_1}_{-1}$ term, the cylinder expansion was trivial because the cylinder contour carried no factor of $(t-t_2)$ which appears in $\mathcal{I}_1$ expression. However, for the $\tilde{J}^{-,t\to t_1}_{-2}$ term, the cylinder contour carried a factor $(t-t_2)^{-1}$ which becomes extremely difficult to expand in terms of the coordinate $z=e^w$ which is necessary in order to write our two deformation operators completely in terms of cylinder modes. To avoid this problem altogether we simply leave our integral expression, $\mathcal{I}_2$, in terms of $G^-$ contours. 

We note here that we have two cases to consider for $\mathcal{I}_2$ which we examine below. The first case is when $\dot{A}=\dot{B}$ and the second case is for general $\dot{A}$ and $\dot{B}$. We note that we compute the special case of $\dot{A}=\dot{B}$ in the $t$ plane as opposed to back on the cylinder.
\subsection{\underline{$\dot{A}=\dot{B}$}}\label{A = B one}
We first begin with the case of  $\dot{A}=\dot{B}$. With this constraint the expression $\mathcal{I}_2$ can be written as:
\bea
\mathcal{I}_{2} &\equiv& CC'(z_1z_2)^{1/2}\sum_{k=0}^{2}(-1)^k(t_1-t_2)^{-k-2} {1\over 2 \pi i}{1\over 2 \pi i} \oint_{t_1}\oint_{t_1,|t-t_1|>|t'-t_1|}(t+a)(t+b)\cr
&& (t-t_1)^{k-1} (t'+a)(t'+b) (t'-t_1)^{-1} G^-_{\dot{A}}(t) G^-_{\dot{A}}(t')\diff t \diff t'
\label{GG two}
\eea
Let us insert the $t$ plane mode expansions given in (\ref{t plane G mode}) into (\ref{GG two}).
Therefore (\ref{GG two}) becomes
\bea
\mathcal{I}_{2} &\to&  CC'(z_1z_2)^{1/2}(t_1-t_2)^{-2}(t_1+a)^2(t_1+b)^2 \tilde{G}^{-,t\to t_1}_{\dot{A},-3/2} \tilde{G}^{-,t\to t_1}_{\dot{A},-3/2}
\cr
\cr
&&\quad - CC'(z_1z_2)^{1/2}(t_1-t_2)^{-3} (t_1 + a)^2(t_1+b)^2\tilde{G}^{-,t\to t_1}_{\dot{A},-1/2} \tilde{G}^{-,t\to t_1}_{\dot{A},-3/2}
\cr
\cr
&&\quad + CC'(z_1z_2)^{1/2}(t_1-t_2)^{-4} (t_1+a)^2(t_1+b)^2\tilde{G}^{-,t\to t_1}_{\dot{A},1/2} \tilde{G}^{-,t\to t_1}_{\dot{A},-3/2}
\eea
Since we have a local $NS$ vacuum at $t_1$ we compute the action of each term on the vacuum. Looking at the first term we find:
\bea
\tilde{G}^{t\to t_1,-}_{\dot{A},-3/2}\tilde{G}^{t\to t_1,-}_{\dot{A},-3/2}|0_{NS}\rangle_{t_1}=0
\eea
since two fermions aren't allowed to be in the same state. Looking at the second and third terms we find:
\bea
&& \tilde{G}^{-,t\to t_1}_{\dot{A},-1/2}\tilde{G}^{-,t\to t_1}_{\dot{A},-3/2}|0_{NS}\rangle_{t_1}=-\tilde{G}^{-,t\to t_1}_{\dot{A},-3/2}\tilde{G}^{-,t\to t_1}_{\dot{A},-1/2}|0_{NS}\rangle_{t_1}=0\cr
&&  \tilde{G}^{-,t\to t_1}_{\dot{A},1/2}\tilde{G}^{-,t\to t_1}_{\dot{A},-3/2}|0_{NS}\rangle_{t_1}=-\tilde{G}^{-,t\to t_1}_{\dot{A},-3/2}\tilde{G}^{-,t\to t_1}_{\dot{A},1/2}|0_{NS}\rangle_{t_1}=0
\eea
where we have used anticommutation relation (\ref{anticommutator}). We see that these terms also vanish and therefore conclude that for $\dot{A} = \dot{B}$ the integral expression $\mathcal{I}_2$ vanishes. This gives a drastic simplification to the full expression of the two deformation operators given in (\ref{full deformation one}).

In the next section we compute the integral expression for $\mathcal{I}_2$ for general $\dot{A}$ and $\dot{B}$.
\subsection{\underline{General $\dot{A}$ and $\dot{B}$}}
We now compute integral expression for $\mathcal{I}_2$ for general $\dot{A}$ and $\dot{B}$. Again writing our expression we have:
\bea
 \mathcal{I}_{2}&\to&CC'(z_1z_2)^{1/2}\sum_{k=0}^{2}{}^{-1}C_{k}(t_1-t_2)^{-k-2} {1\over 2 \pi i}{1\over 2 \pi i} \oint_{t_1}\oint_{t_1,|t-t_1|>|t'-t_1|}(t+a)(t+b)\cr
&& (t-t_1)^{k-1} (t'+a)(t'+b) (t'-t_1)^{-1} G^-_{\dot{B}}(t) G^-_{\dot{A}}(t')\diff t \diff t'
\label{spectral flow expanded GG}
\eea
Taking:
\bea
{}^{-1}C_{k}=(-1)^k
\eea 
gives:
\bea
 \mathcal{I}_{2} &\equiv&  CC'(z_1z_2)^{1/2}\sum_{k=0}^{2}(-1)^k(t_1-t_2)^{-k-2} {1\over 2 \pi i}{1\over 2 \pi i} \oint_{t_1}\oint_{t_1,|t-t_1|>|t'-t_1|}(t+a)(t+b)\cr
&& (t-t_1)^{k-1} (t'+a)(t'+b) (t'-t_1)^{-1} G^-_{\dot{B}}(t) G^-_{\dot{A}}(t')\diff t \diff t'
\eea

Let us now expand both the $t$ and $t'$ contours away from $t_1$ to points corresponding to initial and final points on the cylinder. Since each contour can land on four different points stretching them will produce a total of sixteen terms. However, we must take care to keep up with minus signs as contours will be reversing direction at finite points. Let us tabulate the minus sign changes. We note that stretching the $\dot{B}$ contour to finite points will reverse direction as well as place it on the inside of the $\dot{A}$ contour that also lands on that point however we always leave the $\dot{B}$ supercharge outside of the $\dot{A}$ supercharge. When the $\dot{B}$ and $\dot{A}$ contour both land at $t=\infty$ we get no sign change because the contours are in the right direction and the and $\dot{A}$ contour is still inside of the $\dot{B}$ contour. Proceeding forward, we have
\bea
&&\dot{B}~~\text{at}~~\infty,\quad\dot{A}~~\text{at} ~~\infty   \quad \to \quad (+1)(+1)=+1\cr
&&\dot{B}~~\text{at}~~\infty,\quad\dot{A}~~\text{at} ~~0 \quad \to \quad (+1)(-1)=-1\cr
&&\dot{B}~~\text{at}~~\infty,\quad\dot{A}~~\text{at} ~~a   \quad \to \quad (+1)(-1)=-1\cr
&&\dot{B}~~\text{at}~~\infty,\quad\dot{A}~~\text{at} ~~b   \quad \to \quad (+1)(-1)=-1\cr
\cr
\cr
&&\dot{B}~~\text{at}~~ 0,\quad\dot{A}~~\text{at} ~~\infty   \quad \to \quad (-1)(+1)=-1\cr
&&\dot{B}~~\text{at}~~ 0,\quad\dot{A}~~\text{at} ~~0   \quad \to \quad (-1)(-1)=+1\cr
&&\dot{B}~~\text{at}~~ 0,\quad\dot{A}~~\text{at} ~~-a   \quad \to \quad (-1)(-1)=+1\cr
&&\dot{B}~~\text{at}~~ 0,\quad\dot{A}~~\text{at} ~~-b   \quad \to \quad  (-1)(-1)=+1\cr
\cr
&&\dot{B}~~\text{at}~~ -a,\quad\dot{A}~~\text{at} ~~\infty   \quad \to \quad  (-1)(+1)=-1\cr
&&\dot{B}~~\text{at}~~ -a,\quad\dot{A}~~\text{at} ~~0   \quad \to \quad  (-1)(-1)=+1\cr
&&\dot{B}~~\text{at}~~ -a,\quad\dot{A}~~\text{at} ~~-a   \quad \to \quad  (-1)(-1)=+1\cr
&&\dot{B}~~\text{at}~~ -a,\quad\dot{A}~~\text{at} ~~-b   \quad \to \quad  (-1)(-1)=+1\cr
\cr
\cr
&&\dot{B}~~\text{at}~~ -b,\quad\dot{A}~~\text{at} ~~\infty   \quad \to \quad  (-1)(+1)=-1\cr
&&\dot{B}~~\text{at}~~ -b,\quad\dot{A}~~\text{at} ~~0   \quad \to \quad  (-1)(-1)=+1\cr
&&\dot{B}~~\text{at}~~ -b,\quad\dot{A}~~\text{at} ~~-a   \quad \to \quad  (-1)(-1)=+1\cr
&&\dot{B}~~\text{at}~~ -b,\quad\dot{A}~~\text{at} ~~-b   \quad \to \quad  (-1)(-1)=+1
\eea
Implementing these sign changes as we stretch our contours in (\ref{spectral flow expanded GG}) gives the following:
\bea
 \mathcal{I}_{2}&\to&CC'(z_1z_2)^{1/2}\sum_{k=0}^{2}(-1)^k(t_1-t_2)^{-k-2}\cr
&& {1\over 2 \pi i}{1\over 2 \pi i} \bigg(\oint_{t=\infty}\oint_{t'=\infty, |t|>|t'|} - \oint_{t=\infty}\oint_{t'=0} - \oint_{t=\infty}\oint_{t'=-a} - \oint_{t=\infty}\oint_{t'=-b}\cr
&& - \oint_{t=0}\oint_{t'=\infty} + \oint_{t=0}\oint_{t'=0,|t'|>|t|}  + \oint_{t=0}\oint_{t'=-a} + \oint_{t=0}\oint_{t'=-b} \cr
&& - \oint_{t=-a}\oint_{t'=\infty} + \oint_{t=-a}\oint_{t'=0}  + \oint_{t'=-a} \oint_{t=-a,|t'+a|>|t+a|}   +  \oint_{t=-a}\oint_{t'=-b}\cr
&&  - \oint_{t=-b}\oint_{t'=\infty} + \oint_{t=-b}\oint_{t'=0} + \oint_{t=-b}\oint_{t'=-a}  + \oint_{t'=-b}\oint_{t=-b,|t'+b|>|t+b|}  \bigg )\cr
\cr
&&\qquad (t+a)(t+b) (t-t_1)^{k-1}(t'+a)(t'+b) (t'-t_1)^{-1} G^-_{\dot{B}}(t) G^-_{\dot{A}}(t')\diff t \diff t'
\nn
\label{expanded GG} 
\eea
Let us rearrange (\ref{expanded GG}) into three terms grouped according to the contour locations specified below:
\begin{enumerate}
\item  $t=0,\infty; ~~\quad t'=0,\infty$\\
\item  $t=0,\infty;~~\quad t'=-a,-b~~~~$ and $~~\quad t=-a,-b;\quad t'=0,\infty$\\
\item  $t=-a,-b; ~~\quad t'=-a,-b$ 
\end{enumerate}
Contours circling points  in set 1 are placed on the cylinder  after the two twists. For set 2, each term has one contour before the twists and one after the twists. For set 3, the contours are both placed before the twists. 
 
Therefore (\ref{expanded GG}) becomes:
\bea
 \mathcal{I}_{2}&\to&CC'(z_1z_2)^{1/2}\sum_{k=0}^{2}(-1)^k(t_1-t_2)^{-k-2}\cr
&& {1\over 2 \pi i}{1\over 2 \pi i} \bigg(\oint_{t=\infty}\oint_{t'=\infty, |t|>|t'|} - \oint_{t=\infty}\oint_{t'=0}  - \oint_{t=0}\oint_{t'=\infty} + \oint_{t=0}\oint_{t'=0,|t'|>|t|}\bigg)
\cr
\cr
&&\qquad\qquad\times~(t+a)(t+b) (t-t_1)^{k-1}(t'+a)(t'+b) (t'-t_1)^{-1} G^-_{\dot{B}}(t) G^-_{\dot{A}}(t')\diff t \diff t'
\cr
\cr
&&~ +  ~CC'(z_1z_2)^{1/2}\sum_{k=0}^{2}(-1)^k(t_1-t_2)^{-k-2}\cr
&& \quad\times~{1\over 2 \pi i}{1\over 2 \pi i}\bigg( - \oint_{t=\infty}\oint_{t'=-a} - \oint_{t=\infty}\oint_{t'=-b} + \oint_{t=0}\oint_{t'=-a} + \oint_{t=0}\oint_{t'=-b}\cr
&&\qquad\qquad  - \oint_{t=-a}\oint_{t'=\infty} + \oint_{t=-a}\oint_{t'=0} - \oint_{t=-b}\oint_{t'=\infty} + \oint_{t=-b}\oint_{t'=0}\bigg)
\cr
\cr
&&\qquad\qquad\quad\times~(t+a)(t+b) (t-t_1)^{k-1}(t'+a)(t'+b) (t'-t_1)^{-1} G^-_{\dot{B}}(t) G^-_{\dot{A}}(t')\diff t \diff t'
\cr
\cr
&& ~ + ~ CC'(z_1z_2)^{1/2}\sum_{k=0}^{2}(-1)^k(t_1-t_2)^{-k-2}\cr
&&\quad\times~ {1\over 2 \pi i}{1\over 2 \pi i}\bigg(\oint_{t'=-a} \oint_{t=-a,|t'+a|>|t+a|}   + \oint_{t=-a}\oint_{t'=-b} \cr
&&\qquad\qquad\qquad ~ +  \oint_{t=-b}\oint_{t'=-a} + \oint_{t'=-b}\oint_{t=-b,|t'+b|>|t+b|} \bigg ) \cr
&&\qquad\qquad\qquad\qquad\times(t+a)(t+b) (t-t_1)^{k-1}\cr
&&\qquad\qquad\qquad\qquad\times~(t'+a)(t'+b) (t'-t_1)^{-1} G^-_{\dot{B}}(t) G^-_{\dot{A}}(t')\diff t \diff t' 
\label{expanded GG rearranged prime}
\eea

Let us now reverse spectral flow in exactly the opposite order in order to remove the constants $C$ and $C'$ and then map back to the the cylinder. Even though we don't explicitly write the intermediate step of mapping from the $z$ plane to the $t$ plane we note that there is sign change that occurs for any integral around $t$ or $t'=0$. This is again because in mapping from the $t$ plane to the $z$ plane the contours around $t=0$ reverse the direction of the contour bringing in a minus sign just as they did when mapping from the $z$ plane to the $t$ plane. In the case where both integrals are around $t=t'=0$, we will gain two minus signs for that term giving no overall sign change. We note that the $\dot{B}$ contour maps from inside of the $\dot{A}$ contour in the $t$ plane, to outside the $\dot{A}$ contour in the $z$ plane. Then the same ordering is kept when going to the cylinder. Using the transformation in (\ref{t to w}) and (\ref{spin field t to w}) for both $G$ contours gives the following cylinder expression for $\mathcal{I}_2$:
\bea
 \mathcal{I}_{2}&\to&\sum_{k=0}^{2}(-1)^k(t_1-t_2)^{-k-2}\cr
&&\times~\bigg[{1\over 2 \pi i}\oint_{\s=0,\t>\t_2}^{\s=2\pi} \diff w~ (t-t_1)^k(t-t_2) \bigg( G^{(1)-}_{\dot{B}}(w) +  G^{(2)-}_{\dot{B}}(w)\bigg)\cr
&&\qquad\times{1\over 2 \pi i}\oint_{\s'=0,\t'>\t_2,\t>\t'}^{\s'=2\pi} \diff w'~ (t'-t_2) \bigg( G^{(1)-}_{\dot{A}}(w') +  G^{(2)-}_{\dot{A}}(w')\bigg)\bigg]\s^+_2(w_2)\s^+_2(w_1)
\cr
\cr
\cr
&&+\sum_{k=0}^{2}(-1)^k(t_1-t_2)^{-k-2}\cr
&&\quad\times~\bigg[ - {1\over 2 \pi i}\oint_{\s=0,\t>\t_2}^{\s=2\pi}\diff w~ (t-t_1)^k(t-t_2)\bigg( G^{(1)-}_{\dot{B}}(w) + G^{(2)-}_{\dot{B}}(w)\bigg)\cr
&&\qquad\times\s^+_2(w_2)\s^+_2(w_1)\cr
&&\quad\quad\quad\quad \times{1\over 2 \pi i}\oint_{\s'=0,\t'<\t_1}^{\s'=2\pi} \diff w'~ (t'-t_2)\bigg( G^{(1)-}_{\dot{A}}(w') + G^{(2)-}_{\dot{A}}(w')\bigg)
\cr\cr
&&\quad\quad\quad{}+ {1\over 2 \pi i}\oint_{\s'=0,\t'>\t_2}^{\s'=2\pi}\diff w'~ (t'-t_2)\bigg( G^{(1)-}_{\dot{A}}(w') + G^{(2)-}_{\dot{A}}(w')\bigg)\cr
&&\quad\quad\quad\quad\times\s^+_2(w_2)\s^+_2(w_1)\cr
&&\quad\quad\quad\quad\times ~ {1\over 2 \pi i}\oint_{\s=0,\t<\t_1}^{\s=2\pi} \diff w~ (t-t_1)^k(t-t_2)\bigg( G^{(1)-}_{\dot{B}}(w) + G^{(2)-}_{\dot{B}}(w)\bigg)\bigg]\cr
\cr
\cr
&&- \sum_{k=0}^{2}(-1)^k(t_1-t_2)^{-k-2} \s^+_2(w_2)\s^+_2(w_1)\cr
&&\quad\times ~\bigg[{1\over 2 \pi i}\oint_{\s'=0,\t'<\t_1}^{\s'=2\pi} \diff w'~ (t'-t_2) \bigg( G^{(1)-}_{\dot{A}}(w') +  G^{(2)-}_{\dot{A}}(w')\bigg)\cr
&&\quad\quad\times~{1\over 2 \pi i}\oint_{\s=0,\t<\t_1,\t'>\t}^{\s=2\pi} \diff w~ (t-t_1)^k(t-t_2) \bigg( G^{(1)-}_{\dot{B}}(w) +  G^{(2)-}_{\dot{B}}(w)\bigg)\bigg]\nn
\label{expanded GG rsf z to w}
\eea

We have now computed the integral expression $\mathcal{I}_2$ in terms of contours before and after the twists on the cylinder. Our next goal is to write the full expression of the deformation operator on the cylinder before expanding our coordinate $t$ in terms of the cylinder coordinate $w$.

We again remind the reader that the   general case reduces to the case where $\dot{A}=\dot{B}$ that we also computed. However, it is much easier to evaluate this while still in the $t$ plane as we have done as opposed to waiting until we have reached the cylinder.

\section{Full expression before expanding $t$ in terms of cylinder coordinate $w$}\label{cylinder result}

Here we compute the full deformation operator expressions on the cylinder prior to expanding our coordinate $t$ in terms of $z=e^w$. We compute the two cases below. 

\subsection{\underline{$\dot{A}=\dot{B}$}}

Looking at the case when $\dot{A}=\dot{B}$, in (\ref{A = B one}) we found that $\mathcal{I}_2=0$ which gives a great simplification. Taking $\mathcal{I}_2=0$ in (\ref{full deformation one}) gives the following expression for the two deformation operator:
\bea
\hat{O}_{\dot{A}}\hat{O}_{\dot{A}}&\equiv& \big( G^{(1),-}_{\dot{B},0} + G^{(2),-}_{\dot{B},0}\big)\mathcal{I}_1 + \mathcal{I}_{1}\big( G^{(1),-}_{\dot{B},0} + G^{(2),-}_{\dot{B},0}\big)
\label{A = B cylinder expression}
\eea
Now inserting our expression for $\mathcal{I}_1$ which is given in (\ref{I four}) into (\ref{A = B cylinder expression}) gives us the following:
\bea
\hat{O}_{\dot{A}}\hat{O}_{\dot{A}}\!\!{}&\equiv&{}\!\!- {1\over 2t_2}\big( G^{(1),-}_{\dot{A},0} + G^{(2),-}_{\dot{A},0}\big)\cr
&&\quad\bigg[\bigg( {1\over 2 \pi i}\int_{\s = 0,\t>\t_2}^{\s = 2\pi}\diff w (t-t_2)G^{(1)-}_{\dot{A}}(w)\cr
&&\qquad\qquad + ~ {1\over 2 \pi i}\int_{\s = 0,\t>\t_2}^{\s = 2\pi}\diff w (t-t_2)G^{(2)-}_{\dot{A}}(w) \bigg) \s_2^+(w_2)\s_2^+(w_1)
\cr
\cr
&&\quad~-\s_2^+(w_2)\s_2^+(w_1) ~ \bigg( {1\over 2 \pi i}\int_{\s = 0,\t<\t_1}^{\s = 2\pi}\diff w (t-t_2)G^{(1)-}_{\dot{A}}(w)\cr
&&\qquad\qquad\qquad\qquad\qquad\qquad + {1\over 2 \pi i}\int_{\s = 0,\t<\t_1}^{\s = 2\pi}\diff w (t-t_2)G^{(2)-}_{\dot{A}}(w) \bigg) \bigg]
\cr
\cr
\cr
&&  -{1\over 2t_2}\bigg[\bigg({1\over 2 \pi i}\int_{\s = 0,\t>\t_2}^{\s = 2\pi}\diff w (t-t_2) G^{(1)-}_{\dot{A}}(w)\cr
&&\qquad\qquad+ ~{1\over 2 \pi i}\int_{\s = 0,\t>\t_2}^{\s = 2\pi}\diff w (t-t_2)G^{(2)-}_{\dot{A}}(w) \bigg) ~ \s_2^+(w_2)\s_2^+(w_1)\cr
&&\qquad~~ -~\s_2^+(w_2)\s_2^+(w_1)\bigg( {1\over 2 \pi i}\int_{\s = 0,\t<\t_1}^{\s = 2\pi}\diff w (t-t_2)G^{(1)-}_{\dot{A}}(w)\cr
&&\qquad\qquad\qquad\qquad\qquad\qquad {}+ {1\over 2 \pi i}\int_{\s = 0,\t<\t_1}^{\s = 2\pi}\diff w (t-t_2)G^{(2)-}_{\dot{A}}(w) \bigg) \bigg]\cr
&&\quad\times\big( G^{(1),-}_{\dot{B},0} + G^{(2),-}_{\dot{B},0}\big)
\label{A = B two}
\eea
This is our cylinder expression for the two deformation operators for the case where $\dot{A}=\dot{B}$ before computing the $G^-$ contour integrals in terms of cylinder modes. Next we record the cylinder expression for our two deformation operator for general $\dot{A}$ and $\dot{B}$.
\subsection{\underline{General $\dot{A}$ and $\dot{B}$}}
Now looking at the case for general $\dot{A}$ and $\dot{B}$ we insert (\ref{expanded GG rsf z to w}) and (\ref{I four}) into (\ref{full deformation one}) giving the expression:
\bea
\hat{O}_{\dot{B}}\hat{O}_{\dot{A}}\!\! &=&\!\!  -{1\over  2t_2}\big( G^{(1),-}_{\dot{B},0} + G^{(2),-}_{\dot{B},0}\big)\cr
&&\times~\bigg[ \bigg({1\over 2 \pi i}\int_{\s = 0,\t>\t_2}^{\s = 2\pi}\diff w (t-t_2) G^{(1)-}_{\dot{A}}(w)\cr
&&\quad\quad{} + {1\over 2 \pi i}\int_{\s = 0,\t>\t_2}^{\s = 2\pi}\diff w (t-t_2) G^{(2)-}_{\dot{A}}(w) \bigg)\s_2^+(w_2)\s_2^+(w_1)
\cr
\cr
\cr
&&~ -\s_2^+(w_2)\s_2^+(w_1)\bigg({1\over 2 \pi i}\int_{\s = 0,\t<\t_1}^{\s = 2\pi}\diff w (t-t_2) G^{(1)-}_{\dot{A}}(w)\cr
&&\qquad\qquad\qquad\qquad\qquad{} + {1\over 2 \pi i}\int_{\s = 0,\t<\t_1}^{\s = 2\pi}\diff w (t-t_2)G^{(2)-}_{\dot{A}}(w) \bigg) \bigg]
\cr
\cr
\cr
&&\!\!\!\!\!\! -\sum_{k=0}^{2}(-1)^k(t_1-t_2)^{-k-2}\cr
&&\times~\bigg[\bigg({1\over 2 \pi i}\int_{\s=0,\t>\t_2}^{\s=2\pi} \diff w~ (t-t_1)^k(t-t_2) G^{(1)-}_{\dot{B}}(w)
\cr
&&\quad\quad +  {1\over 2 \pi i}\int_{\s=0,\t>\t_2}^{\s=2\pi} \diff w~(t-t_1)^k (t-t_2)G^{(2)-}_{\dot{B}}(w)\bigg)
\cr
&&\quad\times\bigg({1\over 2 \pi i}\int_{\s'=0,\t'>\t_2,\t>\t'}^{\s'=2\pi} \diff w'~ (t'-t_2) G^{(1)-}_{\dot{A}}(w')\cr
&&\quad\quad{}+  {1\over 2 \pi i}\int_{\s'=0,\t'>\t_2,\t>\t'}^{\s'=2\pi} \diff w'~ (t'-t_2)G^{(2)-}_{\dot{A}}(w')\bigg)\bigg]\s^+_2(w_2)\s^+_2(w_1)
\cr
\cr
\cr
&&\!\!\!\!\!\!+ \sum_{k=0}^{2}(-1)^k(t_1-t_2)^{-k-2}\cr
&&\times~\bigg[  \bigg({1\over 2 \pi i}\oint_{\s=0,\t>\t_2}^{\s=2\pi}\diff w~ (t-t_1)^k(t-t_2) G^{(1)-}_{\dot{B}}(w)
\cr
&&\quad\quad {}+  {1\over 2 \pi i}\oint_{\s=0,\t>\t_2}^{\s=2\pi}\diff w~ (t-t_1)^k(t-t_2)G^{(2)-}_{\dot{B}}(w)\bigg)\s^+_2(w_2)\s^+_2(w_1)
\cr
\cr
&&\quad\times\bigg( {1\over 2 \pi i}\int_{\s'=0,\t'<\t_1}^{\s'=2\pi} \diff w'~ (t'-t_2) G^{(1)-}_{\dot{A}}(w')\cr
&&\quad\quad{}+  {1\over 2 \pi i}\int_{\s'=0,\t'<\t_1}^{\s'=2\pi} \diff w'~ (t'-t_2)G^{(2)-}_{\dot{A}}(w')\bigg)
\cr
\cr
\cr
&&\quad - \bigg({1\over 2 \pi i}\int_{\s'=0,\t'>\t_2}^{\s'=2\pi}\diff w'~ (t'-t_2) G^{(1)-}_{\dot{A}}(w')\cr
&&\quad\quad{} + {1\over 2 \pi i}\int_{\s'=0,\t'>\t_2}^{\s'=2\pi}\diff w'~ (t'-t_2) G^{(2)-}_{\dot{A}}(w')\bigg)\s^+_2(w_2)\s^+_2(w_1)
\cr
&&\quad\quad \times\bigg({1\over 2 \pi i}\int_{\s=2,\pi\t<\t_1}^{\s=2\pi} \diff w~ (t-t_1)^k(t-t_2) G^{(1)-}_{\dot{B}}(w)
\cr
&&\quad\quad\quad+ {1\over 2 \pi i}\int_{\s=2,\pi\t<\t_1}^{\s=2\pi} \diff w~ (t-t_1)^k(t-t_2) G^{(2)-}_{\dot{B}}(w)\bigg)\bigg]
\cr
\cr
\cr
&&\!\!\!\!\!\!+\s^+_2(w_2)\s^+_2(w_1) \sum_{k=0}^{2}(-1)^k(t_1-t_2)^{-k-2}\cr
&&\times~\bigg[\bigg({1\over 2 \pi i}\int_{\s'=0,\t'<\t_1}^{\s'=2\pi} \diff w'~ (t'-t_2)  G^{(1)-}_{\dot{A}}(w')\cr
&&\quad\quad{} + {1\over 2 \pi i}\int_{\s'=0,\t'<\t_1}^{\s'=2\pi} \diff w'~ (t'-t_2) G^{(2)-}_{\dot{A}}(w')\bigg)\cr
&&\quad\times\bigg({1\over 2 \pi i}\int_{\s=0,\t<\t_1,\t'>\t}^{\s=2\pi} \diff w~ (t-t_1)^k(t-t_2) G^{(1)-}_{\dot{B}}(w)
\cr
&&\quad\quad {} +  {1\over 2 \pi i}\int_{\s=0,\t<\t_1,\t'>\t}^{\s=2\pi}\diff w(t-t_1)^k(t-t_2)G^{(2)-}_{\dot{B}}(w)\bigg)\bigg]
\cr
\cr
\cr
&&\!\!\!\!\!\!  -{1\over 2t_2}\bigg[\bigg({1\over 2 \pi i}\int_{\s = 0,\t>\t_2}^{\s = 2\pi}\diff w (t-t_2) G^{(1)-}_{\dot{A}}(w)\cr
&&\quad\quad{} + {1\over 2 \pi i}\int_{\s = 0,\t>\t_2}^{\s = 2\pi}\diff w (t-t_2)G^{(2)-}_{\dot{A}}(w) \bigg)\s_2^+(w_2)\s_2^+(w_1)
\cr
\cr
&&\quad~~ -~\s_2^+(w_2)\s_2^+(w_1)\bigg( {1\over 2 \pi i}\int_{\s = 0,\t<\t_1}^{\s = 2\pi}\diff w (t-t_2)G^{(1)-}_{\dot{A}}(w)\cr
&&\qquad\qquad\qquad\qquad\qquad\qquad{} + {1\over 2 \pi i}\int_{\s = 0,\t<\t_1}^{\s = 2\pi}\diff w (t-t_2)G^{(2)-}_{\dot{A}}(w) \bigg) \bigg]\cr
&&\quad\times\big ( G^{(1),-}_{\dot{B},0} + G^{(2),-}_{\dot{B},0} \big )
\label{full deformation two}
\eea

We have now written the expression for the two deformation operators on the cylinder for both the specific case $\dot{A} = \dot{B}$ and the more general case $\dot{A}\neq\dot{B}$. Remarkably, we were able to write the expression in terms of $G$ contours before and after the twists. Our next step is to expand the coordinate $t$ in terms of $z=e^w$ in order to write the full deformation operator in terms of cylinder modes.  

\section{Expanding $t$ in terms of the cylinder coordinate $w$}\label{map inversion}

Now that we have written our two deformation operator as contours on the cylinder our goal is to write these contours in terms of cylinder modes.

\subsection{Map Inversion}

To do this we must first invert our map so that we can expand the integrands in terms of the cylinder coordinate $e^{w}$. We start with the following:
\bea
&&z={(t+a)(t+b)\over t}\cr
&&\to zt= t^2+(a+b)t +ab\cr
&& \to t^2+t(a+b-z) +ab = 0
\eea
Inserting this result into \textit{Mathematica} and using $t_2=\sqrt{ab}$ gives the solutions
\bea
t_{\pm}&=&{1\over 2}\bigg(z-(a+b)\pm \sqrt{-(2t_2)^2 + (a+b-z)^2}\bigg)\cr
&=&{1\over 2}\bigg(z-(a+b)\pm \sqrt{(z - (a+b+2\sqrt{ab}))( z - (a+b-2\sqrt{ab}))}\bigg)\nn
\label{t solution}
\eea
Let us now look at the limiting behavior in order to determine which solutions will correspond to which initial and final states.

\subsubsection*{ Final Copies $\t\to\infty\Rightarrow z\to \infty$}

Taking the limit as $z\to\infty$ we have:
\bea
t_{\pm}={1\over 2}(z \pm \sqrt{z^2})
\eea
Taking the positive branch of $ \sqrt{z^2}$:
\bea
t_{\pm} = {1\over 2}(z \pm z)
\eea
We there see that the $t_+$ solution corresponds to Copy $1$ final because 
\bea
t\sim z
\eea
and similarly $t_-$ corresponds to Copy $2$ final because
\bea
t\sim 0
\eea
We also perform this same analysis for the initial copies:
\subsubsection*{ Initial Copies $\t \to -\infty \Rightarrow z\to 0$}
Taking the limit $z\to 0$
\bea
t_{\pm}&=&{1\over 2}\bigg(-(a+b)\pm \sqrt{(b-a)^2}\bigg)
\eea
Staying consistent with the final copy notation we want the $+$ solution to correspond to Copy 1 initial. We therefore take the positive branch of $\sqrt{(b-a)^2}$ we get:
\bea
t_{\pm}&=&{1\over 2}\bigg(-(a+b)\pm (b-a)\bigg)
\eea
The $t_+$ therefore corresponds to Copy $1$ because:
\bea
t\sim-a
\eea
and the $t_-$ solution corresponds to Copy $2$ because
\bea
t\sim -b
\eea
Now that we have picked the appropriate solutions for the appropriate initial and final states we proceed forward with evaluating $t_{\pm}$. Substituting
\bea
(a+b+2\sqrt{ab})&=&z_2=e^{{\Delta w\over 2}}\nn
(a+b-2\sqrt{ab})&=&z_1=e^{-{\Delta w\over 2}}\nn
a+b &=& -2t_2 + e^{{\Delta w\over 2}}
\eea
into (\ref{t solution}) gives
\bea
t_{\pm}&=&{1\over 2}\bigg(z-(a+b)\pm \sqrt{(z - e^{{\Delta w \over 2}})(z - e^{-{\Delta w \over 2}})}\bigg)
\label{t pm}
\eea
%
Since we will have to expand $z$ around different points corresponding to initial and final states, let us do each one in turn in the following subsections. 

\subsection{Copy 1 and Copy 2 Final}

Here we expand our $t$ coordinate around $\t = \infty\to z =\infty$ in order to write our $G$ contours in terms of Copy 1 and Copy 2 final modes on the cylinder. Looking at the square root terms in (\ref{t pm}), we have a general expansion of the form 
\bea
\big(z+x\big)^{1/2} &=& z^{1/2}\big(1+xz^{-1}\big)^{1/2}\cr
&=&  z^{1/2}\big(1+xz^{-1}\big)^{1/2}\cr
&=& \sum_{p_1\geq 0}{}^{1/2}C_{p_1} z^{-p_1 + 1/2}x^{p_1}
\eea
Applying the above expansion to (\ref{t pm}) we obtain
\bea
t_{\pm} = {1\over 2} ( z - (a+b)) \pm   {1\over 2}\sum_{p_1\geq 0}\sum_{p_2\geq 0}{}^{1/2}C_{p_1}{}^{1/2}C_{p_2} z^{-p_1-p_2 + 1}(-1)^{p_1+p_2}e^{{\Delta w \over 2}(p_1-p_2)}
\label{solution}
\eea
which are the solutions, $t_{\pm}$, at final points on the cylinder. Let us make the following variable redefinitions:
\bea
k&=&{p_1 - p_2\over 2}\cr
n &=& p_1 + p_2 \cr
\Rightarrow p_1&=& {n\over 2}+k\cr
\Rightarrow p_2&=&{n\over 2} - k
\label{redefinition}
\eea
Evaluating the limits give:
\bea
&& p_1\geq 0 ~~\Rightarrow~~ k+ {n\over 2}\geq 0 ~~\Rightarrow k~~ \geq -  {n\over 2}
\cr
&& p_2\geq 0 ~~\Rightarrow~~  {n\over 2}-k\geq 0 ~~\Rightarrow {n\over 2} \geq k
\label{limits}
\eea
our solutions, (\ref{solution}), become:
\bea
t_{\pm} = {1\over 2} ( z - (a+b)) \pm   {1\over 2}\sum_{n = 0}^{\infty} (-1)^{n} z^{-n + 1}\sum_{k = - {n\over 2}}^{{n \over 2}}{}^{1/2}C_{{n\over 2} + k}{}^{1/2}C_{{n\over 2} - k}e^{k\Delta w}
\label{solution two}
\eea
Looking at our $k$ sum we see for each $k = j$ term  with $j\in \mathbb{Z}_+$ there will be a corresponding $k =-j $ which allows us to make the following replacement:
\bea
e^{k\Delta w}\to \cosh\big( k\Delta w \big)
\label{cosine argument}
\eea
Applying this replacement to (\ref{solution two}) gives:
\bea
t_{\pm} &=& {1\over 2} ( z - (a+b) ) \pm   {1\over 2}\sum_{n = 0}^{\infty} (-1)^{n} z^{-n + 1}\sum_{k = - {n\over 2}}^{{n \over 2}}{}^{1/2}C_{{n\over 2} + k}{}^{1/2}C_{{n\over 2} - k}\cosh\big(k \Delta w\big)\cr
&=& {1\over 2} \big( z - (a+b) \pm  \sum_{n = 0}^{\infty}C_n  z^{1-n}\big)
\label{t plus z}
\eea
where we've made the following definition:
\bea
C_n&\equiv&  (-1)^{n}\sum_{k = - {n\over 2}}^{{n \over 2}}{}^{1/2}C_{{n\over 2} + k}{}^{1/2}C_{{n\over 2} - k}\cosh\big( k \Delta w\big)
\label{C}
\eea
Making the replacement $z=e^w$ in (\ref{t plus z}) gives
\bea
t_{\pm}&=&  {1\over 2} \big( e^w - (a+b) \pm  \sum_{n = 0}^{\infty}C_n  e^{w(1-n)}\big)
\label{t plus}
\eea

Here we expanded our $t$ coordinate in terms of the coordinate $z=e^w$ for Copy 1 and Copy 2 final modes. Next we will expand the $t$ coordinate around Copy 1 and Copy 2 initial states.
 
\subsection{Copy 1 and Copy 2 Initial}

Here we expand our $t$ coordinate around $\t = -\infty\to z =0$. Looking back at $t_{\pm}$ in (\ref{t pm}) we must expand the square root portion around around small $z$. We therefore have:
\bea
\big(z+x\big)^{1/2} &=& x^{1/2}\big(1+x^{-1}z \big)^{1/2}\cr
 &=&\sum_{p_1\geq 0}{}^{1/2}C_{p_1}x^{-p_1 +1/2}z^{p_1}
\eea
Applying this to (\ref{t pm}) gives:
\bea
t_{\pm} &=& {1\over 2} ( z -(a+b)) \mp {1\over 2}\sum_{p_1\geq 0}\sum_{p_2\geq 0}{}^{1/2}C_{p_1}{}^{1/2}C_{p_2} z^{p_1+p_2 }(-1)^{p_1+p_2}e^{-{\Delta w \over 2}(p_1-p_2)}\nn
\label{t pm initial}
\eea
Again making the substitutions given in (\ref{redefinition}) and (\ref{limits}), (\ref{t pm initial}) becomes:
\bea
t_{\pm} = {1\over 2} ( z - (a+b )) \mp  {1\over 2}\sum_{n = 0}^{\infty} (-1)^{n} z^{ n }\sum_{k = - {n\over 2}}^{{n \over 2}}{}^{1/2}C_{{n\over 2} + k}{}^{1/2}C_{{n\over 2} - k}e^{-k\Delta w}
\label{t pm initial two}
\eea
Using the same argument given in (\ref{cosine argument}) we write (\ref{t pm initial two}) as:
\bea
t_{\pm} &=& {1\over 2} \big( z -( a+b ) \mp \sum_{n = 0}^{\infty} C_n z^{ n }\big)\nn
\label{t pm one initial z}
\eea
Again inserting $z=e^w$ in (\ref{t pm one initial z}) gives:
\bea
t_{\pm} &=& {1\over 2} \big( e^w -( a+b ) \mp \sum_{n = 0}^{\infty} C_n e^{ nw }\big)\nn
\label{t pm one initial}
\eea
We see that the only difference between the $t_{\pm}$ expansion for initial and final copies comes in the \textbf{third} term. There is a modification of the mode number as well as a minus sign difference. We summarize this difference below:
\bea
&&\text{Final Copies}: \pm \sum_{n = 0}^{\infty} C_n e^{ (1-n)w }\cr
&&\text{Initial Copies} :\mp \sum_{n = 0}^{\infty} C_n e^{ nw }
\label{third term modification}
\eea

We have now written our $t$ coordinate in terms of expansions around the location of Copy 1 and Copy 2 initial states. 

In the final section we compute the integral expressions on the cylinder by inserting the expansions that we have computed for Copy 1 and Copy 2 final as well as Copy 1 and Copy 2 initial. We then compute the expression for the final result of the two deformation operators on the cylinder.

\section{Computing Supercharge Contours on the Cylinder}\label{cylinder modes} 

Our goal in this section is to compute the expression of the full two deformation operator in terms of cylinder modes using the expansions computed in Section \ref{map inversion}. We first define the supercharge modes on the cylinder. We then compute the necessary contour integrals in terms of cylinder modes at locations corresponding to Copy 1 and Copy 2 final states as well as Copy 1 and Copy 2 initial states that appear in our two deformation operators. 

\subsection{Supercharge modes on the cylinder}

Here we define the supercharge modes on the cylinder. We will have modes before the two twists and modes after the two twists. Our modes after the twist are defined as follows:
\subsubsection*{\underline{$\t>\t_2$}}
\bea
G^{(1),-}_{\dot{C},n}\equiv {1\over 2 \pi i}\int_{\s = 0,\t > \t_2}^{2\pi }  \diff w G^{(1),-}_{\dot{C}}(w) e^{nw}\cr
G^{(2),-}_{\dot{C},n}\equiv {1\over 2 \pi i}\int_{\s = 0,\t > \t_2}^{2\pi }  \diff w G^{(2),-}_{\dot{C}}(w) e^{nw}\cr
\label{cylinder G}
\eea
with $n$ being at integer.
Our modes before the twist are defined as follows:
\subsubsection*{\underline{$\t<\t_1$}}
\bea
G^{(1),-}_{\dot{C},n}\equiv{1\over 2 \pi i}\int_{\s = 0,\t < \t_1}^{2\pi }  \diff w G^{(1),-}_{\dot{C}}(w) e^{nw}\cr
G^{(2),-}_{\dot{C},n}\equiv{1\over 2 \pi i}\int_{\s = 0,\t < \t_1}^{2\pi }  \diff w G^{(2),-}_{\dot{C}}(w) e^{nw}\cr
\label{cylinder G initial}
\eea
Here we have defined our supercharge modes on the cylinder. Next we compute the contour integrals on the cylinder in terms of these cylinder modes.
\subsection{Computing Copy 1 and Copy 2 Final Contours}
Now we compute the necessary contours that appear in (\ref{A = B two}) and (\ref{full deformation two}) in terms of cylinder modes defined in (\ref{cylinder G}). There will be contours of two types in these expressions. We compute the simpler contours first.
\subsubsection{Contours containing $(t-t_2)$} 
The first contours we compute are of the form:
\bea
\text{Copy 1}&:&{1\over 2 \pi i}\int_{\s = 0,\t>\t_2}^{\s = 2\pi}\diff w (t_+-t_2) G^{(1)-}_{\dot{C}}(w)
\cr
\cr
\cr
\text{Copy 2}&:&{1\over 2 \pi i}\int_{\s = 0,\t>\t_2}^{\s = 2\pi}\diff w (t_--t_2) G^{(2)-}_{\dot{C}}(w)
\label{t minus t2 final prime}
\eea
where $t_{\pm}$ correspond to the two roots that were computed in Section \ref{map inversion} with $t_+$ corresponding to Copy 1 and $t_-$ corresponding to Copy 2. Using (\ref{t plus}) we compute the following contribution:
\bea
t_{\pm} - t_2 &=&  {1\over 2} \big( e^w - (a+b) \pm  \sum_{n = 0}^{\infty}C_n  e^{(1-n)w}\big)-t_2\cr
&=&  {1\over 2} \big( e^w - e^{\Delta w \over 2} \pm  \sum_{n = 0}^{\infty}C_n  e^{(1-n)w}\big)
\label{t minus tpm}
\eea
Inserting (\ref{t minus tpm}) into (\ref{t minus t2 final prime}) and using the mode definitions in (\ref{cylinder G}) gives:
\bea
\text{Copy 1}&:&{1\over 2 \pi i}\int_{\s = 0,\t>\t_2}^{\s = 2\pi}\diff w (t_+-t_2) G^{(1)-}_{\dot{C}}(w) \cr
&&\qquad = {1\over 2}\bigg ( G^{(1)-}_{\dot{C},1} - e^{\Delta w\over 2}G^{(1)-}_{\dot{C},0}  +  \sum_{n = 0}^{\infty}  C_n G^{(1)-}_{\dot{C},1-n} \bigg) \cr
\cr
\cr
\text{Copy 2}&:&{1\over 2 \pi i}\int_{\s = 0,\t>\t_2}^{\s = 2\pi}\diff w (t_--t_2) G^{(2)-}_{\dot{C}}(w)\cr
&&\qquad = {1\over 2}\bigg ( G^{(2)-}_{\dot{C},1} - e^{\Delta w\over 2}G^{(2)-}_{\dot{C},0}  -  \sum_{n = 0}^{\infty}  C_n G^{(2)-}_{\dot{C},1-n} \bigg) 
\label{t minus t2 final}
\eea
Let us define the symmetric antisymmtric basis:
\bea
G^{-}_{\dot{C},n} &=&  \big( G^{(1)-}_{\dot{C},n} +  G^{(2)-}_{\dot{C},n}  \big )\cr
\tilde{G}^{-}_{\dot{C},n} &=& \big( G^{(1)-}_{\dot{C},n} - G^{(2)-}_{\dot{C},n}\big)
\label{new basis}
\eea
We do this because computing amplitudes will turn out to be easier when we write all of our modes in terms of the symmetric and antisymmetric basis.

Now combining both results given in (\ref{t minus t2 final}) and using the basis (\ref{new basis}) gives the following expression:
\bea
&&{1\over 2 \pi i}\int_{\s = 0,\t>\t_2}^{\s = 2\pi}\diff w (t_+-t_2) G^{(1)-}_{\dot{C}}(w) + {1\over 2 \pi i}\int_{\s = 0,\t>\t_2}^{\s = 2\pi}\diff w (t_--t_2) G^{(2)-}_{\dot{C}}(w)\cr
&&\qquad =  {1\over 2}\bigg ( G^{-}_{\dot{C},1} - e^{\Delta w\over 2}G^{-}_{\dot{C},0}  + \sum_{n = 0}^{\infty}  C_n \tilde{G}^{-}_{\dot{C},1-n} \bigg)
\label{t minus t2 final 2}
\eea
For this type of term we see a finite number of symmetric annihilation modes \textit{after} the twist and an infinite number of antisymmetric creation modes \textit{after} the twists. Since creation modes annihilate when acting on a state to the left we need not worry about getting an infinite number of terms.
\subsubsection{Contours containing $\sum_{k=0}^{2}(t_1-t_2)^{-k-2}(-1)^k(t-t_2)(t-t_1)^k$} 
Now we compute the more complicated contours of the form:
\bea
\text{Copy 1}&:&{1\over 2 \pi i}\int_{\s = 0,\t>\t_2}^{\s = 2\pi}\diff w \sum_{k=0}^{2}(t_1-t_2)^{-k-2}(-1)^k(t_+-t_2)(t_+-t_1)^k G^{(1),-}_{\dot{C}}(w)
\cr
\cr
\text{Copy 2}&:&{1\over 2 \pi i}\int_{\s = 0,\t>\t_2}^{\s = 2\pi}\diff w \sum_{k=0}^{2}(t_1-t_2)^{-k-2}(-1)^k(t_--t_2)(t_--t_1)^k G^{(2),-}_{\dot{C}}(w)\nn
\label{Copy 1 final}
\eea
Let us simplify the following term that we get coming in the integrand of (\ref{Copy 1 final}):
\bea
&&\sum_{k=0}^{2}(-1)^{k}(t_1-t_2)^{-k-2}(t_{\pm}-t_2)(t_{\pm}-t_1)^k\cr
&&\qquad = {1\over 4 t_2^2}\bigg(  (t_{\pm}-t_2) - {(t_{\pm}-t_2)(t_{\pm}+t_2) \over -2 t_2} +   {(t_{\pm}-t_2)(t_{\pm}+t_2)^2 \over 4t_2^2}  \bigg)\cr
&&\qquad = {1\over 16 t_2^4}\bigg( (t_{\pm}-t_2)(t_{\pm}^2 + 4 t_{\pm} t_2 + 7 t_2^2)  \bigg)\cr
&&\qquad = {1\over 16 t_2^4}\bigg(t_{\pm} (z^2 - z A + B)-zt_2^2 + z_1t_2^2-8t_2^3\bigg)
\label{expansion of integrand}
\eea
where considerable simplification was made in (\ref{expansion of integrand}). We've also made the following definitions:
\bea
A &\equiv&(2z_1 + t_2)\cr
B &\equiv& z_1 (a+b)+ t_2(z_2 - 2(a+b))
\label{A B}
\eea
We note that we used the relation
\bea
z={(t+a)(t+b)\over t}\to t^2=(z-(a+b))t -t_2^2
\eea
to reduce powers of $t$ in (\ref{expansion of integrand}).
First inserting $z=e^w$ into the RHS of (\ref{expansion of integrand}) and then using, (\ref{t plus}) we write (\ref{expansion of integrand}) as: 
\bea
&&\sum_{k=0}^{2}(-1)^{k}(t_1-t_2)^{-k-2}(t_{\pm}-t_2)(t_{\pm}-t_1)^k\cr
&&\qquad =  {1\over 32 t_2^4}\bigg( e^{3w}-e^{2w}\big(A + a+ b\big)+e^w\big(A(a+b)+ B - 2t_2^2 \big)\cr
&&\qquad\qquad \pm\bigg(\sum_{n = 0}^{\infty}C_n  e^{(3-n)w} -   A\sum_{n = 0}^{\infty}C_n  e^{(2-n)w} + B \sum_{n = 0}^{\infty}C_n  e^{(1-n)w}\bigg)\cr
&&\qquad\qquad +\big(2z_1t_2^2 - (a+b)B - 16 t_2^3 \big)\bigg) \nn
\label{expansion of integrand 2}
\eea
Here we have written our expression in terms of the cylinder coordinate $w$. Simplifying (\ref{expansion of integrand 2}) using the definitions of $A$ and $B$ in (\ref{A B}) give:
\bea
&&\sum_{k=0}^{2}(-1)^{k}(t_1-t_2)^{-k-2}(t_{\pm}-t_2)(t_{\pm}-t_1)^k\cr
&&\qquad =  {1\over 32  t_2^4}\bigg[ e^{3w} - A'e^{2w} + B'e^w + C'\cr
&&\qquad\qquad\qquad\qquad \pm \bigg(  \sum_{n = 0}^{\infty}C_n  e^{(3-n)w} -  A\sum_{n = 0}^{\infty}C_n  e^{(2-n)w} +B \sum_{n = 0}^{\infty}C_n  e^{(1-n)w}\bigg)\bigg]
\nn
\label{tpm one}
\eea
where we make the following definitions:
\bea
A'&\equiv&A+a+b\cr
B'&\equiv&A(a+b) + B - 2t_2^2\cr
C'&\equiv& 2z_1t_2^2 -  (a+b)B - 16 t_2^3
\label{A B C prime}
\eea
Let us combine the sums together. To do this let us expand each individual sum in (\ref{tpm one}) appropriately:
\bea
\sum_{n = 0}^{\infty}C_n  e^{(3-n)w} &=&  C_0e^{3w} +  C_1e^{2w} +\sum_{n = 2}^{\infty}C_n e^{(3-n)w}\cr
&=& e^{3w}  -(a+b)e^{2w} +\sum_{n = 2}^{\infty}C_n e^{(3-n)w}
\cr
\cr
A\sum_{n = 0}^{\infty}C_n e^{(2-n)} &=& A e^{2w}  +   A\sum_{n = 1}^{\infty}C_n e^{(2-n)w}
\label{sum post twist}
\eea
Now shifting each sum back to $n=0$ gives:

\bea
\sum_{n = 0}^{\infty}C_n  e^{(3-n)w} &=& e^{3w}  -(a+b)e^{2w} + \sum_{n = 0}^{\infty}C_{n+2} e^{(1-n)w}
\cr
\cr
A\sum_{n = 0}^{\infty}C_n e^{(2-n)w} &=& A e^{2w} + A\sum_{n = 0}^{\infty}C_{n+1} e^{(1-n)w}   
\label{sum two post twist}
\eea
Inserting (\ref{sum two post twist}) into term containing the three summations in (\ref{tpm one}) gives:
\bea
&&  \sum_{n = 0}^{\infty}C_n  e^{(3-n)w} -  A\sum_{n = 0}^{\infty}C_n  e^{(2-n)w} +B \sum_{n = 0}^{\infty}C_n  e^{(1-n)w} 
 \cr
 \cr
 &&\qquad\qquad =  e^{3w} - A'e^{2w} + \sum_{n=0}^{\infty} \big (C_{n+2}- A C_{n+1} + B C_n\big)e^{(1-n)w}
 \cr
 \cr
  &&\qquad\qquad=  e^{3w} - A'e^{2w} + \sum_{n=0}^{\infty}C'_n e^{(1-n)w}
  \label{sum}
\eea
where
\bea
C'_n\equiv (C_{n+2}- A C_{n+1} + B C_n\big)
\label{Cn prime}
\eea
Inserting (\ref{sum}) into  (\ref{tpm one}) gives:
\bea
&&\sum_{k=0}^{2}(-1)^{k}(t_1-t_2)^{-k-2}(t_{\pm}-t_2)(t_{\pm}-t_1)^k\cr
&&\quad\quad =  {1\over 32 t_2^4}\bigg( e^{3w} - A'e^{2w} + B'e^w + C' \pm \bigg( e^{3w} - A'e^{2w} + \sum_{n=0}^{\infty}C'_n e^{(1-n)w}\bigg)\bigg)  \nn
\label{t pm two}
\eea
We have now reduced our term to one infinite sum.
Now inserting (\ref{t pm two}) into (\ref{Copy 1 final}) for both Copy 1 and Copy 2 contours, taking $t=t_+$ for Copy 1 and $t=t_-$ for Copy 2, and using the appropriate mode definitions given in (\ref{cylinder G}), we obtain the following result for the Copy 1 and Copy 2 final contours:
\bea
\text{Copy } 1&:&{1\over 2 \pi i}\int_{\s = 0,\t>\t_2}^{\s = 2\pi}\diff w \sum_{k=0}^{2}(-1)^k(t_1-t_2)^{-k-2}(t_+-t_2)(t_+-t_1)^k G^{(1)-}_{\dot{C}}(w)
\cr
\cr
&&\qquad\qquad = {1\over 32 t_2^4}\bigg[ G_{\dot{C},3}^{(1)-}-A'G_{\dot{C},2}^{(1)-}+B'G_{\dot{C},1}^{(1)-} + C'  G_{\dot{C},0}^{(1)-}\cr
&&\qquad\qquad\qquad\qquad\qquad\qquad{} + \bigg( G_{\dot{C},3}^{(1)-} - A'G_{\dot{C},2}^{(1)-} + \sum_{n=0}^{\infty}C'_n G_{\dot{C},1-n}^{(1)-}\bigg)\bigg] 
\cr
\cr
\cr
\cr
\text{Copy } 2&:&{1\over 2 \pi i}\int_{\s = 0,\t>\t_2}^{\s = 2\pi}\diff w \sum_{k=0}^{2}(-1)^k(t_1-t_2)^{-k-2}(t_--t_2)(t_--t_1)^k G^{(2)-}_{\dot{C}}(w)
\cr
\cr
&&\qquad\qquad = {1\over  32 t_2^4}\bigg[ G_{\dot{C},3}^{(2)-}-A'G_{\dot{C},2}^{(2)-}+B'G_{\dot{C},1}^{(2)-} + C'  G_{\dot{C},0}^{(2)-}\cr
&&\qquad\qquad\qquad\qquad\qquad\qquad{} - \bigg( G_{\dot{C},3}^{(2)-} - A'G_{\dot{C},2}^{(2)-} + \sum_{n=0}^{\infty}C'_n G_{\dot{C},1-n}^{(2)-}\bigg)\bigg]\nn  
\label{cylinder contours 1}
\eea
Using our mode definitions in (\ref{new basis}) we add together both Copy 1 and Copy 2 contours given in (\ref{cylinder contours 1}) to obtain:
\bea
&&{1\over 2 \pi i}\int_{\s = 0,\t>\t_2}^{\s = 2\pi}\diff w \sum_{k=0}^{2}(-1)^k(t_1-t_2)^{-k-2}(t_+-t_2)(t_+-t_1)^k G^{(1)-}_{\dot{C}}(w)
\cr
&& \quad + 
{1\over 2 \pi i}\int_{\s = 0,\t>\t_2}^{\s = 2\pi}\diff w \sum_{k=0}^{2}(-1)^k(t_1-t_2)^{-k-2}(t_--t_2)(t_--t_1)^k G^{(2)-}_{\dot{C}}(w)
\cr
\cr
&&\quad = {1\over  32 t_2^4}\bigg( G_{\dot{C},3}^{-} - A'G_{\dot{C},2}^{-} + B'G_{\dot{C},1}^{-} + C'  G_{\dot{C},0}^{-} +  \tilde{G}_{\dot{C},3}^{-} - A'\tilde{G}_{\dot{C},2}^{-} + \sum_{n=0}^{\infty}C'_n \tilde{G}_{\dot{C},1-n}^{-}\bigg)\nn
\label{combined final}
\eea
We note that there are a finite number of both symmetric and antisymmetric annihilation modes \textit{after} the twist and an infinite number of creation modes \textit{after} the twist. Again, since creation modes annihilate when acting on a state to the left we need not worry about getting an infinite number of terms.

We have now computed the two relevant types contours at locations on the cylinder corresponding to Copy 1 and Copy 2 \textit{final} states. Next, we compute these same contours at locations on the cylinder corresponding to Copy 1 and Copy 2 \textit{initial} states.

\subsection{Integral Expressions for Copy 1 and Copy 2 Initial}

Here we compute the relevant contours corresponding to Copy 1 and Copy 2 \textit{initial} states. We will again have two types of terms analogous to the case for Copy 1 and Copy 2 \textit{final} states.

\subsubsection{Contours containing $(t-t_2)$} 

Just as before we begin with the simpler of the two contours with terms of the following form:
\bea
\text{Copy 1}&:&{1\over 2 \pi i}\int_{\s = 0,\t<\t_1}^{\s = 2\pi}\diff w (t_+-t_2) G^{(1)-}_{\dot{C}}(w)
\cr
\cr
\cr
\text{Copy 2}&:&{1\over 2 \pi i}\int_{\s = 0,\t<\t_1}^{\s = 2\pi}\diff w (t_--t_2) G^{(2)-}_{\dot{C}}(w)
\label{t minus t2 initial prime}
\eea
where again we take $t_+$ to correspond to Copy 1 initial and $t_1$ to correspond to Copy 2 initial.
For these contours we compute the following expression using (\ref{t pm one initial}):
\bea
t_{\pm} - t_2 &=&  {1\over 2} \big( e^w - e^{\Delta w \over 2} \mp  \sum_{n = 0}^{\infty}C_n  e^{nw}\big)
\label{t minus tpm initial}
\eea
Inserting (\ref{t minus tpm initial}) into (\ref{t minus t2 initial prime}) and using the mode definitions given in (\ref{cylinder G initial}), we obtain the following result for Copy 1 and Copy 2 initial contours:
\bea
\text{Copy 1}&:&{1\over 2 \pi i}\int_{\s = 0,\t<\t_1}^{\s = 2\pi}\diff w (t_+-t_2) G^{(1)-}_{\dot{C}}(w) \cr
&&\qquad = {1\over 2}\bigg ( G^{(1)-}_{\dot{C},1} - e^{\Delta w\over 2}G^{(1)-}_{\dot{C},0}  -  \sum_{n = 0}^{\infty}  C_n G^{(1)-}_{\dot{C},n} \bigg) \cr
\cr
\cr
\cr
\text{Copy 2}&:&{1\over 2 \pi i}\int_{\s = 0,\t<\t_1}^{\s = 2\pi}\diff w (t_--t_2) G^{(2)-}_{\dot{C}}(w)\cr
&&\qquad = {1\over 2}\bigg ( G^{(2)-}_{\dot{C},1} - e^{\Delta w\over 2}G^{(2)-}_{\dot{C},0}  +  \sum_{n = 0}^{\infty}  C_n G^{(2)-}_{\dot{C},n} \bigg) 
\label{t minus t2 initial}
\eea
Combining both contours given in (\ref{t minus t2 initial}) and using the basis defined in (\ref{new basis}) gives:
\bea
&&{1\over 2 \pi i}\int_{\s = 0,\t<\t_1}^{\s = 2\pi}\diff w (t_+-t_2) G^{(1)-}_{\dot{C}}(w) + {1\over 2 \pi i}\int_{\s = 0,\t<\t_1}^{\s = 2\pi}\diff w (t_--t_2) G^{(2)-}_{\dot{C}}(w)\cr
&&\qquad =  {1\over 2}\bigg ( G^{-}_{\dot{C},1} - e^{\Delta w\over 2}G^{-}_{\dot{C},0}  -  \sum_{n = 0}^{\infty}  C_n \tilde{G}^{-}_{\dot{C},n} \bigg)
\label{t minus t2 initial 2}
\eea
We note here that we have a finite number of symmetric and antisymmetric annihilation modes \textit{before} the twist and infinite number of antisymmetric annihilation modes \textit{before} the twist. Since annihilation modes annihilate when acting on a state to the right we need not worry about getting an infinite number of terms.

Here we have now computed the simpler of the two relevant contours at cylinder locations of the \textit{inital} copies of the CFT. Next we compute the more complicated contours.

\subsubsection{Contours containing $\sum_{k=0}^{2}(t_1-t_2)^{-k-2}(-1)^k(t-t_2)(t-t_1)^k$}

Here we compute the more complicated contours at initial locations on the cylinder. For this case we again have contours of the form:

\bea
\text{Copy 1}&:&{1\over 2 \pi i}\int_{\s = 0,\t<\t_1}^{\s = 2\pi}\diff w \sum_{k=0}^{2}(t_1-t_2)^{-k-2}(-1)^k(t_+-t_2)(t_+-t_1)^k G^{(1),-}_{\dot{C}}(w)
\cr
\cr
\text{Copy 2}&:&{1\over 2 \pi i}\int_{\s = 0,\t<\t_1}^{\s = 2\pi}\diff w \sum_{k=0}^{2}(t_1-t_2)^{-k-2}(-1)^k(t_--t_2)(t_--t_1)^k G^{(2),-}_{\dot{C}}(w)\nn
\label{Copy 1 initial}
\eea
where we again take the $t_+$ solution for Copy 1 initial and the $t_-$ solution for Copy 2 initial. For the above contours we again need to compute the integrand
\bea
&&\sum_{k=0}^{2}(-1)^k(t_1-t_2)^{-k-2}(t_{\pm}-t_2)(t_\pm-t_1)^k\cr
&&\quad\quad = {1\over 16 t_2^4}\bigg(t_{\pm} (z^2 - z A + B)-zt_2^2 + z_1t_1^2-8t_2^3\bigg)
\label{expansion of integrand two}
\eea
but now using initial state expansions. Inserting the initial state expansions of $t_{\pm}$ given in (\ref{t pm one initial}) into (\ref{expansion of integrand 2}), again with considerable simplification, we obtain:
\bea
&&\sum_{k=0}^{2}(-1)^k(t_1-t_2)^{-k-2}(t_{\pm}-t_2)(t_\pm-t_1)^k\cr
&&\quad\quad  =  {1\over 32  t_2^4}\bigg[ z^3-A'z^2+B'z + C'\cr
&&\qquad\qquad\qquad\qquad{} \mp \bigg(  \sum_{n = 0}^{\infty}C_n  z^{n + 2} -  A\sum_{n = 0}^{\infty}C_n  z^{n + 1} +B \sum_{n = 0}^{\infty}C_n  z^{n}\bigg)\bigg]\nn
\label{t pm two initial z}
\eea
Inserting $z=e^w$ into (\ref{t pm two initial z}) gives the expression:
\bea
&&\sum_{k=0}^{2}(-1)^k(t_1-t_2)^{-k-2}(t_{\pm}-t_2)(t_\pm-t_1)^k\cr
&&\quad\quad  =  {1\over 32  t_2^4}\bigg[ e^{3w} - A'e^{2w} + B'e^w + C'\cr
&&\qquad\qquad\qquad\qquad{} \mp \bigg(  \sum_{n = 0}^{\infty}C_n  e^{(n + 2)w} -  A\sum_{n = 0}^{\infty}C_n  e^{(n + 1)w} +B \sum_{n = 0}^{\infty}C_n  e^{nw}\bigg)\bigg]\nn
\label{t pm two initial}
\eea
where we have again written our expression in terms of the cylinder coordinate $w$.
We again adjust the above sums in order to combine terms. Adjusting all three of the sums in (\ref{t pm two initial}) gives:
\bea
\sum_{n = 0}^{\infty}C_{n}  e^{(n + 2)w } &=& \sum_{n = 2}^{\infty}C_{n-2}  e^{nw} 
\cr
\cr
A\sum_{n = 0}^{\infty}C_n  e^{(n + 1)w} &=& A\sum_{n = 1}^{\infty}C_{n-1}  e^{nw}\cr
&=& AC_0e^w + A\sum_{n = 2}^{\infty}C_{n-1}  e^{nw}\cr
&=& Ae^w + A\sum_{n = 2}^{\infty}C_{n-1}  e^{nw}
\cr
B\sum_{n = 0}^{\infty}C_n  e^{nw}&=&BC_0 + BC_1e^w +  B\sum_{n = 2}^{\infty}C_{n}  e^{nw }\cr
&=& B - B(a+b)e^w +  B\sum_{n = 2}^{\infty}C_{n}  e^{nw }
\label{sum initial}
\eea
Inserting the expansions in (\ref{sum initial}) into (\ref{t pm two initial}) gives:
\bea
&&\sum_{k=0}^{2}(-1)^k(t_1-t_2)^{-k-2}(t_{\pm}-t_2)(t_\pm-t_1)^k\cr
&&\quad\quad=  {1\over 32 t_2^4}\bigg[ e^{3w} - A'e^{2w} + B'e^w + C' \mp \bigg(B - (A + B(a+b))e^w \cr
&&\quad\quad\quad\quad\quad\quad\quad\quad +  \sum_{n = 2}^{\infty}\bigg(C_{n-2} - A C_{n-1} + B C_n \bigg) e^{nw} \bigg) \bigg]
\cr
&&\quad\quad = {1\over 32 t_2^4}\bigg[ e^{3w} - A'e^{2w} + B'e^w + C' \mp  \bigg( B - (A-B(a+b))e^w +  \sum_{n = 2}^{\infty}C''_n e^{nw}\bigg) \bigg]\nn
\label{t pm three initial}
\eea
where we define
\bea
C''_n\equiv C_{n-2} - A C_{n-1} + B C_n
\label{Cn prime prime}
\eea
We see that we have again writtin our expression in terms of a single infinite sum.

Inserting the expression (\ref{t pm three initial}) into (\ref{Copy 1 initial}) and using the mode definitions defined in (\ref{cylinder G initial}) we obtain the following result for Copy 1 and Copy 2 initial contours:
\bea
\text{Copy 1}&:&{1\over 2 \pi i}\int_{\s = 0,\t<\t_1}^{\s = 2\pi}\diff w \sum_{k=0}^{2}(t_1-t_2)^{-k-2}(-1)^{k}(t-t_2)(t-t_1)^k G^{(1)-}_{\dot{C}}(w)\cr
&&\qquad\qquad = {1\over  32 t_2^4}\bigg( G^{(1)-}_{ \dot{C},3} -A'G^{(1)-}_{ \dot{C},2} +B'G^{(1)-}_{ \dot{C},1} + C'G^{(1)-}_{ \dot{C},0} \cr
&&\qquad\qquad\qquad\qquad - BG^{(1)-}_{\dot{C},0} + (A + B(a+b))G^{(1)-}_{\dot{C},1} -  \sum_{n = 0}^{\infty}C''_n G^{(1)-}_{ \dot{C},n} \bigg)
\cr
\cr
\text{Copy 2}&:&{1\over 2 \pi i}\int_{\s = 0,\t<\t_1}^{\s = 2\pi}\diff w \sum_{k=0}^{2}(t_1-t_2)^{-k-2}(-1)^{k}(t-t_2)(t-t_1)^k G^{(2)-}_{\dot{C}}(w)\cr
&&\qquad\qquad = {1\over 32 t_2^4}\bigg( G^{(2)-}_{ \dot{C},3} -A'G^{(2)-}_{ \dot{C},2} +B'G^{(2)-}_{ \dot{C},1} + C'G^{(2)-}_{ \dot{C},0}\cr
&&\qquad\qquad\qquad\qquad + BG^{(2)-}_{\dot{C},0} - (A + B(a+b))G^{(2)-}_{\dot{C},1} + \sum_{n = 0}^{\infty}C''_n G^{(2)-}_{ \dot{C},n} \bigg)\nn
\label{initial contours}
\eea
Combining both contours in (\ref{initial contours}) again using the basis defined in (\ref{new basis}) gives:
\bea
&&{1\over 2 \pi i}\int_{\s = 0,\t<\t_1}^{\s = 2\pi}\diff w \sum_{k=0}^{2}(t_1-t_2)^{-k-2}(-1)^{k}(t-t_2)(t-t_1)^k G^{(1)-}_{\dot{C}}(w)\cr
&&\quad + {1\over 2 \pi i}\int_{\s = 0,\t<\t_1}^{\s = 2\pi}\diff w \sum_{k=0}^{2}(t_1-t_2)^{-k-2}(-1)^{k}(t-t_2)(t-t_1)^k G^{(2)-}_{\dot{C}}(w)
\cr
\cr
&&\quad =  {1\over  32 t_2^4}\bigg( G^{-}_{ \dot{C},3} -A'G^{-}_{ \dot{C},2} + B'G^{-}_{ \dot{C},1} + C'G^{-}_{ \dot{C},0}\cr
&&\qquad\qquad\qquad\qquad{} - B\tilde{G}^{-}_{\dot{C},0} + (A + B(a+b))\tilde{G}^{-}_{\dot{C},1} -   \sum_{n = 0}^{\infty}C''_n \tilde{G}^{-}_{ \dot{C},n} \bigg)\nn
\label{combined initial}
\eea

We note here that our expression contains a finite number of symmetric annihilation modes before the twist as well as an infinite number of antisymmetric annihilation modes before the twist. Again, since annihilation modes annihilate when acting on a state to the right we need not worry about getting an infinite number of terms.
We have now computed the two relevant types contours at locations on the cylinder corresponding to Copy 1 and Copy 2 \textit{initial} states. 

Since we have computed the relevant supercharge modes at initial and final state locations on the cylinder appearing in the expressions for our two deformation operators we are now in position to compute the final result on the cylinder.

\section{Final Result}\label{final result}
We are now able to write the expression of our two deformation operators in terms cylinder modes. Let us now rewrite (\ref{full deformation two}) in terms of $G$ modes on the cylinder which we computed in Section \ref{cylinder modes}. We recall that we have two cases to compute. The first case is when $\dot{A}=\dot{B}$ and the second case for general $\dot{A}$ and $\dot{B}$. We remind the reader that the general case reduces to the case of $\dot{A}=\dot{B}$ but case is much easier computed while in the $t$ plane. We write our two cases below.
\subsection{$\dot{A}=\dot{B}$}
Here we compute the final expression for the case where $\dot{A}=\dot{B}$. 
First let us define a shorthand notation for our twists:
\bea
\hat{\s} = \s_{2}^+(w_2)\s_{2}^+(w_1)
\label{sigma defintion}
\eea
Now inserting (\ref{t minus t2 final 2}), (\ref{t minus t2 initial 2}), and (\ref{sigma defintion}) into (\ref{A = B two}) and using (\ref{redefinition}) we obtain:
\bea
\hat{O}_{\dot{A}}\hat{O}_{\dot{A}}\!\!&=&\!\!  -{1\over 4t_2}\bigg[ G^{-}_{\dot{A},0} \bigg ( G^{-}_{\dot{A},1}   +  \sum_{n = 0}^{\infty}  C_n \tilde{G}^{-}_{\dot{A},1-n} \bigg)~\hat{\s} -  G^{-}_{\dot{A},0}~ \hat{\s}~\bigg ( G^{-}_{\dot{A},1}   -  \sum_{n = 0}^{\infty}  C_n \tilde{G}^{-}_{\dot{A},n} \bigg) 
\cr
\cr
&&\qquad{} + \bigg ( G^{-}_{\dot{A},1}  +  \sum_{n = 0}^{\infty}  C_n \tilde{G}^{-}_{\dot{A},1-n} \bigg)~\hat{\s}~ G^{-}_{\dot{A},0}  - \hat{\s}~\bigg ( G^{-}_{\dot{A},1} -  \sum_{n = 0}^{\infty}  C_n \tilde{G}^{-}_{\dot{A},n} \bigg) G^{-}_{\dot{A},0}\bigg]\nn
\eea
\subsection{General $\dot{A}$ and $\dot{B}$} 
For the case for general $\dot{A}$ and $\dot{B}$, inserting (\ref{t minus t2 final 2}), (\ref{combined final}), (\ref{t minus t2 initial 2}), (\ref{combined initial}) and (\ref{sigma defintion}) into (\ref{full deformation two}) and using the expression
\bea
C' = 2z_1t_2^2 - (a+b)B - 16 t_2^3,
\eea 
gives the final result:
\bea
\hat{O}_{\dot{B}}\hat{O}_{\dot{A}}\!\! &=&\!\! - {1\over  64 t_2^4}\bigg( G_{\dot{B},3}^{-} - A'G_{\dot{B},2}^{-} + B'G_{\dot{B},1}^{-} + \big( 2z_1t_2^2 -B(a+b)\big) G_{\dot{B},0}^{-}\cr
&&\qquad\qquad {}+  \tilde{G}_{\dot{B},3}^{-} - A'\tilde{G}_{\dot{B},2}^{-} + \sum_{n=0}^{\infty}C'_n \tilde{G}_{\dot{B},1-n}^{-}\bigg)\cr
&&\qquad\qquad\!\!\!\!\!\!\!\times ~\bigg ( G^{-}_{\dot{A},1} - e^{\Delta w\over 2}G^{-}_{\dot{A},0}  +  \sum_{n = 0}^{\infty}  C_n \tilde{G}^{-}_{\dot{A},1-n} \bigg)~\hat{\s}
\cr
\cr
\cr
&&\!\!+{1\over  64 t_2^4}\bigg( G_{\dot{B},3}^{-} - A'G_{\dot{B},2}^{-} + B'G_{\dot{B},1}^{-} + \big( 2z_1t_2^2 -B(a+b)\big)G_{\dot{B},0}^{-}\cr
&&\qquad\qquad{} +  \tilde{G}_{\dot{B},3}^{-} - A'\tilde{G}_{\dot{B},2}^{-} + \sum_{n=0}^{\infty}C'_n \tilde{G}_{\dot{B},1-n}^{-}\bigg)\cr
&&\qquad\qquad\!\!\!\!\!\!\!\times ~\hat{\s}~\bigg( G^{-}_{\dot{A},1} - e^{\Delta w\over 2}G^{-}_{\dot{A},0}  -  \sum_{n = 0}^{\infty}  C_n \tilde{G}^{-}_{\dot{A},n} \bigg)
\cr
\cr
\cr
&&\!\!- {1\over  64 t_2^4}\bigg ( G^{-}_{\dot{A},1} - e^{\Delta w\over 2}G^{-}_{\dot{A},0}  +  \sum_{n = 0}^{\infty}  C_n \tilde{G}^{-}_{\dot{A},1-n} \bigg)
~\hat{\s}
\cr
&&\qquad\qquad\!\!\!\!\!\!\!\!\!\times~\bigg( G^{-}_{ \dot{B},3} -A'G^{-}_{ \dot{B},2} + B'G^{-}_{ \dot{B},1} + \big(  2z_1t_2^2 - (a+b)B \big)G^{-}_{ \dot{B},0}\cr
&&\qquad\qquad  - B\tilde{G}^{-}_{\dot{B},0}  + (A + B(a+b))\tilde{G}^{-}_{\dot{B},1} -   \sum_{n = 2}^{\infty}C''_n \tilde{G}^{-}_{ \dot{B},n} \bigg)
\cr
\cr
\cr
&&\!\!+{1\over  64 t_2^4}~\hat{\s}~
 \bigg ( G^{-}_{\dot{A},1} - e^{\Delta w\over 2}G^{-}_{\dot{A},0}  -  \sum_{n = 0}^{\infty}  C_n \tilde{G}^{-}_{\dot{A},n} \bigg)\cr
&&\qquad\qquad\!\!\!\!\!\!\times~\bigg( G^{-}_{ \dot{B},3} -A'G^{-}_{ \dot{B},2} + B'G^{-}_{ \dot{B},1} + \big( 2z_1t_2^2 - (a+b)B \big)G^{-}_{ \dot{B},0}\cr
&&\qquad\qquad - B\tilde{G}^{-}_{\dot{B},0} + (A + B(a+b))\tilde{G}^{-}_{\dot{B},1}  -   \sum_{n = 2}^{\infty}C''_n \tilde{G}^{-}_{ \dot{B},n} \bigg)\nn
\label{final result one}
\eea
Writing our final result in compact notation, (\ref{final result one}) becomes:
\bea
\hat{O}_{\dot{B}}\hat{O}_{\dot{A}} &=& -{1\over 64 t_2^4}\bigg[\bigg(\sum_{n=0}^{3} D_{n}G^{-}_{\dot{B},n} + \sum_{n=-2}^{\infty}E_{n}\tilde{G}^{-}_{\dot{B},1-n} \bigg)\bigg(\sum_{n=0}^1 D'_n G^-_{\dot{A},n} + \sum_{n=0}^{\infty} C_n\tilde{G}^-_{\dot{A},1-n}\bigg)~\hat{\s}\cr
&&\qquad\quad -\bigg(\sum_{n=0}^{3} D_{n}G^{-}_{\dot{B},n} + \sum_{n=-2}^{\infty}E_{n}\tilde{G}^{-}_{\dot{B},1-n} \bigg)~\hat{\s}~\bigg(\sum_{n=0}^1 D'_n G^-_{\dot{A},n} - \sum_{n=0}^{\infty} C_n\tilde{G}^-_{\dot{A},n}\bigg)\cr
&&\qquad\quad + \bigg(\sum_{n=0}^1 D'_n G^-_{\dot{A},n} + \sum_{n=0}^{\infty} C_n\tilde{G}^-_{\dot{A},1-n}\bigg)~\hat{\s}~ \bigg(\sum_{n=0}^{3} D_{n}G^{-}_{\dot{B},n} + \sum_{n=0}^{\infty}E'_{n}\tilde{G}^{-}_{\dot{B},n} \bigg)\cr
&&\qquad\quad -\hat{\s}~\bigg(\sum_{n=0}^1 D'_n G^-_{\dot{A},n} - \sum_{n=0}^{\infty} C_n\tilde{G}^-_{\dot{A},n}\bigg) \bigg(\sum_{n=0}^{3} D_{n}G^{-}_{\dot{B},n} + \sum_{n=0}^{\infty}E'_{n}\tilde{G}^{-}_{\dot{B},n} \bigg)\bigg]\nn
\eea
where we have computed the above coefficients in terms of $\Delta w$ in detail in Appendix \ref{coefficients}. We tabulate the results below:
\bea
C_n &=& (-1)^{n}\sum_{k = - {n\over 2}}^{{n \over 2}}{}^{1/2}C_{{n\over 2} + k}{}^{1/2}C_{{n\over 2} - k}\cosh\big( k \Delta w\big)\cr
D_0 &=& -{1\over 4}e^{-{3\Delta w\over2}}\big( 1 + 3e^{\Delta w} \big)\cr
D_1 &=&{3\over 2}\big( 1 + e^{-\Delta w} \big)\cr
D_2 &=& -{3\over 4}e^{-{\Delta w\over 2}}\big( 3 + e^{\Delta w} \big)\cr
D_3 &=&  1\cr
\cr
D'_0 &=& -e^{\Delta w \over 2}\cr
D'_1 &=& 1\cr
\cr
E_{-2}&=& 1\cr
E_{-1}&=& -{3\over 4}e^{-{\Delta w\over 2}}\big( 3 + e^{\Delta w} \big)\cr
E_{n} &=&  \bigg(C_{n+2}- {1\over 4}e^{-{\Delta w\over 2}}\big( 7 + e^{\Delta w} \big) C_{n+1} +{1\over 4}\big(1 + 3e^{-\Delta w} \big) C_n\bigg),\quad n\geq 0\cr
\cr
E'_0 &=& -{1\over 4}\big(1 + 3e^{-\Delta w} \big)\cr
E'_1 &=& {3\over4} e^{-{\Delta w\over 2}} \big(3 + \cosh\big(\Delta w\big)\big)\cr
E'_n &=& -\bigg( C_{n-2} - {1\over 4}e^{-{\Delta w\over 2}}\big( 7 + e^{\Delta w} \big) C_{n-1} + {1\over 4}\big(1 + 3e^{-\Delta w} \big)C_n\bigg),\quad n\geq 2
\eea
We see that in both cases above, we were able to write the full expression of two deformation operators in terms of modes before and after the twists using the covering space method.

Also, we again note that there are a finite number of symmetric and antisymmetric annihilation modes after the twist and a finite number of symmetric annihilation modes before the twist but an infinite number of antisymmetric creation modes after the twist and an infinite number of antisymmetric annihilation modes before the twist. These terms all give finite contributions when computing an amplitude because creators kill on the left and annihilators kill on the right. We need not worry about an infinite number of contributions when capping each side with an appropriate state.

\subsection{Numerical Plots of $C_{n}$}
Here we numerically plot the coefficient
\bea
C_n = \sum_{k=-{n\over2}}^{n\over2}{}^{\h}C_{{n\over2}+k}{}^{\h}C_{{n\over2}-k}\cosh(k\D w)
\label{C euclidean}
\eea 
to determine its behavior for various mode numbers, $n$.
We can write our twist separation $\Delta w$ as:
\bea
\Delta w = \Delta \t + i\Delta \s
\eea
We wick rotate back to Minkowski signature by taking $\Delta \t\to i\Delta t$ which gives
\bea
\Delta w \to i \Delta w
\label{dw}
\eea
with $\Delta w = \Delta t + \Delta \s$ being completely real. Inserting (\ref{dw}) into (\ref{C euclidean}) gives
\bea
C_n = \sum_{k=-{n\over2}}^{n\over2}{}^{\h}C_{{n\over2}+k}{}^{\h}C_{{n\over2}-k}\cos(k\D w)
\label{C minkowksi}
\eea 
We now plot (\ref{C minkowksi}) in Figure \ref{figtwo} for various values of $n$:
\begin{figure}[tbh]
\begin{center}
\includegraphics[width=0.42\columnwidth]{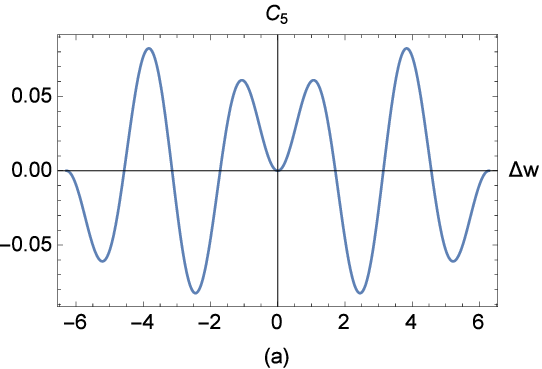} $\qquad\qquad$ \includegraphics[width=0.42\columnwidth]{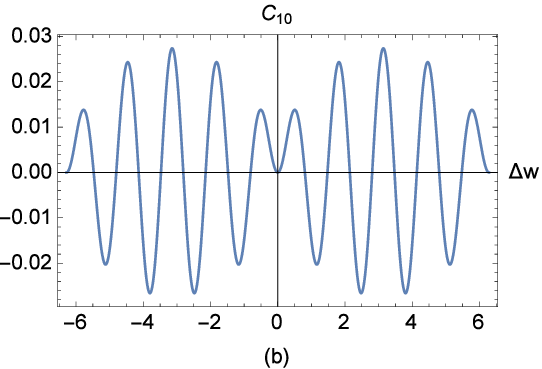} $\qquad\qquad$
\includegraphics[width=0.42\columnwidth]{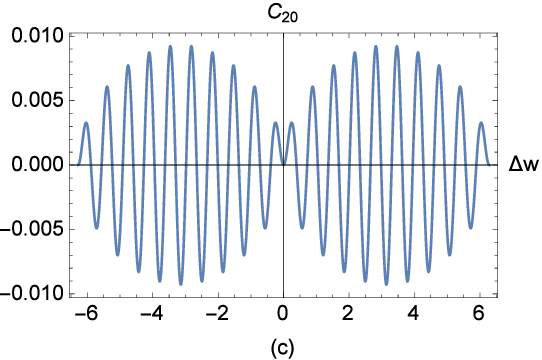}
\end{center}
\caption{We plot $C_n$ for (a) $n=5$ (b) $n=10$ and (c) $n=20$ each from $\Delta w \in [-2\pi,2\pi]$}
\label{figtwo}
\end{figure}

We see an oscillatory behavior as well as an envelope behavior coming from the twist separation, $\Delta w$. The number of peaks from $0$ to $2\pi$ is given by ${ n\over 2}$ for even $n$ and ${n-1\over 2}$ for odd $n$.

\section{Discussion}

Black hole formation is a process that has yet to possess a full quantitative description. Because this process is difficult to study in the gravity setting, it is useful to try to gain some insight using the dual CFT.  Generally, the process of black hole formation in the gravity theory would manifest as the process of thermalization in the CFT. Looking at the D1D5 CFT, we investigate this question of thermalization by considering a deformation away from its conjectured `orbifold point' where the theory is free.\footnote{For another approach to thermalization in the CFT, see \cite{kaplan}.} This deformation, $\hat{O}_{\dot{A}}$, consists of a supercharge $G$ and a twist $\s$. Looking at first order in the deformation operator, one finds no clear evidence of thermalization. We thus extended our analysis to second order in the twist operator where the first twist joins two singly wound copies into one doubly wound copy and then the second twist returns the double wound copy back to two singly wounds copies. One finds  that after applying two twists to the vacuum a squeezed state is produced just as in the one twist case. In prior work, we computed the bosonic and fermionic Bogoliubov coefficients $\g^{B},\g^{F\pm}$ describing this squeezed state. In the large $m,n$ limit we found a decay with $m,n$ similar to the one twist case which was computed in chapter \ref{sec:gamma} which came from \cite{chmt1,cmt}, but also found  an additional oscillatory dependence arising from the twist separation parameter $\Delta w$ found in chapter \ref{two twist gamma} which came from \cite{chm1}. We conjectured that to all orders in the deformation, the Bogoliubov coefficients could be factored into  a part independent of the twist separations, and a part that oscillated with these separations.\footnote{For other computations with the twist deformation in the CFT, see \cite{peet}.}

\chapter{Lifting of D1-D5-P States}\label{lifting chapter}

While investigating the question of thermalization, we realized that we could use the deformation of the CFT to compute the anomalous dimension of states in the CFT. We call this the `lifting' of states. In this chapter we show how to compute such a quantity. Fortunately, some of the methodology used in computing the lifting can also be applied to the question of thermalization which we address in the next chapter.

\section{Introduction}\label{Intro}

One of the most useful examples of a black hole is the hole made with D1, D5, and P charges in string theory. The microscopic entropy for these charges, $S_{micro}$,  agrees with the Bekenstein entropy, $S_{bek}$, obtained from the classical gravity solution with the same charges \cite{sv,cm}. Further, a weak coupling computation of radiation from the branes, $\Gamma_{micro}$,  agrees with the Hawking radiation from the gravitational solution, $\Gamma_{hawking}$ \cite{dmcompare,maldastrom}. 

AdS$_3$/CFT$_2$ duality  \cite{adscft, gkp, witten} states that the near horizon dynamics of the black hole is described by a 1+1 dimensional CFT called the D1-D5 CFT. The momentum charge P is carried by left moving excitations of this CFT. The CFT has a `free point' called the `orbifold CFT', where the theory can be described using free bosons and free fermions on a set of twisted sectors \cite{Vafa:1995bm,Dijkgraaf:1998gf,orbifold2,Larsen:1999uk,Arutyunov:1997gt,Arutyunov:1997gi,Jevicki:1998bm}, see \cite{deformation} for a review of the D1-D5 brane system. 

 At the orbifold point all states which have only left moving excitations are BPS; i.e. they have energy equal to their charge. This need not be true as we deform the theory along some direction in the moduli space of  the D1-D5 CFT. Some of the states which were BPS at the orbifold point will remain BPS, while others can pair up and `lift'.

In this chapter we will look at a specific family of D1-D5-P states which are BPS at the orbifold point but which lift as we move away from this free point towards the supergravity description of the black hole. We use conformal perturbation theory to compute the lifting at quadratic order in the coupling $\lambda$. The form of this lifting will tell us about the behavior of string states in the gravity dual, and shed light on the nature of the fuzzball configurations that describe black hole microstates \cite{fuzzballs_i,fuzzballs_ii,fuzzballs_iii,fuzzballs_iv,fuzzballs_v}. 

We now summarize the set-up of the computation and the main results.

\subsection{The D1-D5 CFT}

As we have noted several times now, consider type IIB string theory compactified as
\be
M_{9,1}\r M_{4,1}\times S^1\times T^4\ .
\ee
We wrap $n_1$ D1 branes on $S^1$ and $n_5$ D5 branes on $S^1\times T^4$. The bound states of these branes generate the D1-D5 CFT,  which is a 1+1 dimensional field theory living on the cylinder made from  the $S^1$ and time directions. This theory is believed to have an orbifold point, where we have
\be
N=n_1n_5
\ee
copies of a $c=6$ free CFT. The free CFT is made of $4$ free bosons and $4$ free fermions in the left-moving sector and likewise in the right-moving sector. The free fields are subject to an orbifold symmetry generated by the  group of permutations $S_N$; this leads to various twisted sectors around the circle $S^1$. The field theory is a CFT with small ${\cal N}=4$ supersymmetry in each of the left and right-moving sectors; thus the left sector has chiral algebra generators $L_n, G^{\pm}_{\dot A,r}, J^a_n$ associated with the stress-energy tensor, the supercurrents, and the $\mathfrak{su}(2)$ R-currents (the right-moving sector has analogous generators). The small $\mathcal N=4$ superconformal algebra and our notations are outlined in appendix \ref{app_cft}.

\subsection{The states of interest}\label{introsec}

\begin{figure}
\begin{center}
\includegraphics[width=67mm]{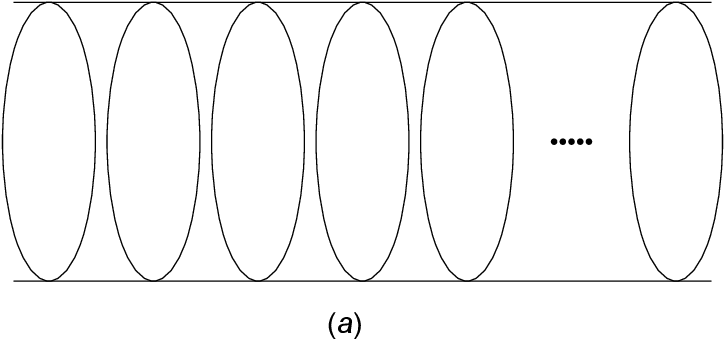}$\qquad\qquad$
\includegraphics[width=67mm]{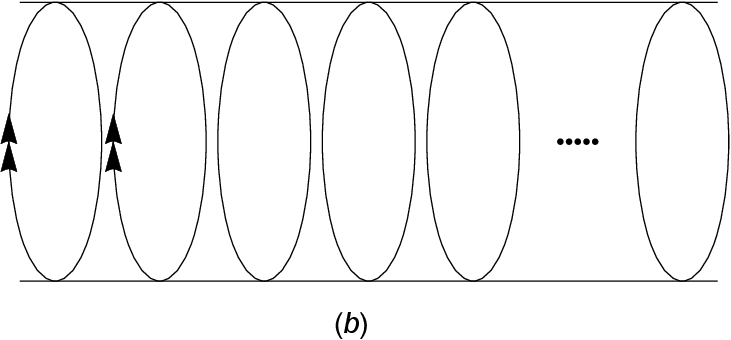}
\end{center}
\label{singlwinding}
\caption{(a)  $N$ singly-wound component strings in their vacuum state wrapping the $S^1$; this gives the vacuum of the theory.  (b) $n$ of these $N$ copies have been excited by the application of current operators.}
\label{fig_singlwindingexcited}
\end{figure}

Consider the untwisted sector, i.e. the sector where each copy of the $c=6$ CFT is singly wound around the $S^1$.  The $N$ copies of the $c=6$ CFT can be depicted  by $N$ separate circles; we sketch this in fig. \ref{fig_singlwindingexcited}. We consider the NS sector. If all the copies are unexcited, we get the vacuum state $|0\rangle$ as shown in the left panel of the figure; the gravity dual of this state is $AdS_3 \times S^3\times T^4$ which was shown in Chapter \ref{D1D5 system}.

We will now consider a set of excited states, proceeding in the following steps:

\b

(i) Consider the state where $N-1$ copies are in the NS vacuum state, and one of the copies is excited by application of the operator
\be
 J^{+}_{-(2m-1)}\dots J^{+}_{-3}J^{+}_{-1}
\label{threeq}
\ee
on the NS vacuum. This is illustrated in the right panel of fig. \ref{fig_singlwindingexcited}. At the orbifold point this operator has quantum numbers
\be
(h,\bar h)=(m^2,0)\ ,\qquad(j,\bar j)=(m,0)\ .
\ee
Thus the energy of the state at the orbifold point  is
\be
E_{\mathrm{orbifold}}=h+\bar h = m^2\ .
\label{four}
\ee
At the orbifold point, the excitation is BPS since the right movers of the CFT are  in the supersymmetric ground state on all copies. 

We will argue that when we move to the supergravity domain, this state can be heuristically described by  a string localized at the center of the AdS space. Because the excitation is a string rather than a supergravity quantum, the energy will change away from the orbifold point (\ref{four}); we write the extra  energy as
\be
E-E_{\mathrm{orbifold}}=\Delta E\ .
\ee

\b

(ii) Suppose we place the excitation (\ref{threeq}) on {\it two} copies of the CFT. At the orbifold point we have
\be
(h,\bar h)=(2m^2,0)\ ,\qquad(j,\bar j)=(2m,0)\ ,
\ee
and
\be
E_{\mathrm{orbifold}}=2m^2\ .
\ee
At the supergravity point in the dual theory, our heuristic picture will  have {\it two} strings placed at the center of AdS. Each string will have an extra energy $\Delta E$ as before. But there will also be some gravitational attraction between these strings, which will lower the energy by some amount $\Delta E_{\mathrm{grav}}$. This suggests that the energy at the supergravity point will have the schematic form
\be
E-E_{\mathrm{orbifold}}=2\Delta E-E_{\mathrm{grav}}\ .
\ee

\b

(iii) Now suppose we place the excitation (\ref{threeq}) on $n$ out the the $N$ copies. In the dual gravity description we have $n$ strings. We get a positive energy from each of the $n$ strings, and a negative contribution to the energy from the attraction between each pair of strings. Thus the total energy has the schematic form
\be
E_n-E_{\mathrm{orbifold},n }=n\Delta E - {n(n-1)\over 2} E_{\mathrm{grav}}\ ,
\label{five}
\ee
where we have added a subscript  $n$ to the energies to indicate the number of copies which have been excited.

\b

(iv) Finally we place the excitation (\ref{threeq}) on {\it all} the $N$ copies of the $c=6$ CFT. In this situation we  know the energy $E$ exactly, because this state is obtained by a spectral flow of the vacuum by $2m$ units. We have
\be
E_N-E_{\mathrm{orbifold},N}=0
\ee

\b

\begin{figure}
\begin{center}
\includegraphics[width=80mm]{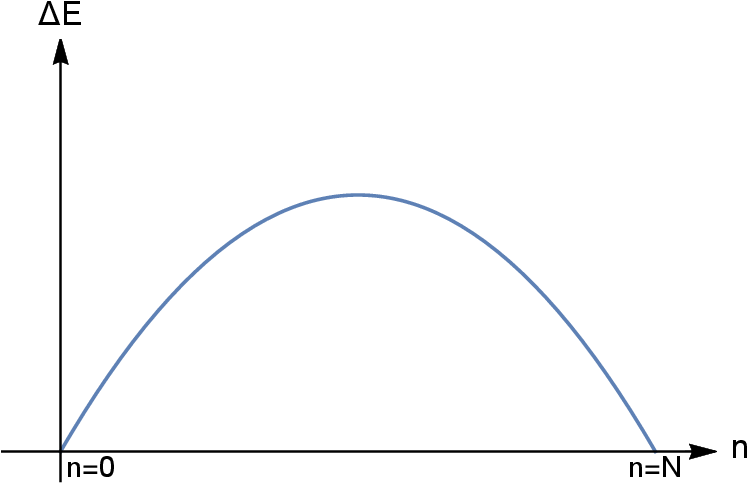}
\end{center}
\caption{A schematic plot showing the energy lift when $n$ out of the $N$ copies of the $c=6$ CFT  are excited.  The lift vanishes when $n=0$ and when $n=N$: the $n=0$ state is the vacuum and the $n=N$ state is a spectral flow of the vacuum.}
\label{fig_deltaE}
\end{figure}

To summarize, the quantity  $E_n-E_{\mathrm{orbifold},n}$ should have the schematic behavior depicted in fig. \ref{fig_deltaE}: it vanishes at $n=0$, rises for low values of $n$, then falls back to zero at $n=N$. 

\subsection{The results}

In this chapter we study the deformation of the D1-D5 CFT off the orbifold point towards the supergravity point, upto second order in the coupling $\lambda$. We show that, at this order, the energy $E_n$ of the family of the states we consider in eq. (\ref{threeq}) indeed has the form depicted in fig.\ref{fig_deltaE}:
\be
\langle ( E_n-E_{\mathrm{orbifold}}) \rangle =\lambda^2 \frac{\pi^{3\over 2}}{2}\,{\Gamma[m^2-{1\over2}]\over \Gamma[m^2-1]}\,n(N-n)\ ,
\label{wwone}
\ee
where $m$ is the number of the R-currents in the initial state (\ref{threeq}), see section \ref{section7} for details.

We also consider the case where the $N$ copies of the CFT are grouped into twist sectors with winding $k$ each, and then excited in a manner similar to that discussed above, We again find an expression of the energy lift of the form (\ref{wwone}). 

Finally, we note a general property of the computation of lifting at second order. If the deformation operators join two component strings and then break them apart, the covering surface arising in the computation has genus 0. If on the other hand  the deformation operators break and then rejoin a component string,  the covering space arising in the computation has genus 1. The maximally twisted sector can only exhibit the second possibility; this suggests that the large class of unlifted states needed to explain black hole entropy may lie in this sector.

There are several earlier works that have studied conformal perturbation theory, the lifting of the states, the acquiring of anomalous dimensions, and the issue of operator mixing,  in particular in the context of the D1-D5 CFT;  see for example \cite{chmt1,cmt,chmt2,chm1,chm2,chm3,Pakman:2009mi,Avery:2010er,Burrington:2012yq,Burrington:2014yia,Gaberdiel:2015uca,Burrington:2017jhh}.

\subsection{The plan of this chapter}
In section \ref{section2} we outline the computation that gives the lift to second order; in particular we explain why the issue of operator renormalization does not arise to this order in our problem.

 In section \ref{section3} we describe the deformation operator and the states whose lift we are interested in.

In section \ref{section4} we compute  the vacuum correlation function of two twist-2 operators; we call this the `base' amplitude, as it appears as a starting element in the computation of all other correlation functions.

In section \ref{section5} we compute the lifting of energies of the states under consideration.

In section \ref{section6} we consider global modes and use our approach to show that they are not lifted under conformal perturbation, to the order we study, as expected.

In section \ref{section7} we perform the needed combinatorics to extend our result to the case where we have an arbitrary number $N$ of component strings.

In section \ref{section8} we compute the lift for the case where the component strings on the initial state are grouped into sets with twist $k$ each.

In section \ref{section9} we analyze the nature of the covering space  in different instances of second order perturbation theory, and find a special role for the maximally twisted sector.

Section  \ref{section10} is a general discussion, where we state the physical implications of our results.

\section{Outline of the method}\label{section2}
In this section we first outline the conformal perturbation theory approach that we will use to compute the lifting of energies. We then derive a general expression for lifting to second order.

\subsection{Conformal perturbation theory on the cylinder}\label{outline}
We proceed in the following steps:

\b

(a) Suppose we have a conformal field $\phi$ with left and right-moving dimensions $(h, 0)$. On the plane, the 2-point function is
\be
\langle \phi(z) \phi(0) \rangle_0 ={1\over z^{2h}}\ ,
\ee
where the subscript $``0"$ corresponds to the unperturbed theory. After a perturbation of the CFT, the conformal dimensions can change to $(h+\delta h, \delta h)$. The left and right dimensions must increase by the same amount, since $h-\bar h$ must always be an integer for the operator to be local. The operator $\tilde\phi$ in the perturbed theory, having a well defined dimension, will also in general be different from $\phi$. So we should write
\be\label{phiphi_i}
\phi=\tilde\phi+\delta \tilde\phi\ .
\ee
We will see, however, that to the order where we will be working, the correction $\delta \tilde\phi$ will not be relevant (see eq.s (\ref{phiphi}), (\ref{bone_i}), (\ref{bone}), and footnote \ref{fn_bone}). We can then write
\be
\langle \phi(z) \phi(0) \rangle_{\mathrm{pert}} ={1\over z^h}{1\over |z|^{2\delta h}}= {1\over z^h}e^{-2\delta h\log |z|}\approx {1\over z^h} ( 1-4\delta h\log |z|)\ ,
\ee
where the subscript ``pert" corresponds to the perturbed theory. Thus the perturbation to the 2-point function has a correction term of the form $ \sim z^{-h}\log|z|$, and the perturbation to the  dimension $\delta h$ can be read off from the coefficient of this term. For more details on the analyses of conformal perturbation theory in two and higher dimensional CFTs see, e.g. \cite{kadanoff,Dijkgraaf:1987jt,Cardy:1987vr,Eberle:2001jq,Gaberdiel:2008fn,Berenstein:2014cia,Berenstein:2016avf}.

\b

(b) We will find it convenient to work on the cylinder rather than   the plane, so let us see how the expressions in (a) change when  we work on the cylinder.  The cylinder coordinate is given by
\be
z=e^w, ~~~w=\tau+i\sigma\ .
\ee
Consider the  state $|\phi\rangle$ corresponding to the operator $\phi$; we assume that this state is normalized as
\be\label{unit_norm}
\langle \phi |\phi\rangle_0 =1\ .
\ee
As will become clear below, we can ignore the change $\delta \tilde\phi$ in this state itself. Let us also assume for the moment  that the energy level of $\phi$ is nondegenerate. We  place the state $|\phi\rangle$ at $\tau=-{T\over 2}$ and the conjugate state $\langle \phi |$ at $\tau={T\over 2}$. We compute the amplitude $A(T)$ for transition between these two states. In the unperturbed theory, we have for our operator $\phi$ with dimensions $(h,0)$:
\be\label{amp_unpert}
A(T)=\langle \phi |e^{-H^{(0)} T}|\phi\rangle = e^{-hT}\ ,
\ee
where the energy of the state is $E=h+\bar h = h$, and $H^{(0)}$ is the Hamiltonian in the unperturbed theory. After the perturbation, we get
\be\label{amp_pert}
A(T)+\delta A(T) = \langle \phi |e^{-(H^{(0)}+\delta H) T}|\phi\rangle= e^{-(h+2\delta h)T}\approx e^{-hT}-2 \delta h\,  T e^{-hT}\ .
\ee
Thus, we can read off $\delta h$ from the coefficient of $Te^{-hT}$ in $\delta A(T)$. 

\b

(c) With these preliminaries, we now set up the formalism for situation that we actually have.  In our problem, the space of operators with dimension $(h, 0)$ is degenerate. Thus, we consider the case where we have operators $\phi_a, a=1, \dots,n$, all with the same dimension $(h,0)$. Let the remaining operators having well defined scaling dimensions be  called $\phi_\mu$; there will in general be an infinite number of the $\phi_\mu(\lambda)$, with dimensions going all the way to infinity. These operators are normalized as
\be
\langle \phi_{a}|\phi_{b}\rangle=\delta_{ab}\ ,\qquad\langle \phi_{\mu}|\phi_{\nu}\rangle=\delta_{\mu\nu}\ ,\qquad\langle \phi_{a}|\phi_{\mu}\rangle=0\ .
\label{none}
\ee
After the perturbations, there will be a different set of operators which have well defined scaling dimensions; let these operators be denoted by a tilde on top. We separate these operators into two classes. The first class is the operators that are deformations of the degenerate set $\phi_a, a=1, \dots n$. We call these deformed operators $\tilde \phi_{a'}(\lambda), a'=1, \dots n$, where we have explicitly noted the dependence of these operators on the coupling $\lambda$.  The second class is comprised of the remaining operators in the deformed theory which have well defined scaling dimensions; let us call these $\tilde\phi_\mu(\lambda)$.  We assume that these operators are normalized
\be
\langle \tilde\phi_{a'}(\lambda)|\tilde\phi_{b'}(\lambda)\rangle=\delta_{a'b'}\ ,\qquad\langle \tilde\phi_{\mu}(\lambda)|\tilde\phi_{\nu}(\lambda)\rangle=\delta_{\mu\nu}\ ,\qquad\langle \tilde\phi_{a'}(\lambda)|\tilde\phi_{\mu}(\lambda)\rangle=0\ .
\ee
The $\tilde\phi_{a'}$ have conformal dimensions of $(h+\delta h_{a'}(\lambda), \delta h_{a'}(\lambda))$.  The energies of the unperturbed states $\phi_{a'}$ and the perturbed states $\tilde\phi_{a'}$ are therefore
\be
E=h+\bar h=h\ , ~~~\tilde E_{a'}= h + 2\delta h_{a'}(\lambda)\ ,
\ee 
We expand the perturbed energies as  
\be
\tilde E_{a'}(\lambda)=E+\lambda E^{(1)}_{a'}+\lambda^2 E^{(2)}_{a'}+\cdots\ ,
\ee
\be
\tilde E_{\mu'}=E_{\mu'}+\lambda E^{(1)}_{\mu'}+\lambda^2 E^{(2)}_{\mu'}+\cdots \ .
\ee

Let us now consider the expansions of operators themselves. We can write
\bea
\tilde\phi_{a'}(\lambda)&=&\tilde C_{a'a}(\lambda)\phi_a+\tilde D_{a'\mu}(\lambda)\phi_\mu\ ,\nn
\tilde\phi_{\mu'}(\lambda)&=&\tilde F_{\mu'a}(\lambda)\phi_a+\tilde G_{\mu'\nu}(\lambda)\phi_\nu\ ,
\eea
where $\tilde C_{a'a}$, $\tilde D_{a'\mu}$, $\tilde F_{\mu'a}$, and $\tilde G_{\mu'\nu}$ are  $\lambda$-dependent expansion coefficients. We can invert these expansions to write
\bea
\phi_{a}&=&C_{aa'}(\lambda)\tilde \phi_{a'}(\lambda)+D_{a\mu'}(\lambda)\tilde \phi_{\mu'}(\lambda)\ ,\nn
\phi_{\mu}&=&F_{\mu a'}(\lambda)\tilde \phi_{a'}(\lambda)+G_{\mu\nu'}(\lambda)\tilde \phi_{\nu'}(\lambda)\ .
\eea
Finally, we expand the coefficients above in powers of $\lambda$: 
\bea
C_{aa'}(\lambda)&=&C^{(0)}_{aa'}+\lambda C^{(1)}_{aa'}+\lambda^2 C^{(2)}_{aa'}+\dots\nn
D_{a\mu'}(\lambda)&=&D^{(0)}_{a\mu'}+\lambda D^{(1)}_{a\mu'}+\lambda^2 D^{(2)}_{a\mu'}+\dots
\eea
Thus, in particular $\phi_a$ can be expanded as
\be\label{phiphi}
\phi_a=C^{(0)}_{aa'}\tilde\phi_{a'}+\lambda C^{(1)}_{aa'} \tilde\phi_{a'}+\lambda^2 C^{(2)}_{aa'}  \tilde \phi_{a'}+\cdots +\lambda D^{(1)}_{a\mu} \tilde \phi_\mu +\cdots\ .
\ee
The condition (\ref{none}) gives at leading order
\be
C^{(0)}_{aa'} C^{(0)*}_{ba'}=\delta_{ab}\ .
\label{nfive}
\ee 

The reason all these preliminaries are needed is that when computing an amplitude in  perturbation theory we find ourselves in the following situation. The operators in the amplitude  are taken to be the unperturbed operators $\phi_a, \phi_\mu$, since these are the ones with well understood and explicit constructions. But the operators that have well defined scaling dimensions are the $\tilde\phi_{a'}, \tilde\phi_{\mu'}$, which are not explicitly known. Thus we would compute an amplitude of the type
\be
A_{ab}(T)\equiv\Big\langle\phi_b({\tfrac T 2})\Big|e^{-(H^{(0)}+\delta H(\lambda)) T}\Big| \phi_a(-{\tfrac T 2})\Big\rangle\ .
\label{ccone}
\ee
Here the operators $\phi_a, \phi_b$ are operators in the unperturbed theory, and therefore explicitly known to us. But these unperturbed operators do not give eigenstates of the full Hamiltonian  $H^{(0)}+\delta H(\lambda)$; the latter eigenstates correspond to the perturbed operators $\tilde\phi_{a'}, \tilde\phi_{\mu'}$. Thus we have
\bea
\langle\tilde\phi_{b'}({\tfrac T 2})\Big|e^{-(H^{(0)}+\delta H(\lambda)) T}\Big| \tilde\phi_{a'}(-{\tfrac T 2})\Big\rangle&=&e^{-\tilde E_{a'} T}\delta_{a'b'}\ ,\nn
\langle\tilde\phi_{\nu'}({\tfrac T 2})\Big|e^{-(H^{(0)}+\delta H(\lambda)) T}\Big| \tilde\phi_{\mu'}(-{\tfrac T 2})\Big\rangle&=&e^{-\tilde E_{\mu'} T}\delta_{\mu'\nu'}\ ,\nn
\langle\tilde\phi_{a'}({\tfrac T 2})\Big|e^{-(H^{(0)}+\delta H(\lambda)) T}\Big| \tilde\phi_{\mu'}(-{\tfrac T 2})\Big\rangle&=&0\ .
\eea

Substituting the expansions (\ref{phiphi}) in eq. (\ref{ccone}), we find
\bea\label{AabT}
&&A_{ab}(T)\equiv\Big\langle\phi_b({\tfrac T 2})\Big|e^{-(H^{(0)}+\delta H(\lambda)) T}\Big| \phi_a(-{\tfrac T 2})\Big\rangle \nn
&&\!\!=\Big(C^{(0)*}_{ba'}+\lambda C^{(1)*}_{ba'}+\lambda^2 C^{(2)*}_{ba'} +\dots\Big)\Big(C^{(0)}_{aa'}+\lambda C^{(1)}_{aa'}+\lambda^2 C^{(2)}_{aa'} +\dots\Big)e^{-(E+\lambda E^{(1)}_{a'}+\lambda^2 E^{(2)}_{a'}+\dots )T}\nn
&&\!\!+\,\lambda^2 D^{(1)*}_{b\mu}D^{(1)}_{a\mu}e^{-(E_\mu+\lambda E^{(1)}_{\mu}+\lambda^2 E^{(2)}_{\mu}+\dots )T} +\cdots\ .
\eea
In general amplitudes like $A_{ab}(T)$ are functions of the fields like $\phi_a, \phi_b$ placed at the upper and lower time slices, the time interval $T$ between the slices, and the coupling $\lambda$. From the set of such amplitudes, we can extract the perturbed dimensions of the theory.  We will do this below, but first we note that it is convenient  to expand the above amplitude  in powers of $\lambda$ 
\be\label{Aab_define}
A_{ab}(T)=A^{(0)}_{ab}+\lambda A^{(1)}_{ab} +\lambda^2 A^{(2)}_{ab}+\cdots\ .
\ee

\b

(d) We first look at the coefficient of $\lambda Te^{-ET}$ in (\ref{AabT}). This coefficient is found to be
\be
-C^{(0)}_{aa'}E^{(1)}_{a'} C^{(0)*}_{ba'}\ .
\ee
We can write the above relation in matrix form, defining $(\hat A^{(1)})_{ab}=A^{(1)}_{ab}$, $(\hat C^{(0)})_{aa'}=C^{(0)}_{aa'}$, and $(\hat{E}^{(1)})_{a'b'}=\delta_{a'b'}  E^{(1)}_{a'}$. This gives
\be
 \hat A^{(1)}  \r - T e^{-ET} \hat C^{(0)} \hat{E}^{(1)} \hat C^{(0)\dagger}\ .
\label{bone_i}
\ee
 where the arrow indicates that we are writing only the coefficient of $Te^{-ET}$ in $ \hat A^{(1)} $.

We now note that, in our problem, the amplitude $A_{ab}(T)$ has no terms at $O(\lambda)$. This is because, as we will see in the next subsection, the deformation operator $D$ which perturbs the theory away from the orbifold point is in the twist 2 sector, while the states $|\phi_a\rangle$ and $|\phi_b\rangle$ are in the untwisted sector. The 3-point function $\langle\phi_b|D|\phi_a\rangle$ then vanishes due to the orbifold group selection rules. From eq. (\ref{nfive}) we see that $\hat C$ is unitary. Thus the vanishing of the above contribution tells us that $\hat{E}^{(1)}=0$; i.e. $E^{(1)}_{a'}=0$ for all $a'\in\{1, \dots,n\}$.

Now we look at  the coefficient of $\lambda^2 Te^{-ET}$ in $ A_{ab}(T)$ in eq. (\ref{AabT}). We find 
\be\label{A2abT}
 A^{(2)}_{ab} \r   - Te^{-ET} \left ( C^{(0)}_{aa'}C^{(0)*}_{ba'} E^{(2)}_{a'}\right )\ .
\ee
In matrix form, this reads 
\be
 \hat A^{(2)}  \r - T e^{-ET} \hat C^{(0)} \hat { E^{(2)}} \hat C^{(0)\dagger}\ .
\label{bone}
\ee
Thus, if we compute the matrix $\hat A^{(2)}$ and look at the coefficient of $-Te^{-ET}$,  then the eigenvalues of this matrix give the corrections to the energies upto $O(\lambda^2)$ :
\be\label{E2T}
 E^{(2)}_{a'}=2\delta h_{a'}\ ,
\ee
 and the eigenvectors give the linear combinations of the $\phi_a$ which correspond to operators with definite conformal dimensions\footnote{We note that, as mentioned below eq. (\ref{unit_norm}), $\delta\tilde \phi$ defined in eq. (\ref{phiphi_i}) does not appear in the expectation value up to second order in perturbation theory. $\delta\tilde\phi$ corresponds to the terms with the $C^{(i)}$ and $D^{(i)}$ ($i\in\mathbb Z_{>0}$) coefficients in eq. (\ref{phiphi}) and do not appear at the first and second order amplitudes in eq.s (\ref{bone_i}) and (\ref{bone}), respectively.\label{fn_bone}}.

\b

(e) In our system, we  have states $|\Phi^{(m)}\rangle $  labelled by a parameter $m\in\mathbb Z_{\ge0}$, see eq. (\ref{threeq}).  As we go to higher $m$, the number of states with the same conformal dimensions as $|\Phi^{(m)}\rangle $ increases; in fact even for the lowest interesting value, $m=2$, the number of degenerate states is large enough to make the computation of the matrix $A^{(2)}_{ab}$ difficult. We will be interested in computing something a bit different. The state $|\Phi^{(m)}\rangle $ of interest to us is one of the states $|\phi_a\rangle$; let us call it $|\phi_1\rangle$. Then we compute the quantity $A^{(2)}_{11}$ in eq. (\ref{A2abT}). From (\ref{bone}) we see that the coefficient of $-Te^{-ET}$ in $\hat A^{(2)}$ is
\be
\sum_{a'} |C_{1a'}|^2 \, E^{(2)}_{a'}= \sum_{a'} |\langle \tilde\phi_{a'}|\phi_1\rangle|^2 E^{(2)}_{a'}\ .
\label{nfour}
\ee
Thus we get the {\it expectation value} of the increase in energy for the  state $|\phi_1\rangle = |\Phi^{(m)}\rangle $.  Computing this quantity will allow us to make our arguments about the nature of lifting of string states.

\subsection{The general expression for lifting at second order}\label{subsec_Aab}
In the above discussion we have expressed the amplitude $A_{ab}$ in eq. (\ref{Aab_define}) in terms of Hamiltonian evolution. But  we will actually compute $A_{ab}$ using path integrals, since the perturbation is known as a change to the   Lagrangian rather than a change to the Hamiltonian:
\be
S_0\r S_{\mathrm{pert}}=S_0+\lambda \int d^2 w D(w, \bar w)\ ,
\label{assevent}
\ee
where $D(w,\bar w)$ is an exactly marginal operator deforming the CFT. As mentioned before, $A^{(1)}_{ab}=0$, see the discussion below eq. (\ref{bone_i}). We will work with the next order, where we have
\be
A^{(2)}_{ab}(T)=\h \bigg\langle \phi_b({\tfrac T2})\bigg|\bigg(\int d^2w_2  D(w_2, \bar w_2)\bigg)\bigg(\int d^2w_1  D(w_1, \bar w_1)\bigg)\bigg|\phi_a(-{\tfrac T2})\bigg\rangle\ ,
\label{wnasel}
\ee
where the range of the $w_i$ integrals are
\be
0\le \sigma_i<2\pi\ , ~~~-{\tfrac T 2} <\tau_i<{\tfrac T 2}\ .
\label{nrange}
\ee

Before proceeding, we write the initial and final states in a convenient form. The local operators like $J^a(w)$ in the CFT depend on time $\tau$. Thus if we create the state $|\psi_a\rangle$ by the application of such an operator, then the value of $\tau$ at the point of application is relevant. But if we expand in modes $J^a(w)=\sum_n J^a_n e^{nw}$, then the operators $J^a_n$ do not have the information about the point of application. It is convenient to write the state in terms of mode operators like $J^a_n$, and so we need to factor out the $\tau$-dependence explicitly. 

For $\tau<-{T\over 2}$, the state is the NS vacuum $|0\rangle$. Suppose that the state created at $\tau=-{T\over 2}$ has energy $E$.  Then we have
\be
\big|\phi(-{\tfrac T2})\big\rangle=e^{-{ET\over 2}}|\Phi\rangle\ ,
\ee
where the state $ |\Phi\rangle$ is written with upper case letters: this will denote the fact that this state is made from modes like $J^a_n$ which have no $\tau$-dependence. 
Similarly, the final state  is
\be
\big\langle \phi({\tfrac T2})\big|=e^{-{ET\over 2}} \langle \Phi|\ .
\ee
Eq. (\ref{wnasel}) then reads:
\be
A^{(2)}_{ab}(T)=\h e^{-ET}\bigg\langle \Phi_b({\tfrac T2})\bigg|\bigg(\int d^2w_2  D(w_2, \bar w_2)\bigg)\bigg(\int d^2w_1  D(w_1, \bar w_1)\bigg)\bigg|\Phi_a({-\tfrac T2})\bigg\rangle\ .
\label{wnasel_ii}
\ee

To compute  $A^{(2)}_{ab}$,  we proceed as follows:

\b

(a) Since $\Phi_a$ has the same energy as $\Phi_b$, the integrand depends only on 
\be
\Delta w = w_2-w_1\ .
\ee
It would be convenient if we could write the integrals over $w_1, w_2$ as an integral over $\Delta w$, and factor out the integral over 
\be
s=\h (w_1+w_2)\ .
\ee
We cannot immediately do this, however, as the ranges of the $\tau_i$ integrals given in  (\ref{nrange}) do not factor into a range for $\Delta w$ and a range for $s$. But for our case, we will see that we {\it can} obtain the needed factorization by taking the limit $T\r \infty$. 

Suppose that $w_1<w_2$. In the region $-{T\over 2}<\tau<\tau_1$, we have the state $\Phi_a$, and Hamiltonian evolution gives the factor $\sim e^{-E\tau}$. Similarly, in the region $\tau_2<\tau<{T\over 2}$, we have the state $\Phi_b$, and Hamiltonian evolution  gives $\sim e^{-E\tau}$. In the region $\tau_1<\tau<\tau_2$, we have a state $\Phi_k$ with energy $E_k$, giving a factor $\sim e^{-E_k\tau}$. As we will show below, we have
\be
E_k\ge E+2\ ,
\label{ntwo}
\ee
so that the integrand in $A^{(2)}_{ab}$ in eq. (\ref{wnasel_ii}) is exponentially suppressed as we increase $\Delta w$. Thus  we can fix $w_1=0$, and integrate over $w_2\equiv w$ to compute
\be
\bigg\langle \phi_b({\tfrac T2})\bigg|\bigg(\int d^2w  D(w, \bar w)\bigg) \,  D(0)\bigg|\phi_a(-{\tfrac T2})\bigg\rangle\ .
\label{wnaselqqpre}
\ee
Here the $w$ integral ranges over $0\le \sigma<2\pi, -{T\over 2}<\tau<{T\over 2}$. The $\tau$  range is large  in the limit $T\r\infty$. But the contributions to the integral die off quickly for $|w|\gg 2\pi$. Integration over $w_1$ then just gives a factor
\be
\int d^2 w_1 \r 2\pi T\ .
\ee
Thus in the limit $T\r\infty$, eq. (\ref{wnasel_ii}) reads
\be
 A^{(2)}_{ab}(T)=(2\pi T)\,\h\,e^{-ET}\bigg\langle\Phi_b({\tfrac T2})\bigg|\bigg(\int d^2w  D(w, \bar w)\bigg)\,D(0)\bigg|\Phi_a(-{\tfrac T2})\bigg\rangle\ .
\label{wnaselqq}
\ee
To prove eq. (\ref{ntwo}), we note that $E_k$ must lie in the conformal block of some primary operator $\chi$ with dimensions $(h_\chi, \bar h_\chi)$. Thus, we need a non-vanishing 3-point function
\be
f=\langle \phi_a (z_1) \, D(z_2, \bar z_2) \, \chi(z_3, \bar z_3)\rangle\ .
\ee
Since $\phi_a$ has dimensions $(h,0)$, there is no power of $\bar z_2-\bar z_1$ in the correlator. Since $D$ has dimensions $(h_D,\bar h_D)=(1,1)$, this implies that $\bar h_\chi=1$. Further, since the $w_i$ are integrated over the spatial coordinates $\sigma_i$ with no phase, the state $\phi_k$ must have the same spin as $\phi_a$; i.e. $h_\chi-\bar h_\chi=h$. Thus we have
\be
(h_\chi, \bar h_\chi)=(h+1,1)\ .
\ee
and the lowest state $E_k$ corresponding to such a primary has $E_k=E+2$. If we have a descendent of this lowest state, then we have $E_k>E+2$. Thus we obtain (\ref{ntwo}).

\b

(b) Now consider the integrand of $A^{(2)}_{ab}$ in eq. (\ref{wnaselqq}). We have the correlation function
\be
\langle \Phi_a | \, D(w, \bar w) \, D(0)\, | \Phi_b\rangle\ .
\ee
The right-moving dimensions of $\Phi_a$ and $\Phi_b$ are zero, so the antiholomorphic part of this correlator is $\langle 0 | \, D( \bar w) \,  D(0)\, | 0\rangle$. Since $\bar h_D=1$, we find
\be\label{rightsigmasigma}
\langle 0 | \, D( \bar w) \,  D(0)\, | 0\rangle= {C_1\over  \sinh^2 ({\bar w\over 2})}
\ee
for some constant $C_1$. The left moving part is more complicated and we will calculate it in later sections. But this part also has the same singularity as the right movers when the two $D$ operators approach. So the full correlator will have the form 
\be
\langle \Phi_a | \, D(w, \bar w) \, D(0)\, | \Phi_b\rangle={Q_{ab}(w)\over  \sinh^2({w\over 2})}\,{1\over \sinh^2({\bar w\over 2})}\ .
\label{bbtwo}
\ee
We define
\be
X_{ab}(T)\equiv\int d^2w  {Q_{ab}(w)\over   \sinh^2  ({w\over 2})}\,{1\over  \sinh^2 ({\bar w\over 2})}\ .
\label{wnaselqa}
\ee
The amplitude (\ref{wnaselqq}) then reads
\be
 A^{(2)}_{ab}(T)=\h\,(2\pi T)\,e^{-ET}\,X_{ab}(T)\ .
\label{nthree}
\ee

\begin{figure}
\begin{center}
\includegraphics[width=35mm]{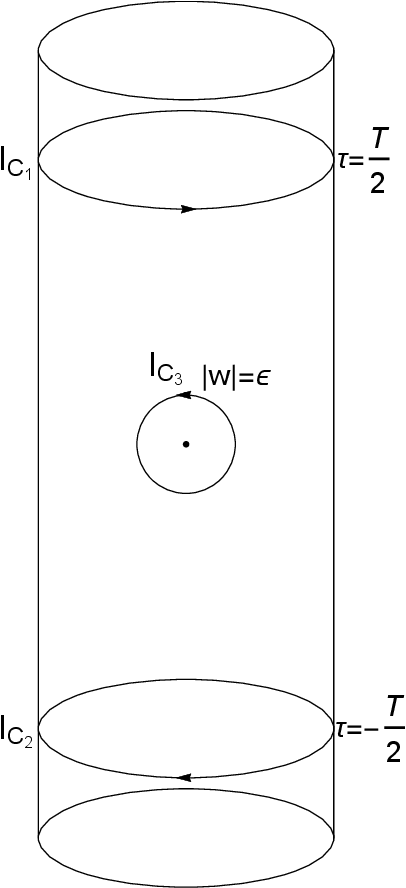}
\end{center}
\caption{Here we show the locations of the three boundary contour integrals, $I_{C_1},I_{C_2}$, and $I_{C_3}$ given in equations (\ref{IC1}), (\ref{IC2}), and (\ref{IC3}) respectively.}
\label{contours}
\end{figure}

\b

(c) To evaluate eq. (\ref{wnaselqa}) we write
\bea
 X_{ab}(T)&=& \int d^2w  {Q_{ab}(w)\over   \sinh^2  ({w\over 2})}\,{1\over  \sinh^2 ({\bar w\over 2})}\nn
&=& -2\int d^2w  {Q_{ab}(w)\over   \sinh^2  ({w\over 2})}\left ( \p_{\bar w}\coth(\tfrac{\bar w}2)\right ) \nn
&=& -2\int d^2w \, \p_{\bar w} \left (  {Q_{ab}(w)\over   \sinh^2  ({w\over 2})}\,  \coth(\tfrac{\bar w}2)\right ) \nn
&=&  i \int_C dw \,  \left (  {Q_{ab}(w)\over   \sinh^2  ({w\over 2})}\,  \coth({\tfrac{\bar w} 2})\right )\ ,
\label{wnaselqb}
\eea
where in the last line we have used the divergence theorem in complex coordinates. The boundary integral is defined over a contour $C$ consisting of three parts:

\b

(i) $C_1$: the upper boundary of the integration range at $\tau={T\over 2}$. This integral, which we call $I_{C_1}$, runs in the direction of positive $\sigma$. Note that $i(dw)=i(id\sigma)=-d\sigma$ and we have
\be
I_{C_1}=-\int_{0}^{2\pi}  d\sigma \left (  {Q_{ab}(w)\over   \sinh^2  ({w\over 2})}\,  \coth({\tfrac{\bar w} 2})\right )\ .
\label{IC1}
\ee

\b

(ii) $C_2$: the lower boundary of the integration range at $\tau=-{T\over 2}$. The contour runs in the direction of negative $\sigma$. The integral is called $I_{C_2}$ and has the form 
\be
I_{C_2}=\int_{0}^{2\pi}  d\sigma \left (  {Q_{ab}(w)\over   \sinh^2  ({w\over 2})}\,  \coth({\tfrac{\bar w} 2})\right)\ .
\label{IC2}
\ee

\b

(iii) $C_3$: An integral over a small circle of radius $\epsilon$ around the origin where we have the operator $D(0)$. The contour here runs clockwise as this is an inner boundary of the integration domain. We therefore write it as
\be
I_{C_3}=-i\int_{|w|=\epsilon} dw \left (  {Q_{ab}(w)\over   \sinh^2  ({w\over 2})}\,  \coth({\tfrac{\bar w} 2})\right )
\label{IC3}
\ee
where now the integral runs in the usual anticlockwise direction. 
The integral over $C_3$ contains divergent contributions from the appearance of operators with dimension $h+\bar h \le 2$ in the OPE $D(w, \bar w) D(0,0)$. These divergences have to be removed by adding counterterms terms to the action. Thus we get
\be
I_{C_{3}}+I_{C_3,\,\mathrm{counterterm}}= I_{C_3,\,\mathrm{renormalized}}\ .
\ee
Then eq. (\ref{wnaselqb}) reads
\be
X_{ab}(T)=I_{C_1}+I_{C_2}+I_{C_3,\,\mathrm{renormalized}}\ .
\label{bsix}
\ee
Fig. \ref{contours} shows the locations of the three contours.
\b

(iv) Let us now summarize the above discussion. As mentioned in section \ref{outline}(d), we compute $A^{(2)}_{11}\equiv A^{(2)}$ for just one state $|\Phi_1\rangle$, see eq. (\ref{nfour}). This will give the expectation value of the increase in energy of $|\Phi_1\rangle$. From eq.s (\ref{nthree}) and (\ref{bsix}) we obtain
\be\label{A2T_i}
A^{(2)}(T)=\pi T e^{-ET} \left ( I_{C_1}+I_{C_2}+I_{C_3,\,\mathrm{renormalized}}\right )\ .
\ee
Finally, for our state $|\Phi_1\rangle$, the lift in the expectation value of the energy is given by the coefficient of $-Te^{-ET}$ in the limit $T\to\infty$, see eq.s (\ref{bone}) and (\ref{E2T}). Thus, we find
\be
\langle E^{(2)}\rangle=-\pi \lim_{T\r\infty} X_{ab}(T)=-\pi \lim_{T\r\infty}\left(I_{C_1}+I_{C_2}+I_{C_3,\,\mathrm{renormalized}}\right)\ .
\label{deltaE}
\ee

\section{Setting up the computation}\label{section3}
\subsection{The deformation operator}\label{sebsec_deformation}
The orbifold CFT describes the system at its `free' point in moduli space. To move towards the supergravity description, we deform the orbifold CFT by adding a deformation operator $D$, as noted in (\ref{assevent}).

To understand the structure of $D$ we recall that the orbifold CFT contains `twist' operators. Twist operators can link any number $k$ out of the $N$ copies of the CFT together to give a $c=6$ CFT living on a circle of length $2\pi k$ rather than $2\pi$.    We will call such a set of linked copies a `component string' with winding number $k$.

The deformation operator contains a twist of order $2$. The twist itself carries left and right charges $j=\pm \h, \bar j=\pm \h$ \cite{lm2}. Suppose we start with both these charges positive; this gives the twist $\sigma_2^{++}$. Then the deformation operators in this twist sector have the form
\be\label{exactlymarginal}
D=P^{\dot A\dot B}\hat O_{\dot A\dot B}= P^{\dot A \dot B}G^-_{\dot A, -\h}\bar G^-_{\dot B, -\h} \sigma^{++}_2\ .
\ee
Here $P^{\dot A \dot B}$ is a polarization. We will later choose
\be
P^{\dot A \dot B}=\epsilon^{\dot A \dot B}
\ee
where $\epsilon^{+-}=-1$. This choice gives a deformation carrying no charges.

We will omit the subscript $2$ on the twist operator from now on, and will also consider its holomorphic and antiholomorphic parts separately. 
We normalize the twist operator as
\be
\sigma^{-}(z )\sigma^{+}(z')\sim {1\over (z-z')}\ .
\label{asone}
\ee
We note that \cite{acm1,Avery:2010qw}
\be
G^-_{\dot A, -\h}\sigma^+=-G^+_{\dot A, -\h} \sigma^-\ .
\label{eeone}
\ee
It will be convenient to write one of the two deformation operators as $G^-_{\dot A, -\h}\sigma^+$ and the other as $-G^+_{\dot C, -\h} \sigma^-$. We will make this choice for both the left and right movers, so  the negative sign in (\ref{eeone}) cancels out. Thus on each of the left and right sides we  write one deformation operator in the form $G^-_{\dot A, -\h}\sigma^+$ and the other in the form  $G^+_{\dot A, -\h} \sigma^-$.

From (\ref{asone}) we find that on the cylinder
\be
\langle 0 | \sigma^-(w_2) \sigma^+(w_1)|0\rangle = {1\over 2 \sinh ({\Delta w\over 2})}
\label{axthree}
\ee
where
\be
\Delta w = w_2-w_1\ .
\ee

\subsection{The states}

We start by looking at a CFT with $N=2$; i.e., we have two copies of the $c=6$ CFT (we will consider general values of $N$ in section \ref{section7}). The vacuum $|0\rangle$ with $h=j=0$ is given by two singly-wound copies of the CFT, i.e. there is no twist linking the copies, and the fermions on each of the copies are in the NS sector. Thus we can write
\be
|0\rangle=|0\rangle^{(1)}\,|0\rangle^{(2)}\ ,
\ee
where the superscripts indicate the copy number.

We consider one of the copies to be excited by the application of $m$ R-current operators. The orbifold symmetry requires that the state be symmetric between the two copies, so the state we take is
\bea
|\Phi^{(m)}\rangle&=&{1\over \sqrt{2}}\,
\Big(J^{+(1)}_{-(2m-1)}\dots J^{+(1)}_{-3}J^{+(1)}_{-1} ~+~  J^{+(2)}_{-(2m-1)}\dots J^{+(2)}_{-3}J^{+(2)}_{-1}\Big)|0\rangle\nn
&\equiv& |\Phi^{(m)}\rangle^{(1)}~+~|\Phi^{(m)}\rangle^{(2)}\ ,
\label{threex}
 \eea
where in $|\Phi^{(m)}\rangle^{(i)}$ the excitations act on copy $i$. This state has
\be
(h,\bar h)=(m^2,0),\qquad(j,\bar j)=(m,0)\ .
\ee
The energy of the state is 
\be
E\equiv h+\bar h  = m^2
\label{assix}
\ee
and its momentum is 
\be
P\equiv h-\bar h =m^2\ .
\label{asfive}
\ee
The final state is the conjugate of the initial state
\be
\langle \Psi^{(m)}|={1\over \sqrt{2}}\;\langle 0|\Big( J^{-(1)}_{1}J^{-(1)}_{3}\dots J^{-(1)}_{(2m-1)}+
J^{-(2)}_{1}J^{-(2)}_{3}\dots J^{-(2)}_{(2m-1)}\Big)\ . 
\label{threexq}
\ee

\section{The vacuum correlator}\label{section4}

\begin{figure}
\begin{center}
\includegraphics[width=38mm]{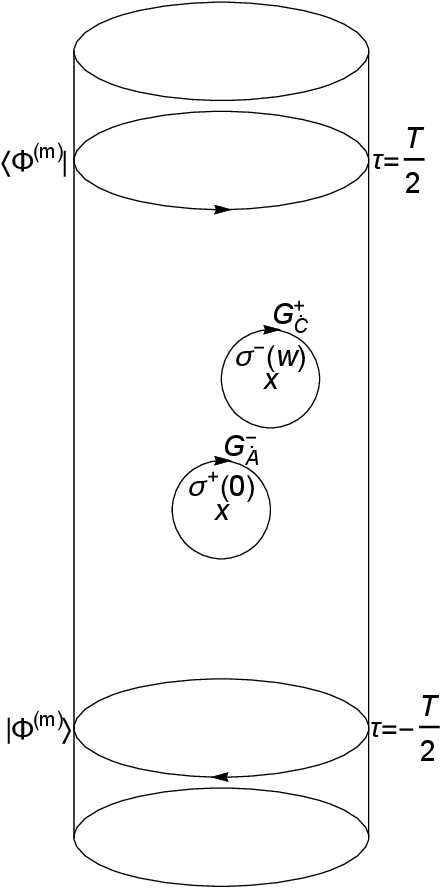}
\end{center}
\caption{The cylinder with the locations of the initial state $|\Phi^{(m)}\rangle$ at $\tau = -{T\over2}$, the final state $\langle\Phi^{(m)}|$ at $\tau={T\over2}$, and the two deformation operators  $G^-_{\dot{A}}\sigma^+$ and $G^+_{\dot{C}}\sigma^-$ at $w_1=0$ and $w_2=w$, respectively.}
\label{cylinderstate}
\end{figure}

As a first step, we compute the vacuum to vacuum correlator
\be
T_{\dot C\dot A}(w_2,w_1)=\langle0|\big(G^+_{\dot C, -\h}\sigma^-(w_2)\big)\,\big(G^-_{\dot A, -\h}\sigma^+(w_1)\big)\big|0\rangle\ .
\label{adoneq}
\ee
The complex conjugate of this correlator will give the right moving part of the correlator of $A^{(2)}(T)$, see eq. (\ref{wnaselqq}). Fig.\ref{cylinderstate} represents the full state on the cylinder.

\subsection{The map to the covering space}\label{map}

To compute the vacuum amplitude $T_{\dot C\dot A}(w_2,w_1)$ in eq. (\ref{adoneq}) we first map the cylinder labeled by $w$ to the complex plane labeled by $z$:
\be
z=e^w\ .
\ee
We then map this plane to its covering space where the twist operators are resolved, see \cite{lm1} for the details of the covering space analyses. We consider the map
 \be
z={(t+a)(t+b)\over t}\ .
\label{amap}
\ee
We have
\be
{dz\over dt} = 1-{ab\over t^2}\ .
\ee
The twist operators correspond to the locations given by ${dz\over dt}=0$; i.e. the points
\bea
&&t_1=-\sqrt{ab}\ ,\qquad z_1=e^{w_1}=(\sqrt{a}-\sqrt{b})^2\ ,\\
&&t_2=\sqrt{ab}\ ,\qquad\;\;\, z_2=e^{w_2}=(\sqrt{a}+\sqrt{b})^2\ .
\eea
Note that
\be
{dz\over dt} = {(t-t_1)(t-t_2)\over t^2}\ .
\ee
We define
\be
\Delta w = w_2-w_1\ ,
\ee
\be
s={\h}(w_1+w_2)\ ,
\ee
Then we find
\be
a=  e^{s} \cosh^2 (\tfrac{\Delta w }{4}), ~~~b=  e^{s} \sinh^2 (\tfrac{\Delta w }{4})\ .
\ee
It will be useful to note the relations
\be
a-b=e^{s}, ~~~z_1z_2=e^{2s}, ~~~z_1-z_2= -2 e^{s} \sinh (\tfrac{\Delta w }{2})\ .
\ee

\subsection{The `base' amplitude}\label{base}

To compute the vacuum correlator $T_{\dot C\dot A}(w_2,w_1)$ in eq. (\ref{adoneq}), we start by computing
\be
U(w_2,w_1)\equiv\langle 0|  \sigma^-(w_2)\,  \sigma^+(w_1) |0\rangle\ .
\label{axone}
\ee
We call this the `base' amplitude since each correlator we compute will have this structure of twist operators, and the only extra elements will be local operators with no twist. 

The computation of correlators like (\ref{axone}) is discussed in \cite{lm1,lm2}. We briefly summarise the computation by proceeding in the following steps:

\b

(i) We already mapped the cylinder labelled by the coordinate $w$ to the complex plane labelled by the coordinate $z$ in subsection \ref{map}, through the map $z=e^w$. We then mapped the plane to the covering space, labelled by the coordinate $t$, though the map (\ref{amap}). These maps generate a Liouville factor since the curvature of the covering space is different from the curvature on the cylinder, and the change of curvature changes the partition function due to the conformal anomaly of the CFT. Let this Liouville factor be
\be
L\,[z_1,z_2]\ .
\label{axtwo}
\ee

\b

(ii) The twist operator $\sigma^+(w_1)$ has left-moving dimension $h=\h$ and transforms to the plane as
\be\label{sigmaplus}
\left ( {dz\over dw}(z_1)\right ) ^\h  \sigma^+(z_1)=z_1^\h \sigma^+(z_1)\ .
\ee
Now consider the map to the cover. A twist $\sigma^+(z=0)$ on the plane $z$ transforms to a spin field ${S}^+(t=0)$ on the covering space\footnote{This is because fermionic fields have different boundary conditions in the odd versus even twisted sectors of the symmetric orbifold. The NS sector fermions have the usual NS-type half-integer modes in the odd twisted sector, whereas in the even twisted sector they have Ramond-type integer modes. The spin fields account for the ground state energy in the Ramond (R) sector, see \cite[section 2.2 ]{lm2} for details.\label{fn_RNS}} under the map $z=t^2$. The spin field has left-moving dimension $h={1\over 4}$. In the map (\ref{amap}) we have
\be
z-z_1\approx C(t-t_1)^2\ ,\qquad C=-{1\over \sqrt{ab}}\ ,
\ee 
so we have an extra scaling factor $\sqrt{C}$ furnishing $t$ compared to the standard map $z-z_1\approx (t-t_1)^2$, see \cite{lm2}. Combining with eq. (\ref{sigmaplus}), we find that the twist 2 operator $\sigma^+(w_1)$ transforms to the spin field ${S}^+(t_1)$ on the covering surface with an overall factor
\be
z_1^\h\big(\sqrt{C}\big)^{1\over 4}=z_1^\h\left ( -{1\over \sqrt{ab}}\right )^{1\over 8}\ .
\ee

\b

(iii) Similarly, the operator $\sigma^-(w_2)$ transforms to the spin field ${S}^-(t_2)$ on the cover acquiring an overall factor
\be
z_2^\h \left ( {1\over \sqrt{ab}}\right )^{1\over 8}\ .
\ee

\b

(iv) At this stage we have on the $t$ plane the amplitude
\be
\langle0|{S}^-(t_2) \, {S}^+(t_1)|0\rangle\ .
\ee
As discussed in footnote \ref{fn_RNS}, the spin field ${S}^+(t_1)$ creates an R vacuum at $t_1$. We can make a spectral flow transformation around the point $t=t_1$ to map this R vacuum to the NS vacuum $|0\rangle$ which we have done several times in the previous chapters. As a reminder to the reader, under spectral flow by $\a$ units, the dimension, $h$, and the charge, $j$, transform like
\bea
h' = h +\alpha j +{c\a^2\over 24}\ , \qquad j' = j +{\a c\over12}\ ,
\eea
where $c$ is the central charge of the CFT. Consider an operator $\mathcal{O}(z)$ of charge $q$. Under spectral flow by $\a$ units at a point $z_0$, the operator transforms as
\bea\label{sf_O}
\mathcal{O}(z)\to (z-z_0)^{-\a q}O(z)\ .
\eea

The NS vacuum is equivalent to no insertion on the covering space at all, so we would have taken all the effects of the twist into account. The spectral flow parameter needed is $\alpha=-1$, and we obtain
\be
{S}^+(t_1)|0\rangle \longmapsto |0\rangle\ .
\ee
Under such a spectral flow transformation, other fields in the $t$ plane pick up a factor as given in eq. (\ref{sf_O}). Thus, the field ${S}^-(t_2)$ (which has R-charge $j=-\h$) acquires a factor $(t_2-t_1)^{-{1\over 2}}$.

\b

(v) Now consider the spin field ${S}^-(t_2)$. We perform a similar spectral flow around the point $t=t_2$ with $\alpha=1$. This gives
\be
{S}^-(t_2)|0\rangle \longmapsto |0\rangle\ .
\ee
There are no other fields in the $t$ plane, so this time we get no additional factors from the spectral flow. 

\b

(vi) We now just have the $t$ plane with no insertions. The amplitude for this vacuum state is unity: it has been set to this value when defining the Liouville factor (\ref{axtwo}). Collecting all the factors (i)-(v) above, we obtain the amplitude $U(w_2,w_1)$ in eq. (\ref{axone}). 

\b

While we can compute $U(w_2,w_1)$ as outlined above, it turns out that we do not need to carry out these steps in this specific example: we already know the result from eq. (\ref{axthree})
\be
U(w_2,w_1)={1\over 2 \sinh({\Delta w\over 2})}\ .
\label{base_amplitude}
\ee
The reason we do not have to carry out the steps (i)-(v) explicitly here is that we have only two twist operators in our correlator; in this situation the factors from steps (i)-(v) can be absorbed in the normalization of the twists. However, if we have more than two twists then we do need to compute all factors explicitly.

Even though we can compute $U(w_2,w_1)$ without carrying out these steps, it is important to list the steps since when we have other excitations in the correlator then we will get additional factors from each of these steps. For later use, it will be helpful to also write the base amplitude (\ref{base_amplitude}) in alternative ways using the relations in section \ref{map}:
\be
U={z_1^\h z_2^\h\over (z_2-z_1)}={(a-b)\over 4 \sqrt{ab}}\ .
\label{bbone}
\ee

\subsection{The complete vacuum amplitude}\label{sect}

We now return to the computation of the vacuum amplitude $T_{\dot C\dot A}(w_2,w_1)$ defined in (\ref{adoneq}). Consider the operator
\be
 G^-_{\dot A, -\h} ={1\over 2\pi i} \int_{w_1} dw'_1 G^-_{\dot A}(w'_1)\ .
 \ee
 We proceed in the following steps:
 
 \b
 
 (i) We have
\bea
 {1\over 2\pi i} \int_{w_1} dw'_1 G^-_{\dot A}(w'_1)&=&{1\over 2\pi i} \int_{z_1} dz'_1 \Big({dz'_1\over dw'_1}\Big)^\h G^-_{\dot A}(z'_1)\\
 &=&{1\over 2\pi i} \int_{t_1} dt'_1\Big({dt'_1\over dz'_1}\Big)^\h\Big({dz'_1\over dw'_1}\Big)^\h G^-_{\dot A}(t'_1)\nn  
 &=&{1\over 2\pi i} \int_{t_1} dt'_1 (t'_1-t_1)^{-\h} (t'_1-t_2)^{-\h} {t'_1}^\h (t'_1+a)^\h (t'_1+b)^\h G^-_{\dot A}(t'_1)\ .\nonumber
\eea

\b

(ii) In section \ref{base} above we have seen that we perform a spectral flow around $t=t_1$ by $\alpha=-1$ and another one around $t=t_2$ by  $\alpha=1$. These flows give the factors
\be
G^-_{\dot A}(t'_1)\longmapsto(t'_1-t_1)^{-\h}(t'_1-t_2)^{\h} G^-_{\dot A}(t'_1)\ ,
\ee
see appendix \ref{app_sf}. Thus, we obtain
\bea
 {1\over 2\pi i} \int_{w_1} dw'_1 G^-_{\dot A}(w'_1) &\longmapsto& {1\over 2\pi i} \int_{t_1} dt'_1 (t'_1-t_1)^{-1} {t'_1}^\h (t'_1+a)^\h (t'_1+b)^\h G^-_{\dot A}(t'_1)\nn
 &=&\!\!\!{t_1}^\h (t_1+a)^\h (t_1+b)^\h G^-_{\dot A}(t_1)\ .
 \eea
 
\b

(iii) Similarly we have
\be
 {1\over 2\pi i} \int_{w_2} dw'_2 G^+_{\dot C}(w'_2)\longmapsto{t_2}^\h (t_2+a)^\h (t_2+b)^\h G^+_{\dot C}(t_2)\ .
 \ee
 
 \b
 
 (iv) Apart from the c-number factors in steps (i)-(iii), we have the $t$ plane correlator
 \be
\langle0|G^+_{\dot C}(t_2)\, G^-_{\dot A}(t_1) |0\rangle~=~\epsilon_{\dot C\dot A}\,{(-2)\over (t_2-t_1)^3}\ .
 \label{axtw}
 \ee
 
\b

(v) Collecting all the factors and noting the base amplitude (\ref{base_amplitude}), we find\footnote{Since there are fractional powers in the expressions here, the overall phase involves a choice of branch. But similar fractional powers appear in the right moving sector, and we can choose the signs as taken here with the understanding that we choose similar signs for the right movers.}
\bea
&&T_{\dot C\dot A}(w_2,w_1)=\left ( {t_1}^\h (t_1+a)^\h (t_1+b)^\h \right )\left ( {t_2}^\h (t_2+a)^\h (t_2+b)^\h \right )\times\nonumber\\
&&\qquad\qquad\qquad\qquad\qquad\qquad\qquad\qquad\qquad\times\left ( \epsilon_{\dot C\dot A}\,{(-2)\over (t_2-t_1)^3}U(w_2,w_1) \right ) \nn
&&\qquad\qquad\quad\,\,=\epsilon_{\dot C\dot A}\,{(a-b)^2\over 16\,ab} \, = \,\epsilon_{\dot C\dot A}\,{1\over 4\sinh^2 ({\Delta w \over 2} )}\ .
\label{axeight}
\eea
The right moving part of the correlator in the integrand of $A^{(2)}(T)$ in eq. ({\ref{A2T_i}) is found by taking the complex conjugate of this expression and taking $\e_{\dot{C}\dot{A}}\to\e_{\dot{D}\dot{B}}$
\be
\big\langle0|\big(G^+_{\dot D, -\h}\sigma^-(\bar w_2)\big)~\big( G^-_{\dot B, -\h} \sigma^+(\bar w_1)\big)\big|0\rangle=
\epsilon_{\dot D\dot B}\,{1\over 4\sinh^2 ({\Delta \bar w \over 2} )}\ .
\ee

\section{Lifting of D1-D5-P states}\label{section5}

In this section we evaluate lifting of the D1-D5-P states (\ref{threex}). We compute the left part of the correlator appearing in the amplitude $A^{(2)}(T)$, see eq.s (\ref{wnaselqq}) and (\ref{A2T_i}). Analogous to (\ref{adoneq}), we define
\be
T^{(j)(i)}_{\dot C\dot A,m}(w_2,w_1)={}^{(j)}\big\langle\Phi^{(m)}\big|\big(G^+_{\dot C, -\h} \sigma^-(w_2)\big)\big(G^-_{\dot A, -\h}\sigma^+(w_1)\big)\big|\Phi^{(m)}\big\rangle^{(i)}\ ,
\label{adoneqq}
\ee
where the superscripts $(i),(j)$ indicate which of the two copies carries the current excitations.

\subsection{Computing $T^{(1)(1)}_{\dot C\dot A,m}(w_2,w_1)$}\label{spectralflowfactors}

Let us start by computing $T^{(1)(1)}_{\dot C\dot A,m}(w_2,w_1)$. We will see that this computation will automatically extend to yield all the $T^{(j)(i)}_{\dot C\dot A,m}(w_2,w_1)$ amplitudes.

The operator
\be
{\cal J}^{+,(m)}\equiv J^{+}_{-(2m-1)}\cdots J^{+}_{-3}J^{+}_{-1} 
\label{calj}
\ee
has quantum numbers $(h,\bar h)=(m^2,0)$ and $(j,\bar j)=(m,0)$. With the commutation relations given in (\ref{commutations_ii}), we find that ${\cal J}^+_m(z=0)$ generates a state with unit norm at $z=0$. The operator conjugate to ${\cal J}^+_m$ is
\be
{\cal J}^{-,(m)}\equiv J^{-}_{1}J^{-}_{3} \cdots J^{-}_{(2m-1)}\ .
\ee

We follow the same process by which we computed the amplitude $T_{\dot C\dot A}(w_2,w_1)$ in subsection \ref{sect}.  The initial and final states have been written in terms of operator modes and can therefore  be assumed to be placed at $\tau\r-\infty$ and $\tau\r \infty$, respectively. We first map the cylinder $w$ to the plane $z$. The currents in the initial state give the operator ${\cal J}^{+,(m)}(z=0)$ on copy $1$. The currents in the final state give the operator ${\cal J}^{-,(m)}(z=\infty)$, again on copy $1$.

Next we map to the $t$ plane via the map (\ref{amap}). The point $z=0$ for copy $1$ maps to $t=-a$. Thus, ${\cal J}^{+,(m)}(z=0)$ maps as
\be
{\cal J}^{+,(m)}(z=0) ~\longmapsto~ \Big({dt\over dz}\Big)^{m^2} {\cal J}^{+,(m)}(t=-a)=\Big({a\over a-b}\Big)^{m^2} {\cal J}^{+,(m)}(t=-a)\ .
\label{axtenq}
\ee
The point $z=\infty$ for copy $1$ maps to $t=\infty$. We have $z\sim t$ at $t=\infty$, so  the operator ${\cal J}^{-,(m)}(z=\infty)$ maps as
\be
{\cal J}^{-,(m)}(z=\infty) ~\longmapsto~ {\cal J}^{-,(m)}(t=\infty)\ .
\label{axelq}
\ee

We note that the state ${\cal J}^{+,(m)}|0\rangle$ is obtained by spectral flow of the vacuum $|0\rangle$ by $\alpha=2m$. Thus we can use the spectral flow by $\alpha=-2m$ to map ${\cal J}^{+,(m)}|0\rangle$ to the vacuum $|0\rangle$. We will use this trick to remove the insertion of ${\cal J}^{+,(m)}$ on the $t$ plane: this reduces the amplitude to the one we had for $T_{\dot C\dot A}(w_2,w_1)$ in eq. (\ref{axeight}). (Note that when we remove ${\cal J}^{+,(m)}$ from any point in the $t$ plane, we remove at the same time the operator ${\cal J}^{-,(m)}$ at infinity.)
\begin{figure}
\begin{center}
\includegraphics[width=69mm]{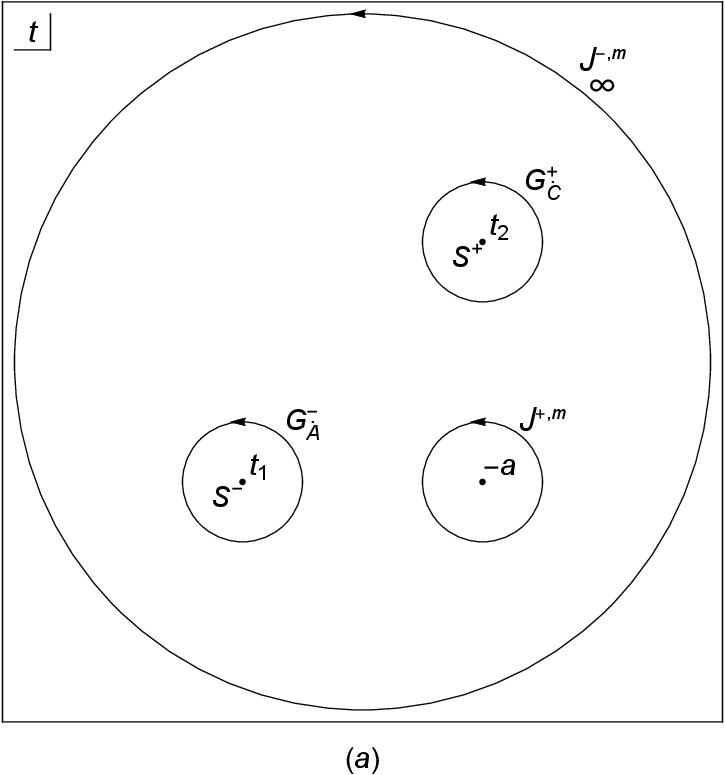}
\vspace{0.3cm}
\hspace{3mm}
\includegraphics[width=69mm]{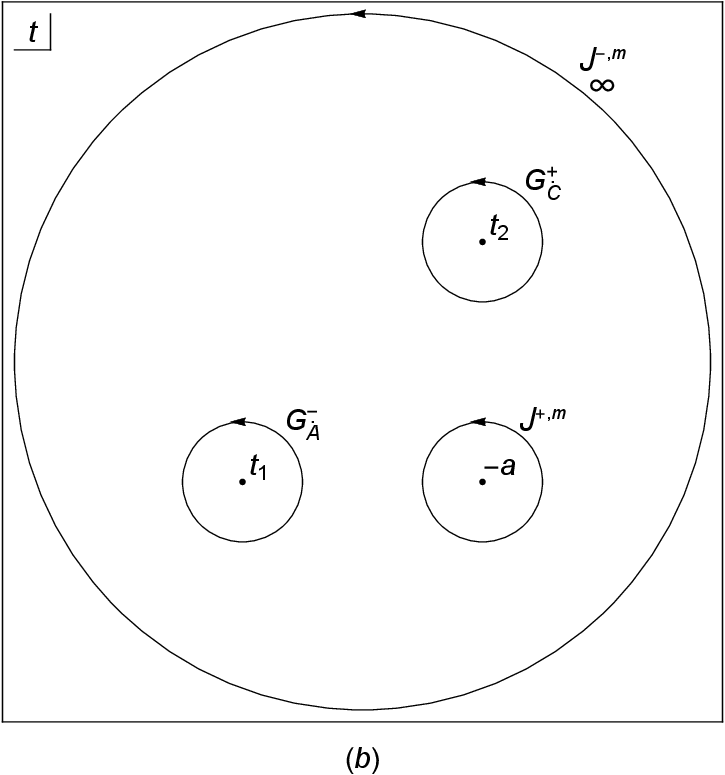}
\includegraphics[width=69mm]{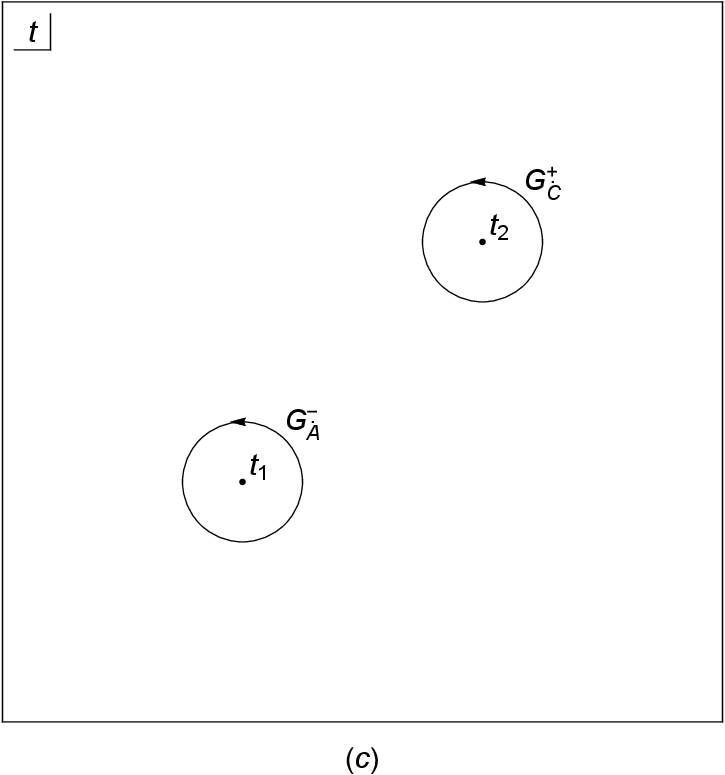}
\end{center}
\caption{(a) The $t$ plane with the spin fields $S^{+}(t_1)$ and $S^{-}(t_2)$, the $G^{-}_{\dot{A}}$ contour circling $t_1$, the $G^{+}_{\dot{C}}$ contour circling $t_2$, the $\mathcal{J}^{+,m}$ contour circling $t=-a$, and the $\mathcal{J}^{-,m}$ contour circling $t=\infty$. (b) The spin fields, $S^{+}(t_1)$ and $S^{-}(t_2)$ spectral flowed away. (c) The spin fields $S^{+}(t_1)$ and $S^{-}(t_2)$ and the currents $\mathcal{J}^{+,m}$ and $\mathcal{J}^{-,m}$ all spectral flowed away. All of the spectral flow factors are given in subsection \ref{spectralflowfactors}.}
\label{fig_spectralflow}
\end{figure}
Fig.\ref{fig_spectralflow} shows the covering space insertions for the unspectral flowed amplitude, the amplitude with only spin fields spectral flowed away, and the amplitude with both spin fields and currents spectral flowed away.

Let us note the extra factors we get in computing $T^{(1)(1)}_{\dot C\dot A,m}(w_2,w_1)$ in eq. (\ref{adoneqq}) as compared to $T_{\dot C\dot A}(w_2,w_1)$ in eq. (\ref{axeight}):

\b

(i) From (\ref{threex}) we see that the initial and final states $|\Psi^{(m)}\rangle^{(1)}$ and  ${}^{(1)}\langle \Psi^{(m)}|$ have a normalization ${1\over \sqrt{2}}$ each; this gives a factor
\be\label{eq_f1}
f_1=\left ( {1\over \sqrt{2}}\right )^2\ .
\ee

(ii) We have the factor  obtained in (\ref{axtenq}) when mapping  ${\cal J}^{+,(m)}$ from the $z$ to the $t$ plane:
\be\label{eq_f2}
f_2=\left({a\over a-b}\right)^{m^2}\ . 
\ee

\b

(iii) We perform a spectral flow by $\alpha=-1$ around the point $t=t_1$. Under this spectral flow, the operator ${\cal J}^{+,(m)}(t=-a)$ picks up the factor
\be\label{eq_f3}
f_3=(-a-t_1)^{m}\ .
\ee

\b

(iv) We perform a spectral flow by $\alpha=1$ around the point $t=t_2$. Under this spectral flow, the operator ${\cal J}^{+,(m)}(t=-a)$ picks up the factor
\be\label{eq_f4}
f_4=(-a-t_2)^{-m}\ .
\ee

\b

(v) We perform a spectral flow around $t=-a$ by $\alpha=-2m$. This gives
\be
{\cal J}^{+,(m)}|0\rangle_{(t=-a)}~\longmapsto~|0\rangle_{(t=-a)}\ .
\ee
We have the operator $G^-_{\dot A}(t=t_1)$; this picks up a factor
\be\label{eq_f5}
f_5=(t_1+a)^{-{m}}\ .
\ee
Likewise, the operator $G^+_{\dot C}(t=t_2)$ picks up a factor
\be\label{eq_f6}
f_6=(t_2+a)^{{m}}\ .
\ee

\b

We are now left with just the amplitude (\ref{axeight}). Combining eq.s (\ref{eq_f1})-(\ref{eq_f6}), we find 
\bea
&&T^{(1)(1)}_{\dot C\dot A,m}(w_2,w_1)=\prod_{i=1}^6f_i\,T_{\dot C\dot A}(w_2,w_1)={\epsilon_{\dot C\dot A}\over 2}\Big({a\over a-b}\Big)^{m^2}  \left [ {(a-b)^2\over 16 ab}\right ] \nn
&&\qquad\qquad\qquad\;=\epsilon_{\dot C\dot A} {\big(\cosh({\Delta w\over 4})\big)^{2m^2}\over 8 \sinh^2 ({\Delta w \over 2})}\ .
\label{axthir}
\eea

\subsection{Computing the remaining $T_{\dot C\dot A}^{(j)(i)}$}

Suppose we fix $\sigma_1$ and consider the shift $\sigma_2\mapsto \sigma_2+2\pi$. This gives $\Delta w \mapsto \Delta w + 2 \pi i$. Under this change $\sinh ({\Delta  w \over 2}) \mapsto -\sinh ({\Delta  w \over 2})$, and so $\sinh^2 ({\Delta  w \over 2})$ is invariant. Similarly, $\sinh^2 ({ \Delta  \bar w \over 2})$ is invariant. But $\cosh({\Delta w\over 4})\mapsto i\sinh({\Delta w\over 4})$. Thus, the integrand in (\ref{axthir}) is not periodic under $\sigma_2\mapsto \sigma_2+2\pi$. The reason is that when we move $\sigma_2$ through $2\pi$, we move from copy $1$ to copy $2$. This implies
\be
T_{\dot C\dot A, m}^{(1)(1)}~\mapsto~ T_{\dot C\dot A,m}^{(2)(1)}\ .
\ee
Under the shift $\sigma_2\mapsto \sigma_2+2\pi$ we find
\be
\big(\cosh(\tfrac{\Delta w}{4})\big)^{2m^2}\mapsto (-1)^m\big(\sinh(\tfrac{\Delta w}{4})\big)^{2m^2}\ .
\ee

Thus, we see that we can take into account all the four terms $T^{(j)(i)}$ by taking $T^{(1)(1)}$ and making the replacement
\be
\big(\cosh(\tfrac{\Delta w}{4})\big)^{2m^2}\mapsto 2\left ( \big(\cosh(\tfrac{\Delta w}{4})\big)^{2m^2}+(-1)^m\big(\sinh(\tfrac{\Delta w}{4})\big)^{2m^2}\right )\ .
\ee
Collecting the left and right parts of the correlator, we find that
\be
\langle \Psi^{(m)}| D(w) D(0)|\Psi^{(m)}\rangle = P^{\dot A\dot B} P^{\dot C \dot D} \epsilon_{\dot C\dot A}\epsilon_{\dot D\dot B} {\left ( \big(\cosh({\Delta w\over 4})\big)^{2m^2}+(-1)^m\big(\sinh({\Delta w\over 4})\big)^{2m^2}\right )\over 16 \sinh^2 ({\Delta w \over 2}) \sinh^2 ({\Delta \bar w \over 2})}\ .
\ee
Comparing with (\ref{bbtwo}), we find that
\be
Q^{(m)}(w)=P^{\dot A\dot B} P^{\dot C \dot D} \epsilon_{\dot C\dot A}\,\epsilon_{\dot D\dot B}\,{\left ( \big(\cosh({\Delta w\over 4})\big)^{2m^2}+(-1)^m
\big(\sinh({\Delta w\over 4})\big)^{2m^2}\right )\over 16 }\ .
\label{bbthree}
\ee

\subsection{Computing $X^{(m)}(T)$ for $m$ even}

Due to the term $(-1)^m$ in (\ref{bbthree}), it is convenient to treat the cases of even and odd $m$ separately. We consider even values of $m$ in this subsection and treat the odd $m$ case in the next subsection.

We first compute the contour integral $I_{C_1}$ in eq. (\ref{IC1}) in the limit $\tau\r\infty$. To do so, we set $w_2\equiv w$, $w_1=0$, and expand the functions in (\ref{bbthree}) in powers of $e^{-w}$. We find:
\bea
\big(\cosh(\tfrac{w}{4})\big)^{2m^2}&=&{1\over 2^{2m^2}} e^{{m^2\over 2} w}\sum_{k=0}^{2m^2}\, {}^{2m^2}C_k e^{-{k\over 2}w}\ ,\nn
\big(\sinh(\tfrac{w}{4})\big)^{2m^2}&=&{1\over 2^{2m^2}} e^{{m^2\over 2} w}\sum_{k=0}^{2m^2}\, {}^{2m^2}C_k (-1)^k e^{-{k\over 2}w}\ ,
\eea
where ${}^mC_n$ are the binomial coefficients. Defining $k=2k'$, $k^\prime\in\mathbb Z$, we find
\be
\big(\cosh(\tfrac{w}{4})\big)^{2m^2}+\big(\sinh(\tfrac{w}{4})\big)^{2m^2}={2\over 2^{2m^2}} e^{{m^2\over 2} w}\sum_{k'=0}^{m^2}\, {}^{2m^2}C_{2k'}  e^{-k'w}\ .
\ee
We also have
\bea
&&{1\over  \sinh^2 ({w\over 2})}=4e^{-w}\sum_{l=0}^\infty (l+1) e^{-l w}\ ,\\
&&\coth(\tfrac{\bar w}{2})=(1+e^{-\bar w})\sum_{n=0}^\infty  e^{-n\bar w}\ .
\eea
We have $I_{C_1}$ in eq. (\ref{IC1}) as an integral over $w$ at $\tau={T\over 2}$:
\bea
I_{C_1}&=&-P^{\dot A \dot B}P^{\dot C\dot D} \epsilon_{\dot C\dot A}\epsilon_{\dot D\dot B}\cr
&&~~~~~~~~\times ~~{1\over 2}\int_{\sigma=0}^{2\pi} d\sigma  \, \left ( {e^{{m^2\over 2} w}\over 2^{2m^2}} \right )
\left ( \sum_{k'=0}^{m^2}\, {}^{2m^2}C_{2k'}  e^{-k'w}\right ) \left (e^{-w}\sum_{l=0}^\infty (l+1) e^{-l w}\right )\times\nn
&&
~~~~~~~~~~~~~~~\times ~~ \left ((1+e^{-\bar w})\sum_{n=0}^\infty  e^{-n\bar w} \right ) \ ,
\label{mone_i}
\eea
where the last bracket contains antiholomorphic factors of the form  $1, e^{-\bar w}, e^{-2\bar w},\cdots$. We will now argue that only the leading term, $1$, survives from this bracket, in the limit $T\r \infty$. To see this, note that the first bracket on the RHS of (\ref{mone_i}) has a power $e^{{m^2\over 2} w}$. From the second and third brackets, we can get a power $e^{-k_1w}$ with $k_1\ge 0$ and from the last bracket we can get a power $e^{-k_2 \bar w}$ with $k_2\ge 0$. These factors give
\be
e^{({m^2\over 2}-k_1-k_2)\tau} e^{i({m^2\over 2}-k_1+k_2)\sigma}\ .
\ee
We have
\be
\int_0^{2\pi} d\sigma e^{i({m^2\over 2}-k_1+k_2)\sigma}= 2\pi \delta_{{m^2\over 2}-k_1+k_2,0}\ ,
\label{mtwo}
\ee
so that we have $k_1={m^2\over 2}+k_2$. The power of $e^{\tau}$ then gives (since $\tau={T\over 2}$)
\be
e^{-k_2{T}}\ .
\ee
Thus in the limit $T\r\infty$, the only surviving term is $k_2=0$, and therefore the last bracket in (\ref{mone_i}) can be replaced by unity. We then get 
\be
k_1=k'+l+1={m^2\over 2}
\ee
which sets $l={m^2\over 2} -k'-1$. The condition $l\ge 0$ then gives
\be
k'\le {m^2\over 2}-1\ .
\ee
Using (\ref{mtwo}), we find
\bea
I_{C_1}&=&-P^{\dot A \dot B}P^{\dot C\dot D} \epsilon_{\dot C\dot A}\epsilon_{\dot D\dot B}\,{\pi\over 2^{2m^2}}\sum_{k'=0}^{{m^2\over 2}-1} \, {}^{2m^2}C_{2k'} \,(\tfrac{m^2}2-k')\cr
&=& -P^{\dot A \dot B}P^{\dot C\dot D} \epsilon_{\dot C\dot A}\epsilon_{\dot D\dot B}\,{\sqrt{\pi} \over 8}\,{\Gamma[m^2-\h]\over \Gamma[m^2-1]}\ .
\eea

We proceed similarly for the contour integral $I_{C_2}$ in eq. (\ref{IC2}) at $\tau=-{T\over 2}$. This time we expand the functions in (\ref{bbthree}) in powers of $e^w$:
\bea
\big(\cosh(\tfrac{w}{4})\big)^{2m^2} &=&{1\over 2^{2m^2}} e^{-{m^2\over 2} w}\sum_{k=0}^{2m^2}\, {}^{2m^2}C_k e^{{k\over 2}w}\ ,\nn
\big(\sinh(\tfrac{w}{4})\big)^{2m^2}&=&{1\over 2^{2m^2}} e^{-{m^2\over 2} w}\sum_{k=0}^{2m^2}\, {}^{2m^2}C_k (-1)^k e^{{k\over 2}w}\ .
\eea
Defining again $k=2k'$, $k'\in\mathbb Z$ in these sums, we find
\be
\big(\cosh(\tfrac{w}{4})\big)^{2m^2}+\big(\sinh(\tfrac{w}{4})\big)^{2m^2}={2\over 2^{2m^2}} e^{-{m^2\over 2} w}\sum_{k'=0}^{m^2}\, {}^{2m^2}C_{2k'}  e^{k'w}\ .
\ee
We further have
\bea
{1\over  \sinh^2 (\tfrac{w}{2})}&=&4e^{w}\sum_{l=0}^\infty (l+1) e^{l w} \label{sinh}\ ,\\
\coth(\tfrac{\bar w}{2})&=&-(1+e^{\bar{w}})\sum_{n=0}^\infty  e^{n\bar w}\ .\label{coth}
\eea

Our integral becomes
\bea
I_{C_2}&=&-P^{\dot A \dot B}P^{\dot C\dot D} \epsilon_{\dot C\dot A}\epsilon_{\dot D\dot B}\cr
&&~~\times ~~{1\over 2}\int_{\sigma=0}^{2\pi} d\sigma  \, \left ( {2e^{-{m^2\over 2} w}\over 2^{2m^2}} \right ) \left ( \sum_{k'=0}^{m^2}\, {}^{2m^2}C_{2k'}  e^{k'w}\right ) \left (e^{w}\sum_{l=0}^\infty (l+1) e^{l w}\right )\nn
&&
~~~~~~~~~~~~~~~\times ~~ \left ((1+e^{\bar w})\sum_{n=0}^\infty  e^{n\bar w} \right )\ ,
\label{mone}
\eea
where the last bracket now contains antiholomorphic factors of the form  $1, e^{\bar w}, e^{2\bar w},\cdots$. Following a reasoning similar to that in the case of the computation for $I_{C_1}$ in eq. (\ref{mone_i}), we find that only the leading term, $1$, survives from the last bracket in the limit $T\to \infty$. We then have
\bea
I_{C_2}&=&-P^{\dot C\dot D} P^{\dot A \dot B}\epsilon_{\dot C\dot A}\epsilon_{\dot D\dot B}\,{\pi\over 2^{2m^2}}\sum_{k'=0}^{{m^2\over 2}-1} \, {}^{2m^2}C_{2k'} \, ({\tfrac{m^2}2}-k')\cr
&=&- P^{\dot C\dot D} P^{\dot A \dot B}\epsilon_{\dot C\dot A}\epsilon_{\dot D\dot B}\,{\sqrt{\pi} \over 8}\,{\Gamma[m^2-{1\over2}]\over \Gamma[m^2-1]}=I_{C_1}\ .
\eea

We next consider $I_{C_3}$ in eq. (\ref{IC3}), the contribution from the contour around $w=0$.  For small $|w|$ we have (with $m\ge 2$)
\bea
&&\big(\cosh(\tfrac{w}{4})\big)^{2m^2}+\big(\sinh(\tfrac{w}{4})\big)^{2m^2}=1+{m^2\over 16} w^2+\cdots\ ,\\
&&{1\over \sinh^2(\tfrac{w}{2})}={4\over w^2}-{1\over 3}+\cdots\ ,\label{IC3sinh}\\
&&\coth (\tfrac{\bar w}{2})={2\over \bar w}+{\bar w\over 6}+\cdots\ .\label{IC3cosh}
\eea
The leading term in the integrand (\ref{IC3}) gives:
\be
I_{C_3}\r -P^{\dot A\dot B} P^{\dot C \dot D} \epsilon_{\dot C\dot A}\epsilon_{\dot D\dot B}\,{i\over 2}\int_{|w|=\epsilon} { dw\over w^2}\,{1\over \bar w} = P^{\dot A\dot B}
P^{\dot C \dot D} \epsilon_{\dot C\dot A}\epsilon_{\dot D\dot B}\;{\pi\over \epsilon^2}\ .
\ee
This is a constant independent of the states at the bottom and top of the cylinder  in the correlator, and will arise even if we replace these states with the vacuum state $|0\rangle$. Thus, to maintain the normalization
\be
\langle 0 | 0 \rangle=1
\ee
of the vacuum at $O(\lambda^2)$ we must add to the Lagrangian  a counterterm proportional to the identity operator, which generates an integral
\be
I_{C_3,\,\mathrm{counterterm}}= -P^{\dot A\dot B} P^{\dot C \dot D} \epsilon_{\dot C\dot A}\epsilon_{\dot D\dot B}\,{\pi\over \epsilon^2}\ .
\label{counterterm}
 \ee
  This cancels the divergent contribution to $I_{C_3}$. The next largest terms in the integrand of (\ref{IC3}) give 
\be
P^{\dot A\dot B} P^{\dot C \dot D} \epsilon_{\dot C\dot A}\epsilon_{\dot D\dot B} \left (  -{ im^2\over 32}\int_{|w|=\epsilon} {dw\over \bar w} +  {i\over 24}\int_{|w|=\epsilon} {dw\over \bar w} - {i\over 24}\int_{|w|=\epsilon} {dw  \over  w^2}\,  \bar w  \right )\ .
\ee
We find that each of these terms vanishes. Higher order terms give contributions having positive powers of $|\epsilon|$ and so for these terms we have $I_{C_3,\,\mathrm{renormalized}} = 0$.

Finally, using (\ref{bsix}), we find that for even $m$, $X^{(m)}(T)$ is of the form
\be
\lim_{T\r\infty} X^{(m)}=- P^{\dot A \dot B}P^{\dot C\dot D} \epsilon_{\dot C\dot A}\epsilon_{\dot D\dot B}\,{\sqrt{\pi} \over 4}\,{\Gamma[m^2-{1\over2}]\over \Gamma[m^2-1]}\ .
\ee

\subsection{Computing $X^{(m)}(T)$ for $m$ odd}

We next compute eq. (\ref{bbthree}) for odd values of $m$. Defining $k=2k'+1$, $k'\in\mathbb Z$ we find that
\be
\big(\cosh(\tfrac w4)\big)^{2m^2}-\big(\sinh(\tfrac w4)\big)^{2m^2}={2\over 2^{2m^2}} e^{{(m^2-1)\over 2} w}\sum_{k'=0}^{m^2-1}\, {}^{2m^2}C_{2k'+1}  e^{-k'w}\ .
\ee
Proceeding just as for the case of even $m$, we find
\bea
I_{C_1}&=&- P^{\dot A \dot B}P^{\dot C\dot D}\epsilon_{\dot C\dot A}\epsilon_{\dot D\dot B}\,
{\pi\over 2^{2m^2}}\sum_{k'=0}^{{(m^2-1)\over 2}-1} \, {}^{2m^2}C_{2k'+1} \, ({(m^2-1)\over 2}-k')\cr
&=& - P^{\dot A \dot B}P^{\dot C\dot D} \epsilon_{\dot C\dot A}\epsilon_{\dot D\dot B}\,{\sqrt{\pi} \over 8}\,{\Gamma[m^2-\h]\over \Gamma[m^2-1]}\ .
\eea
We also find, as before,
\be
I_{C_2}=I_{C_1}, ~~~I_{C_3,\,\mathrm{renormalized}}\r 0\ .
\ee
Thus, for odd $m$ we get the same expression as for even $m$
\be
\lim_{T\r\infty} X^{(m)}=- P^{\dot C\dot D} P^{\dot A \dot B}\epsilon_{\dot C\dot A}\epsilon_{\dot D\dot B}\,{\sqrt{\pi} \over 4}\,{\Gamma[m^2-{1\over2}]\over \Gamma[m^2-1]}\ .
\ee

\subsection{Expectation values}

We shall now compute the expectation value of the energy (\ref{deltaE}) for the states (\ref{threex}):
\be
\langle E^{(2)}\rangle=-\pi \lim_{T\r\infty}X^{(m)}= P^{\dot C\dot D} P^{\dot A \dot B}\epsilon_{\dot C\dot A}\epsilon_{\dot D\dot B}\,{\pi^{3\over 2} \over 4}\,
{\Gamma[m^2-{1\over2}]\over \Gamma[m^2-1]}\ .
\ee
We set the polarization $P^{\dot A\dot B}$ to correspond to a perturbation with no quantum numbers
\be
P^{\dot{A}\dot{B}}=\e^{\dot{A}\dot{B}}\ .
\ee
We expect such a perturbation to correspond to the direction towards the supergravity spacetime $AdS_3\times S^3\times T^4$. Then we obtain
\be\label{expval_i}
\langle E^{(2)}\rangle={\pi^{3\over 2} \over 2}\,{\Gamma[m^2-{1\over2}]\over \Gamma[m^2-1]}\ .
\ee
This is the main result of this section. We will generalise this result to arbitrary values of $N$ in section \ref{section7}.

\section{No lift for global modes}\label{section6}

The chiral algebra generators of the CFT at the orbifold point are described by a sum over the generators of each copy:
\be
J^+_{-n}=J^{+(1)}_{-n}+J^{+(2)}_{-n}+\cdots+J^{+(N)}_{-n}\ .
\ee
If we act with such a current on any state then the dimension of the state will rise as 
\be
h\r h+n\ .
\label{azsix}
\ee
There cannot be any anomalous contribution since the change (\ref{azsix}) is determined by the chiral algebra. We can use this fact as a check on the computations that we have performed; we will perform this check for a simple case in this section. 
 
Let us assume that we have  two copies (as in the above sections); so $N=2$.  First consider the case where we apply $J^+_{-1}$; this gives the state
\be
|\chi_1\rangle={1\over \sqrt{2}}J^+_{-1}\, |0\rangle^{(1)}|0\rangle^{(2)}=
{1\over \sqrt{2}}\left ( J^{+(1)}_{-1}+J^{+(2)}_{-1}\right ) \, |0\rangle^{(1)}|0\rangle^{(2)}\ ,
\ee
where we have added a normalization factor to normalize the state to unity. This is the same as the state $|\Phi^{(1)}\rangle$ defined in (\ref{threex}). From (\ref{expval_i}) we see that the lift vanishes in this case.

Next consider the state
\bea
|\chi_2\rangle&=&{1\over \sqrt{8}}J^+_{-3}J^+_{-1} \, |0\rangle^{(1)}|0\rangle^{(2)}\nn
&=&{1\over \sqrt{8}} \left ( J^{+(1)}_{-3}J^{+(1)}_{-1}+J^{+(2)}_{-3}J^{+(2)}_{-1}\right )  \, |0\rangle^{(1)}|0\rangle^{(2)}\cr
&& + {1\over \sqrt{8}}\left ( J^{+(1)}_{-3}J^{+(2)}_{-1}+J^{+(2)}_{-3}J^{+(1)}_{-1}\right )  \, |0\rangle^{(1)}|0\rangle^{(2)}
\nn
&\equiv& |\chi_{2,1}\rangle+|\chi_{2,2}\rangle.
\label{global mode}
\eea
The conjugate state is written in a similar way in two parts
\be
\langle \chi_2|~=~\langle \chi_{2,1}|+\langle \chi_{2,2}|\ .
\ee
Note that the state $|\chi_{2,1}\rangle$ is (upto a normalization factor) the same as the state $|\Phi^{(2)}\rangle$ defined in (\ref{threex}). 

We now consider the lift of the state $|\chi_2\rangle$. There are four contributions to this lift. The first has $|\chi_{2,1}\rangle$ as the initial and final states, and this is proportional to the lift we have computed for $|\Phi^{(2)}\rangle$, see eq. (\ref{expval_i}). Thus, this contribution is nonzero. But there are three other contributions which involve $|\chi_{2,2}\rangle$. When we add all these contributions and subtract the identity contribution in (\ref{axeight}), the lift is expected vanish as we now check. 

Computing the amplitudes by the same method as in the above sections, we find

\bea
 \langle\chi_{2,1}|\big(G^+_{\dot{C},-{1\over2}}\s^-(w_2)\big)\big( G^-_{\dot{A},-{1\over2}}\s^+(w_1)  \big)|\chi_{2,1}\rangle&=&\e_{\dot{C}\dot{A}}{1\over 4}
 \bigg({\cosh^8({\Delta w\over 4}) \over 4\sinh^2({\Delta w\over 2})}  + {\sinh^8({\Delta w\over 4}) \over 4\sinh^2({\Delta w\over 2})}\bigg)\ ,
\cr
\cr
\cr
 \langle\chi_{2,2}|\big(G^+_{\dot{C},-{1\over2}}\s^-(w_2)\big)\big( G^-_{\dot{A},-{1\over2}}\s^+(w_1)  \big)|\chi_{2,1}\rangle  &=& \e_{\dot{C}\dot{A}} {1\over4}
 \bigg( - {\cosh^8 ({\Delta w\over 4})\over 4\sinh^2({\Delta w\over 2})}  - {\sinh^8 ({\Delta w\over 4})\over 4\sinh^2({\Delta w\over 2})}+ \cr
  &&\qquad~~+ {1\over  4\sinh^2({\Delta w\over 2})}\bigg)\ ,
  \cr
  \cr
  \cr
  \langle\chi_{2,1}|\big(G^+_{\dot{C},-{1\over2}}\s^-(w_2)\big)\big( G^-_{\dot{A},-{1\over2}}\s^+(w_1)  \big)|\chi_{2,2}\rangle&=&  \e_{\dot{C}\dot{A}}{1\over4}
  \bigg(-{\cosh^8 ({\Delta w\over 4})\over 4\sinh^2({\Delta w\over 2})} - {\sinh^8 ({\Delta w\over 4})\over 4\sinh^2({\Delta w\over 2})}+\cr
 &&\qquad~~ + {1\over 4\sinh^2({\Delta w\over 2})}\bigg)\ ,
 \cr
 \cr
 \cr
  \langle\chi_{2,2}|\big(G^+_{\dot{C},-{1\over2}}\s^-(w_2)\big)\big( G^-_{\dot{A},-{1\over2}}\s^+(w_1)  \big)|\chi_{2,2}\rangle  &=&\e_{\dot{C}\dot{A}}{1\over4}
  \bigg( {\cosh^8 ({\Delta w\over 4})\over 4\sinh^2({\Delta w\over 2})} +  {\sinh^8 ({\Delta w\over 4})\over 4\sinh^2({\Delta w\over 2})}+\cr
 &&\qquad ~~+  {2\over  4\sinh^2({\Delta w\over 2})}\bigg)\ .
\eea
We add up all four contributions and obtain
\be
 \langle\chi_{2}|\big(G^+_{\dot{C},-{1\over2}}\s^-(w_2)\big)\big( G^-_{\dot{A},-{1\over2}}\s^+(w_1)  \big)|\chi_{2}\rangle=\e_{\dot{C}\dot{A}}{1\over  4\sinh^2({\Delta w\over 2})}\ .
\ee
Collecting the left at right parts of the correlator we find
\be
\langle \chi_2| D(w,\bar{w}) D(0)|\chi_2\rangle = P^{\dot A\dot B} P^{\dot C \dot D} \epsilon_{\dot C\dot A}\epsilon_{\dot D\dot B}
{1\over  16\sinh^2({\Delta w\over 2})}\,{1\over  \sinh^2({\Delta \bar{w}\over 2})}\ .
\ee
Comparing with (\ref{bbtwo}) we have
\be
Q = P^{\dot A\dot B} P^{\dot C \dot D} \epsilon_{\dot C\dot A}\epsilon_{\dot D\dot B}\,{1\over 16}\ .
\label{Q}
\ee
We want to compute $I_{C_1}, I_{C_2}$ and $I_{C_3,\,\mathrm{renormalized}}$ in eqs. (\ref{IC1})-(\ref{IC3}). For $I_{C_1}$, inserting (\ref{Q}) into (\ref{IC1}) yields
\be
I_{C_1}=-P^{\dot A\dot B} P^{\dot C \dot D} \epsilon_{\dot C\dot A}\epsilon_{\dot D\dot B}\int_{\s=0}^{2\pi} d\s \left (  {1\over   16\sinh^2  ({w\over 2})}\,  \coth(\tfrac{\bar w}{2})\right )\ .
\ee
Inserting the expansions (\ref{sinh}) and (\ref{coth}) give
\bea
I_{C_1}=-{1\over4}\int_{\sigma=0}^{2\pi} d\sigma  \,  \left (e^{-w}\sum_{l=0}^\infty {}^{-2}C_l (-1)^l e^{-l w}\right )
\!\!\left ((1+e^{-\bar w})\sum_{n=0}^\infty {}^{-1}C_n(-1)^n e^{-n\bar w} \right )\ . 
\eea

This contour is evaluated at $\tau= {T\over2}$ with $T\to\infty$. Since our expression only contains negative powers of $w$ and $\bar{w}$, every term vanishes and we find that 
\bea
I_{C_1}=0\ .
\eea
Similarly, for $I_{C_2}$ we find
\be
I_{C_2}\,=\,-{1\over4}\int_{\sigma=0}^{2\pi} d\sigma  \, \left (e^{w}\sum_{l=0}^\infty {}^{-2}C_l (-1)^l e^{l w}\right )\!\! \left ((1+e^{\bar w})\sum_{n=0}^\infty {}^{-1}C_n(-1)^n e^{n\bar w}\right)\ .
\ee
This contour is evaluated at $\tau= -{T\over2}$ with $T\to\infty$. Our expression only contains positive powers of $w$ and $\bar{w}$ and we find 
\bea
I_{C_2} = 0\ .
\eea
For $I_{C_3}$, inserting (\ref{Q}) into (\ref{IC3}) yields
\be
 I_{C_3}=-P^{\dot A\dot B} P^{\dot C \dot D} \epsilon_{\dot C\dot A}\epsilon_{\dot D\dot B}i\int_{|w|=\epsilon} dw \left (  {1\over   16\sinh^2  ({w\over 2})}\,  \coth({\bar w\over 2})\right)\ .
\ee
Inserting the expansions (\ref{IC3sinh}) and (\ref{IC3cosh}) and taking the leading order term in the integrand we have
\be
I_{C_3}\r -P^{\dot A\dot B} P^{\dot C \dot D} \epsilon_{\dot C\dot A}\epsilon_{\dot D\dot B}\,{i\over 2}\int_{|w|=\epsilon} { dw\over w^2\bar w}=
P^{\dot A\dot B} P^{\dot C \dot D} \epsilon_{\dot C\dot A}\epsilon_{\dot D\dot B}\,{\pi\over \epsilon^2}\ .
\ee
Using (\ref{counterterm}), we find
\bea
I_{C_{3,\,\mathrm{renormalized}}}=I_{C_3} + I_{C_{3,\,\mathrm{counterterm}}}=0\ .
\eea
The expectation value (\ref{deltaE}) then reads
\bea
\langle E^{(2)}\rangle = -\pi \lim_{T\r\infty}\left ( I_{C_1}+I_{C_2}+I_{C_{3,\,\mathrm{renormalized}}}\right ) = 0\ ,
\eea
proving the absence of lifting for the global mode in (\ref{global mode}). From (\ref{nfour}) we can now argue that the global mode does not mix with any eigenstate that lifts. We have $E^{(2)}_{a'}\ge 0$ for all $a'$, since the states $\tilde\phi_{a'}$ are chiral primaries in the right-moving sector, and therefore must have $\Delta \bar h =\Delta h \ge 0$. Thus, a vanishing of $\langle E^{(2)}\rangle=0$ means a vanishing overlap of $|\chi_2\rangle$ with each of the $\tilde\phi_{a'}$ which lift; from this it follows that $|\chi_2\rangle$ remains an unlifted eigenstate of the Hamiltonian.

\section{General values of $N$}\label{section7}

So far we have considered two copies of the $c=6$ CFT: $N=2$. The initial state had two singly-wound copies; the twist operators twisted these together and then untwisted them, so that we ended with two singly-wound copies again. In general, the orbifold CFT has an arbitrary number $N=n_1n_5$ of copies of the CFT. But as we will now see, for our situation, the computation with two copies that we have carried out allows us to obtain the expectation values for arbitrary $N$.

\subsection{The initial state}

We have $N$ copies of the $c=6$ CFT and each copy is singly-wound. Each copy is in the NS sector. Out of these copies, we assume that $n$ copies are excited as
\be
J^{+(i)}_{-(2m-1)}\dots J^{+(i)}_{-3}J^{+(i)}_{-1} |0\rangle^{(i)}\ ,\quad i\in\{1,\cdots,n\}
\ee
in the left moving sector, while the right moving sector is in the vacuum state $|0\rangle$. The remaining $N-n$ copies are in the vacuum state $|0\rangle$ on both the left and right sectors. This state is depicted in fig.\ref{fig_singlwindingexcited}.

There are ${}^NC_n$ ways to choose which strings are excited. Thus, the initial state is composed of ${}^NC_n$ different terms, with each term describing one set of possible excitations. The sum of these terms must be multiplied by a factor
\be
{\cal N}= \left ( {}^NC_n\right ) ^{-\h}
\label{axtone}
\ee
in order that the overall state is normalised to unity.

\subsection{Action of the deformation operator}

When we were dealing with just two copies, we denoted the twist operator of  the deformation by $\sigma^\pm$; it was implicit that this operator would twist together the two copies that we had, see subsection \ref{sebsec_deformation}. But when we have $N>2$ copies, then we need to specify which two copies are being twisted. If the $(i)$ and $(j)$ copies are twisted, we denote the twist operator by $\sigma^\pm_{(i)(j)}$. Thus the deformation operator (\ref{exactlymarginal}) now has the form 
\be
D= P^{\dot A \dot B}G^-_{\dot A, -\h}\bar G^-_{\dot B, -\h} \sum_{i<j}\sigma^{++}_{(i)(j)}\ ,
\ee
where $1\le i,j\le N$. The supercurrents $G^-_{\dot A}$ are given by a sum over the contributions from each copy:
\be
G^-_{\dot A}\equiv\sum_{i=1}^N G^{-(i)}_{\dot A}\ .
\ee

\subsection{Expectation values for general values of $N$}

We argue in the following steps:

\b

(i) Consider the action of the first deformation operator on the initial state. Suppose this first deformation operator twists together the copies $(i),(j)$. Since we are computing an expectation value, the final state must be the same as the initial state; thus the final state must also have all copies singly-wound.  So the second deformation operator must twist the same copies $i,j$ to produce a state with all copies singly-wound. 

\b

(ii) Copies {\it other} than the two copies that get twisted act like `spectators'; thus we get an inner product between their initial state and their final state. If the initial state for such a copy is unexcited, then the final state must also be unexcited, and if the initial state is excited then the final state has to be excited.  The inner product between copies with the same initial and final state is unity.

Since the state of the spectator copies does not change, the number of spectator copies which are excited are also the same between the initial and final states. As a consequence, for the pair $i,j$ which do get twisted, the number of excited copies in the initial and final states is the same. 

\b

(iii) We therefore find that there are three possibilities for the excitations among the twisted copies $i,j$:

\b

(a) Neither of the copies $(i),(j)$ are excited. In this case there is no contribution to the lifting, as the vacuum state $|0\rangle^{(i)}|0\rangle^{(j)}$ is not lifted. 

\b

(b) Both the copies $(i),(j)$ are excited. But the state where both copies are excited is a spectral flow of the state where neither copy is excited, as we have seen in section \ref{section6}. So again there is no lift.

\b

(c) One of the copies out of $(i),(j)$ is excited and one is unexcited. There are four contributions here: (1) Copy $(i)$ excited in the initial state, copy $(i)$ excited in the final state; (2) Copy $(j)$ excited in the initial state, copy $(i)$ excited in the final state; (3) Copy $(i)$ excited in the initial state, copy $(j)$ excited in the final state;  (4) Copy $(j)$ excited in the initial state, copy $(j)$ excited in the final state. These are the four contributions we had in eq. (\ref{adoneqq}). Thus summing these four contributions gives the same anomalous dimension $E^{(2)}$ that we computed for the case $N=2$, see eq. (\ref{expval_i}), with an extra factor of $2$ since we do not here have the normalization factors ${1\over \sqrt{2}}$ in the initial and final state. Thus, the pair $(i),(j)$ contribute $2E^{(2)}$, with $E^{(2)}$ given by eq. (\ref{expval_i}).

\b

(iv) Let us now collect combinatoric factors. First we select the pair $(i),(j)$ out of the $N$ copies. This is done in ${}^NC_2$ ways. Out of these two copies, one has  to be excited (as noted in (iii) above). Thus, out of the remaining $N-2$ copies, $n-1$ are excited. These $n-1$ copies can be chosen in ${}^{N-2}C_{n-1}$ ways, so this is the number of terms which have the contribution $2E^{(2)}$. We note that $1\le n\le N-1$.

\b

(v) Let us finally collect all the factors. We have a normalization factor (\ref{axtone}) both from the initial and final configurations, so we get a factor $|{\cal N}|^2$. Together with the combinatorial factors from the previous paragraph, we find that the expectation value is of the form:
\bea
\langle (E-E_{orbifold})\rangle &=&\lambda^2({}^NC_n) ^{-1}\, ({}^NC_2)\, ({}^{N-2}C_{n-1}) \, (2 \langle E^{(2)}\rangle)\nn
&=&\lambda^2n(N-n)\langle E^{(2)}\rangle \nn
&=&\lambda^2\frac12\,\pi^{3\over 2}\,n(N-n)\,{\Gamma[m^2-{1\over2}]\over \Gamma[m^2-1]}\ .
\eea
The above expression gives the lifting to order $O(\lambda^2)$ for the case where we have $N$ copies of the seed $c=6$ CFT, and $n$ of these are excited by application of the operator ${\cal J}^{(+, m)}$ (eq. (\ref{calj})) which is composed of  $m$ currents. We note that $N=2$ and $n=1$ corresponds to the case studied in section \ref{section5}, see eq. (\ref{expval_i}).

\section{Multi-wound initial states}\label{section8}

So far our initial state has consisted of $N$ singly-wound copies of the seed $c=6$ CFT. We now consider the case where we link together $k$ of these copies to make a `multi-wound copy'. We assume that all copies are grouped into such sets; i.e. there are ${N\over k}$ sets of linked copies, with each set having winding $k$. Note that this requires that $N$ be divisible by $k$. This case is depicted in fig.\ref{fig_multiwound}.

\begin{figure}
\begin{center}
\includegraphics[width=100mm]{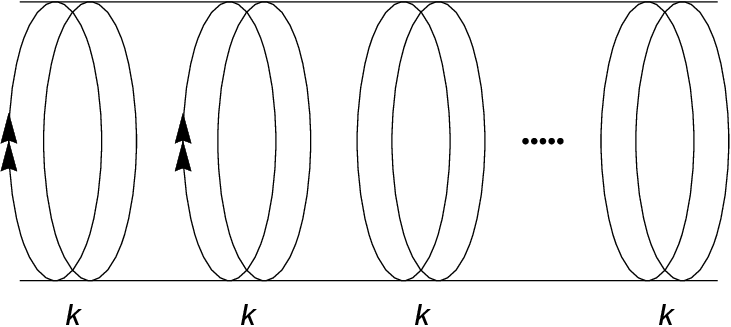}
\end{center}
\caption{ A total of $N$ singly wound copies that have been linked into ${N\over k}$ `multiwound' copies each of winding $k$. A number  $n$ of these twisted sets have been excited by current operators.}
\label{fig_multiwound}
\end{figure}

For the case where the copies are singly-wound, the ground state of each copy was the vacuum $|0\rangle$ with $h=j=0$. A set of linked copies, however, has a  nontrivial dimension, see, e.g. \cite{lm2} for the computation of the ground state energies in both odd and even twisted sectors. We start with each set being in a chiral primary state $|k\rangle$ with
\be\label{cps}
h={k-1\over 2}, ~~~j={k-1\over 2}, ~~~\bar h = {k-1\over 2}, ~~~\bar j = {k-1\over 2}\ .
\ee
We now take $n$ of these linked sets, and excite each of these by the application of current operators:
\be
|k\rangle_{ex}\equiv J^+_{-{(2m-1)\over k}}\dots J^+_{-{3\over k}}J^+_{-{1\over k}}|k\rangle\ .
\label{axttwo}
\ee
This excitation adds a momentum
\be
{1\over k}+{3\over k}+\dots + {(2m-1)\over k}={m^2\over k}
\ee
to this set of linked copies. This momentum must be an integer \cite{dasmathur1}, so $m^2$ should be divisible by $k$.

We wish to find the expectation value of the energy of the state constructed in this way, see eq. (\ref{deltaE}). It turns out that this computation is related to the ones we performed in the above sections by having multiply wound copies in the initial state instead of singly-wound copies. We will now see that we obtain the expectation value for the multi-wound case by going to a covering space of the cylinder, where we undo the multi-winding, and then relating the computation to the singly-wound case.

\subsection{The action of the twist operator}

Consider the action of the first twist operator $\sigma^+_{(i)(j)}$.  There are two possibilities:

\b

(i) The copies $i,j$ are from the same set of linked copies.

\b

(ii) The copies $i,j$ belong to different sets of linked copies.

\b

In case (i), the twist will break up the set of linked copies into two sets with windings $k', k-k'$. Since we are computing an expectation value, the second twist has to link these two sets back to a single set of $k$ linked copies. 

But we can easily see that such an action of twist operators will give no contribution to the expectation value, $\langle E^{(2)}\rangle $. The set of linked copies that we started with could be either unexcited, i.e. in the state $|k\rangle$ (\ref{cps}), or excited, i.e. in the state $|k\rangle_{ex}$ (\ref{axttwo}). The copies other than the ones in our set of $k$ linked copies play no role in the computation. Thus, the action of the two deformation operators tells us the correction $E^{(2)}$ to $|k\rangle$ or $|k\rangle_{ex}$. But $|k\rangle$ is a chiral primary state; this means that its dimension is determined by its charge and so  its anomalous dimension $E^{(2)}$ will have to vanish. The state $|k\rangle_{ex}$ arises from a spectral flow of $|k\rangle$, so again its anomalous dimension will vanish. Thus we get no contributions from the case (i). 

In case (ii), the first twist takes the two sets of $k$ linked copies and  joins them into one set of $2k$ linked copies.  Since we are looking for an expectation value, the second twist must break up this set of $2k$ linked copies back to two sets of copies with linking $k$ each. This is the situation that we will analyze in more detail now.

\subsection{The $k$-fold cover of the cylinder}

Since each set of linked copies has winding number $k$, we can go to a covering space of the cylinder where the spatial coordinate $\tilde\sigma$ runs over the range $0\le \tilde \sigma< 2\pi k$. It is convenient to think of the range of $\tilde\sigma$ to be subdivided into the $k$ intervals
\be
0\le \tilde\sigma < 2\pi\ , ~~2\pi\le \tilde\sigma < 4\pi, ~~\dots ~~ 2\pi (k-1)\le \tilde\sigma < 2\pi k\ .
\ee
We have not changed the time $\tau$ in going to the cover, so we have
\be
\tilde \tau=\tau\ .
\ee

Now we look at the factors emerging from going to this cover:

\b

(i) For the first set of linked copies, we label the copies $1, 2, \cdots,k$. For the second set, we label them $1', 2', \cdots,k'$. Then the first twist operator has the form $\sigma^+_{(i)(j')}$, where $i$ is from the first set and $j'$ is from the second set. Suppose we consider a twist at the point $\sigma=0$. Then there are  $k^2$ ways of joining the two sets into one set of $2k$ linked copies. 

On the covering space $\tilde\sigma$, there are $k$ images of the point $\sigma=0$, and we can apply a twist at any of these points. Thus we get $k$ rather than $k^2$ similar interaction points. Thus we must multiply the result we get from the covering space by an extra factor $k$.(The origin of this factor can be understood alternatively as follows: we can choose any of the copies $i=1, \dots k$ to be the first copy $i=1$, and use this to set the origin $\tilde\sigma=0$; the factor $k$ then describes  the different ways we can choose the  copy $i'=1$ from  the second set of $k$ linked copies.)

\b

(ii) The second twist acts on a set of $2k$ linked copies labeled by $i=1, \cdots,2k$. Suppose this twist is at the location $\sigma=0$ on the cylinder. This time there are only $k$ possible ways for the twist to act: once we choose one of the copies $i$, the second copy must be $i'=i+k$, since otherwise the set will not break up into two sets of winding $k$ each. Because the twist is symmetric  between $i$ and $i'$, the different possibilities are given by choosing $i=1, \dots k$; i.e., there are $k$ possible choices. 

We now see that on the covering space $\tilde\sigma$, these $k$ different choices are accounted for by the $k$ different images of $\sigma=0$ on the space $\tilde\sigma$. Thus there is no additional factor (analogous to case (i)) from the second twist. 

\b

(iii) We now make a conformal map  from the covering space $\tilde\sigma, \tilde\tau$ to a cylinder where the spatial coordinate has the usual range $(0, 2\pi)$:
\be
\sigma'={1\over k}\,{\tilde\sigma}\ , ~~~~\tau'={1\over k}\,\tilde\tau\ .
\ee
Under this map the deformation operators scale as
\be
D(\tilde w_1, \bar {\tilde w}_1) \r  {1\over k^2} D(w'_1, \bar {w'}_1)\ , ~~~D(\tilde w_2, \bar {\tilde w}_2) \r  {1\over k^2} D(w'_2, \bar {w'}_2)\ ,
\ee
so we get an overal factor of ${1\over k^4}$ from this scaling.


\b

(iv) Let us now recall the computation of  the amplitude $A^{(2)}(T)$ in subsection \ref{subsec_Aab}, see eq.s (\ref{wnasel_ii}) and (\ref{A2T_i}). This amplitude involved integrals over the positions $w_1, w_2$  of the two deformation operators. (We later recast these integrals in the form of contour integrals in eq. (\ref{A2T_i}), but it is simpler to see the scalings in terms of the original integrals in eq. (\ref{wnasel_ii})). We have
\be
\int d^2 \tilde w_1 \r k^2 \int  d^2 w'_1\ ,\qquad\int d^2 \tilde w_2 \r k^2 \int  d^2 w'_2\ ,
\ee
so we obtain a overall factor of $k^4$.

\b

(v) The integral over $\Delta \tilde w=(\tilde w_2-\tilde w_1)$ converges, but the integral over $s=\h(\tilde w_1+\tilde w_2)$ gives a factor of $\tilde T$. We need to multiply by a factor ${1\over \tilde T}$, which scales as
\be
{1\over \tilde T} \r {1\over k}\,{1\over T'}\ .
\label{axtfiveqq}
\ee

\b

(vi) We have now mapped the problem to the cylinder $w'$, which is just like the cylinder $w$ which worked with in the situation with $k=1$.  Collecting all the factors we obtained from (i)-(v) above, we find that the factors of $k$ cancel out. We thus find the following: suppose we have $N=n_1n_5$ copies of the CFT. These copies are grouped into ${N\over k}$ sets of copies, with each set having $k$ linked copies. A number $0\le n\le {N\over k}$ of these sets is excited in the form $|k\rangle_{ex}$, while the remainder are in the state $|k\rangle$. Then the lifting of the energy is given by
\be
\langle (E-E_{orbifold})\rangle=\lambda^2\frac{\pi^{3\over 2}}2\,n({N\over k}-n)\,{\Gamma[m^2-{1\over2}]\over \Gamma[m^2-1]}\ .
\ee
The above expression gives the lifting to order $O(\lambda^2)$ for the case where we have $N$ copies of the $c=6$ CFT, with these copies being grouped into ${N\over k}$ sets (with $N$ being divisible by $k$) with each set  having winding $k$. Of these sets,  $n$ are excited in the form (\ref{axttwo}) which describes the action of  $m$ fractionally moded currents.

\section{The maximally wound sector}\label{section9}

We have computed the lifting of certain states which are excited on the left, but are a chiral primary on the right. Apart from special cases this lifting was found to be nonzero. 
In \cite{gavanarain}, the lifting of a more general class of  states was computed in a certain approximation; again it was found that generic states were lifted. 

Let us analyze this lifting in the context of the elliptic genus \cite{Witten:1993jg} which tells us how many unlifted states we expect at a given energy for the left movers. The elliptic genus for the case where the compactification was $K3\times S^1$ was computed in \cite{deBoer:1998kjm,deBoer:1998us}. The elliptic genus vanishes for the compactification $T^4\times S^1$ that we have considered, but a modified index was defined in \cite{Maldacena:1999bp}. This index protects very few states for low levels of the left moving energy. Thus in the Ramond (R) sector, there are very few unlifted states for 
\be
h\le {N\over 4}\ .
\ee
But for $h>{N\over 4}$ the number of states that are unlifted is very large; in fact their number ${\cal N}$ has to reproduce the black hole entropy which behaves as
\be
S=\ln {\cal N} \approx 2\pi \sqrt{Nh}\ .
\label{azone}
\ee
Thus we need to ask: what changes when we cross the threshold $h={N\over 4}$? Since we have been working in the NS sector, let us first spectral flow to the NS sector. Consider the R sector ground state with maximal twist $k=N$. This state has dimension
\be
h={c\over 24}={N\over 4}\ .
\ee
and charge $j=\pm \h$; let us take $j=-\h$. The spectral flow of this state to the NS sector gives a  chiral primary with
\be
h=j={N-1\over 2}\ .
\ee
This is the state $|k\rangle$ we defined above with $k=N$. We can obtain states contributing to the entropy (\ref{azone}) by acting with left moving creation and annihilation operators on $|N\rangle$. 

 Let us now ask if there is a special property shared by states in  the maximally wound sector, which is not present for states in sectors where we do not have maximal winding. We will now argue that there is indeed such a property: the nature of the linkage that is produced by the action of twists.
 
 If a state is not in the maximally wound sector, then the twist $\sigma_{(i)(j)}$ present in the deformation operator can do one of two things: 
 
 \b
 
 (i) It can join two different set of linked copies, with windings $k_1, k_2$,  into one linked copy with winding $k_1+k_2$. The second deformation operator will break this back to two sets with windings $k_1, k_2$, since we are computing an expectation value and so need the final state to be the same as the initial state.
 
 \b
 
 (ii) If we have a subset of strings with winding $k_1+k_2>1$, then it can break this subset into two sets of linked copies, with windings $k_1, k_2$. The second deformation operator will join these sets back to one set with winding $k_1+k_2$. 
 
 \b
 
 If on the other hand we have a state in the maximally wound sector, then there are no other sets of copies in the state; thus we are allowed possibility (ii) but not possibility (i).
 
  We must now ask if there is a difference in the action of the deformation operators in the cases (i) and (ii). We will see that there is indeed a difference:  in case (i), the covering space obtained when we `undo' the twist operators is a sphere (genus $g=0$), while in case (ii) the covering space is a torus (genus $g=1$). 
  
  To see this, we recall how we compute the genus of the covering space obtained from undoing the action of twist operators \cite{lm1}. Suppose we have twists of order $k_i, i=1, \dots i_{max}$. The ramification order at a twist $\sigma_{k_i}$ is $r_i=k_i-1$. Let the number of sheets (i.e. copies) over a generic point be ${ s}$. Then the genus of the covering surface is given by the Riemann-Hurwitz relation
  \be
  g=\h \sum_i r_i -s+1\ .
  \ee
  
  Let us now compute $g$ in the two cases above. We focus only on the copies which are involved in the interaction:
  
  \b
  
  (i') In case (i), we create the initial set of linked strings using twist operators $\sigma_{k_1}, \sigma_{k_2}$. The final state is created by the twists of the same order. The two deformation operators carry twists $\sigma_2$ each. The number of sheets is $s=k_1+k_2$. Thus
  \be
  g=\h [ 2(k_1-1)+2(k_2-1)+2]-(k_1+k_2)+1=0\ .
  \ee

  \b
  
  (ii') In case (ii), we create the initial state by a twist $\sigma_{k_1+k_2}$.  The final state is created by a twist of the same order. The two deformation operators carry twists $\sigma_2$ each. The number of sheets is $s=k_1+k_2$. Thus
  \be
  g=\h [ 2(k_1+k_2-1)+2 ]-(k_1+k_2)+1=1\ .
  \ee

In this chapter we have considered an example of case (i), where the covering space $t$ was a sphere. In this situation we found that the lift $\langle E^{(2)}\rangle$ was nonzero. It is possible that when the covering space is a torus, then the lift vanishes, at least for some class of states. If that happens, then such states in the maximally wound sector will not be lifted. We hope to return to this issue elsewhere.

\section{Discussion}\label{section10}

We have considered the family of states depicted in fig.\ref{fig_singlwindingexcited}, and computed the correction to the expectation value of their  energy - the `lift' - upto second order in the deformation parameter $\lambda$. The results, depicted in fig.\ref{fig_deltaE}, suggest a heuristic picture for this lift; this picture was discussed in section \ref{introsec}. 

We know that the lift vanishes in two extreme cases: (i) when no copies are excited and (ii) when all copies are excited. (The state in (ii) is just a spectral flow of the state in (i).) But in between these two extremes, the energy does rise. The heuristic picture aims to explain this phenomenon as follows.

Each set of linked copies corresponds, in this heuristic picture, to one elementary object in the dual gravity configuration. The excitations (\ref{threex}) for  $m>1$ do not correspond to supergravity quanta; thus we must think of them as `string' states. 
String states will have more  mass  than charge, and will therefore `lift'. But the gravitational attraction between the strings will cause the overall energy to reduce.   If we have enough strings so that all the copies in the CFT are excited, then the negative potential energy cancels the energy from string tension, and we end up with no lift.

We noted in section \ref{section8} that this picture holds also for the case where the copies of the $c=6$ CFT are linked together in sets of $k$ copies each. Thus if none of these sets is excited then we have no lift, and again if all the sets are excited we have no lift. But when some of the sets are excited, then we do have a lift in general. 

Now consider the limiting case where {\it all} the $N$ copies of the CFT are linked into one copy with winding $k=N$. If we excite this multi-wound set, then we have excited {\it all} the sets, since there is only one set to excite. If we extrapolate our heuristic picture above to this limiting case, then we see that the energy lift of this string state will be cancelled by the self-gravitation of the state, and the state will not lift at all. This suggests that states in the maximally wound sector will not lift.\footnote{It has been argued earlier \cite{maldasuss} that states relevant for the dynamics of the {\it near-extremal} hole should be in the highly wound sectors.} This is interesting, because we know that at high energies we have to reproduce the large entropy of the extremal hole \cite{sv}, so we need a large class of states that will not lift.

This picture also tells us how we should think about states in the fuzzball paradigm. There are certainly some states in general winding sectors that are not lifted, and the gravity description of many of these states have been constructed. But as we have seen, many states {\it will} lift. The fuzzball paradigm says that {\it all} states are fuzzballs; i.e., they have no regular horizon. Thus the class of states obtained in the fuzzball construction will in general cover both extremal and non-extremal states. It turns out that it is often easier to take a limit where the non-extremal states are in fact near-extremal. We can then look for the subclass of extremal states as limits of the construction that gives the near-extremal states. 

We also note that one should not make a sharp distinction between `supergarvity' states and
`stringy' states. In fig.\ref{fig_deltaE} the state with $n=0$ is a supergravity state with no strings. As $n$ rises, we add more an more strings, but at $n=N$ this collection of strings again behaves like a supergravity state with no lift.

We have suggested above that a large class of extremal states might lie in the maximally wound sector. We noted in section \ref{section8}  that the $O(\lambda^2)$ lift of such states  has contributions only from  genus $1$ covering surfaces while states with lower winding have both genus $0$ and genus $1$ contributions; this fact may be relevant to the relation between lifting and maximal winding. But maximally wound sectors are difficult to study in the classical limit: the classical limit corresponds to  $N\r\infty$, and a winding $k=N$ will typically produce a conical defect with conical angle $1/k=1/N\r 0$ \cite{bal,mm,glmt}. Thus such states should be thought of as limits of states with finite $k$. If the general $k$ states are near-extremal, and the $k=N$ state is extremal, then the extremal state can be seen as a limit of a family of near-extremal states.

Finally, we note that we have considered only a special family of D1-D5-P states.  It is of interest to ask if there is a general characterization of which D1-D5-P states lift and which do not. We hope to study this issue elsewhere.

\chapter{Computing the Thermalization Vertex}\label{thermalization}

This chapter is perhaps the most important of all the chapters in this work. Everything that we have done up to this point was to position us to answer the question, What is the thermalization vertex in the D1D5 CFT? Which interaction produces thermalization? This would signal black hole formation in the dual gravity picture which, in our setting, would correspond to tunneling into fuzzball states. In this chapter we present the first evidence of this thermalization effect. For this computation we don't directly use the results from the previous chapters, particularly those in Chapter \ref{G contours}. This is because while working on the problem of thermalization, as mentioned last chapter, we shifted our focus to computing the `lifting' of states in the CFT. This allowed us to approach the problem of thermalization through a more direct route. With that being said, we do however use the techniques we have developed through our original approach, which are presented in the chapters prior to Chapter \ref{lifting chapter}, as well as some additional ideas and techniques which we have developed in Chapter \ref{lifting chapter} to address thermalization. We want to compute what we call `splitting amplitudes'. These amplitudes describe the probability of an initial mode or modes with a specific energy or energies to split into lower energy modes in the final state. For thermalization to occur, energy carried by some initial state must be transferred and redistributed amongst many lower energy modes so as to produce a system in the final state which looks thermal. This can only take place through an interaction. In this chapter we compute the interaction which we call the thermalization vertex. Let us begin. 

\section{The deformation operator}\label{sebsec_deformation}
This computation will combine elements from all previous chapters. We begin by writing the action of a perturbed CFT:
\be
S_0\rightarrow S_{\mathrm{pert}}=S_0+\lambda \int d^2 w D(w, \bar w)\ ,
\ee
which was given in the last chapter in (\ref{assevent}), where $D(w,\bar w)$ is an exactly marginal operator deforming the CFT which we also used in the previous chapter. Since we are interested in second order effects. The amplitude we'd like to compute is of the form
\bea
\mathcal{A}^{i\to f}_{\text{int}} = {1\over2}\lambda^2\langle \Psi_f|\bigg(\int d^2w_2 D(w_2,\bar{w}_2)\bigg)\bigg(\int d^2w_1D(w_1,\bar{w}_1)\bigg)|\Phi_i\rangle
\label{thermal amplitude}
\eea
where $|\Psi_i\rangle,\langle\Phi_f|$ represent arbitrary initial and final states respectively, which we choose in the next section, and $i\to f$ represents the transition from some initial state $i$ to some final state $f$. This is has a similar form to that of (\ref{wnasel}). The states $|\Psi_i\rangle,\langle\Phi_f|$ include both holomorphic and antiholomorphic components.

As we have seen in previous chapters we know that the deformation operator of the D1D5 CFT contains a twist of order $2$, $\s_2^+$ for the holomorphic sector. The twist itself carries left and right charges $j=\pm \h, \bar j=\pm \h$ \cite{lm2}. Suppose we start with both these charges positive; this gives the twist $\sigma_2^{++}$. Then the deformation operators in this twist sector have the form
\be\label{exactlymarginal}
D=\e^{\dot A\dot B}\hat O_{\dot A\dot B}= \e^{\dot A \dot B}G^-_{\dot A, -\h}\bar G^-_{\dot B, -\h} \sigma^{++}_2\ .
\ee
where $\epsilon^{+-}=-1$. This choice gives a deformation carrying no charges. We will omit the subscript $2$ on the twist operator from now on, and will also consider its holomorphic and antiholomorphic parts separately. 
 As we have shown previously, we normalize the twist operator as
\be
\sigma^{-}(z )\sigma^{+}(z')\sim {1\over (z-z')}\ .
\label{asone}
\ee
We note that \cite{acm2,acm3}
\be
G^-_{\dot A, -\h}\sigma^+=-G^+_{\dot A, -\h} \sigma^-\ .
\label{eeone}
\ee

Following the conventions of Chapter 8, it will be convenient to write one of the two deformation operators as $G^-_{\dot A, -\h}\sigma^+$ and the other as $-G^+_{\dot C, -\h} \sigma^-$. We will make this choice for both the left and right movers, so the negative sign in (\ref{eeone}) cancels out. Thus on each of the left and right sides we write one deformation operator in the form $G^-_{\dot A, -\h}\sigma^+$ and the other in the form  $G^+_{\dot A, -\h} \sigma^-$. This is different than the charge combination which was used in Chapter 7 where the deformation operators carried the same charge, $G^-\s^+G^-\s^+$. 


From (\ref{asone}) we find that on the cylinder
\be
\langle 0 | \sigma^-(w_2) \sigma^+(w_1)|0\rangle = {1\over 2 \sinh ({\Delta w\over 2})}
\label{axthree}
\ee
where
\be
\Delta w = w_2-w_1\ .
\ee
where 
\bea
w_i = \t_i + i \s_i
\eea
For the antiholomorphic quantities we have the following
\be
\langle\bar 0 | \bar{\sigma}^-(\bar w_2) \bar{\sigma}^+(\bar w_1)|\bar{0}\rangle = {1\over 2 \sinh ({\Delta \bar{w}\over 2})}
\label{axthree}
\ee
where
\be
\Delta \bar{w} = \bar w_2-\bar w_1
\ee
where 
\be
\bar{w}_i = \t_i - i \s_i,\quad i = 1,2
\ee
Given the definition of $D$ in (\ref{exactlymarginal}) our amplitude in (\ref{thermal amplitude}) can be written as
\bea
\mathcal{A}^{i\to f}_{int} &=& {1\over2}\lambda^2\int d^2w_2\int d^2w_1\cr
&&\e^{\dot{C}\dot{D}}\e^{\dot{A}\dot{B}}\langle\Psi_f|\big(G^+_{\dot{C},-{1\over2}}\s^-(w_2)G^-_{\dot{A},-{1\over2}}\s^+(w_1)\big)\big(\bar{G}^+_{\dot{D},-{1\over2}}\bar{\s}^-(\bar{w}_2)\bar{G}^-_{\dot{B},-{1\over2}}\bar{\s}^+(\bar{w}_1)\big)|\Phi_i\rangle\nn
\label{amplitude two}
\eea
where we have factored the two deformation operators into their holomorphic and antiholomorphic components. Since we know that both $|\Psi_i\rangle$ and $\langle\Phi_f|$ factorize, we know that our amplitude will be of the form
\bea
\mathcal{A}^{i\to f}_{int} &=& {1\over2}\lambda^2\int d^2w_2\int d^2w_1\mathcal{A}^{i\to f}(w_1,w_2,\bar{w}_1,\bar{w}_2)\cr
&=& {1\over2}\lambda^2\int d^2w_2\int d^2w_1\e^{\dot{C}\dot{D}}\e^{\dot{A}\dot{B}}\mathcal{A}^{i\to f}_{\dot{C}\dot{A}}(w_2,w_1)\bar{\mathcal{A}}^{\bar{i}\to\bar{f}}_{\dot{D}\dot{B}}(\bar{w}_2,\bar{w}_1)
\eea
where
\bea
\mathcal{A}^{i\to f}(w_1,w_2,\bar{w}_1,\bar{w}_2)\equiv\e^{\dot{C}\dot{D}}\e^{\dot{A}\dot{B}}\mathcal{A}^{i\to f}_{\dot{C}\dot{A}}(w_2,w_1)\bar{\mathcal{A}}^{\bar{i}\to\bar{f}}_{\dot{D}\dot{B}}(\bar{w}_2,\bar{w}_1)
\eea
Now we have derived the form of our amplitude at second order in the deformation. In the next section, we choose specific initial and final states, $|\Psi_i\rangle$ and $\langle\Phi_f|$ to evaluate.

\section{The states}

We start by looking at a CFT with $N=2$ with $c=6$ just as we did in the previous chapter. The vacuum $|0\rangle$ with $h=j=0$ is given by two singly-wound copies of the CFT, i.e. there is no twist linking the copies, and the fermions on each of the copies are in the NS sector. Thus we can write for the holomorphic sector
\be
|0\rangle=|0\rangle^{(1)}\,|0\rangle^{(2)}\ ,
\ee
where the superscripts indicate the copy number. The antiholomorphic sector is identical. We ultimately want to evaluate our amplitudes in the Ramond sector, however it is easier to compute everything in the $NS$ sector first. We can then spectral our result to the Ramond sector. This spectral flow will simply result in a mode shift for all of the fermion indices. Now, let's record all of the in and out states we wish to use.

For the initial state we consider two possibilities. The first possibility is a single excitation in the initial state. Including both left and right movers and symmetrizing over copy 1 and copy 2 indices, the simplest state with a single excitation is given by:
\bea
|\Psi_1\rangle &=& {1\over \sqrt2}{1\over n}\a^{(1)i}_{--,-n}|0\rangle^{(1)}|0\rangle^{(2)}\bar{\a}^{(1)i}_{++,-n}|\bar{0}\rangle^{(1)}|\bar{0}\rangle^{(2)}\cr
&& +  {1\over \sqrt2}{1\over n}\a^{(2)i}_{--,-n}|0\rangle^{(1)}|0\rangle^{(2)}\bar{\a}^{(2)i}_{++,-n}|\bar{0}\rangle^{(1)}|\bar{0}\rangle^{(2)}\nn
\label{single excitation}
\eea
with energy
\bea
h = n,~~j=0,\qquad \bar{h}=n,~~\bar{j}=0
\eea
where the bosonic modes on the cylinder are defined in (\ref{BosonCylinderMode after}). We have also normalized the state to unity as one can check. 
To connect the states on the CFT side to physical states on the gravity side we note that the above graviton states can be written as linear combinations of physical states such as the graviton, the RR two form B-field, and the dilaton as described in Appendix (\ref{Field Definitions}). Any linear combination of physical states are also physical states.



For the second possibility we choose two excitations in the initial state which is just two factors of the single excitation (\ref{single excitation}). Therefore a state with two excitations is also a physical state.
Symmetrizing in a similar manner as for the single excitation, the state with two excitations is given by 
\bea
|\Psi_2\rangle &=& {1\over2}{1\over n_1n_2}\a^{(1)i}_{++,-n_1}\a^{(1)i}_{--,-n_2}|0\rangle^{(1)}|0\rangle^{(2)}\bar{\a}^{(1)i}_{--,-n_1}\bar{\a}^{(1)i}_{++,-n_2}|\bar{0}\rangle^{(1)}|\bar{0}\rangle^{(2)} \cr
&&\!\!\!\!+ {1\over2}{1\over n_1n_2}\a^{(1)i}_{++,-n_1}\a^{(2)i}_{--,-n_2}|0\rangle^{(1)}|0\rangle^{(2)}  \bar{\a}^{(1)i}_{--,-n_1}\bar{\a}^{(2)i}_{++,-n_2}|\bar{0}\rangle^{(1)}|\bar{0}\rangle^{(2)}\cr
&&\!\!\!\!+ {1\over2}{1\over n_1n_2}\a^{(2)i}_{++,-n_1}\a^{(1)i}_{--,-n_2}|0\rangle^{(1)}|0\rangle^{(2)}  \bar{\a}^{(2)i}_{--,-n_1}\bar{\a}^{(1)i}_{++,-n_2}|\bar{0}\rangle^{(1)}|\bar{0}\rangle^{(2)}\cr
&&\!\!\!\!+ {1\over2}{1\over n_1n_2}\a^{(2)i}_{++,-n_1}\a^{(2)i}_{--,-n_2}|0\rangle^{(1)}|0\rangle^{(2)}  \bar{\a}^{(2)i}_{--,-n_1}\bar{\a}^{(2)i}_{++,-n_2}|\bar{0}\rangle^{(1)}|\bar{0}\rangle^{(2)}
\label{initial states}
\eea
with
\bea
h=n_1 + n_2,~~j=0\qquad \bar{h} = n_1 + n_2,~~\bar{j}=0
\eea

We consider the splitting of these initial states into lower energy modes in the final state. The final states we are interested in are the following:

\bea
&&\text{3 Bosons}\cr
\langle \Phi_1|&=&{1\over\sqrt{2}}{1\over pqr}{}^{(1)}\langle \bar{0}|{}^{(2)}\langle \bar{0}|\bar{\a}^{(1)f}_{--,p}\bar{\a}^{(1)f}_{++,q}\bar{\a}^{(1)f}_{--,r}{}^{(1)}\langle 0|{}^{(2)}\langle 0|\a^{(1)f}_{++,p}\a^{(1)f}_{--,q}\a^{(1)f}_{++,r}\cr
&&+{1\over\sqrt{2}}{1\over pqr}{}^{(1)}\langle \bar{0}|{}^{(2)}\langle \bar{0}|\bar{\a}^{(2)f}_{--,p}\bar{\a}^{(2)f}_{++,q}\bar{\a}^{(2)f}_{--,r}{}^{(1)}\langle 0|{}^{(2)}\langle 0|\a^{(2)f}_{++,p}\a^{(2)f}_{--,q}\a^{(2)f}_{++,r}\cr
\cr
\cr
&&\text{1 Boson 2 Fermions}\cr
\langle \Phi_2|&=&{1\over\sqrt{2}}{1\over p}{}^{(1)}\langle \bar{0}|{}^{(2)}\langle \bar{0}|\bar{\a}^{(1)f}_{--,p}\bar{d}^{(1)f,-+}_q\bar{d}^{(1)f,+-}_r{}^{(1)}\langle 0|{}^{(2)}\langle 0|\a^{(1)f}_{++,p}d^{(1)f,+-}_qd^{(1)f,-+}_r\cr
&&+{1\over\sqrt{2}}{1\over p}{}^{(1)}\langle \bar{0}|{}^{(2)}\langle \bar{0}|\bar{\a}^{(2)f}_{--,p}\bar{d}^{(2)f,-+}_q\bar{d}^{(2)f,+-}_r{}^{(1)}\langle 0|{}^{(2)}\langle 0|\a^{(2)f}_{++,p}d^{(1)f,+-}_qd^{(1)f,-+}_r
\cr
\cr
\cr
&&\text{4 Bosons}\cr
\langle \Phi_3|&=&{1\over\sqrt{2}}{1\over pqrs}{}^{(1)}\langle \bar{0}|{}^{(2)}\langle \bar{0}|\bar{\a}^{(1)f}_{--,p}\bar{\a}^{(1)f}_{++,q}\bar{\a}^{(1)f}_{--,r}\bar{\a}^{(1)f}_{++,s}\cr
&&\qquad\qquad\qquad{}^{(1)}\langle 0|{}^{(2)}\langle 0|\a^{(1)f}_{++,p}\a^{(1)f}_{--,q}\a^{(1)f}_{++,r}\a^{(1)f}_{--,s}\cr
&&+{1\over\sqrt{2}}{1\over pqrs}{}^{(1)}\langle \bar{0}|{}^{(2)}\langle \bar{0}|\bar{\a}^{(2)f}_{--,p}\bar{\a}^{(2)f}_{++,q}\bar{\a}^{(2)f}_{--,r}\bar{\a}^{(2)f}_{++,s}\cr
&&\qquad\qquad\qquad{}^{(1)}\langle 0|{}^{(2)}\langle 0|\a^{(2)f}_{++,p}\a^{(2)f}_{--,q}|\a^{(2)f}_{++,p}\a^{(2)f}_{--,q}
\cr
\cr
\cr
&&\text{2 Bosons 2 Fermions}\cr
\langle \Phi_4|&=&{1\over\sqrt{2}}{1\over pq}{}^{(1)}\langle \bar{0}|{}^{(2)}\langle \bar{0}|\bar{\a}^{(1)f}_{--,p}\bar{\a}^{(1)f}_{++,q}\bar{d}^{(1)f,-+}_r\bar{d}^{(1)f,+-}_s\cr
&&\qquad\qquad\qquad{}^{(1)}\langle 0|{}^{(2)}\langle 0|\a^{(1)f}_{++,p}\a^{(1)f}_{--,q}d^{(1)f,+-}_rd^{(1)f,-+}_s\cr
&&+{1\over\sqrt{2}}{1\over pq}{}^{(1)}\langle \bar{0}|{}^{(2)}\langle \bar{0}|\bar{\a}^{(2)f}_{--,p}\bar{\a}^{(2)f}_{++,q}\bar{d}^{(2)f,-+}_q\bar{d}^{(2)f,+-}_r\cr
&&\qquad\qquad\qquad{}^{(1)}\langle 0|{}^{(2)}\langle 0|\a^{(2)f}_{++,p}\a^{(2)f}_{--,q}d^{(1)f,+-}_rd^{(1)f,-+}_s\nn
\label{final states}
\eea
For the case with one boson in the initial state we want to compute the amplitude for splitting into three modes in the final state. For the case with two bosons in the initial state we want to compute the amplitude for splitting into four modes in the final state. The number of modes in each case are chosen so to produce a nonzero amplitude.

Now combining the initial states, (\ref{initial states}), final states, (\ref{final states}), and our deformation operator (\ref{amplitude two}), we obtain the following amplitudes:
\bea
&&\mathcal{A}^{\a\to \a\a\a}(w_2,w_1,\bar{w}_2,\bar{w}_1)\cr
&&\quad=\e^{\dot{C}\dot{D}}\e^{\dot{A}\dot{B}}\langle \Phi_1|\big(G^+_{\dot{C},-{1\over2}}\s^-(w_2)G^-_{\dot{A},-{1\over2}}\s^+(w_1)\big)\big(\bar{G}^+_{\dot{D},-{1\over2}}\bar{\s}^-(\bar{w}_2)\bar{G}^-_{\dot{B},-{1\over2}}\bar{\s}^+(\bar{w}_1)\big)|\Psi_1\rangle\cr
&&\quad ={1\over 2 npqr}\e^{\dot{C}\dot{D}}\e^{\dot{A}\dot{B}}\mathcal{A}^{\a^{(1)}\to \a^{(1)}\a^{(1)}\a^{(1)}}_{\dot{C}\dot{A}}(w_2,w_1)\bar{\mathcal{A}}^{\a^{(1)}\to \a^{(1)}\a^{(1)}\a^{(1)}}_{\dot{D}\dot{B}}(\bar{w}_2,\bar{w}_1)\cr
&&\qquad + {1\over 2 npqr}\e^{\dot{C}\dot{D}}\e^{\dot{A}\dot{B}}\mathcal{A}^{\a^{(2)}\to \a^{(1)}\a^{(1)}\a^{(1)}}_{\dot{C}\dot{A}}(w_2,w_1)\bar{\mathcal{A}}^{\a^{(2)}\to \a^{(1)}\a^{(1)}\a^{(1)}}_{\dot{D}\dot{B}}(\bar{w}_2,\bar{w}_1)\cr
&&\qquad + {1\over 2 npqr}\e^{\dot{C}\dot{D}}\e^{\dot{A}\dot{B}}\mathcal{A}^{\a^{(1)}\to \a^{(2)}\a^{(2)}\a^{(2)}}_{\dot{C}\dot{A}}(w_2,w_1)\bar{\mathcal{A}}^{\a^{(1)}\to \a^{(2)}\a^{(2)}\a^{(2)}}_{\dot{D}\dot{B}}(\bar{w}_2,\bar{w}_1)\cr
&&\qquad + {1\over 2 npqr}\e^{\dot{C}\dot{D}}\e^{\dot{A}\dot{B}}\mathcal{A}^{\a^{(2)}\to \a^{(2)}\a^{(2)}\a^{(2)}}_{\dot{C}\dot{A}}(w_2,w_1)\bar{\mathcal{A}}^{\a^{(2)}\to \a^{(2)}\a^{(2)}\a^{(2)}}_{\dot{D}\dot{B}}(\bar{w}_2,\bar{w}_1)\cr
\cr
&&\mathcal{A}^{\a\to \a dd}(w_2,w_1,\bar{w}_2,\bar{w}_1)\cr
&&\quad=\e^{\dot{C}\dot{D}}\e^{\dot{A}\dot{B}}\langle \Phi_2|\big(G^+_{\dot{C},-{1\over2}}\s^-(w_2)G^-_{\dot{A},-{1\over2}}\s^+(w_1)\big)\big(\bar{G}^+_{\dot{D},-{1\over2}}\bar{\s}^-(\bar{w}_2)\bar{G}^-_{\dot{B},-{1\over2}}\bar{\s}^+(\bar{w}_1)\big)|\Psi_2\rangle\cr
&&\quad ={1\over 2 np}\e^{\dot{C}\dot{D}}\e^{\dot{A}\dot{B}}\mathcal{A}^{\a^{(1)}\to \a^{(1)}d^{(1)}d^{(1)}}_{\dot{C}\dot{A}}(w_2,w_1)\bar{\mathcal{A}}^{\a^{(1)}\to \a^{(1)}d^{(1)}d^{(1)}}_{\dot{D}\dot{B}}(\bar{w}_2,\bar{w}_1)\cr
&&\qquad +{1\over 2 np}\e^{\dot{C}\dot{D}}\e^{\dot{A}\dot{B}}\mathcal{A}^{\a^{(2)}\to \a^{(1)}d^{(1)}d^{(1)}}_{\dot{C}\dot{A}}(w_2,w_1)\bar{\mathcal{A}}^{\a^{(2)}\to \a^{(1)}d^{(1)}d^{(1)}}_{\dot{D}\dot{B}}(\bar{w}_2,\bar{w}_1)\cr
&&\qquad + {1\over 2 np}\e^{\dot{C}\dot{D}}\e^{\dot{A}\dot{B}}\mathcal{A}^{\a^{(1)}\to \a^{(2)}d^{(2)}d^{(2)}}_{\dot{C}\dot{A}}(w_2,w_1)\bar{\mathcal{A}}^{\a^{(1)}\to \a^{(2)}d^{(2)}d^{(2)}}_{\dot{D}\dot{B}}(\bar{w}_2,\bar{w}_1)\cr
&&\qquad + {1\over 2 np}\e^{\dot{C}\dot{D}}\e^{\dot{A}\dot{B}}\mathcal{A}^{\a^{(2)}\to \a^{(2)}d^{(2)}d^{(2)}}_{\dot{C}\dot{A}}(w_2,w_1)\bar{\mathcal{A}}^{\a^{(2)}\to \a^{(2)}d^{(2)}d^{(2)}}_{\dot{D}\dot{B}}(\bar{w}_2,\bar{w}_1)
\cr
\cr
\cr
&&\mathcal{A}^{\a\a\to \a\a \a\a}(w_2,w_1,\bar{w}_2,\bar{w}_1)\cr
&&\quad=\e^{\dot{C}\dot{D}}\e^{\dot{A}\dot{B}}\langle \Phi_3|\big(G^+_{\dot{C},-{1\over2}}\s^-(w_2)G^-_{\dot{A},-{1\over2}}\s^+(w_1)\big)\big(\bar{G}^+_{\dot{D},-{1\over2}}\bar{\s}^-(\bar{w}_2)\bar{G}^-_{\dot{B},-{1\over2}}\bar{\s}^+(\bar{w}_1)\big)|\Psi_2\rangle\cr
&&\quad ={1\over 2\sqrt2 n_1n_2pqrs}\e^{\dot{C}\dot{D}}\e^{\dot{A}\dot{B}}\mathcal{A}^{\a^{(1)}\a^{(1)}\to \a^{(1)}\a^{(1)}\a^{(1)}\a^{(1)}}_{\dot{C}\dot{A}}(w_2,w_1)\bar{\mathcal{A}}^{\a^{(1)}\a^{(1)}\to \a^{(1)}\a^{(1)}\a^{(1)}\a^{(1)}}_{\dot{D}\dot{B}}(\bar{w}_2,\bar{w}_1)\cr
&&\qquad +{1\over 2\sqrt2 n_1n_2pqrs} \e^{\dot{C}\dot{D}}\e^{\dot{A}\dot{B}}\mathcal{A}^{\a^{(1)}\a^{(2)}\to \a^{(1)}\a^{(1)}\a^{(1)}\a^{(1)}}_{\dot{C}\dot{A}}(w_2,w_1)\bar{\mathcal{A}}^{\a^{(1)}\a^{(2)}\to \a^{(1)}\a^{(1)}\a^{(1)}\a^{(1)}}_{\dot{D}\dot{B}}(\bar{w}_2,\bar{w}_1)\cr
&&\qquad + {1\over 2\sqrt2 n_1n_2pqrs}\e^{\dot{C}\dot{D}}\e^{\dot{A}\dot{B}}\mathcal{A}^{\a^{(2)}\a^{(1)}\to \a^{(1)}\a^{(1)}\a^{(1)}\a^{(1)}}_{\dot{C}\dot{A}}(w_2,w_1)\bar{\mathcal{A}}^{\a^{(2)}\a^{(1)}\to \a^{(1)}\a^{(1)}\a^{(1)}\a^{(1)}}_{\dot{D}\dot{B}}(\bar{w}_2,\bar{w}_1)\cr
&&\qquad + {1\over 2\sqrt2 n_1n_2pqrs}\e^{\dot{C}\dot{D}}\e^{\dot{A}\dot{B}}\mathcal{A}^{\a^{(2)}\a^{(2)}\to \a^{(1)}\a^{(1)}\a^{(1)}\a^{(1)}}_{\dot{C}\dot{A}}(w_2,w_1)\bar{\mathcal{A}}^{\a^{(2)}\a^{(2)}\to \a^{(1)}\a^{(1)}\a^{(1)}\a^{(1)}}_{\dot{D}\dot{B}}(\bar{w}_2,\bar{w}_1)\cr
&&\qquad + {1\over 2\sqrt2 n_1n_2pqrs}\e^{\dot{C}\dot{D}}\e^{\dot{A}\dot{B}}\mathcal{A}^{\a^{(1)}\a^{(1)}\to \a^{(2)}\a^{(2)}\a^{(2)}\a^{(2)}}_{\dot{C}\dot{A}}(w_2,w_1)\bar{\mathcal{A}}^{\a^{(1)}\a^{(1)}\to \a^{(2)}\a^{(2)}\a^{(2)}\a^{(2)}}_{\dot{D}\dot{B}}(\bar{w}_2,\bar{w}_1)\cr
&&\qquad + {1\over 2\sqrt2 n_1n_2pqrs}\e^{\dot{C}\dot{D}}\e^{\dot{A}\dot{B}}\mathcal{A}^{\a^{(1)}\a^{(2)}\to \a^{(2)}\a^{(2)}\a^{(2)}\a^{(2)}}_{\dot{C}\dot{A}}(w_2,w_1)\bar{\mathcal{A}}^{\a^{(1)}\a^{(2)}\to \a^{(2)}\a^{(2)}\a^{(2)}\a^{(2)}}_{\dot{D}\dot{B}}(\bar{w}_2,\bar{w}_1)\cr
&&\qquad + {1\over 2\sqrt2 n_1n_2pqrs}\e^{\dot{C}\dot{D}}\e^{\dot{A}\dot{B}}\mathcal{A}^{\a^{(2)}\a^{(1)}\to \a^{(2)}\a^{(2)}\a^{(2)}\a^{(2)}}_{\dot{C}\dot{A}}(w_2,w_1)\bar{\mathcal{A}}^{\a^{(2)}\a^{(1)}\to \a^{(2)}\a^{(2)}\a^{(2)}\a^{(2)}}_{\dot{D}\dot{B}}(\bar{w}_2,\bar{w}_1)\cr
&&\qquad + {1\over 2\sqrt2 n_1n_2pqrs}\e^{\dot{C}\dot{D}}\e^{\dot{A}\dot{B}}\mathcal{A}^{\a^{(2)}\a^{(2)}\to \a^{(2)}\a^{(2)}\a^{(2)}\a^{(2)}}_{\dot{C}\dot{A}}(w_2,w_1)\bar{\mathcal{A}}^{\a^{(2)}\a^{(2)}\to \a^{(2)}\a^{(2)}\a^{(2)}\a^{(2)}}_{\dot{D}\dot{B}}(\bar{w}_2,\bar{w}_1)
\cr
\cr
\cr
&&\mathcal{A}^{\a\a\to \a\a dd}(w_2,w_1,\bar{w}_2,\bar{w}_1)\cr
&&\quad=\e^{\dot{C}\dot{D}}\e^{\dot{A}\dot{B}}\langle \Phi_4|\big(G^+_{\dot{C},-{1\over2}}\s^-(w_2)G^-_{\dot{A},-{1\over2}}\s^+(w_1)\big)\big(\bar{G}^+_{\dot{D},-{1\over2}}\bar{\s}^-(\bar{w}_2)\bar{G}^-_{\dot{B},-{1\over2}}\bar{\s}^+(\bar{w}_1)\big)|\Psi_2\rangle\cr
&&\quad ={1\over 2\sqrt2 n_1n_2pqrs}\e^{\dot{C}\dot{D}}\e^{\dot{A}\dot{B}}\mathcal{A}^{\a^{(1)}\a^{(1)}\to \a^{(1)}\a^{(1)}d^{(1)}d^{(1)}}_{\dot{C}\dot{A}}(w_2,w_1)\bar{\mathcal{A}}^{\a^{(1)}\a^{(1)}\to \a^{(1)}\a^{(1)}d^{(1)}d^{(1)}}_{\dot{D}\dot{B}}(\bar{w}_2,\bar{w}_1)\cr
&&\qquad +{1\over 2\sqrt2 n_1n_2pqrs} \e^{\dot{C}\dot{D}}\e^{\dot{A}\dot{B}}\mathcal{A}^{\a^{(1)}\a^{(2)}\to \a^{(1)}\a^{(1)}d^{(1)}d^{(1)}}_{\dot{C}\dot{A}}(w_2,w_1)\bar{\mathcal{A}}^{\a^{(1)}\a^{(2)}\to \a^{(1)}\a^{(1)}d^{(1)}d^{(1)}}_{\dot{D}\dot{B}}(\bar{w}_2,\bar{w}_1)\cr
&&\qquad + {1\over 2\sqrt2 n_1n_2pqrs}\e^{\dot{C}\dot{D}}\e^{\dot{A}\dot{B}}\mathcal{A}^{\a^{(2)}\a^{(1)}\to \a^{(1)}\a^{(1)}d^{(1)}d^{(1)}}_{\dot{C}\dot{A}}(w_2,w_1)\bar{\mathcal{A}}^{\a^{(2)}\a^{(1)}\to \a^{(1)}\a^{(1)}d^{(1)}d^{(1)}}_{\dot{D}\dot{B}}(\bar{w}_2,\bar{w}_1)\cr
&&\qquad + {1\over 2\sqrt2 n_1n_2pqrs}\e^{\dot{C}\dot{D}}\e^{\dot{A}\dot{B}}\mathcal{A}^{\a^{(2)}\a^{(2)}\to \a^{(1)}\a^{(1)}d^{(1)}d^{(1)}}_{\dot{C}\dot{A}}(w_2,w_1)\bar{\mathcal{A}}^{\a^{(2)}\a^{(2)}\to \a^{(1)}\a^{(1)}d^{(1)}d^{(1)}}_{\dot{D}\dot{B}}(\bar{w}_2,\bar{w}_1)\cr
&&\qquad + {1\over 2\sqrt2 n_1n_2pqrs}\e^{\dot{C}\dot{D}}\e^{\dot{A}\dot{B}}\mathcal{A}^{\a^{(1)}\a^{(1)}\to \a^{(2)}\a^{(2)}d^{(2)}d^{(2)}}_{\dot{C}\dot{A}}(w_2,w_1)\bar{\mathcal{A}}^{\a^{(1)}\a^{(1)}\to \a^{(2)}\a^{(2)}d^{(2)}d^{(2)}}_{\dot{D}\dot{B}}(\bar{w}_2,\bar{w}_1)\cr
&&\qquad + {1\over 2\sqrt2 n_1n_2pqrs}\e^{\dot{C}\dot{D}}\e^{\dot{A}\dot{B}}\mathcal{A}^{\a^{(1)}\a^{(2)}\to \a^{(2)}\a^{(2)}d^{(2)}d^{(2)}}_{\dot{C}\dot{A}}(w_2,w_1)\bar{\mathcal{A}}^{\a^{(1)}\a^{(2)}\to \a^{(2)}\a^{(2)}d^{(2)}d^{(2)}}_{\dot{D}\dot{B}}(\bar{w}_2,\bar{w}_1)\cr
&&\qquad + {1\over 2\sqrt2 n_1n_2pqrs}\e^{\dot{C}\dot{D}}\e^{\dot{A}\dot{B}}\mathcal{A}^{\a^{(2)}\a^{(1)}\to \a^{(2)}\a^{(2)}d^{(2)}d^{(2)}}_{\dot{C}\dot{A}}(w_2,w_1)\bar{\mathcal{A}}^{\a^{(2)}\a^{(1)}\to \a^{(2)}\a^{(2)}d^{(2)}d^{(2)}}_{\dot{D}\dot{B}}(\bar{w}_2,\bar{w}_1)\cr
&&\qquad + {1\over 2\sqrt2 n_1n_2pqrs}\e^{\dot{C}\dot{D}}\e^{\dot{A}\dot{B}}\mathcal{A}^{\a^{(2)}\a^{(2)}\to \a^{(2)}\a^{(2)}d^{(2)}d^{(2)}}_{\dot{C}\dot{A}}(w_2,w_1)\bar{\mathcal{A}}^{\a^{(2)}\a^{(2)}\to \a^{(2)}\a^{(2)}d^{(2)}d^{(2)}}_{\dot{D}\dot{B}}(\bar{w}_2,\bar{w}_1)
\label{all amplitudes}
\eea
Where


\bea
&&\!\!\!\!\!\!\!\!\!\!\mathcal{A}^{\a^{(j)} \to \a^{(k)}\a^{(k)}\a^{(k)}}_{\dot{C}\dot{A}}(w_2,w_1)\cr
&&\!\!\!\!\!\!\!\!\!= {}^{(1)}\langle 0|{}^{(2)}\langle 0|\a^{(k)f}_{++,p}\a^{(k)f}_{--,q}\a^{(k)f}_{++,r}G^+_{\dot{C},-{1\over2}}\s^-_{2}(w_2)G^-_{\dot{A},-{1\over2}}\s^+_{2}(w_1)\a^{(j)i}_{--,-n}|0\rangle^{(1)}|0\rangle^{(2)}
\cr
&&\!\!\!\!\!\!\!\!\!\!\bar{\mathcal{A}}^{\bar{\a}^{(j)} \to \bar{\a}^{(k)}\bar{\a}^{(k)}\bar{\a}^{(k)}}_{\dot{D}\dot{B}}(\bar{w}_2,\bar{w}_1)\cr
&&\!\!\!\!\!\!\!\!\!= {}^{(1)}\langle \bar{0}|{}^{(2)}\langle \bar{0}|\bar{\a}^{(k)f}_{--,p}\bar{\a}^{(k)f}_{++,q}\bar{\a}^{(k)f}_{--,r}\bar{G}^+_{\dot{D},-{1\over2}}\bar{\s}^-_{2}(\bar{w}_2)\bar{G}^-_{\dot{B},-{1\over2}}\bar{\s}^+_{2}(\bar{w}_1)\bar{\a}^{(j)i}_{++,-n}|\bar{0}\rangle^{(1)}|\bar{0}\rangle^{(2)}
\cr
\cr
\cr
&&\!\!\!\!\!\!\!\!\!\!\mathcal{A}^{\a^{(j)} \to \a^{(k)}d^{(k)}d^{(k)}}_{\dot{C}\dot{A}}(w_2,w_1)\cr
&&\!\!\!\!\!\!\!\!\!= {}^{(1)}\langle 0|{}^{(2)}\langle 0|\a^{(k)f}_{++,p}d^{(k)f,+-}_{q}d^{(k)f,-+}_{r}G^+_{\dot{C},-{1\over2}}\s^-_{2}(w_2)G^-_{\dot{A},-{1\over2}}\s^+_{2}(w_1)\a^{(j)i}_{--,-n}|0\rangle^{(1)}|0\rangle^{(2)}
\cr
&&\!\!\!\!\!\!\!\!\!\!\bar{\mathcal{A}}^{\bar{\a}^{(j)} \to \bar{\a}^{(k)}\bar{d}^{(k)}\bar{d}^{(k)}}_{\dot{D}\dot{B}}(\bar{w}_2,\bar{w}_1)\cr
&&\!\!\!\!\!\!\!\!\!= {}^{(1)}\langle \bar{0}|{}^{(2)}\langle \bar{0}|\bar{\a}^{(k)f}_{--,p}\bar{d}^{(k)f,-+}_{q}\bar{d}^{(k)f,+-}_{r}\bar{G}^+_{\dot{D},-{1\over2}}\bar{\s}^-_{2}(\bar{w}_2)\bar{G}^-_{\dot{B},-{1\over2}}\bar{\s}^+_{2}(w_1)\bar{\a}^{(j)i}_{++,-n}|\bar{0}\rangle^{(1)}|\bar{0}\rangle^{(2)}
\cr
\cr
\cr
&&\!\!\!\!\!\!\!\!\!\!\mathcal{A}^{\a^{(j)}\a^{(k)} \to \a^{(l)}\a^{(l)}\a^{(l)}\a^{(l)}}_{\dot{C}\dot{A}}(w_2,w_1)\cr
&&\!\!\!\!\!\!\!\!\!= {}^{(1)}\langle 0|{}^{(2)}\langle 0|\a^{(l)f}_{++,p}\a^{(l)f}_{--,q}\a^{(l)f}_{++,r}\a^{(l)f}_{--,s}G^+_{\dot{C},-{1\over2}}\s^-_{2}(w_2)G^-_{\dot{A},-{1\over2}}\s^+_{2}(w_1)\a^{(j)i}_{++,-n_1}\a^{(k)i}_{--,-n_2}|0\rangle^{(1)}|0\rangle^{(2)}\cr
&&\!\!\!\!\!\!\!\!\!\!\bar{\mathcal{A}}^{\bar{\a}^{(j)}\bar{\a}^{(k)} \to \bar{\a}^{(l)}\bar{\a}^{(l)}\bar{\a}^{(l)}\bar{\a}^{(l)}}_{\dot{C}\dot{A}}(\bar{w}_2,\bar{w}_1)\cr
&&\!\!\!\!\!\!\!\!\!= {}^{(1)}\langle \bar{0}|{}^{(2)}\langle \bar{0}|\bar{\a}^{(l)f}_{--,p}\bar{\a}^{(l)f}_{++,q}\bar{\a}^{(l)f}_{--,r}\bar{\a}^{(l)f}_{++,s}\bar{G}^+_{\dot{D},-{1\over2}}\bar{\s}^-_{2}(\bar{w}_2)\bar{G}^-_{\dot{B},-{1\over2}}\bar{\s}^+_{2}(\bar{w}_1)\bar{\a}^{(j)i}_{--,-n_1}\bar{\a}^{(k)i}_{++,-n_2}|\bar{0}\rangle^{(1)}|\bar{0}\rangle^{(2)}
\cr
\cr
\cr
&&\!\!\!\!\!\!\!\!\!\!\mathcal{A}^{\a^{(j)}\a^{(k)} \to \a^{(l)}\a^{(l)}d^{(l)}d^{(l)}}_{\dot{C}\dot{A}}(w_2,w_1)\cr
&&\!\!\!\!\!\!\!\!\!= {}^{(1)}\langle 0|{}^{(2)}\langle 0|\a^{(l)f}_{++,p}\a^{(l)f}_{--,q}d^{(l)f,+-}_{r}d^{(l)f,-+}_{s}G^+_{\dot{C},-{1\over2}}\s^-_{2}(w_2)G^-_{\dot{A},-{1\over2}}\s^+_{2}(w_1)\a^{(j)i}_{++,-n_1}\a^{(k)i}_{--,-n_2}|0\rangle^{(1)}|0\rangle^{(2)}\cr
&&\!\!\!\!\!\!\!\!\!\!\bar{\mathcal{A}}^{\bar{\a}^{(j)}\bar{\a}^{(k)} \to \bar{\a}^{(l)}\bar{\a}^{(l)}\bar{d}^{(l)}\bar{d}^{(l)}}_{\dot{C}\dot{A}}(\bar{w}_2,\bar{w}_1)\cr
&&\!\!\!\!\!\!\!\!\!= {}^{(1)}\langle \bar{0}|{}^{(2)}\langle \bar{0}|\bar{\a}^{(l)f}_{--,p}\bar{\a}^{(l)f}_{++,q}\bar{d}^{(l)f,-+}_{r}\bar{d}^{(l)f,+-}_{s}\bar{G}^+_{\dot{D},-{1\over2}}\bar{\s}^-_{2}(\bar{w}_2)\bar{G}^-_{\dot{B},-{1\over2}}\bar{\s}^+_{2}(\bar{w}_1)\bar{\a}^{(j)i}_{--,-n_1}\bar{\a}^{(k)i}_{++,-n_2}|\bar{0}\rangle^{(1)}|\bar{0}\rangle^{(2)}\nn
\label{amplitudes}
\eea
where the superscript $\a^{(j)}$ on the left hand side represents a bosonic mode along copy $j$ and the superscript $d^{(k)}$ represents a fermionic mode along copy $k$. The full superscript labels on the amplitudes represent the corresponding scattering processes being computed. For example, the superscript
\bea
\a^{(1)}\to\a^{(1)}\a^{(1)}\a^{(1)}
\eea
indicates that we are looking at the amplitude for one initial boson on copy 1 to split into three final bosons on copy 1 in the left moving sector, etc. 

To compute these amplitudes we will need to map them to the $t$ plane. In doing so, our amplitude will consist of several factors. There will be factors coming from the base amplitude, factors coming from lifting the $G$ contours to the cover, and a $t$ plane amplitude containing the modes which have been mapped from the cylinder to the $t$ plane and undergone the necessary spectral flows to remove all spin fields. Schematically, our result will look like
\bea
\mathcal{A}^{i\to f}_{\dot{C}\dot{A}}=A_{base}f_1f_2\mathcal{A}^{i\to f}_{t,\dot{C}\dot{A}}
\label{holo amplitude}
\eea
where $f_1$ and $f_2$ are the factors coming from lifting the $G$ contours circling $w_1,w_2$ on the cylinder to the $t$ plane where they circle points $t_1,t_2$. In the following subsections we compute these various factors. Similarly, the antiholomorphic component is
\bea
\bar{\mathcal{A}}^{\bar{i}\to \bar{f}}_{\dot{D}\dot{B}}=\bar{A}_{base}\bar{f}_1\bar{f}_2\bar{\mathcal{A}}^{\bar{i}\to \bar{f}}_{\bar t,\dot{D}\dot{B}}
\label{anti holo amplitude}
\eea

\section{The map}\label{map}
Here we recall the map for the two twist case. We first map the cylinder labeled by $w$ to the complex plane labeled by $z$:
\be
z=e^w\ .
\ee
We then map this plane to its covering space where the twist operators are resolved, see \cite{lm1} for the details of the covering space analyses. This is also performed in Chapter's 6 and 7 as well. We consider the map
 \be
z={(t+a)(t+b)\over t}\ .
\label{amap two}
\ee
We have
\be
{dz\over dt} = 1-{ab\over t^2}\ .
\ee
The twist operators correspond to the locations given by ${dz\over dt}=0$; i.e. the points
\bea
&&t_1=-\sqrt{ab}\ ,\qquad z_1=e^{w_1}=(\sqrt{a}-\sqrt{b})^2\ ,\\
&&t_2=\sqrt{ab}\ ,\qquad\;\;\, z_2=e^{w_2}=(\sqrt{a}+\sqrt{b})^2\ .
\eea
Note that
\be
{dz\over dt} = {(t-t_1)(t-t_2)\over t^2}\ .
\ee
We define
\be
\Delta w = w_2-w_1\ ,
\ee
\be
s={\h}(w_1+w_2)\ ,
\ee
Then we find
\be
a=  e^{s} \cosh^2 (\tfrac{\Delta w }{4}), ~~~b=  e^{s} \sinh^2 (\tfrac{\Delta w }{4})\ .
\ee
It will be useful to note the relations
\be
a-b=e^{s}, ~~~z_1z_2=e^{2s}, ~~~z_1-z_2= -2 e^{s} \sinh (\tfrac{\Delta w }{2})\ .
\ee

We have similar relations for antiholomorphic quantities
\be
\bar{z}=e^{\bar{w}}\ .
\ee
\be
\bar{z} = {(\bar t + \bar a)(\bar t + \bar b)\over \bar t}
\ee
\bea
&&\bar{t}_1=-\sqrt{\bar{a}\bar b}\ ,\qquad \bar z_1=e^{\bar w_1}=(\sqrt{\bar {a}}-\sqrt{\bar{b}})^2\ . \cr
&&\bar{t}_2=\sqrt{\bar a\bar b}\ ,\qquad\;\;\, \bar z_2=e^{\bar w_2}=(\sqrt{\bar a}+\sqrt{\bar b})^2\ .
\eea
We have
\be
\Delta \bar{w} = \bar w_2- \bar w_1\ ,
\ee
\be
\bar s={\h}(\bar w_1+ \bar w_2)\ ,
\ee
and
\be
\bar a=  e^{\bar s} \cosh^2 (\tfrac{\Delta \bar{w} }{4}), ~~~\bar b=  e^{\bar s} \sinh^2 (\tfrac{\Delta \bar{w} }{4})\ .
\ee
\be
\bar a-\bar b=e^{\bar s}, ~~~\bar z_1\bar z_2=e^{2\bar s}, ~~~\bar z_1-\bar z_2= -2 e^{\bar s} \sinh (\tfrac{\Delta \bar{w} }{2})\ .
\ee

\section{Base Amplitude}
First, we'll have to compute what we call the base amplitude which comes from the twist insertions alone:
\bea
A_{base} =  {}^{(1)}\langle 0|{}^{(2)}\langle 0|\s^-(w_2)\s^+(w_1)|0\rangle^{(1)}|0\rangle^{(2)}
\eea
We can compute this by going to the $z$ plane.
\bea
A_{base} &=&{}^{(1)}\langle 0|{}^{(2)}\langle 0|({dz\over dw})^{{1\over 2}}(z_2)\s^-(z_2)({dz\over dw})^{{1\over 2}}(z_1)\s^+(z_1)|0\rangle^{(1)}|0\rangle^{(2)}\cr
&=&z_1^{{1\over2}}z_2^{{1\over2}}{1 \over z_2 - z_1}
\label{base}
\eea
Next we compute the contribution of
\bea
G^{+}_{\dot{C},-{1\over 2}}\s^-_2(w_2)G^{-}_{\dot{A},-{1\over 2}}\s^+_2(w_1)
\eea
when lifting to the cover. 
\section{The point $w_1$}
At point the point $w_1$ on the cylinder we have the following contour
\bea
G^{-}_{\dot{A},-{1\over 2}}\s^+_2(w_1) = {1\over 2\pi i}\int_{w_1} dw G^-_{\dot{A}}(w)\s^+_{2}(w_1)
\label{G1 cylinder}
\eea
Mapping (\ref{G1 cylinder}) to the $z$ plane using the map $z=e^w$ yields
\bea
{1\over 2\pi i}\int_{w_1} dw G^-_{\dot{A}}(w)\s^+_{2}(w_1) \to{1\over 2\pi i}\int_{z_1} dz z^{{1\over2}}G^-_{\dot{A}}(z) [z_1^{1/2}\s^+_{2}(z_1)]
\label{G1 z plane}
\eea
Mapping (\ref{G1 z plane}) to the $t$ plane using $z={(t+a)(t+b)\over t}$ yields
\bea
&&{1\over 2\pi i}\int_{z_1} dz z^{{1\over2}}G^-_{\dot{A}}(z) [z_1^{1/2}\s^+_{2}(z_1)]\cr
 &&\qquad\to{1\over 2\pi i}\int_{t_1} dt ({dz\over dt})^{-{1\over2}}z^{{1\over2}}G^-_{\dot{A}}(t) [z_1^{1/2}S^+(t_1)]\cr
&&\qquad={1\over 2\pi i}\int_{t_1} dt (t-t_1)^{-{1\over2}} (t-t_2)^{-{1\over2}}t^{{1\over2}}(t+a)^{{1\over2}}(t+b)^{{1\over2}}G^-_{\dot{A}}(t) [Cz_1^{1/2}S^+(t_1)]
\cr
&&\qquad= (t_1-t_2)^{-{1\over2}}t_1^{{1\over2}}(t_1+a)^{{1\over2}}(t_1+b)^{{1\over2}}{1\over 2\pi i}\int_{t_1} dt (t-t_1)^{-{1\over2}}G^-_{\dot{A}}(t) [Cz_1^{1/2}S^+(t_1)]
\cr
&&\qquad= (t_1-t_2)^{-{1\over2}}t_1^{{1\over2}}(t_1+a)^{{1\over2}}(t_1+b)^{{1\over2}}\tilde{G}^-_{\dot{A},-1}[Cz_1^{1/2}S^+(t_1)]
\eea
where $C$ is a normalization factor acquired by the spin field when mapping to the $t$ plane. 
The term in brackets will contribute to the base amplitude which we've already computed.
Where in the fourth line we've expanded the integrand and only kept the zeroth order term. This gives the factor 
\bea
(t_1-t_2)^{-{1\over2}}t_1^{{1\over2}}(t_1+a)^{{1\over2}}(t_1+b)^{{1\over2}}
\eea
Spectral flowing our result by $\a=-1$ at $t_1$ to remove the spin field $S^+(t_1)$ yields
\bea
&&(t_1-t_2)^{-{1\over2}}t_1^{{1\over2}}(t_1+a)^{{1\over2}}(t_1+b)^{{1\over2}}\tilde{G}^-_{\dot{A},-{3\over2}}[Cz_1^{1/2}|0\rangle_t]\cr
&&=(t_1-t_2)^{-{1\over2}}t_1^{{1\over2}}(t_1+a)^{{1\over2}}(t_1+b)^{{1\over2}}\tilde{G}^-_{\dot{A}}(t_1)[Cz_1^{1/2}]\nn
\eea

At the point $t_2$ where we'll spectral flow by $\a=+1$ to remove the spin field $S^-(t_2)$ we get a contribution for $G^-$ at $t_1$. Our expression becomes

\bea
&&(t_1 - t_2)^{{1\over2}}(t_1-t_2)^{-{1\over2}}t_1^{{1\over2}}(t_1+a)^{{1\over2}}(t_1+b)^{{1\over2}}\tilde{G}^-_{\dot{A}}(t_1)[Cz_1^{1/2}]\cr
&&\quad = t_1^{{1\over2}}(t_1+a)^{{1\over2}}(t_1+b)^{{1\over2}}\tilde{G}^-_{\dot{A}}(t_1)[Cz_1^{1/2}]
\eea

The factor that we'll need is 
\bea
f_1=t_1^{{1\over2}}(t_1+a)^{{1\over2}}(t_1+b)^{{1\over2}}
\label{G factor one}
\eea
\section{The point $w_2$}

At point $w_2$ we have the following
\bea
G^{+}_{\dot{C},-{1\over 2}}\s^-_2(w_2) = {1\over 2\pi i}\int_{w_2} dw G^+_{\dot{A}}(w)\s^-_{2}(w_2)
\label{G2 cylinder}
\eea
Mapping (\ref{G2 cylinder}) to the $z$ plane using $z=e^w$ yields
\bea
{1\over 2\pi i}\int_{w_2} dw G^+_{\dot{C}}(w)\s^-_{2}(w_2) \to{1\over 2\pi i}\int_{z_2} dz z^{{1\over2}}G^+_{\dot{C}}(z) [Cz_2^{1/2}\s^-_{2}(z_2)]
\label{G2 z plane}
\eea
Mapping (\ref{G2 z plane}) to the $t$ plane using $z={(t+a)(t+b)\over t}$ yields
\bea
&&{1\over 2\pi i}\int_{z_2} dz z^{{1\over2}}G^+_{\dot{C}}(z) [z_2^{1/2}\s^-_{2}(z_2)]\cr
 &&\qquad\to{1\over 2\pi i}\int_{t_2} dt ({dz\over dt})^{-{1\over2}}z^{{1\over2}}G^+_{\dot{C}}(t) [Cz_2^{1/2}S^-(t_2)]\cr
&&\qquad={1\over 2\pi i}\int_{t_2} dt (t-t_1)^{-{1\over2}} (t-t_2)^{-{1\over2}}t^{{1\over2}}(t+a)^{{1\over2}}(t+b)^{{1\over2}}G^+_{\dot{C}}(t) [Cz_2^{1/2}S^-(t_2)]
\cr
&&\qquad= (t_2-t_1)^{-{1\over2}} t_2^{{1\over2}}(t_2+a)^{{1\over2}}(t_2+b)^{{1\over2}}{1\over 2\pi i}\int_{t_2} dt (t-t_2)^{-{1\over2}}G^+_{\dot{C}}(t) [Cz_2^{1/2}S^-(t_2)]
\cr
&&\qquad = (t_2-t_1)^{-{1\over2}} t_2^{{1\over2}}(t_2+a)^{{1\over2}}(t_2+b)^{{1\over2}}\tilde{G}^+_{\dot{C},-1}[Cz_2^{1/2}S^-(t_2)]
\eea
where again the term in brackets will contribute to the base amplitude, which we've already computed.
In the fourth line we have again only taken the zeroth order term in the expansion. This gives the factor 
\bea
(t_2-t_1)^{-{1\over2}} t_2^{{1\over2}}(t_2+a)^{{1\over2}}(t_2+b)^{{1\over2}}
\eea

From spectral flowing around $t_1$ by $\a=-1$ the contribution at $t_2$ yields
\bea
&&(t_2-t_1)^{{1\over2}}(t_2-t_1)^{-{1\over2}} t_2^{{1\over2}}(t_2+a)^{{1\over2}}(t_2+b)^{{1\over2}}\tilde{G}^+_{\dot{C},-1}[Cz_2^{1/2}(t_2-t_1)^{-{1\over2}}S^-(t_2)]\cr
&&\quad = t_2^{{1\over2}}(t_2+a)^{{1\over2}}(t_2+b)^{{1\over2}}\tilde{G}^+_{\dot{C},-1}[Cz_2^{1/2}(t_2-t_1)^{-{1\over2}}S^-(t_2)]
\eea
Spectral flowing our result by $\a=1$ at $t_2$ yields
\bea
 t_2^{{1\over2}}(t_2+a)^{{1\over2}}(t_2+b)^{{1\over2}}\tilde{G}^+_{\dot{C}}(t_2)[C(t_2-t_1)^{-{1\over2}}z_2^{1/2}]
\eea

The factor that we'll need is 
\bea
f_2=t_2^{{1\over2}}(t_2+a)^{{1\over2}}(t_2+b)^{{1\over2}}
\label{G factor two}
\eea
\section{Combining $A_{base}$, $f_1$, and $f_2$}
Lets compute the expression for $A_{base}f_1f_2$ in terms of $\Delta w$. First, combining (\ref{base}), (\ref{G factor one}), and (\ref{G factor two}), we get
\bea
A_{base}f_1f_2 &=&{(z_1z_2)^{1\over2}\over z_2-z_1} t_1^{{1\over2}}(t_1+a)^{{1\over2}}(t_1+b)^{{1\over2}}t_2^{{1\over2}}(t_2+a)^{{1\over2}}(t_2+b)^{{1\over2}}
\eea
Using the relations in (\ref{map}) we  get
\bea
A_{base}f_1f_2&=&-{z_1z_2\over z_2-z_1} t_2^2\cr
&=&- {e^{2s} \over 2e^s\sinh({\Delta w\over2}) }e^{2s}\cosh^2({\Delta w\over4}) \sinh^2({\Delta w\over4})\cr
&=&-{e^{3s}\sinh({\Delta w\over2})\over8}
\eea
and similarly for the holomorphic sector
\bea
\bar{A}_{base}\bar{f}_1\bar{f}_2 = -{e^{3\bar{s}}\sinh({\Delta \bar{w}\over2})\over8}
\eea
Inserting these expressions into (\ref{holo amplitude}) and (\ref{anti holo amplitude}) yields
\bea
\mathcal{A}^{i\to f}_{\dot{C}\dot{A}}=-{e^{3s}\sinh({\Delta w\over2})\over8}\mathcal{A}^{i\to f}_{t,\dot{C}\dot{A}}
\label{holo amplitude two}
\eea
and 
\bea
\mathcal{A}^{i\to f}_{\dot{C}\dot{A}}=-{e^{3\bar{s}}\sinh({\Delta \bar{w}\over2})\over8}\mathcal{A}^{\bar{i}\to \bar{f}}_{t,\dot{D}\dot{B}}
\label{anti holo amplitude two}
\eea
Next we map the fermions and bosons to the cover.
\section{Mapping bosons to cover (holomorphic sector)}\label{map bosons}
Here we map the bosons to the cover which we've done several times now. The cylinder modes are
\bea
\a^{(i)f}_{A\dot{A},m} = {1\over 2\pi }\int_{\t>\t_2} dw e^{mw} \partial X_{A\dot{A}}(t),\quad i =1,2\cr
\a^{(i)i}_{A\dot{A},m} = {1\over 2\pi }\int_{\t<\t_1} dw e^{mw} \partial X_{A\dot{A}}(t),\quad i =1,2
\eea
We map to the $z$ and $t$ planes using
\bea
e^w=z={(t+a)(t+b)\over t}
\eea
Since bosons are dimension one, their transformation and jacobian exactly cancel. Therefore
\bea
\a^{(1)f}_{A\dot{A},m}\to \a'^{(1)f}_{A\dot{A},m} = {1\over 2\pi }\int_{t=\infty} dt{(t+a)^m(t+b)\over t^m}^{m}\partial X_{A\dot{A}}(t)\cr
\a^{(2)f}_{A\dot{A},m}\to \a'^{(2)f}_{A\dot{A},m} = -{1\over 2\pi }\int_{t=0} dt{(t+a)^m(t+b)\over t^m}^{m}\partial X_{A\dot{A}}(t)\cr
\a^{(1)i}_{A\dot{A},m}\to \a'^{(1)i}_{A\dot{A},m} = {1\over 2\pi }\int_{t=-a} dt{(t+a)^m(t+b)\over t^m}^{m}\partial X_{A\dot{A}}(t)\cr
\a^{(2)i}_{A\dot{A},m}\to \a'^{(2)i}_{A\dot{A},m} = {1\over 2\pi }\int_{t=-b} dt{(t+a)^m(t+b)\over t^m}^{m}\partial X_{A\dot{A}}(t)
\eea
where the minus sign comes for Copy 2 final because around $t=0$ the map goes like $z\sim {1\over t}$. This reverses the direction of the contour.

Also, bosons are uncharged under the $R$ symmetry and are therefore unaffected by spectral flow. Now we map the fermions to the cover. 
\section{Mapping fermions to cover}\label{map fermions}
Here we map fermions to cover. We first consider fermions in the $NS$ sector on the cylinder
\subsection{$NS$ sector}
Since we don't have any fermions in the initial state we will not consider them. For fermions after the two twist we have
\bea
d^{(i)f,\a A}_r = {1\over 2\pi i}\int_{\t>\t_2}dw \psi^{\a A}(w)e^{rw},\quad i =1,2
\eea
where $r$ is half integer. Mapping to $z$ plane yields
\bea
d^{(i)f,\a A}_r &\to& {1\over 2\pi i}\int_{z=\infty}dz{dw\over dz}({dz\over dw})^{{1\over2}} \psi^{\a A}(z)z^{r}\cr
&=&  {1\over 2\pi i}\int_{z=\infty}dz\psi^{\a A}(z)z^{r-{1\over2}},\quad i=1,2
\eea

Mapping to $t$ plane

\bea
d^{(1)f,\a A}_r \to d'^{(1)f,\a A}_r &=& {1\over 2\pi i}\int_{t=\infty}dt{dz\over dt}({dt\over dz})^{{1\over2}}\psi^{\a A}(t)(t+a)^{r-{1\over2}}(t+b)^{r-{1\over2}}t^{-r+{1\over2}}\cr
&=&{1\over 2\pi i}\int_{t=\infty}dt(t^2 -ab)^{1/2}t^{-1}\psi^{\a A}(t)(t+a)^{r-{1\over2}}(t+b)^{r-{1\over2}}t^{-r+{1\over2}}\cr
&=&{1\over 2\pi i}\int_{t=\infty}dt\psi^{\a A}(t)(t -t_1)^{1/2}(t -t_2)^{1/2}(t+a)^{r-{1\over2}}(t+b)^{r-{1\over2}}t^{-r-{1\over2}}\nn
d^{(2)f,\a A}_r \to d'^{(1)f,\a A}_r &=& -{1\over 2\pi i}\int_{t=0}dt\psi^{\a A}(t)(t -t_1)^{1/2}(t -t_2)^{1/2}(t+a)^{r-{1\over2}}(t+b)^{r-{1\over2}}t^{-r-{1\over2}}\nn
\eea
Now lets apply the spectral flows. For this we will split the charges up into plus and minus

\bea
d'^{(1)f,+ A}_r&=&{1\over 2\pi i}\int_{t=\infty}dt\psi^{+ A}(t)(t -t_1)^{1/2}(t -t_2)^{{1\over2}}(t+a)^{r-{1\over2}}(t+b)^{r-{1\over2}}t^{-r-{1\over2}}\cr
d'^{(2)f,+ A}_r&=&-{1\over 2\pi i}\int_{t=0}dt\psi^{+ A}(t)(t -t_1)^{1/2}(t -t_2)^{{1\over2}}(t+a)^{r-{1\over2}}(t+b)^{r-{1\over2}}t^{-r-{1\over2}}\cr
d'^{(1)f,- A}_r&=&{1\over 2\pi i}\int_{t=\infty}dt\psi^{- A}(t)(t -t_1)^{1/2}(t -t_2)^{{1\over2}}(t+a)^{r-{1\over2}}(t+b)^{r-{1\over2}}t^{-r-{1\over2}}\cr
d'^{(2)f,- A}_r&=&-{1\over 2\pi i}\int_{t=0}dt\psi^{- A}(t)(t -t_1)^{1/2}(t -t_2)^{{1\over2}}(t+a)^{r-{1\over2}}(t+b)^{r-{1\over2}}t^{-r-{1\over2}}
\eea
Where again, the minus sign for Copy 2 comes from reversal of the contour.
Spectral flowing away the spin field $S^+(t_1)$ by $\a=-1$ at $t_1$ yields the transformation

\bea
\psi^{+ A}(t) &\to& (t-t_1)^{{1\over2}}\psi^{+ A}(t)\cr
\psi^{- A}(t) &\to& (t-t_1)^{-{1\over2}}\psi^{- A}(t)
\eea

Applying these transformations yield
\bea
d'^{(1)f,+ A}_r&=&{1\over 2\pi i}\int_{t=\infty}dt\psi^{+ A}(t)(t -t_1)(t -t_2)^{{1\over2}}(t+a)^{r-{1\over2}}(t+b)^{r-{1\over2}}t^{-r-{1\over2}}\cr
d'^{(2)f,+ A}_r&=&-{1\over 2\pi i}\int_{t=0}dt\psi^{+ A}(t)(t -t_1)(t -t_2)^{{1\over2}}(t+a)^{r-{1\over2}}(t+b)^{r-{1\over2}}t^{-r-{1\over2}}\cr
d'^{(1)f,- A}_r&=&{1\over 2\pi i}\int_{t=\infty}dt\psi^{- A}(t)(t -t_2)^{{1\over2}}(t+a)^{r-{1\over2}}(t+b)^{r-{1\over2}}t^{-r-{1\over2}}\cr
d'^{(2)f,- A}_r&=&-{1\over 2\pi i}\int_{t=0}dt\psi^{- A}(t)(t -t_2)^{{1\over2}}(t+a)^{r-{1\over2}}(t+b)^{r-{1\over2}}t^{-r-{1\over2}}
\eea

Spectral flowing away the spin field $S^-(t_2)$ by $\a=+1$ at $t_2$ yields the transformation

\bea
\psi^{+ A}(t) &\to& (t-t_2)^{-{1\over2}}\psi^{+ A}(t)\cr
\psi^{- A}(t) &\to& (t-t_2)^{{1\over2}}\psi^{- A}(t)
\eea

Applying these transformations yield
\bea
d'^{(1)f,+ A}_r&=&{1\over 2\pi i}\int_{t=\infty}dt\psi^{+ A}(t)(t -t_1)(t+a)^{r-{1\over2}}(t+b)^{r-{1\over2}}t^{-r-{1\over2}}\cr
d'^{(2)f,+ A}_r&=&-{1\over 2\pi i}\int_{t=0}dt\psi^{+ A}(t)(t -t_1)(t+a)^{r-{1\over2}}(t+b)^{r-{1\over2}}t^{-r-{1\over2}}\cr
d'^{(1)f,- A}_r&=&{1\over 2\pi i}\int_{t=\infty}dt\psi^{- A}(t)(t -t_2)(t+a)^{r-{1\over2}}(t+b)^{r-{1\over2}}t^{-r-{1\over2}}\cr
d'^{(2)f,- A}_r&=&-{1\over 2\pi i}\int_{t=0}dt\psi^{- A}(t)(t -t_2)(t+a)^{r-{1\over2}}(t+b)^{r-{1\over2}}t^{-r-{1\over2}}
\label{NS cylinder fermions}
\eea

\subsection{Fermions in the Ramond sector: Initial vacuua $|0^-_R\rangle^{(1)}|0^-_R\rangle^{(2)}$, Final vacuua  ${}^{(2)}\langle 0_{R,-}|{}^{(1)}\langle 0_{R,-}|$}
Consider that our amplitudes start in the Ramond sector on the cylinder where we have the vacuua
\bea
\text{Initial State}&&|0^-_R\rangle^{(1)}|0^-_R\rangle^{(2)}\cr
\text{Final State}&&{}^{(2)}\langle 0_{R,-}|{}^{(1)}\langle 0_{R,-}|
\eea
Therefore, in addition to the spectral flows that were made to remove the spin fields $S^+(t_1)$ and $S^-(t_2)$ in the $t$ plane, coming from the twist insertions $\s^+(w_1)$ and $\s^-(w_2)$ on the cylinder, we have additional spectral flows to take $|0^-_R\rangle^{(1)}|0^-_R\rangle^{(2)}\to|0\rangle^{(1)}|0\rangle^{(2)}$ and  ${}^{(2)}\langle 0_{R,-}|{}^{(1)}\langle 0_{R,-}|\to {}^{(2)}\langle 0|{}^{(1)}\langle 0|$. Therefore, any amplitude in the Ramond sector can be obtained from the same amplitude in $NS$ sector with an additional set of spectral flows. These spectral flows can either be performed on the cylinder or in the $t$ plane. We perform them on the cylinder. Let us start with the Ramond sector modes on the cylinder
\bea
d^{(1)f,\a A}_r &=& {1\over 2\pi i}\int_{\t>\t_2}dw \psi^{\a A}(w)e^{rw}\cr
d^{(2)f,\a A}_r &=& {1\over 2\pi i}\int_{\t>\t_2}dw \psi^{\a A}(w)e^{rw}
\eea
were $r$ is now an integer. Spectral flowing by $\a=+1$ to remove the $-$ Ramond vacua, our fermion fields change like
\bea
\psi^{\pm A}(w)\to e^{\mp{1\over2}w}\psi^{\pm A}(w)
\eea
Therefore our modes after the twist become
\bea
d'^{(1)f,+ A}_r &=& {1\over 2\pi i}\int_{\t>\t_2}dw \psi^{+ A}(w)e^{(r-{1\over2})w}\cr
d'^{(2)f,+ A}_r &=& {1\over 2\pi i}\int_{\t>\t_2}dw \psi^{+ A}(w)e^{(r-{1\over2})w}\cr
d'^{(1)f,- A}_r &=& {1\over 2\pi i}\int_{\t>\t_2}dw \psi^{- A}(w)e^{(r+{1\over2})w}\cr
d'^{(2)f,- A}_r &=& {1\over 2\pi i}\int_{\t>\t_2}dw \psi^{- A}(w)e^{(r+{1\over2})w}
\eea
In the $t$ plane our modes become
\bea
d'^{(1)f,+ A}_r&=&{1\over 2\pi i}\int_{t=\infty}dt\psi^{+ A}(t)(t -t_1)(t+a)^{r-1}(t+b)^{r-1}t^{-r}\cr
d'^{(2)f,+ A}_r&=&-{1\over 2\pi i}\int_{t=0}dt\psi^{+ A}(t)(t -t_1)(t+a)^{r-1}(t+b)^{r-1}t^{-r}\cr
d'^{(1)f,- A}_r&=&{1\over 2\pi i}\int_{t=\infty}dt\psi^{- A}(t)(t -t_2)(t+a)^{r}(t+b)^{r}t^{-r-1}\cr
d'^{(2)f,- A}_r&=&-{1\over 2\pi i}\int_{t=0}dt\psi^{- A}(t)(t -t_2)(t+a)^{r}(t+b)^{r}t^{-r-1}
\eea
Therefore we see that we can shift our NS sector modes in (\ref{NS cylinder fermions}) by the following amount
\bea
+&:& r \to r-{1\over2}\cr
-&:& r \to r+{1\over2}
\label{mode shift}
\eea 
where after the shift, $r$ takes on integer values.

\section{Computing the amplitudes}\label{amplitudes}

In this section we compute expressions for the various amplitudes given in (\ref{all amplitudes}). The first amplitude we compute is \textit{one} boson going to \textit{three} bosons. The second is \textit{one} boson going to \textit{one} boson and \textit{two} fermions. The third and final amplitude is \textit{two} bosons going to \textit{two} bosons and \textit{two} fermions.

\subsection{Amplitude for $\a\to\a\a\a$}\label{subsection a to aaa}
Therefore using (\ref{holo amplitude two}) and (\ref{anti holo amplitude two}) we can write a copy generalized version of the individual holomorphic and antiholomorphic terms in the $\a\to\a\a\a$ amplitude in (\ref{all amplitudes}) as
\bea
\mathcal{A}^{\a^{(j)}\to \a^{(j)}\a^{(j)}\a^{(j)}}_{\dot{C}\dot{A}}&=&-{e^{3s}\sinh({\Delta w\over2})\over8}\mathcal{A}^{\a^{(j)}\to \a^{(k)}\a^{(k)}\a^{(k)}}_{t,\dot{C}\dot{A}}
\label{a to aaa  two holo}\\
\bar{\mathcal{A}}^{\bar{\a}^{(j)}\to \bar{\a}^{(j)}\bar{\a}^{(j)}\bar{\a}^{(j)}}_{\dot{D}\dot{B}}&=&-{e^{3\bar{s}}\sinh({\Delta \bar{w}\over2})\over8}\bar{\mathcal{A}}^{\bar{\a}^{(j)}\to \bar{\a}^{(k)}\bar{\a}^{(k)}\bar{\a}^{(k)}}_{\bar{t},\dot{D}\dot{B}}
\label{a to aaa  two antiholo}
\eea 
where $j,k=1,2$. We have also dropped the explicit $w_1,w_2, \bar{w}_1,\bar{w}_2$ dependence for brevity. Now we start with the following $t$ plane amplitude:
\bea
\mathcal{A}^{\a^{(j)} \to \a^{(k)}\a^{(k)}\a^{(k)}}_{t,\dot{C}\dot{A}}(w_2,w_1) &=& \langle 0|\a'^{(k)}_{++,p}\a'^{(k)}_{--,q}\a'^{(k)}_{++,r}\tilde{G}^{+,t_2}_{\dot{C},-{3\over2}}\tilde{G}^{-,t_1}_{\dot{A},-{3\over2}}\a'^{(j)}_{--,-n}|0\rangle
\label{t plane amplitude 2}
\eea
where the primes indicate that the modes have been mapped to the $t$ plane and modified by necessary spectral flows and the $\tilde{G}$ modes a natural $t$ plane at their specified points $t_i$. This was done in section \ref{map bosons} for the bosons. The tildes represent modes that are natural to the $t$ plane defined at the point $t_1$ or $t_2$. Let's write the $G^{\pm}$'s in terms of fermions and bosons using the following relation
\bea
 \tilde{G}^{-,t_i}_{\dot{A},r}=-i\sum_{s}\tilde{d}^{-A,t_i}_{s}\tilde{\a}^{t_i}_{A\dot{A},r-s}
 \eea
where $r,s$ are both half integer modes. Our $t$ plane amplitude becomes.
\bea
\mathcal{A}^{\a^{(j)} \to \a^{(k)}\a^{(k)}\a^{(k)}}_{t,\dot{C}\dot{A}}&=& - \langle 0|\a'^{(k)f}_{++,p}\a'^{(k)f}_{--,q}\a'^{(k)f}_{++,r}\tilde{d}^{+C,t_2}_{-{1\over2}}\tilde{\a}^{t_2}_{C\dot{C},-1}\tilde{d}^{-A,t_1}_{-{1\over2}}\tilde{\a}^{t_1}_{A\dot{A},-1}\a'^{(j)i}_{--,-n}|0\rangle\cr
&=&  - \langle 0|\a'^{(k)f}_{++,p}\a'^{(k)f}_{--,q}\a'^{(k)f}_{++,r}\tilde{d}^{++,t_2}_{-{1\over2}}\tilde{\a}^{t_2}_{+\dot{C},-1}\tilde{d}^{-A,t_1}_{-{1\over2}}\tilde{\a}^{t_1}_{A\dot{A},-1}\a'^{(j)i}_{--,-n}|0\rangle\cr
&& - \langle 0|\a'^{(k)f}_{++,p}\a'^{(k)f}_{--,q}\a'^{(k)f}_{++,r}\tilde{d}^{+-,t_2}_{-{1\over2}}\tilde{\a}^{t_2}_{-\dot{C},-1}\tilde{d}^{-A,t_1}_{-{1\over2}}\tilde{\a}^{t_1}_{A\dot{A},-1}\a'^{(j)i}_{--,-n}|0\rangle\cr
&=&  - \langle 0|\a'^{(k)f}_{++,p}\a'^{(k)f}_{--,q}\a'^{(k)f}_{++,r}\tilde{d}^{++,t_2}_{-{1\over2}}\tilde{\a}^{t_2}_{+\dot{C},-1}\tilde{d}^{-+,t_1}_{-{1\over2}}\tilde{\a}^{t_1}_{+\dot{A},-1}\a'^{(j)i}_{--,-n}|0\rangle\cr
&&- \langle 0|\a'^{(k)f}_{++,p}\a'^{(k)f}_{--,q}\a'^{(k)f}_{++,r}\tilde{d}^{++,t_2}_{-{1\over2}}\tilde{\a}^{t_2}_{+\dot{C},-1}\tilde{d}^{--,t_1}_{-{1\over2}}\tilde{\a}^{t_1}_{-\dot{A},-1}\a'^{(j)i}_{--,-n}|0\rangle\cr
&& - \langle 0|\a'^{(k)f}_{++,p}\a'^{(k)f}_{--,q}\a'^{(k)f}_{++,r}\tilde{d}^{+-,t_2}_{-{1\over2}}\tilde{\a}^{t_2}_{-\dot{C},-1}\tilde{d}^{-+,t_1}_{-{1\over2}}\tilde{\a}^{t_1}_{+\dot{A},-1}\a'^{(j)i}_{--,-n}|0\rangle\cr
&& - \langle 0|\a'^{(k)f}_{++,p}\a'^{(k)f}_{--,q}\a'^{(k)f}_{++,r}\tilde{d}^{+-,t_2}_{-{1\over2}}\tilde{\a}^{t_2}_{-\dot{C},-1}\tilde{d}^{--,t_1}_{-{1\over2}}\tilde{\a}^{t_1}_{-\dot{A},-1}\a'^{(j)i}_{--,-n}|0\rangle
\cr\cr
&=&- \langle 0|\a'^{(k)f}_{++,p}\a'^{(k)f}_{--,q}\a'^{(k)f}_{++,r}\tilde{d}^{++,t_2}_{-{1\over2}}\tilde{\a}^{t_2}_{+\dot{C},-1}\tilde{d}^{--,t_1}_{-{1\over2}}\tilde{\a}^{t_1}_{-\dot{A},-1}\a'^{(j)i}_{--,-n}|0\rangle\cr
&& - \langle 0|\a'^{(k)f}_{++,p}\a'^{(k)f}_{--,q}\a'^{(k)f}_{++,r}\tilde{d}^{+-,t_2}_{-{1\over2}}\tilde{\a}^{t_2}_{-\dot{C},-1}\tilde{d}^{-+,t_1}_{-{1\over2}}\tilde{\a}^{t_1}_{+\dot{A},-1}\a'^{(j)i}_{--,-n}|0\rangle\nn
\label{t plane amplitude 3}
\eea
We note that bosons are unaffected by spectral flow. 
\subsubsection{$\dot{C}=+$ and $\dot{A}=-$}
For the charge combination $\dot{C}=+$ and $\dot{A}=-$. We have
\bea
\mathcal{A}^{\a^{(j)} \to \a^{(k)}\a^{(k)}\a^{(k)}}_t (w_2,w_1)&=&- \langle 0|\a'^{(k)f}_{++,p}\a'^{(k)f}_{--,q}\a'^{(k)f}_{++,r}\tilde{d}^{++,t_2}_{-{1\over2}}\tilde{\a}^{t_2}_{++,-1}\tilde{d}^{--,t_1}_{-{1\over2}}\tilde{\a}^{t_1}_{--,-1}\a'^{(j)i}_{--,-n}|0\rangle\cr
&& - \langle 0|\a'^{(k)f}_{++,p}\a'^{(k)f}_{--,q}\a'^{(k)f}_{++,r}\tilde{d}^{+-,t_2}_{-{1\over2}}\tilde{\a}^{t_2}_{-+,-1}\tilde{d}^{-+,t_1}_{-{1\over2}}\tilde{\a}^{t_1}_{+-,-1}\a'^{(j)i}_{--,-n}|0\rangle\nn
\eea
Wick contracting, we get
\bea
\mathcal{A}^{\a^{(j)} \to \a^{(k)}\a^{(k)}\a^{(k)}}_{t,+-}&=&-\bigg(   (\a'^{(k)f}_{++,p}\a'^{(k)f}_{--,q})(\a'^{(k)f}_{++,r}\tilde{\a}^{t_1}_{--,-1})(\tilde{\a}^{t_2}_{++,-1}\a'^{(j)i}_{--,-n})\cr
&&\quad+(\a'^{(k)f}_{++,p}\a'^{(k)f}_{--,q})(\a'^{(k)f}_{++,r}\a'^{(j)i}_{--,-n})(\tilde{\a}^{t_2}_{++,-1}\tilde{\a}^{t_1}_{--,-1})\cr
&&\quad+(\a'^{(k)f}_{++,p}\tilde{\a}^{t_1}_{--,-1})(\a'^{(k)f}_{--,q}\a'^{(k)f}_{++,r})(\tilde{\a}^{t_2}_{++,-1}\a'^{(j)i}_{--,-n})\cr
&&\quad+(\a'^{(k)f}_{++,p}\tilde{\a}^{t_1}_{--,-1})(\a'^{(k)f}_{--,q}\tilde{\a}^{t_2}_{++,-1})(\a'^{(k)f}_{++,r}\a'^{(j)i}_{--,-n})\cr
&&\quad+(\a'^{(k)f}_{++,p}\a'^{(j)i}_{--,-n})(\a'^{(k)f}_{--,q}\a'^{(k)f}_{++,r})(\tilde{\a}^{t_2}_{++,-1}\tilde{\a}^{t_1}_{--,-1})\cr
&&\quad+(\a'^{(k)f}_{++,p}\a'^{(j)i}_{--,-n})(\a'^{(k)f}_{--,q}\tilde{\a}^{t_2}_{++,-1})(\a'^{(k)f}_{++,r}\tilde{\a}^{t_1}_{--,-1})\bigg)(\tilde{d}^{++,t_2}_{-{1\over2}}\tilde{d}^{--,t_1}_{-{1\over2}})\cr
&&-\bigg((\a'^{(k)f}_{++,p},\a'^{(k)f}_{--,q})(\a'^{(k)f}_{++,r},\a'^{(j)i}_{--,-n}) + (\a'^{(k)f}_{++,p},\a'^{(j)i}_{--,-n})(\a'^{(k)f}_{--,q},\a'^{(k)f}_{++,r})\bigg)\cr
&&\quad (\tilde{\a}^{t_2}_{-+,-1}\tilde{\a}^{t_1}_{+-,-1})(\tilde{d}^{+-,t_2}_{-{1\over2}},\tilde{d}^{-+,t_1}_{-{1\over2}})
\label{ a aaa pm}
\eea

\subsubsection{$\dot{C}=-$ and $\dot{A}=+$}
For $\dot{C}=-$ and $\dot{A}=+$ we get the $t$ plane amplitude
\bea
\mathcal{A}^{\a^{(j)} \to \a^{(k)}\a^{(k)}\a^{(k)}}_{t,-+} &=& - \langle 0|\a'^{(k)f}_{++,p}\a'^{(k)f}_{--,q}\a'^{(k)f}_{++,r}\tilde{d}^{++,t_2}_{-{1\over2}}\tilde{\a}^{t_2}_{+-,-1}\tilde{d}^{--,t_1}_{-{1\over2}}\tilde{\a}^{t_1}_{-+,-1}\a'^{(j)i}_{--,-n}|0\rangle\cr
&& - \langle 0|\a'^{(k)f}_{++,p}\a'^{(k)f}_{--,q}\a'^{(k)f}_{++,r}\tilde{d}^{+-,t_2}_{-{1\over2}}\tilde{\a}^{t_2}_{--,-1}\tilde{d}^{-+,t_1}_{-{1\over2}}\tilde{\a}^{t_1}_{++,-1}\a'^{(j)i}_{--,-n}|0\rangle\nn
\eea
Wick contracting gives
\bea
\mathcal{A}^{\a^{(j)} \to \a^{(k)}\a^{(k)}\a^{(k)}}_{t,-+} &=&-\bigg((\a'^{(k)f}_{++,p},\a'^{(k)f}_{--,q})(\a'^{(k)f}_{++,r},\a'^{(j)i}_{--,-n}) + (\a'^{(k)f}_{++,p},\a'^{(j)i}_{--,-n})(\a'^{(k)f}_{--,q},\a'^{(k)f}_{++,r})\bigg)\cr
&&\quad(\tilde{\a}^{t_2}_{+-,-1},\tilde{\a}^{t_1}_{-+,-1})(\tilde{d}^{++,t_2}_{-{1\over2}},\tilde{d}^{--,t_1}_{-{1\over2}})\cr
&&-\bigg(  (\a'^{(k)f}_{++,p}\a'^{(k)f}_{--,q})(\a'^{(k)f}_{++,r}\tilde{\a}^{t_2}_{--,-1})(\tilde{\a}^{t_1}_{++,-1}\a'^{(j)i}_{--,-n})\cr
&&\quad+(\a'^{(k)f}_{++,p}\a'^{(k)f}_{--,q})(\a'^{(k)f}_{++,r}\a'^{(j)i}_{--,-n})(\tilde{\a}^{t_2}_{--,-1}\tilde{\a}^{t_1}_{++,-1})\cr
&&\quad+(\a'^{(k)f}_{++,p}\tilde{\a}^{t_2}_{--,-1})(\a'^{(k)f}_{--,q}\a'^{(k)f}_{++,r})(\tilde{\a}^{t_1}_{++,-1}\a'^{(j)i}_{--,-n})\cr
&&\quad+(\a'^{(k)f}_{++,p}\tilde{\a}^{t_2}_{--,-1})(\a'^{(k)f}_{--,q}\tilde{\a}^{t_1}_{++,-1})(\a'^{(k)f}_{++,r}\a'^{(j)i}_{--,-n})\cr
&&\quad+(\a'^{(k)f}_{++,p}\a'^{(j)i}_{--,-n})(\a'^{(k)f}_{--,q}\a'^{(k)f}_{++,r})(\tilde{\a}^{t_2}_{--,-1}\tilde{\a}^{t_1}_{++,-1})\cr
&&\quad+(\a'^{(k)f}_{++,p}\a'^{(j)i}_{--,-n})(\a'^{(k)f}_{--,q}\tilde{\a}^{t_1}_{++,-1})(\a'^{(k)f}_{++,r}\tilde{\a}^{t_2}_{--,-1})\bigg)(\tilde{d}^{+-,t_2}_{-{1\over2}}\tilde{d}^{-+,t_1}_{-{1\over2}})\nn
\eea

Following a similar computation, the antiholomorphic $t$ plane amplitude is given by
\subsubsection{$\dot{D}=+$ and $\dot{B}=-$}
\bea
\bar{\mathcal{A}}^{\bar{\a}^{(j)} \to  \bar{\a}^{(k)}\bar{\a}^{(k)}\bar{\a}^{(k)}}_{\bar{t},+-} &=&-\bigg((\bar{\a}'^{(k)f}_{--,p},\bar{\a}'^{(k)f}_{++,q})(\bar{\a}'^{(k)f}_{--,r},\bar{\a}'^{(j)i}_{++,-n}) + (\bar{\a}'^{(k)f}_{--,p},\bar{\a}'^{(j)i}_{++,-n})(\bar{\a}'^{(k)f}_{++,q},\bar{\a}'^{(k)f}_{--,r})\bigg)\cr
&&\quad(\tilde{\bar{\a}}^{\bar{t}_2}_{-+,-1},\tilde{\bar{\a}}^{\bar{t}_1}_{+-,-1})(\tilde{\bar{d}}^{+-,\bar{t}_2}_{-{1\over2}},\tilde{\bar{d}}^{-+,\bar{t}_1}_{-{1\over2}})\cr
&&-\bigg((\bar{\a}'^{(k)f}_{--,p},\bar{\a}'^{(k)f}_{++,q})(\bar{\a}'^{(k)f}_{--,r},\tilde{\bar{\a}}^{\bar{t}_2}_{++,-1})(\tilde{\bar{\a}}^{\bar{t}_1}_{--,-1},\bar{\a}'^{(j)i}_{++,-n})\cr
&&\quad+(\bar{\a}'^{(k)f}_{--,p},\bar{\a}'^{(k)f}_{++,q})(\bar{\a}'^{(k)f}_{--,r},\bar{\a}'^{(j)i}_{++,-n})(\tilde{\bar{\a}}^{\bar{t}_2}_{++,-1},\tilde{\bar{\a}}^{\bar{t}_1}_{--,-1})\cr
&&\quad+(\bar{\a}'^{(k)f}_{--,p},\tilde{\bar{\a}}^{\bar{t}_2}_{++,-1})(\bar{\a}'^{(k)f}_{++,q},\bar{\a}'^{(k)f}_{--,r})(\tilde{\bar{\a}}^{\bar{t}_1}_{--,-1},\bar{\a}'^{(j)i}_{++,-n})\cr
&&\quad+(\bar{\a}'^{(k)f}_{--,p},\tilde{\bar{\a}}^{\bar{t}_2}_{++,-1})(\bar{\a}'^{(k)f}_{++,q},\tilde{\bar{\a}}^{\bar{t}_1}_{--,-1})(\bar{\a}'^{(k)f}_{--,r},\bar{\a}'^{(j)i}_{++,-n})\cr
&&\quad+(\bar{\a}'^{(k)f}_{--,p},\bar{\a}'^{(j)i}_{++,-n})(\bar{\a}'^{(k)f}_{++,q},\bar{\a}'^{(k)f}_{--,r})(\tilde{\bar{\a}}^{\bar{t}_2}_{++,-1},\tilde{\bar{\a}}^{\bar{t}_1}_{--,-1})\cr
&&\quad+(\bar{\a}'^{(k)f}_{--,p},\bar{\a}'^{(j)i}_{++,-n})(\bar{\a}'^{(k)f}_{++,q},\tilde{\bar{\a}}^{\bar{t}_1}_{--,-1})(\bar{\a}'^{(k)f}_{--,r},\tilde{\bar{\a}}^{\bar{t}_2}_{++,-1})\bigg)(\tilde{\bar{d}}^{++,\bar{t}_2}_{-{1\over2}},\tilde{\bar{d}}^{--,\bar{t}_1}_{-{1\over2}})\nn
\eea
and
\subsubsection{$\dot{D}=-$ and $\dot{B}=+$}
\bea
\bar{\mathcal{A}}^{\bar{\a}^{(j)} \to  \bar{\a}^{(k)}\bar{\a}^{(k)}\bar{\a}^{(k)}}_{\bar{t},-+} &=& -\bigg( (\bar{\a}'^{(k)f}_{--,p},\bar{\a}'^{(k)f}_{++,q})(\bar{\a}'^{(k)f}_{--,r},\bar{\a}'^{(j)i}_{++,-n})(\tilde{\bar{\a}}^{\bar{t}_2}_{--,-1},\tilde{\bar{\a}}^{\bar{t}_1}_{++,-1})\cr
&&\quad + (\bar{\a}'^{(k)f}_{--,p},\bar{\a}'^{(k)f}_{++,q})(\bar{\a}'^{(k)f}_{--,r},\tilde{\bar{\a}}^{\bar{t}_1}_{++,-1})(\tilde{\bar{\a}}^{\bar{t}_2}_{--,-1},  \bar{\a}'^{(j)i}_{++,-n})\cr
&&\quad + (\bar{\a}'^{(k)f}_{--,p},\bar{\a}'^{(j)i}_{++,-n})(\bar{\a}'^{(k)f}_{++,q},\bar{\a}'^{(k)f}_{--,r})(\tilde{\bar{\a}}^{\bar{t}_2}_{--,-1},\tilde{\bar{\a}}^{\bar{t}_1}_{++,-1})\cr
&&\quad+ (\bar{\a}'^{(k)f}_{--,p},\bar{\a}'^{(j)i}_{++,-n})(\bar{\a}'^{(k)f}_{++,q},\tilde{\bar{\a}}^{\bar{t}_2}_{--,-1})(\bar{\a}'^{(k)f}_{--,r},\tilde{\bar{\a}}^{\bar{t}_1}_{++,-1})\cr
&&\quad +(\bar{\a}'^{(k)f}_{--,p},\tilde{\bar{\a}}^{\bar{t}_1}_{++,-1})(\bar{\a}'^{(k)f}_{--,r},\bar{\a}'^{(j)i}_{++,-n})(\bar{\a}'^{(k)f}_{++,q},\tilde{\bar{\a}}^{\bar{t}_2}_{--,-1})\cr
&&\quad+(\bar{\a}'^{(k)f}_{--,p},\tilde{\bar{\a}}^{\bar{t}_1}_{++,-1})(\tilde{\bar{\a}}^{\bar{t}_2}_{--,-1},\bar{\a}'^{(j)i}_{++,-n})(\bar{\a}'^{(k)f}_{++,q},\bar{\a}'^{(k)f}_{--,r}) \bigg)(\tilde{\bar{d}}^{+-,\bar{t}_2}_{-{1\over2}},\tilde{\bar{d}}^{-+,\bar{t}_1}_{-{1\over2}})\cr
&& - \bigg( (\bar{\a}'^{(k)f}_{--,p},\bar{\a}'^{(k)f}_{++,q})(\bar{\a}'^{(k)f}_{--,r},\bar{\a}'^{(j)i}_{++,-n})\cr
&&\quad+(\bar{\a}'^{(k)f}_{--,p},\bar{\a}'^{(j)i}_{++,-n})(\bar{\a}'^{(k)f}_{++,q},\bar{\a}'^{(k)f}_{--,r})\bigg)  (\tilde{\bar{d}}^{++,\bar{t}_2}_{-{1\over2}},\tilde{\bar{d}}^{--,\bar{t}_1}_{-{1\over2}})(\tilde{\bar{\a}}^{\bar{t}_2}_{+-,-1},\tilde{\bar{\a}}^{\bar{t}_1}_{-+,-1})\nn
\eea

\subsection{Collecting Results}
Using the four $t$ plane amplitudes computed above, (\ref{a to aaa  two holo}) and (\ref{a to aaa  two antiholo}) become
\bea
&&\!\!\!\!\!\!\!\mathcal{A}^{\a^{(j)} \to \a^{(k)}\a^{(k)}\a^{(k)}}_{+-}\cr
&&={e^{{3\over2}s}\sinh({\Delta w\over2})\over8}\cr
&&\quad\Bigg[  \bigg(   (\a'^{(k)f}_{++,p}\a'^{(k)f}_{--,q})(\a'^{(k)f}_{++,r}\tilde{\a}^{t_2}_{--,-1})(\tilde{\a}^{t_1}_{++,-1}\a'^{(j)i}_{--,-n})\cr
&&\quad+(\a'^{(k)f}_{++,p}\a'^{(k)f}_{--,q})(\a'^{(k)f}_{++,r}\a'^{(j)i}_{--,-n})(\tilde{\a}^{t_2}_{--,-1}\tilde{\a}^{t_1}_{++,-1})\cr
&&\quad+(\a'^{(k)f}_{++,p}\tilde{\a}^{t_2}_{--,-1})(\a'^{(k)f}_{--,q}\a'^{(k)f}_{++,r})(\tilde{\a}^{t_1}_{++,-1}\a'^{(j)i}_{--,-n})\cr
&&\quad+(\a'^{(k)f}_{++,p}\tilde{\a}^{t_2}_{--,-1})(\a'^{(k)f}_{--,q}\tilde{\a}^{t_1}_{++,-1})(\a'^{(k)f}_{++,r}\a'^{(j)i}_{--,-n})\cr
&&\quad+(\a'^{(k)f}_{++,p}\a'^{(j)i}_{--,-n})(\a'^{(k)f}_{--,q}\a'^{(k)f}_{++,r})(\tilde{\a}^{t_2}_{--,-1}\tilde{\a}^{t_1}_{++,-1})\cr
&&\quad+(\a'^{(k)f}_{++,p}\a'^{(j)i}_{--,-n})(\a'^{(k)f}_{--,q}\tilde{\a}^{t_1}_{++,-1})(\a'^{(k)f}_{++,r}\tilde{\a}^{t_2}_{--,-1})\bigg)(\tilde{d}^{++,t_2}_{-{1\over2}}\tilde{d}^{--,t_1}_{-{1\over2}})\cr
&&\quad+\bigg((\a'^{(k)f}_{++,p},\a'^{(k)f}_{--,q})(\a'^{(k)f}_{++,r},\a'^{(j)i}_{--,-n}) + (\a'^{(k)f}_{++,p},\a'^{(j)i}_{--,-n})(\a'^{(k)f}_{--,q},\a'^{(k)f}_{++,r})\bigg)\cr
&&\qquad (\tilde{\a}^{t_2}_{-+,-1}\tilde{\a}^{t_1}_{+-,-1})(\tilde{d}^{+-,t_2}_{-{1\over2}},\tilde{d}^{-+,t_1}_{-{1\over2}})\Bigg]\label{a to aaa pm}\\
\cr
\cr
&&\!\!\!\!\!\!\!\mathcal{A}^{\a^{(j)} \to \a^{(k)}\a^{(k)}\a^{(k)}}_{-+}\cr
&& = {e^{{3\over2}s}\sinh({\Delta w\over2})\over8}\cr
&&\quad\bigg[\bigg((\a'^{(k)f}_{++,p},\a'^{(k)f}_{--,q})(\a'^{(k)f}_{++,r},\a'^{(j)i}_{--,-n}) + (\a'^{(k)f}_{++,p},\a'^{(j)i}_{--,-n})(\a'^{(k)f}_{--,q},\a'^{(k)f}_{++,r})\bigg)\cr
&&\quad(\tilde{\a}^{t_2}_{+-,-1},\tilde{\a}^{t_1}_{-+,-1})(\tilde{d}^{++,t_2}_{-{1\over2}},\tilde{d}^{--,t_1}_{-{1\over2}})\cr
&&\quad+\bigg(  (\a'^{(k)f}_{++,p}\a'^{(k)f}_{--,q})(\a'^{(k)f}_{++,r}\tilde{\a}^{t_2}_{--,-1})(\tilde{\a}^{t_1}_{++,-1}\a'^{(j)i}_{--,-n})\cr
&&\quad+(\a'^{(k)f}_{++,p}\a'^{(k)f}_{--,q})(\a'^{(k)f}_{++,r}\a'^{(j)i}_{--,-n})(\tilde{\a}^{t_2}_{--,-1}\tilde{\a}^{t_1}_{++,-1})\cr
&&\quad+(\a'^{(k)f}_{++,p}\tilde{\a}^{t_2}_{--,-1})(\a'^{(k)f}_{--,q}\a'^{(k)f}_{++,r})(\tilde{\a}^{t_1}_{++,-1}\a'^{(j)i}_{--,-n})\cr
&&\quad+(\a'^{(k)f}_{++,p}\tilde{\a}^{t_2}_{--,-1})(\a'^{(k)f}_{--,q}\tilde{\a}^{t_1}_{++,-1})(\a'^{(k)f}_{++,r}\a'^{(j)i}_{--,-n})\cr
&&\quad+(\a'^{(k)f}_{++,p}\a'^{(j)i}_{--,-n})(\a'^{(k)f}_{--,q}\a'^{(k)f}_{++,r})(\tilde{\a}^{t_2}_{--,-1}\tilde{\a}^{t_1}_{++,-1})\cr
&&\quad+(\a'^{(k)f}_{++,p}\a'^{(j)i}_{--,-n})(\a'^{(k)f}_{--,q}\tilde{\a}^{t_1}_{++,-1})(\a'^{(k)f}_{++,r}\tilde{\a}^{t_2}_{--,-1})\bigg)(\tilde{d}^{+-,t_2}_{-{1\over2}}\tilde{d}^{-+,t_1}_{-{1\over2}})\bigg]\label{a to aaa mp}
\cr
\cr
\cr
&&\!\!\!\!\!\!\!\bar{\mathcal{A}}^{\bar{\a}^{(j)} \to \bar{\a}^{(k)}\bar{\a}^{(k)}\bar{\a}^{(k)}}_{-+}\cr
&&={e^{{3\over2}\bar s}\sinh({\Delta \bar{w}\over2})\over8}\cr
&&\bigg[\bigg( (\bar{\a}'^{(k)f}_{--,p},\bar{\a}'^{(k)f}_{++,q})(\bar{\a}'^{(k)f}_{--,r},\bar{\a}'^{(j)i}_{++,-n})(\tilde{\bar{\a}}^{\bar{t}_2}_{--,-1},\tilde{\bar{\a}}^{\bar{t}_1}_{++,-1})\cr
&&\quad + (\bar{\a}'^{(k)f}_{--,p},\bar{\a}'^{(k)f}_{++,q})(\bar{\a}'^{(k)f}_{--,r},\tilde{\bar{\a}}^{\bar{t}_1}_{++,-1})(\tilde{\bar{\a}}^{\bar{t}_2}_{--,-1},  \bar{\a}'^{(j)i}_{++,-n})\cr
&&\quad + (\bar{\a}'^{(k)f}_{--,p},\bar{\a}'^{(j)i}_{++,-n})(\bar{\a}'^{(k)f}_{++,q},\bar{\a}'^{(k)f}_{--,r})(\tilde{\bar{\a}}^{\bar{t}_2}_{--,-1},\tilde{\bar{\a}}^{\bar{t}_1}_{++,-1})\cr
&&\quad+ (\bar{\a}'^{(k)f}_{--,p},\bar{\a}'^{(j)i}_{++,-n})(\bar{\a}'^{(k)f}_{++,q},\tilde{\bar{\a}}^{\bar{t}_2}_{--,-1})(\bar{\a}'^{(k)f}_{--,r},\tilde{\bar{\a}}^{\bar{t}_1}_{++,-1})\cr
&&\quad +(\bar{\a}'^{(k)f}_{--,p},\tilde{\bar{\a}}^{\bar{t}_1}_{++,-1})(\bar{\a}'^{(k)f}_{--,r},\bar{\a}'^{(j)i}_{++,-n})(\bar{\a}'^{(k)f}_{++,q},\tilde{\bar{\a}}^{\bar{t}_2}_{--,-1})\cr
&&\quad+(\bar{\a}'^{(k)f}_{--,p},\tilde{\bar{\a}}^{\bar{t}_1}_{++,-1})(\tilde{\bar{\a}}^{\bar{t}_2}_{--,-1},\bar{\a}'^{(j)i}_{++,-n})(\bar{\a}'^{(k)f}_{++,q},\bar{\a}'^{(k)f}_{--,r}) \bigg)(\tilde{\bar{d}}^{+-,\bar{t}_2}_{-{1\over2}},\tilde{\bar{d}}^{-+,\bar{t}_1}_{-{1\over2}})\cr
&& + \bigg( (\bar{\a}'^{(k)f}_{--,p},\bar{\a}'^{(k)f}_{++,q})(\bar{\a}'^{(k)f}_{--,r},\bar{\a}'^{(j)i}_{++,-n})+(\bar{\a}'^{(k)f}_{--,p},\bar{\a}'^{(j)i}_{++,-n})(\bar{\a}'^{(k)f}_{--,r},\bar{\a}'^{(k)f}_{++,q})\bigg)\cr
&&\quad  (\tilde{\bar{d}}^{++,\bar{t}_2}_{-{1\over2}},\tilde{\bar{d}}^{--,\bar{t}_1}_{-{1\over2}})(\tilde{\bar{\a}}^{\bar{t}_2}_{+-,-1},\tilde{\bar{\a}}^{\bar{t}_1}_{-+,-1})\bigg]\label{a to aaa pm bar}\\
\cr
\cr
&&\!\!\!\!\!\!\!\bar{\mathcal{A}}^{\bar{\a}^{(j)} \to \bar{\a}^{(k)}\bar{\a}^{(k)}\bar{\a}^{(k)}}_{-+}\cr
&&={e^{{3\over2}\bar s}\sinh({\Delta \bar{w}\over2})\over8}\cr
&&\quad\bigg[\bigg((\bar{\a}'^{(k)f}_{--,p},\bar{\a}'^{(k)f}_{++,q})(\bar{\a}'^{(k)f}_{--,r},\bar{\a}'^{(j)i}_{++,-n}) + (\bar{\a}'^{(k)f}_{--,p},\bar{\a}'^{(j)i}_{++,-n})(\bar{\a}'^{(k)f}_{++,q},\bar{\a}'^{(k)f}_{--,r})\bigg)\cr
&&\quad(\tilde{\bar{\a}}^{\bar{t}_2}_{-+,-1},\tilde{\bar{\a}}^{\bar{t}_1}_{+-,-1})(\tilde{\bar{d}}^{+-,\bar{t}_2}_{-{1\over2}},\tilde{\bar{d}}^{-+,\bar{t}_1}_{-{1\over2}})\cr
&&\quad+\bigg((\bar{\a}'^{(k)f}_{--,p},\bar{\a}'^{(k)f}_{++,q})(\bar{\a}'^{(k)f}_{--,r},\tilde{\bar{\a}}^{\bar{t}_2}_{++,-1})(\tilde{\bar{\a}}^{\bar{t}_1}_{--,-1},\bar{\a}'^{(j)i}_{++,-n})\cr
&&\quad+(\bar{\a}'^{(k)f}_{--,p},\bar{\a}'^{(k)f}_{++,q})(\bar{\a}'^{(k)f}_{--,r},\bar{\a}'^{(j)i}_{++,-n})(\tilde{\bar{\a}}^{\bar{t}_2}_{++,-1},\tilde{\bar{\a}}^{\bar{t}_1}_{--,-1})\cr
&&\quad+(\bar{\a}'^{(k)f}_{--,p},\tilde{\bar{\a}}^{\bar{t}_2}_{++,-1})(\bar{\a}'^{(k)f}_{++,q},\bar{\a}'^{(k)f}_{--,r})(\tilde{\bar{\a}}^{\bar{t}_1}_{--,-1},\bar{\a}'^{(j)i}_{++,-n})\cr
&&\quad+(\bar{\a}'^{(k)f}_{--,p},\tilde{\bar{\a}}^{\bar{t}_2}_{++,-1})(\bar{\a}'^{(k)f}_{++,q},\tilde{\bar{\a}}^{\bar{t}_1}_{--,-1})(\bar{\a}'^{(k)f}_{--,r},\bar{\a}'^{(j)i}_{++,-n})\cr
&&\quad+(\bar{\a}'^{(k)f}_{--,p},\bar{\a}'^{(j)i}_{++,-n})(\bar{\a}'^{(k)f}_{++,q},\bar{\a}'^{(k)f}_{--,r})(\tilde{\bar{\a}}^{\bar{t}_2}_{++,-1},\tilde{\bar{\a}}^{\bar{t}_1}_{--,-1})\cr
&&\quad+(\bar{\a}'^{(k)f}_{--,p},\bar{\a}'^{(j)i}_{++,-n})(\bar{\a}'^{(k)f}_{++,q},\tilde{\bar{\a}}^{\bar{t}_1}_{--,-1})(\bar{\a}'^{(k)f}_{--,r},\tilde{\bar{\a}}^{\bar{t}_2}_{++,-1})\bigg)(\tilde{\bar{d}}^{++,\bar{t}_2}_{-{1\over2}},\tilde{\bar{d}}^{--,\bar{t}_1}_{-{1\over2}})\bigg]\label{a to aaa mp bar}
\eea
Therefore our full amplitude for $\a\to\a\a\a$ given in (\ref{all amplitudes}) is
\bea
&&\mathcal{A}^{\a\to \a\a\a}(w_2,w_1,\bar{w}_2,\bar{w}_1)\cr
&&\quad =\e^{\dot{C}\dot{D}}\e^{\dot{A}\dot{B}}\mathcal{A}^{\a^{(1)}\to \a^{(1)}\a^{(1)}\a^{(1)}}_{\dot{C}\dot{A}}(w_2,w_1)\bar{\mathcal{A}}^{\a^{(1)}\to \a^{(1)}\a^{(1)}\a^{(1)}}_{\dot{D}\dot{B}}(\bar{w}_2,\bar{w}_1)\cr
&&\qquad + \e^{\dot{C}\dot{D}}\e^{\dot{A}\dot{B}}\mathcal{A}^{\a^{(2)}\to \a^{(1)}\a^{(1)}\a^{(1)}}_{\dot{C}\dot{A}}(w_2,w_1)\bar{\mathcal{A}}^{\a^{(2)}\to \a^{(1)}\a^{(1)}\a^{(1)}}_{\dot{D}\dot{B}}(\bar{w}_2,\bar{w}_1)\cr
&&\qquad + \e^{\dot{C}\dot{D}}\e^{\dot{A}\dot{B}}\mathcal{A}^{\a^{(1)}\to \a^{(2)}\a^{(2)}\a^{(2)}}_{\dot{C}\dot{A}}(w_2,w_1)\bar{\mathcal{A}}^{\a^{(1)}\to \a^{(2)}\a^{(2)}\a^{(2)}}_{\dot{D}\dot{B}}(\bar{w}_2,\bar{w}_1)\cr
&&\qquad + \e^{\dot{C}\dot{D}}\e^{\dot{A}\dot{B}}\mathcal{A}^{\a^{(2)}\to \a^{(2)}\a^{(2)}\a^{(2)}}_{\dot{C}\dot{A}}(w_2,w_1)\bar{\mathcal{A}}^{\a^{(2)}\to \a^{(2)}\a^{(2)}\a^{(2)}}_{\dot{D}\dot{B}}(\bar{w}_2,\bar{w}_1)
\cr
\cr
&&\quad =-{1\over 2 npqr}\mathcal{A}^{\a^{(1)} \to \a^{(1)}\a^{(1)}\a^{(1)}}_{+-}(w_2,w_1)\bar{\mathcal{A}}^{\bar{\a}^{(1)} \to \bar{\a}^{(1)}\bar{\a}^{(1)}\bar{\a}^{(1)}}_{-+}(\bar{w}_2,\bar{w}_1)\cr
&& \qquad- {1\over 2 npqr}\mathcal{A}^{\a^{(1)} \to \a^{(1)}\a^{(1)}\a^{(1)}}_{-+}(w_2,w_1)\bar{\mathcal{A}}^{\bar{\a}^{(1)} \to\bar \a^{(1)}\bar\a^{(1)}\bar\a^{(1)}}_{+-}(\bar{w}_2,\bar{w}_1)
\cr\cr
&&\qquad -{1\over 2 npqr}\mathcal{A}^{\a^{(2)} \to \a^{(1)}\a^{(1)}\a^{(1)}}_{+-}(w_2,w_1)\bar{\mathcal{A}}^{\bar\a^{(2)} \to\bar \a^{(1)}\bar\a^{(1)}\bar\a^{(1)}}_{-+}(\bar{w}_2,\bar{w}_1)\cr
&& \qquad- {1\over 2 npqr}\mathcal{A}^{\a^{(2)} \to \a^{(1)}\a^{(1)}\a^{(1)}}_{-+}(w_2,w_1)\bar{\mathcal{A}}^{\bar\a^{(2)} \to \bar\a^{(1)}\bar\a^{(1)}\bar\a^{(1)}}_{+-}(\bar{w}_2,\bar{w}_1)
\cr\cr
&&\qquad -{1\over 2 npqr}\mathcal{A}^{\a^{(1)} \to \a^{(2)}\a^{(2)}\a^{(2)}}_{+-}(w_2,w_1)\bar{\mathcal{A}}^{\bar\a^{(1)} \to \bar\a^{(2)}\bar\a^{(2)}\bar\a^{(2)}}_{-+}(\bar{w}_2,\bar{w}_1)\cr
&& \qquad- {1\over 2 npqr}\mathcal{A}^{\a^{(1)} \to \a^{(2)}\a^{(2)}\a^{(2)}}_{-+}(w_2,w_1)\bar{\mathcal{A}}^{\bar\a^{(1)} \to \bar\a^{(2)}\bar\a^{(2)}\bar\a^{(2)}}_{+-}(\bar{w}_2,\bar{w}_1)
\cr\cr
&&\qquad -{1\over 2 npqr}\mathcal{A}^{\a^{(2)} \to \a^{(2)}\a^{(2)}\a^{(2)}}_{+-}(w_2,w_1)\bar{\mathcal{A}}^{\bar\a^{(2)} \to \bar\a^{(2)}\bar\a^{(2)}\bar\a^{(2)}}_{-+}(\bar{w}_2,\bar{w}_1)\cr
&& \qquad- {1\over 2 npqr}\mathcal{A}^{\a^{(2)} \to \a^{(2)}\a^{(2)}\a^{(2)}}_{-+}(w_2,w_1)\bar{\mathcal{A}}^{\bar\a^{(2)} \to \bar\a^{(2)}\bar\a^{(2)}\bar\a^{(2)}}_{+-}(\bar{w}_2,\bar{w}_1)
\label{a to aaa full}
\eea
where the individual copy amplitudes for each charge combination are in (\ref{a to aaa pm}), (\ref{a to aaa mp}), (\ref{a to aaa pm bar}) and (\ref{a to aaa mp bar}). We have also restored the dependence on the cylinder twist locations and also used the fact that 
\bea
\mathcal{A}^{\a^{(j)} \to \a^{(k)}\a^{(k)}\a^{(k)}}_{++}(w_2,w_1)&=&\mathcal{A}^{\a^{(j)} \to \a^{(k)}\a^{(k)}\a^{(k)}}_{--}(w_2,w_1)=\bar{\mathcal{A}}^{\bar{\a}^{(j)} \to \bar{\a}^{(k)}\bar{\a}^{(k)}\bar{\a}^{(k)}}_{--}(\bar{w}_2,\bar{w}_1)\cr
&=&\bar{\mathcal{A}}^{\bar{\a}^{(j)} \to \bar{\a}^{(k)}\bar{\a}^{(k)}\bar{\a}^{(k)}}_{++}(\bar{w}_2,\bar{w}_1)=0
\eea
We also note the following copy symmetry relations between the various amplitudes
\bea
\mathcal{A}^{\a^{(1)} \to \a^{(1)}\a^{(1)}\a^{(1)}}_{\dot{C}\dot{A}}(w_2,w_1) &=& \mathcal{A}^{\a^{(2)} \to \a^{(2)}\a^{(2)}\a^{(2)}}_{\dot{C}\dot{A}}(w_2,w_1)\cr
\mathcal{A}^{\a^{(1)} \to \a^{(2)}\a^{(2)}\a^{(2)}}_{\dot{C}\dot{A}}(w_2,w_1) &=& \mathcal{A}^{\a^{(2)} \to \a^{(1)}\a^{(1)}\a^{(1)}}_{\dot{C}\dot{A}}(w_2,w_1)\cr
\bar{\mathcal{A}}^{\bar\a^{(1)} \to \bar\a^{(1)}\bar\a^{(1)}\bar\a^{(1)}}_{\dot{D}\dot{B}}(\bar w_2,\bar w_1) &=& \bar{\mathcal{A}}^{\bar\a^{(2)} \to \bar\a^{(2)}\bar\a^{(2)}\bar\a^{(2)}}_{\dot{D}\dot{B}}(\bar w_2,\bar w_1)\cr 
\bar{\mathcal{A}}^{\bar\a^{(1)} \to \bar\a^{(2)}\bar\a^{(2)}\bar\a^{(2)}}_{\dot{D}\dot{B}}(\bar w_2,\bar w_1) &=& \bar{\mathcal{A}}^{\bar\a^{(2)} \to \bar\a^{(1)}\bar\a^{(1)}\bar\a^{(1)}}_{\dot{D}\dot{B}}(\bar w_2,\bar w_1)    
\eea

\subsection{Amplitudes for the process $\a\to\a dd$} \label{subsection a to add}
In this subsection we compute the amplitudes for the process $\a\to \a dd$:
\bea
\mathcal{A}^{\a^{(j)}\to \a^{(j)}d^{(j)}d^{(j)}}_{\dot{C}\dot{A}}&=&-{e^{3s}\sinh({\Delta w\over2})\over8}\mathcal{A}^{\a^{(j)}\to \a^{(k)}d^{(k)}d^{(k)}}_{t,\dot{C}\dot{A}}
\label{a to add  two holo}\\
\bar{\mathcal{A}}^{\bar{\a}^{(j)}\to \bar{\a}^{(j)}\bar{d}^{(j)}\bar{d}^{(j)}}_{\dot{D}\dot{B}}&=&-{e^{3\bar{s}}\sinh({\Delta \bar{w}\over2})\over8}\bar{\mathcal{A}}^{\bar{\a}^{(j)}\to \bar{\a}^{(k)}\bar{d}^{(k)}\bar{d}^{(k)}}_{\bar t,\dot{D}\dot{B}}
\label{a to add  two antiholo}
\eea 
The holomorphic and antiholomorphic $t$ plane amplitudes are
\bea
\mathcal{A}^{\a^{(j)} \to \a^{(k)}d^{(k)}d^{(k)}}_{t,\dot{C}\dot{A}} &=& \langle 0|\a'^{(k)f}_{++,p}d'^{(k)f,+-}_{q}d'^{(k)f,-+}_{r}\tilde{G}^{+,t_2}_{\dot{C},-{3\over2}}\tilde{G}^{-,t_1}_{\dot{A},-{3\over2}}\a'^{(j)i}_{--,-n}|0\rangle\cr
\bar{\mathcal{A}}^{\bar{\a}^{(j)} \to \bar{\a}^{(k)}\bar{d}^{(k)}\bar{d}^{(k)}}_{\bar t,\dot{D}\dot{B}}&=& \langle \bar{0}|\bar{\a}'^{(k)f}_{--,p}\bar{d}'^{(k)f,-+}_{q}\bar{d}'^{(k)f,+-}_{r}\tilde{\bar{G}}^{+, \bar{t}_2}_{\dot{D},-{3\over2}}\tilde{\bar{G}}^{-,\bar t_1}_{\dot{B},-{3\over2}}\bar{\a}^{(j)i}_{++,-n}|\bar{0}\rangle\nn
\eea
The computation is very similar to the one in the previous subsection and so we simply record the final expressions for the generalized amplitudes and the full amplitude. The generalized amplitudes are
\bea
&&\!\!\!\!\!\!\!\mathcal{A}^{\a^{(j)} \to \a^{(k)}d^{(k)}d^{(k)}}_{+-}\cr
&&\quad= {e^{{3\over2}s}\sinh({\Delta w\over2})\over8}\cr
&&\qquad\bigg[\bigg(  (\a'^{(k)f}_{++,p} , \a'^{(j)i}_{--,-n})(\tilde{\a}^{t_2}_{++,-1},\tilde{\a}^{t_1}_{--,-1})(d'^{(k)f,+-}_q ,d'^{(k)f,- +}_r)(\tilde{d}^{++,t_2}_{-{1\over2}},\tilde{d}^{--,t_1}_{-{1\over2}})
\cr
&&\qquad+(\a'^{(k)f}_{++,p} ,\tilde{\a}^{t_1}_{--,-1})(\tilde{\a}^{t_2}_{++,-1}, \a'^{(j)i}_{--,-n})(d'^{(k)f,+-}_q ,d'^{(k)f,- +}_r)(\tilde{d}^{++,t_2}_{-{1\over2}},\tilde{d}^{--,t_1}_{-{1\over2}})
\cr
&&\qquad+(\a'^{(k)f}_{++,p} , \a'^{(j)i}_{--,-n})(\tilde{\a}^{t_2}_{-+,-1},\tilde{\a}^{t_1}_{+-,-1})(d'^{(k)f,+-}_q ,d'^{(k)f,- +}_r)(\tilde{d}^{+-,t_2}_{-{1\over2}},\tilde{d}^{-+,t_1}_{-{1\over2}})\cr
&&\qquad+(\a'^{(k)f}_{++,p} , \a'^{(j)i}_{--,-n})(\tilde{\a}^{t_2}_{-+,-1},\tilde{\a}^{t_1}_{+-,-1})(d'^{(k)f,+-}_q ,\tilde{d}^{-+,t_1}_{-{1\over2}})(d'^{(k)f,- +}_r,\tilde{d}^{+-,t_2}_{-{1\over2}})  \bigg]\nn
\label{a to add pm}\\
\cr
&&\!\!\!\!\!\!\!\mathcal{A}^{\a^{(j)} \to \a^{(k)}d^{(k)}d^{(k)}}_{-+}(w_2,w_1)\cr
&&=  {e^{{3\over2}s}\sinh({\Delta w\over2})\over8}\cr
&&\qquad\bigg[  (\a'^{(k)f}_{++,p},\a'^{(j)i}_{--,-n})(\tilde{\a}^{t_2}_{+-,-1},\tilde{\a}^{t_1}_{-+,-1})(d'^{(k)f,+-}_q ,d'^{(k)f,- +}_r)(\tilde{d}^{++,t_2}_{-{1\over2}},\tilde{d}^{--,t_1}_{-{1\over2}})\cr
&&\qquad+(\a'^{(k)f}_{++,p},\tilde{\a}^{t_2}_{--,-1})(\tilde{\a}^{t_1}_{++,-1},\a'^{(j)i}_{--,-n})(d'^{(k)f,+-}_q ,d'^{(k)f,- +}_r)(\tilde{d}^{+-,t_2}_{-{1\over2}},\tilde{d}^{-+,t_1}_{-{1\over2}})\cr
&&\qquad+(\a'^{(k)f}_{++,p},\tilde{\a}^{t_2}_{--,-1})(\tilde{\a}^{t_1}_{++,-1},\a'^{(j)i}_{--,-n})(d'^{(k)f,+-}_q ,  \tilde{d}^{-+,t_1}_{-{1\over2}})(d'^{(k)f,- +}_r ,\tilde{d}^{+-,t_2}_{-{1\over2}} )\cr
&&\qquad+(\a'^{(k)f}_{++,p},\a'^{(j)i}_{--,-n})(\tilde{\a}^{t_2}_{--,-1},\tilde{\a}^{t_1}_{++,-1})(d'^{(k)f,+-}_q ,d'^{(k)f,- +}_r)(\tilde{d}^{+-,t_2}_{-{1\over2}},\tilde{d}^{-+,t_1}_{-{1\over2}})\cr
&&\qquad+(\a'^{(k)f}_{++,p},\a'^{(j)i}_{--,-n})(\tilde{\a}^{t_2}_{--,-1},\tilde{\a}^{t_1}_{++,-1})(d'^{(k)f,+-}_q ,  \tilde{d}^{-+,t_1}_{-{1\over2}})(d'^{(k)f,- +}_r ,\tilde{d}^{+-,t_2}_{-{1\over2}} )  \bigg]\nn
\label{a to add mp}\\
\cr
&&\!\!\!\!\!\!\! \bar{\mathcal{A}}^{\bar\a^{(j)} \to \bar\a^{(k)}\bar d^{(k)}\bar d^{(k)}}_{+-}\cr
 &&={e^{{3\over2}\bar s}\sinh({\Delta \bar{w}\over2})\over8}\cr
 &&\quad\bigg[ \bigg((\bar{d}^{(k)f,-+}_{q},\bar{d}^{(k)f,+-}_{r})(\tilde{\bar{d}}^{+-,\bar{t}_2}_{-{1\over2}},\tilde{\bar{d}}^{-+,\bar{t}_1}_{-{1\over2}})  -(\bar{d}^{(k)f,-+}_{q},\tilde{\bar{d}}^{+-,\bar{t}_2}_{-{1\over2}})(\bar{d}^{(k)f,+-}_{r},\tilde{\bar{d}}^{-+,\bar{t}_1}_{-{1\over2}}) \bigg)\cr
 &&\quad(\bar{\a}^{(k)f}_{--,p},\bar{\a}'^{(j)i}_{++,-n})(\tilde{\bar{\a}}^{\bar{t}_2}_{-+,-1},\tilde{\bar{\a}}^{\bar{t}_1}_{+-,-1})
 \cr
 &&\quad+\bigg( (\bar{\a}^{(k)f}_{--,p},\bar{\a}'^{(j)i}_{++,-n})(\tilde{\bar{\a}}^{\bar{t}_2}_{++,-1},\tilde{\bar{\a}}^{\bar{t}_1}_{--,-1}) + (\bar{\a}^{(k)f}_{--,p},\tilde{\bar{\a}}^{\bar{t}_2}_{++,-1})(\tilde{\bar{\a}}^{\bar{t}_1}_{--,-1},\bar{\a}'^{(j)i}_{++,-n})\bigg)\cr
 &&\quad (\bar{d}^{(k)f,-+}_{q},\bar{d}^{(k)f,+-}_{r})(\tilde{\bar{d}}^{++,\bar{t}_2}_{-{1\over2}},\tilde{\bar{d}}^{--,\bar{t}_1}_{-{1\over2}}) \bigg]
 \label{a to add pm bar}\\
 \cr
&&\!\!\!\!\!\!\!\bar{\mathcal{A}}^{\bar\a^{(j)} \to \bar\a^{(k)}\bar d^{(k)}\bar d^{(k)}}_{-+} \cr
&&={e^{{3\over2}\bar s}\sinh({\Delta \bar{w}\over2})\over8}\cr
 &&\quad\bigg[ \bigg((\bar{\a}^{(k)f}_{--,p},\bar{\a}'^{(j)i}_{++,-n}) (\tilde{\bar{\a}}^{\bar{t}_2}_{--,-1},\tilde{\bar{\a}}^{\bar{t}_1}_{++,-1}) +(\bar{\a}^{(k)f}_{--,p},\tilde{\bar{\a}}^{\bar{t}_1}_{++,-1}) (\tilde{\bar{\a}}^{\bar{t}_2}_{--,-1},\bar{\a}'^{(j)i}_{++,-n})  \bigg)\cr
 &&\quad\bigg((\bar{d}^{(k)f,-+}_{q},\bar{d}^{(k)f,+-}_{r})(\tilde{\bar{d}}^{+-,\bar{t}_2}_{-{1\over2}},\tilde{\bar{d}}^{-+,\bar{t}_1}_{-{1\over2}}) - (\bar{d}^{(k)f,-+}_{q},\tilde{\bar{d}}^{+-,\bar{t}_2}_{-{1\over2}})(\bar{d}^{(k)f,+-}_{r},\tilde{\bar{d}}^{-+,\bar{t}_1}_{-{1\over2}})\bigg)\cr
 &&\quad + (\bar{\a}^{(k)f}_{--,p},\bar{\a}'^{(j)i}_{++,-n})(\tilde{\bar{\a}}^{\bar{t}_2}_{+-,-1},\tilde{\bar{\a}}^{\bar{t}_1}_{-+,-1})(\bar{d}^{(k)f,-+}_{q},\bar{d}^{(k)f,+-}_{r})(\tilde{\bar{d}}^{++,\bar{t}_2}_{-{1\over2}},\tilde{\bar{d}}^{--,t_1}_{-{1\over2}}) \bigg]
\label{a to add mp bar}
\eea

The full amplitude for $\a\to\a dd$ in (\ref{all amplitudes}) is then
\bea
&&\mathcal{A}^{\a\to \a dd}(w_2,w_1,\bar{w}_2,\bar{w}_1)\cr
&&\quad =-{1\over 2 np}\mathcal{A}^{\a^{(1)} \to \a^{(1)}d^{(1)}d^{(1)}}_{+-}(w_2,w_1)\bar{\mathcal{A}}^{\bar{\a}^{(1)} \to \bar{\a}^{(1)}\bar{d}^{(1)}\bar{d}^{(1)}}_{-+}(\bar{w}_2,\bar{w}_1)\cr
&& \qquad- {1\over 2 np}\mathcal{A}^{\a^{(1)} \to \a^{(1)}d^{(1)}d^{(1)}}_{-+}(w_2,w_1)\bar{\mathcal{A}}^{\bar{\a}^{(1)} \to\bar \a^{(1)}\bar d^{(1)}\bar d^{(1)}}_{+-}(\bar{w}_2,\bar{w}_1)
\cr\cr
&&\qquad -{1\over 2 np}\mathcal{A}^{\a^{(2)} \to \a^{(1)} d^{(1)} d^{(1)}}_{+-}(w_2,w_1)\bar{\mathcal{A}}^{\bar\a^{(2)} \to\bar \a^{(1)}\bar d^{(1)}\bar d^{(1)}}_{-+}(\bar{w}_2,\bar{w}_1)\cr
&& \qquad- {1\over 2 np}\mathcal{A}^{\a^{(2)} \to \a^{(1)} d^{(1)} d^{(1)}}_{-+}(w_2,w_1)\bar{\mathcal{A}}^{\bar\a^{(2)} \to \bar\a^{(1)}\bar d^{(1)}\bar d^{(1)}}_{+-}(\bar{w}_2,\bar{w}_1)
\cr\cr
&&\qquad -{1\over 2 np}\mathcal{A}^{\a^{(1)} \to \a^{(2)} d^{(2)} d^{(2)}}_{+-}(w_2,w_1)\bar{\mathcal{A}}^{\bar\a^{(1)} \to \bar\a^{(2)}\bar d^{(2)}\bar d^{(2)}}_{-+}(\bar{w}_2,\bar{w}_1)\cr
&& \qquad- {1\over 2 np}\mathcal{A}^{\a^{(1)} \to \a^{(2)} d^{(2)} d^{(2)}}_{-+}(w_2,w_1)\bar{\mathcal{A}}^{\bar\a^{(1)} \to \bar\a^{(2)}\bar d^{(2)}\bar d^{(2)}}_{+-}(\bar{w}_2,\bar{w}_1)
\cr\cr
&&\qquad -{1\over 2 np}\mathcal{A}^{\a^{(2)} \to \a^{(2)} d^{(2)} d^{(2)}}_{+-}(w_2,w_1)\bar{\mathcal{A}}^{\bar\a^{(2)} \to \bar\a^{(2)}\bar d^{(2)}\bar d^{(2)}}_{-+}(\bar{w}_2,\bar{w}_1)\cr
&& \qquad- {1\over 2 np}\mathcal{A}^{\a^{(2)} \to \a^{(2)} d^{(2)} d^{(2)}}_{-+}(w_2,w_1)\bar{\mathcal{A}}^{\bar\a^{(2)} \to \bar\a^{(2)}\bar d^{(2)}\bar d^{(2)}}_{+-}(\bar{w}_2,\bar{w}_1)
\label{a to add full}
\eea
where the copy amplitudes for each charge combination are given in (\ref{a to add pm}), (\ref{a to add mp}), (\ref{a to add pm bar}), and (\ref{a to add mp bar}). 
We also note the following copy symmetry relations between the various amplitudes
\bea
\mathcal{A}^{\a^{(1)} \to \a^{(1)}d^{(1)}d^{(1)}}_{\dot{C}\dot{A}}(w_2,w_1) &=& \mathcal{A}^{\a^{(2)} \to \a^{(2)}d^{(2)}d^{(2)}}_{\dot{C}\dot{A}}(w_2,w_1)\cr
\mathcal{A}^{\a^{(1)} \to \a^{(2)}d^{(2)}d^{(2)}}_{\dot{C}\dot{A}}(w_2,w_1) &=& \mathcal{A}^{\a^{(2)} \to \a^{(1)}d^{(1)}d^{(1)}}_{\dot{C}\dot{A}}(w_2,w_1)\cr
\bar{\mathcal{A}}^{\bar\a^{(1)} \to \bar\a^{(1)}\bar d^{(1)}\bar d^{(1)}}_{\dot{D}\dot{B}}(\bar w_2,\bar w_1) &=& \bar{\mathcal{A}}^{\bar\a^{(2)} \to \bar\a^{(2)}\bar d^{(2)}\bar d^{(2)}}_{\dot{D}\dot{B}}(\bar w_2,\bar w_1)\cr 
\bar{\mathcal{A}}^{\bar\a^{(1)} \to \bar\a^{(2)}\bar d^{(2)}\bar d^{(2)}}_{\dot{D}\dot{B}}(\bar w_2,\bar w_1) &=& \bar{\mathcal{A}}^{\bar\a^{(2)} \to \bar\a^{(1)}\bar d^{(1)}\bar d^{(1)}}_{\dot{D}\dot{B}}(\bar w_2,\bar w_1)    
\eea

\subsection{$\a\a\to \a\a \a\a$}\label{subsection aa to aaaa}

In this subsection we compute the amplitudes for the process $\a\a\to \a\a \a\a$:
\bea
\mathcal{A}^{\a^{(j)}\a^{(k)}\to \a^{(l)}\a^{(l)}\a^{(l)}\a^{(l)}}_{\dot{C}\dot{A}}&=&-{e^{3s}\sinh({\Delta w\over2})\over8}\mathcal{A}^{\a^{(j)}\a^{(k)}\to \a^{(l)}\a^{(l)}\a^{(l)}\a^{(l)}}_{t,\dot{C}\dot{A}}
\label{aa to aadd  two holo}\\
\bar{\mathcal{A}}^{\bar{\a}^{(j)}\bar{\a}^{(k)}\to \bar{\a}^{(l)}\bar{\a}^{(l)}\bar{\a}^{(l)}\bar{\a}^{(l)}}_{\dot{D}\dot{B}}&=&-{e^{3\bar{s}}\sinh({\Delta \bar{w}\over2})\over8}\bar{\mathcal{A}}^{\bar{\a}^{(j)}\bar{\a}^{(k)}\to \bar{\a}^{(l)}\bar{\a}^{(l)}\bar{\a}^{(l)}\bar{\a}^{(l)}}_{\bar t,\dot{D}\dot{B}}
\label{aa to aaaa  two antiholo}
\eea 
The holomorphic and antiholomorphic $t$ plane amplitudes are
\bea
\mathcal{A}^{\a^{(j)}\a^{(k)} \to \a^{(l)}\a^{(l)}\a^{(l)}\a^{(l)}}_{t,\dot{C}\dot{A}} &=& \langle 0|\a'^{(l)f}_{++,p}\a'^{(l)f}_{--,q}\a'^{(l)f}_{++,r}\a'^{(l)f}_{--,s}\tilde{G}^{+,t_2}_{\dot{C},-{3\over2}}\tilde{G}^{-,t_1}_{\dot{A},-{3\over2}}\a'^{(j)i}_{++,-n_1}\a'^{(k)i}_{--,-n_2}|0\rangle\cr
\bar{\mathcal{A}}^{\bar{\a}^{(j)}\bar\a^{(k)} \to \bar\a^{(l)}\bar\a^{(l)}\bar d^{(l)}\bar d^{(l)}}_{\bar t,\dot{D}\dot{B}} &=& \langle \bar{0}|\bar{\a}'^{(l)f}_{--,p}\bar{\a}'^{(l)f}_{++,q}\bar{\a}'^{(l)f}_{--,r}\bar{\a}'^{(l)f}_{++,s}\tilde{\bar{G}}^{+,\bar{t}_2}_{\dot{D},-{3\over2}}\tilde{\bar{G}}^{-,\bar{t}_1}_{\dot{B},-{3\over2}}\bar{\a}'^{(j)i}_{--,-n_1}\bar{\a}'^{(k)i}_{++,-n_2}|\bar{0}\rangle\nn
\eea

Again, the computation is very similar to the ones in the previous subsections and so we simply record the final expressions for the generalized amplitudes and the full amplitude. The generalized amplitudes are

\bea
&&\!\!\!\!\!\!\!\mathcal{A}^{\a^{(j)}\a^{(k)} \to \a^{(l)}\a^{(l)}\a^{(l)}\a^{(l)}}_{+-}\cr
&&= {e^{{3\over2}s}\sinh({\Delta w\over2})\over8}\cr
&&\quad\bigg[(\a'^{(l)f}_{++,p},\a'^{(l)f}_{--,q})(\a'^{(l)f}_{++,r},\a'^{(l)f}_{--,s})(\tilde{\a}^{t_2}_{++,-1},\tilde{\a}^{t_1}_{--,-1})(\a'^{(j)i}_{++,-n_1},\a'^{(k)i}_{--,-n_2})\nn
 &&\qquad +(\a'^{(l)f}_{++,p},\a'^{(l)f}_{--,q})(\a'^{(l)f}_{++,r},\a'^{(l)f}_{--,s})(\tilde{\a}^{t_2}_{++,-1},\a'^{(k)i}_{--,-n_2})(\tilde{\a}^{t_1}_{--,-1},\a'^{(j)i}_{++,-n_1})\cr
 &&\qquad +(\a'^{(l)f}_{++,p},\a'^{(l)f}_{--,q})(\a'^{(l)f}_{++,r},\tilde{\a}^{t_1}_{--,-1})(\a'^{(l)f}_{--,s},\tilde{\a}^{t_2}_{++,-1})(\a'^{(j)i}_{++,-n_1},\a'^{(k)i}_{--,-n_2})\cr
 &&\qquad +(\a'^{(l)f}_{++,p},\a'^{(l)f}_{--,q})(\a'^{(l)f}_{++,r},\tilde{\a}^{t_1}_{--,-1})(\a'^{(l)f}_{--,s},\a'^{(j)i}_{++,-n_1})(\tilde{\a}^{t_2}_{++,-1},\a'^{(k)i}_{--,-n_2})\cr
 &&\qquad + (\a'^{(l)f}_{++,p},\a'^{(l)f}_{--,q})(\a'^{(l)f}_{++,r},\a'^{(k)i}_{--,-n_2})(\tilde{\a}^{t_2}_{++,-1},\tilde{\a}^{t_1}_{--,-1})(\a'^{(l)f}_{--,s},\a'^{(j)i}_{++,-n_1})\cr
 &&\qquad + (\a'^{(l)f}_{++,p},\a'^{(l)f}_{--,q})(\a'^{(l)f}_{++,r},\a'^{(k)i}_{--,-n_2})(\a'^{(l)f}_{--,s},\tilde{\a}^{t_2}_{++,-1})(\tilde{\a}^{t_1}_{--,-1},\a'^{(j)i}_{++,-n_1})
 \cr
 \cr
 &&\qquad + (\a'^{(l)f}_{++,p},\a'^{(l)f}_{--,s})(\a'^{(l)f}_{--,q},\a'^{(l)f}_{++,r})(\tilde{\a}^{t_2}_{++,-1},\tilde{\a}^{t_1}_{--,-1})(\a'^{(j)i}_{++,-n_1},\a'^{(k)i}_{--,-n_2})\cr
 &&\qquad + (\a'^{(l)f}_{++,p},\a'^{(l)f}_{--,s})(\a'^{(l)f}_{--,q},\a'^{(l)f}_{++,r})(\tilde{\a}^{t_2}_{++,-1},\a'^{(k)i}_{--,-n_2})(\tilde{\a}^{t_1}_{--,-1},\a'^{(j)i}_{++,-n_1})\cr
 &&\qquad + (\a'^{(l)f}_{++,p},\a'^{(l)f}_{--,s})(\a'^{(l)f}_{--,q},\tilde{\a}^{t_2}_{++,-1})(\a'^{(l)f}_{++,r},\tilde{\a}^{t_1}_{--,-1})(\a'^{(j)i}_{++,-n_1},\a'^{(k)i}_{--,-n_2})\cr
 &&\qquad + (\a'^{(l)f}_{++,p},\a'^{(l)f}_{--,s})(\a'^{(l)f}_{--,q},\tilde{\a}^{t_2}_{++,-1})(\a'^{(l)f}_{++,r},\a'^{(k)i}_{--,-n_2})(\tilde{\a}^{t_1}_{--,-1},\a'^{(j)i}_{++,-n_1})\cr
 &&\qquad + (\a'^{(l)f}_{++,p},\a'^{(l)f}_{--,s})(\a'^{(l)f}_{--,q},\a'^{(j)i}_{++,-n_1})(\tilde{\a}^{t_2}_{++,-1},\tilde{\a}^{t_1}_{--,-1})(\a'^{(l)f}_{++,r},\a'^{(k)i}_{--,-n_2})\cr
 &&\qquad + (\a'^{(l)f}_{++,p},\a'^{(l)f}_{--,s})(\a'^{(l)f}_{--,q},\a'^{(j)i}_{++,-n_1})(\a'^{(l)f}_{++,r},\tilde{\a}^{t_1}_{--,-1})(\tilde{\a}^{t_2}_{++,-1},\a'^{(k)i}_{--,-n_2})
 \cr
 \cr
&&\qquad + (\a'^{(l)f}_{++,p},\tilde{\a}^{t_1}_{--,-1})(\a'^{(l)f}_{++,r},\a'^{(l)f}_{--,s})(\a'^{(l)f}_{--,q},\tilde{\a}^{t_2}_{++,-1})(\a'^{(j)i}_{++,-n_1},\a'^{(k)i}_{--,-n_2})\cr
&&\qquad + (\a'^{(l)f}_{++,p},\tilde{\a}^{t_1}_{--,-1})(\a'^{(l)f}_{++,r},\a'^{(l)f}_{--,s})(\a'^{(l)f}_{--,q},\a'^{(j)i}_{++,-n_1})(\tilde{\a}^{t_2}_{++,-1},\a'^{(k)i}_{--,-n_2})\cr
&&\qquad + (\a'^{(l)f}_{++,p},\tilde{\a}^{t_1}_{--,-1})(\a'^{(l)f}_{++,r},\a'^{(l)f}_{--,q})(\a'^{(l)f}_{--,s},\tilde{\a}^{t_2}_{++,-1})(\a'^{(j)i}_{++,-n_1},\a'^{(k)i}_{--,-n_2})\cr
&&\qquad + (\a'^{(l)f}_{++,p},\tilde{\a}^{t_1}_{--,-1})(\a'^{(l)f}_{++,r},\a'^{(l)f}_{--,q})(\a'^{(l)f}_{--,s},\a'^{(j)i}_{++,-n_1})(\tilde{\a}^{t_2}_{++,-1},\a'^{(k)i}_{--,-n_2})\cr
&&\qquad + (\a'^{(l)f}_{++,p},\tilde{\a}^{t_1}_{--,-1})(\a'^{(l)f}_{++,r},\a'^{(k)i}_{--,-n_2})(\a'^{(l)f}_{--,q},\tilde{\a}^{t_2}_{++,-1})(\a'^{(l)f}_{--,s},\a'^{(j)i}_{++,-n_1})\cr
&&\qquad + (\a'^{(l)f}_{++,p},\tilde{\a}^{t_1}_{--,-1})(\a'^{(l)f}_{++,r},\a'^{(k)i}_{--,-n_2})(\a'^{(l)f}_{--,q},\a'^{(j)i}_{++,-n_1})(\a'^{(l)f}_{--,s},\tilde{\a}^{t_2}_{++,-1})
\cr
\cr
&&\qquad + (\a'^{(l)f}_{++,p},\a'^{(k)i}_{--,-n_2})(\a'^{(l)f}_{++,r},\a'^{(l)f}_{--,s})(\tilde{\a}^{t_2}_{++,-1},\tilde{\a}^{t_1}_{--,-1})(\a'^{(l)f}_{--,q},\a'^{(j)i}_{++,-n_1})\cr
&&\qquad + (\a'^{(l)f}_{++,p},\a'^{(k)i}_{--,-n_2})(\a'^{(l)f}_{++,r},\a'^{(l)f}_{--,s})(\a'^{(l)f}_{--,q},\tilde{\a}^{t_2}_{++,-1})(\tilde{\a}^{t_1}_{--,-1},\a'^{(j)i}_{++,-n_1})\cr
&&\qquad + (\a'^{(l)f}_{++,p},\a'^{(k)i}_{--,-n_2})(\a'^{(l)f}_{++,r},\tilde{\a}^{t_1}_{--,-1})(\a'^{(l)f}_{--,s},\tilde{\a}^{t_2}_{++,-1})(\a'^{(l)f}_{--,q},\a'^{(j)i}_{++,-n_1})\cr
&&\qquad + (\a'^{(l)f}_{++,p},\a'^{(k)i}_{--,-n_2})(\a'^{(l)f}_{++,r},\tilde{\a}^{t_1}_{--,-1})(\a'^{(l)f}_{--,s},\a'^{(j)i}_{++,-n_1})(\tilde{\a}^{t_2}_{++,-1},\a'^{(l)f}_{--,q})\cr
&&\qquad + (\a'^{(l)f}_{++,p},\a'^{(k)i}_{--,-n_2})(\a'^{(l)f}_{--,q},\a'^{(l)f}_{++,r})(\tilde{\a}^{t_2}_{++,-1},\tilde{\a}^{t_1}_{--,-1})(\a'^{(l)f}_{--,s},\a'^{(j)i}_{++,-n_1})\cr
&&\qquad + (\a'^{(l)f}_{++,p},\a'^{(k)i}_{--,-n_2})(\a'^{(l)f}_{--,q},\a'^{(l)f}_{++,r})(\a'^{(l)f}_{--,s},\tilde{\a}^{t_2}_{++,-1})(\tilde{\a}^{t_1}_{--,-1},\a'^{(j)i}_{++,-n_1})
  \bigg)\cr
  &&\quad~~\times(\tilde{d}^{++,t_2}_{-{1\over2}},\tilde{d}^{--,t_1}_{-{1\over2}})
  \cr
  \cr
 &&\qquad\!\!\!\!+ \bigg((\a'^{(l)f}_{++,p},\a'^{(l)f}_{--,q})(\a'^{(l)f}_{++,r},\a'^{(l)f}_{--,s})(\a'^{(j)i}_{++,-n_1},\a'^{(k)i}_{--,-n_2})\cr
 &&\qquad +(\a'^{(l)f}_{++,p},\a'^{(l)f}_{--,q})(\a'^{(l)f}_{++,r},,\a'^{(k)i}_{--,-n_2})(\a'^{(l)f}_{--,s},\a'^{(j)i}_{++,-n_1})\cr
 &&\qquad+(\a'^{(l)f}_{++,p},\a'^{(l)f}_{--,s})(\a'^{(l)f}_{--,q},\a'^{(l)f}_{++,r})(\a'^{(j)i}_{++,-n_1},\a'^{(k)i}_{--,-n_2})\cr
 &&\qquad+(\a'^{(l)f}_{++,p},\a'^{(l)f}_{--,s})(\a'^{(l)f}_{--,q},\a'^{(j)i}_{++,-n_1})(\a'^{(l)f}_{++,r},\a'^{(k)i}_{--,-n_2})\cr
 &&\qquad+(\a'^{(l)f}_{++,p},\a'^{(k)i}_{--,-n_2})(\a'^{(l)f}_{++,r},\a'^{(l)f}_{--,s})(\a'^{(l)f}_{--,q} , \a'^{(j)i}_{++,-n_1})\cr
 &&\qquad+(\a'^{(l)f}_{++,p},\a'^{(k)i}_{--,-n_2})(\a'^{(l)f}_{--,q},\a'^{(l)f}_{++,r})(\a'^{(l)f}_{--,s},\a'^{(j)i}_{++,-n_1})\bigg)\cr
 &&\quad~~\times(\tilde{d}^{+-,t_2}_{-{1\over2}},\tilde{d}^{-+,t_1}_{-{1\over2}})(\tilde{\a}^{t_2}_{-+,-1},\tilde{\a}^{t_1}_{+-,-1})\bigg]
 \label{aa to aaaa pm}
\eea

\bea
&&\!\!\!\!\!\!\!\mathcal{A}^{\a^{(j)}\a^{(k)} \to \a^{(l)}\a^{(l)}\a^{(l)}\a^{(l)}}_{-+}\cr
&&=   {e^{{3\over2}s}\sinh({\Delta w\over2})\over8}\cr
&&\quad\bigg[\bigg( (\a'^{(l)f}_{++,p},\a'^{(l)f}_{--,q})(\a'^{(l)f}_{++,r},\a'^{(l)f}_{--,s})(\a'^{(j)i}_{++,-n_1},\a'^{(k)i}_{--,-n_2})\cr
 &&\qquad +(\a'^{(l)f}_{++,p},\a'^{(l)f}_{--,q})(\a'^{(l)f}_{++,r},,\a'^{(k)i}_{--,-n_2})(\a'^{(l)f}_{--,s},\a'^{(j)i}_{++,-n_1})\cr
 &&\qquad+(\a'^{(l)f}_{++,p},\a'^{(l)f}_{--,s})(\a'^{(l)f}_{--,q},\a'^{(l)f}_{++,r})(\a'^{(j)i}_{++,-n_1},\a'^{(k)i}_{--,-n_2})\cr
 &&\qquad+(\a'^{(l)f}_{++,p},\a'^{(l)f}_{--,s})(\a'^{(l)f}_{--,q},\a'^{(j)i}_{++,-n_1})(\a'^{(l)f}_{++,r},\a'^{(k)i}_{--,-n_2})\cr
 &&\qquad+(\a'^{(l)f}_{++,p},\a'^{(k)i}_{--,-n_2})(\a'^{(l)f}_{++,r},\a'^{(l)f}_{--,s})(\a'^{(l)f}_{--,q} , \a'^{(j)i}_{++,-n_1})\cr
 &&\qquad+(\a'^{(l)f}_{++,p},\a'^{(k)i}_{--,-n_2})(\a'^{(l)f}_{--,q},\a'^{(l)f}_{++,r})(\a'^{(l)f}_{--,s},\a'^{(j)i}_{++,-n_1})\bigg)\cr
 &&\quad~~\times(\tilde{d}^{++,t_2}_{-{1\over2}},\tilde{d}^{--,t_1}_{-{1\over2}})(\tilde{\a}^{t_2}_{+-,-1},\tilde{\a}^{t_1}_{-+,-1})
 \cr
 \cr
 &&\quad+ \bigg((\a'^{(l)f}_{++,p},\a'^{(l)f}_{--,q})(\a'^{(l)f}_{++,r},\a'^{(l)f}_{--,s})(\tilde{\a}^{t_2}_{--,-1},\tilde{\a}^{t_1}_{++,-1})(\a'^{(j)i}_{++,-n_1},\a'^{(k)i}_{--,-n_2})\nn
 &&\qquad +(\a'^{(l)f}_{++,p},\a'^{(l)f}_{--,q})(\a'^{(l)f}_{++,r},\a'^{(l)f}_{--,s})(\tilde{\a}^{t_1}_{++,-1},\a'^{(k)i}_{--,-n_2})(\tilde{\a}^{t_2}_{--,-1},\a'^{(j)i}_{++,-n_1})\cr
 &&\qquad +(\a'^{(l)f}_{++,p},\a'^{(l)f}_{--,q})(\a'^{(l)f}_{++,r},\tilde{\a}^{t_2}_{--,-1})(\a'^{(l)f}_{--,s},\tilde{\a}^{t_1}_{++,-1})(\a'^{(j)i}_{++,-n_1},\a'^{(k)i}_{--,-n_2})\cr
 &&\qquad +(\a'^{(l)f}_{++,p},\a'^{(l)f}_{--,q})(\a'^{(l)f}_{++,r},\tilde{\a}^{t_2}_{--,-1})(\a'^{(l)f}_{--,s},\a'^{(j)i}_{++,-n_1})(\tilde{\a}^{t_1}_{++,-1},\a'^{(k)i}_{--,-n_2})\cr
 &&\qquad + (\a'^{(l)f}_{++,p},\a'^{(l)f}_{--,q})(\a'^{(l)f}_{++,r},\a'^{(k)i}_{--,-n_2})(\tilde{\a}^{t_2}_{--,-1},\tilde{\a}^{t_1}_{++,-1})(\a'^{(l)f}_{--,s},\a'^{(j)i}_{++,-n_1})\cr
 &&\qquad + (\a'^{(l)f}_{++,p},\a'^{(l)f}_{--,q})(\a'^{(l)f}_{++,r},\a'^{(k)i}_{--,-n_2})(\a'^{(l)f}_{--,s},\tilde{\a}^{t_1}_{++,-1})(\tilde{\a}^{t_2}_{--,-1},\a'^{(j)i}_{++,-n_1})
 \cr
 \cr
 &&\qquad + (\a'^{(l)f}_{++,p},\a'^{(l)f}_{--,s})(\a'^{(l)f}_{--,q},\a'^{(l)f}_{++,r})(\tilde{\a}^{t_2}_{--,-1},\tilde{\a}^{t_1}_{++,-1})(\a'^{(j)i}_{++,-n_1},\a'^{(k)i}_{--,-n_2})\cr
 &&\qquad + (\a'^{(l)f}_{++,p},\a'^{(l)f}_{--,s})(\a'^{(l)f}_{--,q},\a'^{(l)f}_{++,r})(\tilde{\a}^{t_1}_{++,-1},\a'^{(k)i}_{--,-n_2})(\tilde{\a}^{t_2}_{--,-1},\a'^{(j)i}_{++,-n_1})\cr
 &&\qquad + (\a'^{(l)f}_{++,p},\a'^{(l)f}_{--,s})(\a'^{(l)f}_{--,q},\tilde{\a}^{t_1}_{++,-1})(\a'^{(l)f}_{++,r},\tilde{\a}^{t_2}_{--,-1})(\a'^{(j)i}_{++,-n_1},\a'^{(k)i}_{--,-n_2})\cr
 &&\qquad + (\a'^{(l)f}_{++,p},\a'^{(l)f}_{--,s})(\a'^{(l)f}_{--,q},\tilde{\a}^{t_1}_{++,-1})(\a'^{(l)f}_{++,r},\a'^{(k)i}_{--,-n_2})(\tilde{\a}^{t_2}_{--,-1},\a'^{(j)i}_{++,-n_1})\cr
 &&\qquad + (\a'^{(l)f}_{++,p},\a'^{(l)f}_{--,s})(\a'^{(l)f}_{--,q},\a'^{(j)i}_{++,-n_1})(\tilde{\a}^{t_2}_{--,-1},\tilde{\a}^{t_1}_{++,-1})(\a'^{(l)f}_{++,r},\a'^{(k)i}_{--,-n_2})\cr
 &&\qquad + (\a'^{(l)f}_{++,p},\a'^{(l)f}_{--,s})(\a'^{(l)f}_{--,q},\a'^{(j)i}_{++,-n_1})(\a'^{(l)f}_{++,r},\tilde{\a}^{t_2}_{--,-1})(\tilde{\a}^{t_1}_{++,-1},\a'^{(k)i}_{--,-n_2})
 \cr
 \cr
&&\qquad + (\a'^{(l)f}_{++,p},\tilde{\a}^{t_2}_{--,-1})(\a'^{(l)f}_{++,r},\a'^{(l)f}_{--,s})(\a'^{(l)f}_{--,q},\tilde{\a}^{t_1}_{++,-1})(\a'^{(j)i}_{++,-n_1},\a'^{(k)i}_{--,-n_2})\cr
&&\qquad + (\a'^{(l)f}_{++,p},\tilde{\a}^{t_2}_{--,-1})(\a'^{(l)f}_{++,r},\a'^{(l)f}_{--,s})(\a'^{(l)f}_{--,q},\a'^{(j)i}_{++,-n_1})(\tilde{\a}^{t_1}_{++,-1},\a'^{(k)i}_{--,-n_2})\cr
&&\qquad + (\a'^{(l)f}_{++,p},\tilde{\a}^{t_2}_{--,-1})(\a'^{(l)f}_{++,r},\a'^{(l)f}_{--,q})(\a'^{(l)f}_{--,s},\tilde{\a}^{t_1}_{++,-1})(\a'^{(j)i}_{++,-n_1},\a'^{(k)i}_{--,-n_2})\cr
&&\qquad + (\a'^{(l)f}_{++,p},\tilde{\a}^{t_2}_{--,-1})(\a'^{(l)f}_{++,r},\a'^{(l)f}_{--,q})(\a'^{(l)f}_{--,s},\a'^{(j)i}_{++,-n_1})(\tilde{\a}^{t_1}_{++,-1},\a'^{(k)i}_{--,-n_2})\cr
&&\qquad + (\a'^{(l)f}_{++,p},\tilde{\a}^{t_2}_{--,-1})(\a'^{(l)f}_{++,r},\a'^{(k)i}_{--,-n_2})(\a'^{(l)f}_{--,q},\tilde{\a}^{t_1}_{++,-1})(\a'^{(l)f}_{--,s},\a'^{(j)i}_{++,-n_1})\cr
&&\qquad + (\a'^{(l)f}_{++,p},\tilde{\a}^{t_2}_{--,-1})(\a'^{(l)f}_{++,r},\a'^{(k)i}_{--,-n_2})(\a'^{(l)f}_{--,q},\a'^{(j)i}_{++,-n_1})(\a'^{(l)f}_{--,s},\tilde{\a}^{t_1}_{++,-1})
\cr
\cr
&&\qquad + (\a'^{(l)f}_{++,p},\a'^{(k)i}_{--,-n_2})(\a'^{(l)f}_{++,r},\a'^{(l)f}_{--,s})(\tilde{\a}^{t_2}_{--,-1},\tilde{\a}^{t_1}_{++,-1})(\a'^{(l)f}_{--,q},\a'^{(j)i}_{++,-n_1})\cr
&&\qquad + (\a'^{(l)f}_{++,p},\a'^{(k)i}_{--,-n_2})(\a'^{(l)f}_{++,r},\a'^{(l)f}_{--,s})(\a'^{(l)f}_{--,q},\tilde{\a}^{t_1}_{++,-1})(\tilde{\a}^{t_2}_{--,-1},\a'^{(j)i}_{++,-n_1})\cr
&&\qquad + (\a'^{(l)f}_{++,p},\a'^{(k)i}_{--,-n_2})(\a'^{(l)f}_{++,r},\tilde{\a}^{t_2}_{--,-1})(\a'^{(l)f}_{--,s},\tilde{\a}^{t_1}_{++,-1})(\a'^{(l)f}_{--,q},\a'^{(j)i}_{++,-n_1})\cr
&&\qquad + (\a'^{(l)f}_{++,p},\a'^{(k)i}_{--,-n_2})(\a'^{(l)f}_{++,r},\tilde{\a}^{t_2}_{--,-1})(\a'^{(l)f}_{--,s},\a'^{(j)i}_{++,-n_1})(\tilde{\a}^{t_1}_{++,-1},\a'^{(l)f}_{--,q})\cr
&&\qquad + (\a'^{(l)f}_{++,p},\a'^{(k)i}_{--,-n_2})(\a'^{(l)f}_{--,q},\a'^{(l)f}_{++,r})(\tilde{\a}^{t_1}_{++,-1},\tilde{\a}^{t_2}_{--,-1})(\a'^{(l)f}_{--,s},\a'^{(j)i}_{++,-n_1})\cr
&&\qquad + (\a'^{(l)f}_{++,p},\a'^{(k)i}_{--,-n_2})(\a'^{(l)f}_{--,q},\a'^{(l)f}_{++,r})(\a'^{(l)f}_{--,s},\tilde{\a}^{t_1}_{++,-1})(\tilde{\a}^{t_2}_{--,-1},\a'^{(j)i}_{++,-n_1})
  \bigg)\cr
  &&\quad~~\times(\tilde{d}^{+-,t_2}_{-{1\over2}},\tilde{d}^{-+,t_1}_{-{1\over2}})\bigg]
   \label{aa to aaaa mp}
\eea

\bea
&&\!\!\!\!\!\!\!\bar{\mathcal{A}}^{\bar\a^{(j)}\bar\a^{(k)} \to \bar\a^{(l)}\bar\a^{(l)}\bar\a^{(l)}\bar\a^{(l)}}_{+-}\cr
&& =   {e^{{3\over2}\bar s}\sinh({\Delta\bar{  w}\over2})\over8}\cr
&&\quad\bigg[ \bigg((\bar{\a}'^{(l)f}_{--,p},\bar{\a}'^{(l)f}_{++,q})(\bar{\a}'^{(l)f}_{--,r},\bar{\a}'^{(l)f}_{++,s})(\tilde{\bar{\a}}^{\bar{t}_2}_{++,-1},\tilde{\bar{\a}}^{\bar{t}_1}_{--,-1})(\bar{\a}'^{(j)i}_{--,-n_1},\bar{\a}'^{(k)i}_{++,-n_2})\cr
 &&\qquad + (\bar{\a}'^{(l)f}_{--,p},\bar{\a}'^{(l)f}_{++,q})(\bar{\a}'^{(l)f}_{--,r},\bar{\a}'^{(l)f}_{++,s})(\tilde{\bar{\a}}^{\bar{t}_2}_{++,-1},\bar{\a}'^{(j)i}_{--,-n_1})(\tilde{\bar{\a}}^{\bar{t}_1}_{--,-1},\bar{\a}'^{(k)i}_{++,-n_2})\cr
 &&\qquad + (\bar{\a}'^{(l)f}_{--,p},\bar{\a}'^{(l)f}_{++,q})(\bar{\a}'^{(l)f}_{--,r},\tilde{\bar{\a}}^{\bar{t}_2}_{++,-1})(\bar{\a}'^{(l)f}_{++,s},\tilde{\bar{\a}}^{\bar{t}_1}_{--,-1})(\bar{\a}'^{(j)i}_{--,-n_1},\bar{\a}'^{(k)i}_{++,-n_2})\cr
 &&\qquad + (\bar{\a}'^{(l)f}_{--,p},\bar{\a}'^{(l)f}_{++,q})(\bar{\a}'^{(l)f}_{--,r},\tilde{\bar{\a}}^{\bar{t}_2}_{++,-1})(\bar{\a}'^{(l)f}_{++,s},\bar{\a}'^{(j)i}_{--,-n_1})(\tilde{\bar{\a}}^{\bar{t}_1}_{--,-1},\bar{\a}'^{(k)i}_{++,-n_2})\cr
 &&\qquad + (\bar{\a}'^{(l)f}_{--,p},\bar{\a}'^{(l)f}_{++,q})(\bar{\a}'^{(l)f}_{--,r},\bar{\a}'^{(k)i}_{++,-n_2})(\tilde{\bar{\a}}^{\bar{t}_2}_{++,-1},\tilde{\bar{\a}}^{\bar{t}_1}_{--,-1})(\bar{\a}'^{(l)f}_{++,s},\bar{\a}'^{(j)i}_{--,-n_1})\cr
 &&\qquad + (\bar{\a}'^{(l)f}_{--,p},\bar{\a}'^{(l)f}_{++,q})(\bar{\a}'^{(l)f}_{--,r},\bar{\a}'^{(k)i}_{++,-n_2})(\bar{\a}'^{(l)f}_{++,s},\tilde{\bar{\a}}^{\bar{t}_1}_{--,-1})(\tilde{\bar{\a}}^{\bar{t}_2}_{++,-1},\bar{\a}'^{(j)i}_{--,-n_1})
 \cr
 \cr
  &&\qquad +(\bar{\a}'^{(l)f}_{--,p},\bar{\a}'^{(l)f}_{++,s})(\bar{\a}'^{(l)f}_{++,q},\bar{\a}'^{(l)f}_{--,r})(\tilde{\bar{\a}}^{\bar{t}_2}_{++,-1},\tilde{\bar{\a}}^{\bar{t}_1}_{--,-1})(\bar{\a}'^{(j)i}_{--,-n_1},\bar{\a}'^{(k)i}_{++,-n_2})\cr
  &&\qquad +(\bar{\a}'^{(l)f}_{--,p},\bar{\a}'^{(l)f}_{++,s})(\bar{\a}'^{(l)f}_{++,q},\bar{\a}'^{(l)f}_{--,r})(\tilde{\bar{\a}}^{\bar{t}_2}_{++,-1}\bar{\a}'^{(j)i}_{--,-n_1})(\tilde{\bar{\a}}^{\bar{t}_1}_{--,-1},\bar{\a}'^{(k)i}_{++,-n_2})\cr
  &&\qquad +(\bar{\a}'^{(l)f}_{--,p},\bar{\a}'^{(l)f}_{++,s})(\bar{\a}'^{(l)f}_{++,q},\tilde{\bar{\a}}^{\bar{t}_1}_{--,-1})(\bar{\a}'^{(l)f}_{--,r},\tilde{\bar{\a}}^{\bar{t}_2}_{++,-1})(\bar{\a}'^{(j)i}_{--,-n_1},\bar{\a}'^{(k)i}_{++,-n_2})\cr
  &&\qquad +(\bar{\a}'^{(l)f}_{--,p},\bar{\a}'^{(l)f}_{++,s})(\bar{\a}'^{(l)f}_{++,q},\tilde{\bar{\a}}^{\bar{t}_1}_{--,-1})(\bar{\a}'^{(l)f}_{--,r},\bar{\a}'^{(k)i}_{++,-n_2})(\tilde{\bar{\a}}^{\bar{t}_2}_{++,-1},\bar{\a}'^{(j)i}_{--,-n_1})\cr
  &&\qquad +(\bar{\a}'^{(l)f}_{--,p},\bar{\a}'^{(l)f}_{++,s})(\bar{\a}'^{(l)f}_{++,q},\bar{\a}'^{(j)i}_{--,-n_1})(\tilde{\bar{\a}}^{\bar{t}_2}_{++,-1},\tilde{\bar{\a}}^{\bar{t}_1}_{--,-1})(\bar{\a}'^{(l)f}_{--,r},\bar{\a}'^{(k)i}_{++,-n_2})\cr
  &&\qquad +(\bar{\a}'^{(l)f}_{--,p},\bar{\a}'^{(l)f}_{++,s})(\bar{\a}'^{(l)f}_{++,q},\bar{\a}'^{(j)i}_{--,-n_1})(\bar{\a}'^{(l)f}_{--,r},\tilde{\bar{\a}}^{\bar{t}_2}_{++,-1})(\tilde{\bar{\a}}^{\bar{t}_1}_{--,-1},\bar{\a}'^{(k)i}_{++,-n_2})\cr
  \cr
  &&\qquad +(\bar{\a}'^{(l)f}_{--,p},\tilde{\bar{\a}}^{\bar{t}_2}_{++,-1})(\bar{\a}'^{(l)f}_{--,r},\bar{\a}'^{(l)f}_{++,s})(\bar{\a}'^{(l)f}_{++,q},\tilde{\bar{\a}}^{\bar{t}_1}_{--,-1})(\bar{\a}'^{(j)i}_{--,-n_1},\bar{\a}'^{(k)i}_{++,-n_2})\cr
   &&\qquad +(\bar{\a}'^{(l)f}_{--,p},\tilde{\bar{\a}}^{\bar{t}_2}_{++,-1})(\bar{\a}'^{(l)f}_{--,r},\bar{\a}'^{(l)f}_{++,s})(\bar{\a}'^{(l)f}_{++,q},\bar{\a}'^{(j)i}_{--,-n_1})(\tilde{\bar{\a}}^{\bar{t}_1}_{--,-1},\bar{\a}'^{(k)i}_{++,-n_2})\cr
    &&\qquad +(\bar{\a}'^{(l)f}_{--,p},\tilde{\bar{\a}}^{\bar{t}_2}_{++,-1})(\bar{\a}'^{(l)f}_{++,q},\bar{\a}'^{(l)f}_{--,r})(\bar{\a}'^{(l)f}_{++,s},\tilde{\bar{\a}}^{\bar{t}_1}_{--,-1})(\bar{\a}'^{(j)i}_{--,-n_1},\bar{\a}'^{(k)i}_{++,-n_2})\cr
    &&\qquad +(\bar{\a}'^{(l)f}_{--,p},\tilde{\bar{\a}}^{\bar{t}_2}_{++,-1})(\bar{\a}'^{(l)f}_{++,q},\bar{\a}'^{(l)f}_{--,r})(\bar{\a}'^{(l)f}_{++,s},\bar{\a}'^{(j)i}_{--,-n_1})(\tilde{\bar{\a}}^{\bar{t}_1}_{--,-1},\bar{\a}'^{(k)i}_{++,-n_2})\cr
    &&\qquad +(\bar{\a}'^{(l)f}_{--,p},\tilde{\bar{\a}}^{\bar{t}_2}_{++,-1})(\bar{\a}'^{(l)f}_{--,r},\bar{\a}'^{(k)i}_{++,-n_2})(\bar{\a}'^{(l)f}_{++,q},\tilde{\bar{\a}}^{\bar{t}_1}_{--,-1})(\bar{\a}'^{(l)f}_{++,s},\bar{\a}'^{(j)i}_{--,-n_1})\cr
     &&\qquad +(\bar{\a}'^{(l)f}_{--,p},\tilde{\bar{\a}}^{\bar{t}_2}_{++,-1})(\bar{\a}'^{(l)f}_{--,r},\bar{\a}'^{(k)i}_{++,-n_2})(\bar{\a}'^{(l)f}_{++,q},\bar{\a}'^{(j)i}_{--,-n_1})(\bar{\a}'^{(l)f}_{++,s},\tilde{\bar{\a}}^{\bar{t}_1}_{--,-1})
     \cr
     \cr
    &&\qquad+ (\bar{\a}'^{(l)f}_{--,p},\bar{\a}'^{(k)i}_{++,-n_2})(\bar{\a}'^{(l)f}_{--,r},\bar{\a}'^{(l)f}_{++,s})(\tilde{\bar{\a}}^{\bar{t}_2}_{++,-1},\tilde{\bar{\a}}^{\bar{t}_1}_{--,-1})(\bar{\a}'^{(l)f}_{++,q},\bar{\a}'^{(j)i}_{--,-n_1})\cr
     &&\qquad+ (\bar{\a}'^{(l)f}_{--,p},\bar{\a}'^{(k)i}_{++,-n_2})(\bar{\a}'^{(l)f}_{--,r},\bar{\a}'^{(l)f}_{++,s})(\bar{\a}'^{(l)f}_{++,q},\tilde{\bar{\a}}^{\bar{t}_1}_{--,-1})(\tilde{\bar{\a}}^{\bar{t}_2}_{++,-1},\bar{\a}'^{(j)i}_{--,-n_1})\cr
     &&\qquad+ (\bar{\a}'^{(l)f}_{--,p},\bar{\a}'^{(k)i}_{++,-n_2})(\bar{\a}'^{(l)f}_{--,r},\tilde{\bar{\a}}^{\bar{t}_2}_{++,-1})(\bar{\a}'^{(l)f}_{++,s},\tilde{\bar{\a}}^{\bar{t}_1}_{--,-1})(\bar{\a}'^{(l)f}_{++,q},\bar{\a}'^{(j)i}_{--,-n_1})\cr
     &&\qquad+ (\bar{\a}'^{(l)f}_{--,p},\bar{\a}'^{(k)i}_{++,-n_2})(\bar{\a}'^{(l)f}_{--,r},\tilde{\bar{\a}}^{\bar{t}_2}_{++,-1})(\bar{\a}'^{(l)f}_{++,q},\tilde{\bar{\a}}^{\bar{t}_1}_{--,-1})(\bar{\a}'^{(l)f}_{++,s},\bar{\a}'^{(j)i}_{--,-n_1})\cr
     &&\qquad+ (\bar{\a}'^{(l)f}_{--,p},\bar{\a}'^{(k)i}_{++,-n_2})(\bar{\a}'^{(l)f}_{++,q},\bar{\a}'^{(l)f}_{--,r})(\tilde{\bar{\a}}^{\bar{t}_2}_{++,-1},\tilde{\bar{\a}}^{\bar{t}_1}_{--,-1})(\bar{\a}'^{(l)f}_{++,s},\bar{\a}'^{(j)i}_{--,-n_1})\cr
     &&\qquad+ (\bar{\a}'^{(l)f}_{--,p},\bar{\a}'^{(k)i}_{++,-n_2})(\bar{\a}'^{(l)f}_{++,q},\bar{\a}'^{(l)f}_{--,r})(\tilde{\bar{\a}}^{\bar{t}_2}_{++,-1},\bar{\a}'^{(j)i}_{--,-n_1})(\bar{\a}'^{(l)f}_{++,s},\tilde{\bar{\a}}^{\bar{t}_1}_{--,-1})
     \bigg)\cr
     &&\quad~~\times(\tilde{\bar{d}}^{++,\bar{t}_2}_{-{1\over2}},\tilde{\bar{d}}^{--,\bar{t}_1}_{-{1\over2}})
     \cr
     \cr
     &&\quad+\bigg((\bar{\a}'^{(l)f}_{--,p},\bar{\a}'^{(l)f}_{++,q})(\bar{\a}'^{(l)f}_{--,r},\bar{\a}'^{(l)f}_{++,s})(\bar{\a}'^{(j)i}_{--,-n_1},\bar{\a}'^{(k)i}_{++,-n_2})\cr
     &&\qquad +(\bar{\a}'^{(l)f}_{--,p},\bar{\a}'^{(l)f}_{++,q})(\bar{\a}'^{(l)f}_{--,r},\bar{\a}'^{(k)i}_{++,-n_2})(\bar{\a}'^{(l)f}_{++,s},\bar{\a}'^{(j)i}_{--,-n_1})\cr
     &&\qquad +(\bar{\a}'^{(l)f}_{--,p},\bar{\a}'^{(l)f}_{++,s})(\bar{\a}'^{(l)f}_{++,q},\bar{\a}'^{(l)f}_{--,r})(\bar{\a}'^{(j)i}_{--,-n_1},\bar{\a}'^{(k)i}_{++,-n_2})\cr
     &&\qquad +(\bar{\a}'^{(l)f}_{--,p},\bar{\a}'^{(l)f}_{++,s})(\bar{\a}'^{(l)f}_{++,q},\bar{\a}'^{(j)i}_{--,-n_1})(\bar{\a}'^{(l)f}_{--,r},\bar{\a}'^{(k)i}_{++,-n_2})\cr
      &&\qquad +(\bar{\a}'^{(l)f}_{--,p},\bar{\a}'^{(k)i}_{++,-n_2})(\bar{\a}'^{(l)f}_{--,r},\bar{\a}'^{(l)f}_{++,s})(\bar{\a}'^{(l)f}_{++,q},\bar{\a}'^{(j)i}_{--,-n_1})\cr
      &&\qquad +(\bar{\a}'^{(l)f}_{--,p},\bar{\a}'^{(k)i}_{++,-n_2})(\bar{\a}'^{(l)f}_{++,q},\bar{\a}'^{(l)f}_{--,r})(\bar{\a}'^{(l)f}_{++,s},\bar{\a}'^{(j)i}_{--,-n_1})
     \bigg)\cr
     &&\quad~~\times(\tilde{\bar{d}}^{+-,\bar{t}_2}_{-{1\over2}},\tilde{\bar{d}}^{-+,\bar{t}_1}_{-{1\over2}})(\tilde{\bar{\a}}^{\bar{t}_2}_{-+,-1},\tilde{\bar{\a}}^{\bar{t}_1}_{+-,-1})\bigg]
      \label{aa to aaaa pm bar}
\eea

\bea
&&\!\!\!\!\!\!\!\bar{\mathcal{A}}^{\bar\a^{(j)}\bar\a^{(k)} \to \bar\a^{(l)}\bar\a^{(l)}\bar\a^{(l)}\bar\a^{(l)}}_{-+}\cr
&& =   {e^{{3\over2}\bar s}\sinh({\Delta  \bar{w}\over2})\over8}\cr
&&\quad\bigg[\bigg((\bar{\a}'^{(l)f}_{--,p},\bar{\a}'^{(l)f}_{++,q})(\bar{\a}'^{(l)f}_{--,r},\bar{\a}'^{(l)f}_{++,s})(\bar{\a}'^{(j)i}_{--,-n_1},\bar{\a}'^{(k)i}_{++,-n_2})\cr
     &&\quad +(\bar{\a}'^{(l)f}_{--,p},\bar{\a}'^{(l)f}_{++,q})(\bar{\a}'^{(l)f}_{--,r},\bar{\a}'^{(k)i}_{++,-n_2})(\bar{\a}'^{(l)f}_{++,s},\bar{\a}'^{(j)i}_{--,-n_1})\cr
     &&\quad +(\bar{\a}'^{(l)f}_{--,p},\bar{\a}'^{(l)f}_{++,s})(\bar{\a}'^{(l)f}_{++,q},\bar{\a}'^{(l)f}_{--,r})(\bar{\a}'^{(j)i}_{--,-n_1},\bar{\a}'^{(k)i}_{++,-n_2})\cr
     &&\quad +(\bar{\a}'^{(l)f}_{--,p},\bar{\a}'^{(l)f}_{++,s})(\bar{\a}'^{(l)f}_{++,q},\bar{\a}'^{(j)i}_{--,-n_1})(\bar{\a}'^{(l)f}_{--,r},\bar{\a}'^{(k)i}_{++,-n_2})\cr
      &&\quad +(\bar{\a}'^{(l)f}_{--,p},\bar{\a}'^{(k)i}_{++,-n_2})(\bar{\a}'^{(l)f}_{--,r},\bar{\a}'^{(l)f}_{++,s})(\bar{\a}'^{(l)f}_{++,q},\bar{\a}'^{(j)i}_{--,-n_1})\cr
      &&\quad +(\bar{\a}'^{(l)f}_{--,p},\bar{\a}'^{(k)i}_{++,-n_2})(\bar{\a}'^{(l)f}_{++,q},\bar{\a}'^{(l)f}_{--,r})(\bar{\a}'^{(l)f}_{++,s},\bar{\a}'^{(j)i}_{--,-n_1})
     \bigg)\cr
     &&\quad~~\times(\tilde{\bar{d}}^{++,\bar{t}_2}_{-{1\over2}},\tilde{\bar{d}}^{--,\bar{t}_1}_{-{1\over2}})(\tilde{\bar{\a}}^{\bar{t}_2}_{+-,-1},\tilde{\bar{\a}}^{\bar{t}_1}_{-+,-1})
     \cr
     \cr
 &&\quad+\bigg((\bar{\a}'^{(l)f}_{--,p},\bar{\a}'^{(l)f}_{++,q})(\bar{\a}'^{(l)f}_{--,r},\bar{\a}'^{(l)f}_{++,s})(\tilde{\bar{\a}}^{\bar{t}_2}_{--,-1},\tilde{\bar{\a}}^{\bar{t}_1}_{++,-1})(\bar{\a}'^{(j)i}_{--,-n_1},\bar{\a}'^{(k)i}_{++,-n_2})\cr
 &&\qquad + (\bar{\a}'^{(l)f}_{--,p},\bar{\a}'^{(l)f}_{++,q})(\bar{\a}'^{(l)f}_{--,r},\bar{\a}'^{(l)f}_{++,s})(\tilde{\bar{\a}}^{\bar{t}_1}_{++,-1},\bar{\a}'^{(j)i}_{--,-n_1})(\tilde{\bar{\a}}^{\bar{t}_2}_{--,-1},\bar{\a}'^{(k)i}_{++,-n_2})\cr
 &&\qquad + (\bar{\a}'^{(l)f}_{--,p},\bar{\a}'^{(l)f}_{++,q})(\bar{\a}'^{(l)f}_{--,r},\tilde{\bar{\a}}^{\bar{t}_1}_{++,-1})(\bar{\a}'^{(l)f}_{++,s},\tilde{\bar{\a}}^{\bar{t}_2}_{--,-1})(\bar{\a}'^{(j)i}_{--,-n_1},\bar{\a}'^{(k)i}_{++,-n_2})\cr
 &&\qquad + (\bar{\a}'^{(l)f}_{--,p},\bar{\a}'^{(l)f}_{++,q})(\bar{\a}'^{(l)f}_{--,r},\tilde{\bar{\a}}^{\bar{t}_1}_{++,-1})(\bar{\a}'^{(l)f}_{++,s},\bar{\a}'^{(j)i}_{--,-n_1})(\tilde{\bar{\a}}^{\bar{t}_2}_{--,-1},\bar{\a}'^{(k)i}_{++,-n_2})\cr
 &&\qquad + (\bar{\a}'^{(l)f}_{--,p},\bar{\a}'^{(l)f}_{++,q})(\bar{\a}'^{(l)f}_{--,r},\bar{\a}'^{(k)i}_{++,-n_2})(\tilde{\bar{\a}}^{\bar{t}_2}_{--,-1},\tilde{\bar{\a}}^{\bar{t}_1}_{++,-1})(\bar{\a}'^{(l)f}_{++,s},\bar{\a}'^{(j)i}_{--,-n_1})\cr
 &&\qquad + (\bar{\a}'^{(l)f}_{--,p},\bar{\a}'^{(l)f}_{++,q})(\bar{\a}'^{(l)f}_{--,r},\bar{\a}'^{(k)i}_{++,-n_2})(\bar{\a}'^{(l)f}_{++,s},\tilde{\bar{\a}}^{\bar{t}_2}_{--,-1})(\tilde{\bar{\a}}^{\bar{t}_1}_{++,-1},\bar{\a}'^{(j)i}_{--,-n_1})
 \cr
 \cr
  &&\qquad +(\bar{\a}'^{(l)f}_{--,p},\bar{\a}'^{(l)f}_{++,s})(\bar{\a}'^{(l)f}_{++,q},\bar{\a}'^{(l)f}_{--,r})(\tilde{\bar{\a}}^{\bar{t}_2}_{--,-1},\tilde{\bar{\a}}^{\bar{t}_1}_{++,-1})(\bar{\a}'^{(j)i}_{--,-n_1},\bar{\a}'^{(k)i}_{++,-n_2})\cr
  &&\qquad +(\bar{\a}'^{(l)f}_{--,p},\bar{\a}'^{(l)f}_{++,s})(\bar{\a}'^{(l)f}_{++,q},\bar{\a}'^{(l)f}_{--,r})(\tilde{\bar{\a}}^{\bar{t}_1}_{++,-1}\bar{\a}'^{(j)i}_{--,-n_1})(\tilde{\bar{\a}}^{\bar{t}_2}_{--,-1},\bar{\a}'^{(k)i}_{++,-n_2})\cr
  &&\qquad +(\bar{\a}'^{(l)f}_{--,p},\bar{\a}'^{(l)f}_{++,s})(\bar{\a}'^{(l)f}_{++,q},\tilde{\bar{\a}}^{\bar{t}_2}_{--,-1})(\bar{\a}'^{(l)f}_{--,r},\tilde{\bar{\a}}^{\bar{t}_1}_{++,-1})(\bar{\a}'^{(j)i}_{--,-n_1},\bar{\a}'^{(k)i}_{++,-n_2})\cr
  &&\qquad +(\bar{\a}'^{(l)f}_{--,p},\bar{\a}'^{(l)f}_{++,s})(\bar{\a}'^{(l)f}_{++,q},\tilde{\bar{\a}}^{\bar{t}_2}_{--,-1})(\bar{\a}'^{(l)f}_{--,r},\bar{\a}'^{(k)i}_{++,-n_2})(\tilde{\bar{\a}}^{\bar{t}_1}_{++,-1},\bar{\a}'^{(j)i}_{--,-n_1})\cr
  &&\qquad +(\bar{\a}'^{(l)f}_{--,p},\bar{\a}'^{(l)f}_{++,s})(\bar{\a}'^{(l)f}_{++,q},\bar{\a}'^{(j)i}_{--,-n_1})(\tilde{\bar{\a}}^{\bar{t}_2}_{--,-1},\tilde{\bar{\a}}^{\bar{t}_1}_{++,-1})(\bar{\a}'^{(l)f}_{--,r},\bar{\a}'^{(k)i}_{++,-n_2})\cr
  &&\qquad +(\bar{\a}'^{(l)f}_{--,p},\bar{\a}'^{(l)f}_{++,s})(\bar{\a}'^{(l)f}_{++,q},\bar{\a}'^{(j)i}_{--,-n_1})(\bar{\a}'^{(l)f}_{--,r},\tilde{\bar{\a}}^{\bar{t}_1}_{++,-1})(\tilde{\bar{\a}}^{\bar{t}_2}_{--,-1},\bar{\a}'^{(k)i}_{++,-n_2})\cr
  \cr
  &&\qquad +(\bar{\a}'^{(l)f}_{--,p},\tilde{\bar{\a}}^{\bar{t}_1}_{++,-1})(\bar{\a}'^{(l)f}_{--,r},\bar{\a}'^{(l)f}_{++,s})(\bar{\a}'^{(l)f}_{++,q},\tilde{\bar{\a}}^{\bar{t}_2}_{--,-1})(\bar{\a}'^{(j)i}_{--,-n_1},\bar{\a}'^{(k)i}_{++,-n_2})\cr
   &&\qquad +(\bar{\a}'^{(l)f}_{--,p},\tilde{\bar{\a}}^{\bar{t}_1}_{++,-1})(\bar{\a}'^{(l)f}_{--,r},\bar{\a}'^{(l)f}_{++,s})(\bar{\a}'^{(l)f}_{++,q},\bar{\a}'^{(j)i}_{--,-n_1})(\tilde{\bar{\a}}^{\bar{t}_2}_{--,-1},\bar{\a}'^{(k)i}_{++,-n_2})\cr
    &&\qquad +(\bar{\a}'^{(l)f}_{--,p},\tilde{\bar{\a}}^{\bar{t}_1}_{++,-1})(\bar{\a}'^{(l)f}_{++,q},\bar{\a}'^{(l)f}_{--,r})(\bar{\a}'^{(l)f}_{++,s},\tilde{\bar{\a}}^{\bar{t}_2}_{--,-1})(\bar{\a}'^{(j)i}_{--,-n_1},\bar{\a}'^{(k)i}_{++,-n_2})\cr
    &&\qquad +(\bar{\a}'^{(l)f}_{--,p},\tilde{\bar{\a}}^{\bar{t}_1}_{++,-1})(\bar{\a}'^{(l)f}_{++,q},\bar{\a}'^{(l)f}_{--,r})(\bar{\a}'^{(l)f}_{++,s},\bar{\a}'^{(j)i}_{--,-n_1})(\tilde{\bar{\a}}^{\bar{t}_2}_{--,-1},\bar{\a}'^{(k)i}_{++,-n_2})\cr
    &&\qquad +(\bar{\a}'^{(l)f}_{--,p},\tilde{\bar{\a}}^{\bar{t}_1}_{++,-1})(\bar{\a}'^{(l)f}_{--,r},\bar{\a}'^{(k)i}_{++,-n_2})(\bar{\a}'^{(l)f}_{++,q},\tilde{\bar{\a}}^{\bar{t}_2}_{--,-1})(\bar{\a}'^{(l)f}_{++,s},\bar{\a}'^{(j)i}_{--,-n_1})\cr
     &&\qquad +(\bar{\a}'^{(l)f}_{--,p},\tilde{\bar{\a}}^{\bar{t}_1}_{++,-1})(\bar{\a}'^{(l)f}_{--,r},\bar{\a}'^{(k)i}_{++,-n_2})(\bar{\a}'^{(l)f}_{++,q},\bar{\a}'^{(j)i}_{--,-n_1})(\bar{\a}'^{(l)f}_{++,s},\tilde{\bar{\a}}^{\bar{t}_2}_{--,-1})
     \cr
     \cr
    &&\qquad+ (\bar{\a}'^{(l)f}_{--,p},\bar{\a}'^{(k)i}_{++,-n_2})(\bar{\a}'^{(l)f}_{--,r},\bar{\a}'^{(l)f}_{++,s})(\tilde{\bar{\a}}^{\bar{t}_2}_{--,-1},\tilde{\bar{\a}}^{\bar{t}_1}_{++,-1})(\bar{\a}'^{(l)f}_{++,q},\bar{\a}'^{(j)i}_{--,-n_1})\cr
     &&\qquad+ (\bar{\a}'^{(l)f}_{--,p},\bar{\a}'^{(k)i}_{++,-n_2})(\bar{\a}'^{(l)f}_{--,r},\bar{\a}'^{(l)f}_{++,s})(\bar{\a}'^{(l)f}_{++,q},\tilde{\bar{\a}}^{\bar{t}_2}_{--,-1})(\tilde{\bar{\a}}^{\bar{t}_1}_{++,-1},\bar{\a}'^{(j)i}_{--,-n_1})\cr
     &&\qquad+ (\bar{\a}'^{(l)f}_{--,p},\bar{\a}'^{(k)i}_{++,-n_2})(\bar{\a}'^{(l)f}_{--,r},\tilde{\bar{\a}}^{\bar{t}_1}_{++,-1})(\bar{\a}'^{(l)f}_{++,s},\tilde{\bar{\a}}^{\bar{t}_2}_{--,-1})(\bar{\a}'^{(l)f}_{++,q},\bar{\a}'^{(j)i}_{--,-n_1})\cr
     &&\qquad+ (\bar{\a}'^{(l)f}_{--,p},\bar{\a}'^{(k)i}_{++,-n_2})(\bar{\a}'^{(l)f}_{--,r},\tilde{\bar{\a}}^{\bar{t}_1}_{++,-1})(\bar{\a}'^{(l)f}_{++,q},\tilde{\bar{\a}}^{\bar{t}_2}_{--,-1})(\bar{\a}'^{(l)f}_{++,s},\bar{\a}'^{(j)i}_{--,-n_1})\cr
     &&\qquad+ (\bar{\a}'^{(l)f}_{--,p},\bar{\a}'^{(k)i}_{++,-n_2})(\bar{\a}'^{(l)f}_{++,q},\bar{\a}'^{(l)f}_{--,r})(\tilde{\bar{\a}}^{\bar{t}_2}_{--,-1},\tilde{\bar{\a}}^{\bar{t}_1}_{++,-1})(\bar{\a}'^{(l)f}_{++,s},\bar{\a}'^{(j)i}_{--,-n_1})\cr
     &&\qquad+ (\bar{\a}'^{(l)f}_{--,p},\bar{\a}'^{(k)i}_{++,-n_2})(\bar{\a}'^{(l)f}_{++,q},\bar{\a}'^{(l)f}_{--,r})(\tilde{\bar{\a}}^{\bar{t}_1}_{++,-1},\bar{\a}'^{(j)i}_{--,-n_1})(\bar{\a}'^{(l)f}_{++,s},\tilde{\bar{\a}}^{\bar{t}_2}_{--,-1})
     \bigg)\cr
     &&\quad~~\times(\tilde{\bar{d}}^{+-,\bar{t}_2}_{-{1\over2}},\tilde{\bar{d}}^{-+,\bar{t}_1}_{-{1\over2}})\bigg]
      \label{aa to aaaa mp bar}
\eea
The full amplitude for $\a\a\to\a\a \a\a$ in (\ref{all amplitudes}) is then
\bea
&&\mathcal{A}^{\a\a\to \a\a \a\a}(w_2,w_1,\bar{w}_2,\bar{w}_1)\cr
&&\quad =-{1\over 2\sqrt2 n_1n_2pqrs}\mathcal{A}^{\a^{(1)}\a^{(1)} \to  \a^{(1)}\a^{(1)}\a^{(1)}\a^{(1)}}_{+-}(w_2,w_1)\bar{\mathcal{A}}^{\bar{\a}^{(1)}\bar{\a}^{(1)} \to  \bar{\a}^{(1)}\bar{\a}^{(1)}\bar{\a}^{(1)}\bar{\a}^{(1)}}_{-+}(\bar{w}_2,\bar{w}_1)\cr
&& \qquad- {1\over 2\sqrt2 n_1n_2pqrs}\mathcal{A}^{\a^{(1)} \a^{(1)} \to \a^{(1)}\a^{(1)}\a^{(1)}\a^{(1)}}_{-+}(w_2,w_1)\bar{\mathcal{A}}^{ \bar{\a}^{(1)}\bar{\a}^{(1)} \to\bar{\a}^{(1)}\bar \a^{(1)}\bar{\a}^{(1)}\bar{\a}^{(1)}}_{+-}(\bar{w}_2,\bar{w}_1)
\cr\cr
&&\qquad -{1\over 2\sqrt2 n_1n_2pqrs}\mathcal{A}^{\a^{(1)}\a^{(2)} \to \a^{(1)} \a^{(1)} \a^{(1)}\a^{(1)}}_{+-}(w_2,w_1)\bar{\mathcal{A}}^{ \bar{\a}^{(1)}\bar\a^{(2)} \to  \bar{\a}^{(1)}\bar\a^{(1)}\bar{\a}^{(1)}\bar{\a}^{(1)}}_{-+}(\bar{w}_2,\bar{w}_1)\cr
&& \qquad- {1\over 2\sqrt2 n_1n_2pqrs}\mathcal{A}^{\a^{(1)}\a^{(2)} \to \a^{(1)}\a^{(1)} \a^{(1)}\a^{(1)}}_{-+}(w_2,w_1)\bar{\mathcal{A}}^{ \bar{\a}^{(1)}\bar\a^{(2)} \to  \bar{\a}^{(1)}\bar\a^{(1)}\bar{\a}^{(1)}\bar{\a}^{(1)}}_{+-}(\bar{w}_2,\bar{w}_1)
\cr\cr
&&\qquad -{1\over 2\sqrt2 n_1n_2pqrs}\mathcal{A}^{\a^{(2)}\a^{(1)} \to \a^{(1)} \a^{(1)} \a^{(1)}\a^{(1)}}_{+-}(w_2,w_1)\bar{\mathcal{A}}^{\bar\a^{(2)}  \bar{\a}^{(1)}\to  \bar{\a}^{(1)}\bar\a^{(1)}\bar{\a}^{(1)}\bar{\a}^{(1)}}_{-+}(\bar{w}_2,\bar{w}_1)\cr
&& \qquad- {1\over 2\sqrt2 n_1n_2pqrs}\mathcal{A}^{\a^{(2)}\a^{(1)} \to  \a^{(1)}\a^{(1)} \a^{(1)}\a^{(1)}}_{-+}(w_2,w_1)\bar{\mathcal{A}}^{\bar\a^{(2)}  \bar{\a}^{(1)}\to  \bar{\a}^{(1)}\bar\a^{(1)}\bar{\a}^{(1)}\bar{\a}^{(1)}}_{+-}(\bar{w}_2,\bar{w}_1)
\cr\cr
&&\qquad -{1\over 2\sqrt2 n_1n_2pqrs}\mathcal{A}^{\a^{(2)}\a^{(2)} \to \a^{(1)} \a^{(1)} \a^{(1)}\a^{(1)}}_{+-}(w_2,w_1)\bar{\mathcal{A}}^{\bar\a^{(2)} \bar{\a}^{(2)} \to  \bar{\a}^{(1)}\bar\a^{(1)}\bar{\a}^{(1)}\bar{\a}^{(1)}}_{-+}(\bar{w}_2,\bar{w}_1)\cr
&& \qquad- {1\over 2\sqrt2 n_1n_2pqrs}\mathcal{A}^{\a^{(2)}\a^{(2)} \to  \a^{(1)}\a^{(1)} \a^{(1)}\a^{(1)}}_{-+}(w_2,w_1)\bar{\mathcal{A}}^{\bar\a^{(2)}  \bar{\a}^{(2)}\to \bar{\a}^{(1)}\bar\a^{(1)}\bar{\a}^{(1)}\bar{\a}^{(1)}}_{+-}(\bar{w}_2,\bar{w}_1)
\cr\cr
&&\qquad-{1\over 2\sqrt2 n_1n_2pqrs}\mathcal{A}^{\a^{(2)}\a^{(2)} \to  \a^{(2)}\a^{(2)}\a^{(2)}\a^{(2)}}_{+-}(w_2,w_1)\bar{\mathcal{A}}^{\bar{\a}^{(1)}\bar{\a}^{(1)} \to  \bar{\a}^{(2)}\bar{\a}^{(2)}\bar{\a}^{(2)}\bar{\a}^{(2)}}_{-+}(\bar{w}_2,\bar{w}_1)\cr
&& \qquad- {1\over 2\sqrt2 n_1n_2pqrs}\mathcal{A}^{\a^{(1)} \a^{(1)} \to\a^{(2)}\a^{(2)}\a^{(2)}\a^{(2)}}_{-+}(w_2,w_1)\bar{\mathcal{A}}^{ \bar{\a}^{(1)}\bar{\a}^{(1)} \to\bar{\a}^{(2)}\bar{\a}^{(2)}\bar{\a}^{(2)}\bar{\a}^{(2)}}_{+-}(\bar{w}_2,\bar{w}_1)
\cr\cr
&&\qquad -{1\over 2\sqrt2 n_1n_2pqrs}\mathcal{A}^{\a^{(1)}\a^{(2)} \to \a^{(2)}\a^{(2)}\a^{(2)}\a^{(2)}}_{+-}(w_2,w_1)\bar{\mathcal{A}}^{ \bar{\a}^{(1)}\bar\a^{(2)} \to\bar{\a}^{(2)}\bar{\a}^{(2)}\bar{\a}^{(2)}\bar{\a}^{(2)}}_{-+}(\bar{w}_2,\bar{w}_1)\cr
&& \qquad- {1\over 2\sqrt2 n_1n_2pqrs}\mathcal{A}^{\a^{(1)}\a^{(2)} \to \a^{(2)}\a^{(2)}\a^{(2)}\a^{(2)}}_{-+}(w_2,w_1)\bar{\mathcal{A}}^{ \bar{\a}^{(1)}\bar\a^{(2)} \to \bar{\a}^{(2)}\bar{\a}^{(2)}\bar{\a}^{(2)}\bar{\a}^{(2)}}_{+-}(\bar{w}_2,\bar{w}_1)
\cr\cr
&&\qquad -{1\over 2\sqrt2 n_1n_2pqrs}\mathcal{A}^{\a^{(2)}\a^{(1)} \to \a^{(2)}\a^{(2)}\a^{(2)}\a^{(2)}}_{+-}(w_2,w_1)\bar{\mathcal{A}}^{\bar\a^{(2)}  \bar{\a}^{(1)}\to \bar{\a}^{(2)}\bar{\a}^{(2)}\bar{\a}^{(2)}\bar{\a}^{(2)}}_{-+}(\bar{w}_2,\bar{w}_1)\cr
&& \qquad-{1\over 2\sqrt2 n_1n_2pqrs} \mathcal{A}^{\a^{(2)}\a^{(1)} \to  \a^{(2)}\a^{(2)}\a^{(2)}\a^{(2)}}_{-+}(w_2,w_1)\bar{\mathcal{A}}^{\bar\a^{(2)}  \bar{\a}^{(1)}\to\bar{\a}^{(2)}\bar{\a}^{(2)}\bar{\a}^{(2)}\bar{\a}^{(2)}}_{+-}(\bar{w}_2,\bar{w}_1)
\cr\cr
&&\qquad -{1\over 2\sqrt2 n_1n_2pqrs}\mathcal{A}^{\a^{(2)}\a^{(2)} \to \a^{(2)}\a^{(2)}\a^{(2)}\a^{(2)}}_{+-}(w_2,w_1)\bar{\mathcal{A}}^{\bar\a^{(2)} \bar{\a}^{(2)} \to\bar{\a}^{(2)}\bar{\a}^{(2)}\bar{\a}^{(2)}\bar{\a}^{(2)}}_{-+}(\bar{w}_2,\bar{w}_1)\cr
&& \qquad-{1\over 2\sqrt2 n_1n_2pqrs} \mathcal{A}^{\a^{(2)}\a^{(2)} \to \a^{(2)}\a^{(2)}\a^{(2)}\a^{(2)}}_{-+}(w_2,w_1)\bar{\mathcal{A}}^{\bar\a^{(2)}  \bar{\a}^{(2)}\to \bar{\a}^{(2)}\bar{\a}^{(2)}\bar{\a}^{(2)}\bar{\a}^{(2)}}_{+-}(\bar{w}_2,\bar{w}_1)\nn
\label{aa to aaaa full}
\eea
where the copy amplitudes for each charge combination are given in equations (\ref{aa to aaaa pm}), (\ref{aa to aaaa mp}), (\ref{aa to aaaa pm bar}), and (\ref{aa to aaaa mp bar}).

\subsection{$\a\a\to \a\a dd$}\label{subsection aa to aadd}

In this subsection we compute the amplitudes for the process $\a\a\to \a\a dd$:
\bea
\mathcal{A}^{\a^{(j)}\a^{(k)}\to \a^{(l)}\a^{(l)}d^{(l)}d^{(l)}}_{\dot{C}\dot{A}}&=&-{e^{3s}\sinh({\Delta w\over2})\over8}\mathcal{A}^{\a^{(j)}\a^{(k)}\to \a^{(l)}\a^{(l)}d^{(l)}d^{(l)}}_{t,\dot{C}\dot{A}}
\label{aa to aadd  two holo}\\
\bar{\mathcal{A}}^{\bar{\a}^{(j)}\bar{\a}^{(k)}\to \bar{\a}^{(l)}\bar{\a}^{(l)}\bar{d}^{(l)}\bar{d}^{(l)}}_{\dot{D}\dot{B}}&=&-{e^{3\bar{s}}\sinh({\Delta \bar{w}\over2})\over8}\bar{\mathcal{A}}^{\bar{\a}^{(j)}\bar{\a}^{(k)}\to \bar{\a}^{(l)}\bar{\a}^{(l)}\bar{d}^{(l)}\bar{d}^{(l)}}_{\bar{t},\dot{D}\dot{B}}
\label{aa to aadd  two antiholo}
\eea 
The holomorphic and antiholomorphic $t$ plane amplitudes are
\bea
\mathcal{A}^{\a^{(j)}\a^{(k)} \to \a^{(l)}\a^{(l)}d^{(l)}d^{(l)}}_{t,\dot{C}\dot{A}} &=& \langle 0|\a'^{(l)}_{++,p}\a'^{(l)}_{--,q}d'^{(l)f,+-}_{r}d'^{(l)f,-+}_{s}\tilde{G}^{+,t_2}_{\dot{C},-{3\over2}}\tilde{G}^{-,t_1}_{\dot{A},-{3\over2}}\a'^{(j)}_{++,-n_1}\a'^{(k)}_{--,-n_2}|0\rangle\cr
\bar{\mathcal{A}}^{\bar{\a}^{(j)}\bar\a^{(k)} \to \bar\a^{(l)}\bar\a^{(l)}\bar d^{(l)}\bar d^{(l)}}_{\bar t,\dot{D}\dot{B}} &=& \langle \bar{0}|\bar{\a}^{(l)f}_{--,p}\bar{\a}^{(l)f}_{++,q}\bar{d}^{(l)f,-+}_{r}\bar{d}^{(l)f,+-}_{s}\tilde{\bar{G}}^{+,\bar{t}_2}_{\dot{D},-{3\over2}}\tilde{\bar{G}}^{-,\bar{t}_1}_{\dot{B},-{3\over2}}\bar{\a}^{(j)i}_{--,-n_1}\bar{\a}^{(k)i}_{++,-n_2}|\bar{0}\rangle\nn
\eea
Again, the computation is very similar to the ones in the previous subsections and so we simply record the final expressions for the generalized amplitudes and the full amplitude. The generalized amplitudes are

\bea
&&\!\!\!\!\!\!\!\mathcal{A}^{\a^{(j)}\a^{(k)} \to \a^{(l)}\a^{(l)}d^{(l)}d^{(l)}}_{+-}\cr
&&= {e^{{3\over2}s}\sinh({\Delta w\over2})\over8}\cr
&&\quad\Bigg[ \bigg((\a'^{(l)f}_{++,p},\a'^{(l)}_{--,q})(\tilde{\a}^{t_2}_{++,-1},\tilde{\a}^{t_1}_{--,-1})(\a'^{(j)}_{++,-n_1},\a'^{(k)}_{--,-n_2}) \cr
&&\quad + (\a'^{(l)f}_{++,p},\a'^{(l)}_{--,q})(\tilde{\a}^{t_2}_{++,-1},\a'^{(k)}_{--,-n_2})(\tilde{\a}^{t_1}_{--,-1},\a'^{(j)}_{++,-n_1}) \cr
&&\quad+(\a'^{(l)f}_{++,p},\tilde{\a}^{t_1}_{--,-1})(\a'^{(l)}_{--,q},\tilde{\a}^{t_2}_{++,-1})(\a'^{(j)}_{++,-n_1},\a'^{(k)}_{--,-n_2})\cr
&&\quad+ (\a'^{(l)f}_{++,p},\tilde{\a}^{t_1}_{--,-1})(\a'^{(l)}_{--,q},\a'^{(j)}_{++,-n_1})(\tilde{\a}^{t_2}_{++,-1},\a'^{(k)}_{--,-n_2})\cr
&&\quad+ (\a'^{(l)f}_{++,p},\a'^{(k)}_{--,-n_2})(\a'^{(l)}_{--,q},\tilde{\a}^{t_2}_{++,-1})(\tilde{\a}^{t_1}_{--,-1},\a'^{(j)}_{++,-n_1})\cr
&&\quad+ (\a'^{(l)f}_{++,p},\a'^{(k)}_{--,-n_2})(\a'^{(l)}_{--,q},\a'^{(j)}_{++,-n_1})(\tilde{\a}^{t_2}_{++,-1},\tilde{\a}^{t_1}_{--,-1})   \bigg)\cr
&&\qquad (d'^{(l)f,+-}_r ,d'^{(l)f,- +}_s)(\tilde{d}^{++,t_2}_{-{1\over2}},\tilde{d}^{--,t_1}_{-{1\over2}})
\cr
&&+\bigg((\a'^{(l)f}_{++,p},\a'^{(l)}_{--,q})(\a'^{(j)}_{++,-n_1},\a'^{(k)}_{--,-n_2} ) + (\a'^{(l)f}_{++,p},\a'^{(k)}_{--,-n_2})(\a'^{(l)}_{--,q},\a'^{(j)}_{++,-n_1})\bigg)\cr
&&\quad\bigg((d'^{(l)f,+-}_r ,d'^{(l)f,- +}_s)(\tilde{d}^{+-,t_2}_{-{1\over2}},\tilde{d}^{-+,t_1}_{-{1\over2}})  + (d'^{(l)f,+-}_r ,\tilde{d}^{-+,t_1}_{-{1\over2}})(d'^{(l)f,- +}_s,\tilde{d}^{+-,t_2}_{-{1\over2}}) \bigg)\cr
&&\quad(\tilde{\a}^{t_2}_{-+,-1}, \tilde{\a}^{t_1}_{+-,-1})\Bigg]
\label{aa to aadd pm}\\
\cr
&&\!\!\!\!\!\!\!\mathcal{A}^{\a^{(j)}\a^{(k)} \to \a^{(l)}\a^{(l)}d^{(l)}d^{(l)}}_{-+} \cr
&&=  {e^{{3\over2}s}\sinh({\Delta w\over2})\over8}\cr
&&\quad\bigg[\bigg((\a'^{(l)f}_{++,p},\a'^{(l)}_{--,q})(\tilde{\a}^{t_2}_{+-,-1},\tilde{\a}^{t_1}_{-+,-1})(\a'^{(j)}_{++,-n_1},\a'^{(k)}_{--,-n_2}) \cr
&&\quad+ (\a'^{(l)f}_{++,p},\a'^{(k)}_{--,-n_2})(\a'^{(l)}_{--,q},\a'^{(j)}_{++,-n_1})(\tilde{\a}^{t_2}_{+-,-1},\tilde{\a}^{t_1}_{-+,-1})   \bigg)\cr
&&\quad(d'^{(l)f,+-}_r ,d'^{(l)f,- +}_s)(\tilde{d}^{++,t_2}_{-{1\over2}},\tilde{d}^{--,t_1}_{-{1\over2}})
\cr
&&+\bigg((\a'^{(l)f}_{++,p},\a'^{(l)}_{--,q})(\tilde{\a}^{t_2}_{--,-1},\tilde{\a}^{t_1}_{++,-1})(\a'^{(j)}_{++,-n_1},\a'^{(k)}_{--,-n_2} )\cr
&&\quad+ (\a'^{(l)f}_{++,p},\a'^{(l)}_{--,q})(\tilde{\a}^{t_2}_{--,-1},\a'^{(j)}_{++,-n_1})(\tilde{\a}^{t_1}_{++,-1},\a'^{(k)}_{--,-n_2} ) \cr
&&\quad+ (\a'^{(l)f}_{++,p},\tilde{\a}^{t_2}_{--,-1})(\a'^{(l)}_{--,q},\tilde{\a}^{t_1}_{++,-1})(\a'^{(j)}_{++,-n_1},\a'^{(k)}_{--,-n_2} )\cr
&&\quad  +  (\a'^{(l)f}_{++,p},\tilde{\a}^{t_2}_{--,-1})(\a'^{(l)}_{--,q},\a'^{(j)}_{++,-n_1})(\tilde{\a}^{t_1}_{++,-1},\a'^{(k)}_{--,-n_2} )\cr
&&\quad + (\a'^{(l)f}_{++,p},\a'^{(k)}_{--,-n_2})(\a'^{(l)}_{--,q},\tilde{\a}^{t_1}_{++,-1})(\tilde{\a}^{t_2}_{--,-1},\a'^{(j)}_{++,-n_1} )\cr
&&\quad + (\a'^{(l)f}_{++,p},\a'^{(k)}_{--,-n_2})(\a'^{(l)}_{--,q},\a'^{(j)}_{++,-n_1})(\tilde{\a}^{t_2}_{--,-1}, \tilde{\a}^{t_1}_{++,-1})\bigg)\cr
&&\quad\bigg((d'^{(l)f,+-}_r ,d'^{(l)f,- +}_s)(\tilde{d}^{+-,t_2}_{-{1\over2}},\tilde{d}^{-+,t_1}_{-{1\over2}})  + (d'^{(l)f,+-}_r ,\tilde{d}^{-+,t_1}_{-{1\over2}})(d'^{(l)f,- +}_s,\tilde{d}^{+-,t_2}_{-{1\over2}}) \bigg)\bigg]\nn
\label{aa to aadd mp}\\
\cr
&&\!\!\!\!\!\!\!\!\!\!\bar{\mathcal{A}}^{\bar{\a}^{(j)}\bar{\a}^{(k)} \to \bar{\a}^{(l)}\bar{\a}^{(l)}\bar{d}^{(l)}\bar{d}^{(l)}}_{+-}\cr
&&={e^{{3\over2}\bar s}\sinh({\Delta \bar{w}\over2})\over8}\cr
&&\quad\Bigg[\bigg((\bar{\a}^{(l)f}_{--,p},\bar{\a}^{(l)f}_{++,q})(\bar{\a}^{(j)i}_{--,-n_1},\bar{\a}^{(k)i}_{++,-n_2}) + (\bar{\a}^{(l)f}_{--,p},\bar{\a}^{(k)i}_{++,-n_2})(\bar{\a}^{(l)f}_{++,q},\bar{\a}^{(j)i}_{--,-n_1})\bigg)\cr
&&\qquad\bigg((\bar{d}^{(l)f,-+}_{r},\bar{d}^{(l)f,+-}_{s})(\tilde{\bar{d}}^{+-,\bar{t}_2}_{-{1\over2}},\tilde{\bar{d}}^{-+,\bar{t}_1}_{-{1\over2}}) - (\bar{d}^{(l)f,-+}_{r},\tilde{\bar{d}}^{+-,\bar{t}_2}_{-{1\over2}})(\bar{d}^{(l)f,+-}_{s},\tilde{\bar{d}}^{-+,\bar{t}_1}_{-{1\over2}})\bigg)\cr
&&\qquad(\tilde{\bar{\a}}^{\bar{t}_2}_{-+,-1},\tilde{\bar{\a}}^{\bar{t}_1}_{+-,-1})\cr
&&\qquad + \bigg((\bar{\a}^{(l)f}_{--,p},\bar{\a}^{(l)f}_{++,q})(\tilde{\bar{\a}}^{\bar{t}_2}_{++,-1},\tilde{\bar{\a}}^{\bar{t}_1}_{--,-1})(\bar{\a}^{(j)i}_{--,-n_1},\bar{\a}^{(k)i}_{++,-n_2})\cr
&&\qquad~~ + (\bar{\a}^{(l)f}_{--,p},\bar{\a}^{(l)f}_{++,q})(\tilde{\bar{\a}}^{\bar{t}_2}_{++,-1},\bar{\a}^{(j)i}_{--,-n_1})(\tilde{\bar{\a}}^{\bar{t}_1}_{--,-1},\bar{\a}^{(k)i}_{++,-n_2})\cr
&&\qquad~~ + (\bar{\a}^{(l)f}_{--,p},\tilde{\bar{\a}}^{\bar{t}_2}_{++,-1})(\bar{\a}^{(l)f}_{++,q},\tilde{\bar{\a}}^{\bar{t}_1}_{--,-1})(\bar{\a}^{(j)i}_{--,-n_1},\bar{\a}^{(k)i}_{++,-n_2})\cr
&&\qquad~~ + (\bar{\a}^{(l)f}_{--,p},\tilde{\bar{\a}}^{\bar{t}_2}_{++,-1})(\bar{\a}^{(l)f}_{++,q},\bar{\a}^{(j)i}_{--,-n_1})(\tilde{\bar{\a}}^{\bar{t}_1}_{--,-1},\bar{\a}^{(k)i}_{++,-n_2})\cr
&&\qquad~~ +(\bar{\a}^{(l)f}_{--,p},\bar{\a}^{(k)i}_{++,-n_2})(\bar{\a}^{(l)f}_{++,q},\bar{\a}^{(j)i}_{--,-n_1})(\tilde{\bar{\a}}^{\bar{t}_2}_{++,-1},\tilde{\bar{\a}}^{\bar{t}_1}_{--,-1})\cr
&&\qquad~~ +(\bar{\a}^{(l)f}_{--,p},\bar{\a}^{(k)i}_{++,-n_2})(\bar{\a}^{(l)f}_{++,q},\tilde{\bar{\a}}^{\bar{t}_1}_{--,-1})(\tilde{\bar{\a}}^{\bar{t}_2}_{++,-1},\bar{\a}^{(j)i}_{--,-n_1})\bigg)\cr
&&\qquad(\bar{d}^{(l)f,-+}_{r},\bar{d}^{(l)f,+-}_{s})(\tilde{\bar{d}}^{++,\bar{t}_2}_{-{1\over2}},\tilde{\bar{d}}^{--,\bar{t}_1}_{-{1\over2}})\Bigg]\nn
\label{aa to aadd pm bar}
\eea
\bea
&&\!\!\!\!\!\!\!\!\!\!\bar{\mathcal{A}}^{\bar{\a}^{(j)}\bar{\a}^{(k)} \to \bar{\a}^{(l)}\bar{\a}^{(l)}\bar{d}^{(l)}\bar{d}^{(l)}}_{-+}\cr
&&={e^{{3\over2}\bar s}\sinh({\Delta \bar{w}\over2})\over8}\cr
&&\quad\Bigg[\bigg((\bar{\a}^{(l)f}_{--,p},\bar{\a}^{(l)f}_{++,q})(\tilde{\bar{\a}}^{\bar{t}_2}_{--,-1},\tilde{\bar{\a}}^{\bar{t}_1}_{++,-1})(\bar{\a}^{(j)i}_{--,-n_1},\bar{\a}^{(k)i}_{++,-n_2})\cr
&&\quad~~+ (\bar{\a}^{(l)f}_{--,p},\bar{\a}^{(l)f}_{++,q})(\tilde{\bar{\a}}^{\bar{t}_2}_{--,-1},\bar{\a}^{(k)i}_{++,-n_2})(\tilde{\bar{\a}}^{\bar{t}_1}_{++,-1},\bar{\a}^{(j)i}_{--,-n_1})\cr
&&\quad~~+(\bar{\a}^{(l)f}_{--,p},\bar{\a}^{(k)i}_{++,-n_2})(\bar{\a}^{(l)f}_{++,q},\tilde{\bar{\a}}^{\bar{t}_2}_{--,-1})(\tilde{\bar{\a}}^{\bar{t}_1}_{++,-1},\bar{\a}^{(j)i}_{--,-n_1})\cr
&&\quad~~+(\bar{\a}^{(l)f}_{--,p},\bar{\a}^{(k)i}_{++,-n_2})(\bar{\a}^{(l)f}_{++,q},\bar{\a}^{(j)i}_{--,-n_1})(\tilde{\bar{\a}}^{\bar{t}_2}_{--,-1},\tilde{\bar{\a}}^{\bar{t}_1}_{++,-1})\cr
&&\quad~~+(\bar{\a}^{(l)f}_{--,p},\tilde{\bar{\a}}^{\bar{t}_1}_{++,-1})(\bar{\a}^{(l)f}_{++,q},\tilde{\bar{\a}}^{\bar{t}_2}_{--,-1})(\bar{\a}^{(j)i}_{--,-n_1},\bar{\a}^{(k)i}_{++,-n_2})\cr
&&\quad~~+(\bar{\a}^{(l)f}_{--,p},\tilde{\bar{\a}}^{\bar{t}_1}_{++,-1})(\bar{\a}^{(l)f}_{++,q},\bar{\a}^{(j)i}_{--,-n_1})(\tilde{\bar{\a}}^{\bar{t}_2}_{--,-1},\bar{\a}^{(k)i}_{++,-n_2})\bigg)\cr
&&\quad~~\bigg((\bar{d}^{(l)f,-+}_{r},\bar{d}^{(l)f,+-}_{s})(\tilde{\bar{d}}^{+-,\bar{t}_2}_{-{1\over2}},\tilde{\bar{d}}^{-+,\bar{t}_1}_{-{1\over2}}) - (\bar{d}^{(l)f,-+}_{r},\tilde{\bar{d}}^{+-,\bar{t}_2}_{-{1\over2}})(\bar{d}^{(l)f,+-}_{s},\tilde{\bar{d}}^{-+,\bar{t}_1}_{-{1\over2}})\bigg)\cr
&&\qquad + \bigg((\bar{\a}^{(l)f}_{--,p},\bar{\a}^{(l)f}_{++,q})(\bar{\a}^{(j)i}_{++,-n_1},\bar{\a}^{(k)i}_{--,-n_2}) + (\bar{\a}^{(l)f}_{--,p},\bar{\a}^{(k)i}_{++,-n_2})(\bar{\a}^{(l)f}_{++,q},\bar{\a}^{(j)i}_{--,-n_1})\bigg)\cr
&&\qquad~~(\tilde{\bar{\a}}^{\bar{t}_2}_{+-,-1},\tilde{\bar{\a}}^{\bar{t}_1}_{-+,-1})(\bar{d}^{(l)f,-+}_{r},\bar{d}^{(l)f,+-}_{s})(\tilde{\bar{d}}^{++,\bar{t}_2}_{-{1\over2}},\tilde{\bar{d}}^{--,\bar{t}_1}_{-{1\over2}})\Bigg]\nn
\label{aa to aadd mp bar}
\eea

The full amplitude for $\a\a\to\a\a dd$ in (\ref{all amplitudes}) is then
\bea
&&\mathcal{A}^{\a\a\to \a\a dd}(w_2,w_1,\bar{w}_2,\bar{w}_1)\cr
&&\quad =-{1\over 2\sqrt2 n_1n_2pq}\mathcal{A}^{\a^{(1)}\a^{(1)} \to  \a^{(1)}\a^{(1)}\a^{(1)}\a^{(1)}}_{+-}(w_2,w_1)\bar{\mathcal{A}}^{\bar{\a}^{(1)}\bar{\a}^{(1)} \to  \bar{\a}^{(1)}\bar{\a}^{(1)}\bar{d}^{(1)}\bar{d}^{(1)}}_{-+}(\bar{w}_2,\bar{w}_1)\cr
&& \qquad- {1\over 2\sqrt2 n_1n_2pq}\mathcal{A}^{\a^{(1)} \a^{(1)} \to \a^{(1)}\a^{(1)}\a^{(1)}\a^{(1)}}_{-+}(w_2,w_1)\bar{\mathcal{A}}^{ \bar{\a}^{(1)}\bar{\a}^{(1)} \to\bar{\a}^{(1)}\bar \a^{(1)}\bar d^{(1)}\bar d^{(1)}}_{+-}(\bar{w}_2,\bar{w}_1)
\cr\cr
&&\qquad -{1\over 2\sqrt2 n_1n_2pq}\mathcal{A}^{\a^{(1)}\a^{(2)} \to \a^{(1)} \a^{(1)} \a^{(1)}\a^{(1)}}_{+-}(w_2,w_1)\bar{\mathcal{A}}^{ \bar{\a}^{(1)}\bar\a^{(2)} \to  \bar{\a}^{(1)}\bar\a^{(1)}\bar d^{(1)}\bar d^{(1)}}_{-+}(\bar{w}_2,\bar{w}_1)\cr
&& \qquad- {1\over 2\sqrt2 n_1n_2pq}\mathcal{A}^{\a^{(1)}\a^{(2)} \to \a^{(1)}\a^{(1)} \a^{(1)}\a^{(1)}}_{-+}(w_2,w_1)\bar{\mathcal{A}}^{ \bar{\a}^{(1)}\bar\a^{(2)} \to  \bar{\a}^{(1)}\bar\a^{(1)}\bar d^{(1)}\bar d^{(1)}}_{+-}(\bar{w}_2,\bar{w}_1)
\cr\cr
&&\qquad -{1\over 2\sqrt2 n_1n_2pq}\mathcal{A}^{\a^{(2)}\a^{(1)} \to \a^{(1)} \a^{(1)} \a^{(1)}\a^{(1)}}_{+-}(w_2,w_1)\bar{\mathcal{A}}^{\bar\a^{(2)}  \bar{\a}^{(1)}\to  \bar{\a}^{(1)}\bar\a^{(1)}\bar d^{(1)}\bar d^{(1)}}_{-+}(\bar{w}_2,\bar{w}_1)\cr
&& \qquad- {1\over 2\sqrt2 n_1n_2pq}\mathcal{A}^{\a^{(2)}\a^{(1)} \to  \a^{(1)}\a^{(1)} \a^{(1)}\a^{(1)}}_{-+}(w_2,w_1)\bar{\mathcal{A}}^{\bar\a^{(2)}  \bar{\a}^{(1)}\to  \bar{\a}^{(1)}\bar\a^{(1)}\bar d^{(1)}\bar d^{(1)}}_{+-}(\bar{w}_2,\bar{w}_1)
\cr\cr
&&\qquad -{1\over 2\sqrt2 n_1n_2pq}\mathcal{A}^{\a^{(2)}\a^{(2)} \to \a^{(1)} \a^{(1)} \a^{(1)}\a^{(1)}}_{+-}(w_2,w_1)\bar{\mathcal{A}}^{\bar\a^{(2)} \bar{\a}^{(2)} \to  \bar{\a}^{(1)}\bar\a^{(1)}\bar d^{(1)}\bar d^{(1)}}_{-+}(\bar{w}_2,\bar{w}_1)\cr
&& \qquad- {1\over 2\sqrt2 n_1n_2pq}\mathcal{A}^{\a^{(2)}\a^{(2)} \to  \a^{(1)}\a^{(1)} \a^{(1)}\a^{(1)}}_{-+}(w_2,w_1)\bar{\mathcal{A}}^{\bar\a^{(2)}  \bar{\a}^{(2)}\to \bar{\a}^{(1)}\bar\a^{(1)}\bar d^{(1)}\bar d^{(1)}}_{+-}(\bar{w}_2,\bar{w}_1)
\cr\cr
&&\qquad-{1\over 2\sqrt2 n_1n_2pq}\mathcal{A}^{\a^{(2)}\a^{(2)} \to  \a^{(2)}\a^{(2)}d^{(2)}d^{(2)}}_{+-}(w_2,w_1)\bar{\mathcal{A}}^{\bar{\a}^{(1)}\bar{\a}^{(1)} \to  \bar{\a}^{(2)}\bar{\a}^{(2)}\bar{d}^{(2)}\bar{d}^{(2)}}_{-+}(\bar{w}_2,\bar{w}_1)\cr
&& \qquad- {1\over 2\sqrt2 n_1n_2pq}\mathcal{A}^{\a^{(1)} \a^{(1)} \to\a^{(2)}\a^{(2)}d^{(2)}d^{(2)}}_{-+}(w_2,w_1)\bar{\mathcal{A}}^{ \bar{\a}^{(1)}\bar{\a}^{(1)} \to\bar{\a}^{(2)}\bar{\a}^{(2)}\bar{d}^{(2)}\bar{d}^{(2)}}_{+-}(\bar{w}_2,\bar{w}_1)
\cr\cr
&&\qquad -{1\over 2\sqrt2 n_1n_2pq}\mathcal{A}^{\a^{(1)}\a^{(2)} \to \a^{(2)}\a^{(2)}d^{(2)}d^{(2)}}_{+-}(w_2,w_1)\bar{\mathcal{A}}^{ \bar{\a}^{(1)}\bar\a^{(2)} \to\bar{\a}^{(2)}\bar{\a}^{(2)}\bar{d}^{(2)}\bar{d}^{(2)}}_{-+}(\bar{w}_2,\bar{w}_1)\cr
&& \qquad- {1\over 2\sqrt2 n_1n_2pq}\mathcal{A}^{\a^{(1)}\a^{(2)} \to \a^{(2)}\a^{(2)}d^{(2)}d^{(2)}}_{-+}(w_2,w_1)\bar{\mathcal{A}}^{ \bar{\a}^{(1)}\bar\a^{(2)} \to \bar{\a}^{(2)}\bar{\a}^{(2)}\bar{d}^{(2)}\bar{d}^{(2)}}_{+-}(\bar{w}_2,\bar{w}_1)
\cr\cr
&&\qquad -{1\over 2\sqrt2 n_1n_2pq}\mathcal{A}^{\a^{(2)}\a^{(1)} \to \a^{(2)}\a^{(2)}d^{(2)}d^{(2)}}_{+-}(w_2,w_1)\bar{\mathcal{A}}^{\bar\a^{(2)}  \bar{\a}^{(1)}\to \bar{\a}^{(2)}\bar{\a}^{(2)}\bar{d}^{(2)}\bar{d}^{(2)}}_{-+}(\bar{w}_2,\bar{w}_1)\cr
&& \qquad-{1\over 2\sqrt2 n_1n_2pq} \mathcal{A}^{\a^{(2)}\a^{(1)} \to  \a^{(2)}\a^{(2)}d^{(2)}d^{(2)}}_{-+}(w_2,w_1)\bar{\mathcal{A}}^{\bar\a^{(2)}  \bar{\a}^{(1)}\to\bar{\a}^{(2)}\bar{\a}^{(2)}\bar{d}^{(2)}\bar{d}^{(2)}}_{+-}(\bar{w}_2,\bar{w}_1)
\cr\cr
&&\qquad -{1\over 2\sqrt2 n_1n_2pq}\mathcal{A}^{\a^{(2)}\a^{(2)} \to \a^{(2)}\a^{(2)}d^{(2)}d^{(2)}}_{+-}(w_2,w_1)\bar{\mathcal{A}}^{\bar\a^{(2)} \bar{\a}^{(2)} \to\bar{\a}^{(2)}\bar{\a}^{(2)}\bar{d}^{(2)}\bar{d}^{(2)}}_{-+}(\bar{w}_2,\bar{w}_1)\cr
&& \qquad-{1\over 2\sqrt2 n_1n_2pq} \mathcal{A}^{\a^{(2)}\a^{(2)} \to \a^{(2)}\a^{(2)}d^{(2)}d^{(2)}}_{-+}(w_2,w_1)\bar{\mathcal{A}}^{\bar\a^{(2)}  \bar{\a}^{(2)}\to \bar{\a}^{(2)}\bar{\a}^{(2)}\bar{d}^{(2)}\bar{d}^{(2)}}_{+-}(\bar{w}_2,\bar{w}_1)\nn
\label{aa to aadd full}
\eea
where the copy amplitudes for each charge combination are given in equations (\ref{aa to aadd pm}), (\ref{aa to aadd mp}), (\ref{aa to aadd pm bar}), and (\ref{aa to aadd mp bar}).
Thus far we have computed all of the necessary amplitudes. Now we must compute all of the necessary wick contraction terms contained within the amplitudes. We do this in the next section.

\section{Wick Contraction Terms}

In this section we perform all possible wick contractions that we'll need to compute the various amplitudes. We remind the reader of the following $t$-plane locations for various copies
\bea
\text{Copy 1 initial}&:& t=-a\cr
\text{Copy 2 initial}  &:& t=-b\cr
\text{Copy 1 final} &:& t=\infty\cr
\text{Copy 2 final} &:& t=0
\eea
and of course $t_1$ and $t_2$ correspond to the $t$-plane images of the twist insertions. We have $\tilde{G}^-$ at $t_1$ and $\tilde{G}^+$ at $t_2$. Next we list all pairs of $t$-plane locations for which we have a wick contraction between the modes located there.
\bea
&&\infty,\infty\cr
&&\infty,t_i\cr
&&\infty,-a\cr
&&\infty,-b\cr
\cr
&&0,0\cr
&&0,t_i\cr
&&0,-a\cr
&&0,-b\cr
\cr
&&-a,-a\cr
&&-b,-b\cr
&&-a,-b
\cr
\cr
&&t_i,-a\cr
&&t_i,-b\cr
\cr
&&t_1,t_2
\eea
Let's compute expressions for each type of contraction. We first record the various boson and fermion $t$-plane contours below

\bea
\a'^{(1)f}_{A\dot{A},p} &=& {1\over 2\pi}\int_{t=\infty} dt {(t+a)^p(t+b)^p\over t^p}\partial X_{A\dot{A}}(t),\qquad \cr
\a'^{(2)f}_{A\dot{A},p} &=& -{1\over 2\pi}\int_{t=0} dt {(t+a)^p(t+b)^p\over t^p}\partial X_{A\dot{A}}(t)\cr
d'^{(1)f,+ A}_r&=&{1\over 2\pi i}\int_{t=\infty}dt\psi^{+ A}(t)(t -t_1)(t+a)^{r-{1\over2}}(t+b)^{r-{1\over2}}t^{-r-{1\over2}}\cr
d'^{(1)f,- A}_r&=&{1\over 2\pi i}\int_{t=\infty}dt\psi^{- A}(t)(t -t_2)(t+a)^{r-{1\over2}}(t+b)^{r-{1\over2}}t^{-r-{1\over2}}\cr
d'^{(2)f,+ A}_r&=&-{1\over 2\pi i}\int_{t=0}dt\psi^{+ A}(t)(t -t_1)(t+a)^{r-{1\over2}}(t+b)^{r-{1\over2}}t^{-r-{1\over2}}\cr
d'^{(2)f,- A}_r&=&-{1\over 2\pi i}\int_{t=0}dt\psi^{- A}(t)(t -t_2)(t+a)^{r-{1\over2}}(t+b)^{r-{1\over2}}t^{-r-{1\over2}}\cr
\a'^{(1)i}_{A\dot{A},-n}&=&  {1\over 2\pi}\int_{t=-a} dt {t^n\over   (t+a)^n(t+b)^n}\partial X_{A\dot{A}}(t)\cr
\a'^{(2)i}_{A\dot{A},-n}&=&  {1\over 2\pi}\int_{t=-b} dt {t^n\over   (t+a)^n(t+b)^n}\partial X_{A\dot{A}}(t)\cr
\tilde{\a}^{t_i}_{A\dot{A},-1}&=&i \partial X_{A\dot{A}}(t_i)\cr
\tilde{d}^{\a A,t_i}_{-{1\over2}} &=&\psi^{\a A}(t_i)
\eea
where the minus sign around all copy two final modes comes because locally at $t=0$ the map has the behavior $z\sim {1\over t}$ and therefore reverses direction.
Since there are no fermion modes in the initial state, we do not include their $t$-plane contours above.

\subsection{Contraction between bosons at $\infty$ and $\infty$}
Here we compute the contraction between two modes at $t=\infty$ (both modes are Copy 1 final).
Our boson contour is
\bea
\a'^{(1)f}_{A\dot{A},p}&=&{1\over 2\pi}\int_{t=\infty} dt {(t+a)^p(t+b)^p\over t^p}\partial X_{A\dot{A}}(t)\cr
&=&\sum_{k,k'\geq0}{}^{p}C_{k}{}^{p}C_{k'}a^kb^{k'}\tilde{\a}_{A\dot{A},p-k-k'}
\eea
where we have expanded our integrand at $t=\infty$ and modes natural to $t=\infty$ are defined as
\bea
\tilde{\a}_{A\dot{A},n}={1\over 2\pi}\int_{t=\infty}dt t^n\partial X_{A\dot{A}}(t)
\eea
The contraction is
\bea
&&(\a'^{(1)f}_{A\dot{A},p}\a'^{(1)f}_{B\dot{B},q})_{\text{contr.}}\cr
&&\qquad = \sum_{k,k',j,j'\geq0}{}^{p}C_{k}{}^{p}C_{k'}{}^{q}C_{j}{}^{q}C_{j'}a^{k+j}b^{k'+j'}[\tilde{\a}_{A\dot{A},p-k-k'},\tilde{\a}_{B\dot{B},q-j-j'}]\cr
&&\qquad = (-\e_{AB}\e_{\dot{A}\dot{B}})\sum_{k,k',j,j'\geq0}(p-k-k'){}^{p}C_{k}{}^{p}C_{k'}{}^{q}C_{j}{}^{q}C_{j'}a^{k+j}b^{k'+j'}\d_{p-k-k' + q-j-j',0}
\cr
&&\qquad = (-\e_{AB}\e_{\dot{A}\dot{B}})\sum_{k,k',j,j'\geq0}(p-k-k'){}^{p}C_{k}{}^{p}C_{k'}{}^{q}C_{j}{}^{q}C_{p+q-k-k'-j}a^{k+j}b^{p+q-k-j}\nn
\eea
Using 
\bea
&&p-k-k' + q-j-j'=0\to p-k-k' + q-j=j\cr
&& j'\geq 0 \to p-k-k' + q\geq j\cr
&&p-k-k'>0\to p - k > k' \cr
&&k'\geq 0 \to p>k
\eea
gives
\bea
(\a'^{(1)f}_{A\dot{A},p}\a'^{(1)f}_{B\dot{B},q})_{\text{contr.}}\!\!&=&\!\! (-\e_{AB}\e_{\dot{A}\dot{B}})\sum_{k=0}^{p-1}\sum_{k'=0}^{p-k-1}\sum_{j=0}^{p+q-k-k'}(p-k-k'){}^{p}C_{k}{}^{p}C_{k'}{}^{q}C_{j}{}^{q}C_{p+q-k-k'-j}\cr
&&\qquad\qquad\quad\times a^{k+j}b^{p+q-k-j}\nn
\eea

\subsection{Contraction between bosons at $\infty$ and $t_i$}
Here we compute the contraction between a mode at $t=\infty$ (Copy 1 final) and a mode at $t_i$ (Boson coming from $G$). From the above we know that the expansion at $t=\infty$ is
\bea
\a'^{(1)f}_{A\dot{A},p}&=&\sum_{k,k'\geq0}{}^{p}C_{k}{}^{p}C_{k'}a^kb^{k'}\tilde{\a}_{A\dot{A},p-k-k'}\cr
&=& \sum_{k=0}^{p-1}\sum_{k'=0}^{p-k-1}{}^{p}C_{k}{}^{p}C_{k'}a^kb^{k'}{1\over 2\pi}\int_{t=\infty} dt_{\infty}t_{\infty}^{p-k-k'}\partial X_{\dot{A}A}(t_{\infty})
\eea
Now the contraction gives
\bea
&&(\a'^{(1)f}_{B\dot{B},p}\tilde{\a}^{t_i}_{C\dot{C},-1})_{\text{contr.}}\cr
&&\qquad = i \sum_{k=0}^{p-1}\sum_{k'=0}^{p-k-1}{}^{p}C_{k}{}^{p}C_{k'}a^kb^{k'}{1\over 2\pi}\int_{t=\infty} dt_{\infty}t_{\infty}^{p-k-k'}\partial X_{B\dot{B}}(t_{\infty})\partial X_{C\dot{C}}(t_i)\cr
&&\qquad =  i\sum_{k=0}^{p-1}\sum_{k'=0}^{p-k-1}{}^{p}C_{k}{}^{p}C_{k'}a^kb^{k'}{1\over 2\pi}\int_{t=\infty} dt_{\infty}t_{\infty}^{p-k-k'}{1\over( t_{\infty}-t_i)^2}\cr
&&\qquad=-\e_{BC}\e_{\dot{B}\dot{C }}\sum_{k=0}^{p-1}\sum_{k'=0}^{p-k-1}(p-k-k'){}^{p}C_{k}{}^{p}C_{k'}a^kb^{k'}t_i^{p-k-k'-1}
\eea
Lets also expand the mode at $t_i$ around infinity. We have
\bea
\tilde{\a}^{t_i}_{C\dot{C},-1} = {1\over2\pi}\int_{t_i} dt (t-t_i)^{-1}\partial X_{C\dot{C}}(t)
\eea
Expanding around $\infty$ yields
\bea
\tilde{\a}^{t_i}_{C\dot{C},-1} &=&\sum_{k\geq0}t_i^k\tilde{\a}_{C\dot{C},-k-1}
\eea

\subsection{Contraction between bosons at $\infty$ and $-a$}
Here we compute the contraction between a mode at $t=\infty$ (Copy 1 final) and a mode at $t=-a$ (Copy 1 initial). From the above we know that the expansion at $t=\infty$ is 
\bea
\a'^{(1)f}_{A\dot{A},p}&=&\sum_{k=0}^{p-1}\sum_{k'=0}^{p-k-1}{}^{p}C_{k}{}^{p}C_{k'}a^kb^{k'}\tilde{\a}_{A\dot{A},p-k-k'}
\eea
For a mode $t=-a$ we have the following expansion
\bea
\a'^{(1)i}_{A\dot{A},-n}&=&  {1\over 2\pi}\int_{t=-a} dt {t^n\over   (t+a)^n(t+b)^n}\partial X_{A\dot{A}}(t)\cr
&=&\sum_{j,j'\geq 0}{}^nC_{j}{}^{-n}C_{j'}(-a)^{n-j}(b-a)^{-n-j'}\tilde{\a}^{-a}_{A\dot{A},-n+j+j'}
\eea
where modes natural to $t=-a$ are defined as
\bea
\tilde{\a}^{-a}_{A\dot{A},m}={1\over 2\pi}\int_{t=-a} dt(t+a)^m\partial X_{A\dot{A}}(t)
\eea
In order to not annihilate the local vacuum we require
\bea
-n+j+j'<0&\to& j'<n-j\cr
j'\geq0&\to& j<n
\eea
This gives
\bea
\a'^{(1)i}_{A\dot{A},-n}&=&\sum_{j= 0}^{n-1}\sum_{j'= 0}^{n-j-1}{}^nC_{j}{}^{-n}C_{j'}(-a)^{n-j}(b-a)^{-n-j'}\tilde{\a}^{-a}_{A\dot{A},-n+j+j'}
\label{copy one initial b}
\eea
Now we expand the mode natural to $t=-a$ around $t=\infty$. We have
\bea
\tilde{\a}^{-a}_{A\dot{A},-n+j+j'}  &=&   {1\over 2\pi}\int_{t=-a} dt(t+a)^{-n + j + j'}\partial X_{A\dot{A}}(t)\cr
&=& \sum_{j''\geq 0}{}^{-n+j+j'}C_{j''}a^{j''}\tilde{\a}_{A\dot{A},-n+j+j'-j''}
\eea
Inserting this into the expression $\a'^{(1)i}_{A\dot{A},-n}$ yields
\bea
\a'^{(1)i}_{A\dot{A},-n}&=&\sum_{j= 0}^{n-1}\sum_{j'= 0}^{n-j-1}\sum_{j''\geq 0}{}^nC_{j}{}^{-n}C_{j'}{}^{-n+j+j'}C_{j''}(-a)^{n-j}(b-a)^{-n-j'}a^{j''}\tilde{\a}_{A\dot{A},-n+j+j'-j''}\nn
\eea
Now we compute the contraction
\bea
&&\!\!\!\!\!\!\!\!\!\!\!\!\!\!\!(\a'^{(1)f}_{B\dot{B},p},\a'^{(1)i}_{A\dot{A},-n})\cr
&&\!\!\!\!\!\!\!\!\!\! =  \sum_{k=0}^{p-1}\sum_{k'=0}^{p-k-1}\sum_{j= 0}^{n-1}\sum_{j'= 0}^{n-j-1}\sum_{j''\geq 0}{}^{p}C_{k}{}^{p}C_{k'}{}^nC_{j}{}^{-n}C_{j'}{}^{-n+j+j'}C_{j''}a^kb^{k'}(-a)^{n-j}(b-a)^{-n-j'}a^{j''}\cr
&&(\tilde{\a}_{A\dot{A},p-k-k'},\tilde{\a}_{B\dot{B},-n+j+j'-j''})_{\text{contr.}}\cr
&&\!\!\!\!\!\!\!\!\!\!= \sum_{k=0}^{p-1}\sum_{k'=0}^{p-k-1}\sum_{j= 0}^{n-1}\sum_{j'= 0}^{n-j-1}\sum_{j''\geq 0}{}^{p}C_{k}{}^{p}C_{k'}{}^nC_{j}{}^{-n}C_{j'}{}^{-n+j+j'}C_{j''}a^kb^{k'}(-a)^{n-j}(b-a)^{-n-j'}a^{j''}\cr
&&(-1)(p-k-k')\e_{AB}\e_{\dot{A}\dot{B}}\d_{p-k-k' -n+j+j'-j'' ,0}
\cr
&& \!\!\!\!\!\!\!\!\!\!= - \e_{AB}\e_{\dot{A}\dot{B}}\sum_{k=0}^{p-1}\sum_{k'=0}^{p-k-1}\sum_{j= 0}^{n-1}\sum_{j'= 0}^{n-j-1}(p-k-k'){}^{p}C_{k}{}^{p}C_{k'}{}^nC_{j}{}^{-n}C_{j'}{}^{-n+j+j'}C_{p-k-k' -n+j+j'}\cr
&&(-1)^{n-j}a^{p-k' +j'}b^{k'}(b-a)^{-n-j'}
\eea
Using creation and annihilation constraints on the modes indices yield 
\bea
&&j''\geq 0 \quad\to  \quad p-k-k' -n+j+j'\geq 0  \quad\to\quad k'\leq p-n+j+j'-k \cr
&&k'\geq 0 \to k\leq p-n+j+j'
\eea
Also, we have the additional constraint into the expression for that
\bea
k+k'\leq p - n +j +j' \text{  and  } k+k'\geq 0 \quad\to\quad j+j' \geq n-p \quad\to\quad j'\geq n-p-j\nn
\eea
Inserting these constraints into the expression for $(\a'^{(1)f}_{B\dot{B},p},\a'^{(1)i}_{A\dot{A},-n})$ yield
\bea
&&\!\!\!\!\!\!\!(\a'^{(1)f}_{B\dot{B},p}\a'^{(1)i}_{A\dot{A},-n})_{\text{contr.}}\cr
&&= - \e_{AB}\e_{\dot{A}\dot{B}}\sum_{j= 0}^{n-1}\sum_{j'= \max{[0,n-p -j]}}^{n-j-1}\sum_{k=0}^{p-n +j+j'}\sum_{k'=0}^{p-n +j+j'-k}(p-k-k'){}^{p}C_{k}{}^{p}C_{k'}{}^nC_{j}{}^{-n}C_{j'}\cr
&&\quad{}^{-n+j+j'}C_{p-k-k' -n+j+j'}(-1)^{n-j}a^{p-k' +j'}b^{k'}(b-a)^{-n-j'}
\label{init}
\eea
For an initial boson on Copy $2$ we just make the switch $a\leftrightarrow b$ in for the expression (\ref{init}). This yields
\bea
&&(\a'^{(1)f}_{B\dot{B},p},\a'^{(2)i}_{A\dot{A},-n})\cr
&& =  - \e_{AB}\e_{\dot{A}\dot{B}}\sum_{j= 0}^{n-1}\sum_{j'= \max{[0,n-p -j]}}^{n-j-1}\sum_{k=0}^{p-n +j+j'}\sum_{k'=0}^{p-n +j+j'-k}(p-k-k'){}^{p}C_{k}{}^{p}C_{k'}{}^nC_{j}{}^{-n}C_{j'}\cr
&&\quad{}^{-n+j+j'}C_{p-k-k' -n+j+j'}(-1)^{n-j}b^{p-k' +j'}a^{k'}(a-b)^{-n-j'}
\eea

\subsection{Contraction between bosons at $\infty$ and $0$}
Here we compute the contraction between a boson at $t=\infty$ (Copy 1 final) and $t=0$ (Copy 2 final). The expansion at $t=\infty$ was shown to be
\bea
\a'^{(1)f}_{A\dot{A},p} &=& \sum_{k,k'\geq0}{}^{p}C_{k}{}^{p}C_{k'}a^kb^{k'}\tilde{\a}_{A\dot{A},p-k-k'}
\eea
The expansion at $t=0$ is
\bea
\a'^{(2)f}_{A\dot{A},p} &=& - {1\over 2\pi}\int_{t=0} dt {(t+a)^p(t+b)^p\over t^p}\partial X_{A\dot{A}}(t)\cr
&=&-\sum_{j,j'\geq 0}{}^pC_j{}^pC_{j'} a^{p-j}b^{p-j'}\tilde{\a}^0_{A\dot{A},j+j'-p}
\label{copy two final}
\eea
where modes at $t=0$ are defined as 
\bea
\tilde{\a}_{A\dot{A},n}={1\over 2\pi}\int_{t=\infty}dt t^n\partial X_{A\dot{A}}(t)
\eea
The contraction is 
\bea
&&\!\!\!\!\!\!\!\!\!\!(\a'^{(1)f}_{B\dot{B},p},\a'^{(2)f}_{A\dot{A},q})\cr
&&\!\!\!\!\!= -\sum_{k,k',j,j'\geq0}{}^{p}C_{k}{}^{p}C_{k'}{}^qC_j{}^qC_{j'} a^{q-j+k}b^{q-j'+k'}[\tilde{\a}_{B\dot{B},p-k-k'},\tilde{\a}^0_{A\dot{A},j+j'-q}]\cr
&&\!\!\!\!\!= -\sum_{k,k',j,j'\geq0}{}^{p}C_{k}{}^{p}C_{k'}{}^qC_j{}^qC_{j'} a^{q-j+k}b^{q-j'+k'}(-(p-k-k')\e_{BA}\e_{\dot{B}\dot{A}}\d_{p-k-k' + j+j'-q,0})\cr
&&\!\!\!\!\!= \e_{BA}\e_{\dot{B}\dot{A}}\sum_{k,k',j\geq0}(p-k-k'){}^{p}C_{k}{}^{p}C_{k'}{}^qC_j{}^qC_{q-p-j+k+k'} a^{q-j+k}b^{p+j-k}
\eea
Constraints on indices give
\bea
&&\!\!\!\!\!(\a'^{(1)f}_{B\dot{B},p},\a'^{(2)f}_{A\dot{A},q})\cr
 &&=  \e_{BA}\e_{\dot{B}\dot{A}}\sum_{k=0}^{p-1}\sum_{k'=0}^{p-k-1}\sum_{j=0}^{q-p+k+k'}(p-k-k'){}^{p}C_{k}{}^{p}C_{k'}{}^qC_j{}^qC_{q-p-j+k+k'} a^{q-j+k}b^{p+j-k}\nn
\eea
Similarly, for the other combination we have
\bea
(\a'^{(2)f}_{B\dot{B},p},\a'^{(1)f}_{A\dot{A},q}) 
&=&\e_{BA}\e_{\dot{B}\dot{A}}\sum_{k=0}^{p-1}\sum_{k'=0}^{p-k-1}\sum_{j=0}^{q-p+k+k'}(p-k-k'){}^{p}C_{k}{}^{p}C_{k'}{}^qC_j{}^qC_{q-p-j+k+k'}\cr
&&\qquad\quad\times a^{p+j-k}b^{q-j+k}\nn
\eea

\subsection{Contraction between bosons at $-a$ and $-a$}
Here we compute the contraction between two bosons both at $t=-a$ (Both Copy 1 initial). Using the expansions recorded in (\ref{copy one initial b}) we have the contraction
\bea
&&\!\!\!\!\!\!\!\!\!\!(\a'^{(1)i}_{B\dot{B},-n_1},\a'^{(1)i}_{A\dot{A},-n_2})\cr
&&\!\!\!\!\!= \sum_{j,j',j'',j'''= 0}{}^{n_1}C_{j}{}^{-n_1}C_{j'}{}^{n_2}C_{j''}{}^{-n_2}C_{j'''}(-a)^{n_1+n_2-j-j''}(b-a)^{-n_1-n_2-j'-j'''}\cr
&&\quad(\tilde{\a}^{-a}_{B\dot{B},-n_1+j+j'},\tilde{\a}^{-a}_{A\dot{A},-n_2+j''+j'''})\cr
&&\!\!\!\!\!=\sum_{j,j',j'',j'''= 0}{}^{n_1}C_{j}{}^{-n_1}C_{j'}{}^{n_2}C_{j''}{}^{-n_2}C_{j'''}(-a)^{n_1+n_2-j-j''}(b-a)^{-n_1-n_2-j'-j'''}\cr
&&(-(-n_1+j+j')\e_{BA}\e_{\dot{B}\dot{A}})\d_{-n_1-n_2+j+j'+j''+j''',0}\cr
&&\!\!\!\!\!=-\e_{BA}\e_{\dot{B}\dot{A}}\sum_{j,j',j'',j'''= 0}(n_2-j''-j'''){}^{n_1}C_{n_1 + n _2 -j'-j''-j'''}{}^{-n_1}C_{j'}{}^{n_2}C_{j''}{}^{-n_2}C_{j'''}\cr
&&\quad(-a)^{j'+j'''}(b-a)^{-n_1-n_2-j'-j'''}
\eea

We have the following constraints on the mode indices
\bea
j &=& n_1 + n_2 - j'-j''-j'''\cr
-n_2 + j'' + j''' &<&0\to j'''<n_2-j''\cr
j'''&\geq&0\to j''<n_2
\cr
j&\geq& 0 \to n_1 + n _2 -j'-j''-j'''\geq 0\to n_1 + n _2 -j''-j'''\geq j'\nn
\eea
Therefore the contraction becomes
\bea
&&\!\!\!\!\!(\a'^{(1)i}_{B\dot{B},-n_1},\a'^{(1)i}_{A\dot{A},-n_2})\cr
&&=-\e_{BA}\e_{\dot{B}\dot{A}}\sum_{j''= 0}^{n_2-1}\sum_{j'''= 0}^{n_2-j''-1} \sum_{j'= 0}^{n_1 + n _2 -j''-j'''}(n_2-j''-j'''){}^{n_1}C_{n_1 + n _2 -j'-j''-j'''}\cr
&&{}^{-n_1}C_{j'}{}^{n_2}C_{j''}{}^{-n_2}C_{j'''}(-a)^{j'+j'''}(b-a)^{-n_1-n_2-j'-j'''}
\eea
For both excitations on Copy 2 $(t=-b)$ we make the switch $a\leftrightarrow b$
\bea
&&\!\!\!\!\!(\a'^{(2)i}_{B\dot{B},-n_1},\a'^{(2)i}_{A\dot{A},-n_2})\cr
&&=-\e_{BA}\e_{\dot{B}\dot{A}}\sum_{j''= 0}^{n_2-1}\sum_{j'''= 0}^{n_2-j''-1} \sum_{j'= 0}^{n_1 + n _2 -j''-j'''}(n_2-j''-j'''){}^{n_1}C_{n_1 + n _2 -j'-j''-j'''}\cr
&&{}^{-n_1}C_{j'}{}^{n_2}C_{j''}{}^{-n_2}C_{j'''}(-b)^{j'+j'''}(a-b)^{-n_1-n_2-j'-j'''}
\eea

For one excitation on copy 1 and another on copy 2 we must expand one of the copies around the other one. Lets first expand copy 2 around copy 1. We have
\bea
\a'^{(2)i}_{A\dot{A},-n}&=&\sum_{j= 0}^{n-1}\sum_{j'= 0}^{n-j-1}{}^nC_{j}{}^{-n}C_{j'}(-b)^{n-j}(a-b)^{-n-j'}\tilde{\a}^{-b}_{A\dot{A},-n+j+j'}
\eea
\bea
\tilde{\a}^{-b}_{A\dot{A},-n+j+j'} = {1\over 2\pi}\int_{t=-b}dt (t+b)^{-n+j+j'}\partial X_{A\dot{A}}(t)
\eea
Expanding the integrand around $t=-a$ we have
\bea
(t+b)^{-n+j+j'} &=& (t+a +b-a)^{-n+j+j'}\cr
&=&\sum_{j''\geq 0}{}^{-n+j+j'}C_{j''}(b-a)^{-n+j+j'-j''}(t+a)^{j''}
\eea
Inserting this into the above yields
\bea
&&\!\!\!\!\!\a'^{(2)i}_{A\dot{A},-n}\cr
&&=-\sum_{j= 0}^{n-1}\sum_{j'= 0}^{n-j-1}\sum_{j''\geq 0}{}^nC_{j}{}^{-n}C_{j'}{}^{-n+j+j'}C_{j''}(-b)^{n-j}(a-b)^{-n-j'}(b-a)^{-n+j+j'-j''}\tilde{\a}^{-a}_{A\dot{A},j''}\nn
\eea
Where the minus comes because we have reversed the direction of the contour when expanding around $t=-a$. Our contraction becomes
\bea
&&\!\!\!\!\!\!\!\!\!\!(\a'^{(2)i}_{B\dot{B},-n_1},\a'^{(1)i}_{A\dot{A},-n_2})\cr
&&\!\!\!\!\!= -\sum_{j= 0}^{n_1-1}\sum_{j'= 0}^{n_1-j-1}\sum_{j''\geq 0}\sum_{j'''= 0}^{n_2-1}\sum_{j''''= 0}^{n_2-j'''-1}{}^{n_1}C_{j}{}^{-n_1}C_{j'}{}^{-n_1+j+j'}C_{j''}{}^{n_2}C_{j'''}{}^{-n_2}C_{j''''}\cr
&&(-b)^{n_1-j}(a-b)^{-n_1-j'}(b-a)^{-n_1+j+j'-j''}(-a)^{n_2-j'''}(b-a)^{-n_2-j''''}\cr
&&[\tilde{\a}^{-a}_{A\dot{A},j''}\tilde{\a}^{-a}_{A\dot{A},-n_2+j'''+j''''}]
\cr
\cr
&&\!\!\!\!\!= -\sum_{j= 0}^{n_1-1}\sum_{j'= 0}^{n_1-j-1}\sum_{j''\geq 0}\sum_{j'''= 0}^{n_2-1}\sum_{j''''= 0}^{n_2-j'''-1}{}^{n_1}C_{j}{}^{-n_1}C_{j'}{}^{-n_1+j+j'}C_{j''}{}^{n_2}C_{j'''}{}^{-n_2}C_{j''''}\cr
&&(-b)^{n_1-j}(a-b)^{-n_1-j'}(b-a)^{-n_1+j+j'-j''}(-a)^{n_2-j'''}(b-a)^{-n_2-j''''}\cr
&&(-j''\e_{BA}\e_{\dot{B}\dot{A}}\d_{j''-n_2+j'''+j'''',0})
\cr
\cr
&&\!\!\!\!\!= \e_{BA}\e_{\dot{B}\dot{A}}\sum_{j= 0}^{n_1-1}\sum_{j'= 0}^{n_1-j-1}\sum_{j'''= 0}^{n_2-1}\sum_{j''''= 0}^{n_2-j'''-1}(n_2-j'''-j''''){}^{n_1}C_{j}{}^{-n_1}C_{j'}\cr
&&{}^{-n_1+j+j'}C_{n_2-j'''-j''''}{}^{n_2}C_{j'''}{}^{-n_2}C_{j''''}(-b)^{n_1-j}(a-b)^{-n_1-j'}\cr
&&(b-a)^{-n_1+j+j'-(n_2-j'''-j'''')}(-a)^{n_2-j'''}(b-a)^{-n_2-j''''}\nn
\cr
\cr
&&\!\!\!\!\!= \e_{BA}\e_{\dot{B}\dot{A}}\sum_{j= 0}^{n_1-1}\sum_{j'= 0}^{n_1-j-1}\sum_{j'''= 0}^{n_2-1}\sum_{j''''= 0}^{n_2-j'''-1}(n_2-j'''-j''''){}^{n_1}C_{j}{}^{-n_1}C_{j'}\cr
&&{}^{-n_1+j+j'}C_{n_2-j'''-j''''}{}^{n_2}C_{j'''}{}^{-n_2}C_{j''''}\cr
&&(-b)^{n_1-j}(a-b)^{-n_1-j'}(b-a)^{-n_1-2n_2+j+j'+j'''}(-a)^{n_2-j'''}
\cr
\cr
&&\!\!\!\!\!= \e_{BA}\e_{\dot{B}\dot{A}}\sum_{j= 0}^{n_1-1}\sum_{j'= 0}^{n_1-j-1}\sum_{j'''= 0}^{n_2-1}\sum_{j''''= 0}^{n_2-j'''-1}(n_2-j'''-j''''){}^{n_1}C_{j}{}^{-n_1}C_{j'}\cr
&&{}^{-n_1+j+j'}C_{n_2-j'''-j''''}{}^{n_2}C_{j'''}{}^{-n_2}C_{j''''}\cr
&&(-b)^{n_1-j}(a-b)^{-n_1-j'}(b-a)^{-n_1-2n_2+j+j'+j'''}(-a)^{n_2-j'''}\nn
\eea
For the other combination, we make the switch $a$ and $b$
\bea
(\a'^{(1)i}_{B\dot{B},-n_1},\a'^{(2)i}_{A\dot{A},-n_2})&=& \e_{BA}\e_{\dot{B}\dot{A}}\sum_{j= 0}^{n_1-1}\sum_{j'= 0}^{n_1-j-1}\sum_{j'''= 0}^{n_2-1}\sum_{j''''= 0}^{n_2-j'''-1}(n_2-j'''-j''''){}^{n_1}C_{j}{}^{-n_1}C_{j'}\cr
&&{}^{-n_1+j+j'}C_{n_2-j'''-j''''}{}^{n_2}C_{j'''}{}^{-n_2}C_{j''''}\cr
&&(-a)^{n_1-j}(b-a)^{-n_1-j'}(a-b)^{-n_1-2n_2+j+j'+j'''}(-b)^{n_2-j'''}\nn
\eea

Because of Copy symmetry we have the relations
\bea
(\a'^{(1)i}_{B\dot{B},-n_1},\a'^{(1)i}_{A\dot{A},-n_2})&=&(\a'^{(2)i}_{B\dot{B},-n_1},\a'^{(2)i}_{A\dot{A},-n_2})\cr
(\a'^{(1)i}_{B\dot{B},-n_1},\a'^{(2)i}_{A\dot{A},-n_2})&=&(\a'^{(2)i}_{B\dot{B},-n_1},\a'^{(1)i}_{A\dot{A},-n_2})\cr
(\a'^{(1)i}_{B\dot{B},-n_1},\a'^{(1)i}_{A\dot{A},-n_2})&=&-(\a'^{(1)i}_{B\dot{B},-n_1},\a'^{(2)i}_{A\dot{A},-n_2})
\eea

\subsection{Contraction between bosons at $0$ and $0$}
Here we compute the contraction of two modes at $t=0$ (Both Copy 2 final). From (\ref{copy two final}) we have the expansion
\bea
\a'^{(2)f}_{A\dot{A},p} &=&-\sum_{j,j'\geq 0}{}^pC_j{}^pC_{j'} a^{p-j}b^{p-j'}\tilde{\a}^0_{A\dot{A},j+j'-p}
\eea

We also note that when we perform the wick contraction with another Copy 2 final mode, the outer mode on the cylinder gets mapped to the inner mode in the $t$ plane, again because of the behavior of the map. The larger $|z|$ is, the smaller $|t|$ is. We have:
\bea
(\a'^{(2)f}_{B\dot{B},p},\a'^{(2)f}_{A\dot{A},q}) &=& \sum_{j,j',k,k'\geq 0}{}^qC_j{}^qC_{j'}{}^pC_k{}^pC_{k'} a^{p+q-j-k}b^{p+q-j'-k'}[\tilde{\a}^0_{A\dot{A},j+j'-q},\tilde{\a}^0_{B\dot{B},k+k'-p}]\cr
&=& \sum_{j,j',k,k'\geq 0}{}^qC_j{}^qC_{j'}{}^pC_k{}^pC_{k'} a^{p+q-j-k}b^{p+q-j'-k'}\cr
&&\qquad(-(j+j'-q)\e_{AB}\e_{\dot{A}\dot{B}}\d_{j+j'-q + k +k'-p,0})\cr
&=&-\e_{AB}\e_{\dot{A}\dot{B}}\sum_{j,k,k'\geq 0}(p-k-k'){}^pC_k{}^pC_{k'}{}^qC_j{}^qC_{p+q-j-k-k'} a^{p+q-j-k}b^{j+k}\nn
\eea
Applying constraints on the sums yields
\bea
k+k'-p &<& 0 \to k'< p-k\cr
k'&\geq& 0\to k<p\cr
j' &=& p+q-k-k-j\cr
j'&\geq& 0\to p+q-k-k'\geq j
\eea
Implementing these changes yield
\bea
(\a'^{(2)f}_{B\dot{B},p},\a'^{(2)f}_{A\dot{A},q}) &=& -\e_{AB}\e_{\dot{A}\dot{B}}\sum_{k=0}^{p-1}\sum_{k'=0}^{p-k-1}\sum_{j=0}^{p+q-k-k'}(p-k-k'){}^pC_k{}^pC_{k'}{}^qC_j{}^qC_{p+q-j-k-k'}\cr
&&\qquad\qquad\qquad\qquad a^{p+q-j-k}b^{j+k}\nn
\eea

\subsection{Contraction between bosons at $0$ and $t_i$}
Here we compute the contraction between a mode at $t=0$ (Copy 2 final) and a mode at $t=t_i$ where $i=1,2$.
Let's expand the mode that is around $t=0$ around $t_i$ (Boson coming from $G$). First we have the expansion
\bea
\a'^{(2)f}_{A\dot{A},p}&=&-\sum_{k,k'\geq 0}{}^pC_k{}^pC_{k'} a^{p-k}b^{p-k'}\tilde{\a}^0_{A\dot{A},k+k'-p}
\eea 
To not annihilate the local vacuum requires
\bea
k+k'-p &<& 0 \to k'< p-k\cr
k'&\geq& 0\to k<p
\eea
Implementing these constraints yield
\bea
\a'^{(2)f}_{A\dot{A},p}&=&-\sum_{k=0}^{p-1}\sum_{k'=0}^{p-k-1}{}^pC_k{}^pC_{k'} a^{p-k}b^{p-k'}\tilde{\a}^0_{A\dot{A},k+k'-p}
\eea
We now from previous cases that the mode expansion at $t=0$ is
\bea
\tilde{\a}^0_{A\dot{A},k+k'-p} = {1\over 2\pi}\int dt t^{k+k'-p}\partial X_{A\dot{A}}(t)
\eea
Let's expand the above integrand around $t=t_i$
\bea
t^{k+k'-p}&=&(t-t_i +t_i)^{k+k'-p}\cr
&=&t_i^{k+k'-p}(1+t_i^{-1}(t-t_i))^{k+k'-p}\cr
&=&\sum_{l\geq0}{}^{k+k'-p}C_lt_i^{k+k'-p-l}(t-t_i)^l
\label{ti expansion}
\eea
Where modes natural to $t_i,\quad i=1,2$ are defined as
\bea
\tilde{\a}^{t_i}_{A\dot{A},n}={1\over 2\pi}\int dt (t-t_i)^n\partial X_{A\dot{A}}(t)
\eea
Inserting the above expansion into our contour yields
\bea
\a'^{(2)f}_{A\dot{A},p}=\sum_{k=0}^{p-1}\sum_{k'=0}^{p-k-1}\sum_{l\geq0}{}^pC_k{}^pC_{k'} {}^{k+k'-p}C_la^{p-k}b^{p-k'}t_i^{k+k'-p-l}\tilde{\a}^{t_i}_{A\dot{A},l}
\eea
where the minus sign again comes from reversing the contour. Since there was already a minus there, an additional minus sign gives an overall plus sign.
Now computing the wick contraction we have:
\bea
(\a'^{(2)f}_{B\dot{B},p},\tilde{\a}^{t_i}_{A\dot{A},-1}) &=&  \sum_{k=0}^{p-1}\sum_{k'=0}^{p-k-1}\sum_{l\geq0}{}^pC_k{}^pC_{k'} {}^{k+k'-p}C_la^{p-k}b^{p-k'}t_i^{k+k'-p-l}[\tilde{\a}^{t_i}_{B\dot{B},l},\tilde{\a}^{t_i}_{A\dot{A},-1}]\nn
&=&\sum_{k=0}^{p-1}\sum_{k'=0}^{p-k-1}\sum_{l\geq0}{}^pC_k{}^pC_{k'} {}^{k+k'-p}C_la^{p-k}b^{p-k'}t_i^{k+k'-p-l}(-\e_{\dot{B}\dot{A}}\e_{BA}l\d_{l-1,0})\cr
&=&-\e_{\dot{B}\dot{A}}\e_{BA}\sum_{k=0}^{p-1}\sum_{k'=0}^{p-k-1}{}^pC_k{}^pC_{k'} {}^{k+k'-p}C_1a^{p-k}b^{p-k'}t_i^{k+k'-p-1}\cr
&=&-\e_{\dot{B}\dot{A}}\e_{BA}\sum_{k=0}^{p-1}\sum_{k'=0}^{p-k-1}(k+k'-p){}^pC_k{}^pC_{k'} a^{p-k}b^{p-k'}t_i^{k+k'-p-1}
\eea

\subsection{Contraction between bosons at $0$ and $-a$}
Here we compute the contraction between a mode at $t=0$ (Copy 2 final) and $t=-a$ (Copy 1 initial). The expansion for Copy 2 final is 
\bea
\a'^{(2)f}_{A\dot{A},p}&=&-\sum_{k=0}^{p-1}\sum_{k'=0}^{p-k-1}{}^pC_k{}^pC_{k'} a^{p-k}b^{p-k'}\tilde{\a}^0_{A\dot{A},k+k'-p}
\eea
and Copy 1 initial is
\bea
\a'^{(1)i}_{A\dot{A},-n}&=&\sum_{j= 0}^{n-1}\sum_{j'= 0}^{n-j-1}{}^nC_{j}{}^{-n}C_{j'}(-a)^{n-j}(b-a)^{-n-j'}\tilde{\a}^{-a}_{A\dot{A},-n+j+j'}
\eea
We'll do the computation for Copy 1 initial and then note that the result for Copy 2 initial can be obtained simply making the switch $a\leftrightarrow b$. Proceeding, we'd like to expand the contour circling $t=0$ around $t=-a$ so we have

\bea
\tilde{\a}^0_{A\dot{A},k+k'-p} &=& {1\over 2\pi}\int_{t=0} dt t^{k+k'-p}\partial X_{A\dot{A}}(t)
\eea
Expanding the integrand around $t=-a$ yields 
\bea
 t^{k+k'-p}=\sum_{l\geq0}{}^{k+k'-p}C_l(-a)^{k+k'-p-l}(t+a)^l
\eea
Inserting this into the above contour yields
\bea
\tilde{\a}^0_{A\dot{A},k+k'-p}  &=&-\sum_{l\geq0}{}^{k+k'-p}C_l(-a)^{k+k'-p-l}\tilde{\a}^{-a}_{A\dot{A},l}
\eea
where the minus sign comes from reversing the direction of the contour.
Therefore $\a'^{(2)f}$ becomes
\bea
\a'^{(2)f}_{A\dot{A},p}&=&\sum_{k=0}^{p-1}\sum_{k'=0}^{p-k-1}\sum_{l\geq0}{}^pC_k{}^pC_{k'} {}^{k+k'-p}C_l (-a)^{k+k'-p-l}a^{p-k}b^{p-k'}\tilde{\a}^{-a}_{A\dot{A},l}
\eea

Now computing the wick contraction, we have
\bea
(\a'^{(2)f}_{B\dot{B},p},\a'^{(1)i}_{A\dot{A},-n})&=&\sum_{k=0}^{p-1}\sum_{k'=0}^{p-k-1}\sum_{j= 0}^{n-1}\sum_{j'= 0}^{n-j-1}\sum_{l\geq0}{}^pC_k{}^pC_{k'} {}^{k+k'-p}C_l {}^nC_{j}{}^{-n}C_{j'}\cr
&&(-a)^{n-j+k+k'-p-l}a^{p-k}(b-a)^{-n-j'}b^{p-k'}[\tilde{\a}^{-a}_{B\dot{B},l},\tilde{\a}^{-a}_{A\dot{A},-n+j+j'}]
\cr
\cr
&=&\sum_{k=0}^{p-1}\sum_{k'=0}^{p-k-1}\sum_{j= 0}^{n-1}\sum_{j'= 0}^{n-j-1}\sum_{l\geq0}{}^pC_k{}^pC_{k'} \cr
&&{}^{k+k'-p}C_l {}^nC_{j}{}^{-n}C_{j'}(-a)^{n-j+k+k'-p-l}a^{p-k}(b-a)^{-n-j'}b^{p-k'}\cr
&&(-l\e_{BA}\e_{\dot{B}\dot{A}}\d_{l-n+j+j',0})
\cr
\cr
&=&-\e_{BA}\e_{\dot{B}\dot{A}}\sum_{k=0}^{p-1}\sum_{k'=0}^{p-k-1}\sum_{j= 0}^{n-1}\sum_{j'= 0}^{n-j-1}(n-j-j'){}^pC_k{}^pC_{k'} {}^{k+k'-p}C_{n-j-j'} \cr
&&{}^nC_{j}{}^{-n}C_{j'}(-a)^{j'+k+k'-p}a^{p-k}b^{p-k'}(b-a)^{-n-j'}
\cr
\cr
&=&-\e_{BA}\e_{\dot{B}\dot{A}}\sum_{k=0}^{p-1}\sum_{k'=0}^{p-k-1}\sum_{j= 0}^{n-1}\sum_{j'= 0}^{n-j-1}(n-j-j'){}^pC_k{}^pC_{k'} {}^{k+k'-p}C_{n-j-j'} \cr
&&{}^nC_{j}{}^{-n}C_{j'}(-1)^{p+j'+k+k'}a^{j'+k'}b^{p-k'}(b-a)^{-n-j'}
\eea
To give the contraction with a Copy 2 initial state we switch $a$ and $b$ in the above expression giving.
\bea
(\a'^{(2)f}_{B\dot{B},p},\a'^{(2)i}_{A\dot{A},-n}) &=& -\e_{BA}\e_{\dot{B}\dot{A}}\sum_{k=0}^{p-1}\sum_{k'=0}^{p-k-1}\sum_{j= 0}^{n-1}\sum_{j'= 0}^{n-j-1}(n-j-j'){}^pC_k{}^pC_{k'} {}^{k+k'-p}C_{n-j-j'} \cr
&&{}^nC_{j}{}^{-n}C_{j'}(-1)^{p+j'+k+k'}b^{j'+k'}a^{p-k'}(a-b)^{-n-j'}
\eea
\subsection{Contraction between bosons at $t_1,t_2$}
Here we compute the contraction between modes at $t_1$ (Boson from $G^-$) and $t_2$ (Boson from $G^+$). We have
\bea
(\tilde{\a}^{t_2}_{C\dot{C},-1}\tilde{\a}^{t_1}_{A\dot{A},-1})_{\text{contr.}}  &=&  i\partial X_{C\dot{C}}(t_2)i\partial X_{A\dot{A}}(t_1)\cr
&=&-\e_{CA}\e_{\dot{C}\dot{A}}{1\over (t_2-t_1)^2}
\eea
\subsection{Contraction between bosons at $t_i$ and $-a$}

Here we compute the contraction between a mode at $t_i$ (Boson from $G$) and $t=-a$ (Copy 1 initial). As we have seen, the expansion for a Copy 1 initial boson is
\bea
\a'^{(1)i}_{A\dot{A},-n}&=&\sum_{j= 0}^{n-1}\sum_{j'= 0}^{n-j-1}{}^nC_{j}{}^{-n}C_{j'}(-a)^{n-j}(b-a)^{-n-j'}\tilde{\a}^{-a}_{A\dot{A},-n+j+j'}
\eea
Now the modes natural to the point $t=-a$
\bea
\tilde{\a}^{-a}_{A\dot{A},-n+j+j'}  &=&   {1\over 2\pi}\int_{t=-a} dt(t+a)^{-n + j + j'}\partial X_{A\dot{A}}(t)
\eea
We want to expand these modes around $t=t_i$. Expanding the integrand yields.

\bea
(t+a)^{-n + j + j'} &=&(t-t_i + t_i + a)^{-n + j + j'}\cr
&=&\sum_{k\geq0} {}^{-n+j+j'}C_k (t_i+a)^{-n + j + j'-k}(t-t_i)^k
\eea

Therefore

\bea
\a'^{(1)i}_{A\dot{A},-n}&=&-\sum_{j= 0}^{n-1}\sum_{j'= 0}^{n-j-1}\sum_{k\geq0}{}^nC_{j}{}^{-n}C_{j'}{}^{-n+j+j'}C_k(-a)^{n-j}(b-a)^{-n-j'}(t_i+a)^{-n + j + j'-k}\tilde{\a}^{t_i}_{A\dot{A},k}\nn
\eea

Where the minus sign comes from reversing the direction of the contour. Therefore we have

\bea
(\tilde{\a}^{t_i}_{C\dot{C},-1}\a'^{(1)i}_{A\dot{A},-n})_{\text{contr.}} &=& -\sum_{j= 0}^{n-1}\sum_{j'= 0}^{n-j-1}\sum_{k\geq0}{}^nC_{j}{}^{-n}C_{j'}{}^{-n+j+j'}C_k(-a)^{n-j}(b-a)^{-n-j'}\cr
&&(t_i+a)^{-n + j + j'-k}[\tilde{\a}^{t_i}_{A\dot{A},k},\tilde{\a}^{t_i}_{C\dot{C},-1}]\cr
&=&-\sum_{j= 0}^{n-1}\sum_{j'= 0}^{n-j-1}\sum_{k\geq0}{}^nC_{j}{}^{-n}C_{j'}{}^{-n+j+j'}C_k(-a)^{n-j}(b-a)^{-n-j'}\cr
&&(t_i+a)^{-n + j + j'-k}(-k\e_{CA}\e_{\dot{C}\dot{A}}\d_{k-1,0})\cr
&=&\e_{CA}\e_{\dot{C}\dot{A}}\sum_{j= 0}^{n-1}\sum_{j'= 0}^{n-j-1}{}^nC_{j}{}^{-n}C_{j'}{}^{-n+j+j'}C_1(-a)^{n-j}(b-a)^{-n-j'}\cr
&&\qquad (t_i+a)^{-n + j + j'-1}
\eea
For contractions between modes located at $t_i$ and $-b$ we just make the following switch in the above equation $a\leftrightarrow b$. This yields

\bea
&&\!\!\!\!\!(\tilde{\a}^{t_i}_{C\dot{C},-1}\a'^{(2)i}_{A\dot{A},-n})\cr
&&=\e_{CA}\e_{\dot{C}\dot{A}}\sum_{j= 0}^{n-1}\sum_{j'= 0}^{n-j-1}{}^nC_{j}{}^{-n}C_{j'}{}^{-n+j+j'}C_1(-b)^{n-j}(a-b)^{-n-j'}(t_i+b)^{-n + j + j'-1}\nn
\eea

\subsection{Contraction between fermions at $\infty$ and $\infty$}
Here we compute the contraction between two fermions at $t=\infty$ (Both Copy 1). Lets write the mode expansions for the fermions at $t=\infty$. We begin with the contours 
\bea
d'^{(1)f,+ A}_r&=&{1\over 2\pi i}\int_{|t|=\infty}dt\psi^{+ A}(t)(t -t_1)(t+a)^{r-{1\over2}}(t+b)^{r-{1\over2}}t^{-r-{1\over2}}\cr
d'^{(1)f,- A}_r&=&{1\over 2\pi i}\int_{|t|=\infty}dt\psi^{+ A}(t)(t -t_2)(t+a)^{r-{1\over2}}(t+b)^{r-{1\over2}}t^{-r-{1\over2}}
\eea
Expanding the integrands, we have
\bea
(t+a)^{r-{1\over2}}&=&t^{r-{1\over2}}(1+at^{-1})^{r-{1\over2}}\cr
&=&\sum_{k\geq 0}{}^{r-{1\over2}}C_ka^kt^{r-k-{1\over2}}
\eea
and similarly
\bea
(t+b)^{r-{1\over2}}=\sum_{k'\geq 0}{}^{r-{1\over2}}C_{k'}b^{k'}t^{r-k'-{1\over2}}
\eea
Inserting these expansions into the contour we get for the $+$ charge
\bea
d'^{(1)f,+ A}_r&=&\sum_{k,k'\geq 0}{}^{r-{1\over2}}C_k{}^{r-{1\over2}}C_{k'}a^kb^{k'}\tilde{d}^{+A}_{r-k-k'}-\sum_{k,k'\geq 0}{}^{r-{1\over2}}C_k{}^{r-{1\over2}}C_{k'}a^kb^{k'}t_1\tilde{d}^{+A}_{r-k-k'-1}\cr
&=&\sum_{k,k'\geq 0}{}^{r-{1\over2}}C_k{}^{r-{1\over2}}C_{k'}a^kb^{k'}\tilde{d}^{+A}_{r-k-k'}+\sum_{k,k'\geq 0}{}^{r-{1\over2}}C_k{}^{r-{1\over2}}C_{k'}a^kb^{k'}t_2\tilde{d}^{+A}_{r-k-k'-1}\nn
\eea
and similarly for the $-$ charge
\bea
d'^{(1)f,- A}_r&=&\sum_{k,k'\geq 0}{}^{r-{1\over2}}C_k{}^{r-{1\over2}}C_{k'}a^kb^k\tilde{d}^{-A}_{r-k-k'}-\sum_{k,k'\geq 0}{}^{r-{1\over2}}C_k{}^{r-{1\over2}}C_{k'}a^kb^kt_2\tilde{d}^{-A}_{r-k-k'-1}\nn
\eea
Therefore our contraction becomes
\bea
&&\!\!\!\!(d'^{(1)f,+-}_q ,d'^{(1)f,- +}_r)\cr
&&= \bigg(\sum_{j,j'\geq 0}{}^{q-{1\over2}}C_j{}^{q-{1\over2}}C_{j'}a^jb^{j'}\tilde{d}^{+-}_{q-j-j'}+\sum_{j,j'\geq 0}{}^{q-{1\over2}}C_j{}^{q-{1\over2}}C_{j'}a^jb^{j'}t_2\tilde{d}^{+-}_{q-j-j'-1},\cr
&&\quad~~\sum_{k,k'\geq 0}{}^{r-{1\over2}}C_k{}^{r-{1\over2}}C_{k'}a^kb^{k'}\tilde{d}^{-+}_{r-k-k'}-\sum_{k,k'\geq 0}{}^{r-{1\over2}}C_k{}^{r-{1\over2}}C_{k'}a^kb^{k'}t_2\tilde{d}^{-+}_{r-k-k'-1}\bigg)\cr
&&=\sum_{j,j'\geq 0}\sum_{k,k'\geq 0}{}^{q-{1\over2}}C_j{}^{q-{1\over2}}C_{j'}{}^{r-{1\over2}}C_k{}^{r-{1\over2}}C_{k'}a^{j+k}b^{j'+k'}\lbrace\tilde{d}^{+-}_{q-j-j'},\tilde{d}^{-+}_{r-k-k'}\rbrace\cr
&&\quad-\sum_{j,j'\geq 0}\sum_{k,k'\geq 0}{}^{q-{1\over2}}C_j{}^{q-{1\over2}}C_{j'}{}^{r-{1\over2}}C_k{}^{r-{1\over2}}C_{k'}a^{j+k}b^{j'+k'}t_2\lbrace\tilde{d}^{+-}_{q-j-j'},\tilde{d}^{-+}_{r-k-k'-1}\rbrace\cr
&&\quad+\sum_{j,j'\geq 0}\sum_{k,k'\geq 0}{}^{q-{1\over2}}C_j{}^{q-{1\over2}}C_{j'}{}^{r-{1\over2}}C_k{}^{r-{1\over2}}C_{k'}a^{j+k}b^{j'+k'}t_2\lbrace\tilde{d}^{+-}_{q-j-j'-1},\tilde{d}^{-+}_{r-k-k'}\rbrace\cr
&&\quad-\sum_{j,j'\geq 0}\sum_{k,k'\geq 0}{}^{q-{1\over2}}C_j{}^{q-{1\over2}}C_{j'}{}^{r-{1\over2}}C_k{}^{r-{1\over2}}C_{k'}a^{j+k}b^{j'+k'}t_2^2\lbrace\tilde{d}^{+-}_{q-j-j'-1},\tilde{d}^{-+}_{r-k-k'-1}\rbrace\nn
\eea
First we impose constraints on the $j,j'$ sums. Since they are acting from the left they should be annihilation operators. This yields
\bea
q-j-j'>0 &\to& q-j>j'\cr
j'\geq0 &\to& q>j
\cr
\cr 
q-j-j'-1>0 &\to& q-j-1>j'\cr
j'\geq0 &\to& q-1>j
\eea
Implementing these constraints yield
\bea
&&(d'^{(1)f,+-}_q ,d'^{(1)f,- +}_r)\cr
&&=\sum_{j = 0}^{\lfloor q\rfloor}\sum_{j' = 0}^{\lfloor q-j\rfloor}\sum_{k,k'\geq 0}{}^{q-{1\over2}}C_j{}^{q-{1\over2}}C_{j'}{}^{r-{1\over2}}C_k{}^{r-{1\over2}}C_{k'}a^{j+k}b^{j'+k'}\lbrace\tilde{d}^{+-}_{q-j-j'},\tilde{d}^{-+}_{r-k-k'}\rbrace\cr
&&\quad-\sum_{j = 0}^{\lfloor q\rfloor}\sum_{j' = 0}^{\lfloor q-j\rfloor}\sum_{k,k'\geq 0}{}^{q-{1\over2}}C_j{}^{q-{1\over2}}C_{j'}{}^{r-{1\over2}}C_k{}^{r-{1\over2}}C_{k'}a^{j+k}b^{j'+k'}t_2\lbrace\tilde{d}^{+-}_{q-j-j'},\tilde{d}^{-+}_{r-k-k'-1}\rbrace\cr
&&\quad+\sum_{j = 0}^{\lfloor q-1\rfloor}\sum_{j' = 0}^{\lfloor q-j-1\rfloor}\sum_{k,k'\geq 0}{}^{q-{1\over2}}C_j{}^{q-{1\over2}}C_{j'}{}^{r-{1\over2}}C_k{}^{r-{1\over2}}C_{k'}a^{j+k}b^{j'+k'}t_2\lbrace\tilde{d}^{+-}_{q-j-j'-1},\tilde{d}^{-+}_{r-k-k'}\rbrace\cr
&&\quad-\sum_{j = 0}^{\lfloor q-1\rfloor}\sum_{j' = 0}^{\lfloor q-j-1\rfloor}\sum_{k,k'\geq 0}{}^{q-{1\over2}}C_j{}^{q-{1\over2}}C_{j'}{}^{r-{1\over2}}C_k{}^{r-{1\over2}}C_{k'}a^{j+k}b^{j'+k'}t_2^2\lbrace\tilde{d}^{+-}_{q-j-j'-1},\tilde{d}^{-+}_{r-k-k'-1}\rbrace\nn
\eea
Where $\lfloor x\rfloor$ is the floor function which takes the lowest integer value for a number $x$. We do this because $q,r$ are half integer since we are in the $NS$ sector on the cylinder.
Evaluating the anticommutation relations yield
\bea
&&(d'^{(1)f,+-}_q ,d'^{(1)f,- +}_r)\cr
&&=\sum_{j = 0}^{\lfloor q\rfloor}\sum_{j' = 0}^{\lfloor q-j\rfloor}\sum_{k,k'\geq 0}{}^{q-{1\over2}}C_j{}^{q-{1\over2}}C_{j'}{}^{r-{1\over2}}C_k{}^{r-{1\over2}}C_{k'}a^{j+k}b^{j'+k'}(-\e^{+-}\e^{-+}\d_{q-j-j'+r-k-k',0})\cr
&&\quad-\sum_{j = 0}^{\lfloor q\rfloor}\sum_{j' = 0}^{\lfloor q-j\rfloor}\sum_{k,k'\geq 0}{}^{q-{1\over2}}C_j{}^{q-{1\over2}}C_{j'}{}^{r-{1\over2}}C_k{}^{r-{1\over2}}C_{k'}a^{j+k}b^{j'+k'}t_2(-\e^{+-}\e^{-+}\d_{q-j-j'+r-k-k'-1,0})\cr
&&\quad+\sum_{j = 0}^{\lfloor q-1\rfloor}\sum_{j' = 0}^{\lfloor q-j-1\rfloor}\sum_{k,k'\geq 0}{}^{q-{1\over2}}C_j{}^{q-{1\over2}}C_{j'}{}^{r-{1\over2}}C_k{}^{r-{1\over2}}C_{k'}a^{j+k}b^{j'+k'}t_2(-\e^{+-}\e^{-+}\d_{q-j-j'+r-k-k'-1})\cr
&&\quad-\sum_{j = 0}^{\lfloor q-1\rfloor}\sum_{j' = 0}^{\lfloor q-j-1\rfloor}\sum_{k,k'\geq 0}{}^{q-{1\over2}}C_j{}^{q-{1\over2}}C_{j'}{}^{r-{1\over2}}C_k{}^{r-{1\over2}}C_{k'}a^{j+k}b^{j'+k'}t_2^2(-\e^{+-}\e^{-+}\d_{q-j-j'+r-k-k'-2})
\cr
\cr
\cr
&&=\sum_{j = 0}^{\lfloor q\rfloor}\sum_{j' = 0}^{\lfloor q-j\rfloor}\sum_{k= 0}^{q+r-j-j'}{}^{q-{1\over2}}C_j{}^{q-{1\over2}}C_{j'}{}^{r-{1\over2}}C_k{}^{r-{1\over2}}C_{q-j-j'+r-k}a^{j+k}b^{q+r-j-k}\cr
&&\quad-\sum_{j = 0}^{\lfloor q\rfloor}\sum_{j' = 0}^{\lfloor q-j\rfloor}\sum_{k= 0}^{q+r-j-j'-1}{}^{q-{1\over2}}C_j{}^{q-{1\over2}}C_{j'}{}^{r-{1\over2}}C_k{}^{r-{1\over2}}C_{q-j-j'+r-k-1}a^{j+k}b^{q+r-j-k-1}t_2\cr
&&\quad+\sum_{j = 0}^{\lfloor q-1\rfloor}\sum_{j' = 0}^{\lfloor q-j-1\rfloor}\sum_{k=0}^{q+r-j-j'-1}{}^{q-{1\over2}}C_j{}^{q-{1\over2}}C_{j'}{}^{r-{1\over2}}C_k{}^{r-{1\over2}}C_{q-j-j'+r-k-1}a^{j+k}b^{q+r-j-k-1}t_2\cr
&&\quad-\sum_{j = 0}^{\lfloor q-1\rfloor}\sum_{j' = 0}^{\lfloor q-j-1\rfloor}\sum_{k= 0}^{q+r-j-j'-2}{}^{q-{1\over2}}C_j{}^{q-{1\over2}}C_{j'}{}^{r-{1\over2}}C_k{}^{r-{1\over2}}C_{q-j-j'+r-k-2}a^{j+k}b^{q+r-j-k-2}t_2^2\nn
\eea
and
\bea
&&(d'^{(1)f,-+}_q ,d'^{(1)f,+-}_r)\cr
&&=\sum_{j = 0}^{\lfloor q\rfloor}\sum_{j' = 0}^{\lfloor q-j\rfloor}\sum_{k= 0}^{q+r-j-j'}{}^{q-{1\over2}}C_j{}^{q-{1\over2}}C_{j'}{}^{r-{1\over2}}C_k{}^{r-{1\over2}}C_{q-j-j'+r-k}a^{j+k}b^{q+r-j-k}\cr
&&\quad+\sum_{j = 0}^{\lfloor q\rfloor}\sum_{j' = 0}^{\lfloor q-j\rfloor}\sum_{k= 0}^{q+r-j-j'-1}{}^{q-{1\over2}}C_j{}^{q-{1\over2}}C_{j'}{}^{r-{1\over2}}C_k{}^{r-{1\over2}}C_{q-j-j'+r-k-1}a^{j+k}b^{q+r-j-k-1}t_2\cr
&&\quad-\sum_{j = 0}^{\lfloor q-1\rfloor}\sum_{j' = 0}^{\lfloor q-j-1\rfloor}\sum_{k=0}^{q+r-j-j'-1}{}^{q-{1\over2}}C_j{}^{q-{1\over2}}C_{j'}{}^{r-{1\over2}}C_k{}^{r-{1\over2}}C_{q-j-j'+r-k-1}a^{j+k}b^{q+r-j-k-1}t_2\cr
&&\quad-\sum_{j = 0}^{\lfloor q-1\rfloor}\sum_{j' = 0}^{\lfloor q-j-1\rfloor}\sum_{k= 0}^{q+r-j-j'-2}{}^{q-{1\over2}}C_j{}^{q-{1\over2}}C_{j'}{}^{r-{1\over2}}C_k{}^{r-{1\over2}}C_{q-j-j'+r-k-2}a^{j+k}b^{q+r-j-k-2}t_2^2\nn
\eea

For the Ramond sector we shift our modes according to (\ref{mode shift})
\bea
&&\!\!\!\!\!\!\!\!(d'^{(1)f,+-}_q ,d'^{(1)f,- +}_r)\cr
&&=\sum_{j = 0}^{ q-1}\sum_{j' = 0}^{ q-j-1}\sum_{k= 0}^{q+r-j-j'}{}^{q-1}C_j{}^{q-1}C_{j'}{}^{r}C_k{}^{r}C_{q-j-j'+r-k}a^{j+k}b^{q+r-j-k}\cr
&&\quad-\sum_{j = 0}^{ q-1}\sum_{j' = 0}^{ q-j-1}\sum_{k= 0}^{q+r-j-j'-1}{}^{q-1}C_j{}^{q-1}C_{j'}{}^{r}C_k{}^{r}C_{q-j-j'+r-k-1}a^{j+k}b^{q+r-j-k-1}t_2\cr
&&\quad+\sum_{j = 0}^{q-2}\sum_{j' = 0}^{q-j-2}\sum_{k=0}^{q+r-j-j'-2}{}^{q-1}C_j{}^{q-1}C_{j'}{}^{r}C_k{}^{r}C_{q-j-j'+r-k-1}a^{j+k}b^{q+r-j-k-1}t_2\cr
&&\quad-\sum_{j = 0}^{q-2}\sum_{j' = 0}^{q-j-2}\sum_{k= 0}^{q+r-j-j'-2}{}^{q-1}C_j{}^{q-1}C_{j'}{}^{r}C_k{}^{r}C_{q-j-j'+r-k-2}a^{j+k}b^{q+r-j-k-2}t_2^2\nn
\eea
and
\bea
&&\!\!\!\!\!\!\!\!(d'^{(1)f,-+}_q ,d'^{(1)f,+-}_r)\cr
&&=\sum_{j = 0}^{ q}\sum_{j' = 0}^{ q-j}\sum_{k= 0}^{q+r-j-j'}{}^{q}C_j{}^{q}C_{j'}{}^{r-1}C_k{}^{r-1}C_{q-j-j'+r-k}a^{j+k}b^{q+r-j-k}\cr
&&\quad+\sum_{j = 0}^{ q}\sum_{j' = 0}^{ q-j}\sum_{k= 0}^{q+r-j-j'-1}{}^{q}C_j{}^{q}C_{j'}{}^{r-1}C_k{}^{r-1}C_{q-j-j'+r-k-1}a^{j+k}b^{q+r-j-k-1}t_2\cr
&&\quad-\sum_{j = 0}^{q-1}\sum_{j' = 0}^{ q-j-1}\sum_{k=0}^{q+r-j-j'-1}{}^{q}C_j{}^{q}C_{j'}{}^{r-1}C_k{}^{r-1}C_{q-j-j'+r-k-1}a^{j+k}b^{q+r-j-k-1}t_2\cr
&&\quad-\sum_{j = 0}^{q-1}\sum_{j' = 0}^{ q-j-1}\sum_{k= 0}^{q+r-j-j'-2}{}^{q}C_j{}^{q}C_{j'}{}^{r-1}C_k{}^{r-1}C_{q-j-j'+r-k-2}a^{j+k}b^{q+r-j-k-2}t_2^2\nn
\eea

\subsection{Contraction between fermions at $\infty$ and $t_i$}
Here we compute the contraction between a fermion at $t=\infty$ (Copy 1 initial) and $t_i$ (Fermion from $G$). First we record the mode expansion at $t=\infty$
\bea
d'^{(1)f,+ A}_r&=&\sum_{k,k'\geq 0}{}^{r-{1\over2}}C_k{}^{r-{1\over2}}C_{k'}a^kb^{k'}\tilde{d}^{+A}_{r-k-k'}+\sum_{k,k'\geq 0}{}^{r-{1\over2}}C_k{}^{r-{1\over2}}C_{k'}a^kb^{k'}t_2\tilde{d}^{+A}_{r-k-k'-1}\nn
d'^{(1)f,- A}_r&=&\sum_{k,k'\geq 0}{}^{r-{1\over2}}C_k{}^{r-{1\over2}}C_{k'}a^kb^k\tilde{d}^{-A}_{r-k-k'}-\sum_{k,k'\geq 0}{}^{r-{1\over2}}C_k{}^{r-{1\over2}}C_{k'}a^kb^kt_2\tilde{d}^{-A}_{r-k-k'-1}\nn
\eea
We want to expand the mode at $t_i$ around $t=\infty$. We have 
\bea
 \tilde{d}^{\a A,t_i}_{-{1\over2}} &=& {1\over 2\pi i}\int_{t_i}dt (t-t_i)^{-1}\psi^{\a A}(t )\cr
 &=&\sum_{k\geq0}t_i^k\tilde{d}^{\a A}_{-k-{1\over2}}
\eea
Therefore the wick contraction becomes
\bea
&&\!\!\!\!\!\!\!\!\!\!\!\!(d'^{(1)f,- +}_r ,\tilde{d}^{+-,t_2}_{-{1\over2}} )\cr
&&\!\!\!\!\!\!\!\!\!\!\!= \big( \sum_{j = 0}^{\lfloor q\rfloor}\sum_{j' = 0}^{\lfloor q-j\rfloor}{}^{q-{1\over2}}C_j{}^{q-{1\over2}}C_{j'}a^jb^{j'}\tilde{d}^{-+}_{q-j-j'}\cr
&&\qquad\qquad\qquad  -  \sum_{j = 0}^{\lfloor q-1\rfloor}\sum_{j' = 0}^{\lfloor q-j-1\rfloor}{}^{q-{1\over2}}C_j{}^{q-{1\over2}}C_{j'}a^jb^{j'}t_2\tilde{d}^{-+}_{q-j-j'-1},\sum_{k\geq0}t_2^k\tilde{d}^{+- }_{-k-{1\over2}}\big)\nn
&&\!\!\!\!\!\!\!\!\!\!\!=\sum_{j = 0}^{\lfloor q\rfloor}\sum_{j' = 0}^{\lfloor q-j\rfloor}\sum_{k\geq0}{}^{q-{1\over2}}C_j{}^{q-{1\over2}}C_{j'}a^jb^{j'}t_2^k(-\e^{-+}\e^{+-}\d_{q-j-j'-k-{1\over2}}) \cr
&&\!\!\!\!\!\!\!\!\!\!\!\qquad  -  \sum_{j = 0}^{\lfloor q-1\rfloor}\sum_{j' = 0}^{\lfloor q-j-1\rfloor}\sum_{k\geq0}{}^{q-{1\over2}}C_j{}^{q-{1\over2}}C_{j'}a^jb^{j'}t_2^{k+1}(-\e^{-+}\e^{+-}\d_{q-j-j'-k-{3\over2}})\cr
&&\!\!\!\!\!\!\!\!\!\!\!=\sum_{j = 0}^{\lfloor q\rfloor}\sum_{j' = 0}^{\lfloor q-j\rfloor}{}^{q-{1\over2}}C_j{}^{q-{1\over2}}C_{j'}a^jb^{j'}t_2^{q-j-j'-{1\over2}}  -  \sum_{j = 0}^{\lfloor q-1\rfloor}\sum_{j' = 0}^{\lfloor q-j-1\rfloor}{}^{q-{1\over2}}C_j{}^{q-{1\over2}}C_{j'}a^jb^{j'}t_2^{q-j-j'-{1\over2}}\nn
\eea
where we've specified the mode $t_2$ because it will aways carry a $+$ R charge and will thus only contract with a $-$ R charge. Similarly for the other charge combination we have
\bea
&&(d'^{(1)f,+-}_q ,  \tilde{d}^{-+,t_1}_{-{1\over2}})\cr
&&=\sum_{j = 0}^{\lfloor q\rfloor}\sum_{j' = 0}^{\lfloor q-j\rfloor}{}^{q-{1\over2}}C_j{}^{q-{1\over2}}C_{j'}a^jb^{j'}t_1^{q-j-j'-{1\over2}}  -  \sum_{j = 0}^{\lfloor q-1\rfloor}\sum_{j' = 0}^{\lfloor q-j-1\rfloor}{}^{q-{1\over2}}C_j{}^{q-{1\over2}}C_{j'}a^jb^{j'}t_1t_1^{q-j-j'-{3\over2}}\cr
&&=\sum_{j = 0}^{\lfloor q\rfloor}\sum_{j' = 0}^{\lfloor q-j\rfloor}{}^{q-{1\over2}}C_j{}^{q-{1\over2}}C_{j'}a^jb^{j'}t_1^{q-j-j'-{1\over2}}  -  \sum_{j = 0}^{\lfloor q-1\rfloor}\sum_{j' = 0}^{\lfloor q-j-1\rfloor}{}^{q-{1\over2}}C_j{}^{q-{1\over2}}C_{j'}a^jb^{j'}t_1^{q-j-j'-{1\over2}}\nn
\eea
where we have specified the mode at $t_1$ for similar reasons.
Writing the above contractions in the $R$ sector we again shift our mode indices according to (\ref{mode shift})
\bea
&&\!\!\!\!\!\!\!\!\!\!\!\!\!\!\!(d'^{(1)f,- +}_r ,\tilde{d}^{+-,t_2}_{-{1\over2}} ) \cr
&&\!\!\!\!\!\!\!\!\!\!\!=\sum_{j = 0}^{ r }\sum_{j' = 0}^{r-j}{}^{r}C_j{}^{r}C_{j'}a^jb^{j'}t_2^{r-j-j'}  -  \sum_{j = 0}^{ r-1}\sum_{j' = 0}^{ r-j-1}{}^{r}C_j{}^{r}C_{j'}a^jb^{j'}t_2^{r-j-j'}\nn
\eea
For the other charge combination we have 
\bea
&&(d'^{(1)f,+-}_q ,  \tilde{d}^{-+,t_1}_{-{1\over2}})\cr
&&=\sum_{j = 0}^{ q -1}\sum_{j' = 0}^{q-j-1}{}^{q-1}C_j{}^{q-1}C_{j'}a^jb^{j'}t_1^{q-j-j'-1}  -  \sum_{j = 0}^{ q-2}\sum_{j' = 0}^{ q-j-2}{}^{q-1}C_j{}^{q-1}C_{j'}a^jb^{j'}t_1^{q-j-j'-1}\nn
\eea

\subsection{Contraction between fermions at $0$ and $0$}
Here compute the contraction between two fermions at $t=0$ (Both Copy 2 final). We expand both $+$ and $-$ charge fermions at $t=0$. We begin with
\bea
d'^{(2)f,+ A}_r&=&-{1\over 2\pi i}\int_{t=0}dt\psi^{+ A}(t)(t -t_1)(t+a)^{r-{1\over2}}(t+b)^{r-{1\over2}}t^{-r-{1\over2}}\cr
d'^{(2)f,- A}_r&=&-{1\over 2\pi i}\int_{t=0}dt\psi^{- A}(t)(t -t_2)(t+a)^{r-{1\over2}}(t+b)^{r-{1\over2}}t^{-r-{1\over2}}
\eea
We expand the integrands around $t=0$
\bea
(t+a)^{r-{1\over2}}&=&\sum_{j\geq 0}{}^{r-{1\over2}}C_j a^{r-k-{1\over2}}t^j\cr
(t+b)^{r-{1\over2}}&=&\sum_{j\geq 0}{}^{r-{1\over2}}C_j b^{r-k-{1\over2}}t^j
\eea
Inserting these expansions in the above contours yield the mode expansions
\bea
d'^{(2)f,+ A}_r&=&-(\sum_{j,j'\geq 0}{}^{r-{1\over2}}C_j{}^{r-{1\over2}}C_{j'}a^{r-j-{1\over2}}b^{r-j'-{1\over2}}\tilde{d}^{+A,0}_{j+j'-r+1}\cr
&&\quad - \sum_{j,j'\geq 0}{}^{r-{1\over2}}C_j{}^{r-{1\over2}}C_{j'}a^{r-j-{1\over2}}b^{r-j'-{1\over2}}t_1\tilde{d}^{+A,0}_{j+j'-r})\cr
&=&(\sum_{j,j'\geq 0}{}^{r-{1\over2}}C_j{}^{r-{1\over2}}C_{j'}a^{r-j-{1\over2}}b^{r-j'-{1\over2}}\tilde{d}^{+A,0}_{j+j'-r+1}\cr
&&\quad +\sum_{j,j'\geq 0}{}^{r-{1\over2}}C_j{}^{r-{1\over2}}C_{j'}a^{r-j-{1\over2}}b^{r-j'-{1\over2}}t_2\tilde{d}^{+A,0}_{j+j'-r})\cr
d'^{(2)f,- A}_r&=&-(\sum_{j,j'\geq 0}{}^{r-{1\over2}}C_j{}^{r-{1\over2}}C_{j'}a^{r-j-{1\over2}}b^{r-j'-{1\over2}}\tilde{d}^{-A,0}_{j+j'-r+1}\cr
&&\quad - \sum_{j,j'\geq 0}{}^{r-{1\over2}}C_j{}^{r-{1\over2}}C_{j'}a^{r-j-{1\over2}}b^{r-j'-{1\over2}}t_2\tilde{d}^{-A,0}_{j+j'-r})
\eea
where in the second equality in the second term we've made the replacement $t_1=-t_2$. Now when we compute the contraction the outer fermion on the cylinder will map to the inner fermion on the $t$ plane. This time however there will be an additional minus sign because fermions anticommute. Our contraction becomes
\bea
&&\!\!\!\!\!\!\!\!\!\!\!\!(d'^{(2)f,+ B}_q,d'^{(2)f,- A}_r)\cr
&&\!\!\!\!\!\!\!\!\!\!=-\bigg(  \sum_{k,k'\geq 0}{}^{r-{1\over2}}C_k{}^{r-{1\over2}}C_{k'}a^{r-k-{1\over2}}b^{r-k'-{1\over2}}\tilde{d}^{-A,0}_{k+k'-r+1}\cr
&&\qquad\qquad - \sum_{k,k'\geq 0}{}^{r-{1\over2}}C_k{}^{r-{1\over2}}C_{k'}a^{r-k-{1\over2}}b^{r-k'-{1\over2}}t_2\tilde{d}^{-A,0}_{k+k'-r} , \cr
&&\quad \sum_{j,j'\geq 0}{}^{q-{1\over2}}C_j{}^{q-{1\over2}}C_{j'}a^{q-j-{1\over2}}b^{q-j'-{1\over2}}\tilde{d}^{+B,0}_{j+j'-q+1}\cr
&&\qquad\qquad +\sum_{j,j'\geq 0}{}^{q-{1\over2}}C_j{}^{q-{1\over2}}C_{j'}a^{q-j-{1\over2}}b^{q-j'-{1\over2}}t_2\tilde{d}^{+B,0}_{j+j'-q} \bigg)\cr
&&\!\!\!\!\!\!\!\!\!\!=- \sum_{k,k',j,j'\geq 0}{}^{r-{1\over2}}C_k{}^{r-{1\over2}}C_{k'}{}^{q-{1\over2}}C_j{}^{q-{1\over2}}C_{j'}a^{q+r-j-k-1}b^{q+r-j'-k'-1}\lbrace\tilde{d}^{-A,0}_{k+k'-r+1,0},\tilde{d}^{+B,0}_{j+j'-q+1}\rbrace\cr
&&- \sum_{k,k',j,j'\geq 0}{}^{r-{1\over2}}C_k{}^{r-{1\over2}}C_{k'}{}^{q-{1\over2}}C_j{}^{q-{1\over2}}C_{j'}a^{q+r-j-k-1}b^{q+r-j'-k'-1}t_2\lbrace\tilde{d}^{-A,0}_{k+k'-r+1},\tilde{d}^{+B,0}_{j+j'-q}\rbrace\cr
&&+ \sum_{k,k',j,j'\geq 0}{}^{r-{1\over2}}C_k{}^{r-{1\over2}}C_{k'}{}^{q-{1\over2}}C_j{}^{q-{1\over2}}C_{j'}a^{q+r-j-k-1}b^{q+r-j'-k'-1}t_2\lbrace\tilde{d}^{-A,0}_{k+k'-r},\tilde{d}^{+B,0}_{j+j'-q+1}\rbrace\cr
&&+ \sum_{k,k',j,j'\geq 0}{}^{r-{1\over2}}C_k{}^{r-{1\over2}}C_{k'}{}^{q-{1\over2}}C_j{}^{q-{1\over2}}C_{j'}a^{q+r-j-k-1}b^{q+r-j'-k'-1}t^2_2\lbrace\tilde{d}^{-A,0}_{k+k'-r},\tilde{d}^{+B,0}_{j+j'-q}\rbrace
\cr
\cr
\cr
&&\!\!\!\!\!\!=- \sum_{k,k',j,j'\geq 0}{}^{r-{1\over2}}C_k{}^{r-{1\over2}}C_{k'}{}^{q-{1\over2}}C_j{}^{q-{1\over2}}C_{j'}a^{q+r-j-k-1}b^{q+r-j'-k'-1}\cr
&&\qquad(-\e^{-+}\e^{AB}\d_{k+k'-r+1+j+j'-q+1,0})\cr
&&- \sum_{k,k',j,j'\geq 0}{}^{r-{1\over2}}C_k{}^{r-{1\over2}}C_{k'}{}^{q-{1\over2}}C_j{}^{q-{1\over2}}C_{j'}a^{q+r-j-k-1}b^{q+r-j'-k'-1}t_2\cr
&&\qquad(-\e^{-+}\e^{AB}\d_{k+k'-r+1+j+j'-q,0})\cr
&&+ \sum_{k,k',j,j'\geq 0}{}^{r-{1\over2}}C_k{}^{r-{1\over2}}C_{k'}{}^{q-{1\over2}}C_j{}^{q-{1\over2}}C_{j'}a^{q+r-j-k-1}b^{q+r-j'-k'-1}t_2\cr
&&\qquad(-\e^{-+}\e^{AB}\d_{k+k'-r+j+j'-q+1,0})\cr
&&+ \sum_{k,k',j,j'\geq 0}{}^{r-{1\over2}}C_k{}^{r-{1\over2}}C_{k'}{}^{q-{1\over2}}C_j{}^{q-{1\over2}}C_{j'}a^{q+r-j-k-1}b^{q+r-j'-k'-1}t^2_2\cr
&&\qquad(-\e^{-+}\e^{AB}\d_{k+k'-r+j+j'-q,0})
\cr
\cr
\cr
&&\!\!\!\!\!\!=\e^{AB} \sum_{j,j',k\geq 0}{}^{q-{1\over2}}C_j{}^{q-{1\over2}}C_{j'}{}^{r-{1\over2}}C_k{}^{r-{1\over2}}C_{q+r-j-j'-k-2}a^{q+r-j-k-1}b^{k+j+1}\cr
&& \e^{AB}\sum_{j,j',k\geq 0}{}^{q-{1\over2}}C_j{}^{q-{1\over2}}C_{j'}{}^{r-{1\over2}}C_k{}^{r-{1\over2}}C_{q+r-j-j'-k-1}a^{q+r-j-k-1}b^{k+j}t_2\cr
&&- \e^{AB}\sum_{j,j',k\geq 0}{}^{q-{1\over2}}C_j{}^{q-{1\over2}}C_{j'}{}^{r-{1\over2}}C_k{}^{r-{1\over2}}C_{q+r-j-j'-k-1}a^{q+r-j-k-1}b^{k+j}t_2\cr
&&- \e^{AB}\sum_{j,j',k\geq 0}{}^{q-{1\over2}}C_j{}^{q-{1\over2}}C_{j'}{}^{r-{1\over2}}C_k{}^{r-{1\over2}}C_{q+r-j-j'-k}a^{q+r-j-k-1}b^{j+k-1}t^2_2
\eea
Imposing constraints on the indices we have
\bea
\text{Term 1}&:&  j+j'-q+1 < 0\to j' <  q-j-1\cr
&&j'\geq 0 \to j < q - 1\cr
&&k' = q+r-j-j'-k-2\cr
&&k'\geq 0 \to k\leq q+r-j-j'-2
\cr\cr
\text{Term 2}&:&  j+j'-q < 0\to j' <  q-j\cr
&&j'\geq 0 \to j < q \cr
&&k' = q+r-j-j'-k-1\cr
&&k'\geq 0 \to k\leq q+r-j-j'-1
\cr\cr
\text{Term 3}&:& j+j'-q+1 < 0\to j' <  q-j-1\cr
&&j'\geq 0 \to j < q - 1\cr
&&k' = q+r-j-j'-k-1\cr
&&k'\geq 0 \to k\leq q-r-j-j'-1
\cr\cr
\text{Term 4}&:& j+j'-q < 0\to j' <  q-j\cr
&&j'\geq 0 \to j < q \cr
&&k' = q+r-j-j'-k\cr
&&k'\geq 0 \to k\leq q+r-j-j'
\eea
Implementing these constraints and taking the charge combination $B=-$ and $A=+$ yield the expression
\bea
&&\!\!\!\!\!\!\!\!\!\!\!\!(d'^{(2)f,+ -}_q,d'^{(2)f,- +}_r)\cr
&&\!\!\!\!\!\!=- \sum_{j=0}^{\lfloor q-1\rfloor}\sum_{j'=0}^{\lfloor q-j-1\rfloor}\sum_{k = 0}^{q+r-j-j'-2}{}^{q-{1\over2}}C_j{}^{q-{1\over2}}C_{j'}{}^{r-{1\over2}}C_k{}^{r-{1\over2}}C_{q+r-j-j'-k-2}a^{q+r-j-k-1}b^{k+j+1}\cr
&& -\sum_{j=0}^{\lfloor q\rfloor}\sum_{j'=0}^{\lfloor q-j\rfloor}\sum_{k = 0}^{q+r-j-j'-1}{}^{q-{1\over2}}C_j{}^{q-{1\over2}}C_{j'}{}^{r-{1\over2}}C_k{}^{r-{1\over2}}C_{q+r-j-j'-k-1}a^{q+r-j-k-1}b^{k+j}t_2\cr
&&+ \sum_{j=0}^{\lfloor q-1\rfloor}\sum_{j'=0}^{\lfloor q-j-1\rfloor}\sum_{k = 0}^{q+r-j-j'-1}{}^{q-{1\over2}}C_j{}^{q-{1\over2}}C_{j'}{}^{r-{1\over2}}C_k{}^{r-{1\over2}}C_{q+r-j-j'-k-1}a^{q+r-j-k-1}b^{k+j}t_2\cr
&&+ \sum_{j=0}^{\lfloor q\rfloor}\sum_{j'=0}^{\lfloor q-j\rfloor}\sum_{k = 0}^{q+r-j-j'}{}^{q-{1\over2}}C_j{}^{q-{1\over2}}C_{j'}{}^{r-{1\over2}}C_k{}^{r-{1\over2}}C_{q+r-j-j'-k}a^{q+r-j-k-1}b^{j+k-1}t^2_2\nn
\eea
and similarly
\bea
&&\!\!\!\!\!\!\!\!\!\!\!\!(d'^{(2)f,- +}_q,d'^{(2)f,+ -}_r)\cr
&&\!\!\!\!\!\!=-\ \sum_{j=0}^{\lfloor q-1\rfloor}\sum_{j'=0}^{\lfloor q-j-1\rfloor}\sum_{k = 0}^{q+r-j-j'-2}{}^{q-{1\over2}}C_j{}^{q-{1\over2}}C_{j'}{}^{r-{1\over2}}C_k{}^{r-{1\over2}}C_{q+r-j-j'-k-2}a^{q+r-j-k-1}b^{k+j+1}\cr
&&+ \sum_{j=0}^{\lfloor q\rfloor}\sum_{j'=0}^{\lfloor q-j\rfloor}\sum_{k = 0}^{q+r-j-j'-1}{}^{q-{1\over2}}C_j{}^{q-{1\over2}}C_{j'}{}^{r-{1\over2}}C_k{}^{r-{1\over2}}C_{q+r-j-j'-k-1}a^{q+r-j-k-1}b^{k+j}t_2\cr
&&- \sum_{j=0}^{\lfloor q-1\rfloor}\sum_{j'=0}^{\lfloor q-j-1\rfloor}\sum_{k = 0}^{q+r-j-j'-1}{}^{q-{1\over2}}C_j{}^{q-{1\over2}}C_{j'}{}^{r-{1\over2}}C_k{}^{r-{1\over2}}C_{q+r-j-j'-k-1}a^{q+r-j-k-1}b^{k+j}t_2\cr
&&+\sum_{j=0}^{\lfloor q\rfloor}\sum_{j'=0}^{\lfloor q-j\rfloor}\sum_{k = 0}^{q+r-j-j'}{}^{q-{1\over2}}C_j{}^{q-{1\over2}}C_{j'}{}^{r-{1\over2}}C_k{}^{r-{1\over2}}C_{q+r-j-j'-k}a^{q+r-j-k-1}b^{j+k-1}t^2_2
\eea
Mapping to the Ramond sector we shift our variables appropriately according to (\ref{mode shift}). This gives
\bea
&&\!\!\!\!\!\!\!\!\!\!\!\!(d'^{(2)f,+ -}_q,d'^{(2)f,- +}_r)\cr
&&\!\!\!\!\!\!=- \sum_{j=0}^{ q-2}\sum_{j'=0}^{ q-j-2}\sum_{k = 0}^{q+r-j-j'-2}{}^{q-1}C_j{}^{q-1}C_{j'}{}^{r}C_k{}^{r}C_{q+r-j-j'-k-2}a^{q+r-j-k-1}b^{k+j+1}\cr
&& -\sum_{j=0}^{ q-1}\sum_{j'=0}^{ q-j-1}\sum_{k = 0}^{q+r-j-j'-1}{}^{q-1}C_j{}^{q-1}C_{j'}{}^{r}C_k{}^{r}C_{q+r-j-j'-k-1}a^{q+r-j-k-1}b^{k+j}t_2\cr
&&+ \sum_{j=0}^{ q-2}\sum_{j'=0}^{ q-j-2}\sum_{k = 0}^{q+r-j-j'-1}{}^{q-1}C_j{}^{q-1}C_{j'}{}^{r}C_k{}^{r}C_{q+r-j-j'-k-1}a^{q+r-j-k-1}b^{k+j}t_2\cr
&&+ \sum_{j=0}^{ q-1}\sum_{j'=0}^{q-j-1}\sum_{k = 0}^{q+r-j-j'}{}^{q-1}C_j{}^{q-1}C_{j'}{}^{r}C_k{}^{r}C_{q+r-j-j'-k}a^{q+r-j-k-1}b^{j+k-1}t^2_2
\eea
and
\bea
&&\!\!\!\!\!\!\!\!\!\!\!\!(d'^{(2)f,- +}_q,d'^{(2)f,+ -}_r)\cr
&&\!\!\!\!\!\!=- \sum_{j=0}^{ q-1}\sum_{j'=0}^{ q-j-1}\sum_{k = 0}^{q+r-j-j'-2}{}^{q}C_j{}^{q}C_{j'}{}^{r-1}C_k{}^{r-1}C_{q+r-j-j'-k-2}a^{q+r-j-k-1}b^{k+j+1}\cr
&&+ \sum_{j=0}^{ q}\sum_{j'=0}^{ q-j}\sum_{k = 0}^{q+r-j-j'-1}{}^{q}C_j{}^{q}C_{j'}{}^{r-1}C_k{}^{r-1}C_{q+r-j-j'-k-1}a^{q+r-j-k-1}b^{k+j}t_2\cr
&&- \sum_{j=0}^{q-1}\sum_{j'=0}^{q-j-1}\sum_{k = 0}^{q+r-j-j'-1}{}^{q}C_j{}^{q}C_{j'}{}^{r-1}C_k{}^{r-1}C_{q+r-j-j'-k-1}a^{q+r-j-k-1}b^{k+j}t_2\cr
&&+ \sum_{j=0}^{q}\sum_{j'=0}^{ q-j}\sum_{k = 0}^{q+r-j-j'}{}^{q}C_j{}^{q}C_{j'}{}^{r-1}C_k{}^{r-1}C_{q+r-j-j'-k}a^{q+r-j-k-1}b^{j+k-1}t^2_2
\eea

Because of copy symmetry we have the relations
\bea
(d'^{(1)f,-+}_q ,d'^{(1)f,+-}_r) = (d'^{(2)f,-+}_q ,d'^{(2)f,+-}_r)
\eea

\subsection{Contraction between fermions at $0$ and $t_i$}
Here we compute the contraction between a point at $t=0$ (Copy 2 final) and $t_i$ (Fermion from $G$). We record the fermion expansions at $t=0$ below
\bea
d'^{(2)f,+ A}_r&=&-(\sum_{j,j'\geq 0}{}^{r-{1\over2}}C_j{}^{r-{1\over2}}C_{j'}a^{r-j-{1\over2}}b^{r-j'-{1\over2}}\tilde{d}^{+A,0}_{j+j'-r+1}\cr
&&\quad -\sum_{j,j'\geq 0}{}^{r-{1\over2}}C_j{}^{r-{1\over2}}C_{j'}a^{r-j-{1\over2}}b^{r-j'-{1\over2}}t_1\tilde{d}^{+A,0}_{j+j'-r})\cr
d'^{(2)f,- A}_r&=&-(\sum_{j,j'\geq 0}{}^{r-{1\over2}}C_j{}^{r-{1\over2}}C_{j'}a^{r-j-{1\over2}}b^{r-j'-{1\over2}}\tilde{d}^{-A,0}_{j+j'-r+1}\cr
&&\quad - \sum_{j,j'\geq 0}{}^{r-{1\over2}}C_j{}^{r-{1\over2}}C_{j'}a^{r-j-{1\over2}}b^{r-j'-{1\over2}}t_2\tilde{d}^{-A,0}_{j+j'-r})
\eea
In order to not annihilate the local vacuum, we must require that the mode indices obey the following relations
\bea
d'^{(2)f,+ A}_r,d'^{(2)f,- A}_r:~~\text{Term 1}&:& \,\,  j+j'-r+1<0\to j'< r-j-1\cr
&&j'\geq 0 \to j<r-1\cr
\cr
\text{Term 2 }&:& \,\,  j+j'-r<0\to j'< r-j\cr
&&j'\geq 0 \to j<r
\eea
Implementing these changes yield
\bea
d'^{(2)f,+ A}_r&=&-(\sum_{j = 0}^{\lfloor r-1\rfloor}\sum_{j' = 0}^{\lfloor r-j-1\rfloor}{}^{r-{1\over2}}C_j{}^{r-{1\over2}}C_{j'}a^{r-j-{1\over2}}b^{r-j'-{1\over2}}\tilde{d}^{+A,0}_{j+j'-r+1}\cr
&&\quad -\sum_{j = 0}^{\lfloor r\rfloor}\sum_{j' = 0}^{\lfloor r-j\rfloor}{}^{r-{1\over2}}C_j{}^{r-{1\over2}}C_{j'}a^{r-j-{1\over2}}b^{r-j'-{1\over2}}t_1\tilde{d}^{+A,0}_{j+j'-r})\cr
d'^{(2)f,- A}_r&=&-(\sum_{j = 0}^{\lfloor r-1\rfloor}\sum_{j' = 0}^{\lfloor r-j-1\rfloor}{}^{r-{1\over2}}C_j{}^{r-{1\over2}}C_{j'}a^{r-j-{1\over2}}b^{r-j'-{1\over2}}\tilde{d}^{-A,0}_{j+j'-r+1}\cr
&&\quad - \sum_{j = 0}^{\lfloor r\rfloor}\sum_{j' = 0}^{\lfloor r-j\rfloor}{}^{r-{1\over2}}C_j{}^{r-{1\over2}}C_{j'}a^{r-j-{1\over2}}b^{r-j'-{1\over2}}t_2\tilde{d}^{-A,0}_{j+j'-r})
\eea
We want to expand each of these modes around $t=t_i$. We first write modes natural to $t=0$ as
\bea
\tilde{d}^{-A,0}_{n} = {1\over 2\pi i}\int dt t^{n-{1\over2}}\psi^{-A}(t)
\eea
Expanding the integrand around $t=t_i$ gives
\bea
t^{n-{1\over2}}&=&(t-t_i +t_i)^{n-{1\over2}}\cr
&=&t_i^{n-{1\over2}}(1+t_i^{-1}(t-t_i))^{n-{1\over2}}\cr
&=&\sum_{l\geq0}{}^{n-{1\over2}}C_lt_i^{n-{1\over2}-l}(t-t_i)^l
\label{ti expansion}
\eea
Inserting these modes into the expansions above yield
\bea
d'^{(2)f,+ A}_r&=&\sum_{j = 0}^{\lfloor r-1\rfloor}\sum_{j' = 0}^{\lfloor r-j-1\rfloor}\sum_{l\geq0}{}^{r-{1\over2}}C_j{}^{r-{1\over2}}C_{j'}{}^{j+j'-r+{1\over2}}C_la^{r-j-{1\over2}}b^{r-j'-{1\over2}}t_i^{j+j'-r-l+{1\over2}}\tilde{d}^{+A,t_i}_{l+{1\over2}}\cr
&&\quad -\sum_{j = 0}^{\lfloor r\rfloor}\sum_{j' = 0}^{\lfloor r-j\rfloor}{}^{r-{1\over2}}C_j{}^{r-{1\over2}}C_{j'}{}^{j+j'-r-{1\over2}}C_la^{r-j-{1\over2}}b^{r-j'-{1\over2}}t_1t_i^{j+j'-r-l-{1\over2}}\tilde{d}^{+A,t_i}_{l+{1\over2}}\cr
d'^{(2)f,- A}_r&=&\sum_{j = 0}^{\lfloor r-1\rfloor}\sum_{j' = 0}^{\lfloor r-j-1\rfloor}\sum_{l\geq0}{}^{r-{1\over2}}C_j{}^{r-{1\over2}}C_{j'}{}^{j+j'-r+{1\over2}}C_la^{r-j-{1\over2}}b^{r-j'-{1\over2}}t_i^{j+j'-r-l+{1\over2}}\tilde{d}^{-A,t_i}_{l+{1\over2}}\cr
&&\quad -\sum_{j = 0}^{\lfloor r\rfloor}\sum_{j' = 0}^{\lfloor r-j\rfloor}{}^{r-{1\over2}}C_j{}^{r-{1\over2}}C_{j'}{}^{j+j'-r-{1\over2}}C_la^{r-j-{1\over2}}b^{r-j'-{1\over2}}t_2t_i^{j+j'-r-l-{1\over2}}\tilde{d}^{-A,t_i}_{l+{1\over2}}\nn
\eea
Where the minus sign comes from reversing the direction contour. Now we compute the contraction with the mode at $t = t_i$. We have
\bea
(d'^{(2)f,+ B}_r, \tilde{d}^{-A,t_i}_{-{1\over2}} ) &=& \sum_{j = 0}^{\lfloor r-1\rfloor}\sum_{j' = 0}^{\lfloor r-j-1\rfloor}{}^{r-{1\over2}}C_j{}^{r-{1\over2}}C_{j'}{}^{j+j'-r+{1\over2}}C_la^{r-j-{1\over2}}b^{r-j'-{1\over2}}t_i^{j+j'-r-l + {1\over2}}\cr
&&\quad\lbrace\tilde{d}^{+B,t_i}_{l+{1\over2}},\tilde{d}^{-A,t_i}_{-{1\over2}}\rbrace\cr
&& -\sum_{j = 0}^{\lfloor r\rfloor}\sum_{j' = 0}^{\lfloor r-j\rfloor}{}^{r-{1\over2}}C_j{}^{r-{1\over2}}C_{j'}{}^{j+j'-r-{1\over2}}C_la^{r-j-{1\over2}}b^{r-j'-{1\over2}}t_1t_i^{j+j'-r-l-{1\over2}}\cr
&&\quad\lbrace\tilde{d}^{+B,t_i}_{l+{1\over2}},\tilde{d}^{-A,t_i}_{-{1\over2}}\rbrace \cr
&=&\sum_{j = 0}^{\lfloor r-1\rfloor}\sum_{j' = 0}^{\lfloor r-j-1\rfloor}{}^{r-{1\over2}}C_j{}^{r-{1\over2}}C_{j'}{}^{j+j'-r+{1\over2}}C_la^{r-j-{1\over2}}b^{r-j'-{1\over2}}t_i^{j+j'-r-l+{1\over2}}\cr
&&\quad(-\e^{+-}\e^{BA}\d_{l,0})\cr
&& -\sum_{j = 0}^{\lfloor r\rfloor}\sum_{j' = 0}^{\lfloor r-j\rfloor}{}^{r-{1\over2}}C_j{}^{r-{1\over2}}C_{j'}{}^{j+j'-r-{1\over2}}C_la^{r-j-{1\over2}}b^{r-j'-{1\over2}}t_1t_i^{j+j'-r-l-{1\over2}}\cr
&&\quad(-\e^{+-}\e^{BA}\d_{l,0})\cr
&=&\e^{BA}\sum_{j = 0}^{\lfloor r-1\rfloor}\sum_{j' = 0}^{\lfloor r-j-1\rfloor}{}^{r-{1\over2}}C_j{}^{r-{1\over2}}C_{j'}a^{r-j-{1\over2}}b^{r-j'-{1\over2}}t_i^{j+j'-r+{1\over2}}\cr
&&\quad -\e^{BA}\sum_{j = 0}^{\lfloor r\rfloor}\sum_{j' = 0}^{\lfloor r-j\rfloor}{}^{r-{1\over2}}C_j{}^{r-{1\over2}}C_{j'}a^{r-j-{1\over2}}b^{r-j'-{1\over2}}t_1t_i^{j+j'-r-{1\over2}}\nn
\eea
Picking $B=-$ and $A=+$ yields and selecting $t_1$ yields
\bea
(d'^{(2)f,+ -}_r, \tilde{d}^{-+,t_1}_{-{1\over2}} )&=&\sum_{j = 0}^{\lfloor r-1\rfloor}\sum_{j' = 0}^{\lfloor r-j-1\rfloor}{}^{r-{1\over2}}C_j{}^{r-{1\over2}}C_{j'}a^{r-j-{1\over2}}b^{r-j'-{1\over2}}t_i^{j+j'-r+{1\over2}}\cr
&&\quad -\sum_{j = 0}^{\lfloor r\rfloor}\sum_{j' = 0}^{\lfloor r-j\rfloor}{}^{r-{1\over2}}C_j{}^{r-{1\over2}}C_{j'}a^{r-j-{1\over2}}b^{r-j'-{1\over2}}t_1t_i^{j+j'-r-{1\over2}}\nn
\eea
Similarly, for the other charge combination we have and selecting $t_2$ yields
\bea
(d'^{(2)f,- B}_r, \tilde{d}^{+A,t_2}_{-{1\over2}} ) &=& -\e^{BA}\sum_{j = 0}^{\lfloor r-1\rfloor}\sum_{j' = 0}^{\lfloor r-j-1\rfloor}{}^{r-{1\over2}}C_j{}^{r-{1\over2}}C_{j'}a^{r-j-{1\over2}}b^{r-j'-{1\over2}}t_i^{j+j'-r+{1\over2}}\cr
&&\quad + \e^{BA}\sum_{j = 0}^{\lfloor r\rfloor}\sum_{j' = 0}^{\lfloor r-j\rfloor}{}^{r-{1\over2}}C_j{}^{r-{1\over2}}C_{j'}a^{r-j-{1\over2}}b^{r-j'-{1\over2}}t_2t_i^{j+j'-r-{1\over2}}\nn
\eea
Picking $B=+$ and $A=-$ and picking $t_2$ yields
\bea
(d'^{(2)f,- +}_r, \tilde{d}^{+-,t_2}_{-{1\over2}} ) &=& \sum_{j = 0}^{\lfloor r-1\rfloor}\sum_{j' = 0}^{\lfloor r-j-1\rfloor}{}^{r-{1\over2}}C_j{}^{r-{1\over2}}C_{j'}a^{r-j-{1\over2}}b^{r-j'-{1\over2}}t_i^{j+j'-r+{1\over2}}\cr
&&\quad - \sum_{j = 0}^{\lfloor r\rfloor}\sum_{j' = 0}^{\lfloor r-j\rfloor}{}^{r-{1\over2}}C_j{}^{r-{1\over2}}C_{j'}a^{r-j-{1\over2}}b^{r-j'-{1\over2}}t_2t_i^{j+j'-r-{1\over2}}\nn
\eea

For the Ramond sector we shift our mode indices according to (\ref{mode shift})
\bea
(d'^{(2)f,+ -}_r, \tilde{d}^{-+,t_1}_{-{1\over2}} )&=&\sum_{j = 0}^{r-2}\sum_{j' = 0}^{ r-j-2}  {}^{r-1}C_j{}^{r-1}C_{j'}a^{r-j-1}b^{r-j'-1} t_1^{j+j'-r+1} \cr
&&\quad-\sum_{j = 0}^{ r-1 }\sum_{j' = 0}^{ r-j-1} {}^{r-1}C_j{}^{r-1}C_{j'}a^{r-j-1}b^{r-j'-1} t_1^{j+j'-r+1}\nn    
\eea
Similarly for the other charge combination we have
\bea
(d'^{(2)f,- +}_r, \tilde{d}^{+-,t_2}_{-{1\over2}} ) &=& \sum_{j = 0}^{r-1}\sum_{j' = 0}^{r-j-1}  {}^{r}C_j{}^{r}C_{j'}a^{r-j}b^{r-j'} t_2^{j+j'-r}- \sum_{j = 0}^{ r}\sum_{j' = 0}^{ r-j} {}^{r}C_j{}^{r}C_{j'}a^{r-j}b^{r-j'} t_2^{j+j'-r}\nn
\eea
Because of copy symmetry we have the relation
\bea
(d'^{(1)f,+-}_q ,  \tilde{d}^{-+,t_1}_{-{1\over2}}) &=& -(d'^{(2)f,+-}_q ,  \tilde{d}^{-+,t_1}_{-{1\over2}})
\eea

\subsection{Contraction between fermions at $t_1$ and $t_2$}
Here we compute the contraction between a fermion at $t_1$ (Fermion from $\tilde{G}^-$) and $t_2$ (Fermion from $\tilde{G}^+$). We obtain
\bea
(\tilde{d}^{+C,t_2}_{-{1\over2}}\tilde{d}^{-A,t_1}_{-{1\over2}}) &=& \psi^{+C}(t_2)\psi^{-A}(t_1)\cr
&=&\e^{CA}{1\over t_2-t_1}
\eea

\subsection{Contractions tabulated}\label{all contractions}All fermion and boson wick contractions are tabulated below
\subsubsection{Bosons}
\bea
(\a'^{(1)f}_{A\dot{A},p},\a'^{(1)f}_{B\dot{B},q})&=& -\e_{AB}\e_{\dot{A}\dot{B}}\sum_{k=0}^{p-1}\sum_{k'=0}^{p-k-1}\sum_{j=0}^{p+q-k-k'}(p-k-k'){}^{p}C_{k}{}^{p}C_{k'}{}^{q}C_{j}{}^{q}C_{p+q-k-k'-j}\cr
&&\qquad\qquad a^{k+j}b^{p+q-k-j}
\cr
\cr
(\a'^{(1)f}_{B\dot{B},p},\tilde{\a}^{t_i}_{C\dot{C},-1})&=&-\e_{BC}\e_{\dot{B}\dot{C }}\sum_{k=0}^{p-1}\sum_{k'=0}^{p-k-1}(p-k-k'){}^{p}C_{k}{}^{p}C_{k'}a^kb^{k'}t_i^{p-k-k'-1}
\cr
\cr
(\a'^{(1)f}_{B\dot{B},p},\a'^{(1)i}_{A\dot{A},-n})&=&- \e_{AB}\e_{\dot{A}\dot{B}}\sum_{j= 0}^{n-1}\sum_{j'= \max{[0,n-p -j]}}^{n-j-1}\sum_{k=0}^{p-n +j+j'}\sum_{k'=0}^{p-n +j+j'-k}(p-k-k'){}^{p}C_{k}{}^{p}C_{k'}\cr
&&\qquad\quad{}^nC_{j}{}^{-n}C_{j'}{}^{-n+j+j'}C_{p-k-k' -n+j+j'}(-1)^{n-j}a^{p-k' +j'}b^{k'}(b-a)^{-n-j'}
\cr
\cr
(\a'^{(2)f}_{B\dot{B},p},\a'^{(2)f}_{A\dot{A},q}) &=& -\e_{AB}\e_{\dot{A}\dot{B}}\sum_{k=0}^{p-1}\sum_{k'=0}^{p-k-1}\sum_{j=0}^{p+q-k-k'}(p-k-k'){}^pC_k{}^pC_{k'}{}^qC_j{}^qC_{p+q-j-k-k'}\cr
&&\qquad\qquad a^{p+q-j-k}b^{j+k}
\cr\cr
(\a'^{(2)f}_{B\dot{B},p},\tilde{\a}^{t_i}_{A\dot{A},-1}) &=&-\e_{\dot{B}\dot{A}}\e_{BA}\sum_{k=0}^{p-1}\sum_{k'=0}^{p-k-1}{}^pC_k{}^pC_{k'} {}^{k+k'-p}C_1a^{p-k}b^{p-k'}t_i^{k+k'-p-1}
\cr\cr
(\a'^{(2)f}_{B\dot{B},p},\a'^{(2)i}_{A\dot{A},-n}) &=&-\e_{BA}\e_{\dot{B}\dot{A}}\sum_{k=0}^{p-1}\sum_{k'=0}^{p-k-1}\sum_{j= 0}^{n-1}\sum_{j'= 0}^{n-j-1}(n-j-j'){}^pC_k{}^pC_{k'} {}^{k+k'-p}C_{n-j-j'} \cr
&&\qquad\qquad{}^nC_{j}{}^{-n}C_{j'}(-1)^{p+j'+k+k'}b^{j'+k'}a^{p-k'}(a-b)^{-n-j'}
\cr
\cr
(\a'^{(2)f}_{B\dot{B},p},\a'^{(1)i}_{A\dot{A},-n})&=&-\e_{BA}\e_{\dot{B}\dot{A}}\sum_{k=0}^{p-1}\sum_{k'=0}^{p-k-1}\sum_{j= 0}^{n-1}\sum_{j'= 0}^{n-j-1}(n-j-j'){}^pC_k{}^pC_{k'} {}^{k+k'-p}C_{n-j-j'} \cr
&&\qquad\qquad{}^nC_{j}{}^{-n}C_{j'}(-1)^{p+j'+k+k'}a^{j'+k'}b^{p-k'}(b-a)^{-n-j'}
\cr\cr
(\a'^{(1)f}_{B\dot{B},p}\a'^{(2)i}_{A\dot{A},-n})&=&  - \e_{AB}\e_{\dot{A}\dot{B}}\sum_{j= 0}^{n-1}\sum_{j'= \max{[0,n-p -j]}}^{n-j-1}\sum_{k=0}^{p-n +j+j'}\sum_{k'=0}^{p-n +j+j'-k}(p-k-k'){}^{p}C_{k}{}^{p}C_{k'}\cr
&&\qquad\qquad{}^nC_{j}{}^{-n}C_{j'}{}^{-n+j+j'}C_{p-k-k' -n+j+j'}(-1)^{n-j}b^{p-k' +j'}a^{k'}\cr
&&\qquad\qquad(a-b)^{-n-j'}
\cr\cr
(\a'^{(1)i}_{B\dot{B},-n_1},\a'^{(1)i}_{A\dot{A},-n_2}) &=&-\e_{BA}\e_{\dot{B}\dot{A}}\sum_{j''= 0}^{n_2-1}\sum_{j'''= 0}^{n_2-j''-1} \sum_{j'= 0}^{n_1 + n _2 -j''-j'''}(n_2-j''-j'''){}^{n_1}C_{n_1 + n _2 -j'-j''-j'''}\cr
&&\qquad\qquad{}^{-n_1}C_{j'}{}^{n_2}C_{j''}{}^{-n_2}C_{j'''}(-a)^{j'+j'''}(b-a)^{-n_1-n_2-j'-j'''}
\cr
\cr
(\a'^{(2)i}_{B\dot{B},-n_1},\a'^{(2)i}_{A\dot{A},-n_2})&=&-\e_{BA}\e_{\dot{B}\dot{A}}\sum_{j''= 0}^{n_2-1}\sum_{j'''= 0}^{n_2-j''-1} \sum_{j'= 0}^{n_1 + n _2 -j''-j''}(n_2-j''-j'''){}^{n_1}C_{n_1 + n _2 -j'-j''-j'''}\cr
&&\qquad\qquad{}^{-n_1}C_{j'}{}^{n_2}C_{j''}{}^{-n_2}C_{j'''}(-b)^{j'+j'''}(a-b)^{-n_1-n_2-j'-j'''}
\cr
\cr
(\a'^{(2)i}_{B\dot{B},-n_1},\a'^{(1)i}_{A\dot{A},-n_2}) &=&  \e_{BA}\e_{\dot{B}\dot{A}}\sum_{j= 0}^{n_1-1}\sum_{j'= 0}^{n_1-j-1}\sum_{j'''= 0}^{n_2-1}\sum_{j''''= 0}^{n_2-j'''-1}(n_2-j'''-j''''){}^{n_1}C_{j}{}^{-n_1}C_{j'}\cr
&&\qquad\quad{}^{-n_1+j+j'}C_{n_2-j'''-j''''}{}^{n_2}C_{j'''}{}^{-n_2}C_{j''''}(-b)^{n_1-j}(a-b)^{-n_1-j'}\cr
&&\qquad\quad(b-a)^{-n_1-2n_2+j+j'+j'''}(-a)^{n_2-j'''}
\cr
\cr
(\a'^{(1)i}_{B\dot{B},-n_1},\a'^{(2)i}_{A\dot{A},-n_2})&=& \e_{BA}\e_{\dot{B}\dot{A}}\sum_{j= 0}^{n_1-1}\sum_{j'= 0}^{n_1-j-1}\sum_{j'''= 0}^{n_2-1}\sum_{j''''= 0}^{n_2-j'''-1}(n_2-j'''-j''''){}^{n_1}C_{j}{}^{-n_1}C_{j'}\cr
&&\qquad\quad{}^{-n_1+j+j'}C_{n_2-j'''-j''''}{}^{n_2}C_{j'''}{}^{-n_2}C_{j''''}(-a)^{n_1-j}\cr
&&\qquad\quad(b-a)^{-n_1-j'}(a-b)^{-n_1-2n_2+j+j'+j'''}(-b)^{n_2-j'''}
\cr
\cr
(\tilde{\a}^{t_i}_{C\dot{C},-1}\a'^{(1)i}_{A\dot{A},-n}) &=&\e_{CA}\e_{\dot{C}\dot{A}}\sum_{j= 0}^{n-1}\sum_{j'= 0}^{n-j-1}{}^nC_{j}{}^{-n}C_{j'}{}^{-n+j+j'}C_1(-a)^{n-j}(b-a)^{-n-j'}\cr
&&\qquad\quad(t_i+a)^{-n + j + j'-1}
\cr
\cr
(\tilde{\a}^{t_i}_{C\dot{C},-1}\a'^{(2)i}_{A\dot{A},-n})&=&\e_{CA}\e_{\dot{C}\dot{A}}\sum_{j= 0}^{n-1}\sum_{j'= 0}^{n-j-1}{}^nC_{j}{}^{-n}C_{j'}{}^{-n+j+j'}C_1(-b)^{n-j}(a-b)^{-n-j'}\cr
&&\qquad\quad(t_i+b)^{-n + j + j'-1}
\cr
\cr
(\tilde{\a}^{t_2}_{C\dot{C},-1}\tilde{\a}^{t_1}_{A\dot{A},-1})&=&-\e_{CA}\e_{\dot{C}\dot{A}}{1\over (t_2-t_1)^2}
\eea
The computation for the antiholomorphic contractions follow through in a similar way. The antiholomorphic expressions amount to changing $\Delta w\to \Delta \bar w$ in the above equations.

\subsubsection{Fermions}
The fermion wick contractions are given by
\bea
&&\!\!\!\!\!\!\!\!\!\!\!\!\!\!\!\!\!\!\!\!\!\!\!\!\!\!\!\!\!\!\!\!\!\!\!\!\!\!\!\!\!\!\!\!\!\!\!\!\!\!\!\!\!\!\!\!\!\!\!\!\!\!\!\!\!\!\text{\underline{NS sector}}\cr
&&\!\!\!\!\!\!\!\!\!\!\!\!\!\!\!\!\!\!\!\!\!\!\!\!\!\!\!\!\!\!\!\!\!\!\!\!\!\!\!\!\!\!\!\!\!\!\!\!\!\!\!\!\!\!\!\!\!\!\!\!\!\!\!\!\!\!(d'^{(1)f,+-}_q ,d'^{(1)f,- +}_r)\cr
&&\!\!\!\!\!\!\!\!\!\!\!\!\!\!\!\!\!\!\!\!\!\!\!\!\!\!\!\!\!\!\!\!\!\!\!\!\!\!\!\!\!\!\!\!\!\!\!\!\!\!\!\!\!\!\!\!\!\!=\sum_{j = 0}^{\lfloor q\rfloor}\sum_{j' = 0}^{\lfloor q-j\rfloor}\sum_{k= 0}^{q+r-j-j'}{}^{q-{1\over2}}C_j{}^{q-{1\over2}}C_{j'}{}^{r-{1\over2}}C_k{}^{r-{1\over2}}C_{q-j-j'+r-k}a^{j+k}b^{q+r-j-k}\cr
&&\!\!\!\!\!\!\!\!\!\!\!\!\!\!\!\!\!\!\!\!\!\!\!\!\!\!\!\!\!\!\!\!\!\!\!\!\!\!\!\!\!\!\!\!\!\!\!\!\!\!\!\!-\sum_{j = 0}^{\lfloor q\rfloor}\sum_{j' = 0}^{\lfloor q-j\rfloor}\sum_{k= 0}^{q+r-j-j'-1}{}^{q-{1\over2}}C_j{}^{q-{1\over2}}C_{j'}{}^{r-{1\over2}}C_k{}^{r-{1\over2}}C_{q-j-j'+r-k-1}a^{j+k}b^{q+r-j-k-1}t_2\cr
&&\!\!\!\!\!\!\!\!\!\!\!\!\!\!\!\!\!\!\!\!\!\!\!\!\!\!\!\!\!\!\!\!\!\!\!\!\!\!\!\!\!\!\!\!\!\!\!\!\!\!\!\!+\sum_{j = 0}^{\lfloor q-1\rfloor}\sum_{j' = 0}^{\lfloor q-j-1\rfloor}\sum_{k=0}^{q+r-j-j'-1}{}^{q-{1\over2}}C_j{}^{q-{1\over2}}C_{j'}{}^{r-{1\over2}}C_k{}^{r-{1\over2}}C_{q-j-j'+r-k-1}a^{j+k}b^{q+r-j-k-1}t_2\cr
&&\!\!\!\!\!\!\!\!\!\!\!\!\!\!\!\!\!\!\!\!\!\!\!\!\!\!\!\!\!\!\!\!\!\!\!\!\!\!\!\!\!\!\!\!\!\!\!\!\!\!\!\!-\sum_{j = 0}^{\lfloor q-1\rfloor}\sum_{j' = 0}^{\lfloor q-j-1\rfloor}\sum_{k= 0}^{q+r-j-j'-2}{}^{q-{1\over2}}C_j{}^{q-{1\over2}}C_{j'}{}^{r-{1\over2}}C_k{}^{r-{1\over2}}C_{q-j-j'+r-k-2}a^{j+k}b^{q+r-j-k-2}t_2^2
\cr
\cr
&&\!\!\!\!\!\!\!\!\!\!\!\!\!\!\!\!\!\!\!\!\!\!\!\!\!\!\!\!\!\!\!\!\!\!\!\!\!\!\!\!\!\!\!\!\!\!\!\!\!\!\!\!\!\!\!\!\!\!\!\!\!\!\!\!\!\!\text{\underline{R sector}}\cr
&&\!\!\!\!\!\!\!\!\!\!\!\!\!\!\!\!\!\!\!\!\!\!\!\!\!\!\!\!\!\!\!\!\!\!\!\!\!\!\!\!\!\!\!\!\!\!\!\!\!\!\!\!\!\!\!\!\!\!\!\!\!\!\!\!\!\!(d'^{(1)f,+-}_q ,d'^{(1)f,- +}_r)\cr
&&\!\!\!\!\!\!\!\!\!\!\!\!\!\!\!\!\!\!\!\!\!\!\!\!\!\!\!\!\!\!\!\!\!\!\!\!\!\!\!\!\!\!\!\!\!\!\!\!\!\!\!\!\!\!\!\!\!\!=\sum_{j = 0}^{q-1}\sum_{j' = 0}^{ q-j-1}\sum_{k= 0}^{q+r-j-j'}{}^{q-1}C_j{}^{q-1}C_{j'}{}^{r}C_k{}^{r}C_{q-j-j'+r-k}a^{j+k}b^{q+r-j-k}\cr
&&\!\!\!\!\!\!\!\!\!\!\!\!\!\!\!\!\!\!\!\!\!\!\!\!\!\!\!\!\!\!\!\!\!\!\!\!\!\!\!\!\!\!\!\!\!\!\!\!\!\!\!\!-\sum_{j = 0}^{ q-1}\sum_{j' = 0}^{ q-j-1}\sum_{k= 0}^{q+r-j-j'-1}{}^{q-1}C_j{}^{q-1}C_{j'}{}^{r}C_k{}^{r}C_{q-j-j'+r-k-1}a^{j+k}b^{q+r-j-k-1}t_2\cr
&&\!\!\!\!\!\!\!\!\!\!\!\!\!\!\!\!\!\!\!\!\!\!\!\!\!\!\!\!\!\!\!\!\!\!\!\!\!\!\!\!\!\!\!\!\!\!\!\!\!\!\!\!+\sum_{j = 0}^{ q-2}\sum_{j' = 0}^{q-j-2}\sum_{k=0}^{q+r-j-j'-1}{}^{q-1}C_j{}^{q-1}C_{j'}{}^{r}C_k{}^{r}C_{q-j-j'+r-k-1}a^{j+k}b^{q+r-j-k-1}t_2\cr
&&\!\!\!\!\!\!\!\!\!\!\!\!\!\!\!\!\!\!\!\!\!\!\!\!\!\!\!\!\!\!\!\!\!\!\!\!\!\!\!\!\!\!\!\!\!\!\!\!\!\!\!\!-\sum_{j = 0}^{q-2}\sum_{j' = 0}^{ q-j-2}\sum_{k= 0}^{q+r-j-j'-2}{}^{q-1}C_j{}^{q-1}C_{j'}{}^{r}C_k{}^{r}C_{q-j-j'+r-k-2}a^{j+k}b^{q+r-j-k-2}t_2^2
\cr
\cr
\cr
&&\!\!\!\!\!\!\!\!\!\!\!\!\!\!\!\!\!\!\!\!\!\!\!\!\!\!\!\!\!\!\!\!\!\!\!\!\!\!\!\!\!\!\!\!\!\!\!\!\!\!\!\!\!\!\!\!\!\!\!\!\!\!\!\!\!\!\text{\underline{NS sector}}\cr
&&\!\!\!\!\!\!\!\!\!\!\!\!\!\!\!\!\!\!\!\!\!\!\!\!\!\!\!\!\!\!\!\!\!\!\!\!\!\!\!\!\!\!\!\!\!\!\!\!\!\!\!\!\!\!\!\!\!\!\!\!\!\!\!\!\!\!(d'^{(1)f,-+}_q ,d'^{(1)f,+-}_r)\cr
&&\!\!\!\!\!\!\!\!\!\!\!\!\!\!\!\!\!\!\!\!\!\!\!\!\!\!\!\!\!\!\!\!\!\!\!\!\!\!\!\!\!\!\!\!\!\!\!\!\!\!\!\!=\sum_{j = 0}^{\lfloor q\rfloor}\sum_{j' = 0}^{\lfloor q-j\rfloor}\sum_{k= 0}^{q+r-j-j'}{}^{q-{1\over2}}C_j{}^{q-{1\over2}}C_{j'}{}^{r-{1\over2}}C_k{}^{r-{1\over2}}C_{q-j-j'+r-k}a^{j+k}b^{q+r-j-k}\cr
&&\!\!\!\!\!\!\!\!\!\!\!\!\!\!\!\!\!\!\!\!\!\!\!\!\!\!\!\!\!\!\!\!\!\!\!\!\!\!\!\!\!\!\!\!\!\!\!\!\!\!\!\!+\sum_{j = 0}^{\lfloor q\rfloor}\sum_{j' = 0}^{\lfloor q-j\rfloor}\sum_{k= 0}^{q+r-j-j'-1}{}^{q-{1\over2}}C_j{}^{q-{1\over2}}C_{j'}{}^{r-{1\over2}}C_k{}^{r-{1\over2}}C_{q-j-j'+r-k-1}a^{j+k}b^{q+r-j-k-1}t_2\cr
&&\!\!\!\!\!\!\!\!\!\!\!\!\!\!\!\!\!\!\!\!\!\!\!\!\!\!\!\!\!\!\!\!\!\!\!\!\!\!\!\!\!\!\!\!\!\!\!\!\!\!\!\!-\sum_{j = 0}^{\lfloor q-1\rfloor}\sum_{j' = 0}^{\lfloor q-j-1\rfloor}\sum_{k=0}^{q+r-j-j'-1}{}^{q-{1\over2}}C_j{}^{q-{1\over2}}C_{j'}{}^{r-{1\over2}}C_k{}^{r-{1\over2}}C_{q-j-j'+r-k-1}a^{j+k}b^{q+r-j-k-1}t_2\cr
&&\!\!\!\!\!\!\!\!\!\!\!\!\!\!\!\!\!\!\!\!\!\!\!\!\!\!\!\!\!\!\!\!\!\!\!\!\!\!\!\!\!\!\!\!\!\!\!\!\!\!\!\!-\sum_{j = 0}^{\lfloor q-1\rfloor}\sum_{j' = 0}^{\lfloor q-j-1\rfloor}\sum_{k= 0}^{q+r-j-j'-2}{}^{q-{1\over2}}C_j{}^{q-{1\over2}}C_{j'}{}^{r-{1\over2}}C_k{}^{r-{1\over2}}C_{q-j-j'+r-k-2}a^{j+k}b^{q+r-j-k-2}t_2^2
\cr
\cr
&&\!\!\!\!\!\!\!\!\!\!\!\!\!\!\!\!\!\!\!\!\!\!\!\!\!\!\!\!\!\!\!\!\!\!\!\!\!\!\!\!\!\!\!\!\!\!\!\!\!\!\!\!\!\!\!\!\!\!\!\!\!\!\!\!\!\!\text{\underline{R sector}}\cr
&&\!\!\!\!\!\!\!\!\!\!\!\!\!\!\!\!\!\!\!\!\!\!\!\!\!\!\!\!\!\!\!\!\!\!\!\!\!\!\!\!\!\!\!\!\!\!\!\!\!\!\!\!\!\!\!\!\!\!\!\!\!\!\!\!\!\!(d'^{(1)f,-+}_q ,d'^{(1)f,+-}_r)\cr
&&\!\!\!\!\!\!\!\!\!\!\!\!\!\!\!\!\!\!\!\!\!\!\!\!\!\!\!\!\!\!\!\!\!\!\!\!\!\!\!\!\!\!\!\!\!\!\!\!\!\!\!\!=\sum_{j = 0}^{ q}\sum_{j' = 0}^{ q-j}\sum_{k= 0}^{q+r-j-j'}{}^{q}C_j{}^{q}C_{j'}{}^{r-1}C_k{}^{r-1}C_{q-j-j'+r-k}a^{j+k}b^{q+r-j-k}\cr
&&\!\!\!\!\!\!\!\!\!\!\!\!\!\!\!\!\!\!\!\!\!\!\!\!\!\!\!\!\!\!\!\!\!\!\!\!\!\!\!\!\!\!\!\!\!\!\!\!\!\!\!\!+\sum_{j = 0}^{ q}\sum_{j' = 0}^{ q-j}\sum_{k= 0}^{q+r-j-j'-1}{}^{q}C_j{}^{q}C_{j'}{}^{r-1}C_k{}^{r-1}C_{q-j-j'+r-k-1}a^{j+k}b^{q+r-j-k-1}t_2\cr
&&\!\!\!\!\!\!\!\!\!\!\!\!\!\!\!\!\!\!\!\!\!\!\!\!\!\!\!\!\!\!\!\!\!\!\!\!\!\!\!\!\!\!\!\!\!\!\!\!\!\!\!\!-\sum_{j = 0}^{q-1}\sum_{j' = 0}^{ q-j-1}\sum_{k=0}^{q+r-j-j'-1}{}^{q}C_j{}^{q}C_{j'}{}^{r-1}C_k{}^{r-1}C_{q-j-j'+r-k-1}a^{j+k}b^{q+r-j-k-1}t_2\cr
&&\!\!\!\!\!\!\!\!\!\!\!\!\!\!\!\!\!\!\!\!\!\!\!\!\!\!\!\!\!\!\!\!\!\!\!\!\!\!\!\!\!\!\!\!\!\!\!\!\!\!\!\!-\sum_{j = 0}^{q-1}\sum_{j' = 0}^{ q-j-1}\sum_{k= 0}^{q+r-j-j'-2}{}^{q}C_j{}^{q}C_{j'}{}^{r-1}C_k{}^{r-1}C_{q-j-j'+r-k-2}a^{j+k}b^{q+r-j-k-2}t_2^2
\cr
\cr
&&\!\!\!\!\!\!\!\!\!\!\!\!\!\!\!\!\!\!\!\!\!\!\!\!\!\!\!\!\!\!\!\!\!\!\!\!\!\!\!\!\!\!\!\!\!\!\!\!\!\!\!\!\!\!\!\!\!\!\!\!\!\!\!\!\!\!\text{\underline{NS sector}}\cr
(d'^{(1)f,- +}_r ,\tilde{d}^{+-,t_2}_{-{1\over2}} ) &=&\sum_{j = 0}^{\lfloor r\rfloor}\sum_{j' = 0}^{\lfloor r-j\rfloor}{}^{r-{1\over2}}C_j{}^{r-{1\over2}}C_{j'}a^jb^{j'}t_2^{r-j-j'-{1\over2}}\cr
&&\qquad  -  \sum_{j = 0}^{\lfloor r-1\rfloor}\sum_{j' = 0}^{\lfloor r-j-1\rfloor}{}^{q-{1\over2}}C_j{}^{r-{1\over2}}C_{j'}a^jb^{j'}t_2^{r-j-j'-{1\over2}}
\cr
\cr
&&\!\!\!\!\!\!\!\!\!\!\!\!\!\!\!\!\!\!\!\!\!\!\!\!\!\!\!\!\!\!\!\!\!\!\!\!\!\!\!\!\!\!\!\!\!\!\!\!\!\!\!\!\!\!\!\!\!\!\!\!\!\!\!\!\!\!\text{\underline{R sector}}\cr
(d'^{(1)f,- +}_r ,\tilde{d}^{+-,t_2}_{-{1\over2}} )&=&\sum_{j = 0}^{ r}\sum_{j' = 0}^{ r-j}{}^{r}C_j{}^{r}C_{j'}a^jb^{j'}t_2^{r-j-j'}  -  \sum_{j = 0}^{ r-1}\sum_{j' = 0}^{r-j-1}{}^{r}C_j{}^{r}C_{j'}a^jb^{j'}t_2^{r-j-j'}
\cr
\cr
&&\!\!\!\!\!\!\!\!\!\!\!\!\!\!\!\!\!\!\!\!\!\!\!\!\!\!\!\!\!\!\!\!\!\!\!\!\!\!\!\!\!\!\!\!\!\!\!\!\!\!\!\!\!\!\!\!\!\!\!\!\!\!\!\!\!\!\text{\underline{NS sector}}\cr
(d'^{(1)f,+-}_q ,  \tilde{d}^{-+,t_1}_{-{1\over2}})&=&\sum_{j = 0}^{\lfloor q\rfloor}\sum_{j' = 0}^{\lfloor q-j\rfloor}{}^{q-{1\over2}}C_j{}^{q-{1\over2}}C_{j'}a^jb^{j'}t_1^{q-j-j'-{1\over2}}\cr
&&\qquad  -  \sum_{j = 0}^{\lfloor q-1\rfloor}\sum_{j' = 0}^{\lfloor q-j-1\rfloor}{}^{q-{1\over2}}C_j{}^{q-{1\over2}}C_{j'}a^jb^{j'}t_1^{q-j-j'-{1\over2}}
\cr
\cr
&&\!\!\!\!\!\!\!\!\!\!\!\!\!\!\!\!\!\!\!\!\!\!\!\!\!\!\!\!\!\!\!\!\!\!\!\!\!\!\!\!\!\!\!\!\!\!\!\!\!\!\!\!\!\!\!\!\!\!\!\!\!\!\!\!\!\!\text{\underline{R sector}}\cr
(d'^{(1)f,+-}_q ,  \tilde{d}^{-+,t_1}_{-{1\over2}})&=&\sum_{j = 0}^{q-1}\sum_{j' = 0}^{ q-j-1}{}^{q-1}C_j{}^{q-1}C_{j'}a^jb^{j'}t_1^{q-j-j'-1} \cr
&&\qquad -  \sum_{j = 0}^{q-2 }\sum_{j' = 0}^{q-j-2}{}^{q-1}C_j{}^{q-1}C_{j'}a^jb^{j'}t_1^{q-j-j'-1}
\cr
\cr
&&\!\!\!\!\!\!\!\!\!\!\!\!\!\!\!\!\!\!\!\!\!\!\!\!\!\!\!\!\!\!\!\!\!\!\!\!\!\!\!\!\!\!\!\!\!\!\!\!\!\!\!\!\!\!\!\!\!\!\!\!\!\!\!\!\!\!\text{\underline{NS sector}}\cr
&&\!\!\!\!\!\!\!\!\!\!\!\!\!\!\!\!\!\!\!\!\!\!\!\!\!\!\!\!\!\!\!\!\!\!\!\!\!\!\!\!\!\!\!\!\!\!\!\!\!\!\!\!\!\!\!\!\!\!\!\!\!\!\!\!\!\!(d'^{(2)f,+ -}_q,d'^{(2)f,- +}_r)\cr
&&\!\!\!\!\!\!\!\!\!\!\!\!\!\!\!\!\!\!\!\!\!\!\!\!\!\!\!\!\!\!\!\!\!\!\!\!\!\!\!\!\!\!\!\!\!\!\!\!\!\!\!\!\!\!\!\!\!\!=- \sum_{j=0}^{\lfloor q-1\rfloor}\sum_{j'=0}^{\lfloor q-j-1\rfloor}\sum_{k = 0}^{q+r-j-j'-2}{}^{q-{1\over2}}C_j{}^{q-{1\over2}}C_{j'}{}^{r-{1\over2}}C_k{}^{r-{1\over2}}C_{q+r-j-j'-k-2}a^{q+r-j-k-1}b^{k+j+1}\cr
&&\!\!\!\!\!\!\!\!\!\!\!\!\!\!\!\!\!\!\!\!\!\!\!\!\!\!\!\!\!\!\!\!\!\!\!\!\!\!\!\!\!\!\!\!\!\!\!\!\!\!\!\!-\sum_{j=0}^{\lfloor q\rfloor}\sum_{j'=0}^{\lfloor q-j\rfloor}\sum_{k = 0}^{q+r-j-j'-1}{}^{q-{1\over2}}C_j{}^{q-{1\over2}}C_{j'}{}^{r-{1\over2}}C_k{}^{r-{1\over2}}C_{q+r-j-j'-k-1}a^{q+r-j-k-1}b^{k+j}t_2\cr
&&\!\!\!\!\!\!\!\!\!\!\!\!\!\!\!\!\!\!\!\!\!\!\!\!\!\!\!\!\!\!\!\!\!\!\!\!\!\!\!\!\!\!\!\!\!\!\!\!\!\!\!\!+ \sum_{j=0}^{\lfloor q-1\rfloor}\sum_{j'=0}^{\lfloor q-j-1\rfloor}\sum_{k = 0}^{q+r-j-j'-1}{}^{q-{1\over2}}C_j{}^{q-{1\over2}}C_{j'}{}^{r-{1\over2}}C_k{}^{r-{1\over2}}C_{q+r-j-j'-k-1}a^{q+r-j-k-1}b^{k+j}t_2\cr
&&\!\!\!\!\!\!\!\!\!\!\!\!\!\!\!\!\!\!\!\!\!\!\!\!\!\!\!\!\!\!\!\!\!\!\!\!\!\!\!\!\!\!\!\!\!\!\!\!\!\!\!\!+ \sum_{j=0}^{\lfloor q\rfloor}\sum_{j'=0}^{\lfloor q-j\rfloor}\sum_{k = 0}^{q+r-j-j'}{}^{q-{1\over2}}C_j{}^{q-{1\over2}}C_{j'}{}^{r-{1\over2}}C_k{}^{r-{1\over2}}C_{q+r-j-j'-k}a^{q+r-j-k-1}b^{j+k-1}t^2_2
\cr
\cr
&&\!\!\!\!\!\!\!\!\!\!\!\!\!\!\!\!\!\!\!\!\!\!\!\!\!\!\!\!\!\!\!\!\!\!\!\!\!\!\!\!\!\!\!\!\!\!\!\!\!\!\!\!\!\!\!\!\!\!\!\!\!\!\!\!\!\!\text{\underline{R sector}}\cr
&&\!\!\!\!\!\!\!\!\!\!\!\!\!\!\!\!\!\!\!\!\!\!\!\!\!\!\!\!\!\!\!\!\!\!\!\!\!\!\!\!\!\!\!\!\!\!\!\!\!\!\!\!\!\!\!\!\!\!\!\!\!\!\!\!\!\!(d'^{(2)f,+ -}_q,d'^{(2)f,- +}_r)\cr
&&\!\!\!\!\!\!\!\!\!\!\!\!\!\!\!\!\!\!\!\!\!\!\!\!\!\!\!\!\!\!\!\!\!\!\!\!\!\!\!\!\!\!\!\!\!\!\!\!\!\!\!\!\!\!\!\!\!\!=- \sum_{j=0}^{ q-2}\sum_{j'=0}^{ q-j-2}\sum_{k = 0}^{q+r-j-j'-2}{}^{q-1}C_j{}^{q-1}C_{j'}{}^{r}C_k{}^{r}C_{q+r-j-j'-k-2}a^{q+r-j-k-1}b^{k+j+1}\cr
&&\!\!\!\!\!\!\!\!\!\!\!\!\!\!\!\!\!\!\!\!\!\!\!\!\!\!\!\!\!\!\!\!\!\!\!\!\!\!\!\!\!\!\!\!\!\!\!\!\!\!\!\!  -\sum_{j=0}^{ q-1}\sum_{j'=0}^{ q-j-1}\sum_{k = 0}^{q+r-j-j'-1}{}^{q-1}C_j{}^{q-1}C_{j'}{}^{r}C_k{}^{r}C_{q+r-j-j'-k-1}a^{q+r-j-k-1}b^{k+j}t_2\cr
&&\!\!\!\!\!\!\!\!\!\!\!\!\!\!\!\!\!\!\!\!\!\!\!\!\!\!\!\!\!\!\!\!\!\!\!\!\!\!\!\!\!\!\!\!\!\!\!\!\!\!\!\! + \sum_{j=0}^{ q-2}\sum_{j'=0}^{ q-j-2}\sum_{k = 0}^{q+r-j-j'-1}{}^{q-1}C_j{}^{q-1}C_{j'}{}^{r}C_k{}^{r}C_{q+r-j-j'-k-1}a^{q+r-j-k-1}b^{k+j}t_2\cr
&&\!\!\!\!\!\!\!\!\!\!\!\!\!\!\!\!\!\!\!\!\!\!\!\!\!\!\!\!\!\!\!\!\!\!\!\!\!\!\!\!\!\!\!\!\!\!\!\!\!\!\!\! + \sum_{j=0}^{ q-1}\sum_{j'=0}^{q-j-1}\sum_{k = 0}^{q+r-j-j'}{}^{q-1}C_j{}^{q-1}C_{j'}{}^{r}C_k{}^{r}C_{q+r-j-j'-k}a^{q+r-j-k-1}b^{j+k-1}t^2_2\cr
\cr
&&\!\!\!\!\!\!\!\!\!\!\!\!\!\!\!\!\!\!\!\!\!\!\!\!\!\!\!\!\!\!\!\!\!\!\!\!\!\!\!\!\!\!\!\!\!\!\!\!\!\!\!\!\!\!\!\!\!\!\!\!\!\!\!\!\!\!\text{\underline{NS sector}}\cr
&&\!\!\!\!\!\!\!\!\!\!\!\!\!\!\!\!\!\!\!\!\!\!\!\!\!\!\!\!\!\!\!\!\!\!\!\!\!\!\!\!\!\!\!\!\!\!\!\!\!\!\!\!\!\!\!\!\!\!\!\!\!\!\!\!\!\!(d'^{(2)f,- +}_q,d'^{(2)f,+ -}_r)\cr
&&\!\!\!\!\!\!\!\!\!\!\!\!\!\!\!\!\!\!\!\!\!\!\!\!\!\!\!\!\!\!\!\!\!\!\!\!\!\!\!\!\!\!\!\!\!\!\!\!\!\!\!\!\!\!\!\!\!\!=-\ \sum_{j=0}^{\lfloor q-1\rfloor}\sum_{j'=0}^{\lfloor q-j-1\rfloor}\sum_{k = 0}^{q+r-j-j'-2}{}^{q-{1\over2}}C_j{}^{q-{1\over2}}C_{j'}{}^{r-{1\over2}}C_k{}^{r-{1\over2}}C_{q+r-j-j'-k-2}a^{q+r-j-k-1}b^{k+j+1}\cr
&&\!\!\!\!\!\!\!\!\!\!\!\!\!\!\!\!\!\!\!\!\!\!\!\!\!\!\!\!\!\!\!\!\!\!\!\!\!\!\!\!\!\!\!\!\!\!\!\!\!\!\!\!+ \sum_{j=0}^{\lfloor q\rfloor}\sum_{j'=0}^{\lfloor q-j\rfloor}\sum_{k = 0}^{q+r-j-j'-1}{}^{q-{1\over2}}C_j{}^{q-{1\over2}}C_{j'}{}^{r-{1\over2}}C_k{}^{r-{1\over2}}C_{q+r-j-j'-k-1}a^{q+r-j-k-1}b^{k+j}t_2\cr
&&\!\!\!\!\!\!\!\!\!\!\!\!\!\!\!\!\!\!\!\!\!\!\!\!\!\!\!\!\!\!\!\!\!\!\!\!\!\!\!\!\!\!\!\!\!\!\!\!\!\!\!\!- \sum_{j=0}^{\lfloor q-1\rfloor}\sum_{j'=0}^{\lfloor q-j-1\rfloor}\sum_{k = 0}^{q+r-j-j'-1}{}^{q-{1\over2}}C_j{}^{q-{1\over2}}C_{j'}{}^{r-{1\over2}}C_k{}^{r-{1\over2}}C_{q+r-j-j'-k-1}a^{q+r-j-k-1}b^{k+j}t_2\cr
&&\!\!\!\!\!\!\!\!\!\!\!\!\!\!\!\!\!\!\!\!\!\!\!\!\!\!\!\!\!\!\!\!\!\!\!\!\!\!\!\!\!\!\!\!\!\!\!\!\!\!\!\!+\sum_{j=0}^{\lfloor q\rfloor}\sum_{j'=0}^{\lfloor q-j\rfloor}\sum_{k = 0}^{q+r-j-j'}{}^{q-{1\over2}}C_j{}^{q-{1\over2}}C_{j'}{}^{r-{1\over2}}C_k{}^{r-{1\over2}}C_{q+r-j-j'-k}a^{q+r-j-k-1}b^{j+k-1}t^2_2
\cr\cr
&&\!\!\!\!\!\!\!\!\!\!\!\!\!\!\!\!\!\!\!\!\!\!\!\!\!\!\!\!\!\!\!\!\!\!\!\!\!\!\!\!\!\!\!\!\!\!\!\!\!\!\!\!\!\!\!\!\!\!\!\!\!\!\!\!\!\!\text{\underline{R sector}}\cr
&&\!\!\!\!\!\!\!\!\!\!\!\!\!\!\!\!\!\!\!\!\!\!\!\!\!\!\!\!\!\!\!\!\!\!\!\!\!\!\!\!\!\!\!\!\!\!\!\!\!\!\!\!\!\!\!\!\!\!\!\!\!\!\!\!\!\!(d'^{(2)f,- +}_q,d'^{(2)f,+ -}_r)\cr
&&\!\!\!\!\!\!\!\!\!\!\!\!\!\!\!\!\!\!\!\!\!\!\!\!\!\!\!\!\!\!\!\!\!\!\!\!\!\!\!\!\!\!\!\!\!\!\!\!\!\!\!\!\!\!\!\!\!\!=- \sum_{j=0}^{ q-1}\sum_{j'=0}^{ q-j-1}\sum_{k = 0}^{q+r-j-j'-2}{}^{q}C_j{}^{q}C_{j'}{}^{r-1}C_k{}^{r-1}C_{q+r-j-j'-k-2}a^{q+r-j-k-1}b^{k+j+1}\cr
&&\!\!\!\!\!\!\!\!\!\!\!\!\!\!\!\!\!\!\!\!\!\!\!\!\!\!\!\!\!\!\!\!\!\!\!\!\!\!\!\!\!\!\!\!\!\!\!\!\!\!\!\!+ \sum_{j=0}^{ q}\sum_{j'=0}^{ q-j}\sum_{k = 0}^{q+r-j-j'-1}{}^{q}C_j{}^{q}C_{j'}{}^{r-1}C_k{}^{r-1}C_{q+r-j-j'-k-1}a^{q+r-j-k-1}b^{k+j}t_2\cr
&&\!\!\!\!\!\!\!\!\!\!\!\!\!\!\!\!\!\!\!\!\!\!\!\!\!\!\!\!\!\!\!\!\!\!\!\!\!\!\!\!\!\!\!\!\!\!\!\!\!\!\!\! - \sum_{j=0}^{q-1}\sum_{j'=0}^{q-j-1}\sum_{k = 0}^{q+r-j-j'-1}{}^{q}C_j{}^{q}C_{j'}{}^{r-1}C_k{}^{r-1}C_{q+r-j-j'-k-1}a^{q+r-j-k-1}b^{k+j}t_2\cr
&&\!\!\!\!\!\!\!\!\!\!\!\!\!\!\!\!\!\!\!\!\!\!\!\!\!\!\!\!\!\!\!\!\!\!\!\!\!\!\!\!\!\!\!\!\!\!\!\!\!\!\!\! + \sum_{j=0}^{q}\sum_{j'=0}^{ q-j}\sum_{k = 0}^{q+r-j-j'}{}^{q}C_j{}^{q}C_{j'}{}^{r-1}C_k{}^{r-1}C_{q+r-j-j'-k}a^{q+r-j-k-1}b^{j+k-1}t^2_2
\cr
\cr
&&\!\!\!\!\!\!\!\!\!\!\!\!\!\!\!\!\!\!\!\!\!\!\!\!\!\!\!\!\!\!\!\!\!\!\!\!\!\!\!\!\!\!\!\!\!\!\!\!\!\!\!\!\!\!\!\!\!\!\!\!\!\!\!\!\!\!\text{\underline{NS sector}}\cr
(d'^{(2)f,+ -}_r, \tilde{d}^{-+,t_1}_{-{1\over2}} )&=&\sum_{j = 0}^{\lfloor r-1\rfloor}\sum_{j' = 0}^{\lfloor r-j-1\rfloor}{}^{r-{1\over2}}C_j{}^{r-{1\over2}}C_{j'}a^{r-j-{1\over2}}b^{r-j'-{1\over2}}t_1^{j+j'-r+{1\over2}}\cr
&&\quad -\sum_{j = 0}^{\lfloor r\rfloor}\sum_{j' = 0}^{\lfloor r-j\rfloor}{}^{r-{1\over2}}C_j{}^{r-{1\over2}}C_{j'}a^{r-j-{1\over2}}b^{r-j'-{1\over2}}t_1^{j+j'-r+{1\over2}}\cr
&&\!\!\!\!\!\!\!\!\!\!\!\!\!\!\!\!\!\!\!\!\!\!\!\!\!\!\!\!\!\!\!\!\!\!\!\!\!\!\!\!\!\!\!\!\!\!\!\!\!\!\!\!\!\!\!\!\!\!\!\!\!\!\!\!\!\!\text{\underline{R sector}}\cr
(d'^{(2)f,+ -}_r, \tilde{d}^{-+,t_1}_{-{1\over2}} )&=&\sum_{j = 0}^{r-2}\sum_{j' = 0}^{ r-j-2}  {}^{r-1}C_j{}^{r-1}C_{j'}a^{r-j-1}b^{r-j'-1} t_1^{j+j'-r+1} \cr
&&\quad-\sum_{j = 0}^{ r-1 }\sum_{j' = 0}^{ r-j-1} {}^{r-1}C_j{}^{r-1}C_{j'}a^{r-j-1}b^{r-j'-1} t_1^{j+j'-r+1}\nn    
\cr\cr
&&\!\!\!\!\!\!\!\!\!\!\!\!\!\!\!\!\!\!\!\!\!\!\!\!\!\!\!\!\!\!\!\!\!\!\!\!\!\!\!\!\!\!\!\!\!\!\!\!\!\!\!\!\!\!\!\!\!\!\!\!\!\!\!\!\!\!\text{\underline{NS sector}}\cr
(d'^{(2)f,- +}_r, \tilde{d}^{+-,t_2}_{-{1\over2}} ) &=&\sum_{j = 0}^{\lfloor r-1\rfloor}\sum_{j' = 0}^{\lfloor r-j-1\rfloor}{}^{r-{1\over2}}C_j{}^{r-{1\over2}}C_{j'}a^{r-j-{1\over2}}b^{r-j'-{1\over2}}t_2^{j+j'-r+{1\over2}}\cr
&&\quad - \sum_{j = 0}^{\lfloor r\rfloor}\sum_{j' = 0}^{\lfloor r-j\rfloor}{}^{r-{1\over2}}C_j{}^{r-{1\over2}}C_{j'}a^{r-j-{1\over2}}b^{r-j'-{1\over2}}t_2^{j+j'-r+{1\over2}}\cr
&&\!\!\!\!\!\!\!\!\!\!\!\!\!\!\!\!\!\!\!\!\!\!\!\!\!\!\!\!\!\!\!\!\!\!\!\!\!\!\!\!\!\!\!\!\!\!\!\!\!\!\!\!\!\!\!\!\!\!\!\!\!\!\!\!\!\!\text{\underline{R sector}}\cr
(d'^{(2)f,- +}_r, \tilde{d}^{+-,t_2}_{-{1\over2}} ) &=&  \sum_{j = 0}^{r-1}\sum_{j' = 0}^{r-j-1}  {}^{r}C_j{}^{r}C_{j'}a^{r-j}b^{r-j'} t_2^{j+j'-r}- \sum_{j = 0}^{ r}\sum_{j' = 0}^{ r-j} {}^{r}C_j{}^{r}C_{j'}a^{r-j}b^{r-j'} t_2^{j+j'-r}\nn
\cr\cr
(\tilde{d}^{+C,t_2}_{-{1\over2}}\tilde{d}^{-A,t_1}_{-{1\over2}}) &=&\e^{CA}{1\over t_2-t_1}
\eea

Again, the computation for the antiholomorphic contractions follow through in a similar way. The antiholomorphic expressions amount to changing $\Delta w\to \Delta \bar w$ in the above expressions.
\subsubsection{Quantities which depend on the twist seperation's $\Delta w$ and $\Delta \bar w$}
We remind the reader that $a$, $b$, $t_1$ and $t_2$ are all functions of the twist separation $\Delta w$. Their expressions are
\bea
a &=& \cosh^2 ({\Delta w \over 4})\cr
b &=& \sinh^2 ({\Delta w \over 4})\cr
t_1 &=& -\sqrt{ab}= - \cosh ({\Delta w \over 4}) \sinh({\Delta w \over 4})\cr
t_2 &=& \sqrt{ab}= \cosh ({\Delta w \over 4}) \sinh({\Delta w \over 4})
\eea
Let's express the above quantities in terms of exponentials. This form will be useful in the following sections:
\bea
a &=& {1\over4}(e^{\Delta w\over4} + e^{-{\Delta w\over4}})^2= {1\over4}(e^{\Delta w\over2}+e^{-{\Delta w\over2}}+2)\cr
b &=& {1\over4}(e^{\Delta w\over4} - e^{-{\Delta w\over4}})^2={1\over4}(e^{\Delta w\over2}+e^{-{\Delta w\over2}}-2)\cr
t_1 &=& - {1\over4}(e^{\Delta w\over2} - e^{-{\Delta w\over2}} )\cr
t_2 &=& {1\over4}(e^{\Delta w\over2} - e^{-{\Delta w\over2}} )
\label{exponentials}
\eea
Similarly, the right moving quantities are
\bea
\bar{a} &=& {1\over4}(e^{\Delta \bar w\over4} + e^{-{\Delta \bar w\over4}})^2= {1\over4}(e^{\Delta \bar w\over2}+e^{-{\Delta \bar w\over2}}+2)\cr
\bar{b} &=& {1\over4}(e^{\Delta \bar w\over4} - e^{-{\Delta \bar w\over4}})^2={1\over4}(e^{\Delta \bar w\over2}+e^{-{\Delta \bar w\over2}}-2)\cr
\bar{t}_1 &=& - {1\over4}(e^{\Delta \bar w\over2} - e^{-{\Delta \bar w\over2}} )\cr
\bar{t}_2 &=& {1\over4}(e^{\Delta \bar w\over2} - e^{-{\Delta \bar w\over2}} )
\label{bar exponentials}
\eea

\section{Re-expressing our amplitudes}

Every amplitude recorded in section \ref{amplitudes} contains functions of the variables $a,b,t_1,t_2$ which are expressed in terms of exponentials of $\Delta w$ and $\bar a,\bar b,\bar t_1,\bar t_2$ which are expressed in terms of $\Delta \bar w$ as shown in (\ref{bar exponentials}). If we choose initial and final mode indices, enforce energy conservation, and perform the appropriate sums contained within the wick contractions recorded in subsection (\ref{all contractions}), the amplitudes computed in subsection's (\ref{subsection a to aaa}), (\ref{subsection a to add}), and (\ref{subsection aa to aadd}) will take the schematic form

\bea
\mathcal{A}^{i\to f}(w_1,w_2,\bar w_1,\bar w_2)=\sum_{m=m_{min}(n_{i};n_{f})}^{m_{max}(n_{i};n_{f})}\sum_{m'=m'_{min}(n_{i};n_{f})}^{m'_{max}(n_{i};n_{f})}B^{i\to f}_{m,m'}(n_i;n_f) e^{{m\Delta \bar w\over2}+{m'\Delta w\over2}}
\eea
where $i\to f$ represents the specific splitting process in consideration and $n_{initial},n_{final}$ represent the set of mode indices labeling final and initial states respectively. The coefficients, $B$ as indicated above, are dependent on the initial and final mode numbers. All of our amplitudes can be expressed in terms of sum over various powers of $e^{\Delta w\over2}$ and $e^{\Delta \bar w\over2}$ with coefficients. The upside is that this form simplifies the integration which we show in the next section. The downside is that we are required to pick specific values for the mode numbers in order to do the sums. As of yet, we have not been able to find a closed form solution for general initial and final mode numbers. To compute the above coefficients, $B^{i\to f}_{m,m'}(n_i;n_f)$, we use \textit{Mathematica} expand the various sums for specific values of the mode numbers. Lets do the simplest example for each of the three splitting processes.
\subsection{Computing the amplitude for the process $\a\to \a\a\a$ for specific indices}
Let us compute the simplest splitting amplitude for \textit{one} boson going to \textit{three} bosons. Taking the amplitude in (\ref{a to aaa full}) which is expressed in terms of wick contraction terms tabulated in (\ref{all contractions}), and selecting
\bea
n_f &=& \lbrace p=1,q=1,r=1\rbrace\cr
n_i&=&\lbrace n=3\rbrace
\eea
our amplitude becomes
\bea
&&\!\!\!\!\!\!\!\!\!\!\!\!\mathcal{A}^{\a\to \a\a\a}(w_1,w_2,\bar w_1,\bar w_2)\cr 
&&\!\!\!\!\!\!= \frac{75 e^{-\frac{5 \Delta w}{2}-\frac{5 \Delta \bar w}{2}}}{131072}-\frac{15 e^{-\frac{3 \Delta w}{2}-\frac{5 \Delta \bar w}{2}}}{65536}-\frac{15e^{-\frac{\Delta w}{2}-\frac{5 \Delta \bar w}{2}}}{32768}-\frac{45 e^{\frac{\Delta w}{2}-\frac{5 \Delta \bar w}{2}}}{65536}+\frac{105 e^{\frac{3 \Delta w}{2}-\frac{5\Delta \bar w}{2}}}{131072}
   \cr
   &&-\frac{15 e^{-\frac{5 \Delta w}{2}-\frac{3 \Delta \bar w}{2}}}{65536}+\frac{159 e^{-\frac{3 \Delta w}{2}-\frac{3 \Delta \bar w}{2}}}{131072}-\frac{51e^{-\frac{\Delta w}{2}-\frac{3 \Delta \bar w}{2}}}{65536}-\frac{3 e^{\frac{\Delta w}{2}-\frac{3 \Delta \bar w}{2}}}{8192}-\frac{21 e^{\frac{3 \Delta w}{2}-\frac{3\Delta \bar w}{2}}}{32768}
   \cr
   &&+\frac{9e^{\frac{\Delta w}{2}-\frac{\Delta \bar w}{2}}}{8192}-\frac{3 e^{\frac{3 \Delta w}{2}-\frac{\Delta \bar w}{2}}}{8192}-\frac{45 e^{\frac{5
   \Delta w}{2}-\frac{\Delta \bar w}{2}}}{65536}-\frac{45 e^{-\frac{5 \Delta w}{2}+\frac{\Delta \bar w}{2}}}{65536}-\frac{3 e^{-\frac{3 \Delta w}{2}+\frac{\Delta \bar w}{2}}}{8192}+\frac{9
   e^{-\frac{\Delta w}{2}+\frac{\Delta \bar w}{2}}}{8192}\cr
   &&+\frac{39 e^{\frac{\Delta w}{2}+\frac{\Delta \bar w}{2}}}{32768}+\frac{105 e^{\frac{5 \Delta w}{2}-\frac{3 \Delta \bar w}{2}}}{131072}-\frac{15 e^{-\frac{5 \Delta w}{2}-\frac{\Delta \bar w}{2}}}{32768} -\frac{51e^{-\frac{3 \Delta w}{2}-\frac{\Delta \bar w}{2}}}{65536}+\frac{39 e^{-\frac{\Delta w}{2}-\frac{\Delta \bar w}{2}}}{32768}
   \cr
   &&-\frac{51 e^{\frac{3\Delta w}{2}+\frac{\Delta \bar w}{2}}}{65536}-\frac{15 e^{\frac{5 \Delta w}{2}+\frac{\Delta \bar w}{2}}}{32768}+\frac{105 e^{-\frac{5
   \Delta w}{2}+\frac{3 \Delta \bar w}{2}}}{131072}-\frac{21 e^{-\frac{3 \Delta w}{2}+\frac{3 \Delta \bar w}{2}}}{32768}-\frac{3 e^{-\frac{\Delta w}{2}+\frac{3 \Delta \bar w}{2}}}{8192}
   \cr
   &&-\frac{51e^{\frac{\Delta w}{2}+\frac{3 \Delta \bar w}{2}}}{65536} + \frac{159 e^{\frac{3 \Delta w}{2}+\frac{3 \Delta \bar w}{2}}}{131072}-\frac{15 e^{\frac{5 \Delta w}{2}+\frac{3\Delta \bar w}{2}}}{65536}+\frac{105 e^{-\frac{3 \Delta w}{2}+\frac{5 \Delta \bar w}{2}}}{131072}-\frac{45 e^{-\frac{\Delta w}{2}+\frac{5 \Delta \bar w}{2}}}{65536}\cr
   &&-\frac{15e^{\frac{\Delta w}{2}+\frac{5 \Delta \bar w}{2}}}{32768}-\frac{15 e^{\frac{3 \Delta w}{2}+\frac{5 \Delta \bar w}{2}}}{65536}+\frac{75 e^{\frac{5 \Delta w}{2}+\frac{5
   \Delta \bar w}{2}}}{131072}
   \label{a to aaa 3 to 111}
\eea
where
\bea
m_{min}(n=3; p=1,q=1,r=1)=m'_{min}(n=3; p=1,q=1,r=1)&=&-5\cr
m_{max}(n=3; p=1,q=1,r=1)=m'_{max}(n=3; p=1,q=1,r=1)&=&5\nn
\eea
Writing (\ref{a to aaa 3 to 111}) as a sum over coefficients we have
\bea
\mathcal{A}^{\a\to \a\a\a}(w_1,w_2,\bar w_1,\bar w_2) = \sum_{m=-5}^{5} \sum_{m'=-5}^{5}B^{\a\to \a\a\a}_{m,m'}(n=3; p=1,q=1,r=1)e^{{m\Delta w\over2}+{m'\Delta \bar w\over2}}\nn
\eea
where
\bea
{39\over 32768}&=&B^{\a\to \a\a\a}_{1,1}(n=3; p=1,q=1,r=1)\cr
&=&B^{\a\to \a\a\a}_{-1,-1}(n=3; p=1,q=1,r=1)
\cr
\cr
\frac{9}{8192}&=&B^{\a\to \a\a\a}_{1,-1}(n=3; p=1,q=1,r=1)\cr
&=&B^{\a\to \a\a\a}_{-1,1}(n=3; p=1,q=1,r=1)
\cr
\cr
\frac{159}{131072}&=&B^{\a\to \a\a\a}_{3,3}(n=3; p=1,q=1,r=1)\cr
&=&B^{\a\to \a\a\a}_{-3,-3}(n=3; p=1,q=1,r=1)
\cr
\cr
-\frac{21}{32768}&=&B^{\a\to \a\a\a}_{3,-3}(n=3; p=1,q=1,r=1)\cr
&=&B^{\a\to \a\a\a}_{-3,3}(n=3; p=1,q=1,r=1)
\cr
\cr
{75\over131072}&=&B^{\a\to \a\a\a}_{5,5}(n=3; p=1,q=1,r=1)\cr
&=&B^{\a\to \a\a\a}_{-5,-5}(n=3; p=1,q=1,r=1)
\cr
\cr
-\frac{15 }{65536}&=&B^{\a\to \a\a\a}_{3,5}(n=3; p=1,q=1,r=1)\cr
&=&B^{\a\to \a\a\a}_{5,3}(n=3; p=1,q=1,r=1)\cr
&=&B^{\a\to \a\a\a}_{-3,-5}(n=3; p=1,q=1,r=1)\cr
&=&B^{\a\to \a\a\a}_{-5,-3}(n=3; p=1,q=1,r=1)
\cr
\cr
-\frac{15}{32768}&=&B^{\a\to \a\a\a}_{1,5}(n=3; p=1,q=1,r=1)\cr
&=&B^{\a\to \a\a\a}_{5,1}(n=3; p=1,q=1,r=1)\cr
&=&B^{\a\to \a\a\a}_{-1,-5}(n=3; p=1,q=1,r=1)\cr
&=&B^{\a\to \a\a\a}_{-5,-1}(n=3; p=1,q=1,r=1)
\cr
\cr
-\frac{45}{65536}&=&B^{\a\to \a\a\a}_{1,-5}(n=3; p=1,q=1,r=1)\cr
&=&B^{\a\to \a\a\a}_{-1,5}(n=3; p=1,q=1,r=1)\cr
&=&B^{\a\to \a\a\a}_{5,-1}(n=3; p=1,q=1,r=1)\cr
&=&B^{\a\to \a\a\a}_{-5,1}(n=3; p=1,q=1,r=1)
\cr
\cr
\frac{105}{131072}&=&B^{\a\to \a\a\a}_{3,-5}(n=3; p=1,q=1,r=1)\cr
&=&B^{\a\to \a\a\a}_{5,-3}(n=3; p=1,q=1,r=1)\cr
&=&B^{\a\to \a\a\a}_{-5,3}(n=3; p=1,q=1,r=1)\cr
&=&B^{\a\to \a\a\a}_{-3,5}(n=3; p=1,q=1,r=1)
\cr
\cr
-\frac{51}{65536}&=&B^{\a\to \a\a\a}_{-1,-3}(n=3; p=1,q=1,r=1)\cr
&=&B^{\a\to \a\a\a}_{-3,-1}(n=3; p=1,q=1,r=1)\cr
&=&B^{\a\to \a\a\a}_{3,1}(n=3; p=1,q=1,r=1)\cr
&=&B^{\a\to \a\a\a}_{1,3}(n=3; p=1,q=1,r=1)
\cr
\cr
-\frac{3}{8192}&=&B^{\a\to \a\a\a}_{1,-3}(n=3; p=1,q=1,r=1)\cr
&=&B^{\a\to \a\a\a}_{-1,3}(n=3; p=1,q=1,r=1)\cr
&=&B^{\a\to \a\a\a}_{3,-1}(n=3; p=1,q=1,r=1)\cr
&=&B^{\a\to \a\a\a}_{-3,1}(n=3; p=1,q=1,r=1)
\eea
We note that there are no terms for which $m,m'\in \mathbb{Z}_{\text{even}}$ so
\bea
B^{\a\to \a\a\a}_{m,m'}(n=3; p=1,q=1,r=1) = 0,\quad m,m'\in \mathbb{Z}_{\text{even}}
\eea
We also note the symmetry relations
\bea
B^{\a\to \a\a\a}_{m,-m}(n=3; p=1,q=1,r=1) &=&B^{\a\to \a\a\a}_{-m,m}(n=3; p=1,q=1,r=1)\cr
B^{\a\to \a\a\a}_{m,m}(n=3; p=1,q=1,r=1)&=&B^{\a\to \a\a\a}_{-m,-m}(n=3; p=1,q=1,r=1)\nn
\eea

\subsection{Computing the amplitude $\a\to \a dd$ for specific modes}
Lets compute the simplest splitting amplitude for \textit{one} boson going to \textit{one} boson and \textit{two} fermions. 

\subsubsection{\underline{$NS$ Sector Fermions}}
Taking the amplitude in (\ref{a to aaa full}) which is expressed in terms of wick contraction terms tabulated in (\ref{all contractions}), and selecting
\bea
n_f &=& \lbrace p=1,q={1\over2},r={1\over2}\rbrace\cr
n_i&=&\lbrace n=2\rbrace
\eea
our amplitude becomes
\bea
&&\!\!\!\!\!\!\!\!\!\!\!\!\mathcal{A}^{\a\to \a dd}(w_1,w_2,\bar w_1,\bar w_2)\cr
&& = \frac{9 e^{-\frac{3\Delta w}{2}-\frac{3 \Delta  \bar{w}}{2}}}{4096}-\frac{3e^{-\frac{\Delta w}{2}-\frac{3 \Delta  \bar{w}}{2}}}{2048}-\frac{3 e^{\frac{\Delta w}{2}-\frac{3 \Delta  \bar{w}}{2}}}{4096}-\frac{3 e^{-\frac{3 \Delta w}{2}-\frac{\Delta  \bar{w}}{2}}}{2048}+\frac{5 e^{-\frac{\Delta w}{2}-\frac{\Delta  \bar{w}}{2}}}{4096}+\frac{e^{\frac{\Delta w}{2}-\frac{\Delta  \bar{w}}{2}}}{1024}
\cr
&&\quad -\frac{3e^{\frac{3\Delta w}{2}-\frac{\Delta  \bar{w}}{2}}}{4096}-\frac{3 e^{-\frac{3\Delta w}{2}+\frac{\Delta \bar{w}}{2}}}{4096}+\frac{e^{-\frac{\Delta w}{2}+\frac{\Delta \bar{w}}{2}}}{1024}+\frac{5 e^{\frac{\Delta w}{2}+\frac{\Delta \bar{w}}{2}}}{4096}-\frac{3 e^{\frac{3\Delta w}{2}+\frac{\Delta  \bar{w}}{2}}}{2048} -\frac{3 e^{-\frac{\Delta w}{2}+\frac{3 \Delta  \bar{w}}{2}}}{4096}\cr
&&\quad-\frac{3 e^{\frac{\Delta w}{2}+\frac{3 \Delta  \bar{w}}{2}}}{2048}+\frac{9e^{\frac{3\Delta w}{2} + \frac{3 \Delta  \bar{w}}{2}}}{4096} 
\eea
where
\bea
m_{min}(n=2; p=1,q={1\over2},r={1\over2}) = m'_{min}(n=2; p=1,q={1\over2},r={1\over2})&=&-3\cr
m_{max}(n=2; p=1,q={1\over2},r={1\over2}) = m'_{max}(n=2; p=1,q={1\over2},r={1\over2})&=&3\nn
\eea
Writing our amplitude in condensed notation we have
\bea
\mathcal{A}^{\a\to \a dd}(w_1,w_2,\bar w_1,\bar w_2) = \sum_{m=-3}^{3} \sum_{m'=-3}^{3}B^{\a\to \a dd}_{m,m'}(n=2; p=1,q={1\over2},r={1\over2})e^{{m  \Delta  w\over2}+{m'\Delta \bar w\over2}}\nn
\label{a to add condensed}
\eea
where
\bea
{5\over 4096}&=&B^{\a\to \a dd}_{1,1}(n=2; p=1,q={1\over2},r={1\over2})\cr
&=&B^{\a\to \a dd}_{-1,-1}(n=2; p=1,q={1\over2},r={1\over2})
\cr
\cr
{1\over 1024}&=&B^{\a\to \a dd}_{1,-1}(n=2; p=1,q={1\over2},r={1\over2})\cr
&=&B^{\a\to \a dd}_{-1,1}(n=2; p=1,q={1\over2},r={1\over2})
\cr
\cr
{9\over 4096}&=&B^{\a\to \a dd}_{3,3}(n=2; p=1,q={1\over2},r={1\over2})\cr
&=&B^{\a\to \a dd}_{-3,-3}(n=2; p=1,q={1\over2},r={1\over2})
\cr
\cr
-{3\over 2048}&=&B^{\a\to \a dd}_{1,3}(n=2; p=1,q={1\over2},r={1\over2})\cr
&=&B^{\a\to \a dd}_{3,1}(n=2; p=1,q={1\over2},r={1\over2})\cr
&=&B^{\a\to \a dd}_{-1,-3}(n=2; p=1,q={1\over2},r={1\over2})\cr
&=&B^{\a\to \a dd}_{-3,-1}(n=2; p=1,q={1\over2},r={1\over2})
\cr
\cr
-{3\over 4096}&=&B^{\a\to \a dd}_{1,-3}(n=2; p=1,q={1\over2},r={1\over2})\cr
&=&B^{\a\to \a dd}_{-3,1}(n=2; p=1,q={1\over2},r={1\over2})\cr
&=&B^{\a\to \a dd}_{-1,3}(n=2; p=1,q={1\over2},r={1\over2})\cr
&=&B^{\a\to \a dd}_{3,-1}(n=2; p=1,q={1\over2},r={1\over2})
\eea

\subsection{Computing the amplitude for the process $\a\a\to \a\a \a\a$ for specific indices}
Lets compute the simplest splitting amplitude for \textit{two} bosons going to \textit{four} bosons. Taking the amplitude (\ref{aa to aaaa full}) which is expressed in terms of wick contraction terms in (\ref{all contractions}), and selecting 
\bea
n_f &=& \lbrace p=1,q=1,r=1,s=1\rbrace\cr
n_i&=&\lbrace n_1=2,n_2=2\rbrace
\eea
our amplitude becomes
\bea
&&\!\!\!\!\!\!\!\!\!\!\!\!\mathcal{A}^{\a\a\to \a\a \a\a}(w_1,w_2,\bar w_1,\bar w_2)\cr
&&\!\!\!\!\!\!\!\!\!\!=\frac{59049}{8388608 \sqrt{2}}+\frac{37179 e^{-3 \Delta w}}{33554432\sqrt{2}}-\frac{29889 e^{-2 \Delta w}}{16777216 \sqrt{2}}-\frac{95499 e^{-\Delta w}}{33554432 \sqrt{2}}-\frac{95499 e^{\Delta w}}{33554432 \sqrt{2}}\cr
&&-\frac{29889 e^{2\Delta w}}{16777216 \sqrt{2}}+\frac{37179 e^{3 \Delta w}}{33554432\sqrt{2}}+\frac{37179 e^{-3 \Delta  \bar{w}}}{33554432 \sqrt{2}}-\frac{29889 e^{-2 \Delta \bar{w}}}{16777216 \sqrt{2}}-\frac{95499 e^{-\Delta  \bar{w}}}{33554432 \sqrt{2}}\cr
&&-\frac{95499e^{\Delta  \bar{w}}}{33554432 \sqrt{2}}-\frac{29889 e^{2 \Delta  \bar{w}}}{16777216\sqrt{2}}+\frac{37179 e^{3 \Delta  \bar{w}}}{33554432 \sqrt{2}}+\frac{23409 e^{-3 \Delta w-3 \Delta  \bar{w}}}{134217728 \sqrt{2}}\cr
&&-\frac{18819 e^{-2 \Delta w-3 \Delta\bar{w}}}{67108864 \sqrt{2}}-\frac{60129 e^{-\Delta w-3 \Delta  \bar{w}}}{134217728\sqrt{2}}-\frac{60129 e^{\Delta w-3 \Delta  \bar{w}}}{134217728 \sqrt{2}}-\frac{18819 e^{2\Delta w-3 \Delta  \bar{w}}}{67108864 \sqrt{2}}\cr
&&+\frac{23409 e^{3 \Delta w-3 \Delta\bar{w}}}{134217728 \sqrt{2}}-\frac{18819 e^{-3 \Delta w-2 \Delta  \bar{w}}}{67108864\sqrt{2}}+\frac{15129 e^{-2 \Delta w-2 \Delta  \bar{w}}}{33554432 \sqrt{2}}+\frac{48339e^{-\Delta w-2 \Delta  \bar{w}}}{67108864 \sqrt{2}}\cr
&&+\frac{48339 e^{\Delta w-2\Delta  \bar{w}}}{67108864 \sqrt{2}}+\frac{15129 e^{2 \Delta w-2 \Delta  \bar{w}}}{33554432\sqrt{2}}-\frac{18819 e^{3 \Delta w-2 \Delta  \bar{w}}}{67108864 \sqrt{2}}-\frac{60129e^{-3 \Delta w-\Delta  \bar{w}}}{134217728 \sqrt{2}}\cr
&&+\frac{48339 e^{-2 \Delta w-\Delta  \bar{w}}}{67108864 \sqrt{2}}+\frac{154449 e^{-\Delta w-\Delta \bar{w}}}{134217728 \sqrt{2}}+\frac{154449 e^{\Delta w-\Delta  \bar{w}}}{134217728\sqrt{2}}+\frac{48339 e^{2 \Delta w-\Delta  \bar{w}}}{67108864 \sqrt{2}}\cr
&&-\frac{60129 e^{3\Delta w-\Delta  \bar{w}}}{134217728 \sqrt{2}}-\frac{60129 e^{\Delta  \bar{w}-3\Delta w}}{134217728 \sqrt{2}}+\frac{48339 e^{\Delta  \bar{w}-2 \Delta w}}{67108864 \sqrt{2}}+\frac{154449 e^{\Delta  \bar{w}-\Delta w}}{134217728\sqrt{2}}\cr
&&+\frac{154449 e^{\Delta w+\Delta  \bar{w}}}{134217728 \sqrt{2}}+\frac{48339 e^{2\Delta w+\Delta  \bar{w}}}{67108864 \sqrt{2}}-\frac{60129 e^{3 \Delta w+\Delta\bar{w}}}{134217728 \sqrt{2}}-\frac{18819 e^{2 \Delta  \bar{w}-3 \Delta w}}{67108864\sqrt{2}}\cr
&&+\frac{15129 e^{2 \Delta  \bar{w}-2 \Delta w}}{33554432 \sqrt{2}}+\frac{48339 e^{2\Delta  \bar{w}-\Delta w}}{67108864 \sqrt{2}}+\frac{48339 e^{\Delta w+2 \Delta\bar{w}}}{67108864 \sqrt{2}}+\frac{15129 e^{2 \Delta w+2 \Delta  \bar{w}}}{33554432\sqrt{2}}\cr
&&-\frac{18819 e^{3 \Delta w+2 \Delta  \bar{w}}}{67108864 \sqrt{2}}+\frac{23409 e^{3\Delta  \bar{w}-3 \Delta w}}{134217728 \sqrt{2}}-\frac{18819 e^{3 \Delta  \bar{w}-2\Delta w}}{67108864 \sqrt{2}}-\frac{60129 e^{3 \Delta  \bar{w}-\Delta w}}{134217728 \sqrt{2}}\cr
&&-\frac{60129 e^{\Delta w+3 \Delta  \bar{w}}}{134217728\sqrt{2}}-\frac{18819 e^{2 \Delta w+3 \Delta  \bar{w}}}{67108864 \sqrt{2}}+\frac{23409 e^{3\Delta w+3 \Delta  \bar{w}}}{134217728 \sqrt{2}}
\eea

Writing our amplitude in condensed notation we have
\bea
&&\mathcal{A}^{\a\a\to \a \a\a\a}(w_1,w_2,\bar w_1,\bar w_2)\cr
&&\qquad = \sum_{m=-6}^{6} \sum_{m'=-6}^{6}B^{\a\a\to \a\a\a\a}_{m,m'}(n_1=2,n_2=2; p=1,q=1,r=1,s=1)e^{{m  \Delta  w\over2}+{m'\Delta \bar w\over2}}\nn
\label{aa to aaaa condensed}
\eea
where the nonzero coefficients are given by 
\bea
\frac{59049}{8388608 \sqrt{2}}&=&B^{\a\a\to \a\a \a\a}_{0,0}(n_1=2,n_2=2; p=1,q=1,r=1,s=1)
\cr
\cr
\frac{37179}{33554432\sqrt{2}}&=&B^{\a\a\to \a\a \a\a}_{-6,0}(n_1=2,n_2=2; p=1,q=1,r=1,s=1)\cr
&=&B^{\a\a\to \a\a \a\a}_{0,-6}(n_1=2,n_2=2; p=1,q=1,r=1,s=1)\cr
&=&B^{\a\a\to \a\a \a\a}_{6,0}(n_1=2,n_2=2; p=1,q=1,r=1,s=1)\cr
&=&B^{\a\a\to \a\a \a\a}_{0,6}(n_1=2,n_2=2; p=1,q=1,r=1,s=1)
\cr
\cr
-\frac{29889 }{16777216 \sqrt{2}}&=&B^{\a\a\to \a\a \a\a}_{-4,0}(n_1=2,n_2=2; p=1,q=1,r=1,s=1)\cr
&=&B^{\a\a\to \a\a \a\a}_{0,-4}(n_1=2,n_2=2; p=1,q=1,r=1,s=1)\cr
&=&B^{\a\a\to \a\a \a\a}_{4,0}(n_1=2,n_2=2; p=1,q=1,r=1,s=1)\cr
&=&B^{\a\a\to \a\a \a\a}_{0,4}(n_1=2,n_2=2; p=1,q=1,r=1,s=1)
\cr
\cr
-\frac{95499 }{33554432 \sqrt{2}}&=&B^{\a\a\to \a\a \a\a}_{-2,0}(n_1=2,n_2=2; p=1,q=1,r=1,s=1)\cr
&=&B^{\a\a\to \a\a \a\a}_{0,-2}(n_1=2,n_2=2; p=1,q=1,r=1,s=1)\cr
&=&B^{\a\a\to \a\a \a\a}_{2,0}(n_1=2,n_2=2; p=1,q=1,r=1,s=1)\cr
&=&B^{\a\a\to \a\a \a\a}_{0,2}(n_1=2,n_2=2; p=1,q=1,r=1,s=1)
\cr
\cr
-\frac{18819}{67108864 \sqrt{2}}&=&B^{\a\a\to \a\a \a\a}_{-4,-6}(n_1=2,n_2=2; p=1,q=1,r=1,s=1)\cr
&=&B^{\a\a\to \a\a \a\a}_{-6,-4}(n_1=2,n_2=2; p=1,q=1,r=1,s=1)\cr
&=&B^{\a\a\to \a\a \a\a}_{4,6}(n_1=2,n_2=2; p=1,q=1,r=1,s=1)\cr
&=&B^{\a\a\to \a\a \a\a}_{6,4}(n_1=2,n_2=2; p=1,q=1,r=1,s=1)\cr
&=&B^{\a\a\to \a\a \a\a}_{4,-6}(n_1=2,n_2=2; p=1,q=1,r=1,s=1)\cr
&=&B^{\a\a\to \a\a \a\a}_{6,-4}(n_1=2,n_2=2; p=1,q=1,r=1,s=1)\cr
&=&B^{\a\a\to \a\a \a\a}_{-4,6}(n_1=2,n_2=2; p=1,q=1,r=1,s=1)\cr
&=&B^{\a\a\to \a\a \a\a}_{-6,4}(n_1=2,n_2=2; p=1,q=1,r=1,s=1)
\cr
\cr
-\frac{60129}{134217728\sqrt{2}}&=&B^{\a\a\to \a\a \a\a}_{-2,-6}(n_1=2,n_2=2; p=1,q=1,r=1,s=1)\cr
&=&B^{\a\a\to \a\a \a\a}_{-6,-2}(n_1=2,n_2=2; p=1,q=1,r=1,s=1)\cr
&=&B^{\a\a\to \a\a \a\a}_{2,6}(n_1=2,n_2=2; p=1,q=1,r=1,s=1)\cr
&=&B^{\a\a\to \a\a \a\a}_{6,2}(n_1=2,n_2=2; p=1,q=1,r=1,s=1)\cr
&=&B^{\a\a\to \a\a \a\a}_{2,-6}(n_1=2,n_2=2; p=1,q=1,r=1,s=1)\cr
&=&B^{\a\a\to \a\a \a\a}_{6,-2}(n_1=2,n_2=2; p=1,q=1,r=1,s=1)\cr
&=&B^{\a\a\to \a\a \a\a}_{-2,6}(n_1=2,n_2=2; p=1,q=1,r=1,s=1)\cr
&=&B^{\a\a\to \a\a \a\a}_{-6,-2}(n_1=2,n_2=2; p=1,q=1,r=1,s=1)
\cr
\cr
\frac{48339}{67108864 \sqrt{2}}&=&B^{\a\a\to \a\a \a\a}_{-2,-4}(n_1=2,n_2=2; p=1,q=1,r=1,s=1)\cr
&=&B^{\a\a\to \a\a \a\a}_{-4,-2}(n_1=2,n_2=2; p=1,q=1,r=1,s=1)\cr
&=&B^{\a\a\to \a\a \a\a}_{2,4}(n_1=2,n_2=2; p=1,q=1,r=1,s=1)\cr
&=&B^{\a\a\to \a\a \a\a}_{4,2}(n_1=2,n_2=2; p=1,q=1,r=1,s=1)\cr
&=&B^{\a\a\to \a\a \a\a}_{-2,4}(n_1=2,n_2=2; p=1,q=1,r=1,s=1)\cr
&=&B^{\a\a\to \a\a \a\a}_{-4,2}(n_1=2,n_2=2; p=1,q=1,r=1,s=1)\cr
&=&B^{\a\a\to \a\a \a\a}_{2,-4}(n_1=2,n_2=2; p=1,q=1,r=1,s=1)\cr
&=&B^{\a\a\to \a\a \a\a}_{4,-2}(n_1=2,n_2=2; p=1,q=1,r=1,s=1)\cr
\cr
\cr
\frac{154449}{134217728 \sqrt{2}}&=&B^{\a\a\to \a\a \a\a}_{2,2}(n_1=2,n_2=2; p=1,q=1,r=1,s=1)\cr
&=&B^{\a\a\to \a\a \a\a}_{-2,-2}(n_1=2,n_2=2; p=1,q=1,r=1,s=1)\cr
&=&B^{\a\a\to \a\a \a\a}_{2,-2}(n_1=2,n_2=2; p=1,q=1,r=1,s=1)\cr
&=&B^{\a\a\to \a\a \a\a}_{-2,2}(n_1=2,n_2=2; p=1,q=1,r=1,s=1)\cr
\cr
\cr
\frac{15129}{33554432\sqrt{2}}&=&B^{\a\a\to \a\a \a\a}_{-4,-4}(n_1=2,n_2=2; p=1,q=1,r=1,s=1)\cr
&=&B^{\a\a\to \a\a \a\a}_{4,4}(n_1=2,n_2=2; p=1,q=1,r=1,s=1)\cr
&=&B^{\a\a\to \a\a \a\a}_{-4,4}(n_1=2,n_2=2; p=1,q=1,r=1,s=1)\cr
&=&B^{\a\a\to \a\a \a\a}_{4,-4}(n_1=2,n_2=2; p=1,q=1,r=1,s=1)
\cr
\cr
\cr
\frac{23409}{134217728 \sqrt{2}}&=&B^{\a\a\to \a\a \a\a}_{-6,-6}(n_1=2,n_2=2; p=1,q=1,r=1,s=1)\cr
&=&B^{\a\a\to \a\a \a\a}_{6,6}(n_1=2,n_2=2; p=1,q=1,r=1,s=1)\cr
&=&B^{\a\a\to \a\a \a\a}_{-6,6}(n_1=2,n_2=2; p=1,q=1,r=1,s=1)\cr
&=&B^{\a\a\to \a\a \a\a}_{6,-6}(n_1=2,n_2=2; p=1,q=1,r=1,s=1)\nn
\eea
\subsection{Computing the amplitude for the process $\a\a\to \a\a dd$ in the $NS$ sector for specific indices}
Lets compute the simplest splitting amplitude for \textit{two} bosons going to \textit{two} boson and \textit{two} fermions. 
\subsubsection{\underline{$NS$ Sector Fermions}}
Taking the amplitude in (\ref{aa to aadd full}) which is expressed in terms of wick contraction terms tabulated in (\ref{all contractions}), and selecting
\bea
n_f &=& \lbrace p=1,q=1,r={1\over2},s={1\over2}\rbrace\cr
n_i&=&\lbrace n_1=1,n_2=2\rbrace
\eea
our amplitude becomes
\bea
&&\!\!\!\!\!\!\!\!\!\!\!\!\mathcal{A}^{\a\a\to \a\a dd}(w_1,w_2,\bar w_1,\bar w_2)\cr
&&\!\!\!\!\!\!\!\!\!=\frac{81}{65536 \sqrt{2}}+\frac{169 e^{-2 \Delta w-2 \Delta  \bar{w}}}{262144 \sqrt{2}}-\frac{19 e^{-\Delta w-2\Delta  \bar{w}}}{65536 \sqrt{2}}+\frac{29 e^{\Delta w-2 \Delta\bar{w}}}{65536 \sqrt{2}}-\frac{119 e^{2 \Delta w-2 \Delta  \bar{w}}}{262144\sqrt{2}}
\cr
&&\!\!\!\!-\frac{3 e^{-\frac{\Delta w}{2}-\frac{3 \Delta  \bar{w}}{2}}}{4096\sqrt{2}}-\frac{3 e^{\frac{\Delta w}{2}-\frac{3 \Delta  \bar{w}}{2}}}{8192\sqrt{2}}-\frac{19 e^{-2 \Delta w-\Delta  \bar{w}}}{65536 \sqrt{2}}+\frac{5 e^{-\Delta w-\Delta\bar{w}}}{16384 \sqrt{2}}-\frac{3 e^{\Delta w-\Delta  \bar{w}}}{16384\sqrt{2}}+\frac{29 e^{2 \Delta w-\Delta  \bar{w}}}{65536 \sqrt{2}}
\cr
&&\!\!\!\!-\frac{3 e^{-\frac{3 \Delta w}{2}-\frac{\Delta\bar{w}}{2}}}{4096 \sqrt{2}}+\frac{5 e^{-\frac{\Delta w}{2}-\frac{\Delta \bar{w}}{2}}}{8192 \sqrt{2}}+\frac{e^{\frac{\Delta w}{2}-\frac{\Delta  \bar{w}}{2}}}{2048 \sqrt{2}}-\frac{3 e^{\frac{3 \Delta w}{2}-\frac{\Delta  \bar{w}}{2}}}{8192 \sqrt{2}}-\frac{3 e^{-\frac{3\Delta w}{2}+\frac{\Delta  \bar{w}}{2}}}{8192 \sqrt{2}}+\frac{e^{-\frac{\Delta w}{2}+\frac{\Delta  \bar{w}}{2}}}{2048 \sqrt{2}}\cr
&&\!\!\!\!+\frac{5 e^{\frac{\Delta w}{2}+\frac{\Delta  \bar{w}}{2}}}{8192\sqrt{2}}-\frac{3 e^{\frac{3 \Delta w}{2}+\frac{\Delta  \bar{w}}{2}}}{4096\sqrt{2}}+\frac{29 e^{-2 \Delta w+\Delta  \bar{w}}}{65536 \sqrt{2}}-\frac{3 e^{-\Delta w+\Delta \bar{w}}}{16384 \sqrt{2}}+\frac{5 e^{\Delta w+\Delta  \bar{w}}}{16384\sqrt{2}}-\frac{19 e^{2 \Delta w+\Delta  \bar{w}}}{65536 \sqrt{2}}
\cr
&&\!\!\!\!-\frac{3 e^{-\frac{\Delta w}{2}+\frac{3\Delta  \bar{w}}{2}}}{8192 \sqrt{2}}-\frac{3 e^{\frac{\Delta w}{2}+\frac{3 \Delta \bar{w}}{2}}}{4096 \sqrt{2}}+\frac{9 e^{\frac{3 \Delta w}{2}+\frac{3 \Delta \bar{w}}{2}}}{8192 \sqrt{2}}-\frac{119 e^{-2\Delta w+2 \Delta  \bar{w}}}{262144 \sqrt{2}}+\frac{29 e^{-\Delta w+2 \Delta  \bar{w}}}{65536\sqrt{2}}
\cr
&&\!\!\!\!-\frac{19 e^{\Delta w+2 \Delta  \bar{w}}}{65536 \sqrt{2}}+\frac{169 e^{2 \Delta w+2 \Delta \bar{w}}}{262144 \sqrt{2}}-\frac{45 e^{-2 \Delta  \bar{w}}}{131072\sqrt{2}}-\frac{9 e^{-\Delta  \bar{w}}}{32768\sqrt{2}}-\frac{9 e^{\Delta  \bar{w}}}{32768\sqrt{2}}-\frac{45 e^{2 \Delta  \bar{w}}}{131072 \sqrt{2}}\cr
&&\!\!\!\!-\frac{45 e^{-2 \Delta w}}{131072\sqrt{2}}-\frac{9 e^{-\Delta w}}{32768 \sqrt{2}}-\frac{9 e^{\Delta w}}{32768\sqrt{2}}-\frac{45 e^{2 \Delta w}}{131072 \sqrt{2}}+\frac{9 e^{-\frac{3 \Delta w}{2}-\frac{3 \Delta  \bar{w}}{2}}}{8192\sqrt{2}}
\eea
where
\bea
-4&=&m_{min}(n_1=1,n_2=2; p=1,q=1,r={1\over2},s={1\over2})\cr
&=& m'_{min}(n_1=1,n_2=2; p=1,q=1,r={1\over2},s={1\over2})
\cr
\cr
4&=&m_{max}(n_1=1,n_2=2; p=1,q=1,r={1\over2},s={1\over2})\cr
&=&m'_{max}(n_1=1,n_2=2; p=1,q=1,r={1\over2},s={1\over2})
\eea
Writing our amplitude in condensed notation we have
\bea
&&\mathcal{A}^{\a\a\to \a\a dd}(w_1,w_2,\bar w_1,\bar w_2)\cr
&&\qquad = \sum_{m=-4}^{4} \sum_{m'=-4}^{4}B^{\a\a\to \a\a dd}_{m,m'}(n_1=1,n_2=2; p=1,q=1,r={1\over2},s={1\over2})e^{{m  \Delta  w\over2}+{m'\Delta \bar w\over2}}\nn
\label{aa to aadd condensed}
\eea
where our coefficients are given by
\bea
\frac{81}{65536 \sqrt{2}}&=&B^{\a\a\to \a\a dd}_{0,0}(n_1=1,n_2=2; p=1,q=1,r={1\over2},s={1\over2})
\cr
\cr
\frac{5 }{8192 \sqrt{2}}&=&B^{\a\a\to \a\a dd}_{1,1}(n_1=1,n_2=2; p=1,q=1,r={1\over2},s={1\over2})\cr
&=&B^{\a\a\to \a\a dd}_{-1,-1}(n_1=1,n_2=2; p=1,q=1,r={1\over2},s={1\over2})
\cr
\cr
\frac{1}{2048 \sqrt{2}}&=&B^{\a\a\to \a\a dd}_{1,-1}(n_1=1,n_2=2; p=1,q=1,r={1\over2},s={1\over2})\cr
&=&B^{\a\a\to \a\a dd}_{-1,1}(n_1=1,n_2=2; p=1,q=1,r={1\over2},s={1\over2})
\cr
\cr
\frac{5}{16384 \sqrt{2}}&=&B^{\a\a\to \a\a dd}_{2,2}(n_1=1,n_2=2; p=1,q=1,r={1\over2},s={1\over2})\cr
&=&B^{\a\a\to \a\a dd}_{-2,-2}(n_1=1,n_2=2; p=1,q=1,r={1\over2},s={1\over2})
\cr
\cr
-\frac{3}{16384\sqrt{2}}&=&B^{\a\a\to \a\a dd}_{2,-2}(n_1=1,n_2=2; p=1,q=1,r={1\over2},s={1\over2})\cr
&=&B^{\a\a\to \a\a dd}_{-2,2}(n_1=1,n_2=2; p=1,q=1,r={1\over2},s={1\over2})
\cr
\cr
\frac{9 }{8192\sqrt{2}}&=&B^{\a\a\to \a\a dd}_{3,3}(n_1=1,n_2=2; p=1,q=1,r={1\over2},s={1\over2})\cr
&=&B^{\a\a\to \a\a dd}_{-3,-3}(n_1=1,n_2=2; p=1,q=1,r={1\over2},s={1\over2})
\cr
\cr
\frac{169}{262144 \sqrt{2}}&=&B^{\a\a\to \a\a dd}_{4,4}(n_1=1,n_2=2; p=1,q=1,r={1\over2},s={1\over2})\cr
&=&B^{\a\a\to \a\a dd}_{-4,-4}(n_1=1,n_2=2; p=1,q=1,r={1\over2},s={1\over2})
\cr
\cr
-\frac{119 }{262144\sqrt{2}}&=&B^{\a\a\to \a\a dd}_{4,-4}(n_1=1,n_2=2; p=1,q=1,r={1\over2},s={1\over2})\cr
&=&B^{\a\a\to \a\a dd}_{-4,4}(n_1=1,n_2=2; p=1,q=1,r={1\over2},s={1\over2})
\cr
\cr
-\frac{9}{32768\sqrt{2}}&=&B^{\a\a\to \a\a dd}_{0,2}(n_1=1,n_2=2; p=1,q=1,r={1\over2},s={1\over2})\cr
&=&B^{\a\a\to \a\a dd}_{2,0}(n_1=1,n_2=2; p=1,q=1,r={1\over2},s={1\over2})\cr
&=&B^{\a\a\to \a\a dd}_{0,-2}(n_1=1,n_2=2; p=1,q=1,r={1\over2},s={1\over2})\cr
&=&B^{\a\a\to \a\a dd}_{-2,0}(n_1=1,n_2=2; p=1,q=1,r={1\over2},s={1\over2})
\cr
\cr
-\frac{45}{131072\sqrt{2}}&=&B^{\a\a\to \a\a dd}_{0,4}(n_1=1,n_2=2; p=1,q=1,r={1\over2},s={1\over2})\cr
&=&B^{\a\a\to \a\a dd}_{4,0}(n_1=1,n_2=2; p=1,q=1,r={1\over2},s={1\over2})\cr
&=&B^{\a\a\to \a\a dd}_{0,-4}(n_1=1,n_2=2; p=1,q=1,r={1\over2},s={1\over2})\cr
&=&B^{\a\a\to \a\a dd}_{-4,0}(n_1=1,n_2=2; p=1,q=1,r={1\over2},s={1\over2})
\cr
\cr
-\frac{3}{4096\sqrt{2}}&=&B^{\a\a\to \a\a dd}_{1,3}(n_1=1,n_2=2; p=1,q=1,r={1\over2},s={1\over2})\cr
&=&B^{\a\a\to \a\a dd}_{3,1}(n_1=1,n_2=2; p=1,q=1,r={1\over2},s={1\over2})\cr
&=&B^{\a\a\to \a\a dd}_{-1,-3}(n_1=1,n_2=2; p=1,q=1,r={1\over2},s={1\over2})\cr
&=&B^{\a\a\to \a\a dd}_{-3,-1}(n_1=1,n_2=2; p=1,q=1,r={1\over2},s={1\over2})
\cr
\cr
-\frac{3}{8192\sqrt{2}}&=&B^{\a\a\to \a\a dd}_{1,-3}(n_1=1,n_2=2; p=1,q=1,r={1\over2},s={1\over2})\cr
&=&B^{\a\a\to \a\a dd}_{-1,3}(n_1=1,n_2=2; p=1,q=1,r={1\over2},s={1\over2})\cr
&=&B^{\a\a\to \a\a dd}_{3,-1}(n_1=1,n_2=2; p=1,q=1,r={1\over2},s={1\over2})\cr
&=&B^{\a\a\to \a\a dd}_{-3,1}(n_1=1,n_2=2; p=1,q=1,r={1\over2},s={1\over2})
\cr
\cr
-\frac{19 }{65536 \sqrt{2}}&=&B^{\a\a\to \a\a dd}_{2,4}(n_1=1,n_2=2; p=1,q=1,r={1\over2},s={1\over2})\cr
&=&B^{\a\a\to \a\a dd}_{-2,-4}(n_1=1,n_2=2; p=1,q=1,r={1\over2},s={1\over2})\cr
&=&B^{\a\a\to \a\a dd}_{4,2}(n_1=1,n_2=2; p=1,q=1,r={1\over2},s={1\over2})\cr
&=&B^{\a\a\to \a\a dd}_{-4,-2}(n_1=1,n_2=2; p=1,q=1,r={1\over2},s={1\over2})
\cr
\cr
\frac{29 }{65536 \sqrt{2}}&=&B^{\a\a\to \a\a dd}_{2,-4}(n_1=1,n_2=2; p=1,q=1,r={1\over2},s={1\over2})\cr
&=&B^{\a\a\to \a\a dd}_{-2,4}(n_1=1,n_2=2; p=1,q=1,r={1\over2},s={1\over2})\cr
&=&B^{\a\a\to \a\a dd}_{4,-2}(n_1=1,n_2=2; p=1,q=1,r={1\over2},s={1\over2})\cr
&=&B^{\a\a\to \a\a dd}_{-4,2}(n_1=1,n_2=2; p=1,q=1,r={1\over2},s={1\over2})\nn
\eea

\section{Amplitude Integration}
In this section we perform the integration over the twist insertion points. Our full integrated amplitude will be of the form
\bea
\mathcal{A}^{i\to f}_{\text{int}} &=&{1\over2}\lambda^2\int_{-{\t\over2}}^{{\t\over2}}\int_{-{\t\over2}}^{\t_2}\int_{\s_2=0}^{2\pi}\int_{\s_1=0}^{2\pi}d^2w_2d^2w_1\mathcal{A}^{i\to f}(w_1,w_2,\bar w_1,\bar w_2) \cr
&=&{1\over2}\lambda^2\sum_{m=m_{min}(n_{i};n_{f})}^{m_{max}(n_{i};n_{f})}\sum_{m'=m'_{min}(n_{i};n_{f})}^{m'_{max}(n_{i};n_{f})}B^{i\to f}_{m,m'}(n_i;n_f)\cr
&&\quad \int_{-{\t\over2}}^{{\t\over2}}\int_{-{\t\over2}}^{\t_2}\int_{\s_2=0}^{2\pi}\int_{\s_1=0}^{2\pi}d^2w_2d^2w_1 e^{{m\Delta w\over2}} e^{{m'\Delta \bar{w}\over2}}\cr
&\equiv&{1\over2}\lambda^2\sum_{m=m_{min}(n_{i};n_{f})}^{m_{max}(n_{i};n_{f})}\sum_{m'=m'_{min}(n_{i};n_{f})}^{m'_{max}(n_{i};n_{f})}B^{i\to f}_{m,m'}(n_i;n_f) I_{mm'}(t)
\eea
where
\bea
I_{mm'}(t)\equiv \int_{-{\t\over2}}^{{\t\over2}}\int_{-{\t\over2}}^{\t_2}\int_{\s_2=0}^{2\pi}\int_{\s_1=0}^{2\pi}d^2w_2d^2w_1 e^{{m\Delta w\over2}} e^{{m'\Delta \bar{w}\over2}}
\eea
Writing our cylinder coordinates below
\bea
w_i &=& \t_i + i\s_i, \qquad i=1,2\cr
\bar{w}_i&=& \t_i - i\s_i, \qquad i=1,2
\eea
our measure is given by 
\bea
\diff^2w_2\diff^2w_1=\diff \t_2\diff \t_1\diff \s_2\diff \s_1
\eea
Wick rotating $\t_i$ to the Minkowski time coordinate, $t_i$, through $\t_i\to it_i$, our measure becomes
\bea
\diff \t_2\diff \t_1\diff \s_2\diff \s_1    \to    \diff t_2\diff t_1\diff \s_2\diff \s_1
\label{measure}
\eea
Writing the exponential factors in terms of $\s_i,t_i$, our integral becomes
\bea
I_{mm'}(t)&=&\int_{-{t\over2}}^{{t\over2}}dt_2\int_{-{t\over2}}^{t_2}dt_1\int_{\s_2=0}^{2\pi}\diff \s_2\int_{\s_1=0}^{2\pi}\diff \s_1 e^{{i(m+m')\over2}t'}e^{{i(m-m')\over2}\s'}
\eea
where $t'=t_2-t_1$ and $\s'=\s_2-\s_1$.
When evaluating the above integral, we have four cases
\subsubsection{\underline{$m-m'\neq 0,\quad m+m' \neq 0$}}
\bea
I_{m-m'\neq 0,m+m' \neq 0}(t)&=&\int_{-{t\over2}}^{{t\over2}}dt_2\int_{-{t\over2}}^{t_2}dt_1\int_{\s=0}^{2\pi}d\s_2\int_{\s=0}^{2\pi}d\s_1 e^{{i(m+m')\over2}t'}e^{{i(m-m')\over2}\s'}\cr
&=&{32i\sin^2\big({(m-m')\over2}\pi\big)\bigg((m+m')t - 4e^{i{(m+m')t\over 4}}\sin\big({(m+m')t\over 4}\big)\bigg)\over (m^2 - m'^2)^2}\nn
\eea

\subsubsection{\underline{$m-m'\neq 0, m+m' = 0$}}

\bea
I_{m-m'\neq 0, m+m' = 0}(t)&=&\int_{-{t\over2}}^{{t\over2}}dt_2\int_{-{t\over2}}^{t_2}dt_1\int_{\s=0}^{2\pi}d\s_2\int_{\s=0}^{2\pi}d\s_1e^{{i(m-m')\over2}\s'}\d_{m+m',0}\cr
&&=-{t^2(\cos(2m\pi)-1)\over m^2}\cr
&&=0
\label{integral two}
\eea

\subsubsection{\underline{$m-m'=0 , m+m' \neq0$}}
\bea
I_{m-m'=0 , m+m' \neq0}(t)&=&\int_{-{t\over2}}^{{t\over2}}dt_2\int_{-{t\over2}}^{t_2}dt_1\int_{\s=0}^{2\pi}d\s_2\int_{\s=0}^{2\pi}d\s_1 e^{{i(m+m')\over2}t'}\d_{m-m',0}\cr
&=& {4i \pi^2\bigg(mt  -  2e^{i {mt\over2}}\sin(  {mt\over 2})  \bigg)\over m ^2}
\label{integral three}
\eea

\subsubsection{\underline{$m=m'=0$}}

\bea
I_{m=m'=0}(t)&=&\int_{-{t\over2}}^{{t\over2}}dt_2\int_{-{t\over2}}^{t_2}dt_1\int_{\s=0}^{2\pi}d\s_2\int_{\s=0}^{2\pi}d\s_1\cr
&=& 2\pi^2t^2
\label{integral four}
\eea
Then
\bea
I_{mm'}(t) = I_{mm';m-m'\neq 0,m+m' \neq 0}(t) + I_{mm';m=m' , m+m' \neq0}(t) + I_{m=m'=0}(t)
\eea
So our integrated amplitude becomes
\bea
\mathcal{A}^{i\to f}_{\text{int}}&=&{1\over2}\lambda^2\sum_{m=m_{min}(p,q,r,n)}^{m_{max}(p,q,r,n)}\sum_{m'=m'_{min}(p,q,r,n)}^{m'_{max}(p,q,r,n)}B^{i\to f}_{m,m'}(n_i;n_f)\cr
&&\qquad\bigg(I_{mm';m-m'\neq 0,m+m' \neq 0}(t) + I_{mm';m-m'=0 , m+m' \neq0}(t) + I_{mm';m=m'=0}(t)\bigg)\nn
\label{integrated amplitude}
\eea
In the following subsections we integrate the amplitudes for specific processes.
\subsection{Integrated Amplitude for $\a\to\a\a\a$}
In this subsection we compute the full integrated amplitude for the process where \textit{one} boson with energy $n=3$ splits into \textit{three} bosons, each with energy $p=q=r=1$ by integrating the amplitude in (\ref{a to aaa 3 to 111}). Looking at (\ref{a to aaa 3 to 111}) we note that the indices $m,m'\in \mathbb{Z}_{\text{odd}}$ which means that $m-m'\in \mathbb{Z}_{\text{even}}$. This results in
\bea
I_{mm',m-m'\neq 0,m+m' \neq 0}(t) &=& {32i\sin^2\big({(m-m')\over2}\pi\big)\bigg((m+m')t - 4e^{i{(m+m')t\over 4}}\sin\big({(m+m')t\over 4}\big)\bigg)\over (m^2 - m'^2)^2}
\cr
\cr
&=&0,\quad m,m'\in\mathbb{Z}_{\text{odd}},\quad m-m'\neq 0 \neq m+m'
\eea
We also note that the terms with $m+m'=0$ vanish according to (\ref{integral two}). Also, since there are no $m=m'=0$ terms in (\ref{a to aaa 3 to 111}), we have no contribution from (\ref{integral four}). We only obtain a contribution from the integral (\ref{integral three}). Therefore our integrated amplitude, (\ref{integrated amplitude}), for this process is given by the expression
\bea
\mathcal{A}^{\a\to\a\a\a }_{\text{int}}
&=&{1\over2}\lambda^2  \sum_{m=-5}^{5}B^{\a\to \a\a\a}_{m,m}(n=3; p=1,q=1,r=1)\bigg( {4i \pi^2\big(mt  -  2e^{i {mt\over2}}\sin(  {mt\over 2})  \big)\over m ^2}\bigg)\nn
\eea
Inserting values of the coefficients and performing the sum yields
\bea
\mathcal{A}^{\a\to\a\a\a }_{\text{int}} &=& 8\lambda^2\pi^2  \bigg({39\over 32768}\sin^2(  {t\over 2})  + \frac{159}{131072}{\sin^2(  {3t\over 2})  \over 9} + {3\over131072}\sin^2(  {5t\over 2})   \bigg)\nn
\eea
We see that the amplitude only contains terms which oscillate with time, $t$. We will discuss the implications of this in the next section when compared to the processes where two particles split into four particles.

\subsection{Integrated Amplitude for $\a\to\a dd$}
Applying the same analysis to integrate the amplitude (\ref{a to add condensed}) for the process $\a\to \a dd$ (for an initial energy of $n=2$ and final energies $p=1,q={1\over2},r={1\over2}$) we obtain
\bea
\mathcal{A}^{\a\to \a dd}_{\text{int}} 
&=& 8\lambda^2  \pi^2\bigg( {5\over 4096}\sin^2(  {t\over 2}) +   {1\over 4096}\sin^2(  {3t\over 2}) \bigg)
\eea
Here we see very similar behavior to the integrated amplitude for $\a\to\a\a\a$. The amplitude only oscillates in $t$.
\subsection{Integrated Amplitude for $\a\a\to\a\a \a\a$}
Applying the same analysis to integrate the amplitude (\ref{aa to aaaa condensed}) for the process $\a\a\to \a\a \a\a$ as we did for the processes in the previous subsections, we obtain
\bea
\mathcal{A}^{\a\a\to \a \a\a\a}_{\text{int}} &=& \lambda^2 \pi^2t^2\frac{59049}{8388608 \sqrt{2}} 
\cr
&&+ 8  \lambda^2 \pi^2\Bigg( \frac{154449}{134217728 \sqrt{2}}{ \sin^2(t)   \over 4} + \frac{15129}{33554432\sqrt{2}}{ \sin^2(  2t) \over 16} +  \frac{23409}{134217728 \sqrt{2}}{\sin^2(  3t)   \over 36}\Bigg)\nn
\eea
In this case we see that the amplitude now includes a $t^2$ term in addition to oscillatory terms which are present in the  `1 to 3' processes. This suggests that in the limit of large $t$, the `2 to 4' splitting is preferred over the `1 to 3' splitting. We will discuss this more in depth at the end of the chapter. 
\subsection{Integrated Amplitude for $\a\a\to\a\a dd$}
Applying the same analysis to integrate the amplitude (\ref{aa to aadd condensed}) for the process $\a\a\to \a\a dd$, we obtain
\bea
\mathcal{A}^{\a\a\to \a\a dd}_{\text{int}} 
&=&\lambda^2\pi^2\frac{81}{65536 \sqrt{2}}t^2 \cr
&& + 8 \pi^2\lambda^2 \bigg(\frac{5}{8192 \sqrt{2}}   \sin^2(  {t\over 2})    + \frac{5}{16384 \sqrt{2}} { \sin^2(t)   \over 4}  \cr
&&\qquad\qquad+ \frac{1}{8192\sqrt{2}}  \sin^2(  {3t\over 2})    + \frac{169}{262144 \sqrt{2}}  {  \sin^2(  2t)  \over 16}   \bigg)\nn
\eea

Similar to the process  $\a\a\to\a\a\a\a$, we find that this amplitude has terms which oscillate in $t$ and a term which is grows like $t^2$. We have shown that having a mixture of bosons and fermions in the final state give a similar result as having only bosons in the final state. There is an additional $t^2$ dependence that wasn't present in the `1 to 3' process. This is expected because the D1D5 system is a superconformal field theory.

\section{Coefficients for Larger Energies for $\a\a\to\a\a\a\a$}

In this section we tabulate the $t^2$ coefficients for bosonic splitting amplitudes in the $2\to4$ process, $B^{\a\a\to \a\a\a\a}_{0,0}\big(n_1,n_2; p,q,r,s\big)$, for increasing values of total initial energy $E_{\text{total}}$. For long time scales, $t>>1$, these terms will dominate the amplitude. Our computations in previous sections were performed for simple mode numbers. This was to demonstrate differing behavior between the $1\to3$ versus the $2\to4$ splitting process, namely the appearance of the secular term proportional to $t^2$. However, to address the problem of black hole formation, we must consider higher energies. This ensures that we are above the threshold for black hole formation. We are interested in comparing different splitting processes at each energy level to determine the preferred splitting pattern. For example we would like to know if the energy prefers to split equally or unequally amongst the final modes. Equal splitting would serve as a stronger signal of thermalization as opposed to unequal splitting. This is because equal splitting could be thought of as an equipartioning of energy, a common feature of thermal systems. We tabulate our results below.

\begin{table}[H]
\centering
 \begin{tabular}{||c   | c||} 
 \hline
 $E_{\text{total}} = 2(n_1+n_2) $ & \underline{Equal Splitting}\\ 
  $( n_1 = n_2)$ & $B^{\a\a\to \a\a\a\a}_{0,0}\big(n_1,n_2; p=q=r=s={n_1 + n_2\over4}\big)$ \\[0.5ex] 
 \hline\hline
 $2(n_1+n_2)=2(2+2)=8$ & $4.97746\times 10^{-3}$  \\ 
 \hline
 $2(n_1+n_2)=2(6+6)=24$ &  $7.59583\times 10^{-5}$\\
 \hline
 $2(n_1+n_2)=2(10+10)=40$ & $1.15986\times 10^{-5}$  \\
 \hline
$2(n_1+n_2)=2(14+14)=56$ & $3.36791\times10^{-6}$  \\
 \hline
$2(n_1+n_2)=2(18+18)=72$ & $1.33531\times10^{-6}$  \\ [1ex] 
 \hline
\end{tabular}
\caption{\label{equalsplitting} $t^2$ coefficients for $2\to 4$ splitting for equal energies in the final state: $ p=q=r=s={n_1 + n_2\over4}$. The factor of $2$ in the total energy comes because we must add together the left and right moving energies which are chosen to be the same.}
\end{table}

\begin{table}[H]
\centering
 \begin{tabular}{||c   | c||} 
 \hline
$E_{\text{total}} = 2(n_1+n_2) $ & \underline{Asymmetric Splitting}\\ 
  $( n_1 = n_2)$ & $B^{\a\a\to \a\a\a\a}_{0,0}\big(n_1,n_2; p=n_1 + n_2-3,q=r=s=1\big)$ \\[0.5ex] 
 \hline\hline
 $2(n_1+n_2)=2(2+2)=8$ & $4.97746\times 10^{-3}$\\ 
 \hline
 $2(n_1+n_2)=2(6+6)=24$ & $1.28367\times10^{-8}$  \\
 \hline
 $2(n_1+n_2)=2(10+10)=40$ & $2.33709\times10^{-10}$  \\
 \hline
$2(n_1+n_2)=2(14+14)=56$ & $1.86499\times10^{-11}$  \\
 \hline
$2(n_1+n_2)=2(18+18)=72$ & $2.92488\times10^{-12}$  \\ [1ex] 
 \hline
\end{tabular}
\caption{\label{asymmetricsplitting} $t^2$ coefficients for $2\to 4$ splitting for unequal energies in the final state: $ p=n_1 + n_2-3, q=r=s=1$. Again, the factor of $2$ in the total energy comes because we must add together the left and right moving energies which are chosen to be the same.}
\end{table}

We see that at each energy level, $E_{\text{total}}$, (except for the lowest level due to fact that there is only one way for the final state to split, in which case the coefficients are equal) the $t^2$ coefficient for \textit{equal} splitting is much greater than the $t^2$ coefficient for \textit{asymmetric} splitting. This tells us that the probability for the total energy of two initial modes to split equally amongst four final modes is much higher than to split asymmetrically. This serves as a strong indication of a thermalization process. We will return to this in a future work to analyze the splitting mechanism in detail. 

\section{Discussion}
In order for a system to thermalize we need an initial state with a given energy to break apart into several low energy states which look thermal. Therefore, it must be favorable (high probability) for an initial state to split into lower energy states. In order to compute the full process of thermalization, it is necessary to first compute the fundamental interaction which produces this splitting behavior. We have done this at second order the deformation operator.
We have computed various splitting amplitudes under the application of two deformations of the CFT. We found that each amplitude carries a dependence on the time, $t$, which is the duration of the two deformations. The amplitudes for the `1 to 3' processes contain terms which oscillate in $t$. The amplitudes for the `2 to 4' processes contain terms which oscillate in $t$, just like the `1 to 3' process, but also additional terms which grow like $t^2$. The oscillations are around the leading order state but the $t^2$ terms are `secular' terms. For short time scales, $t<<1$, all four of the amplitudes are of similar order while for long time scales, $t>>1$, the amplitude which dominates is the one where two particles split into four particles. 
Therefore, at large timescales, it is preferable to have two particles in the initial state than to have one. The reason is the following; one $graviton$ in the initial state in the CFT corresponds to a single scalar moving in AdS space. We are in the NS sector in the CFT. This gives a throat and cap region which looks like global AdS. If we want, we can spectral flow this state to the $R$ sector with vacuum spins aligned. This just corresponds to a coordinate transformation in the gravity dual and again yields global AdS. We don't expect this scalar to spontanteously break apart to form something like a black hole as moves through this spacetime. This is because in the gravity theory the scalar moves along a geodesic path. By symmetry all points along its path are equivalent. However, if we collide two scalars in AdS then it is conceivable that a black hole could form if sufficient energies are reached in the collision. In the CFT, this corresponds to applying two deformation operators to two gravitons in the initial state. Therefore we want the `2 to 4' process to dominate over the `1 to 3' process which does indeed happen at long timescales.

This `2 to 4' process in the CFT corresponds to two scalars moving and colliding in the AdS dual. We note that the energies we have used in computing our amplitudes are quite low. Since the mass of the hole grows like $N$, for $N=2$, quite low energy excitations are sufficient to form a black hole.  Additionaly, in Table's \ref{equalsplitting},\ref{asymmetricsplitting} we record coefficients of the bosonic `2 to 4' splitting amplitude for large energies compared to the AdS scale. This ensures that we are above the black hole formation threshold which for $N=2$, is small. We show that equal (\textit{symmetric}) energy splitting is more favored than (\textit{asymmetric}) energy splitting. This is more indicative of a thermal system which is characterized by equipartitioning of energies.

So far, we have shown preliminary evidence that we have found the thermalization vertex in the D1D5 CFT. However, more work is needed is solidify this claim. Because $N=2$ in our computations, this is not a regime which describes a `good' black hole. To make a `good' black hole we need $N$ to be large.  Also, we have only computed one level of splitting with the two deformation operators. We argue, qualitatively, that repeated applications of two deformation operators will move the initial state into the space of all states and thus a thermal state, however we would like to show this in more detail. Though we considered fairly large energies in Table's  \ref{equalsplitting}, \ref{asymmetricsplitting} we would like to consider even larger energies in the initial state in order to see exactly how the CFT thermalizes. It appears that the preferred splitting mechanism is amongst equal energies in the final state. A detailed analysis of the splitting mechanism will help us to determine the phase space evolution of the initial state. It is this phase space evolution which should yield an increase in the entropy of the system, driving it towards thermalization and thus black hole formation in the AdS dual. We hope to address this in future work.

\chapter{Conclusion and Future Outlook}
In this report we reviewed many things. Our primary goal was to understand black hole formation in string theory. To do this we utilized the AdS/CFT correspondence. Black hole formation in a gravity theory corresponds to the thermalization of the dual CFT. The CFT dual is where we performed all of our computations. In order to form a black hole in the bulk either 1) an initial shell of energy, which we identify as a scalar particle, must collapse due to gravitational interactions to form a black hole or 2) two scalar particles must collide with sufficient energy to form a black hole. In either case, the formation process corresponds to a strongly coupled regime in which the gravitational interactions become nonperturbative. In the CFT dual, however, this corresponds to a weakly interacting field theory of `effective strings.' We therefore turned on a perturbative deformation of the theory, the twist operator. This twist operator joined together and split apart `effective strings.'  In our scenario we show 
that case 2) is more likely. We compute the vertex in the CFT which should eventually lead to thermalization. We highlight the major results from each chapter, putting emphasis on Chapter \ref{thermalization}.

In Chapter \ref{D1D5 system} we introduced the D1D5 supergravity solution, the D1D5 CFT, and the twist operator.

In Chapter \ref{sec:gamma} we computed the state obtained from applying a single twist to two initial strings of arbitrary winding. This gave a single multiwound string excited by an exponential of bilinear excitations weighted by Bogoluibov coefficients. This is the single twist $|\chi\rangle$ state.

In Chapter \ref{f function} we computed the state obtained from applying the full deformation, the twist and the supercharge, to a single excitation in the initial state. The final states were characterized by a linear combination of single excitations weighted by a transition function along with bilinear excitations weighted by a different set of coefficients, both applied on the single twist $|\chi\rangle$  state.

In Chapter \ref{two twist gamma chapter} we computed the two twist $|\chi\rangle$ state. The first twist joined two singly wound strings in the initial state into a doubly wound string. The second twist split the doubly wound string back into two singly wound strings with an exponential of bilinear excitations weighted by Bogoluibov coefficients for two twists.

In Chapter \ref{two twist f} we applied two twist operators to an initial excitation. We computed the resulting transition functions. These functions explicitly show the transition of an initial mode from one copy to the other as time is evolved on the cylinder.

In Chapter \ref{G contours} we computed the action of the supercharges in the two twist case. This was done separately in contrast to the one twist case because the two twist case was significantly more complicated. We could stretch part of the $G$ contours to initial and final states on the cylinder but in doing so, we were left with contours sitting between the two twist operators. This is a very complicated state on the cylinder. This required us to map the $G$ contours to the $t$ plane, move them to initial and final states, and then map back to the cylinder. We therefore computed expressions for the contours at initial and final states on the cylinder using this method.

In Chapter \ref{lifting chapter} we computed the lifting of the conformal dimension for a set of non chiral primary states under the application of two deformation operators, $D$, of the theory. To do this, we used conformal pertubation theory. These states were constructed by applications of the current operators $J^+$. We found a simple expression for the lifting of these states.

In Chapter \ref{thermalization} we identified the thermalization vertex of the D1D5 CFT. A signal of thermalization is if an initial excitation breaks apart into lower energy excitations in the final state. For the single twist case the final results were unclear. This is because the box size was changing. We started with two multiwound strings and twisted them into a single multiwound string. Since the vacuum was changing size, it wasn't clear if this was affecting the splitting process. We couldn't effectively separate the vacuum effects from an actual thermalization process. Also, in the single twist case, energy conservation could not be enforced while simultaneously enforcing momentum conservation.  This drastically affected the splitting behavior of initial states into lower energy states. Both of these consequences prompted us to analyze the second order case where we twisted two strings together and then untwisted them. Here, the final box size is the same as the initial box size. Two singly wound strings go to one doubly wound string and back to two singly wound strings. In this scenario we considered 1) a single excitation in the initial state splitting into three excitations and 2) two excitations in the initial state splitting into four excitations. We found that these two cases differed. The splitting amplitude corresponding to one excitation in the initial state had a term which \textit{oscillated} with time, $t$ where $t$ measures the duration of the deformation. The splitting amplitude corresponding to two excitations in the initial state had components which oscillate with $t$ and grow with $t^2$. This `secular' term arose as a consequence of allowed intermediate states which were degenerate with initial and final state energies for the `2 to 4' splitting process. This is not allowed in the `1 to 3' splitting process which is why no secular term appeared. 


For long timescales the `2 to 4' splitting amplitude dominated the `1 to 3' splitting amplitude. Since the amplitude is related to the splitting probability, we see that $two$ excitations are  more likely to split into four excitations. Two excitations in the initial state in the CFT correspond to two scalars moving in AdS. Turning on the deformation in the presence of the two excitations in the CFT correspond to the two scalars interacting in AdS. We've computed the thermalization vertex in the CFT which is necessary to prove thermalization. However, more work is still needed.
We took $N=2$ which doesn't produce a 'good' black hole but provides a low energy threshold for formation. We tabulated values of coefficients corresponding to `1 to 3' splitting and `2 to 4' splitting. At large $t$ `2 to 4' spitting clearly dominated. But we would like to understand this mechanism in greater detail. For example, we would like to know exactly how the initial states split in order to reach a thermal state. Preliminary computations indicate that the preferred splitting mechanism is into equal energies in the final state, however we would like to investigate this in more detail.
Also to produce a `good' black hole in AdS, we need $N$ to be large. With $N$ large the number of ways for an initial state to split apart drastically increases. This will produce a large entropy in the final state. The large entropy comes from how many ways an initial mode or modes can split into lower energy modes, and the size of the hole is related to the entropy of the hole.
Additionally, we can think of this `2 to 4' process being iterated many times to produce a final state with many low energy modes. We argue qualitatively that given enough energy in the initial state, the phase space for the final state is large enough to produce a thermal state which corresponds to a black hole forming in the dual picture. In our scenario, this formation process, corresponds tunneling into a new fuzzball state. We would like to show this more concretely in future work.

\backmatter

\begin{appendix}
\chapter{Notation and conventions} \label{app_cft}
\section{Field Definitions}\label{Field Definitions}
Here we give the notation and conventions used in our computations.
We have 4 real left moving fermions $\psi_1, \psi_2, \psi_3, \psi_4$ which are grouped into doublets $\psi^{\alpha A}$ as follows:
\be
\begin{pmatrix}
\psi^{++} \cr \psi^{-+}
\end{pmatrix}
={1\over\sqrt{2}}
\begin{pmatrix}
\psi_1+i\psi_2 \cr \psi_3+i\psi_4
\end{pmatrix}
\ee
\be
\begin{pmatrix}
\psi^{+-} \cr \psi^{--}
\end{pmatrix}
={1\over\sqrt{2}}
\begin{pmatrix}
\psi_3-i\psi_4 \cr -(\psi_1-i\psi_2)
\end{pmatrix}.
\ee
The index $\alpha=(+,-)$ corresponds to the subgroup $SU(2)_L$ of rotations on $S^3$ and the index $A=(+,-)$ corresponds to the subgroup $SU(2)_1$ from rotations in $T^4$. The 2-point functions read
\be
\langle\psi^{\alpha A}(z)\psi^{\beta B}(w)\rangle=-\epsilon^{\alpha\beta}\epsilon^{AB}{1\over z-w}
\ee
where we have 
\be
\epsilon_{+-}=1, ~~~\epsilon^{+-}=-1
\ee
The 4 real left-moving bosons $X_1, X_2, X_3, X_4$ are grouped into a matrix 
\be
X_{A\dot A}= {1\over\sqrt{2}} X_i \sigma_i
={1\over\sqrt{2}}
\begin{pmatrix}
X_3+iX_4 & X_1-iX_2 \cr X_1+iX_2&-X_3+iX_4
\end{pmatrix}
\ee
where $\sigma_i=(\sigma_a, iI)$. The bosonic field 2-point functions are then of the form
\be
\langle\partial X_{A\dot A}(z) \partial X_{B\dot B}(w)\rangle=\epsilon_{AB}\epsilon_{\dot A\dot B}{1\over (z-w)^2} \,.
\ee
The chiral algebra is generated by the R-currents, supercurrents, and the stress-energy tensor:
\bea
J^a&=& {1\over 4}\e_{\a \g}\e_{AC}\psi^{\g C} (\sigma^{Ta})^\alpha{}_\beta \psi^{\beta A},\qquad a=1,2,3
\cr\cr
G^\alpha_{\dot A}&=& \psi^{\alpha A} \pa X_{A\dot A},\qquad \a = +,-
\cr\cr
T&=& {1\over 2} \e^{AB}\e^{\dot{A}\dot{B}}\pa X_{B\dot B}\pa X_{A\dot A} + {1\over 2} \e_{\a\beta}\e_{AB}\psi^{\beta B} \pa \psi^{\alpha A}
\eea

\section{OPE Algebra}
We note the OPEs between the various operators of interest.

\subsection{OPE's of currents with $\partial X_{A\dot{A}}(z)$ and $\psi^{\a A}(z)$ }

\bea
T(z)\partial X_{A\dot{A}}(w) &\sim& {\partial X_{A \dot A}(w)\over (z-w)^2} + { \partial^2 X_{A \dot A}(w)\over z-w} \cr
T(z)\psi^{\a A}(w) &\sim&  {{1\over 2}\psi^{\a A}(w)\over (z-w)^2}+{\partial \psi^{\a A}(w)\over z-w} \cr
G^{\a}_{\dot{A}}(z)\psi^{\beta B}(w) &\sim& \e^{\a\beta}\e^{BA} {\partial X_{A\dot{A}}(w) \over z-w}\cr
G^{\a}_{\dot{A}}(z)\partial X_{ B\dot{B}}(w)&\sim& \e_{AB}\e_{\dot{A}\dot{B}}{ \psi^{\a A}(w)\over  (z-w)^2} + \e_{AB}\e_{\dot{A}\dot{B}}{ \partial \psi^{\a A}(w)\over  z-w}\cr
J^a(z)\psi^{\a A}(w) &\sim&  {1\over 2} {1\over z-w}(\s^{Ta})^{\a}_{\beta}\psi^{\beta A}(w)\cr
J^{+}(z)\psi^{+ A}(w) &=& 0,\qquad J^{-}(z)\psi^{+ A}(w) =   {\psi^{- A}(w)\over z-w} \cr
J^{+}(z)\psi^{- A}(w) &=&   {\psi^{+ A}(w)\over z-w} ,\qquad  J^{-}(z)\psi^{- A}(w) = 0
\eea

\subsection{OPE's of currents with currents }

\bea
T(z)T(w)&\sim&{{c\over2}\over (z-w)^4} + {2T(w)\over (z-w)^2}  + {\partial T(w)\over z-w}
\cr
J^a(z)J^b(w)&\sim&  {{c\over 12}\delta^{ab}\over(z-w)^2} + {i\e^{ab}_{\,\,\,\,c}J^c(w)\over z-w}
\cr
G^{\a}_{\dot{A}}(z)G^{\beta}_{\dot{B}}(w) &\sim& -\e_{\dot{A}\dot{B}}\bigg[\e^{\beta\a}{{c\over3}\over (z-w)^3}  + \e^{\beta\g}(\s^{aT})^{\a}_{\g}\bigg({2J^a(w)\over (z-w)^2} + {\partial J^a(w)\over z-w}\bigg)  + \e^{\beta\a}{1\over z-w}T(w)   \bigg]
\cr
J^a(z)G^{\a}_{\dot{A}}(w)&\sim& {1\over z-w}{1\over2}(\s^{aT})^{\a}_{\beta}G^{\beta}_{\dot{A}}(w)
\cr
T(z)J^a(w) &\sim&  {J^a(w)\over (z-w)^2} + {\partial J^a(w)\over z-w}
\cr
T(z)G^{\a}_{\dot{A}}(w)&\sim& {{3\over2}G^{\a}_{\dot{A}}(w)\over (z-w)^2} + { \partial G^{\a}_{\dot{A}}(w)\over z-w}
\eea
We convert the relations involving $J^1, J^2$ to those involving $J^+, J^-$. Defining $J^{+}, J^-$ as
\bea
J^+ &=& J^1 + i J^2\cr
J^-&=& J^1 - i J^2
\eea
yield the following OPE's
\begin{align}
J^{+}(z)J^{-}(w)&\sim  {{c\over6}\over (z-w)^2} +  {2J^3(w)\over z-w},& J^{-}(z)J^{+}(w)&\sim  {{c\over6}\over (z-w)^2} -  {2J^3(w)\over z-w}\cr
J^3(z)J^{+}(w)&\sim  {J^{+}(w)\over z-w}, & J^3(z)J^{-}(w)&\sim- {J^{-}(w)\over z-w}\cr
J^+(z)J^3(w)&\sim- {J^+(w)\over z-w}, &J^-(z)J^3(w)&\sim {J^-(w)\over z-w}\cr
T(z)J^{+}(w)&\sim{J^{+}(w)\over (z-w)^2} + {\partial J^{+}(w)\over z-w},&T(z)J^{-}(w)&\sim {J^{-}(w)\over (z-w)^2} + {\partial J^{-}(w)\over z-w}\cr
J^{+}(z)G^{-}_{\dot{A}}(w) &\sim {G^{+}_{\dot{A}}(w)\over z-w},& J^{-}(z)G^{+}_{\dot{A}}(w)& \sim {G^{-}_{\dot{A}}(w)\over z-w}
\end{align}

\section{Mode and contour definitions of the fields} 

The modes are defined in terms of contours through
\bea
L_m&=&\oint {dz\over 2\pi i}z^{m+1}T(z)\cr
J^a_m&=&\oint {dz\over 2\pi i} z^{m}J^a(z)\cr
G^{\a}_{\dot{A},r}&=&\oint {dz\over 2\pi i} z^{r+{1\over2}}G^{\a}_{\dot{A}}(z)\cr
\a_{A\dot{A},m}&=&i\oint {dz\over 2\pi i} z^{m}\partial X_{A\dot{A}}(z)\cr
d^{\a A}_r&=& \oint {dz\over 2\pi i} z^{r-{1\over2}}\psi^{\a A}(z)
\eea 

The inverse relations are
\bea
T(z) &=& \sum_{m}z^{-m-2}L_m\cr
J^a(z) &=& \sum_{m}z^{-m-1}J^a_m\cr
G^{\a}_{\dot{A}}(z) &=& \sum_{r}z^{-r-{3\over2}}G^{\a}_{\dot{A},r}\cr
\partial X_{A\dot{A}}(z) &=& -i\sum_m z^{-m-1}\a_{A\dot{A},m}\cr
\psi^{\a A}(z) &=& \sum_m z^{-m-{1\over2}}d^ {\a A}_m
\eea

\section{Commutation relations}
\subsection{Commutators of $\alpha_{A\dot{A},m}$ and $d^{\a A}_r$}

\bea
[\a_{A\dot{A},m},\a_{B\dot{B},n}] &=& -m\e_{A\dot{A}}\e_{B\dot{B}}\delta_{m+n,0}\cr
\lbrace d^{\alpha A}_r , d^{\beta B}_s\rbrace  &=&-\e^{\alpha\beta}\e^{AB}\delta_{r+s,0}
\eea

\subsection{Commutators of currents with $\alpha_{A\dot{A},m}$ and $d^{\a A}_r$}
\bea\label{commutations}
[L_m,\a_{A\dot{A},n}] &=&-n\a_{A\dot{A},m+n} \cr
[L_m ,d^{\a A}_r] &=&-({m\over2}+r)d^{\a A}_{m+r}\cr
\lbrace G^{\a}_{\dot{A},r} ,  d^{\beta B}_{s} \rbrace&=&i\e^{\a\beta}\e^{AB}\a_{A\dot{A},r+s}\cr
[G^{\a}_{\dot{A},r} , \a_{B \dot{B},m}]&=&  -im\e_{AB}\e_{\dot{A}\dot{B}}d^{\a A}_{r+m}\cr
[J^a_m,d^{\a A}_r] &=&{1\over 2}(\s^{Ta})^{\a}_{\beta}d^{\beta A}_{m+r}\cr
[J^{+}_m,d^{+ A}_r] &=& 0,\qquad~~~~~ [J^{-}_m,d^{+ A}_r] ~=~ d^{-A}_{m+r}\cr
[J^{-}_m,d^{+ A}_r] &=& d^{-A}_{m+r},\qquad [J^{+}_m,d^{+ A}_r] ~=~ 0
\eea

\subsection{Commutators of currents with currents}
\bea\label{commutations_ii}
[L_m,L_n] &=& {c\over12}m(m^2-1)\delta_{m+n,0}+ (m-n)L_{m+n}\cr
[J^a_{m},J^b_{n}] &=&{c\over12}m\delta^{ab}\delta_{m+n,0} +  i\e^{ab}_{\,\,\,\,c}J^c_{m+n}\cr
\lbrace G^{\a}_{\dot{A},r} , G^{\beta}_{\dot{B},s} \rbrace&=&  \e_{\dot{A}\dot{B}}\bigg[\e^{\a\beta}{c\over6}(r^2-{1\over4})\delta_{r+s,0}  + (\s^{aT})^{\a}_{\g}\e^{\g\beta}(r-s)J^a_{r+s}  + \e^{\a\beta}L_{r+s}  \bigg]\cr
[J^a_{m},G^{\a}_{\dot{A},r}] &=&{1\over2}(\s^{aT})^{\a}_{\beta} G^{\beta}_{\dot{A},m+r}\cr
[L_{m},J^a_n]&=& -nJ^a_{m+n}\cr
[L_{m},G^{\a}_{\dot{A},r}] &=& ({m\over2}  -r)G^{\a}_{\dot{A},m+r}\cr
[J^+_{m},J^-_{n}]&=&{c\over6}m\delta_{m+n,0} + 2J^3_{m+n}\cr
[L_m,J^{+}_n] &=& -nJ^{+}_{m+n},\qquad ~[L_m,J^{-}_n] ~=~ -nJ^{-}_{m+n}\cr
[J^{+}_{m},G^{+}_{\dot{A},r}]  &=& 0 ,\qquad\qquad ~~~[J^{-}_{m},G^{+}_{\dot{A},r}]  ~=~ G^{-}_{\dot{A},m+r}\cr
[J^{+}_{m},G^{-}_{\dot{A},r}]  &=&G^{+}_{\dot{A},m+r},\qquad ~[J^{-}_{m},G^{-}_{\dot{A},r}]  ~=~ 0 \cr
[J^3_m , J^{+}_n] &=& J^{+}_{m+n},\qquad\qquad [J^3_m , J^{-}_n] ~=~ -J^{-}_{m+n}
\eea

\section{Current modes written in terms of $\alpha_{A\dot{A},m}$ and $d^{\a A}_r$}
\bea
J^a_m &=& {1\over 4}\sum_{r}\epsilon_{AB}d^ {\g B}_r\epsilon_{\alpha\gamma}(\s^{aT})^{\a}_{\beta}d^ {\beta A}_{m-r},\qquad a=1,2,3\cr
J^3_m &=&  - {1\over 2}\sum_{r} d^ {+ +}_{r}d^ {- -}_{m-r} - {1\over 2}\sum_{r}d^ {- +}_r d^ {+ -}_{m-r}\cr
J^{+}_m&=&\sum_{r}d^ {+ +}_rd^ {+ -}_{m-r} ,\qquad J^{-}_m=\sum_{r}d^ {--}_rd^ {- +}_{m-r}\cr
G^{\a}_{\dot{A},r} &=& -i\sum_{n}d^ {\a A}_{r-n} \a_{A\dot{A},n}\cr
L_m&=& -{1\over 2}\sum_{n} \e^{AB}\e^{\dot A \dot B}\a_{A\dot{A},n}\a_{B\dot{B},m-n}- {1\over 2}\sum_{r}(m-r+{1\over2})\epsilon_{\alpha\beta}\epsilon_{AB}d^ {\a A}_r d^ {\beta B}_{m-r}\nn
\eea

\section{Right moving algebra}
We note that the right moving algebra is identical to all of the algebra presented above with $z\to \bar z$.
\chapter{Physical States of the D1D5 CFT}\label{physical states}
In this appendix we convert the notation for bosonic fields $\partial X_{A\dot{A}},\bar\partial \bar X_{A\dot{A}}$, which are written in the $SU(2)_1\times SU(2)_2$ language, into the notation for physical states, which correspond to excitations along the physical directions of the torus. Let us start with the definitions of $X_{A\dot{A}}$. Using Appendix \ref{Field Definitions} we have

\bea
X_{++}&=&{1\over\sqrt 2}(X_3 + iX_4),\quad \bar{X}_{++} = {1\over \sqrt 2}(\bar X_3 + i\bar X_4)\cr
X_{--}&=&{1\over\sqrt 2}(-X_3 + iX_4),\quad \bar X_{--}={1\over\sqrt2}(-\bar X_3 + i\bar X_4)\cr
X_{+-}&=&{1\over\sqrt 2}(X_1 - iX_2),\quad \bar X_{+-}={1\over\sqrt2}(\bar X_1 - i\bar X_2)\cr
X_{-+}&=&{1\over\sqrt 2}(X_1 + iX_2),\quad \bar X_{-+}={1\over\sqrt2}(\bar X_1 + i\bar X_2)
\label{X}
\eea
Where $X_{A\dot{A}}$ and $\bar{X}_{A\dot{A}}$ are defined independently from one another which is why there is no conjugation of $i$. Their algebras commute.
Inverting the above equations, we get
\bea
X_1&=&{1\over\sqrt 2}(X_{+-}+X_{-+}),\quad \bar X_1={1\over\sqrt 2}(\bar X_{+-}+\bar X_{-+})\cr
X_2&=&{1\over\sqrt 2i}(X_{-+} - X_{+-}),\quad \bar X_2=-{1\over\sqrt 2i}(X_{-+} - X_{+-})\cr
X_3 &=& {1\over \sqrt 2}(X_{++} - X_{--}),\quad \bar X_{3}  = {1\over\sqrt 2}(\bar X_{++} - \bar X_{--}) \cr
X_4 &=& {1\over \sqrt 2i}(X_{++} + X_{--}),\quad \bar X_{4}  = -{1\over\sqrt 2i}(\bar X_{++} +\bar  X_{--})
\eea
We now borrow notation from excitations of a closed string as it is analogous to the situation of a left and right moving open string excitation along a Dbrane. We have four transverse polarization directions which are $X_1,X_2,X_3,X_4$, the directions of $T^4$. The state space at the massless level is given by
\bea
\sum_{i,j=1}^4R^{ij}\partial X_{i}\bar \partial X_j
\eea
where $R^{ij}$ is an arbitrary $4\times 4$ matrix. Now we can decompose $R_{ij}$ as follows
\bea
R^{ij} = S^{ij} + A^{ij}
\eea 
where $S^{ij}$ is a symmetric matrix and $A^{ij}$ is an antisymmetric matrix. The symmetric part can be further decomposed into a symmetric, traceless part, $\hat{S}^{ij}$, and a trace part ${1\over 6-2}S$ where
\bea
\hat{S}^{ij} = S^{ij} - {1\over 6-2}\d^{ij}S
\eea
where $D=6$ because we have the CFT coordinates $t,y$ and the torus directions $x_1,x_2,x_3,x_4$ which gives a total of $6$ dimensions. We note that $\hat{S}^{ij}$ is arbitrary and could be chosen to be any such symmetric, traceless matrix. Looking at the symmetric traceless part we have the possible states
\bea
\sum_{i,j=1}^4\hat{S}^{ij}\partial X_{i}\bar \partial X_j
\label{graviton states}
\eea
These states are in one to one correspondence with graviton states and $\hat{S}^{ij}$ is analogous to the polarizations of the graviton states. Since $\hat{S}^{ij}$ is symmetric and traceless lets write an example matrix which fits these properties
\bea
\hat{S}^{ij}=\begin{pmatrix}
1&1&1&1\\
1&1&1&1\\
1&1&-1&1\\
1&1&1&-1
\end{pmatrix}
\label{symmetric matrix}
\eea
Defining the components of the graviton states, $h_{ij}$, below we have
\bea
h_{12} &=& h_{21} = {1\over\sqrt2}(\partial X_1\bar\partial \bar X_2 + \partial X_2\bar\partial\bar X_1)\cr
h_{13} &=& h_{31} = {1\over\sqrt2}(\partial X_1\bar\partial \bar X_3 + \partial X_3\bar\partial\bar X_1)\cr
h_{14} &=& h_{41} = {1\over\sqrt2}(\partial X_1\bar\partial \bar X_4 + \partial X_4\bar\partial\bar X_1)\cr
h_{23}&=& h_{32} = {1\over\sqrt2}(\partial X_2\bar\partial \bar X_3 + \partial X_3\bar\partial\bar X_2)\cr
h_{24} &=& h_{42} = {1\over\sqrt2}(\partial X_2\bar\partial \bar X_4 + \partial X_4\bar\partial\bar X_2)\cr
h_{34} &=& h_{43} = {1\over\sqrt2}(\partial X_3\bar\partial \bar X_4 + \partial X_4\bar\partial\bar X_3)\cr
h_{11}&=& \partial X_1\bar\partial \bar X_1\cr
h_{22}&=& \partial X_2\bar\partial\bar X_2\cr
h_{33}&=& -\partial X_3\bar\partial \bar X_3\cr
h_{44}&=&-\partial X_4\bar\partial \bar X_4= -(\phi-h_{11} - h_{22} + h_{33})
\eea
For the antisymmetric part we have
\bea
\sum_{i,j=1}^4A^{ij}\partial X_{i}\bar \partial X_j
\label{B state}
\eea
The antisymmetric states are in one to one correspondence with the RR 2 form $B^{RR}_{ij}$ which couples to the $D1$ brane.
We write the antisymmetric matrix as
\bea
A^{ij} = \begin{pmatrix}
0&1&1&1\\
-1&0&1&1\\
-1&-1&0&1\\
-1&-1&-1&0
\end{pmatrix}
\label{antisymmetric matrix}
\eea
Writing the states as components of $B^{RR}_{ij}$, we have
\bea
B^{RR}_{12}&=&-B^{RR}_{21} = {1\over \sqrt 2} ( \partial X_1\bar\partial \bar X_2 - \partial X_2\bar\partial\bar X_1 )\cr
B^{RR}_{13}&=&-B^{RR}_{31} = {1\over \sqrt 2} ( \partial X_1\bar\partial \bar X_3 - \partial X_3\bar\partial\bar X_1 )\cr
B^{RR}_{14}&=&-B^{RR}_{41} = {1\over \sqrt 2} ( \partial X_1\bar\partial \bar X_4 - \partial X_4\bar\partial\bar X_1 )\cr
B^{RR}_{23}&=&-B^{RR}_{32} = {1\over \sqrt 2} ( \partial X_2\bar\partial \bar X_3 - \partial X_3\bar\partial\bar X_2 )\cr
B^{RR}_{24}&=&-B^{RR}_{42} = {1\over \sqrt 2} ( \partial X_2\bar\partial \bar X_4 - \partial X_4\bar\partial\bar X_2 )\cr
B^{RR}_{34}&=&-B^{RR}_{43} = {1\over \sqrt 2} ( \partial X_3\bar\partial \bar X_4 - \partial X_4\bar\partial\bar X_3 )
\eea
The trace part corresponds to the $6D$ dilaton
\bea
{1\over 6-2}S\phi
\label{dilaton}
\eea
which is given by
\bea
\phi = {1\over 2} ( \partial X_1 \bar\partial \bar X_1  +  \partial X_2 \bar\partial \bar X_2 + \partial X_3 \bar\partial \bar X_3 + \partial X_4 \bar\partial \bar X_4) 
\eea

Now lets compute possible combinations of states which will be useful for our purposes. Using (\ref{X}) we have
\bea
\partial{X}_{--}\bar \partial\bar X_{++} &=&{1\over2}(-\partial X_3 + i\partial X_4)(\bar\partial \bar X_3 + i\bar\partial\bar X_4)\cr
&=&{1\over 2}(-\partial X_3\bar\partial \bar X_3 -\partial X_4 \bar\partial\bar X_4  - i \partial X_3\bar\partial\bar X_4 +  i\partial X_4\bar\partial \bar X_3)\cr
&=&{1\over 2}(h_{11} + h_{22}-2\phi-i\sqrt{2}B^{RR}_{34})
\eea
where we see that the above state is formed from a linear combination of linearly independent graviton, antisymmetric tensor and dilaton states given in (\ref{graviton states}),(\ref{B state}) and (\ref{dilaton}).
Similarly for other charge combinations we have
\bea
\partial{X}_{++}\bar \partial\bar X_{--} &=&  {1\over2}(\partial X_3 + i \partial X_4) (-\bar\partial \bar X_3 + i \bar\partial \bar X_4)\cr
&=& {1\over2}(-\partial X_3\bar\partial \bar X_3 - \partial X_4\bar\partial \bar X_4 + i(\partial X_3\bar \partial \bar X_4 - \partial X_4\bar \partial \bar X_3))\cr
&=&{1\over2}(h_{11} + h_{22}-2\phi+i\sqrt{2}B^{RR}_{34})
\cr
\cr
\partial{X}_{+-}\bar \partial\bar X_{-+}&=& {1\over2}(\partial X_1 - i\partial X_2)(\bar \partial \bar X_1 + i\bar \partial \bar X_2)\cr
&=&{1\over2}(\partial X_1 \bar \partial \bar X_1 + \partial X_2\bar \partial \bar X_2 + i(\partial X_1\bar \partial \bar X_2 - \partial X_2\bar \partial \bar X_1))\cr
&=&{1\over2}(h_{11} + h_{22}+i\sqrt{2}B^{RR}_{12})
\cr
\cr
\partial{X}_{-+}\bar \partial\bar X_{+-}&=& {1\over2}(\partial X_1 + i\partial X_2)(\bar \partial \bar X_1 - i\bar \partial \bar X_2)\cr
&=&{1\over2}(\partial X_1 \bar \partial \bar X_1 + \partial X_2\bar \partial \bar X_2 - i\sqrt{2}(\partial X_1\bar \partial \bar X_2 - \partial X_2\bar \partial \bar X_1))\cr
&=&{1\over2}(h_{11} + h_{22}-i\sqrt{2}B^{RR}_{12})
\cr
\cr
\partial{X}_{++}\bar \partial\bar X_{++}&=&{1\over2}(\partial X_3 + i\partial X_4)(\bar\partial \bar X_3 + i\bar\partial\bar X_4)\cr
&=&{1\over2}(\partial X_3\bar\partial \bar X_3 - \partial X_4\bar\partial\bar X_4 + i(\partial X_3\bar \partial \bar X_4 + \partial X_4\bar \partial \bar X_3))\cr
&=&{1\over2}(h_{11} + h_{22}-2\phi-2h_{33}+i\sqrt{2}h_{34})
\cr
\cr
\partial{X}_{--}\bar \partial\bar X_{--}&=&{1\over2}(-\partial X_3 + i\partial X_4)(-\bar\partial \bar X_3 + i\bar\partial\bar X_4)\cr
&=&{1\over2}(\partial X_3\bar\partial \bar X_3 - \partial X_4\bar\partial\bar X_4 - i(\partial X_3\bar \partial \bar X_4 + \partial X_4\bar \partial \bar X_3))\cr
&=&{1\over2}(h_{11} + h_{22}-2\phi-2h_{33} - i\sqrt{2}h_{34})
\cr
\cr
\partial{X}_{+-}\bar \partial\bar X_{+-}&=&{1\over2}(\partial X_1 - i\partial X_2)(\bar \partial \bar X_1 - i\bar \partial \bar X_2)\cr
&=&{1\over2}(\partial X_1\bar\partial \bar X_1 - \partial X_2\bar\partial\bar X_2 - i(\partial X_1\bar \partial \bar X_2 + \partial X_2\bar \partial \bar X_1))\cr
&=&{1\over2}(h_{11} - h_{22} - i\sqrt{2}h_{12})
\cr
\cr
\partial{X}_{-+}\bar \partial\bar X_{-+}&=&{1\over2}(\partial X_1 + i\partial X_2)(\bar \partial \bar X_1 + i\bar \partial \bar X_2)\cr
&=&{1\over2}(\partial X_1\bar\partial \bar X_1 - \partial X_2\bar\partial\bar X_2 + i(\partial X_1\bar \partial \bar X_2 + \partial X_2\bar \partial \bar X_1))\cr
&=&{1\over2}(h_{11} - h_{22} + i\sqrt{2}h_{12})\nn
\eea

\chapter{Mapping modes to the covering plane for the single twist case}\label{sec:mapping}

In this appendix we compute the transformations of the fermion modes for the single twist under the sequence of spectral flow transformations and coordinate changes described in Section \ref{sec:sf}. 

\section{Modes on the cylinder}

We start by recalling from Section \ref{sec:expansions} the mode expansions on the cylinder. 

Below the twist, we have:
\begin{eqnarray}
d_m^{(1)\alpha A} &=& \frac{1}{2\pi i\sqrt M} \int\limits_{\sigma=0}^{2\pi M} \psi^{(1)\alpha A} (w) e^{\frac{m}{M}w} dw\\
d_m^{(2)\alpha A} &=&  \frac{1}{2\pi i\sqrt N} \int\limits_{\sigma=0}^{2\pi N} \psi^{(2)\alpha A} (w) e^{\frac{m}{N}w} dw,
\end{eqnarray}
while above the twist, we have:
\begin{eqnarray}
d_k^{\alpha A} &=& \frac{1}{2\pi i\sqrt{M+N}} \int\limits_{\sigma=0}^{2\pi (M+N)} \psi^{\alpha A} (w) e^{\frac{k}{M+N}w} dw,
\end{eqnarray}
which gives
\begin{eqnarray}
\psi^{\alpha A}\left(w\right)=\frac{1}{\sqrt{M+N}}\sum_{k}d_{k}^{\alpha A}e^{-\frac{k}{M+N}w}.
\end{eqnarray}
The anti-commutation relations are
\begin{eqnarray}
\left\{d_k^{\alpha A}, d_l^{\beta B}\right\} &=& -\varepsilon^{\alpha\beta} \varepsilon^{AB} \delta_{k+l,0}.
\end{eqnarray}

\section{Modes on the $z$ plane}

We map the cylinder to the plane with coordinate $z$ via
\begin{eqnarray}
z &=& e^w.
\end{eqnarray}
The operator modes transform as follows. Before the twist, i.e. $|z| < e^{\tau_0}$, using a contour circling $z=0$, we have
\begin{eqnarray}
d_m^{(1)\alpha A} & \to & \frac{1}{2\pi i \sqrt{M}} \int\limits^{2\pi M}_{\arg (z) = 0} \psi ^{(1) \alpha A} (z) z^{\frac{m}{M} - \frac12} dz\cr
d_m^{(2)\alpha A} & \to & \frac{1}{2\pi i\sqrt{N}} \int\limits_{\arg (z) = 0}^{2\pi N} \psi^{(2)\alpha A} (z) z^{\frac{m}{N}-\frac12} dz.
\end{eqnarray}
After the twist, i.e. $|z| > e^{\tau_0}$, using a contour circling $z=\infty$, we have
\begin{eqnarray}
d_k^{\alpha A} &\to& \frac{1}{2\pi i \sqrt{M + N}} \int\limits_{\arg (z) = 0}^{2\pi (M+N)} \psi ^{\alpha A}(z) z^{\frac{k}{M+N} - \frac{1}{2}} dz.
\end{eqnarray}

\section{Modes on the covering space}

We now map the problem to the covering space with coordinate $t$, where the fields are single-valued.  We use the map
\begin{eqnarray}
z &=& t^M (t-a)^N.
\end{eqnarray}
It is convenient to write the derivative as
\begin{eqnarray}
\frac{dz}{dt} &=& (M+N) \frac{z}{t(t-a)} \left(t -\tfrac{M}{M+N}a\right).
\end{eqnarray}
Under this map, the modes before the twist become:
\begin{eqnarray}
d_m^{(1)\alpha A} &\to & \frac{\sqrt{M+N}}{2\pi i\sqrt{M}} \oint\limits_{t = 0} dt \, \psi^{\alpha A} (t) \left[t^{m-\frac12} \left(t - \tfrac{M}{M+N}a\right)^{\frac12} (t-a)^{\frac{Nm}{M} - \frac12}\right]\cr
d_m^{(2)\alpha A} &\to & \frac{\sqrt{M+N}}{2\pi i\sqrt{N}} \oint\limits_{t = a} dt \, \psi^{\alpha A} (t) \left[t^{\frac{Mm}{N}-\frac12} \left(t - \tfrac{M}{M+N}a\right)^{\frac12} (t-a)^{m-\frac12}\right]
\end{eqnarray}
and after the twist, we have:
\begin{eqnarray}
d_k^{\alpha A} &\to&  \frac{1}{2\pi i} \oint\limits_{t = \infty} dt \, \psi^{\alpha A} (t) \left[t^{\frac{Mk}{M+N}-\frac12} \left(t - \tfrac{M}{M+N}a\right)^{\frac12} (t-a)^{\frac{Nk}{M+N} - \frac12}\right].
\end{eqnarray}

\section{Modes after the first spectral flow}

We now perform the sequence of spectral flow transformations and coordinate changes described in Section \ref{sec:sf}.
As explained there, at this point we ignore normalization factors of spin fields and transformation properties of spin fields under spectral flow. We spectral flow in the same way on left- and right-moving sectors,  but write only the holomorphic expressions. 

We first spectral flow by $\alpha = 1$ in the $t$ plane. Before the spectral flow, we have the amplitude
\begin{eqnarray}
&&{}_t \bra{0_{R,-}} S^- (a) S^+ \left(\tfrac{M}{M+N}a\right)\ket{0_R^-}_t.
\end{eqnarray}
The spectral flow has the following effects:
\begin{enumerate}
\item[(a)] The R ground states $\ket{0_R^-}_t$ and ${}_t\bra{0_{R,-}}$ map to the NS ground states in the $t$ plane:
\be
\ket{0_R^-}_t \to \ket{0_{NS}}_t, \qquad {}_t\bra{0_{R,-}} \to {}_t\bra{0_{NS}}
\ee
\item[(b)] The modes of the fermion fields change as follows (the bosonic modes are unaffected):
    \begin{eqnarray}
    \psi^{\pm A} (t)&\to &t^{\mp\frac12} \psi^{\pm A}(t).
    \end{eqnarray}
\end{enumerate}
Then the modes before the twist become:
    \begin{eqnarray}
    d_m^{(1) + A} &\to& \frac{\sqrt{M+N}}{2\pi i \sqrt M} \oint\limits_{t = 0} {\rm dt} \psi^{(1)+A} (t) \left[t^{m-1} \left(t - \tfrac{M}{M+N}a\right) ^{\frac12} (t-a)^{\frac{Nm}{M} - \frac12}\right]\cr
        d_m^{(2) + A} &\to& \frac{\sqrt{M+N}}{2\pi i \sqrt N} \oint\limits_{t = a} {\rm dt} \psi^{(2)+A} (t) \left[t^{\frac{Mm}{N}-1} \left(t - \tfrac{M}{M+N}a\right) ^{\frac12} (t-a)^{m - \frac12}\right]\cr
    d_m^{(1) - A} &\to& \frac{\sqrt{M+N}}{2\pi i \sqrt M} \oint\limits_{t = 0} {\rm dt} \psi^{(1)-A} (t) \left[t^{m} \left(t - \tfrac{M}{M+N}a\right) ^{\frac12} (t-a)^{\frac{Nm}{M} - \frac12}\right]        \cr
    d_m^{(2) - A} &\to& \frac{\sqrt{M+N}}{2\pi i \sqrt N} \oint\limits_{t = a} {\rm dt} \psi^{(2)-A} (t) \left[t^{\frac{Mm}{N}} \left(t - \tfrac{M}{M+N}a\right) ^{\frac12} (t-a)^{m - \frac12}\right]
    \end{eqnarray}
and the modes after the twist become:
    \begin{eqnarray}
        d_k^{ + A} &\to& \frac{1}{2\pi i } \oint\limits_{t = \infty} {\rm dt} \psi^{+A} (t) \left[t^{\frac{Mk}{M+N}-1} \left(t - \tfrac{M}{M+N}a\right) ^{\frac12} (t-a)^{\frac{Nk}{M+N} - \frac12}\right]\cr
         d_k^{ - A} &\to & \frac{1}{2\pi i} \oint\limits_{t = \infty} {\rm dt} \psi^{-A} (t) \left[t^{\frac{Mk}{M+N}} \left(t - \tfrac{M}{M+N}a\right) ^{\frac12} (t-a)^{\frac{Nk}{M+N} - \frac12}\right].
\end{eqnarray}

\section{Modes after the second spectral flow}
Next, we change coordinate to
\begin{eqnarray}
t'& =& t - \tfrac{M}{M+N}a,
\end{eqnarray}
and we spectral flow by $\alpha = -1$ in the $t'$ plane. Before this second spectral flow, we have the amplitude
\begin{eqnarray}
&&_{t'} \bra{0_{NS}} S^- \left(\tfrac{N}{M+N} a\right) \ket{0_R^+} _{t'}.
\end{eqnarray}
The spectral flow has the following effects:
\begin{enumerate}
\item[(a)] The R ground state $\ket{0_R^+}_{t'}$ maps to the NS vacuum in the $t'$ plane:
\begin{eqnarray}
\ket{0_R^+}_{t'} &\to &\ket{0_{NS}}_{t'}.
\end{eqnarray}
\item[(b)] The modes of the fermions change as follows:
\begin{eqnarray}
\psi^{\pm A} (t') &\to& (t')^{\pm \frac12} \psi^{\pm A} (t').
\end{eqnarray}
\end{enumerate}
Then the modes before the twist become:
  \begin{eqnarray}
  d_{m}^{\left(1\right)+ A} &\to& \frac{\sqrt{M+N}}{2\pi i \sqrt{M}}\oint\limits_{t'=-\frac{Ma}{M+N}} dt' \psi^{\left(1\right)+A}\left[\left(t'+\tfrac{M}{M+N}a\right)^{m-1}t'\left(t'-\tfrac{N}{M+N}a\right)^{\frac{Nm}{M}-\frac{1}{2}}\right]\cr
    	d_{m}^{\left(2\right)+ A} &\to& \frac{\sqrt{M+N}}{2\pi i \sqrt{N}}\oint\limits_{t'=\frac{Na}{M+N}} dt' \psi^{\left(2\right)+A}\left[\left(t'+\tfrac{M}{M+N}a\right)^{\frac{Mm}{N}-1}t'\left(t'-\tfrac{N}{M+N}a\right)^{m-\frac{1}{2}}\right]\cr
    	d_{m}^{\left(1\right)-A}&\to&\frac{\sqrt{M+N}}{2\pi i \sqrt{M}}\oint\limits_{t'=-\frac{Ma}{M+N}} dt'\psi^{\left(1\right)-A}\left[\left(t'-\tfrac{M}{M+N}a\right)^{m}\left(t'-\tfrac{M}{M+N}a\right)^{\frac{Nm}{M}-\frac{1}{2}}\right]\cr
    	d_{m}^{\left(2\right)-A}&\to&\frac{\sqrt{M+N}}{2\pi i \sqrt{N}}\oint\limits_{t'=\frac{Na}{M+N}} dt'\psi^{\left(2\right)-A}\left[\left(t'-\tfrac{M}{M+N}a\right)^{\frac{Mm}{N}}\left(t'-\tfrac{M}{M+N}a\right)^{m-\frac{1}{2}}\right]
  \end{eqnarray}
and the modes after the twist become:
\begin{eqnarray}
       d_{k}^{+A}&\to& \frac{1}{2 \pi i}\oint\limits_{t'=\infty}\psi^{+A}\left(t'\right)\left[\left(t'+\tfrac{M}{M+N}a\right)^{\frac{Mk}{M+N}-1}t'\left(t'-\tfrac{N}{M+N}a\right)^{\frac{Nk}{M+N}-\frac12}\right]\cr
      d_{k}^{-A}&\to& \frac{1}{2 \pi i}\oint\limits_{t'=\infty}\psi^{-A}\left(t'\right)\left[\left(t'+\tfrac{M}{M+N}a\right)^{\frac{Mk}{M+N}}\left(t'-\tfrac{N}{M+N}a\right)^{\frac{Nk}{M+N}-\frac12}\right].
\end{eqnarray}

\section{Modes after the third spectral flow}
Next, we change the coordinate to
\begin{eqnarray}
&&\hat{t}=t'-\tfrac{N}{M+N}a,
\end{eqnarray}
and we spectral flow by $\alpha =1$ in the $\hat t$ plane. The spectral flow has the following effects
\begin{enumerate}
\item[(a)] The R ground state maps to the NS vacuum in the $\hat{t}$ plane:
\begin{eqnarray}
\ket{0_{R}^{-}}_{\hat{t}}&\to& \ket{0_{NS}}_{\hat{t}}.
\end{eqnarray}
\item[(b)] The modes of the fermions change as follows:
\begin{eqnarray}
\psi^{\pm A}\left(\hat{t}\;\!\right)&\to&\left(\hat{t}\;\!\right)^{\mp\frac12}\psi^{\pm A}\left(\hat{t}\;\!\right)
\end{eqnarray}
\end{enumerate}
Then the modes before the twist become:
\begin{eqnarray}
    d_{m}^{\left(1\right)+A}&\to& \hat{d}_{m}'^{\left(1\right)+A}
		~=~\frac{\sqrt{M+N}}{2\pi i\sqrt{M}}
		\oint\limits_{\hat{t}=-a}d\hat{t}\,\psi^{\left(1\right)+A}\left(\hat{t}\;\!\right)\left[\left(\hat{t}+a\right)^{m-1}\left(\hat{t}+\tfrac{Na}{M+N}\right)\hat{t}^{\frac{Nm}{M}-1}\right]\cr
     d_{m}^{\left(2\right)+A}&\to& \hat{d}_{m}'^{\left(2\right)+A}
		~=~\frac{\sqrt{M+N}}{2\pi i\sqrt{N}}
		\oint\limits_{\hat{t}=0}d\hat{t}\,\psi^{\left(2\right)+A}\left(\hat{t}\;\!\right)\left[\left(\hat{t}+a\right)^{\frac{Mm}{N}-1}\left(\hat{t}+\tfrac{Na}{M+N}\right)\hat{t}^{m-1}\right]\cr
     d_{m}^{\left(1\right)-A}&\to& \hat{d}_{m}'^{\left(1\right)-A}
		~=~\frac{\sqrt{M+N}}{2\pi i\sqrt{M}}
		\oint\limits_{\hat{t}=-a}d\hat{t}\,\psi^{\left(1\right)-A}\left(\hat{t}\;\!\right)\left[\left(\hat{t}+a\right)^{m}\hat{t}^{\frac{Nm}{M}}\right]\cr
     d_{m}^{\left(2\right)-A}&\to& \hat{d}_{m}'^{\left(2\right)-A}
		~=~\frac{\sqrt{M+N}}{2\pi i\sqrt{N}}
		\oint\limits_{\hat{t}=0}d\hat{t}\,\psi^{\left(2\right)-A}\left(\hat{t}\;\!\right)\left[\left(\hat{t}+a\right)^{\frac{Mm}{N}}\hat{t}^{m}\right]
\end{eqnarray}
After the twist, we obtain
\begin{eqnarray}
   d_{k}^{+A}&\to&\hat{d}_{k}'^{+A}~=~\frac{1}{2\pi i}\oint\limits_{\hat{t}=\infty}d\hat{t}\,\psi^{+A}\left(\hat{t}\;\!\right)\left[\left(\hat{t}+a\right)^{\tfrac{Mk}{M+N}-1}\left(\hat{t}+\tfrac{N}{M+N}a\right)\hat{t}^{\frac{Nk}{M+N}-1}\right]\cr
   d_{k}^{-A}&\to&\hat{d}_{k}'^{-A}~=~\frac{1}{2\pi i}\oint\limits_{\hat{t}=\infty}d\hat{t}\,\psi^{-A}\left(\hat{t}\;\!\right)\left[\left(\hat{t}+a\right)^{\tfrac{Mk}{M+N}}\hat{t}^{\frac{Nk}{M+N}}\right].
\label{dhat-a}
\end{eqnarray}

\chapter{Calculation of the overall prefactor $C_{MN}$ for the single twist case} \label{app:cmn}

In this appendix we compute the prefactor $C_{MN}$ given by
\bea
C_{MN}&=&\bra{0_{R,--}}
\sigma_{2}^{++}\left(w_{0}\right)
\ket{0_{R}^{--}}^{\left(1\right)}\ket{0_{R}^{--}}^{\left(2\right)} \,.
\label{eq:cmn-a}
\eea
We will have four contributions to our factor of $C_{MN}$. 

First we lift the correlator to the $ z $ plane. The conformal weight of the twist operator then gives a Jacobian factor, which is our first contribution. 

We then have the $ z $ plane correlator
\begin{eqnarray}
\bra{0_{R,--}}
\sigma_{2}^{++}\left(z_{0}\right)
\ket{0_{R}^{--}}^{\left(1\right)}\ket{0_{R}^{--}}^{\left(2\right)} 
&=&
\langle \sigma_{M+N}^{++}(\infty)\sigma_{2}^{++}(z_{0})\sigma_{N}^{--}(0)\sigma_{M}^{--}(0)\rangle\,.
\end{eqnarray}
We employ the methods developed in \cite{lm1,lm2} to compute this correlator.
Introducing the notation that $\sigma_{M+N}^{++}$ has dimension $(\Delta_{{M+N}}^{++},\Delta_{{M+N}}^{++})$, we have
\begin{eqnarray}
\langle\sigma_{M+N}^{++}(\infty)\sigma_{2}^{++}(z_{0})\sigma_{N}^{--}(0)\sigma_{M}^{--}(0)\rangle
&\equiv&
\lim_{|z|\to\infty}|z|^{4\Delta_{{M+N}}^{++}}
\langle\sigma_{M+N}^{++}(z)\sigma_{2}^{++}(z_{0})\sigma_{N}^{--}(0)\sigma_{M}^{--}(0)\rangle
\cr
&=& \lim_{|z|\to\infty}
\frac{
\langle\sigma_{M+N}^{++}(z)\sigma_{2}^{++}(z_{0})\sigma_{N}^{--}(0)\sigma_{M}^{--}(0)\rangle
}
{
\langle\sigma_{M+N}^{++}(z)\sigma_{M+N}^{--}(0)\rangle}\,.
\eea
The computation of this ratio of correlators factorizes~\cite{lm2} into a product of terms coming from the Liouville action, which depend only on the `bare twist' part of the above spin-twist fields, and 
$t$ plane correlators of appropriately normalized spin fields. We denote the image of $z$ and $z_0$ in the $t$ plane by $t(z)$ and $t_0$ respectively. Then we have
\begin{eqnarray}
\langle \sigma_{M+N}^{++}(\infty)\sigma_{2}^{++}(z_{0})\sigma_{N}^{--}(0)\sigma_{M}^{--}(0)\rangle
&=& \lim_{|z|\to\infty}
\frac{\langle\sigma_{M+N}(z)\sigma_{2}(z_{0})\sigma_{N}(0)\sigma_{M}(0)\rangle}
{\langle\sigma_{M+N}(z)\sigma_{M+N}(0)\rangle}
\frac{\langle \mathcal{S}_4 (t(z),t_0) \rangle}{\langle \mathcal{S}_2 (t(z))\rangle} \cr &&
\label{eq:main}
\end{eqnarray}
where $\langle \mathcal{S}_4 (t(z),t_0) \rangle$ and $\langle \mathcal{S}_2 (t(z))\rangle$ are  $t$ plane correlators (four-point and two-point respectively) of appropriately normalized spin fields.

Our second contribution will be the ratio of correlators of bare twist fields in \eq{eq:main}; our third contribution will be the various normalization factors for the $t$ plane spin field correlators, and our final contribution will be the ratio of $t$ plane spin field correlators.

We now compute these four contributions  in turn.

\section{Jacobian factor for lifting from cylinder to plane}

We first lift the correlator to the $z$ plane. Since $\sigma_{2}^{++}$ has weight (1/2,1/2), we obtain the Jacobian factor contribution
\begin{eqnarray}
\left|\frac{dz}{dw}\right|_{z=z_{0}}&=&|a|^{M+N}\frac{M^{M}N^{N}}{(M+N)^{M+N}} \,.
\label{Jacobian}
\end{eqnarray}

\section{Liouville action terms}

In this subsection we deal only with the contribution from the bare twist fields, i.e. we compute
\bea
\langle\sigma_{M+N}(\infty)\sigma_{2}(z_{0})\sigma_{N}(0)\sigma_{M}(0)\rangle
&=&
\lim_{|z|\to\infty}
\frac{\langle\sigma_{M+N}(z)\sigma_{2}(z_{0})\sigma_{N}(0)\sigma_{M}(0)\rangle}
{\langle\sigma_{M+N}(z)\sigma_{M+N}(0)\rangle}
\label{eq:cmn-a3}
\eea
where the bare twist operators are normalized as
\begin{eqnarray}
\langle\sigma_{M}(z)\sigma_{M}(0)\rangle=\frac{1}{|z|^{4\Delta_{M}}} \,.
\label{norm}
\end{eqnarray}

\subsection{Defining regulated twist operators}

We work in a path integral formulation, and we
define regularized bare twist operators by cutting a small circular hole of radius $\epsilon \ll 1$ around each finite insertion point,
\begin{eqnarray} 
\sigma^{\epsilon}_{2}(z_{0}) \,, ~~\sigma^{\epsilon}_{M}(0) \,,~~\sigma^{\epsilon}_{N}(0) \,.
\end{eqnarray}
From \eq{norm} we have the following relationship between the regularized twist operators (whose normalization depends on the cutoff) and the canonically normalized twist operators:
\begin{eqnarray}
\sigma_{M}=\frac{1}{\sqrt{\langle\sigma_{M}^{\epsilon}(0)\sigma_{M}^{\epsilon}(1)\rangle}}\sigma_{M}^{\epsilon} \,.
\label{twist_norm}
\end{eqnarray}
We also define a regularized twist operator at infinity by cutting a large circular hole of radius $1/\tilde\delta$ with $\tilde\delta \ll 1$. This operator is denoted
\begin{eqnarray} 
\sigma^{\tilde\delta}_{M+N}(\infty) \,.
\end{eqnarray}
Since in \eq{eq:cmn-a3} we have a ratio of amplitudes, we do not need to worry about the normalization of $\sigma^{\tilde\delta}_{M+N}(\infty) $, and we obtain
\begin{eqnarray}
\lim_{|z|\to\infty}
\frac{\langle\sigma_{M+N}(z)\sigma_{2}(z_{0})\sigma_{N}(0)\sigma_{M}(0)\rangle}
{\langle\sigma_{M+N}(z)\sigma_{M+N}(0)\rangle}
&=&\frac{\langle\sigma^{\tilde{\delta}}_{M+N}(\infty)\sigma_{2}(z_{0})\sigma_{N}(0)\sigma_{M}(0)\rangle}{\langle\sigma_{M+N}^{\tilde{\delta}}(\infty)\sigma_{M+N}(0)\rangle} \,.
\end{eqnarray}
From (\ref{twist_norm}) we then obtain
\begin{eqnarray}
&&\langle\sigma_{M+N}(\infty)\sigma_{2}(z_{0})\sigma_{N}(0)\sigma_{M}(0)\rangle=\cr
&&\quad\frac{\langle\sigma^{\tilde{\delta}}_{M+N}(\infty)\sigma_{2}^{\epsilon}(z_{0})\sigma_{N}^{\epsilon}(0)\sigma_{M}^{\epsilon}(0)\rangle}{\langle\sigma_{M+N}^{\tilde{\delta}}(\infty)\sigma_{M+N}^{\epsilon}(0)\rangle}
\sqrt{
\frac{\langle\sigma_{M+N}^{\epsilon}(0)\sigma_{M+N}^{\epsilon}(1)\rangle}
{\langle\sigma_{2}^{\epsilon}(0)\sigma_{2}^{\epsilon}(1)\rangle
\langle\sigma_{N}^{\epsilon}(0)\sigma_{N}^{\epsilon}(1)\rangle
\langle\sigma_{M}^{\epsilon}(0)\sigma_{M}^{\epsilon}(1)\rangle}
} \, \qquad\nn
\label{eq:4pt-1}
\end{eqnarray}
and so our four-point function is now expressed entirely in terms of correlators of regularized twist operators, which we now compute.

\subsection{Lifting to the $t$ plane}

We now lift the problem to the $t$-plane using the map
\bea\label{eq:covermap-a}
z &=& t^M(t-a)^N \,.
\eea
In the $ z $ plane we have cut out various circular holes, leaving a path integral over an open set. The image of this open set in the $ t $ plane will be denoted by $\Sigma$. In the $ t $ plane,  we have single-valued fields, with no twist operators, but with appropriately normalized spin field insertions (which we deal with later). There is also a non-trivial metric, which we deal with in the present section.

To take account of the non-trivial metric on the $ t $ plane, we define a fiducial metric on the cover space $\Sigma$ and compute the Liouville action. Replacing the $z$-plane by a closed surface (a sphere), we define the $z$-plane metric $g$ via
\begin{eqnarray}
ds^{2}&=&\begin{cases}
dzd\bar{z} & |z|<\frac{1}{\delta}\\
d\tilde{z}d\bar{\tilde{z}} & |\tilde{z}|<\frac{1}{\delta}
\end{cases} \nn
\tilde{z}&=&\frac{1}{\delta^{2}}\frac{1}{z}
\label{z_fid}
\end{eqnarray}
where $\delta$ is a cutoff which will not play a role in our computation.

We define the fiducial metric on the $ t $ plane $\hat{g}$, to be flat so the curvature term in the Liouville action vanishes. We define $\hat{g}$ via 
\begin{eqnarray}
d\hat{s}^{2}&=& \begin{cases}
dtd\bar{t} & |t|< \frac{1}{\delta'}\\
d\tilde{t}d\bar{\tilde{t}} & |\tilde{t}|< \frac{1}{\delta'}
\end{cases}\nn
\tilde{t}&=&\frac{1}{\delta'^{2}}\frac{1}{t} \,.
\end{eqnarray} 
where $\delta'$ is another cutoff which will not play a role in our computation.

In lifting to the $t$-plane from the $z$-plane our metric will change by
\begin{eqnarray}
ds^{2}=e^{\phi}d\hat{s}^{2}
\label{metric_anomaly}
\end{eqnarray}
This corresponds to a transformation of our path integral of the form
\begin{eqnarray}
Z^{(g)}=e^{S_{L}}Z^{(\hat{g})}
\label{partition_trans}
\end{eqnarray}
where $S_{L}$ is the Liouville action given by (see e.g.~\cite{Friedan:1982is})
\begin{eqnarray}
S_{L}=\frac{c}{96\pi}\int d^{2}t\sqrt{-\hat g}\left[\partial_{\mu}\phi\partial_{\nu}\phi \,{\hat g}^{\m\n}+2R^{(\hat{g})}\phi\right].
\end{eqnarray}
Since we map from the $z$ space to the $t$ space we have an induced metric given by
\begin{eqnarray}
ds^{2}=dzd\bar{z}=\frac{dz}{dt}\frac{d\bar{z}}{d\bar{t}}dtd\bar{t}
\end{eqnarray}
Comparing this equation with (\ref{metric_anomaly}) we find that the Liouville field is given by
\begin{eqnarray}
\phi=\log\frac{dz}{dt}+\log\frac{d\bar{z}}{d\bar{t}} = \log \left ( \left | {dz\over dt} \right |^2 \right ).
\end{eqnarray}

Since the fiducial metric is flat and $\partial_{\mu}\partial^{\mu}\phi=0$, the Liouville action becomes a boundary term, which can be written as
\begin{eqnarray}
S_{L}=\frac{c}{96\pi}\left[i\int\limits_{\partial} dt\phi\partial_{t}\phi+c.c\right]
\label{eq:sl}
\end{eqnarray}
where $c.c.$ denotes the complex conjugate and where $\partial$ is the boundary of $\Sigma$, comprising the images in the $t$ plane of the circular holes defined in the $z$ plane.

The contribution of correlation function reduces to calculating the contributions to the Liouville action from the various holes comprising the boundary of $\Sigma$, which in our problem are:
\begin{itemize}
	\item \underline{Type 1}: The holes whose images are at finite points in the $t$-plane.  In our problem these holes are from finite points within the $z$-plane. 
	\item \underline{Type 2}: The hole whose image is at $t=\infty$.  This hole comes from a cut in the $z$-plane at $\frac{1}{\tilde{\delta}}$ upon taking $\tilde{\delta}\to 0$.  Since all $M+N$ copies of the CFT are twisted together at $z=\infty$, this is the only hole that appears at $t = \infty$.
\end{itemize}

Then our four-point correlation function \eq{eq:4pt-1} can be written as
\begin{eqnarray}
&&\langle\sigma_{M+N}(\infty)\sigma_{2}(z_{0})\sigma_{N}(0)\sigma_{M}(0)\rangle
=\cr
&&\qquad\frac{e^{S_{L_{1}}[\sigma_{M+N}^{\tilde{\delta}}\sigma_{2}^{\epsilon}\sigma_{N}^{\epsilon}\sigma_{M}^{\epsilon}]
+S_{L_{2}}[\sigma_{M+N}^{\tilde{\delta}}\sigma_{2}^{\epsilon}\sigma_{N}^{\epsilon}\sigma_{M}^{\epsilon}]}}
{e^{S_{L_{1}}[\sigma_{M+N}^{\tilde{\delta}}\sigma_{M+N}^{\epsilon}]
+S_{L_{2}}[\sigma_{M+N}^{\tilde{\delta}}\sigma_{M+N}^{\epsilon}]}}
\sqrt{
\frac{\langle\sigma_{M+N}^{\epsilon}(0)\sigma_{M+N}^{\epsilon}(1)\rangle}
{\langle\sigma_{2}^{\epsilon}(0)\sigma_{2}^{\epsilon}(1)\rangle
\langle\sigma_{N}^{\epsilon}(0)\sigma_{N}^{\epsilon}(1)\rangle
\langle\sigma_{M}^{\epsilon}(0)\sigma_{M}^{\epsilon}(1)\rangle}
}\qquad\quad
\label{eq:4pt-2}
\end{eqnarray}
where $S_{L_{1}}[\sigma_{M+N}^{\tilde{\delta}}\sigma_{2}^{\epsilon}\sigma_{N}^{\epsilon}\sigma_{M}^{\epsilon}]$ and $S_{L_{2}}[\sigma^{\tilde{\delta}}_{M+N}\sigma_{2}^{\epsilon}\sigma_{N}^{\epsilon}\sigma_{N}^{\epsilon}]$ are the Liouville contributions to the four-point function from holes of Type 1 and Type 2 respectively, and similarly for the two-point function.

\subsection{The four-point function of regularized twist operators}

We now calculate the contribution to the Liouville action from the four-point function of regularized twist operators,
\begin{equation}
S_{L,4} ~\equiv~S_{L_{1}}[\sigma_{M+N}^{\tilde{\delta}}\sigma_{2}^{\epsilon}\sigma_{N}^{\epsilon}\sigma_{M}^{\epsilon}]
+S_{L_{2}}[\sigma_{M+N}^{\tilde{\delta}}\sigma_{2}^{\epsilon}\sigma_{N}^{\epsilon}\sigma_{M}^{\epsilon}]\,.
\end{equation}

Let us begin by looking at the contribution to the Liouville action from the branch point at $z=0$, $t=0$.  Near this point we have:
\begin{equation}
z=t^{M}(t-a)^{N}\approx t^{M}(-a)^{N};\,\,\,\,\,\; \frac{dz}{dt}\approx Mt^{M-1}(-a)^{N};\,\,\,\,\,\,t\approx \left (\frac{z}{(-a)^{N}}\right )^{1/M};
\end{equation}
\begin{equation}
\phi = \log\left [ \frac{dz}{dt}\right ] +c.c. \approx 2\log \left [ M|t|^{M-1}|a|^N \right ]; \,\,\,\,\,\, \partial_{t}\phi \approx \frac{M-1}{t}
\end{equation}
From \eq{eq:sl}, the contribution to the Liouville action from this point is given by (we note that the central charge is $c=6$)
\begin{equation}
S_{L_1}(z=0,t=0)=\frac{6}{96\pi}\left [ i\int dt \phi\partial_{t}\phi + c.c. \right ]
\end{equation}
We perform branch point integration introducing the following coordinates
\begin{equation}
z \approx \epsilon e^{i\theta}; \qquad t\approx \left (\frac{\epsilon}{(-a)^N}\right )^{1/M}e^{i\theta'}; \qquad \theta' = {\theta \over M}
\end{equation}
Making this change of variables for the integral, we obtain the four-point function's contribution to the Liouville action from this region:
\begin{equation}
\begin{split}
S_{L_1,4}(z=0,t=0)&=\frac{6}{96\pi}\left [ i\int\limits_{0}^{2\pi}d\theta' it\,\frac{M-1}{t}2\log[M|t|^{M-1}|a|^N]+c.c. \right ]\\
&=-\frac{1}{2}(M-1)\left (\log \left [ M\epsilon^{\frac{M-1}{M}}|a|^{\frac{N}{M}}\right ] \right )
\end{split}
\end{equation}

The contributions from the other branch points in the four-point function are calculated in the same manner.  Here we present only the results.
\begin{eqnarray}
S_{L_1,4}(z=0,t=a)&=&-\frac{1}{2}(N-1)\left ( \log \left [ N\epsilon^{\frac{N-1}{N}}|a|^{\frac{M}{N}}\right ] \right )\nn
S_{L_1,4}(z=z_{0},t=t_{0})&=&-\frac{1}{4}\log\left [ 4\left | b_{t_0}\right | \epsilon \right ] \nn
S_{L_2,4}(z=\infty,t=\infty)&=&\frac{1}{2}(M+N-1)\log\left [ (M+N)\tilde{\delta}^{-\frac{M+N-1}{M+N}} \right ]
\end{eqnarray}
where 
\begin{equation}\label{bt0}
\left | b_{t_0}\right |=\frac{|a|^{M+N-2}}{2}\frac{M^{M}N^{N}}{(M+N)^{M+N}}\left ( \frac{(M+N)^{3}}{MN}\right ).
\end{equation}

Combining the Liouville terms for the regularized four-point function we find a total contribution of
\begin{eqnarray} \label{Liouville4}
S_{L,4}&=&-\frac{1}{2}(M-1)\log \left [ M\epsilon^{\frac{M-1}{M}}|a|^{\frac{N}{M}} \right ]-\frac{1}{2}(N-1)\log \left [ N\epsilon^{\frac{N-1}{N}}|a|^{\frac{M}{N}} \right ]-\frac{1}{4}\log\left [ 4\left | b_{t_0}\right | \epsilon\right ]\nn
&&\quad{}+\frac{1}{2}(M+N-1)\log \left [ (M+N)\tilde{\delta}^{-\frac{M+N-1}{M+N}}\right ].
\end{eqnarray}

\subsection{The two-point function of regularized twist operators}

Let us also calculate the contribution to the Liouville action from the regularized two-point function, 
\begin{eqnarray}
S_{L,2}&=&S_{L_{1}}[\sigma_{M+N}^{\tilde{\delta}}\sigma_{M+N}^{\epsilon}]
+S_{L_{2}}[\sigma_{M+N}^{\tilde{\delta}}\sigma_{M+N}^{\epsilon}]
\end{eqnarray}

We use the map
\begin{equation}
\begin{split}
z=t^{M+N},\,\,\,\,\,\,\,\,\, t=z^{\frac{1}{M+N}},\,\,\,\,\,\,\,\, \frac{dz}{dt}=(M+N)t^{M+N-1}\\
\phi\approx \log\left [ (M+N)t^{M+N-1} \right ] +c.c.\,\,\,\,\,\,\,\,\,\, \partial_{t}\phi=\frac{M+N-1}{t}
\end{split}
\end{equation}
We have branch points at $z=0,t=0$ and $z=\infty,t=\infty$. The contribution to the Liouville action from each of these points is calculated in the same manner as the contribution from the four-point function.  We thus find:
\begin{eqnarray}
S_{L_1,2}(t=0,z=0)&=&-\frac{1}{2}(M+N-1)\log\left [ (M+N)\epsilon^{\frac{M+N-1}{M+N}}\right ]\nn
S_{L_2,2}(t=\infty,z=\infty)&=&\frac{1}{2}(M+N-1)\log\left [(M+N)\tilde{\delta}^{\frac{1-M-N}{M+N}} \right ]
\end{eqnarray}

Combining the Liouville terms for both branch points of the two-point function then gives
\begin{equation}\label{Liouville2}
S_{L,2}=\frac{M+N-1}{2}\log \left [ \tilde{\delta}^{\frac{1-M-N}{M+N}} \right ] - \frac{M+N-1}{2}\log \left [ \epsilon^{\frac{M+N-1}{M+N}} \right ].
\end{equation}

\subsection{Normalization of the twist operators}

In \cite{lm1}, it was shown that the two-point function of regularized twist operators at finite separation is given by
\begin{eqnarray}
\langle \sigma_{n}^{\epsilon}(0)\sigma_{n}^{\epsilon}(a)\rangle=a^{-4\Delta_{n}} \left(n^2\epsilon^{A_{n}}Q^{B_{n}}\right)
\end{eqnarray}
where 
\begin{eqnarray}
\Delta_{n}&=&\frac{c}{24}\left(n-\frac{1}{n}\right) \,, \quad
A_{n}~=~-\frac{(n-1)^{2}}{n}\,,\quad B_{n}~=~1-n
\label{norm_terms}
\end{eqnarray}
and where $Q$ is a quantity that is regularization-dependent and which  cancels out.

The contribution from normalization term is then
\begin{equation}
\sqrt{
\frac{\langle\sigma_{M+N}^{\epsilon}(0)\sigma_{M+N}^{\epsilon}(1)\rangle}
{\langle\sigma_{2}^{\epsilon}(0)\sigma_{2}^{\epsilon}(1)\rangle
\langle\sigma_{N}^{\epsilon}(0)\sigma_{N}^{\epsilon}(1)\rangle
\langle\sigma_{M}^{\epsilon}(0)\sigma_{M}^{\epsilon}(1)\rangle}
}
=\sqrt{\frac{(M+N)^{2}\epsilon^{-\frac{(M+N-1)^{2}}{(M+N)}}}{(2MN)^{2}\epsilon^{-\frac{(N-1)^{2}}{N}-\frac{(M-1)^{2}}{M}-\frac{1}{2}}}}
\label{norm_terms-2}
\end{equation}
Combining (\ref{Liouville4}), (\ref{Liouville2}) and (\ref{norm_terms-2}), one can check that the regularization terms cancel, and so \eq{eq:4pt-2} becomes
\begin{eqnarray}\label{LiouvilleNormalization}
&&
\langle\sigma_{M+N}(\infty)\sigma_{2}(z_{0})\sigma_{N}(0)\sigma_{M}(0)\rangle 
\cr
&& {} \qquad =~
2^{-\frac{5}{4}}|a|^{-\frac{3}{4}\left(M+N\right)+\frac{1}{2}\left(\frac{M}{N}+\frac{N}{M}+1\right)}M^{-\frac{3}{4}M-\frac{1}{4}}N^{-\frac{3}{4}N-\frac{1}{4}}(M+N)^{\frac{3}{4}(M+N)-\frac{1}{4}}\,. \qquad
\end{eqnarray}

\section{Normalization factors for spin field insertions}

We next turn to the contributions to our amplitude (\ref{eq:main}) coming from the spin fields, 
\begin{eqnarray}
\lim_{|z|\to\infty}
\frac{\langle \mathcal{S}_4 (t(z),t_0) \rangle}{\langle \mathcal{S}_2 (t(z))\rangle} \quad
\label{eq:main-2}
\end{eqnarray}
where $\langle \mathcal{S}_4 (t(z),t_0) \rangle$ and $\langle \mathcal{S}_2 (t(z))\rangle$ are  $t$ plane  correlators (four-point and two-point respectively) of appropriately normalized spin fields, which we now define.

The normalization coefficient for a spin field $S_n^{\pm}(z_*)$ depends on the local form of the cover map at the insertion point,
\bea
\left(z-z_{*}\right)\approx b_* \left(t-{t_{*}}\right)^{n} \,.
\eea
The corresponding spin field insertion in the $t$ plane is  given by \cite{lm2}
\begin{eqnarray}
b_{*}^{-\frac{1}{4n}}S^{\pm}(t_*) \,.
\end{eqnarray} 
We shall first collect all the normalization terms $b_{*}$, and we shall compute the spin field correlator in the next subsection. 

For the two-point function in the denominator, we use the map
\begin{eqnarray}
z=t^{M+N}
\end{eqnarray}
and so all the normalization factors are trivial. 

For the four-point function, taking the appropriate limits of our map
\begin{eqnarray}
z=t^{M}(t-a)^{N}
\end{eqnarray}
we find the normalization coefficients $b_{*}$ to be
\begin{equation}
\begin{split}
&z=0, t=0;\quad  z \approx (-a)^{N}t^{M} \rightarrow\quad b_{0}=(-a)^{N}\rightarrow\quad b_{0}^{-\frac{1}{4M}}=(-a)^{-\frac{N}{4M}}\\
&z=0,t=a;\quad  z\approx a^{M}(t-a)^{N}\rightarrow\quad b_{a}=a^{M}\rightarrow\quad b_{a}^{-\frac{1}{4N}}=a^{-\frac{M}{4N}}\\
&z=z_{0}, t=t_{0}=\frac{aM}{M+N};\quad z-z_{0}\approx b_{t_0} (t-t_{0})^{2}\rightarrow\quad b_{t_{0}}^{-\frac{1}{8}}\\
&z=\infty, t=\infty;\quad  z\approx t^{M+N}\rightarrow\quad b_{\infty}=1\rightarrow\quad b_{\infty}^{-\frac{1}{4(M+N)}}=1
\label{spin_norm_coeff}
\end{split}
\end{equation}
where $b_{t_0}$ was given in (\ref{bt0}).  We can therefore write the normalized spin field correlator as (writing only holomorphic parts ) 
\begin{eqnarray}
\lim_{|z|\to\infty}
\frac{\langle \mathcal{S}_4 (t(z),t_0) \rangle}{\langle \mathcal{S}_2 (t(z))\rangle}
&=&\left(b_{\infty}^{-\frac{1}{4(M+N)}}b_{t_{0}}^{-\frac{1}{8}}b_{a}^{-\frac{1}{4N}}b_{0}^{-\frac{1}{4M}}\right)\left(\frac{\langle S^{+}(\infty)S^{+}(t_{0})S^{-}(a)S^{-}(0)\rangle}{\langle S^{+}(\infty)S^{-}(0)\rangle}\right). \qquad \quad
\label{total_spin_corr}
\end{eqnarray}
Combining the holomorphic and antiholomorphic parts, the product of all spin field normalization factors is
\begin{eqnarray} \label{SpinFieldNormalization}
&&|b_{\infty}|^{-\frac{1}{2(M+N)}}|b_{t_{0}}|^{-\frac{1}{4}}|b_{a}|^{-\frac{1}{2N}}|b_{0}|^{-\frac{1}{2M}}=\cr
&&\qquad 2^{\frac{1}{4}}|a|^{-\frac{1}{2}\left(\frac{M}{N}+\frac{N}{M}+\frac{M}{2}+\frac{N}{2}-1\right)}M^{-\frac{1}{4}(M-1)}N^{-\frac{1}{4}(N-1)}(M+N)^{\frac{1}{4}(M+N-3)}\,.
\end{eqnarray}

\section{Spin field correlator}
Here we show the calculation of the spin field correlator term, using two different methods: firstly via bosonization  and secondly via spectral flow.
\subsubsection{Method 1: Bosonization}
Here we shall use bosonization to calculate the spin field correlators given in (\ref{total_spin_corr}). We define the bosonized fermion fields as
\begin{eqnarray}
\psi_{1}=e^{i\phi_{5}},\qquad \psi_{2}=e^{i\phi_{6}}
\end{eqnarray}
We can therefore write the spin fields as
\begin{eqnarray}
S^{\pm}(z)=e^{\pm \frac{i}{2}e_{a}\Phi^{a}(z)}
\label{spin_fields}
\end{eqnarray}
where 
\begin{eqnarray}
e_{a}\Phi^{a}(z)=\phi_{5}(z)-\phi_{6}(z) \,.
\end{eqnarray}
The OPE is
\begin{equation}
e^{i\alpha\phi(z)}e^{i\beta\phi(w)}\sim e^{i(\alpha\phi(z)+\beta\phi(w))}(z-w)^{\alpha\beta}\,.
\end{equation}
Thus we have
\begin{eqnarray}
&&\frac{\langle S^{+}(\infty)S^{+}(t_{0})S^{-}(a)S^{-}(0)\rangle}{\langle S^{+}(\infty)S^{-}(0)\rangle}=\lim_{t\to\infty}\frac{\langle S^{+}(t)S^{+}(t_{0})S^{-}(a)S^{-}(0)\rangle}{\langle S^{+}(t)S^{-}(0)\rangle}\cr
&&\quad=\lim_{t\rightarrow \infty}
\frac{
\langle :\!\exp\left(\frac{i}{2}e_{a}\Phi^{a}(t)\right)\!\!:\,:\!\exp\left(\frac{i}{2}e_{a}\Phi^{a}(t_{0})\right)\!\!:\,:\!\exp\left(-\frac{i}{2}e_{a}\Phi^{a}(a)\right)\!\!:\,:\!\exp\left(-\frac{i}{2}e_{a}\Phi^{a}(0)\right)\!\!:\rangle
}{
\langle :\! \exp\left(\frac{i}{2}e_{a}\Phi^{a}(t)\right)\!\!:\,:\!\exp\left(-\frac{i}{2}e_{a}\Phi^{a}(0)\right)\!\!:\rangle 
}
\,.\cr
&&
\label{spin_corr}
\end{eqnarray}

For the four-point correlator, we obtain
\bea
&&\langle :\!\exp\left(\frac{i}{2}e_{a}\Phi^{a}(t)\right)\!\!:\,:\!\exp\left(\frac{i}{2}e_{a}\Phi^{a}(t_{0})\right)\!\!:\,:\!\exp\left(-\frac{i}{2}e_{a}\Phi^{a}(a)\right)\!\!:\,:\!\exp\left(-\frac{i}{2}e_{a}\Phi^{a}(0)\right)\!\!:\rangle \cr
&&{}\qquad =~(t-t_{0})^{\frac{1}{2}}(t-a)^{-\frac{1}{2}}(t_{0}-a)^{-\frac{1}{2}}t^{-\frac{1}{2}}t_{0}^{-\frac{1}{2}}a^{\frac{1}{2}}\,.
\label{4pt_spin_contr}
\eea
For the two-point correlator, we obtain
\bea
\langle :\! \exp\left(\frac{i}{2}e_{a}\Phi^{a}(t)\right)\!\!:\,:\!\exp\left(-\frac{i}{2}e_{a}\Phi^{a}(0)\right)\!\!:\rangle &=& t^{-\frac12} \,.
\label{2pt_spin_contr}
\eea
Combining (\ref{4pt_spin_contr}) and (\ref{2pt_spin_contr}), we find that (\ref{spin_corr}) is given by
\begin{eqnarray}
\frac{\langle S^{+}(\infty)S^{+}(t_{0})S^{-}(a)S^{-}(0)\rangle}{\langle S^{+}(\infty)S^{-}(0)\rangle}&=&\lim_{t\to\infty}(t-t_{0})^{\frac{1}{2}}(t-a)^{-\frac{1}{2}}(t_{0}-a)^{-\frac{1}{2}}t_{0}^{-\frac{1}{2}}a^{\frac{1}{2}}\cr
&=&(t_{0}-a)^{-\frac{1}{2}}t_{0}^{-\frac{1}{2}}a^{\frac{1}{2}}\cr
&=&(-a)^{-\frac{1}{2}}N^{-\frac{1}{2}}M^{-\frac{1}{2}}(M+N)\,.
\label{spin_contr_calc}
\end{eqnarray}
Combining (\ref{spin_contr_calc}) with the antiholomorphic part, we obtain
\bea
\frac{\langle S^{+}(\infty)S^{+}(t_{0})S^{-}(a)S^{-}(0)\rangle}{\langle S^{+}(\infty)S^{-}(0)\rangle}=\frac{(M+N)^2}{MN} \frac{1}{|a|}\,.
\label{eq:spinfieldans2}
\end{eqnarray}

\subsection{Method 2: Spectral Flow}

We now compute the spin field correlator using the sequence of spectral flow transformations and coordinate changes described in Section \ref{sec:sf}. This serves as a cross-check of the above calculation via bosonization.

Having lifted to the $t$ plane, and taken care of the normalization factors of the spin field insertions, we are left with the $t$ plane spin field correlator (writing holomorphic fields only)
\begin{eqnarray}
&&{}_t \bra{0_{R,-}} S^- (a) S^+ \left(\tfrac{M}{M+N}a\right)\ket{0_R^-}_t \,.
\end{eqnarray}
We recall that the action of spectral flow is straightforward for operators where the fermion content may be expressed as a simple exponential in the language in which the fermions are bosonized. For such operators with charge $j$, spectral flow with parameter $\alpha$ gives rise to the transformation
$$\hat O_j (t)\to t^{-\alpha j}\hat O_j (t).$$
We first spectral flow by $\alpha = 1$ in the $t$ plane. 
This gives
\begin{eqnarray}
\sqrt{\frac{M+N}{M}}~
{}_t \bra{0_{NS}} S^- (a) S^+ \left(\tfrac{M}{M+N}a\right)\ket{0_{NS}}_t.
\end{eqnarray}
Next, we change coordinate to
\begin{eqnarray}
t'& =& t - \tfrac{M}{M+N}a \,.
\end{eqnarray}
This gives the $t'$ plane correlator
\begin{eqnarray}
\sqrt{\frac{M+N}{M}}~
{}_{t'} \bra{0_{NS}} S^- \left(\tfrac{N}{M+N} a\right) \ket{0_R^+}_{t'} \;.
\end{eqnarray}
We then spectral flow by $\alpha = -1$ in the $t'$ plane, yielding
\begin{eqnarray}
&&\sqrt{\frac{M+N}{M}} \left(\frac{N}{M+N} a \right)^{-\frac{1}{2}}
{}_{t'} \bra{0_{R, -}} S^- \left(\tfrac{N}{M+N} a\right) \ket{0_{NS}}_{t'} \\
&=&\left(\frac{M+N}{\sqrt{MN}} a^{-\frac{1}{2}}\right) \,
{}_{t'} \bra{0_{R,-}} S^- \left(\tfrac{N}{M+N} a\right) \ket{0_{NS}}_{t'}.
\end{eqnarray}
Next, we change the coordinate to
\begin{eqnarray}
&&\hat{t}=t'-\tfrac{N}{M+N}a \,.
\end{eqnarray}
This gives
\begin{eqnarray}
\left(\frac{M+N}{\sqrt{MN}} a^{-\frac{1}{2}} \right)
{}_{\hat{t}} \braket{0_{R,-}}{0_{R}^{-}}_{\hat{t}} &=& \frac{M+N}{\sqrt{MN}} a^{-\frac{1}{2}} \,.
\end{eqnarray}
The final spectral flow by $\alpha =1$ in the $\hat t$ plane has no effect.

Adding in the anti-holomorphic factors, we obtain
\bea\label{SpinFieldCorrelator}
\frac{(M+N)^2}{MN} \frac{1}{|a|},
\eea
in agreement with \eq{eq:spinfieldans2}.

\bigskip

Combining the Jacobian factor (\ref{Jacobian}), the Liouville action contribution (\ref{LiouvilleNormalization}), the spin field normalization factors (\ref{SpinFieldNormalization}), and the spin field correlator (\ref{SpinFieldCorrelator}), we find
\begin{equation}
C_{MN}=\frac{M+N}{2MN} \,.
\end{equation}

\chapter{Ramond vacua notation}\label{RVN}
Here we define our notation for the various Ramond vacua in the untwisted sector.  There are two copies, which are not technically separate Hilbert spaces.  We start with the vacuum
\be
\rmutvket \equiv |v\rangle
\ee
and act on it with various fermion zero modes to construct the other Ramond vacua.  In order to be consistent with notation from \cite{acm1}, we also require something along the lines of
\be
\rptvket^{(i)} = d_0^{(i)++}d_0^{(i)+-}\rmtvket^{(i)},
\ee
though we do not actually have states containing only one of the two copies.

We now present a table defining our notation for the various vacua.
\bea
|v\rangle & = & \rmutvket \nn
d_0^{(1)+-}|v\rangle &=& |0_R\rangle^{(1)} \otimes |0_R^-\rangle^{(2)}\nn
d_0^{(1)++}|v\rangle &=& |\tilde{0}_R\rangle^{(1)} \otimes |0_R^-\rangle^{(2)}\nn
d_0^{(1)++}d_0^{(1)+-}|v\rangle &=& |0_R^+\rangle^{(1)} \otimes |0_R^-\rangle^{(2)}
\eea
\bea
d_0^{(2)+-}|v\rangle &=& |0_R^-\rangle^{(1)} \otimes |0_R\rangle^{(2)}\nn
d_0^{(1)+-}d_0^{(2)+-}|v\rangle &=& |0_R\rangle^{(1)} \otimes |0_R\rangle^{(2)}\nn
d_0^{(1)++}d_0^{(2)+-}|v\rangle &=& |\tilde{0}_R\rangle^{(1)} \otimes |0_R\rangle^{(2)}\nn
d_0^{(1)++}d_0^{(1)+-}d_0^{(2)+-}|v\rangle &=& |0_R^+\rangle^{(1)} \otimes |0_R\rangle^{(2)}
\eea
\bea
d_0^{(2)++}|v\rangle &=& |0_R^-\rangle^{(1)} \otimes |\tilde{0}_R\rangle^{(2)}\nn
d_0^{(1)+-}d_0^{(2)++}|v\rangle &=& |0_R\rangle^{(1)} \otimes |\tilde{0}_R\rangle^{(2)}\nn
d_0^{(1)++}d_0^{(2)++}|v\rangle &=& |\tilde{0}_R\rangle^{(1)} \otimes |\tilde{0}_R\rangle^{(2)}\nn
d_0^{(1)++}d_0^{(1)+-}d_0^{(2)++}|v\rangle &=& |0_R^+\rangle^{(1)} \otimes |\tilde{0}_R\rangle^{(2)}
\eea
\bea
d_0^{(2)++}d_0^{(2)+-}|v\rangle &=& |0_R^-\rangle^{(1)} \otimes |0_R^+\rangle^{(2)}\nn
d_0^{(1)+-}d_0^{(2)++}d_0^{(2)+-}|v\rangle &=& |0_R\rangle^{(1)} \otimes |0_R^+\rangle^{(2)}\nn
d_0^{(1)++}d_0^{(2)++}d_0^{(2)+-}|v\rangle &=& |\tilde{0}_R\rangle^{(1)} \otimes |0_R^+\rangle^{(2)}\nn
d_0^{(1)++}d_0^{(1)+-}d_0^{(2)++}d_0^{(2)+-}|v\rangle &=& |0_R^+\rangle^{(1)} \otimes |0_R^+\rangle^{(2)}.
\eea
These relations hold for both the initial (pre-twists) and final (post-twists) sectors.

\chapter{Proof of the exponential form for the Two Twist Case}\label{GeneralFormAppendix}
In this section we prove the exponential form of $|\chi(w_1,w_2)\rangle$ for the case of fermions in the NS sector.  The other cases may be treated similarly.

The method of the proof is straightforward.  We look at capping the state $|\chi(w_1,w_2)\rangle$ with a general femionic NS state $\hat{Q}^{\dagger} \nsnsket$.  This provides the relationship
\bea
\mathcal{A}\left ( \hat{Q} \right ) &\equiv&{\nsnsbra \hat{Q} |\chi(w_1,w_2)\rangle \over \nsnsbra \chi(w_1,w_2)\rangle} ~ = ~ {\nstbra \hat{Q}' \nstket \over \nstbra \nstclose}.\label{GeneralMappingRelation}
\eea
The rightmost expression can be calculated from the behaior of the fermion modes under coordinate transformations and spectral flows, while the middle expression can be calculated from our guess for the form of $|\chi(w_1,w_2)\rangle$.  Showing that these two methods are consistent demonstrates that we have the correct form for $|\chi(w_1,w_2)\rangle$.

\section{Middle expression}
Let us begin by taking a close look at the form presented in (\ref{generalform}).  Because we are working with pairs of fermion creation operators in the exponent, all of the terms in the exponent commute.  We can thus rewrite the fermion portion in the NS sector as
\bea
|\chi(w_1,w_2)\rangle^{F}_{NS} & = & e^{\sum\limits_{(i),(j)}\sum\limits_{r,s > 0}\g^{F(i)(j)}_{NS,rs} \left ( d^{(i)f,++}_{-r}d^{(j)f,--}_{-s} - d^{(i)f,+-}_{-r}d^{(j)f,-+}_{-s} \right )}\nsnsket\nn
&=& \prod_{(i),(j)}\prod_{r,s>0}e^{\g^{F(i)(j)}_{NS,rs} \left ( d^{(i)f,++}_{-r}d^{(j)f,--}_{-s} - d^{(i)f,+-}_{-r}d^{(j)f,-+}_{-s} \right )}\nsnsket \nn
&=& \prod_{A,B}\prod_{(i),(j)}\prod_{r,s>0}\left [ 1 + \e_{AB}\g^{F(i)(j)}_{NS,rs}  d^{(i)f,+A}_{-r}d^{(j)f,-B}_{-s} \right ]\nsnsket.\qquad
\eea

From this form, we can see clearly that any state with an odd number of fermion excitations will have no overlap with $|\chi(w_1,w_2)\rangle$.  So let us start with a two-excitation state.  This case is quite simple:
\bea
\mathcal{A}\left(d^{(i)f,+A}_r d^{(j)f,-B}_s\right) &=& {\nsnsbra d^{(i)f,+A}_r d^{(j)f,-B}_s |\chi(w_1,w_2)\rangle \over \nsnsbra \chi(w_1,w_2)\rangle} \nn
&=& \e^{AC}\e^{BD}\e_{CD}\g^{F(i)(j)}_{NS,rs}\nn
&=& \e^{AB}\g^{F(i)(j)}_{NS,rs},\label{2PointAmplitude}
\eea
where the fact that the coefficient is proportional to $\e^{AB}$ simply means that the combinations $d^{++}d^{--}$ and $d^{+-}d^{-+}$ yield coefficients which differ only by an overall sign.

Now what happens when we cap with a state that contains more fermion excitations?  We know that we must have an even number of excitations, so they come in pairs.  We can thus calculate $\mathcal{A}$ in two steps.  First, we write out all possible ways to group the fermion excitations into pairs, accounting for the overal sign required to anticommute the operators into the appropriate pairings.  Then each pairing combination provides a contribution equal to the product of the amplitude for each individual pair within that combination.  Adding the contributions for each combination then gives us the total amplitude.  We can write this schematically as:
\bea
\mathcal{A}\left ( d_1 d_2 d_3 \ldots \right ) &=& \sum_{\text{pairing combinations}} (-1)^p\left (\prod_{\{d_i d_j \}}\mathcal{A}\left ( d_i d_j \right ) \right ), \label{4PointAmplitude}
\eea
where $p$ is the number of anticommutations we need to perform to achieve the pairing configuration.

\section{Right expression}
Let us compare the previous result to the calculation in the $t$ plane.  In Section \ref{Outline} we outlined a series of coordinate maps and spectral flows which bring $|\chi(w_1,w_2)\rangle$ to an empty NS vacuum in the $t$ plane.  Under any combination of such transformations, the modes $d^{(i)f,\a A}_r$ behave in general as:
\bea
d^{(1)f,\a A}_r & \to & d'^{(1)f,\a A}_r ~=~ {1\over 2\pi i}\oint\limits_{t=\infty} \psi^{\a A}(t) h^{\a}_r(t) \diff t\\
d^{(2)f,\a A}_r & \to & d'^{(2)f,\a A}_r ~=~  {1\over 2\pi i}\oint\limits_{t=0} \psi^{\a A}(t) h^{\a}_r(t) \diff t,
\eea
where the function $h_r^{\a}$ is identical for each copy but will in general depend on the original mode number $r$ and the relevant spectral flow charge $\a$.

Now we expand the function $h_r^{\a}$ as a polynomial in $t$, which can in general include negative powers of $t$.  It turns out that the function $h_r^{\a}$ consists of a product of powers of $t$ and binomials in $t$, which makes such an expansion straightforward.  We perform the expansion with large $t$ for Copy 1 and small $t$ for copy 2.  We then find:
\bea
d'^{(1)f,\a A}_r & = & \sum_{p=-\infty}^{\infty}C^{(1)\a}_{rp}{1\over 2\pi i}\oint\limits_{t=\infty} \psi^{\a A}(t) t^p \diff t\\
d'^{(2)f,\a A}_r & = & \sum_{p=-\infty}^{\infty}C^{(2)\a}_{rp}{1\over 2\pi i}\oint\limits_{t=0} \psi^{\a A}(t) t^p \diff t,
\eea
where the coefficients are not in general equal because we are expanding the function in two different regions.

Recall that after these coordinate maps and spectral flows, we have removed all insertions in the $t$ plane.  We can thus close all punctures with the NS vacuum and smoothly deform our contours.  This allows us to deform the contour around $t = \infty$ for the copy 1 mode into a contour around $t=0$.  We then find that in general:
\bea
d'^{(i)f,\a A}_r & = & \sum_{p=-\infty}^{\infty}C^{(i)\a}_{rp}{1\over 2\pi i}\oint\limits_{t=0} \psi^{\a A}(t) t^p \diff t\\
&=& \sum_{p=-\infty}^{\infty} C^{(i)\a}_{rp} \tilde{d}^{\a A}_p. \label{GeneralFermionExpansion}
\eea

From this expression we see that when capping with a two-excitation state, the right side of (\ref{GeneralMappingRelation}) becomes:
\bea\label{A2tPlane}
\mathcal{A}\left(d^{(i)f,+A}_r d^{(j)f,-B}_s\right) &=& \sum_{p,q=-\infty}^{\infty}C^{(i)+}_{rp}C^{(j)-}_{sq}{\nstbra : \tilde{d}^{+A}_p \tilde{d}^{-B}_q :\nstket \over \nstbra\nstclose},
\eea
where the colons indicate radial ordering of the contours in the modes, which may not be the same as the initial $\t$ ordering on the cylinder, as copy 1 modes always map to contours at larger radii than copy 2 modes.

The amplitude in the numerator on the right of (\ref{A2tPlane}) is only nonzero when the two fermion modes can contract together, with the outer mode annihilating the inner mode.  However, we know the anticommutator is:
\bea
\left \{ \tilde{d}^{+A}_p , \tilde{d}^{-B}_q \right \} &=& -\e^{+-}\e^{AB} \d_{p+q,0},
\eea
which is proportional to $\e^{AB}$.  We thus find that for any fermion pair,
\bea
\mathcal{A}\left(d^{(i)f,+A}_r d^{(j)f,-B}_s\right) &=& \e^{AB}D^{(i)(j)}_{rs},
\eea
which matches the result from (\ref{2PointAmplitude}).

Now what happens when we cap with a state that contains more fermion excitations?  In general, we have:
\bea
\mathcal{A}\left( d_1 d_2 d_3 \ldots \right) &=& \sum_{p_1,p_2,p_3,\ldots} \left (C_{p_1} C_{p_2} C_{p_3} \ldots \right ) {\nstbra : \tilde{d}_{p_1} \tilde{d}_{p_2}\tilde{d}_{p_3}\ldots :\nstket \over \nstbra\nstclose}.
\eea
It is clear that the right side vanishes when there are an odd number of fermion modes, as such a case will always leave at least one uncontracted mode to annihilate the vacuum bra or ket.  With an even number of modes, the amplitude can be calculated by performing all possible combinations of contractions in the expression on the right side and adding each contribution together.  However, we also have the relation
\bea
{\nstbra \contraction{}{\tilde{d}}{{}_{p_1}}{\tilde{d}} \tilde{d}_{p_1}\tilde{d}_{p_2}\contraction{}{\tilde{d}}{{}_{p_3}}{\tilde{d}} \tilde{d}_{p_3}\tilde{d}_{p_4}\nstket\over \nstbra\nstclose} &=& {\nstbra :\tilde{d}_{p_1}\tilde{d}_{p_2}:\nstket \over \nstbra\nstclose} {\nstbra:\tilde{d}_{p_3}\tilde{d}_{p_4}:\nstket\over \nstbra\nstclose}.\qquad
\eea
We thus find:
\bea
\mathcal{A}\left (d_1 d_2 d_3 \ldots \right )&=& \sum_{\text{pairing combinations}} (-1)^p\left [ \prod_{\{p_i,p_j\}} \left (\sum_{p_i,p_j}C_{p_i}C_{p_j}{\nstbra : \tilde{d}_{p_i} \tilde{d}_{p_j} :\nstket \over \nstbra\nstclose}\right )\right ] \nn
&=& \sum_{\text{pairing combinations}} (-1)^p\left (\prod_{\{d_i d_j \}}\mathcal{A}\left ( d_i d_j \right ) \right ), \label{4PointTAmplitude}
\eea
where $p$ is again the number of anticommutations we must perform to reach the particular pair combination.  Since this overall sign depends only on how we group the operators in $\hat{Q}$ and not on any of the specifics of the coordinate maps or spectral flows, each $p$ that appears here takes the same value as in in (\ref{4PointAmplitude}).  Thus the two relations are identical, and we see that the claimed form for $|\chi(w_1,w_2)\rangle$ is correct.

\chapter{Proof of copy 1-copy 2 interchange symmetry for $|\chi\rangle$ for the two twist case}\label{CopySymmetry}
Using the vacua listed in Appendix (\ref{RVN}), we can look at the effects of $\g^{F(i)(j)}_{R+-,mn}$ and check the desired $(1) \leftrightarrow (2)$ symmetry.  We first note that $\g^{F(2)(1)}_{R+-;0,0}=1$.  We will also introduce the shorthand
\bea
\g^{F(2)(1)}_{R+-;m,0} ~\equiv~ A_m \implies \g^{F(1)(1)}_{R+-;m,0}~=~ -A_m.
\eea
With this notation, we find:
\bea
|\chi(w_1,w_1)\rangle &=& \left (1+d^{(2)f,++}_{0}d^{(1)f,--}_{0}\right)\left(1-d^{(2)f,+-}_{0}d^{(1)f,-+}_{0}\right)\left(1+A_md^{(2)f,++}_{-m}d^{(1)f,--}_0\right)\nn
&&{}\times\left(1-A_{m'}d^{(2)f,+-}_{-m'}d^{(1)f,-+}_0\right)\left(1-A_nd^{(1)f,++}_{-n}d^{(1)f,--}_0\right)\nn
&&{}\times\left(1+A_{n'}d^{(1)f,+-}_{-n'}d^{(1)f,-+}_0\right)d_0^{(1)f,++}d_0^{(1)f,+-}e^{\hat{Q}}|0_R^-\rangle^{(1)}|0_R^-\rangle^{(2)},\label{ChiZeroes}
\eea
where $\hat{Q}$ now contains no zero modes and the repeated mode indices are summed over positive integers

Noting the relations (\ref{BosonRelation1}), (\ref{BosonRelation2}) and (\ref{symmetryrequirements}), we see that the operator $\hat{Q}$ is symmetric under $(1) \leftrightarrow (2)$.  We must then inqure as to the symmetry of the rest of (\ref{ChiZeroes}).  In doing so, let us write $P$ as the operation which swaps copies $(1)$ and $(2)$.  We then note that the notation outlined in (\ref{RVN}) implies:
\bea
P\left ( |0_R\rangle^{(1)}|\tilde{0}_R\rangle^{(2)} \right ) &=& P \left ( d^{(1)f,+-}_0d^{(2)f,++}_0|0_R^-\rangle^{(1)}|0_R^-\rangle^{(2)}\right ) \nn
&=& d^{(2)f,+-}_0d^{(1)f,++}_0|0_R^-\rangle^{(1)}|0_R^-\rangle^{(2)} \nn
&=& -d^{(1)f,++}_0d^{(2)f,+-}_0|0_R^-\rangle^{(1)}|0_R^-\rangle^{(2)}\nn
&=&-|\tilde{0}_R\rangle^{(1)}|0_R\rangle^{(2)}.
\eea
And since $P^2=1$, we also have:
\bea
P\left ( |\tilde{0}_R\rangle^{(1)}|0_R\rangle^{(2)} \right ) &=& -|0_R\rangle^{(1)}|\tilde{0}_R\rangle^{(2)}.
\eea
On the other hand,
\bea
P\left ( |0_R^+\rangle^{(1)}|0_R^-\rangle^{(2)} \right ) &=& |0_R^-\rangle^{(1)}|0_R^+\rangle^{(2)}\\
P\left(|0_R^-\rangle^{(1)}|0_R^+\rangle^{(2)}\right)&=&|0_R^+\rangle^{(1)}|0_R^-\rangle^{(2)}.
\eea
Using these relations, we find:
\bea
|v\rangle &\equiv& \left(1+d^{(2)f,++}_{0}d^{(1),f--}_{0}\right)\left(1-d^{(2)f,+-}_{0}d^{(1)f,-+}_{0}\right)d_0^{(1)f,++}d_0^{(1)f,+-}|0_R^-\rangle^{(1)}|0_R^-\rangle^{(2)}\nn
&=& |0_R^+\rangle^{(1)}|0_R^-\rangle^{(2)} + |0_R\rangle^{(1)}|\tilde{0}_R\rangle^{(2)} - |\tilde{0}_R\rangle^{(1)}|0_R\rangle^{(2)} + |0_R^-\rangle^{(1)}|0_R^+\rangle^{(2)},
\eea
which is symmetric under $P$.  We now write:
\bea
|\chi(w_1,w_1)\rangle &=& \left(1+A_md^{(2)f,++}_{-m}d^{(1)f,--}_0\right)\left(1-A_{m'}d^{(2)f,+-}_{-m'}d^{(1)f,-+}_0\right)\nn
&&{}\times\left(1-A_nd^{(1)f,++}_{-n}d^{(1)f,--}_0\right)\left(1+A_{n'}d^{(1)f,+-}_{-n'}d^{(1)f,-+}_0\right)e^{\hat{Q}}|v\rangle.\nn\label{Intermediate}
\eea

At this point it is necessary to make use of an interesting property of the state $|v\rangle$.  When applying zero modes to this state, we notice:
\bea
d^{(1)f,--}_0 |v\rangle &=& -|0_R\rangle^{(1)}|0_R^-\rangle^{(2)} + |0_R^-\rangle^{(1)}|0_R\rangle^{(2)}\nn
d^{(2)f,--}_0 |v\rangle &=& |0_R\rangle^{(1)}|0_R^-\rangle^{(2)} - |0_R^-\rangle^{(1)}|0_R\rangle^{(2)}\nn
d^{(1)f,-+}_0 |v\rangle &=&  - |\tilde{0}_R\rangle^{(1)}|0_R^-\rangle^{(2)} + |0_R^-\rangle^{(1)}|\tilde{0}_R\rangle^{(2)}\nn
d^{(2)f,-+}_0 |v\rangle &=&  |\tilde{0}_R\rangle^{(1)}|0_R^-\rangle^{(2)} - |0_R^-\rangle^{(1)}|\tilde{0}_R\rangle^{(2)}.
\eea
Thus whenever we have a copy (1) zero mode with negative charge acting on $|v\rangle$ we can substitute a copy (2) zero mode in its place:
\bea
d^{(1)f,-\a}_0|v\rangle &=& -d^{(2)f,-\a}_0|v\rangle.
\eea
Applying this substitution to the second line of (\ref{Intermediate}), we have:
\bea
|\chi(w_1,w_1)\rangle &=& \left(1+A_md^{(2)f,++}_{-m}d^{(1)f,--}_0\right)\left(1-A_{m'}d^{(2)f,+-}_{-m'}d^{(1)f,-+}_0\right)\nn
&&{}\times\left(1+A_nd^{(1)f,++}_{-n}d^{(2)f,--}_0\right)\left(1-A_{n'}d^{(1)f,+-}_{-n'}d^{(2)f,-+}_0\right)e^{\hat{Q}}|v\rangle,\qquad\nn
\eea
which is manifestly symmetric.

\chapter{Proof of copy symmetry relations for $\g^{B}$ and $\g^{F}$ for the two twist case}\label{GammaRelations}
In this section we analytically prove the relations:
\bea
\g^{B(1)(1)}_{mn}&=&-\g^{B(2)(1)}_{mn}\cr\cr
\left [ \g^{F(1)(1)}_{R+-,mn}\right ]_{n>0}&=&-\left[\g^{F(2)(1)}_{R+-,mn}\right]_{m,n>0}.
\eea

\section{Proof of $\g^{B(1)(1)}_{mn}=-\g^{B(2)(1)}_{mn}$}
Beginning with $\g^{B(i)(j)}_{mn}$, we have the relation
\bea
\g^{B(1)(1)}_{mn}=-\g^{B(2)(1)}_{mn}
\eea
for $a\neq 0$. 
Let us write the expressions for both functions:
\bea
\g^{B(1)(1)}_{mn}&=&\frac{1}{mn}\sum_{l=1}^{n}l{}^{n}C_{l}{}^{m}C_{l}a^{l}b^{m+n-l}{}_2F_1\left(-n,l-n;l+1;\frac{a}{b}\right){}_2F_1\left(-m,l-m;l+1;\frac{a}{b}\right),\nn
\eea
and
\bea
\g^{B(2)(1)}_{mn}&=&-\frac{1}{mn}\sum_{l=1}^{n}l{}^{m}C_{l}{}^{n}C_{l}a^{m}b^{n}{}_2F_1\left(-m,l-m;l+1;\frac{b}{a}\right){}_2F_1\left(-n,l-n;l+1;\frac{a}{b}\right).\nn
\label{gammab12}
\eea
Let us investigate $\g^{B(2)(1)}$. Considering the \emph{first} hypergeometric function with the identity:
\bea
{}_{2}F_{1}(a,b,c;z)&=&\frac{\G(1-a)\G(c)}{\G(b)\G(c-a-b+1)}\left(-1\over z\right)^{a}\left(1-\frac{1}{z}\right)^{c-a-b}\nn
&&~\times~ {}_{2}F_{1}\left(1-b,c-b;c-a-b+1;1-\frac{1}{z}\right)\nn
&& +~ {\G(1-a)\G(c)\over \G(b-a+1)\G(c-b)}\left({1\over z}\right)^{b}{}_{2}F_{1}(b,b-c+1;b-a+1;{1\over z}),\qquad
\label{identity}
\eea
we find:
\bea
{}_2F_1\left(-m,l-m;l+1;\frac{b}{a}\right)&=&\frac{\G(m+1)\G(l+\h)}{\G(l-m)\G(2m+2)}\left(-a\over b\right)^{-m}\left(1-\frac{a}{b}\right)^{2m+1}\nn
&&~\times~{}_{2}F_{1}\left(m-l+1,m+1;2m+2;1-\frac{a}{b}\right)\nn
&& +~ {\G(m+1)\G(l+1)\over \G(l+1)\G(m+1)}\left({a\over b}\right)^{l-m}{}_{2}F_{1}(l-m,-m;l+1;{a\over b}).\nn
\eea
The first term vanishes because 
\bea
\frac{1}{\G(l-m)}=0~~ \text{when} ~~  l-m \leq 0,
\eea
which is always the case. We also note that the ratio of gamma functions for the second term is one. Using the symmetry relation
\bea
{}_{2}F_{1}(a,b,c,z)={}_{2}F_{1}(b,a,c,z),
\label{interchange symmetry}
\eea
we find that
\bea
{}_{2}F_{1}(l-m,-m;l+1;{a\over b})={}_{2}F_{1}(-m,l-m;l+1;{a\over b}).
\eea
Inserting this into (\ref{gammab12}),  we obtain:
\bea
\g^{B(2)(1)}_{mn}&=&-\frac{1}{mn}\sum_{l=1}^{n}l{}^{m}C_{l}{}^{n}C_{l}a^{l}b^{m+n-l}{}_2F_1\left(-m,l-m;l+1;\frac{a}{b}\right){}_2F_1\left(-n,l-n;l+1;\frac{a}{b}\right)\cr\cr
&=&-\g^{B(1)(1)}_{mn}.
\eea

\section{Proof of $\left [ \g^{F(1)(1)}_{R+-,mn}\right ]_{n>0}=-\left[\g^{F(2)(1)}_{R+-,mn}\right]_{m,n>0}$}
Here we prove the $\g^{F(i)(j)}_{R+-,mn}$ relation
\bea
\left [ \g^{F(1)(1)}_{R+-,mn}\right ]_{n>0}=-\left[\g^{F(2)(1)}_{R+-,mn}\right]_{m,n>0}
\eea
for $a\neq 0$.  Let us first write expressions for both functions. We have:
\bea
\left[\g^{F(1)(1)}_{R+-,mn}\right]_{n>0}&=& -\sum_{p = {3\over 2}}^{n+\h}{}^{n-1}C_{n-p+\h}{}^m C_{p-\h}a^{p-\h}b^{m+n-p+\h}\nn
&&\quad{}\times {}_2 F_1 \left ( 1-n, p-n-\h; p-\h;{a\over b}\right ) {}_2 F_1 \left ( p-m-\h,-m;p+\h;{a\over b}\right ) \nn
&&{}+\sum_{p = {3\over 2}}^{n-{3\over2}}{}^{n-1}C_{n-p-{3\over2}}{}^m C_{p-\h}a^{p+\h}b^{m+n-p-\h}\nn
&&\quad{}\times {}_2 F_1 \left ( 1-n, p-n+{3\over2}; p+{3\over2};{a\over b}\right ) {}_2 F_1 \left ( p-m-\h,-m;p+\h;{a\over b}\right ), \nn
\eea
and 
\bea
\left[\g^{F(2)(1)}_{R+-,mn}\right]_{m,n>0}&=& \sum_{p={1\over 2}}^{\min\left(m-\h,n-\h\right)}{}^{n-1}C_{n-p-\h}{}^m C_{m-p-\h}a^m b^n\nn
&&{}\times {}_2 F_1 \left(1-n,p-n+\h;p+\h;{a\over b}\right){}_2 F_1 \left(p-m+\h,-m;p+\frac{3}{2};{b\over a}\right) \nn
&&{}-\sum_{p={1\over2}}^{\min\left(m-\h,n-{5\over2}\right)}{}^{n-1}C_{n-p-{5\over2}}{}^m C_{m-p-\h}a^{m+1}b^{n-1}\nn
&&{}\times {}_2 F_1 \left(1-n,p-n+{5\over2};p+{5\over2};{a\over b}\right){}_2 F_1 \left(p-m+\h,-m;p+{3\over2};{b\over a}\right).\nn
\eea
We shall now investigate $\g^{F(2)(1)}$. First let us make the index shift
\bea
p\to p-1.
\eea
For $\g^{F(2)(1)}$ this gives
\bea
\left[\g^{F(2)(1)}_{R+-,mn}\right]_{m,n>0}&=& \sum_{p={3\over 2}}^{\min\left(m+\h,n+\h\right)}{}^{n-1}C_{n-p+\h}{}^m C_{m-p+\h}a^m b^n\nn
&&{}\times {}_2 F_1 \left(1-n,p-n-\h;p-\h;{a\over b}\right){}_2 F_1 \left(p-m-\h,-m;p+\frac{1}{2};{b\over a}\right) \nn
&&{}-\sum_{p={3\over2}}^{\min\left(m+\h,n-{3\over2}\right)}{}^{n-1}C_{n-p-{3\over2}}{}^m C_{m-p+\h}a^{m+1}b^{n-1}\nn
&&{}\times {}_2 F_1 \left(1-n,p-n+{3\over2};p+{3\over2};{a\over b}\right){}_2 F_1 \left(p-m-\h,-m;p+{1\over2};{b\over a}\right).\nn
\label{gamma21}
\eea
Now using the identity given in (\ref{identity}) to transform the second hypergeometric function of each term as well as the interchange symmetry given in (\ref{interchange symmetry}), we obtain:
\bea
{}_{2}F_{1}\left(p-m-\h,-m,p-\h;{b\over a}\right)&=&{}_{2}F_{1}\left(-m,p-m-\h,p-\h;{b\over a}\right)\cr\cr
&=&\frac{\G(m+1)\G(p+1/2)}{\G(p-m-1/2)\G(2m+2)}\left(-a\over b\right)^{-m}\left(1-\frac{a}{b}\right)^{2m+1}\cr\cr
&&\quad\times{}_{2}F_{1}\left(m-p+3/2,m;2m+2;1-\frac{a}{b}\right)\cr\cr
&&\quad +~ \frac{\G(m+1)\G(p+1/2)}{\G(m+1)\G(p+1/2)}\cr\cr
&&\quad~~~\left({a\over b}\right)^{p-m-1/2}{}_{2}F_{1}\left(p-m-{1\over 2},-m;p+{1\over 2};{a\over b}\right).
\nn
\label{identity}
\eea
We see that the first term vanishes because
\bea
\frac{1}{\G\left(p-m-1/2\right)}=0~~\text{when}~~p\leq m+1/2,
\eea
which is always the case.  Additionally, the ratio of gamma functions in the second term is one. Therefore we have
\bea
{}_{2}F_{1}\left(p-m-\h,-m,p-\h;{b\over a}\right)=\left({a\over b}\right)^{p-m-1/2}{}_{2}F_{1}\left(p-m-{1\over 2},-m;p+{1\over 2};{a\over b}\right).\nn
\eea
Inserting this term into (\ref{gamma21}), we obtain:
\bea
\left[\g^{F(2)(1)}_{R+-,mn}\right]_{m,n>0}&=& \sum_{p={3\over 2}}^{\min\left(m+\h,n+\h\right)}{}^{n-1}C_{n-p+\h}{}^m C_{m-p+\h}a^{p-{1\over 2}} b^{m+n-p+\h}\nn
&&{}\times {}_2 F_1 \left(1-n,p-n-\h;p-\h;{a\over b}\right){}_2 F_1 \left(p-m-\h,-m;p+\frac{1}{2};{a\over b}\right) \nn
&&{}-\sum_{p={3\over2}}^{\min\left(m+\h,n-{3\over2}\right)}{}^{n-1}C_{n-p-{3\over2}}{}^m C_{m-p+\h}a^{p+\h} b^{m+n-p-\h}\nn
&&{}\times {}_2 F_1 \left(1-n,p-n+{3\over 2};p+{3\over 2};{a\over b}\right){}_2 F_1 \left(p-m-\h,-m;p+{1\over 2};{a\over b}\right)\crcr
&=&-\left[\g^{F(1)(1)}_{R+-,mn}\right]_{n>0}.
\eea

\chapter{Proof of supersymmetry relationship between $\g^{B(i)(j)}_{mn}$ and $\g^{F(i)(j)}_{R+-,mn}$ for the two twist case}\label{boson fermion relation}
Here we prove the relationship 
\bea
\g^{F(i)(j)}_{R+-,mn}=-m\g^{B(i)(j)}_{mn}.
\eea
We start by applying $G^{+(p)}_{0,+}$ to $|\chi(w_{1},w_{2})\rangle$.
\bea
\sum_{(p)}G^{(p)+}_{0,+}|\chi(w_{1},w_{2})\rangle&=&\sum_{(p)}G^{(p)+}_{0,+}\s_{2}^{+}(w_{2})\s_{2}^{+}(w_{1})|0_{R}^{-}\rangle^{(1)}|0_{R}^{-}\rangle^{(2)}\nn
&=&\left({1\over 2\pi i}\int_{w_{2}}dwG^{+}_{+}(w)\s_{2}^{+}(w_{2})\right)\s_{2}^{+}(w_{1})|0_{R}^{-}\rangle^{(1)}|0_{R}^{-}\rangle^{(2)}\nn
&&~ + ~ \s_{2}^{+}(w_{2})\left({1\over 2\pi i}\int_{w_{1}}dwG^{+}_{+}(w)\s_{2}^{+}(w_{1})\right)|0_{R}^{-}\rangle^{(1)}|0_{R}^{-}\rangle^{(2)}\cr\cr
&&~ + ~ \s_{2}(w_{1})\s_{2}^{+}(w_{1})\sum_{(j)}G^{(j)+}_{0,+}|0_{R}^{-}\rangle^{(1)}|0_{R}^{-}\rangle^{(2)}\nn
&=& 0.
\label{supercharge chi relation}
\eea
where $j,p\in\lbrace 1,2\rbrace$. We have used the fact that $G^{+}_{0,\dot{A}}$ acting on a single $\s_{2}^{+}(w_{i})$ is which is proven in Appendix \ref{GPlusProof} 
, as well as the fact that $G^{+}_{0,+}$ annihilates the negative Ramond vacuum. More explicitly:

\bea
&&{1\over 2\pi i}\int_{w_{i}}dwG^{+}_{+}(w)\s_{2}^{+}(w_{i})=0\nn
&&G^{(j)+}_{0,+}|0_{R}^{-}\rangle^{(1)}|0_{R}^{-}\rangle^{(2)}=0.
\eea

Using the form of $|\chi(w_{1},w_{2})\rangle$ given in (\ref{chi state}), where we have excluded the $NS$ sector states, we find:
\bea
0 &=&\sum_{(p)}G^{(p)+}_{0,+}|\chi(w_1,w_2)\rangle\nn
&=& \sum_{(p)}G^{(p)+}_{0,+}C_{R+-}\text{exp}\left [ \sum_{(i),(j)}\sum_{k,l > 0}\g^{B(i)(j)}_{kl} \left ( -\a^{(i)f}_{++,-k}\a^{(j)f}_{--,-l} + \a^{(i)f}_{+-,-k}\a^{(j)f}_{-+,-l} \right ) \right ] \nn
&&\quad{}\times \text{exp}\left [ \sum_{(i),(j)}\sum_{k,l \geq 0}\g^{F(i)(j)}_{R+-,kl} \left ( d^{(i)f,++}_{-k}d^{(j)f,--}_{-l} - d^{(i)f,+-}_{-k}d^{(j)f,-+}_{-l} \right ) \right ] \rpmket \ \nn
&=&C_{R+-}\sum_{(i),(j),(p)}\left[\,\sum_{k,l > 0}\g^{B(i)(j)}_{kl} \left[G^{(p)+}_{0,+},  -\a^{(i)f}_{++,-k}\a^{(j)f}_{--,-l} + \a^{(i)f}_{+-,-k}\a^{(j)f}_{-+,-l} \right]\right.\nn
&&\left.\quad{}+\sum_{k,l \geq 0}\g^{F(i)(j)}_{R+-,kl} \left\{ G^{(p)+}_{0,+},  d^{(i)f,++}_{-k}d^{(j)f,--}_{-l} - d^{(i)f,+-}_{-k}d^{(j)f,-+}_{-l} \right\}\right]|\chi(w_1,w_2)\rangle  \nn
&=&C_{R+-}\sum_{(i),(j),(p)}\left[\,\sum_{k,l > 0}\g^{B(i)(j)}_{kl}\left(-\a^{(i)f}_{++,-k}\left[G^{(p)+}_{0,+},\a^{(j)f}_{--,-l}\right]+\left[G^{(p)+}_{0,+},\a^{(i)f}_{+-,-k}\right]\a^{(j)f}_{-+,-l}\right)\right.\nn
&&\left.\quad{}+\sum_{k,l \geq 0}\g^{B(i)(j)}_{kl}\left( - d^{(i)f,++}_{-k}\lbrace G^{(p)+}_{0,+}, d^{(j)f,--}_{-l}\rbrace + \lbrace G^{(p)+}_{0,+},d^{(j)f,-+}_{-l}\rbrace d^{(i)f,+-}_{-k}  \right)\right]|\chi(w_1,w_2)\rangle.\nn
\eea

We now apply the (anti)commutation relations:
\bea
\left[G^{(i)\a}_{\dot{A},0},\a^{(j)}_{B\dot{B},m}\right]&=&-im\e_{AB}\e_{\dot{A}\dot{B}}\d^{(i)(j)}d^{(i)\a A}_{m}\cr
\lbrace G^{(i)\a}_{\dot{A},0},d^{(j),\b B}_{m} \rbrace&=&i\e^{\a\b}\e^{AB}\d^{(i)(j)}\a^{(i)}_{A\dot{A},m}.
\eea
This gives:

\bea
0 &=& C_{R+-} \sum_{(i),(j),(p)}\left[\,\sum_{k,l > 0}\g^{B(i)(j)}_{kl}\left(  \a^{(i)f}_{++,-k}\left(-il\right)\d^{(p)(j)}d^{(p)f,++}_{-l}+(-ik)\d^{(p)(i)}d^{(p)f,+-}_{-k}\a^{(j)f}_{-+,-l}  \right)\right.\nn
&&\left.\quad{}+\sum_{k,l \geq 0}\g^{F(i)(j)}_{kl}\left(  -d^{(i)f,++}_{-k}(i)\d^{(p)(j)}\a_{++,-l}^{(p)f}+(-i)\d^{(p)(j)}\a_{-+,-l}^{(p)f}d^{(i)f,+-}_{-k}\right)\right]|\chi(w_1,w_2)\rangle \nn
&=&-iC_{R+-}\sum_{(i),(p)}\sum_{k,l>0}\left(l\g^{B(i)(p)}_{kl}\a_{++,-k}^{(i)f}d^{(p)f,++}_{-l}+k\g^{B(p)(i)}_{kl}\a_{-+,-l}^{(i)f}d^{(p)f,+-}_{-k}\right.\nn
&&\qquad\qquad\left.{}+\g_{kl}^{F(i)(p)}\a_{++,-l}^{(p)f}d^{(i)f,++}_{-k}+\g^{F(i)(p)}_{kl}\a_{-+,-l}^{(p)f}d^{(i)f,+-}_{-k}\right)|\chi(w_{1},w_{2})\rangle\nn
&&\quad{}-iC_{R+-}\sum_{(i),(p)}\sum_{l > 0}\left(\g^{F(i)(p)}_{0,l}\a_{++,-l}^{(p)f}d^{(i)f,++}_{0} +\g^{F(i)(p)}_{0,l}\a_{-+,-l}^{(p)f}d_{0}^{(i)f,+-}\right)|\chi(w_{1},w_{2})\rangle,\nn
\label{Gplus1}
\eea
where we have used the fact that all bosonic zero modes annihilate $|\chi(w_1,w_2)\rangle$.  The fermionic zero modes do not annihilate this state, but from (\ref{symmetryrequirements}) we know that
\bea
\left[\g^{F(p)(i)}_{R+-;0,l}\right]_{l>0}=0, \quad i, j\in {1,2}.
\eea
Thus the last line of (\ref{Gplus1}) vanishes, and we are left with
\bea
0&=&-iC_{R+-}\sum_{(i),(p)}\sum_{k,l>0}\left(l\g^{B(i)(p)}_{kl}\a_{++,-k}^{(i)f}d^{(p)f,++}_{-l}+k\g^{B(p)(i)}_{kl}\a_{-+,-l}^{(i)f}d^{(p)f,+-}_{-k}\right.\nn
&&\qquad\left. {}+\g_{R+-,kl}^{F(i)(p)}\a_{++,-l}^{(p)f}d^{(i)f,++}_{-k}+\g^{F(i)(p)}_{R+-,kl}\a_{-+,-l}^{(p)f}d^{(i)f,+-}_{-k}\right)|\chi(w_{1},w_{2})\rangle.
\label{Gplus2}
\eea

We now make the following convenient modifications to (\ref{Gplus2}):
\bea
\text{First Term}&:& k\leftrightarrow l\nn
&&\g^{B}_{kl}=\g^{B}_{lk}\nn
&&(i)\leftrightarrow (p)\nn\nn
\text{Fourth Term}&:&(i)\leftrightarrow (p).
\eea
This gives
\bea
0&=&-iC_{R+-}\sum_{(i),(p)}\sum_{k,l>0}\a_{++,-l}^{(p)f}d^{(i)f,++}_{-k}\left(k\g^{B(p)(i)}_{kl}+\g^{F(i)(p)}_{R+-,kl}\right)\nn
&&-iC_{R+-}\sum_{(i),(p)}\sum_{k,l>0}\a_{-+,-l}^{(i)f}d^{(p)f,+-}_{-k}\left(k\g^{B(p)(i)}_{kl}+\g^{F(p)(i)}_{R+-,kl}\right),
\eea
which implies
\bea
\g^{F(p)(i)}_{R+-,kl}&=&-k\g^{B(p)(i)}_{kl}\nn
\g^{F(p)(i)}_{R+-,kl}&=&\g^{F(i)(p)}_{R+-,kl},
\eea
for all $k,l>0$. The second was seen in (\ref{symmetryrequirements}).  The first is a manifestation of the supersymmetry in the theory.

We will note here that we were not required to map to a cover space beyond verifying that $G^+_{\dot A,0}\s_2^+$ vanishes. Since this vanishing should be a local property of the operators, independent of our winding configurations, we expect the relation proved here to extend to arbitrary twist configurations.

\chapter{Computation of Two Twist Wick Contraction Term}
In this section we compute the two twist boson wick contraction term, $C^{B,2}_{++--}$.
\section{Computing $C^{B,2}_{++--}$} 
Before computing $C^{B,2}_{++--}$ let us first define some useful relations that we will need. Consider the wick contraction term for a single untwisting:
\bea
\nn
C^{B,1,\text{unt.}}_{mn,++--}&=&\langle 0|\s_{2}^{+}(w_0)\a_{++,-m}\a_{--,-n}|0\rangle\nn
&=&\left(\langle 0|\a_{++,n}\a_{--,m}\s_{2}^{+}(w_0)|0\rangle\right)^{\dagger}\nn
&=&-mn\left(\g^{B}_{mn}\right)^*\nn
\label{single untwisting wick contraction}
\eea 
where $\left(\g^{B}_{mn}\right)^{\dagger}=\left(\g^{B}_{mn}\right)^{*}$. Even though $\left(\s_{2}^{+}\right)^{\dagger}=\s_{2}^{-}$ we neglect the conjugation because the bosons are unaffected by spectral flow.
We see that the single untwisting wick contraction term is equal the conjugate of the single twist $\g^{B}$. There is one other relation we must consider before we tackle the two twist wick contraction term. Consider the single untwisting $f^{B}$ which we'll write as $f^{B,\text{unt.}}$:
 \bea
 \nn
 f^{B,\text{unt.}}_{pn} &=&{1\over n} \langle 0|\a^{(1)}_{++,n}\s_{2}^{+}(w_0)\a_{--,-p}|0\rangle \nn
 &=&{1\over n} \left(   \langle 0|\a_{++,p}\s_{2}^{+}(w_0)\a^{(1)}_{--,-n}|0\rangle  \right)^{\dagger}\nn
 &=&{p\over n}\left(f^{B}_{np}\right)^{*}\nn
 \label{twist untwist f relation}
 \eea
where $\left(f^{B}_{np}\right)^{\dagger} = \left(f^{B}_{np}\right)^{*}$. We see that the $f^{B,\text{unt.}}$ is just the conjugate transpose of the single twist $f^{B}$. Now that we have these two relations, let us compute the two twist wick contraction term. We begin with the following amplitude
\bea
\nn
C^{B,2}_{mn,++--} &=& {}^{(1)}\langle 0 |  {}^{(2)}\langle 0 | \s_{2}^{+}(w_2)\s_{2}^{+}(w_1)\a_{++,-m}^{(1)}\a_{--,-n}^{(1)}|0\rangle^{(1)}|0\rangle^{(2)}\nn
\eea
where $m,n>0$.
Now, bringing both $\a$'s through the first twist gives the following:
\bea
&&C^{B,2}_{++--,mn} \nn
&&\quad = {}^{(1)}\langle 0 |  {}^{(2)}\langle 0 | \s_{2}^{+}(w_2)\left(\sum_{p\geq 0}f_{mp}^{B}(w_1)\a_{++,-p}\right)\left(\sum_{p'}f_{np'}^{B}(w_1)\a_{--,-p'} \right)\s_{2}^{+}(w_1)|0\rangle^{(1)}|0\rangle^{(2)}\nn
&&\quad~~  + ~ C^{1}_{mn,++--}\nn
\eea
This step was computed in \cite{acm2}.
Let us now pull the two $\a$'s through the second twist. Doing this gives:
\bea
\nn
C^{B,2}_{mn,++--} &=& \sum_{p,p'\in \mathbb{Z}^{+}_{\text{odd}},q,q' > 0}f_{mp}^{B}(w_1)f_{np'}^{B}(w_1)f^{B,\text{unt.}}_{pq}(w_2)f^{B,\text{unt.}}_{p'q'}(w_2)\nn\nn
&&\quad\times {}^{(1)}\langle 0 | {}^{(2)}\langle 0 |\a_{++,-q}^{(1)}\a_{--,-q'}^{(1)}\s_{2}^{+}(w_2)\s_{2}^{+}(w_1)|0\rangle^{(1)}|0\rangle^{(2)}\nn\nn
&&\quad + \sum_{p,p' \in \mathbb{Z}^{+}_{\text{odd}} }f_{mp}^{B}(w_1)C^{B,1,\text{unt.}}_{pp',++--} f_{np'}^{B}(w_1) +  C^{1}_{mn,++--}\nn\nn
&=&\sum_{p,p' \in \mathbb{Z}^{+}_{\text{odd}} }f_{mp}^{B}(w_1)C^{B,1,\text{unt.}}_{pp',++--} f_{np'}^{B}(w_1) +  C^{1}_{mn,++--}\nn
\label{two twist wick contraction}
\eea
where again we use the result computed in \cite{acm2} and also the relation:
\bea
\nn
{}^{(1)}\langle 0 | {}^{(2)}\langle 0 |\a_{++,-q}^{(1)}\a_{--,-q'}^{(1)}\s_{2}^{+}(w_2)\s_{2}^{+}(w_1)|0\rangle^{(1)}|0\rangle^{(2)}=0\nn
\eea
since the two $\a$'s annihilate on the left.

Now using the relations found in (\ref{single untwisting wick contraction}) and (\ref{twist untwist f relation}), (\ref{two twist wick contraction}) becomes:
\bea
\nn
C^{B,2}_{mn,++--} &=& -\sum_{p,p' \in \mathbb{Z}^{+}_{\text{odd}}}pp'f_{mp}^{B}(w_1)\left(\g^{B}_{pp'}(w_2)\right)^* f_{np'}^{B}(w_1) +  C^{1}_{mn,++--}.\nn
\label{two twist wick final}
\eea
\begin{figure}
\includegraphics[width=0.5\columnwidth]{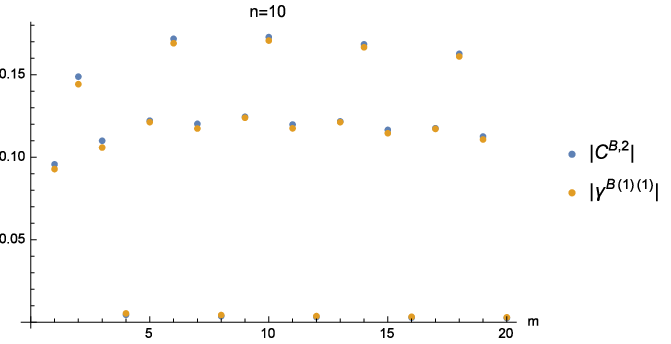}
\includegraphics[width=0.5\columnwidth]{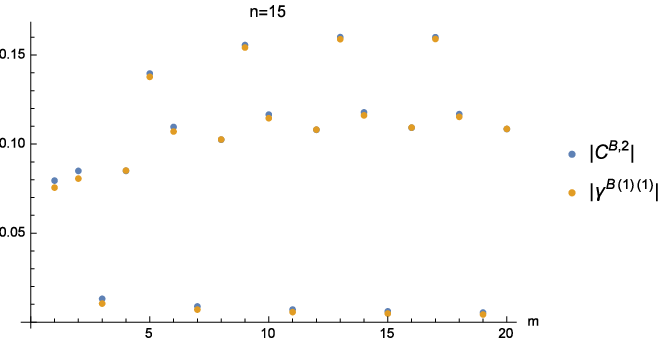}
\caption{Figure 1: Here we plot numerical approximations for both $|C^{B,2}_{mn,++--}|$ and $|\g^{B(1)(1)}_{mn}|$ with $n=10$ (left) and $n=15$ (right) at $\Delta w = {i \pi\over 2}$, for various values of $m$. We see good agreement between the two numerical quantities for each case, as expected.  The small discrepancies for lower $n$ arise from the fact that we have truncated the two infinite sums for some $p,p'>> m,n$.  At low $n$ the convergence takes longer and thus the truncation produces larger discrepancies.}
\label{figure one}
\end{figure}
We can also show that the two twist wick contraction term is related to the complex conjugate of the two twist $\g^{B}$. To do this we again write the two twist wick contraction term:
\bea
C^{B,2}_{mn,++--} &=& {}^{(1)}\langle 0 |  {}^{(2)}\langle 0 | \s_{2}^{+}(w_2)\s_{2}^{+}(w_1)\a_{++,-m}^{(1)}\a_{--,-n}^{(1)}|0\rangle^{(1)}|0\rangle^{(2)}\nn
&=&\left({}^{(1)}\langle 0 |  {}^{(2)}\langle 0 |\a_{++,n}^{(1)}\a_{--,m}^{(1)} \s_{2}^{+}(w_1)\s_{2}^{+}(w_2)|0\rangle^{(1)}|0\rangle^{(2)}\right)^{\dagger}\nn
&=&-mn\left(\g^{B(1)(1)}_{mn,w_{1}\leftrightarrow w_2}\right)^{*}\nn
&=&-mn\g^{B(1)(1)}_{mn}\nn
\label{two twist wick gamma relation}
\eea
where the fourth equality arises because of the reality of $\g^{B(1)(1)}_{mn}$ and the invariance of $a$ and $b$ under the interchange $w_{1}\leftrightarrow w_2$, both conditions of which are a result of our definition of $z_{1}$ and $z_{2}$. In Figure (\ref{figure one}) we present a scatter plot of (\ref{two twist wick final}) and (\ref{two twist wick gamma relation}) where $\g^{B}$ was computed in \cite{acm1}, $f^{B}$ was computed in \cite{acm2}, and $\g^{B(1)(1)}$ was computed in chapter \label{two twist gamma} whic was computed in \cite{chm1}. 

Following the same procedure as for the bosons one can compute the fermion two twist wick contraction term, but there are additional complications one must be aware of. The conjugate relations used in the wick contraction computation change nontrivially for fermions because of the fermion modes actually notice the change in charge of the twist ooperator. Essentially, the conjugate relation is the $\g^{F}$ for a state, $|\chi'\rangle$ built from negative twist operators.  This state has not been computed in any previous work, and we will not do so here. However, the relation does provide an easier way for computing the fermion two twist wick contraction term if necessary.

\chapter{Proof of $G^{+}_{\dot{A},0}\s_{2}^{+}(w_{0}) = 0$}\label{GPlusProof}
Here we prove the relation that a $G^{+}_{\dot{A},0}$ acting on a single $\s^{+}_{2}$ is equal to zero which is necessary for proving the two twist $\g^{B},\g^{F}$ relations found in \cite{chm1} as well as the $f^{B},f^{F\pm}$ relations given by Equations (\ref{NegativeRelation}) and (\ref{PositiveRelation}) . 
\subsection*{Cylinder}
Let us consider the following state on the cylinder:
\bea
{1\over 2 \pi i}\int_{w_0} G^{+}_{\dot{A}}(w)\s_{2}^{+}(w_0) \diff w
\label{cylinder state}
\eea
where $w_0$ is the location of the twist operator. 
\subsubsection*{$z$ plane}
Mapping to the $z$ plane with:
\bea
z=e^{w},\qquad z_0\equiv e^{w_0}
\label{z plane map}
\eea
we find that (\ref{cylinder state}) becomes:
\bea
 z_{0}^{1/2} {1\over 2 \pi i}\int_{z_0} \diff z\, z^{1/2}  G^{+}_{\dot{A}}(z)\s_{2}^{+}(z_0)
 \label{z plane state}
\eea
 where $z_0^{1/2}$ is a Jacobian factor coming from $\s_{2}^{+}(z_0)$.
\subsection*{$t$ plane}
Let us map (\ref{z plane map}) to the $t$ plane. Using the single twist map defined in \cite{acm1}
\bea
z = z_0 + t^{2},
\label{t plane map two}
\eea
(\ref{z plane state}) becomes:
\bea
&&\!\!\!\!\!\!\!\!z_{0}^{1/2} {1\over 2 \pi i}\int_{t=0} \diff t\, z^{1/2} \left({\diff z\over \diff t}\right)^{-1/2}G^{+}_{\dot{A}}(t)S^{+}(0)| 0_{NS} \rangle_{t=0} \nn
&& \!\!\!\!\!\!\!\!\quad = 2^{-1/2} z_{0}^{1/2} {1\over 2 \pi i}\int_{t_0} \diff t\, \left(z_0 + t^2\right)^{1/2}t^{-1/2}G^{+}_{\dot{A}}(t)S^{+}(0)| 0_{NS} \rangle_{t=0}\nn
\label{t plane state one}
\eea
\subsection*{Spectral Flowing away $S^{+}(0)$}
We now perform a single spectral flow to remove $S^{+}(0)$. The change in the fields are as follows:
\subsection*{$\a = -1$ $\text{around}$ $t=0$ }
\bea
S^{+}(0) &\to& 1\nn
G^{+}_{\dot{A}}(t)&\to& G^{+}_{\dot{A}}(t)t^{1/2}
\label{sf t zero}
\eea
Applying (\ref{sf t zero}) to (\ref{t plane state one}) gives:
\bea
2^{-1/2}Cz_{0}^{1/2}{1\over 2 \pi i}\int_{z_0} \diff t\, \left(z_0 +t^{2}\right)^{1/2}G^{+}_{\dot{A}}(t)| 0_{NS} \rangle_{t=0}
\label{t plane state two}
\eea
Let us now expand $\left(z_0 +t^{2}\right)^{1/2}$ around $t=0$:
\bea
 \left(z_0 +t^{2} \right)^{1/2}&=&z_{0}^{1/2}\left(1-z_0^{-1/2} t^{2}\right)^{1/2}\nn
 &=&\sum_{k\geq 0}{}^{1/2}C_{k}z_{0}^{-k/2+1/2}t^{2k}
 \label{expansion}
\eea
Inserting (\ref{expansion}) back into (\ref{t plane state two}) gives:
\bea
2^{-1/2}C\sum_{k\geq 0}{}^{1/2}C_{k}z_{0}^{-k/2+1}{1\over 2 \pi i}\int_{z_0} \diff t\, G^{+}_{\dot{A}}(t)t^{2k} | 0_{NS} \rangle_{t=0}
\label{t plane state three}
\eea
Let us now define $G^{+}_{\dot{A}}$ natural to the $t$ plane at $t=0$:
\bea
\tilde{G}^{+,t\to 0}_{\dot{A},r}&=&{1\over 2 \pi i}\int_{0} \diff t\,  G^{+}_{\dot{A}}(t)t^{r+1/2}, \qquad r \in \mathbb{Z} + 1/2
\label{t plane G modes}
\eea
Rewriting (\ref{t plane state three}) in terms of (\ref{t plane G modes}) gives:
\bea
&&2^{-1/2}C \sum_{k\geq 0}{}^{1/2}C_{k}z_{0}^{-k/2+1} \tilde{G}^{+,t\to 0}_{\dot{A},2k-1/2}| 0_{NS} \rangle_{t=0}
\label{t plane state five}
\eea
It is clear that (\ref{t plane state five}) vanishes when $k\geq 1$:
\bea
\tilde{G}_{\dot{A},k-1/2}^{+,t\to 0}|0_{NS}\rangle_{t=0} = 0,\qquad\qquad k\geq 1
\eea
Let us check the case where $k=0$. Only writing the $G^{+}_{\dot{A}}$ part of (\ref{t plane state five}) we have:
\bea
\tilde{G}_{\dot{A},-1/2}^{+,t\to 0}|0_{NS}\rangle_{t=0}
\label{G on NS}
\eea
This result also vanishes but let us explicitly show this. Splitting our $\tilde{G}^{+}_{\dot{A}}$ into bosons and fermions, a general mode can be written as:
\bea
\tilde{G}^{+,t\to 0}_{\dot{A},r} =-i\sum_{p\in \mathbb{Z}} \tilde{d}^{+A,t\to t_0}_{r-p}\a_{A\dot{A},p}^{t \to 0}
\eea
Applying this to (\ref{G on NS}) we get:
\bea
-i\sum_{p\in \mathbb{Z}} \tilde{d}^{+A,t\to 0}_{1/2-p}\a_{A\dot{A},p}^{t \to 0}|0_{NS}\rangle_{t=0} &=& -i\sum_{p > 0}\tilde{d}^{+A,t\to 0}_{-p+1/2}\a_{A\dot{A},p}^{t \to 0}|0_{NS}\rangle_{t=0}\nn
&&- i\sum_{p > 0}\a_{A\dot{A},-p}^{t \to 0}\tilde{d}^{+A,t\to 0}_{p+1/2}|0_{NS}\rangle_{t=0}\nn
&& -i \tilde{d}^{+A,t\to 0}_{1/2 }\a_{A\dot{A},0}^{t \to 0}|0_{NS}\rangle_{t=0}\nn
&=&0
\eea
Therefore, we have proven that for $k\geq 0$:
\bea
&&\!\!\!\!\!\! 2^{-1/2}C\sum_{k\geq 0}{}^{1/2}C_{k}z_{0}^{-k/2+1} \tilde{G}^{+,t\to 0}_{\dot{A},2k-1/2}| 0_{NS} \rangle_{t=0}=0
\eea
which of course implies that on the cylinder:
\bea
{1\over 2 \pi i}\int_{w_0}G^{+}_{\dot{A}}(w)\s_{2}^{+}(w_0) \diff w =0,
\eea
which is what we wanted to show.

\chapter{Rewriting the expressions for two deformation operators in terms of $\Delta w$}\label{coefficients}
Here we compute the expressions for all of the various coefficients in our final expression for the full second order deformation operator, $\hat{O}_{\dot{B}}\hat{O}_{\dot{A}}$, in terms of $\Delta w$ which is the only free parameter. First let us remind the reader of the following parameters
\bea
z_1&=&e^{-{\Delta w\over 2}}\cr
z_2&=&e^{\Delta w\over 2} \cr
a&=&\cosh^2\bigg({\Delta w\over 4}\bigg)\cr
b&=&\sinh^2\bigg({\Delta w\over 4}\bigg)\cr
t_2&=&\sqrt{ab}\cr
&=&{1\over 2}\sinh\bigg({\Delta w\over 2}\bigg)
\label{coefficients one}
\eea
Now let us write our final expression below:
\bea
\hat{O}_{\dot{B}}\hat{O}_{\dot{A}} &=& -{1\over 64 t_2^4}\bigg[\bigg(\sum_{n=0}^{3} D_{n}G^{-}_{\dot{B},n} + \sum_{n=-2}^{\infty}E_{n}\tilde{G}^{-}_{\dot{B},1-n} \bigg)\bigg(\sum_{n=0}^1 D'_n G^-_{\dot{A},n} + \sum_{n=0}^{\infty} C_n\tilde{G}^-_{\dot{A},1-n}\bigg)~\hat{\s}\cr
&&\qquad\quad -\bigg(\sum_{n=0}^{3} D_{n}G^{-}_{\dot{B},n} + \sum_{n=-2}^{\infty}E_{n}\tilde{G}^{-}_{\dot{B},1-n} \bigg)~\hat{\s}~\bigg(\sum_{n=0}^1 D'_n G^-_{\dot{A},n} - \sum_{n=0}^{\infty} C_n\tilde{G}^-_{\dot{A},n}\bigg)\cr
&&\qquad\quad + \bigg(\sum_{n=0}^1 D'_n G^-_{\dot{A},n} + \sum_{n=0}^{\infty} C_n\tilde{G}^-_{\dot{A},1-n}\bigg)~\hat{\s}~ \bigg(\sum_{n=0}^{3} D_{n}G^{-}_{\dot{B},n} + \sum_{n=0}^{\infty}E'_{n}\tilde{G}^{-}_{\dot{B},n} \bigg)\cr
&&\qquad\quad -\hat{\s}~\bigg(\sum_{n=0}^1 D'_n G^-_{\dot{A},n} - \sum_{n=0}^{\infty} C_n\tilde{G}^-_{\dot{A},n}\bigg) \bigg(\sum_{n=0}^{3} D_{n}G^{-}_{\dot{B},n} + \sum_{n=0}^{\infty}E'_{n}\tilde{G}^{-}_{\dot{B},n} \bigg)\bigg]\nn
\eea
With the following coefficient definitions:
\bea
D_0 &\equiv& \big( 2z_1t_2^2 -B(a+b)\big)\cr
D_1 &\equiv& B'\cr
D_2 &\equiv& -A'\cr
D_3&\equiv&  1\cr
\cr
D'_0 &\equiv& -e^{\Delta w \over 2}\cr
D'_1 &\equiv& 1\cr
\cr
E_{-2}&\equiv& 1\cr
E_{-1}&\equiv& -A'\cr
E_{n} &\equiv& C'_n,\quad n\geq 0\cr
\cr
E'_0 &\equiv& -B\cr
E'_1 &\equiv& A+ B(a+b)\cr
E'_n &\equiv& - C''_n,\quad n\geq 2
\label{coefficients four}
\eea
and 
\bea
C_n &\equiv& (-1)^{n}\sum_{k = - {n\over 2}}^{{n \over 2}}{}^{1/2}C_{{n\over 2} + k}{}^{1/2}C_{{n\over 2} - k}\cosh\big( k \Delta w\big)\cr
C'_n &\equiv&  (C_{n+2}- A C_{n+1} + B C_n\big)\cr
C''_n&\equiv&C_{n-2} - A C_{n-1} + B C_n
\label{coefficients C}
\eea
Where 
\bea
A &\equiv&(2z_1 + t_2)\cr
B &\equiv& z_1 (a+b)+ t_2(z_2 - 2(a+b))
\cr
A'&\equiv&A+a+b\cr
B'&\equiv&A(a+b) + B - 2t_2^2
\label{coefficients two}
\eea

Now evaluating (\ref{coefficients two}) in terms of $\Delta w$ using (\ref{coefficients one}) gives
\bea
A &=&{1\over 4}e^{-{\Delta w\over 2}}\big( 7 + e^{\Delta w} \big)\cr
\cr
B &=&{1\over 4}\big(1 + 3e^{-\Delta w} \big)
\cr
\cr
A'&=& {3\over 4}e^{-{\Delta w\over 2}}\big( 3 + e^{\Delta w} \big)\cr
\cr
B' &=& {3\over 2}\big( 1 + e^{-\Delta w} \big)
\label{coefficients three}
\eea
Inserting (\ref{coefficients three}) into (\ref{coefficients C}) where appropriate gives:
\bea
C_n &=& (-1)^{n}\sum_{k = - {n\over 2}}^{{n \over 2}}{}^{1/2}C_{{n\over 2} + k}{}^{1/2}C_{{n\over 2} - k}\cosh\big( k \Delta w\big)\cr
C'_n &=&  \bigg(C_{n+2}- {1\over 4}e^{-{\Delta w\over 2}}\big( 7 + e^{\Delta w} \big) C_{n+1} +  {1\over 4}\big(1 + 3e^{-\Delta w} \big) C_n\bigg)\cr
C''_n&=&\bigg( C_{n-2} - {1\over 4}e^{-{\Delta w\over 2}}\big( 7 + e^{\Delta w} \big) C_{n-1} + {1\over 4}\big(1 + 3e^{-\Delta w} \big)C_n\bigg)
\label{coefficients C two}
\eea
Now inserting (\ref{coefficients one}), (\ref{coefficients three}), and (\ref{coefficients C two}) into (\ref{coefficients four}) where appropriate gives:
\bea
D_0 &=& -{1\over 4}e^{-{3\Delta w\over2}}\big( 1 + 3e^{\Delta w} \big)\cr
D_1 &=&{3\over 2}\big( 1 + e^{-\Delta w} \big)\cr
D_2 &=& -{3\over 4}e^{-{\Delta w\over 2}}\big( 3 + e^{\Delta w} \big)\cr
D_3 &=&  1\cr
\cr
D'_0 &=& -e^{\Delta w \over 2}\cr
D'_1 &=& 1\cr
\cr
E_{-2}&=& 1\cr
E_{-1}&=& -{3\over 4}e^{-{\Delta w\over 2}}\big( 3 + e^{\Delta w} \big)\cr
E_{n} &=&  \bigg(C_{n+2}- {1\over 4}e^{-{\Delta w\over 2}}\big( 7 + e^{\Delta w} \big) C_{n+1} +  {1\over 4}\big(1 + 3e^{-\Delta w} \big) C_n\bigg),\quad n\geq 0\cr
\cr
E'_0 &=& -{1\over 4}\big(1 + 3e^{-\Delta w} \big)\cr
E'_1 &=& {3\over4} e^{-{\Delta w\over 2}} \big(3 + \cosh\big(\Delta w\big)\big)\cr
E'_n &=& -\bigg( C_{n-2} - {1\over 4}e^{-{\Delta w\over 2}}\big( 7 + e^{\Delta w} \big) C_{n-1} + {1\over 4}\big(1 + 3e^{-\Delta w} \big)C_n\bigg),\quad n\geq 2\nn
\label{coefficients five}
\eea

\end{appendix}

\end{document}